\documentclass[journal=chreay,manuscript=review,layout=traditional]{achemso}

\usepackage{chemformula} 
\usepackage[T1]{fontenc} 

\usepackage{url}

\usepackage[breaklinks=true]{hyperref}

\usepackage{enumitem}

\newcommand{\subsubsubsection}[1]{\paragraph{#1}\mbox{}\\}
\author{\normalsize{Andrey V. Solov'yov}}
\email{solovyov@mbnresearch.com}
\affiliation{\scriptsize{MBN Research Center, Altenh\"oferallee 3, 60438 Frankfurt am Main, Germany}}

\author{Alexey V. Verkhovtsev}
\affiliation{\scriptsize{MBN Research Center, Altenh\"oferallee 3, 60438 Frankfurt am Main, Germany}}

\author{Nigel J. Mason}
\affiliation{School of Physics and Astronomy, University of Kent, Canterbury CT2 7NH, UK}

\author{Richard A. Amos}
\affiliation{Department of Medical Physics and Biomedical Engineering, University College London, London, WC1E 6BT, UK}

\author{Ilko Bald}
\affiliation{Institute of Chemistry, University of Potsdam, Karl-Liebknecht-Str. 24-25, 14476 Potsdam, Germany}

\author{G\'{e}rard Baldacchino}
\affiliation{Universit\'{e} Paris-Saclay, CEA, LIDYL, 91191 Gif-sur-Yvette, France}

\author{Brendan Dromey}
\affiliation{Centre for Light Matter Interactions, School of Mathematics and Physics, Queen's University Belfast, BT7 1NN, United Kingdom}

\author{Martin Falk}
\affiliation{Institute of Biophysics of the Czech Academy of Sciences, Brno, Czech Republic}
\alsoaffiliation{Kirchhoff-Institute for Physics, Heidelberg University, Germany}

\author{Juraj Fedor}
\affiliation{J. Heyrovsk\'{y} Institute of Physical Chemistry, Czech Academy of Sciences, Dolej\v{s}kova 3, 18223 Prague, Czech Republic}

\author{Luca Gerhards}
\affiliation{Institute of Physics, Carl von Ossietzky University, Carl-von-Ossietzky-Str. 9-11, 26129 Oldenburg, Germany}

\author{Michael Hausmann}
\affiliation{Kirchhoff-Institute for Physics, Heidelberg University, Germany}

\author{Georg Hildenbrand}
\affiliation{Kirchhoff-Institute for Physics, Heidelberg University, Germany}
\alsoaffiliation{Faculty of Engineering, University of Applied Sciences Aschaffenburg, Germany}

\author{Milo\v{s} Hrabovsk\'{y}}
\affiliation{TESCAN GROUP, Brno, 62300, Czech Republic}

\author{Stanislav Kadlec}
\affiliation{Eaton European Innovation Center, Bo\v{r}ivojova 2380, 252 63 Roztoky, Czech Republic}

\author{Jaroslav Ko\v{c}i\v{s}ek}
\affiliation{J. Heyrovsk\'{y} Institute of Physical Chemistry, Czech Academy of Sciences, Dolej\v{s}kova 3, 18223 Prague, Czech Republic}

\author{Franck L\'{e}pine}
\affiliation{Universit\'{e} Claude Bernard Lyon 1, CNRS, Institut Lumi\`{e}re Mati\`{e}re, F-69622, Villeurbanne, France}

\author{Siyi Ming}
\affiliation{Yusuf Hamied Department of Chemistry, University of Cambridge, Lensfield Road, Cambridge, CB2 1EW, United Kingdom}

\author{Andrew Nisbet}
\affiliation{Department of Medical Physics and Biomedical Engineering, University College London, United Kingdom}

\author{Kate Ricketts}
\affiliation{Department of Targeted Intervention, University College London, Gower Street, London WC1E 6BT, United Kingdom}

\author{Leo Sala}
\affiliation{J. Heyrovsk\'{y} Institute of Physical Chemistry, Czech Academy of Sciences, Dolej\v{s}kova 3, 18223 Prague, Czech Republic}

\author{Thomas Schlath\"{o}lter}
\affiliation{Zernike Institute for Advanced Materials and University College Groningen, University of Groningen, The Netherlands}

\author{Andrew Wheatley}
\affiliation{Yusuf Hamied Department of Chemistry, University of Cambridge, Lensfield Road, Cambridge, CB2 1EW, United Kingdom}

\author{Ilia A. Solov'yov}
\email{ilia.solovyov@uni-oldenburg.de}
\affiliation{Institute of Physics, Carl von Ossietzky University, Carl-von-Ossietzky-Str. 9-11, 26129 Oldenburg, Germany}

\title{Condensed Matter Systems Exposed to Radiation: Multiscale Theory, Simulations, and Experiment}

\begin{document}

\begin{tocentry}

\includegraphics[width=5.5cm]{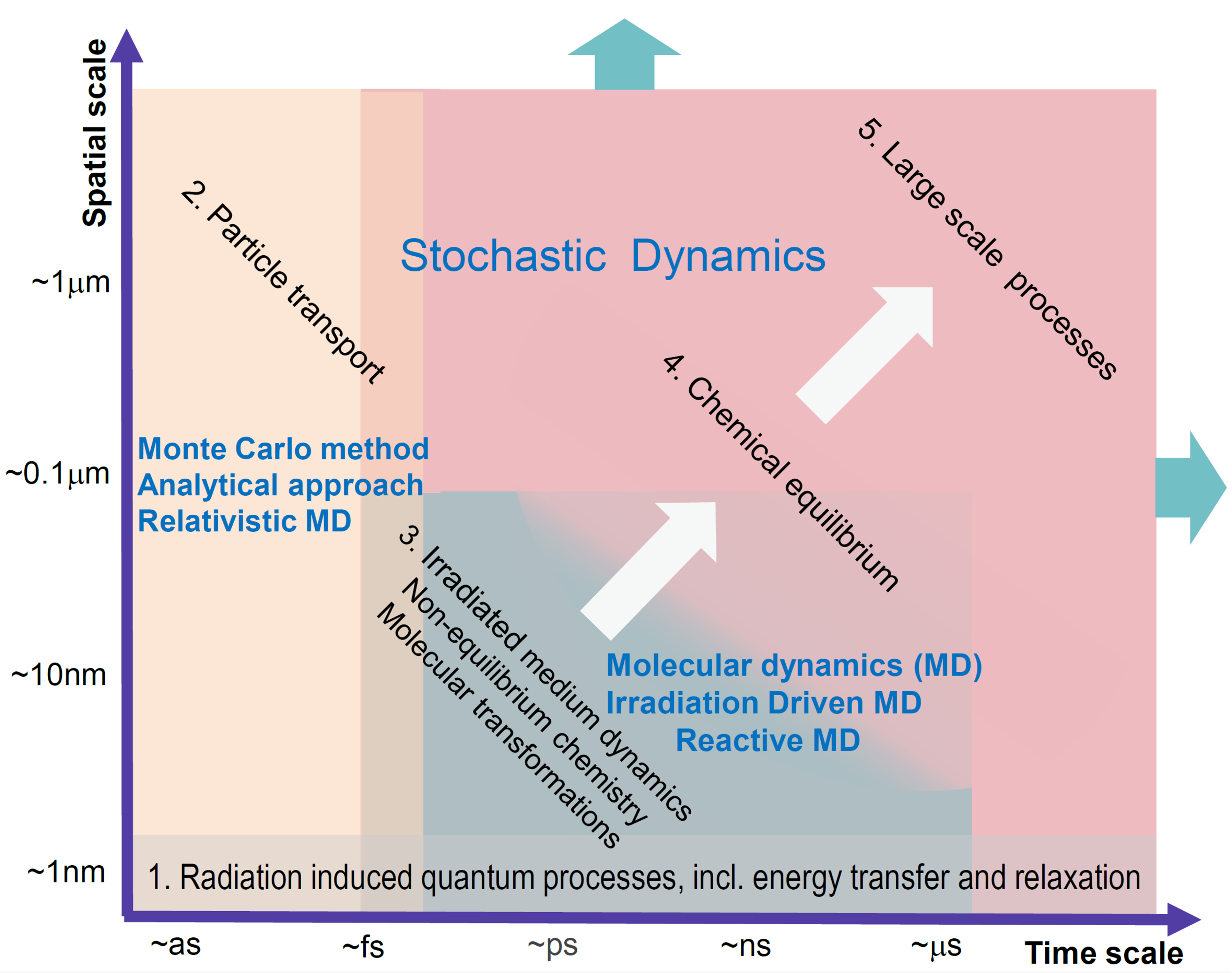}





\end{tocentry}

\begin{abstract}
This paper reviews the new highly interdisciplinary research field studying the behavior of condensed matter systems exposed to radiation. The paper highlights several relevant examples of recent advances in the field and provides a roadmap for the development of the field in the next decade. Condensed matter systems exposed to radiation may have very different natures, being inorganic, organic or biological, finite or infinite, be composed of many different molecular species or materials, existing in different phases (solid, liquid, gaseous or plasma) and operating under different thermodynamic conditions. The essential and novel element of this research is that, despite the vast diversity of such systems, many of the key phenomena related to the behavior of irradiated systems (such as radiation-induced damage, mechanisms of damage repair and control, radiation protection, etc.) are very similar and can be understood based on the same fundamental theoretical principles and computational approaches. One of the essential features of the aforementioned phenomena concerns their multiscale nature as the manifestation of the radiation-induced effects occurring at different spatial and temporal scales ranging from the atomic to the macroscopic. The multiscale nature of the effects and similarity of their manifestation in systems of different origins necessarily brings together different disciplines, such as physics, chemistry, biology, materials and nano-science, and biomedical research, demonstrating numerous interlinks and commonalities between them. This research field is highly relevant to many novel and emerging technologies and medical applications.



\end{abstract}

\tableofcontents

\newpage

\section{Introduction}
\label{sec:Intro}

\textbf{\textit{Condensed Matter Systems Exposed to Radiation.}}
Condensed matter systems represent the states of matter typically in either solid or liquid phases. These states of matter are formed due to the interatomic forces acting within the system \cite{LL3, LL5, LL9, LL10}. The nature of the forces can be different. Often, they have an electromagnetic origin, although due to the quantum motion of atoms and electrons therein, quantum phenomena in interatomic interactions play an essential role. The quantum nature of interatomic interactions under certain conditions can lead to the formation of quantum states of the entire condensed matter system, such as superconducting phases that can be observed in certain materials at low temperatures, the ferromagnetic and antiferromagnetic phases of spins on crystal lattices of atoms, and the Bose-Einstein condensate observed in ultracold atomic systems.

The nature of condensed matter systems can be very different, representing animate and inanimate matter, organic and inorganic, crystal and amorphous materials, glasses, etc. Biological condensed matter systems include large biomolecules, cells, biological media and entire biological systems, including organisms. These large groups of condensed matter systems have different natures, areas of applications and properties but have many similarities from a theoretical and computational point of view \cite{MBNbook_Springer_2017, DySoN_book_Springer_2022}. This makes unraveling the material properties of condensed matter systems a rather general and fundamental task.

Condensed matter systems can be finite and exist in the form of clusters, nanoparticles (NPs), and droplets. They can also be macroscopically large and, in such cases, they can be considered infinite. However, even for an infinite system, one should distinguish its bulk properties and system properties arising near the system surface. Interesting phenomena arise at the interfaces of different condensed matter systems in the same or different phases.

Condensed matter systems can have different and rather district geometries, being zero-, one-, two- or three-dimensional. Concrete examples of the systems include atomic and molecular clusters, fullerenes, nanotubes, nanowires, graphene, monoatomic layers, nanofilms, etc. A condensed matter system often possesses some specific properties (thermomechanical, electromagnetic, optical, etc.) that make it functional and useful for applications \cite{MBNbook_Springer_2017, DySoN_book_Springer_2022}.

The above descriptions demonstrate the existence of many very different condensed matter systems in nature. Their number is enormous, indeed practically infinite. However, despite a large diversity and multiplicity, the interaction of the condensed matter systems with radiation and radiation-induced phenomena in condensed matter systems have many features in common. This roadmap paper reviews the state-of-the-art advances in theoretical and computational methods enabling an accurate description of numerous condensed matter systems exposed to radiation, and related experimental and technological developments. Many of the observed phenomena in this research area are multiscale by nature, i.e. originate from the interconnected processes occurring at very different temporal and spatial scales. Hence, their understanding requires the adoption of appropriate multiscale theories and related computational tools, which are discussed below.

\textbf{\textit{Radiation Modalities.}}
Radiation effects in condensed mater systems can be induced by the radiation of different modalities, which include photons, electrons, positrons, ions, neutrons and other elementary particles. Typically, these particles are delivered in the form of a beam. The physical characteristics of the particle beams, such as energy, intensity, size, emittance, divergence, etc. as well as duration, bunching, fluence, duty cycle, etc., can be very different. This variety of options for particle beams requires various approaches to study radiation-induced effects in different irradiation regimes \cite{Workman_PDG_2022_review}.

\textbf{\textit{Radiation Conditions.}}
Irradiation of condensed matter systems might occur naturally, e.g. by the cosmic rays or the presence of a natural radiation background. In laboratories, it is usually delivered by beams of particles of different energy, size, intensity, etc.

At elevated radiation intensities, the condensed matter systems become highly excited and extreme conditions of their state (high temperature, pressure, ionization state) can be created. If a system in such a state is brought into a gas phase, it turns into a plasma. A detailed discussion of plasma and related phenomena is beyond the scope of this roadmap, although many relevant topics and interdisciplinary connections can be identified between the physics discussed in this roadmap paper and plasma physics. However, the high-intensity irradiation regimes resulting in the delivery of large doses of radiation to the systems are still in the focus of this roadmap.

\textbf{\textit{Common Features in the Response of Condensed Matter Systems to Radiation.}}
Despite a huge diversity of condensed matter systems, their interaction with radiation and related radiation-induced phenomena have many features in common. The reason for this is that the number of different radiation modalities is not so large and the radiation-induced phenomena, defined by the atomic and molecular processes occurring in different condensed matter systems, including biological ones, operate similarly. In this context, one should also mention similarities in the behavior of the cross sections of various elementary radiation-induced quantum processes in all the condensed matter systems. Such similarities arise due to generally a relatively weak dependence of cross sections upon the environment in which radiation-induced quantum processes occur. Typically, the cross sections depend only on a limited number of the relevant physical parameters and dependences of the cross section upon these parameters can be established \cite{LL3}. It should also be noted that the fundamental physical principles and the related theories describing the radiation-induced processes are equally applicable to all the variety of condensed matter systems and that the number of such theories is negligibly small in comparison with the number of different condensed matter systems \cite{LL8}.

\textbf{\textit{The Role of Quantum Processes in the Multiscale Scenario of Radiation-Induced Processes.}}
The interaction of radiation with matter takes place on time scales determined by the characteristic interaction time of a projectile particle with an atom or a molecule \cite{LL3, LL4}. It might vary significantly from attoseconds to femtoseconds depending on the velocity of a projectile and the size of an atom or a molecule. The interaction of projectile particles with atoms and molecules induces their quantum transformations, such as excitations, ionizations, dissociation, electron attachment, etc. These transformations occur via corresponding irradiation-induced quantum processes, the duration of which is relatively short compared to the entire duration of the irradiation-induced processes in the condensed matter systems. However, the initial radiation-induced quantum transformations are followed by further processes that might span over much larger spatial and temporal scales than those typical for the initial quantum processes. For example, irradiation of biological systems may lead to lasting radiobiological effects, such as cell damage, DNA repair processes, mutations, cell apoptosis, etc. Irradiation of organometallic molecules deposited on a surface with electrons or ions results in their degradation and formation of metal clusters, crystals or nanostructures.

It is therefore rather common that exposure of condensed matter systems to radiation triggers a sequence of interconnected processes (physical, chemical and biological) manifesting themselves on different temporal and spatial scales that create a multiscale scenario for the final observables induced by the initial irradiation. This multiscale feature of radiation-induced processes in condensed matter systems was only understood relatively recently and elucidated in a number of concrete case studies
\cite{AVS2017nanoscaleIBCT, Surdutovich_AVS_2014_EPJD.68.353, surdutovich2019multiscale, Sushko_IS_AS_FEBID_2016, DeVera2020, Solovyov_2014_PhysStatSolB.251.609, Stochastic_2022_JCC.43.1442, NovelLSs_Springer_book, AVK_AVS_2020_EPJD.74.201_review}. However, this feature is general and common for very different types of condensed matter systems and radiation modalities.

Therefore, understanding radiation effects in the condensed matter systems requires understanding both the immediate acts of the interaction of radiation with matter and the behavior of the matter (in particular its dynamics) over the long periods after its irradiation. This second and essential part of the problem is far from being understood and is the subject of current intensive research and development. Its thorough discussion is the focus of this roadmap paper.

\textbf{\textit{Characteristic Stages in the Multiscale Scenario of Radiation-Induced Processes.}}
One can find many similarities in the multiscale scenarios for very different condensed matter systems and radiation modalities. Indeed, there are characteristic stages that appear in most of them. Thus, there is a \textbf{stage~(i)} of initial quantum interactions of radiation with atoms and molecules within a system. This is followed by \textbf{stage~(ii)} in which transport of the primary radiation occurs through a condensed matter system with the generation and transport of secondary (as well as ternary, quaternary, etc.) particles created in the system. Transport of the primary radiation and the created (secondary, etc.) particles through a system results in the energy deposition into the system, leading to the next \textbf{stage~(iii)} of the multiscale scenario. This stage is characterized by specific dynamical and thermo-dynamical effects in the medium and chemical transformations. After some period of time, the irradiated system may reach the chemical equilibrium, \textbf{stage~(iv)}. Nevertheless, the multiscale scenario may continue even further, manifesting larger temporal and spatial scale phenomena such as biological (e.g. radiation damage repair, cell apoptosis, mutations, etc), structure formation and evolution, material ageing, morphological transitions, etc. Typically, each of such processes represents a separate case study or even a focused field of research, but together they can be attributed to \textbf{stage~(v)} of the multiscale scenario. All these stages are discussed in detail in this Roadmap and are defined in Figure~\ref{fig:MM_diagram} below. The fundamental understanding of all these complex phenomena is only possible via the inclusive multiscale approaches including all the initial, intermediate and final stage processes involved. Currently, only a few advanced multiscale scenarios have been developed in different research fields \cite{AVS2017nanoscaleIBCT, Surdutovich_AVS_2014_EPJD.68.353, surdutovich2019multiscale, Sushko_IS_AS_FEBID_2016, Solovyov_2014_PhysStatSolB.251.609, MBNbook_Springer_2017, DySoN_book_Springer_2022, Channeling_book, NovelLSs_Springer_book}. Thus, this roadmap paper aims to harmonize these advances and facilitate similar development for many other case studies within the research area defined above. Below, for the sake of concreteness and illustration, let us briefly discuss the three representative examples of multiscale scenarios.

\subsection{Multiscale Scenarios for Clustering, Self-Assembly, Structure Formation and Growth Processes in Condensed Matter Systems}
\label{sec:Intro_Ex_Fractals}

Clustering, self-assembly, structure formation and growth processes represent a group of key phenomena taking place in nearly all kinds of condensed matter systems, including biological ones \cite{MBNbook_Springer_2017, DySoN_book_Springer_2022}.

The processes of aggregation of atoms and small molecules into clusters, NPs and macromolecules, clustering (or coalescence) of NPs and biomolecules into nanostructures, nanostructured materials, biomolecular complexes, and hybrid systems possessing different morphologies permit the creation of a wide range of condensed matter systems \cite{JPC_AVS_WG_EurophysNews.33.200}. Examples of the self-organized systems may include aggregates of metal NPs \cite{Kim_2001_JACS.123.7955, Eichhorn_SelfOrg_MetalNPs}, nanofilms \cite{Kim_2007_JPCC.111.11252, Zhang_2018_IJMS.19.1641}, nanotubes \cite{Mae_2017_SciRep.7.5267, Macak_2007_CurrOpinSolidState.11.3, Shimizu_2014_PolymJ.46.831}, nanowires \cite{Fan_2006_Small.2.700, Yi_2023_NanoRes.16.1606, Zhang_2023_CrystGrowthDes}, functional nanocoatings \cite{Zhang_2018_IJMS.19.1641}, nanofractals \cite{Jensen_1999_RMP.71.1695, Lando_2006_PRL.97.133402, Dick_2010_JPCS.248.012025, Dick_2011_PRB.84.115408, Panshenskov_2014_JCC.35.1317}, and many other ordered or disordered nanostructures with characteristic structure and properties. Some of these systems have been synthesized only recently and have become the subject of intensive investigations due to their unique structural, optical, magnetic, thermomechanical, or thermo-electrical properties and can be utilized in a variety of important applications \cite{JPC_AVS_WG_EurophysNews.33.200}. Clustering, self-organization, and structure formation are general phenomena manifesting themselves over very different levels and scales of matter organization or self-organization thus they are also relevant to numerous dynamical systems studied by other natural sciences. They appear in many different areas of research: astrophysics, physics, chemistry, biology, materials science, nanoscience, neuroscience, and even in technology (clustering in the wireless, computer, or windmill networks, etc). Although there are many examples of these processes, their mechanisms and driving forces are often not well understood \cite{Brechignac_2007_NanomaterNanochem, MBNbook_Springer_2017, DySoN_book_Springer_2022}.

Apart from the fundamental value of understanding the aforementioned processes, this knowledge is also highly relevant to the key problems of modern technology. An important example of the related technological process concerns the manufacturing of nanostructures, nanosystems and nanomaterials. The goal can be achieved by two conceptually different approaches \cite{Yu_2013_ChemSocRev.42.6006}. The ``top-down'' approach deals with different techniques enabling the production of smaller nanostructures by breaking down larger pieces of material. The ``bottom-up'' approach typically relies on self-assembling of atoms, molecules and clusters into larger nanostructures, nanosystems, and nanomaterials.

The so-called ``bottom-up'' approach is driven mainly by the diffusion, reaction, and self-organization processes involving dynamics of the system on rather different spatial and temporal scales. These processes originate from the diffusion of individual atoms and molecules within the system. The quantum interactions of atoms and molecules with their neighbors occur at the characteristic distances of several angstroms and on sub-femtosecond to femtosecond temporal scales. However, the characteristic timescale for the diffusion processes of atoms and small molecules in a condensed matter system occurs on orders of magnitude longer time scales, which depend on the phase, atomic composition and temperature of the condensed matter system or interphase considered. The time scale for the self-assembly processes in a condensed matter system involves much longer time scales, reaching seconds, minutes and longer. The spatial scales of such processes are comparable with the size of an entire system that can be macroscopically large.

The ``bottom-up'' approach for the controllable, reproducible and industrially viable fabrication of nanosystems with desirable morphology and properties can be realized through various atomic, molecular or cluster deposition processes with the follow-up self-assembly of deposited species into the desired nanostructures \cite{Biswas_2012_AdvColloidInterfaceSci, Shimomura_2001_BottomUp}. Among many different deposition techniques and characteristic patterns of deposited species on surfaces, the nanofractal shapes are among the most studied ones \cite{Jensen_1999_RMP.71.1695, Meiwes-Broer_ClustersSurfaces, Brechignac_2007_NanomaterNanochem, MBNbook_Springer_2017, DySoN_book_Springer_2022} because the physics of fractals is of general nature, research interest, and technological importance. Here, we briefly discuss the formation, growth and evolution of nanofractal structures that can be created on surfaces in the course of atomic, molecular or cluster deposition processes as an exemplar case study elucidating the multiscale nature of structure formation and dynamics of such systems.

\begin{figure}[t!]
\includegraphics[width=0.75\textwidth]{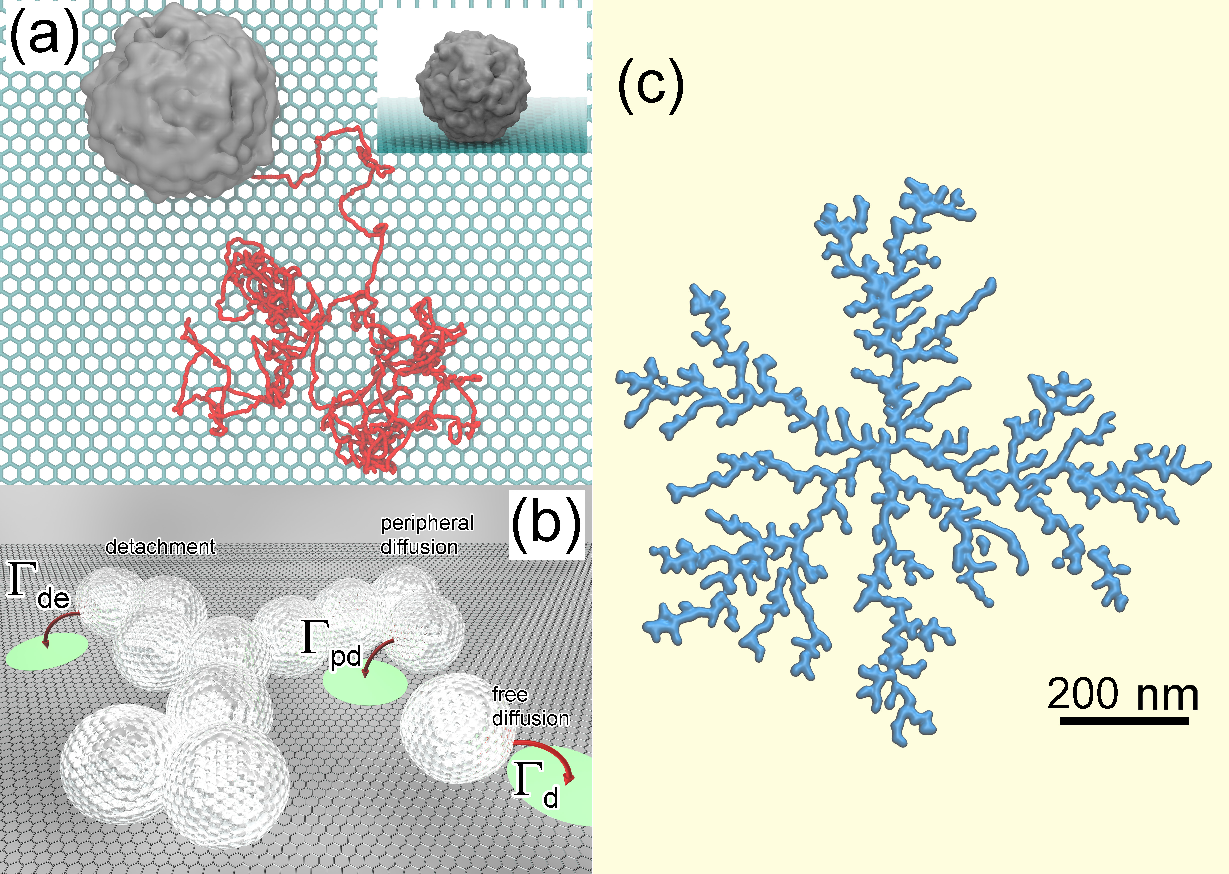}
\caption{Main elementary stochastic processes involving atomic clusters deposited on a surface. Panel~(a) shows the diffusion of a Ag$_{488}$ silver cluster deposited on a graphite surface, plotting the trajectory of the cluster center of mass derived from MD simulations. The figure illustrates that the deposited silver cluster experiences a random, Brownian-like motion, which can be parameterized by the corresponding rate of cluster diffusion $\Gamma_d$, being one of the input parameters for the stochastic dynamics simulation of such a process. Panel~(b) explains that the long time-scale stochastic motion of an ensemble of deposited clusters can be parameterized by rates $\Gamma_d$, $\Gamma_{pd}$, and $\Gamma_{de}$, of the three elementary stochastic processes describing (i) diffusion of a cluster over a surface ($\Gamma_d$), (ii) diffusion of a cluster along the periphery of an island on the surface ($\Gamma_{pd}$), and (iii) detachment of a cluster from an island ($\Gamma_{de}$). Panel~(c) demonstrates that random deposition of new particles on the surface and accounting for the aforementioned stochastic processes leads to the formation of the shown fractal structure. Adapted from ref~\citenum{Solovyov_2014_PhysStatSolB.251.609}.}
\label{fig:Intro_Fractals}
\end{figure}

As explained above and observed in numerous experiments \cite{Lando_2006_PRL.97.133402, Lando_2007_EPJD.43.151, Brechignac_2003_EPJD.24.265, Liu_2006_JCP.124.164707}, nanofractal formation in the course of the surface deposition processes, growth and evolution involves the dynamics of an enormous number of atoms on time scales which are far beyond the current limits of quantum mechanical and even classical Molecular Dynamics (MD) simulations. However, the fractal dynamics can be understood through the multiscale approach. Such an approach invokes quantum mechanics to describe parameters of interatomic potentials utilized as inputs for MD simulations. The MD simulations provide a set of validated parameters for models based on stochastic dynamics \cite{Stochastic_2022_JCC.43.1442}. Using stochastic dynamics, one can simulate the structure formation and dynamics of fractal systems up to the macroscopically large spatial scales and the related temporal scales defined by the performed experiments \cite{Stochastic_2022_JCC.43.1442, Dick_2011_PRB.84.115408, Panshenskov_2014_JCC.35.1317, Solovyov_2014_PhysStatSolB.251.609}.

Figure~\ref{fig:Intro_Fractals} depicts the key elementary processes that are crucial for the formation, evolution and fragmentation of nanofractals.
The described multiscale scenario and the corresponding multiscale approach can explain and quantify the major experimental observations on the formation and evolution of fractal systems, their morphology and properties \cite{Dick_2010_JPCS.248.012025, Dick_2011_PRB.84.115408, Panshenskov_2014_JCC.35.1317, Solovyov_2014_PhysStatSolB.251.609, Stochastic_2022_JCC.43.1442}.

This illustrative example represents a large class of systems and dynamical processes that can be understood by employing simulations carried out on the basis of the multiscale approach interfacing quantum and classical molecular dynamics with stochastic dynamics.

\subsection{Multiscale Scenario for Focused Electron Beam-Induced Deposition}
\label{sec:Intro_Ex_FEBID}

The second illustrative example of a multiscale scenario concerns Focused Electron Beam-Induced Deposition (FEBID). It continues the discussion of the ``bottom-up'' approach for nanofabrication, clustering, self-assembly, structure formation, and growth processes in condensed matter systems started in the previous section. However, let us now consider additionally the possibility of guiding the structure formation of condensed matter systems through their irradiation during this process with focused electron beams. Other radiation modalities (ions, photons) can also be utilized \cite{DeTeresa-book2020, Utke2008} for such applications. They have many similarities with the example considered here and thus will not be discussed below in any detail.

As mentioned in Section~\ref{sec:Intro_Ex_Fractals}, controllable fabrication of condensed matter systems with nanoscale resolution remains a considerable scientific and technological challenge \cite{Plant_2014_JACS.136.7559}. The manufacturing of smaller and smaller structures has been the aim of the electronics industry for several decades since the smaller the fabricated systems, the stronger the operational capacity of the nanodevices produced on their basis. Until recently, the well-known Moore's law has been delivered within the semiconductor industry, enabling smaller and smaller devices to be produced (mainly by the ``top-down'' approach), thus improving operational power within a fixed-size device. However, when the structure size drops below 30~nm, traditional manufacturing methods (e.g. ultraviolet (UV) lithography, plasma etching, plasma-enhanced chemical vapor deposition) struggle to meet Moore's law. Therefore, there is an urgent need to develop new nanofabrication methods (which should be based on the ``bottom-up'' approach), among which FEBID is one of the most promising, allowing controlled creation of nanostructures with nanometer resolution \cite{Huth_2012_BJN.3.597}. Such methods \cite{Utke_book_2012, Cui_Nanofabrication_book} exploit the irradiation of nanosystems with collimated electron beams. These can be used to create specific structural motifs of metal NPs for catalytic and nanoelectrochemistry applications \cite{Xu_2008_NatMater.7.992, Murray_2008_ChemRev.108.2688}, to fabricate metal nanostructures for sensors, nanoantennas and magnetic devices, surface coatings, and prepare thin films with tailored properties for electronic devices and other applications.

The key element of the FEBID multiscale scenario refers to the irradiation of precursor molecules (usually organometallics, below called precursors) \cite{Barth2020_JMaterChemC} by keV electron beams while they are being deposited onto a substrate. Electron-induced fragmentation of the irradiated precursors releases metal atoms, forming a metal-rich deposit on the surface with a size similar to that of the incident electron beam (a few nanometers) \cite{Utke2008}. This phenomenon is multiscale and rather complex. It involves precursor deposition, diffusion, aggregation, clustering, and self-organization processes already discussed in Section~\ref{sec:Intro_Ex_Fractals}. In addition, the FEBID multiscale scenario must take into consideration adsorption and desorption processes; ionization, excitation, and fragmentation of precursors induced by the primary as well as backscattered and secondary electrons emitted from the irradiated surface; electron transport in the substrate; processes of system relaxation after its electronic excitation including energy transfer from electronic to vibrational degrees of freedom and resulting thermomechanical effects; as well as chemical reactions between various reactive species produced upon the system irradiation \cite{Sushko_IS_AS_FEBID_2016, DeVera2020, Prosvetov2021_BJN, Prosvetov2022_PCCP}. Moreover, the formation of 2D and 3D FEBID structures on larger spatial scales involves movement of the electron beam (patterning) and multiple irradiations of the already created structures. Naturally, such processes involve larger temporal scales. The quantitative description and characterization of the entire FEBID scenario can only be achieved within the multiscale approach accounting for a complex interplay of the phenomena mentioned above, taking place on many different temporal and spatial scales \cite{MBNbook_Springer_2017}. Although the essential initial steps in this direction have been recently made \cite{MBNbook_Springer_2017, Sushko_IS_AS_FEBID_2016, DeVera2020}, its complete development remains an important task for the field. The importance of this task is rather obvious. The multiscale approach for FEBID may instruct the technology about choosing optimal nanofabrication regimes of the FEBID systems with desired properties, e.g. metal content, mechanical, electric and magnetic properties.

Typically, FEBID is processed via successive cycles of precursor replenishment and irradiation stages. A popular class of FEBID precursors is represented by metal carbonyls Me$_m$(CO)$_n$ \cite{Huth_2012_BJN.3.597, Kumar2018} consisting of one or more metal atoms (Me) chemically bound to several carbonyl ligands. Metal carbonyls have been widely investigated experimentally, and much information on their thermal and electron-induced fragmentation has been collected \cite{Wysocki_1987_IJMSIP.75.181, Beranova_1994_JAMS.5.1093, Cooks_1990_JASMS.1.16, Wnorowski_2012_IJMS.314.42, Wnorowski_2012_RCMS.26.2093, Neustetter_2016_JCP.145.054301, Lacko_2015_EPJD.69.84, Lengyel2016_JPCC_2, Lengyel2021, Massey_Sanche2015, Bilgilisoy_Fairbrother2021}. Particular attention to the structure and properties of these precursors has been devoted to their peculiar structure with strong C–O bonds and relatively weak Me–C bonds. While the former are relatively hard to cleave, the latter break easily, typically by a sequential loss of CO groups once a sufficient amount of energy is deposited into the molecule.

Until recently, theoretical analysis of FEBID was based on the Monte Carlo (MC) approach and numerical solutions of the diffusion-reaction equation. These methodologies provide the average characteristics of the FEBID structures but cannot give any molecular-level details. The atomistic approach for FEBID simulations, capable of providing the atomistic insights of created FEBID nanostructures, was developed in ref~\citenum{Sushko_IS_AS_FEBID_2016}. This approach is based on the Reactive Molecular Dynamics (RMD) \cite{Sushko2016_rCHARMM} and Irradiation-Driven Molecular Dynamics (IDMD) \cite{Sushko_IS_AS_FEBID_2016} accounting for the quantum and chemical transformations within the absorbed molecular system. These methods are described in Section~\ref{sec:Methods}. The exemplar case study performed in ref~\citenum{Sushko_IS_AS_FEBID_2016} considered the FEBID process of tungsten hexacarbonyl W(CO)$_6$ on a hydroxylated SiO$_2$ surface. It was performed using MBN Explorer \cite{Solovyov_2012_JCC_MBNExplorer}, the software package in which RMD and IDMD were implemented. The simulation results were verified through their comparison with experimental data \cite{Fowlkes2010}.

\begin{figure}[t!]
\includegraphics[width=0.75\textwidth]{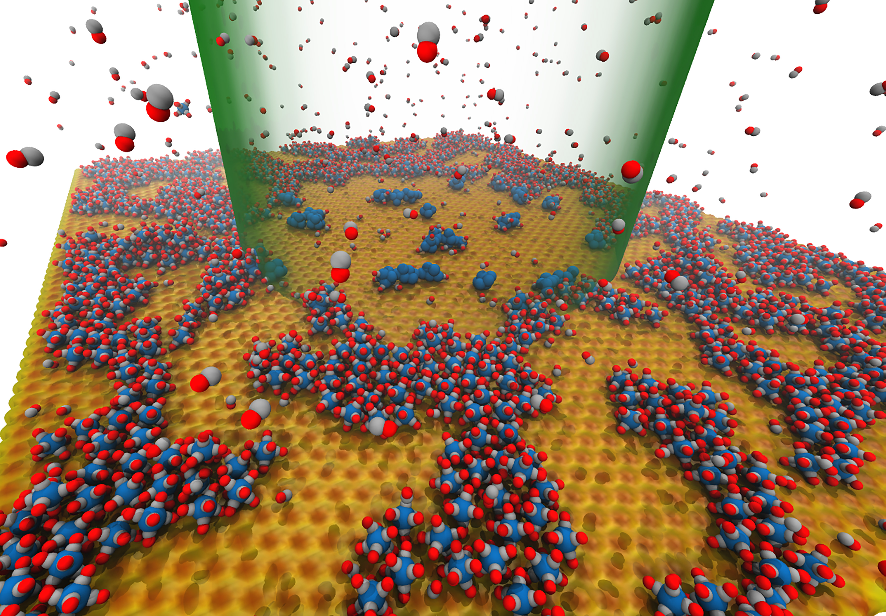}
\caption{An example of MD simulation \cite{Sushko_IS_AS_FEBID_2016} of deposition of W(CO)$_6$ precursors atop the SiO$_2$ substrate depicting initial stages of the irradiation process by an electronic beam (a green semi-transparent cylinder). The interaction of adsorbed precursors with the primary and secondary electrons emitted from the substrate leads to precursor fragmentation and the formation of clusters of tungsten atoms, shown by blue spheres. Adapted from ref~\citenum{Sushko_IS_AS_FEBID_2016}.}
\label{fig:Intro_FEBID}
\end{figure}

Figure~\ref{fig:Intro_FEBID} shows a snapshot of MD simulation \cite{Sushko_IS_AS_FEBID_2016} of deposition of W(CO)$_6$ precursors atop the SiO$_2$ substrate depicting the initial stages of the irradiation process by an electronic beam (a green semi-transparent cylinder). It is seen that most precursors located inside the irradiated area experience fragmentation. The rate of this process was evaluated from the experimental data \cite{Fowlkes2010}.

The atomistic approach for FEBID, RMD and IDMD involves the recently developed rCHARMM force field \cite{Sushko2016_rCHARMM}, which is discussed in detail in Section~\ref{sec:Methods}. The rCHARMM force field requires the specification of several parameters, such as the equilibrium bond lengths, bond stiffness and dissociation energies. Additionally, the dissociative chemistry of precursors should be defined, specifically including the definition of the molecular fragments and atomic valences.

\subsection{Multiscale Scenario for the Radiation Damage of Biological Systems with Ions}
\label{sec:Intro_Ex_IBCT}

The third illustrative example concerns the multiscale scenario for the radiation damage (RADAM) caused by irradiation of condensed matter systems with ions. Particular attention is devoted to the analysis of irradiation of biological systems. This example opens an important topic on how the physical radiation-induced processes are coupled to biological systems and may lead to large scale biological effects, such as cell deactivation, cell mutation, bystander effect, etc. \cite{AVS2017nanoscaleIBCT}. This topic is closely linked with biomedical applications of radiation, such as different kinds of radiotherapies, aiming at the tumor cells irradiation for their subsequent destruction and exploiting different radiation modalities for such purposes.


The MultiScale Approach (MSA) to ion-beam therapy was suggested more than a decade ago \cite{solov2009physics} and the first steps towards developing this approach have been made. The cited work initiated the development of a phenomenon-based approach to the assessment of RADAM with ions, being fundamentally different from any other methods utilised in the field. The primary MSA goal was to understand the multiscale scenario of RADAM with ions in the language of physical, chemical, and biological effects. It was aimed to relate initial physical effects of energy loss by projectiles to the biological effects defining cell inactivation. It is worth to stress that by its principle a MSA is a non-dosimetric approach meaning that in the MSA no damage is solely defined by the locally deposited dose.

The multiscale scenario involves several temporal and spatial scales. It was shown that the ion energy and, consequently, the secondary electrons' energy are essential for defining the concrete realization of the multiscale scenario. The MSA treats the relevant physical, chemical, and biological effects within an inclusive single framework. The scenario begins with the propagation of ions through the tissue, which is substituted in most studies with liquid water since water constitutes about 75\% of tissue mass. The dominating process accompanying the propagation of ions is ionization of molecules of the medium. It is characterized by the depth-dose curve possessing a prominent feature known as a Bragg peak. The position of the Bragg peak depends on the initial energy of ions. In radiotherapy application, the initial energy of ions can be varied so that the Bragg peak is placed into the tumor. The position and profile of the Bragg peak as a function of initial energy can be obtained analytically \cite{Surdutovich_2009_EPJD.51.63, Scifoni_2010_PRE.81.021903, Surdutovich_AVS_2014_EPJD.68.353, AVS2017nanoscaleIBCT} based on the singly-differentiated cross sections (SDCSs) of ionization of water molecules with ions. This analytical approach has been successfully validated by comparing the calculated depth-dose curve with the results of Monte Carlo (MC) simulations and experiment \cite{Pshenichnov_2008_NIMB.266.1094}. The developed methodology is practical for the fast evaluation of the depth-dose curves analysis and can be adopted for treatment planning.

Further analysis of SDCSs of ionization \cite{Scifoni_2010_PRE.81.021903, deVera_2013_PRL.110.148104} has revealed essential features in the energy spectrum of ejected secondary electrons on the temporal scales $10^{-18} - 10^{-17}$~s after ion's passage. It was demonstrated that the dominating fraction of ionized secondary electrons emitted from molecules into the medium in collisions with ions have energies below $\sim$50~eV. More energetic $\delta$-electrons are kinematically suppressed in the Bragg peak region. They can be emitted with smaller probabilities in the plateau region preceding the peak. At energies of $\sim$50~eV, electronic transport through the medium occurs ballistically. At such energies, the cross sections of electronic collisions with molecules of the medium are nearly isotropic \cite{Nikjoo_2006_RadiatMeas.41.1052}. These facts justified the use of the random walk approximation (i.e., diffusion mechanism) to describe the secondary electron transport. The diffusion equation-based approach was successfully developed in refs~\citenum{Surdutovich_AVS_2014_EPJD.68.353, solov2009physics, ES_AVS_2012_EPJD.66.206, ES_AVS_2015_EPJD.69.193, Bug_2012_EPJD.66.291}.

There are several features of secondary electron transport relevant for the multiscale scenario of ion-induced RADAM.

\begin{enumerate}[label=(\roman*)]

\item
In the vicinity of the Bragg peak the secondary electrons lose most of their energy within $1 - 1.5$~nm of the ion's path. This process ends within 50~fs of the ion's track \cite{ES_AVS_2015_EPJD.69.193}. The biologically relevant radiation-induced damage, such as the single- and double-strand breaks (SSBs and DSBs) in the nuclear DNA can be caused by inelastic collisions of secondary electrons with DNA. Low-energy electrons can also create these lesions via their dissociative attachment to DNA. All these processes take place within $3-5$~nm of the ion's track.

\item
The average energy of secondary electrons in the vicinity of the Bragg peak is nearly independent of the projectile's energy and does not depend on the linear energy transfer (LET) of projectiles. Most of these electrons can ionize a few molecules of the medium \cite{Surdutovich_2009_EPJD.51.63}. Therefore, the number of secondary electrons in the vicinity of the Bragg peak is roughly proportional to the LET.

\item
Due to the energy loss by secondary electrons within 50~fs and within $1 - 1.5$~nm of the ion's track, the so-called ``hot'' cylinder is created. There are no means of immediate or fast transport of this energy away from the cylinder because heat conductivity and diffusion occur slowly on the picosecond timescale. Therefore, the pressure rises within the hot cylinder during the $50 - 1000$~fs period, being the characteristic duration of the electron-phonon coupling processes responsible for the energy transfer from electronic degrees of freedom in the system to its vibrational degrees of freedom. The maximum value of pressure is proportional to LET. By the end of this period, a significant collective flow associated with an induced shock wave starts, given a sufficiently large LET. Ion-induced shock waves were predicted by the MSA and have been thoroughly studied in a series of works, both analytically and computationally \cite{surdutovich2010shock, Surdutovich_AVS_2014_EPJD.68.353, surdutovich2013biodamage, Yakubovich_2012_NIMB.279.135, Yakubovich_2011_AIP.1344.230, deVera2016molecular, deVera2017radial, deVera_2018_EPJD.72.147, Friis_2020_JCC, Friis_2021_PRE}.

\item
A manifold of reactive species is formed from the molecules ionized by primary projectiles and secondary electrons. The effect of these reactive species on DNA is deemed to be more significant than the direct effect of secondary electrons. Therefore, understanding their production and transport is vital for the assessment of RADAM. The initial fraction of reactive species is formed within $1-2$~ps of the ion's passage, i.e. during the transport of secondary electrons followed by the relaxation of molecular excitations and energy transfer to vibrational degrees of freedom of the system. The number density of such reactive species may be large and, in the first approximation, grow linear with LET. However, their recombination rates are proportional to the square of their number density. At large LET values, the recombination of reactive species may dominate the transport by diffusion, resulting in the suppression of the number of species that diffuse out of ion tracks. On the other hand, a strong collective flow due to an ion-induced shock wave initiated during $1-2$~ps after the ion's passage can propagate reactive species away from  the tracks more effectively than diffusion. This process reduces the recombination rate of the initially created reactive species, thus affecting the initial conditions for the chemical phase \cite{Surdutovich_AVS_2014_EPJD.68.353, ES_AVS_2015_EPJD.69.193, deVera_2018_EPJD.72.147}.

\item
For larger LETs, corresponding to carbon or heavier ions, the nanoscopic cylindrical shock waves created in the vicinity of ion tracks become sufficiently strong to create breakages of molecular covalent bonds, including those in the DNA strands \cite{surdutovich2013biodamage}. This process provides an essential, and at very high LET the dominating, contribution to the nuclear DNA damage in cells irradiated with ions \cite{Friis_2021_PRE}.
\end{enumerate}

The elements of the MSA described above are related to its physical part. The analytical methods developed for analyzing the physical part of the MSA also provide an efficient methodology for assessing chemical effects. On this basis, a biological model for cell inactivation, which involves the concept of a lethal DNA lesion, has been developed. Introducing the cell lethality criterion, the number of such lesions per unit length of the ion's path can be calculated, and the cell survival probability can be obtained. Two hypotheses underly the concept of lethal damage within the MSA: (i) the inactivation of cells irradiated with ions occurs mainly due to nuclear DNA damage and (ii) a DNA lesion of a certain complexity is lethal. The second hypothesis originates from a series of papers \cite{Ward_1988_ProgNuclAcid.35.95, Ward_1995_RadiatRes.142.362, Malyarchuk_2008_NuclAcidsRes.36.4872, Malyarchuk_2009_DNARepair.8.1343, Sage_2011_MutatRes.711.123} spanning over three decades. These papers analyzed simple DNA lesions (such as SSBs or base damage), DSBs, and complex lesions consisting of a DSB and several simple lesions. After a series of investigations, it was postulated that complex lesions consisting of a DSB and at least two more simple lesions within a length of two DNA twists are lethal, at least for a normal cell \cite{Surdutovich_AVS_2014_EPJD.68.353, verkhovtsev2016multiscale}. This criterion for cell lethality implicitly includes the probability of enzymatic DNA repair. This criterion may be modified for different cancerous cells and some special cell lines \cite{verkhovtsev2016multiscale}. Even more important is that within the MSA, in contrast to other approaches, each lesion has been associated with an action of an agent such as a primary particle, secondary electron, or a reactive species, or can be caused by high pressure gradients arising in the medium at the fronts of the nanoscopic cylindrical shock waves induced by propagating ions. An `action' here means that a lesion is caused by one of the aforementioned processes treated within the MSA that are not necessarily determined by a particular energy deposition but also by many other factors. This feature significantly differs the MSA from the nano- and micro-dosimetric approaches used previously and in most treatment planning models.

\begin{figure}[t!]
\includegraphics[width=1.0\textwidth]{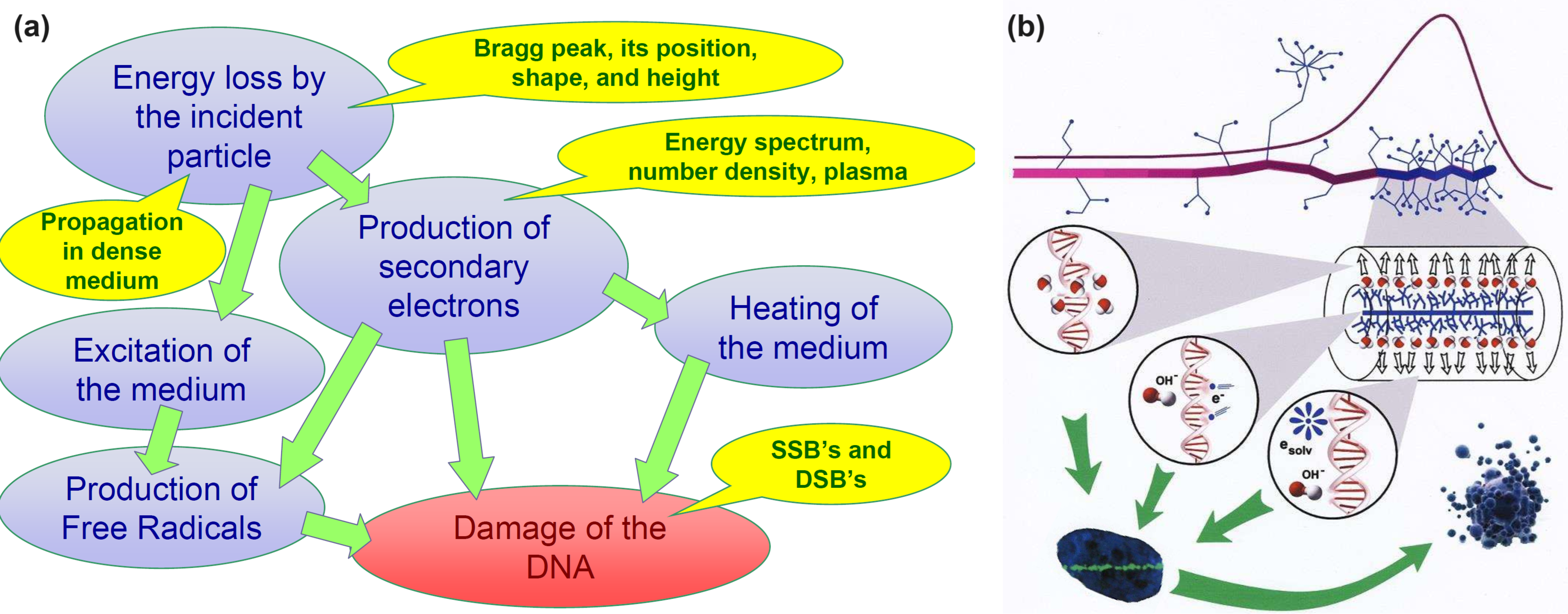}
\caption{The scenario of biological damage with ions: (a) a schematic representation \cite{solov2009physics} and (b) an artistic view \cite{Surdutovich_AVS_2014_EPJD.68.353}. Ion propagation ends with a Bragg peak, shown in the top right corner of panel~(b). Panel~(b) also shows an ion track segment at the Bragg peak in more detail. Secondary electrons and radicals propagate radially from the ion's path, damaging biomolecules (central circle). These reactive species transfer the energy to the medium within the hot cylinder \cite{surdutovich2010shock}, which causes a rapid temperature and pressure increase inside this cylinder. The emerging shock wave (shown in the expanding cylinder) due to this local pressure increase may damage biomolecules by stress (left circle) \cite{surdutovich2013biodamage, Yakubovich_2012_NIMB.279.135, deVera2016molecular, Friis_2021_PRE}. Moreover, the shock wave also effectively propagates reactive species (radicals and solvated electrons) to larger distances (right circle) \cite{Surdutovich_AVS_2014_EPJD.68.353, deVera_2018_EPJD.72.147, Friis_2021_PRE}. The low left corner of panel~(b) shows an image of a cell nucleus, which is crossed by an ion track visualised through foci (visible in the stained cells). The foci arise where DNA lesions are created and then repaired by enzymes carrying luminescent markers. Unsuccessful repair efforts lead to an eminent cell death; an apoptotic cell is shown in the lower right corner of panel~(b). Panel~(a) is adapted from ref~\citenum{solov2009physics}; panel~(b) is adapted from ref~\citenum{Surdutovich_AVS_2014_EPJD.68.353}.}
\label{fig:Intro_IBCT}
\end{figure}

With the defined criterion of lethality, the fluence of agents on a specific DNA segment (located at a given distance from an ion's path) is calculated in accordance with the transport mechanism, taking into account collective flows due to ion-induced shock waves. The fluences are weighted with probabilities of chemical processes leading to lesions. The number of SSBs caused by the direct action of ion-induced shock waves at a given LET is derived from reactive MD simulations. Then the probability of SSBs caused by each particular mechanism leading to such events and a cumulative probability of SSBs, is derived based on Poisson statistics. The yield of lethal lesions per unit length of an ion's path is then also calculated using Poisson statistics \cite{Surdutovich_AVS_2014_EPJD.68.353, verkhovtsev2016multiscale}. It turns out that this quantity depends on three physical characteristics, namely ion fluence, LET, and the dose deposited in the cell nucleus, as well as on the biological characteristics -- the genome size in an irradiated cell. The average length of all tracks through the cell nucleus can be calculated if two of these physical characteristics are treated independently, e.g. the LET and the dose. The yield of lethal lesions per cell could then be calculated as the product of this length with the yield of lethal lesions per unit length of the ion's path. This analysis enables to explain the ``overkill'' effect manifesting itself in a decrease of the biological effectiveness of ionizing radiation as high LET values. The explanation of this effect is that at high LET, the energy is deposited into a target cell nucleus
by a small number of ions, and this energy is larger than that needed for cell inactivation. As a result, high-LET irradiation produces higher DNA damage than actually
required, which leads to a reduction in biological effectiveness \cite{IBT_book_Linz}. The analysis carried out using MSA also demonstrates that the yield of lethal lesions depends on the dose, LET, and oxygen concentration in the medium. The relative biological effectiveness (RBE) can also be derived from the calculated cell survival curves.
In ref~\citenum{verkhovtsev2016multiscale}, the theoretically established survival curves were successfully compared with those experimentally obtained for several cell lines.

The introduced multiscale scenario of the RADAM by ions \cite{solov2009physics, Surdutovich_AVS_2014_EPJD.68.353, AVS2017nanoscaleIBCT, DySoN_book_Springer_2022} is schematically presented in Figure~\ref{fig:Intro_IBCT}, where panel~(a) depicts a schematic representation of the scenario \cite{solov2009physics} and panel~(b) shows an artistic view \cite{Surdutovich_AVS_2014_EPJD.68.353}. The figure shows several pathways leading from the energy loss by a propagating ion to the cell apoptosis.

\textbf{\textit{Important Applications in Technology and Medicine.}}
These three illustrative examples demonstrate that multiscale scenarios can be developed to describe quantitatively the multiscale radiation-induced processes. From these descriptions it is evident that the research field is open to many more investigations of condensed matter systems exposed to radiation of different modalities aiming at unravelling different phenomena and their links to relevant applications in technologies or medicine. Technological applications within this field of research include such important tasks as optimization of ion-beam cancer therapy, and radiotherapies in general, advancing 3D-nanoprinting / controlled nanofabrication, space technologies for radiation protection, atomistic analysis of RADAM and degradation of materials, new materials design, revealing the nature of radiation-induced biological effects, plasma technologies, and many more.

The number of concrete case studies in this field of research and related technologies is rapidly growing. This roadmap paper attempts to describe the state-of-the-art achievements in the field, the main direction of its development paving the way to numerous novel challenging multiscale case studies and their linkage to important applications in technology and medicine. Section~\ref{sec:Case_studies} presents several case studies where the multiscale approach may be applied.

\section{Multiscale Modeling of Condensed Matter Systems Exposed to Radiation: The Key Definitions and Main Problems}
\label{sec:MM_key-definitions}

Let us introduce the main theoretical concepts utilized to study various stages of the multiscale scenario of the radiation-induced processes in condensed matter systems and related phenomena. The description of the entire multiscale scenarios requires the utilization of several methodologies, including those that enable interfacing of the methodologies that operate in different spatial and temporal regimes.

\textbf{\textit{Radiation-Induced Multiscale Processes.}}
As discussed in Section~\ref{sec:Intro}, the dynamical response of condensed matter systems upon their irradiation typically involves a cascade of processes, such as  radiation-induced quantum processes within the system, transport of the primary particles/radiation through the system and production of secondary particles, their propagation in a medium and interaction of projectile particles and secondary electrons with surrounding molecules, energy transfer and relaxation processes in the medium, chemical transformations of molecules and their reactions, thermomechanical and biological transformations of the medium, its dynamics, various many-body/collective effects, biological effects, aging, etc, which are triggered by the initial quantum interactions of radiation with the system. The temporal evolution of such cascades of processes involve variations of different characteristics of the system, such as distributions of particles, energy, thermodynamic variables (temperature, pressure, etc) within a system. This evolution may have very different stages before the system reaches equilibrium. These stages and the corresponding states of the system can be studied experimentally and a set of experiments for the characterization of different stages of the scenario are typically stage specific.

\begin{figure}[t!]
\includegraphics[width=0.9\textwidth]{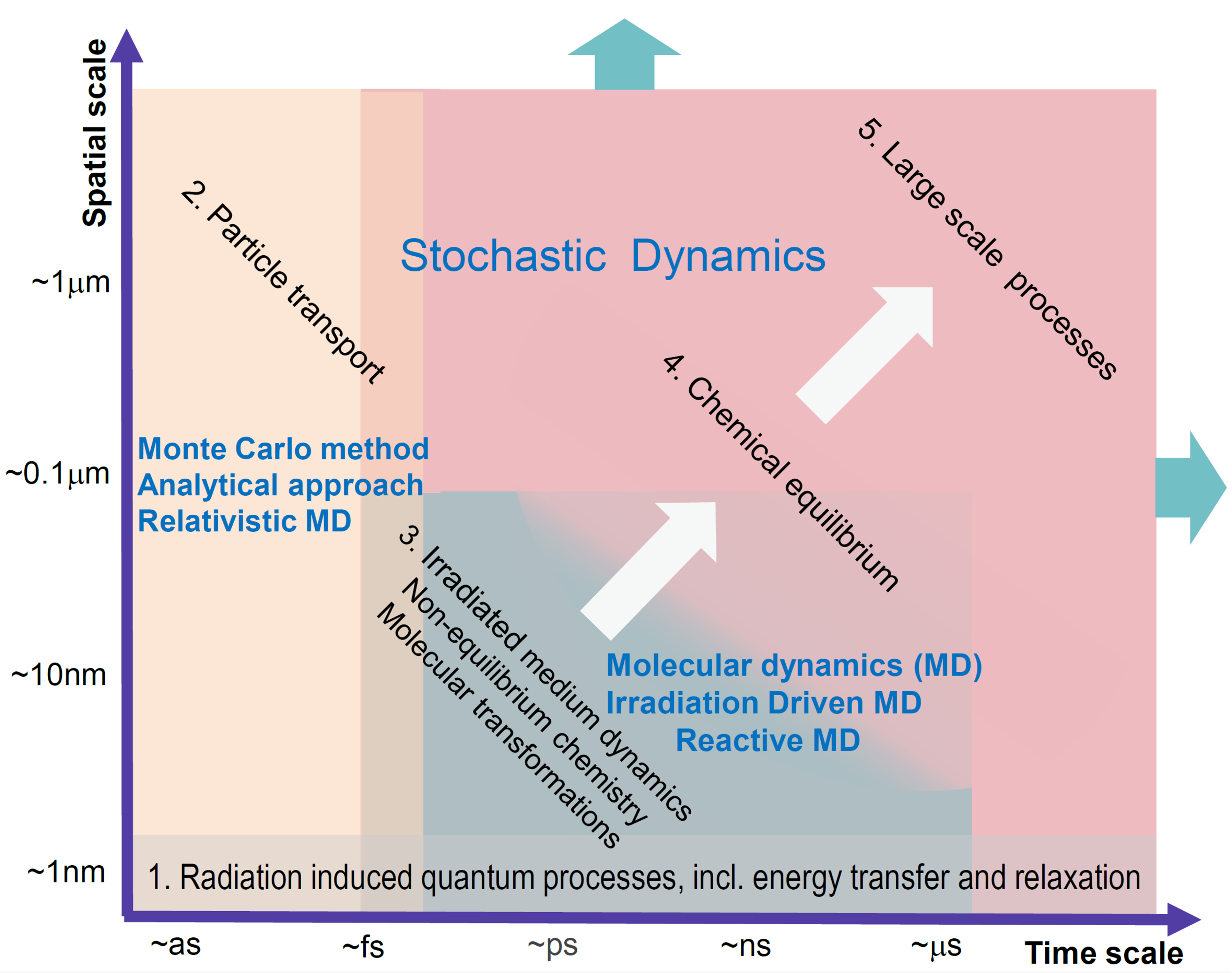}
\caption{A schematic space-time representation of the main stages of the multiscale scenario of radiation-induced processes in condensed matter systems with the corresponding methodologies for their description and limits.}
\label{fig:MM_diagram}
\end{figure}

The whole cascade of processes is essentially multiscale and requires a multiscale theory for its understanding and description, as illustrated in Figure~\ref{fig:MM_diagram}. The diagram schematically represents the main stages of a cascade of processes happening in a condensed matter system after irradiation. The stages correspond to different areas of the temporal and spatial scales characteristic of their manifestation. In Figure~\ref{fig:MM_diagram}, they are entitled with black letters and numbered according to their temporal sequence. The colored areas in Figure~\ref{fig:MM_diagram} introduce the fundamental theoretical approaches and methods (entitled in blue letters) used to describe the corresponding stages of the radiation-induced cascades. The borders of the colored areas indicate the limits of applicability of the corresponding theoretical and computational approaches and methods. The origin in Figure~\ref{fig:MM_diagram} is placed at the beginning of the multiscale cascade of the considered irradiation-induced processes and is typically associated with the primary/initial radiation event. Note here that Figure~\ref{fig:MM_diagram} may correspond to a scenario arising from a single initial radiation event, e.g. passage of an ion through a condensed matter system but, in the case of the multiple events, their statistical analysis and overlap should be considered and analysed. This may result in the dispersion of the interfaces between different areas shown in Figure~\ref{fig:MM_diagram}.

Let us now discuss each element of Figure~\ref{fig:MM_diagram}. The rectangular areas indicated with different colors correspond to the ranges of applicability of different key theoretical and computational methodologies utilised for the multiscale simulations of radiation-induced processes. The ranges for each colored area may indicate the physical limitation for the corresponding processes considered by the method or may correspond to the computational limits of the indicated methodology with the nowadays most powerful computational tools.

\subsection{Radiation-Induced Quantum Processes, Including Energy and Momentum Transfer and Relaxation}
\label{sec:MM_stage1}

The interaction of radiation with a condensed matter system occurs via the radiation-induced elementary quantum processes, see the horizontal greyish rectangular area at the bottom of Figure~\ref{fig:MM_diagram}. The most characteristic example of such a process is atomic or molecular ionization. It can be induced by the radiation of any modality, although the ionization cross section depends on the type of ionized atoms, the type of projectile particles and their energy. In the case of molecular ionization, the ionized molecule can experience fragmentation or a bond cleavage in the course of the process or after it. Electronic excitation of atoms and molecules is another quantum process that can be induced by radiation of any modality in any irradiated system. Electronic excitations in a molecular target may also lead to molecular fragmentation or a bond cleavage. Both excitation and ionization of atoms and molecules may involve different electronic shells. Atomic and molecular excitations may have different mechanisms of energy relaxation, e.g. via the Auger process, radiative de-excitation, electron--phonon coupling, etc.

There are also other important atomic and molecular processes that may play an important role during the interaction of radiation with condensed matter systems, such as electron capture, dissociative electron attachment (DEA), and many others. Here, we discuss only some of them. However, it is important to emphasize that the temporal and spatial scales for the most (if not all) radiation-induced atomic and molecular elementary quantum processes vary within the ranges indicated in Figure~\ref{fig:MM_diagram}. The lower temporal and spatial limits are determined by the characteristic time and distance in a single particle collision with an atom. The upper spatial limit is determined by the collision kinematics of energetic particles with atoms and the upper temporal limit is related to the characteristic relaxation times of various excitation processes in the system (electronic or nuclear) that may proceed via emission of photons, electrons, or other particles (atoms, ions, elementary particles), tunnelling effects, as well as energy transfer to the vibrational degrees of freedom of the system via electron--phonon coupling of the electronic and ionic subsystems.

One should also note that irradiation of condensed matter systems with gamma rays or particles of sufficiently large energy may induce nuclear reactions and related transformations of irradiated condensed matter systems. If these nuclear processes occur, they take place at much smaller spatial scales, although they may last periods orders of magnitude larger than those indicated in Figure~\ref{fig:MM_diagram}. Although they have a different nature than atomic and molecular ones, we do not distinguish them from other quantum processes indicated in Figure~\ref{fig:MM_diagram}. The formal separation of nuclear processes from atomic and molecular would not add any additional aspects to the follow-up considerations of this paper.

The quantum processes are characterised by the corresponding probabilities, which can be derived from quantum mechanics \cite{LL3}. In the case of collision processes, the probability of a process is determined by the product of its cross section and the flux density of projectile particles. The cross sections of collision processes can be calculated using collision theory \cite{Taylor_ScatTheory_book, Newton_ScatTheory_book}. Depending on the studied problem, the flux density of projectile particles can be determined by the primary radiation or the primary radiation together with the radiation of backscattered and secondary particles. The latter can be derived from the particle transport theory.

An important characteristic of radiation-induced quantum processes is the energy and the momentum transferred from the projectile particles into the system. These characteristics can be derived from quantum mechanics and collision theory. The mechanisms of the energy and the momentum transfer, their analysis and quantification are essential for understanding radiation-induced phenomena in condensed matter systems and their efficient computational simulations and quantitative description.
Elementary quantum processes involving nuclei, atoms, and molecules are usually treated by theoretical and computational methods of quantum mechanics and collision theory, including many-body theory, density-functional theory (DFT), and time-dependent density-functional theory (TDDFT). These theoretical methodologies, as well as their computational realizations and limitations, are not indicated in Figure~\ref{fig:MM_diagram} due to the lack of space. However, all these aspects are discussed in Section~\ref{sec:Methods} of this roadmap paper, along with all other theoretical and computational methodologies relevant to the multiscale approach illustrated by Figure~\ref{fig:MM_diagram}.

\subsection{Particle Transport Theories}
\label{sec:MM_stage2}

Let us consider the vertically oriented, rectangular, brown area in Figure~\ref{fig:MM_diagram} corresponding to the particle transport theories: \textit{Monte Carlo method}, \textit{Analytical approach} and \textit{Relativistic Molecular Dynamics (MD)}.

These methodologies describe the two important phenomena related to particles transport through a condensed matter system, namely (i) the primary particle propagation and energy loss and (ii) the creation and propagation of the secondary and follow-up generations of particles. All these phenomena arise in the multiscale scenario of radiation-induced processes presented in Figure~\ref{fig:MM_diagram}.

The first phenomenon deals with the transport of primary particles / radiation through a condensed matter system. The primary particles experience multiple collisions with atoms and molecules during their motion through the system. During these interactions, particles change their direction of motion and lose energy and momentum, transferring them to the medium. The dynamics of such particles is typically considered in a static medium of a given density. Knowing the density of the medium and the cross sections of collision processes of incident particles with atoms in the medium, one can simulate the propagation of incident primary particles through the medium using the \textit{Monte Carlo (MC) method}. Alternatively, one can simulate the propagation of particles through a medium employing \textit{Relativisic MD approach} \cite{RelMD_2013_JCompPhys.252.404} or utilize various \textit{analytical approaches} to describe the propagation of particles through a condensed matter. These methodologies are discussed in more detail in Section~\ref{sec:Methods}. The distances to which ultra-relativistic particles can propagate through the medium can be macroscopically large. Therefore, the upper spatial limit for the brown rectangular area in Figure~\ref{fig:MM_diagram} is indicated as open, stretching towards the larger scales. The temporal scale of primary particle propagation through a medium is determined by the time needed for a particle to pass through a medium of a given size. Primary particles propagation for most of their trajectories is a fast process due to the fast, if not relativistic or ultra-relativistic, character of their motion. For system sizes below $\sim$1~$\mu$m and relativistic velocities of particles ($v \sim c$, where $c$ is the speed of light) the particle propagation times become shorter than $\sim$10~fs.

The second important phenomenon related to the transport of particles through a condensed matter system deals with the process of propagation of secondary and the follow-up generations of particles. The secondary generation of particles is produced by the primary particles/radiation. These secondary particles typically carry less energy than the primary particles. In condensed matter systems, the ionized electrons play a significant role of the secondary particles. However, at sufficiently large energy of the primary radiation, the secondary particles can be produced due to the nuclear reactions induced by the primary particles/ radiation. Secondary particles can in turn create ternary, quaternary, etc. generations of particles before particle propagation stops. The energy of the secondary particles and the ranges of their propagation are less than that of the primary particles and depend on the energy of the primary particles. The primary particles' energy lowers with the propagation of primary particles into a medium.

The production of secondary particles and the follow-up processes induced by them in the medium can be simulated using the MC method, similar to the propagation of primary particles. An alternative to the aforementioned MC approach in simulations of propagation of radiation-induced secondary and follow-up generations of particles is based on the continuous transport theories, e.g. diffusion equation, diffusion-reaction equation, kinetic equation, etc. When justified, these methods can provide faster and reliable solutions for the cumulative dynamics of propagating secondary and higher generations of particles. Thus, in the vicinity of the Bragg peak, the process of propagation of secondary and all the follow-up generations of particles in the medium stops on time scales $\sim 10^1 - 10^2$~fs \cite{Surdutovich_AVS_2014_EPJD.68.353, ES_AVS_2015_EPJD.69.193}. The spatial scales characteristic for the propagation of secondary particles are much smaller than those for the primary particles and depend on the energies at which they are emitted. Thus, $\delta$-electrons emitted after ionization of a medium by relativistic heavy ions can propagate over the micrometer scale, while electrons emitted in the vicinity of the Bragg peak for an ion can only move a few nanometers away from the ion track \cite{Surdutovich_AVS_2014_EPJD.68.353, ES_AVS_2015_EPJD.69.193}. Employing these arguments, the temporal limit of the particle transport domain in Figure~\ref{fig:MM_diagram} has been established \cite{ES_AVS_2015_EPJD.69.193}.

These considerations explain the main features of the particle transport domain presented in Figure~\ref{fig:MM_diagram}. Further details about the main methodologies, their computational realizations and limitations are given in Section~\ref{sec:Methods}. The discussion of the interfaces of this domain with the neighboring domains presented in Figure~\ref{fig:MM_diagram} is given in Section~\ref{sec:Interlinks}.

\subsection{Irradiated Medium Dynamics, Non-Equilibrium Chemistry, and Molecular Transformations}
\label{sec:MM_stage3}

\begin{sloppypar}
The next stage of the multiscale scenario involves \textit{Irradiated medium dynamics}, \textit{non-equilibrium chemistry and molecular transformations}, which arise in most of the case studies considered. This stage is introduced in the central part of Figure~\ref{fig:MM_diagram}. It is seen that the phenomena mentioned above can be simulated using various types of \textit{Molecular Dynamics (MD)}, including the standard \textit{classical MD}, \textit{Reactive MD} \cite{Sushko2016_rCHARMM} and \textit{Irradiation-Driven MD} \cite{Sushko_IS_AS_FEBID_2016}. The temporal and spatial ranges accessible for these methods are introduced in Figure~\ref{fig:MM_diagram} by the blueish rectangular area in the middle part of the figure. These methodologies and their limitations are discussed in detail in Section~\ref{sec:Methods}. Here, let us only mention that the ranges of the blueish rectangular region correspond to the maximum temporal and spatial scales, which can be accessed in simulations of condensed matter systems with the aforementioned computational methods using the most powerful supercomputers. The transparency level of a part of the blueish rectangular region is linked to the feasibility of the simulations; the larger system sizes can be simulated on much shorter temporal scales and the longer temporal behavior can be explored for relatively small systems only.
\end{sloppypar}

On the left-hand side, this blueish region overlaps with stage~\textbf{(ii)} corresponding to \textit{Particle Transport}, as it is seen from a different color of the area where the overlapping takes place. The bottom of the blueish area overlaps with the area of \textit{Radiation-induced quantum processes}.

An alternative for MD based simulations within the range of their feasibility is based on the \textit{Stochastic Dynamics (SD)} approach \cite{Stochastic_2022_JCC.43.1442}, see reddish area in Figure~\ref{fig:MM_diagram}. The SD methodology goes far beyond the capabilities of the MD approach. It is discussed below in this section and the follow-up sections of this roadmap paper.

Stages~\textbf{(i)} and \textbf{(ii)} of the multiscale scenario (see Sections~\ref{sec:MM_stage1} and \ref{sec:MM_stage2}) represent numerous radiation-induced quantum processes in the medium resulting in its transformations such as the breakage of molecular bonds and creation of molecular fragments, formation of defects, ionized centers, free solvated electrons and holes. Such an excited medium is created in a state far from its equilibrium and evolves towards equilibrium through a cascade of follow-up processes. At the end of the particle transport stage, a significant part of the energy of the primary radiation is transferred to the electronic degrees of freedom of the system. Therefore, stage~\textbf{(iii)} of the multiscale scenario continues with the energy transfer processes from the electronic to ionic degrees of freedom in a condensed matter system. This involves various quantum mechanisms, including electron--phonon coupling leading to chemical transformations in the system, such as the formation and subsequent closure of dangling molecular bonds, the creation of new chemical species being products of the irradiated molecules within the system, or their follow-up chemical reactions. At the end of this intermediate state, most of the energy delivered to the medium by the primary radiation is transferred to the ionic degrees of freedom in the system, but the system is still not thermalized. These processes typically occur on a picosecond time scale.

The dynamic response of the medium upon its irradiation does not end with the energy transfer to the ionic degrees of freedom in the system because, at this stage, the energy is distributed non-homogeneously across the system. Therefore, the follow-up process within the multiscale scenario describes the redistribution of the energy deposited into the medium in the vicinity of the particle tracks over the entire volume of the system and amongst all its degrees of freedom. The details of this process depend on the irradiation conditions and the amount of energy deposited. At sufficiently large LET values, it initiates the strong medium dynamics. Under certain conditions, it may lead to a severe distortion of the medium due to a significant increase of temperature and pressure in relatively small volumes where the energy deposition took place and it has been demonstrated that such conditions lead to the formation of nanoscopic shock waves. The strength of these shock waves may be sufficient to create irreversible transformations of the medium, such as molecular bond breakages, defects, formation of craters on surfaces, lethal damages in cells (see the corresponding example in Section~\ref{sec:Intro_Ex_IBCT}), etc.

The dynamical response of the medium starts on the picosecond time scale after the redistribution of energy deposited into the ionic degrees of freedom. The dynamical response of the excited medium leads to its relaxation on the timescale from tens to hundreds of picoseconds, resulting in a more homogeneous distribution of the deposited energy among the ionic degrees of freedom of the system.

The complete thermodynamic equilibration of the system may last up to the nanosecond time scale and even longer, depending on the size of the system and the amount of energy deposited into it. During this period, the distribution of various quantities, such as particle velocities, vibrational excitations, etc., evolve and attain their equilibrated forms consistent with those following from the statistical mechanics for a system being at the thermal equilibrium. Various physicochemical characteristics also evolve towards their equilibrated values during this equilibration process. This affects, for example, the diffusion coefficients of atoms, molecules and other molecular species present in the system after its irradiation. The diffusion coefficients of atoms and molecules determine the resulting chemical transformations of the irradiated medium.

It is important to emphasize that chemical transformations in the system continue during this entire stage and their outcomes are strongly affected by the dynamics of the medium described above. Therefore, such phenomena within the multiscale scenario are characterized as the `non-equilibrium chemistry'. The spatial scales characterizing the non-equilibrium chemistry domain are determined by the spatial distances at which a complete set of possible chemical reactions within the system may take place and reach chemical equilibrium during the period of the medium dynamics and the follow-up thermal relaxation.

Typically, stages~(i) and (ii) adopted for a specific system geometry, utilized radiation sources, radiation modalities, irradiation conditions and characteristics form the initial and boundary conditions for the follow-up multiscale scenario of irradiation-induced/driven processes in each case study. Examples of such case studies have already been presented in Section~\ref{sec:Intro} and will be further provided and discussed in Section~\ref{sec:Case_studies}.

This non-equilibrium chemistry can be studied using RMD and IDMD introduced above and further discussed in Section~\ref{sec:Methods} within the ranges indicated for these methods in Figure~\ref{fig:MM_diagram}. For the larger system sizes, MD simulations of the medium dynamical response upon its irradiation and induced non-equilibrium chemistry processes become more and more challenging or even impossible. The dynamic behavior of such systems can nevertheless be studied employing SD introduced above and further discussed below. The discussion of this approach and its interfaces with other methodologies mentioned above and corresponding stages of the multiscale scenario shown in Figure~\ref{fig:MM_diagram} are given in Section~\ref{sec:Interlinks}.

\subsection{Chemical (and Thermodynamic) Equilibrium}
\label{sec:MM_stage4}

The characteristic size of an irradiated system at which it can be chemically equilibrated is determined by the chemical species present in the system with the smallest concentration. Its concrete value depends on the atomic and molecular composition of the system, the utilized irradiation modality, radiation dose, LET, temperature, etc. It can vary from tens of nanometers to micrometers and above.

The chemical and thermodynamic equilibrium stage is shown in Figure~\ref{fig:MM_diagram} as the fourth stage of the multiscale scenario following the stage of irradiated medium dynamics, non-equilibrium chemistry and molecular transformations. This stage describes the system at the state at which all possible chemical transformations are in balance, and thus all the chemical products exist at specific equilibrated concentrations. Once achieved, the chemical equilibrium can, in principle, last infinitely long, providing that the state of the system is not affected by any further external factors.

It is seen from Figure~\ref{fig:MM_diagram} that simulations of a system in its chemical equilibrium using RMD or IDMD might be challenging because of the relatively large system sizes involved. However, these simulations can be efficiently performed employing SD due to the significant computational advantages gained with the MC approach being a basis for SD.

SD describes dynamical processes in nearly all complex systems, including condensed matter systems, having a probabilistic nature. Such processes may take place at large ranges of temporal and spatial scales. Therefore, the reddish area representing the SD methodology and related processes in Figure~\ref{fig:MM_diagram} occupies the largest part of the figure. The concept of SD and its implementation in the popular software package MBN Explorer are discussed in Sections~\ref{sec:Methods} and \ref{sec:Interlinks}. Here, let us only mention that SD permits simulations of physical, chemical, and biological processes. It is relevant to modeling the multiscale phenomena presented above and to most other case studies in the field.

\subsection{Large-Scale Processes}
\label{sec:MM_stage5}

The last (fifth) stage of the multiscale scenario depicted in Figure~\ref{fig:MM_diagram} corresponds to the large-scale processes that arise in an irradiated system after it reaches the chemical equilibrium. This stage of the multiscale scenario typically involves macroscopic observables characterizing irradiated systems that emerge/originate from the radiation-induced processes/transformations occurring in the system in all the preceding stages of the multiscale scenario. The macroscopic observables usually involve larger spatial and/or longer temporal scales. These large-scale processes arise because the irradiation-induced phenomena occurring within the system and passing through the multiscale scenario described above are usually embedded into larger scale environments. This can be a bulk material or a biological medium being part of a cell, tissue, organ or whole organism; this can be a gas or a plasma volume surrounded by a wall being part of a device, etc.

There are many examples of large-scale processes triggered in systems by their irradiation. For biological systems, this could be most of the radiobiological phenomena. On the cellular level, this could be, for instance, repair mechanisms of the DNA damage caused by irradiation of a cell nucleus, cell apoptosis upon irradiation, or changes in the chromatin properties after its irradiation. On the intercellular level, this could be, for instance, a bystander effect demonstrating the response of a cell to the irradiation of a neighboring cell. Reactions of the immune system on irradiation of some part of the organism could be a representative example of a large-scale process on the level of the organism. There are also many examples of large-scale processes in non-biological systems. For instance, irradiation of materials leads to the alteration of their various bulk properties, such as elasticity, hardness, ability of further disposal, etc., due to the formation of defects in the materials. Radiation-induced defects in electronic chips may lead to errors in the operation of electronic devices, etc.

In order to gain an understanding and quantitative assessment of the large-scale effects, one needs to establish a relationship between the radiation-induced phenomena / transformations emerging from the post-irradiation equilibrium state of the system and the macroscopic observables. This goal has been achieved in the exemplar case studies introduced in Section~\ref{sec:Intro}. For instance, a link was established between the complexity of the DNA damage at which it becomes irreparable and thus, it was possible to determine the cell survival probability. The formation of such complex damages can be quantified within the multiscale approach and on this basis an important macroscopic observable/characteristic, the probability of the cell survival, can be calculated. It is also worth mentioning that for many systems the links between microscopic effects of radiation with macroscopic observables and the large-scale processes are not yet established and it is a topic of intensive current investigations in many different areas of research.

It is obvious that the lower limits for the temporal and spatial scales of large-scale processes should correspond to the scales at which such processes emerge. Often this happens on the scales at which the irradiated parts of the system become equilibrated in the thermodynamic and/or chemical sense. As mentioned above, these scales vary in each case study. The relevant upper spatial and temporal limits for the large-scale processes often become macroscopic. Therefore, the ranges of the SD, being the most suitable approach for simulations of the large-scale processes, are indicated in Figure~\ref{fig:MM_diagram} as extendable with the large arrows directed parallel to the temporal and spatial axes.

Large-scale processes can often be modeled by means of SD because they happen probabilistically. The probabilistic nature of large-scale processes is related to the fact that they are not fully controlled and their key characteristics may depend, for example, on the environment in which processes occur or involve some other known or even as yet unknown phenomena. Establishing probabilities for large-scale processes is often a difficult task. Therefore, there is no general recipe for their determination except experimental measurement of probabilities of relevant events occurring in a system during its SD dynamics. Theoretical or computational derivation of such probabilities can also be achieved, although this is usually done for each particular case study individually.


\textbf{\textit{In conclusion to this section}} let us state that Multiscale Modeling (MM) of condensed matter systems exposed to radiation is represented by a set of complementary and interlinked theoretical and computational methods enabling simulation of condensed matter systems of different origins, their atomistic interactions with radiation and a subsequent cascade of the key processes and phenomena resulting in the formation of specific macroscopic observables relevant to experimental measurements, technological applications and medicine.

The main stages of the multiscale scenario for the dynamical response of condensed matter systems exposed to radiation are presented in Figure~\ref{fig:MM_diagram}, as well as the main theoretical and computational methods relevant to each stage. The presented multiscale scenario is general and applicable to all the systems mentioned above and many different case studies, although it should be adjusted to account for specific, relevant details for each particular case study (see Section~\ref{sec:Case_studies}). In recent years, the MM approach has been developed and validated for a number of case studies, some of which are discussed below in detail. However, there are still many systems and identified problems in the field for which the MM is in its infancy and requires further research.

Therefore, the main challenge of the MM for the next 5--10 years will be devoted to establishing the standards for the methodologies enabling the complete/inclusive MM (i.e. accounting for all relevant phenomena at all temporal and spatial scales involved) of condensed matter systems exposed to radiation in a robust and reliable manner. Different key aspects of this goal are discussed below in this Roadmap paper.

\section{Existing Theoretical and Computational Methodologies, Their Interlinks and Limitations}
\label{sec:Methods}

This section briefly overviews existing theoretical and computational methodologies utilized to study different radiation-induced processes in molecular and condensed matter systems. The existing methods usually focus on particular systems, a particular range of system sizes, and selected phenomena involved. Therefore, these methods cannot model irradiation-induced processes and related phenomena across the different temporal and spatial scales shown in Figure~\ref{fig:MM_diagram}.

In this section, we do not aim to describe the formulation of the existing methods in any detail. Instead, we briefly overview these methods, their areas of application, and their limitations in connection with the overall MM approach depicted in Figure~\ref{fig:MM_diagram}. The interlinks between the different methods are discussed in Section~\ref{sec:Interlinks}.

\subsection{Quantum Processes}
\label{sec:Methods_Quantum-proc}

The interaction of radiation with a molecular or a condensed matter system takes place via the radiation-induced elementary quantum processes, such as ionization, electronic excitation, electron attachment, charge transfer, energy relaxation, and other processes discussed in Section~\ref{sec:MM_key-definitions}. These processes occur on the atomic/sub-nano- and nano scales, see stage~(i) in Figure~\ref{fig:MM_diagram}. Among other possible transformations in atomic and molecular systems, these processes may lead to the cleavage of covalent bonds or the formation of defects in an irradiated system. The quantitative description of these processes is achieved using theoretical and computational methods based on quantum mechanics (\textit{ab initio} methods), such as the Hartree-Fock (HF) method \cite{FroeseFischer_HF_book} and the density-functional theory (DFT) \cite{Kohn_Sham_1965}. These methods have been widely utilized for decades to calculate the electronic properties of many-body systems, such as atoms and molecules \cite{FroeseFischer_HF_book, Parr_Yang_DFT_book}. Since the 1980s, these methods have also been increasingly and successfully applied to larger systems, including atomic and molecular clusters \cite{Ekardt_MetalClusters, LatestAdv_ISAAC_2004, Mariscal_2013_book} and biomolecular systems \cite{Jalkanen_2009_DFT_molbiol}. There are numerous books and reviews \cite{FroeseFischer_HF_book, Parr_Yang_DFT_book, Becke_2014_JCP.140.18A301, Scuseria_XC_funct_2005, Jones_2015_RMP.87.897, Handbook_CompChem} (with the most recent ones \cite{Kaplan_2023_AnnuRevPhysChem, Cances_DFT_book_2023}) devoted to HF and DFT methods; therefore, only the key ideas behind these methods are outlined below.

\subsubsection{Many-Body Theory}
\label{sec:Methods_Many-body_theory}

\textbf{The Hartree-Fock (HF) method} \cite{FroeseFischer_HF_book, Handbook_CompChem} is a computational physics and chemistry tool to solve the time-independent Schr\"{o}dinger equation for a many-body electronic system. Since there are no known analytical solutions for many-electron systems, the problem is solved numerically. The basic idea of the HF method is that the total $N$-body electronic wavefunction of a many-electron system is approximated as an antisymmetrized product of $N$ one-electron wavefunctions characterized by a set of quantum numbers. By invoking the variational method, one can derive a set of $N$-coupled one-electron equations, a solution of which yields an HF wavefunction and the energy of a many-electron system.

\begin{sloppypar}
The \textbf{Dirac-Hartree-Fock (DHF) method} is a well-established method \cite{Grant_AdvPhys1970, Mohanty_1991_IJQC.39.487} for atomic, molecular, cluster and condensed matter systems containing heavy elements that require accurate treatment of relativistic effects \cite{Pyykko_1988_ChemRev.88.563}, such as core orbital contraction, spin-orbit coupling, and spin-spin interactions. In this method, a many-electron wavefunction, the solution of the relativistic Dirac's equation, is constructed as an antisymmetrized product of molecular spinors.
\end{sloppypar}

The HF method does not account for the dynamical electron correlation due to using the closed-shell Slater determinant as a ground-state electronic configuration where electrons are forced to be confined in particular orbitals. In order to calculate the dynamical electron correlation energy, three different classes of \textit{ab initio} methods \cite{MBPT_methods_book_Shavitt} have been developed to permit studies of excitation energies, transition states, spectroscopic properties, etc. In these methods, the HF electronic configuration is used as a starting point for the calculations.

In \textbf{Coupled Cluster (CC)} \cite{Cizek_1980_PhysScr.21.251, Bartlett_2007_RMP.79.291, Zhang_2019_FrontMater.6.123} and \textbf{Configuration Interaction (CI)} \cite{Szalay_2012_ChemRev.112.108} methods, excited determinants are generated from the HF determinant and added to the Slater determinant with appropriate coefficients to improve the wavefunction, keeping the Hamiltonian fixed. In the CC approach, the correlated wavefunction of a many-body system is defined as an exponential of cluster operator $T$ acting on a ``reference'' HF state. In CI the electronic ground-state wavefunction is constructed as a linear combination of configuration-state functions (CSFs). Full CI \cite{Knowles_1989_CPC.54.75} provides a numerically exact solution (within the complete CSF basis set that includes all Slater determinants obtained by exciting all possible electrons to all possible virtual orbitals) to the time-independent, non-relativistic Schr\"{o}dinger equation. However, due to the complexity of the full CI method, it has been applied only to a few-electron systems \cite{Rontani_2006_JCP.124.124102, Joecker_2021_NJP.23.073007}. Full CC is equivalent to full CI, but is applicable only for small-size systems (containing up to a few tens of atoms) due to high computational costs \cite{Xu_Uejima_2018_PRL.121.113001}.

The most common approximation is the truncation of the CC operator or CI space expansion according to the excitation level relative to the ``reference'' HF state. The commonly used truncated CC methods include (i) the CC singles and doubles (CCSD) approach \cite{Purvis_1982_JCP.76.1910, Cullen_1982_JCP.77.4088}, where the cluster operator $T$ is truncated at the two-body component $T_2$, and (ii) the quasi-perturbative correction to CCSD due to the three-body component $T_3$, defining the widely used CCSD(T) approximation \cite{Raghavachari_1989_CPL.157.479}. For the truncated CI method, the widely employed CI singles and doubles (CISD) wavefunction includes only those $N$-electron basis functions representing single or double excitations relative to the reference state.

The other class of post-HF methods relies on the \textbf{many-body perturbation theory (MBPT)}, also known as the \textbf{M{\o}ller-Plesset (MP) perturbation theory} \cite{Moller_Plesset_1934, Cremer_2011_WIREsCMS.1.509}. It improves on the HF method by adding electron correlation effects to the second (MP2), third (MP3), fourth (MP4) or higher order. In this approach, the unperturbed Hamiltonian is replaced by a perturbed Hamiltonian with the constraint that perturbation must be small. The perturbed wavefunction and perturbed energy of a many-body system are expressed as a power series of a parameter that controls the size of the perturbation. MP methods are computationally less demanding than the CC- and CI-based methods, with MP2 being the least costly and most widely used \textit{ab initio} method to correct HF results for correlation effects. MP2, MP3 and MP4 are standard computational methods in many quantum chemistry codes (see below). Higher-level MP calculations, mainly employing MP5, can be performed using some codes (e.g. Gaussian), but these methods are rarely used because of their high computational cost.

\textbf{Software Tools}: Many existing quantum chemistry software tools enable \cite{Wiki_QC_codes} calculations using HF and post-HF (including MP$n$) methods, for instance, CP2K \cite{CP2K_Kuhne_2020_JCP.152.194103}, Dalton \cite{DALTON_Aidas_2014_WIREsCMS.4.269}, GAMESS \cite{GAMESS_Schmidt_1993_JCC.14.1347}, Gaussian \cite{GAUSSIAN_g16}, NWChem \cite{NWChem_Valiev_2010_CPC.181.1477}, and ORCA \cite{ORCA_Neese_2020_JCP.152.224108}. The DHF method has been realized in several relativistic quantum chemistry programs, such as GRASP \cite{GRASP_Grant_2022_Atoms.10.108},  MOLFDIR \cite{MOLFDIR_Visscher_1994_CPC.81.120}, DIRAC \cite{DIRAC_Saue_1997_MolPhys.91.937} and BERTHA \cite{BERTHA_Grant_2000_IJQC.80.283, BERTHA_Belpassi_2020_JCP.152.164118}. The latter program has been utilized for relativistic electronic structure calculations for atoms, diatomic and polyatomic molecules, and atomic clusters \cite{BERTHA_Belpassi_2020_JCP.152.164118, Grant_BERTHA_chapter_2006}. Wavefunctions and energies of the ground and excited states in many-electron atoms have been calculated in the HF and DHF approximations using the ATOM computer program system (see the review \cite{ATOM_Chernysheva_VK_2022_Atoms.10.52} and references therein).

\textbf{Areas of Application}: The HF and DHF methods have been commonly used to calculate the electronic structure of atoms, various molecules, atomic clusters, and periodic systems \cite{March_2003_HF_beyond, Evarestov_QC_Solids}. The HF method also often serves as a starting point for more sophisticated \textit{ab initio} methods, such as the MBPT or the random phase approximation \cite{Ren_2012_JMaterSci.47.7447}. The ground- and excited-state electronic wavefunctions for atoms, molecules and atomic clusters, calculated using the HF and DHF method, have been commonly used to calculate matrix elements for photo- and electron-impact ionization and other processes and their corresponding cross sections \cite{Starace_PI_atoms_2006, Solovyov_2005_IJMPB.19.4143} (see also Section~\ref{sec:Methods_Cross_sections} below).

CI and CC methods require substantial computational resources compared to HF or DFT calculations for the same number of electrons. Therefore, CC- and CI-based methods are not practical for large-scale systems and are often used to verify the quantum many-body theory in experiments or for testing the quality of numerous DFT functionals that are often based on various empirical approaches and assumptions (see Section~\ref{sec:Methods_DFT}).

\textbf{Limitations and Challenges}: It has been commonly discussed in the literature \cite{Strout_1995_JCP.102.8448} that the computational cost of a HF calculation scales as $N^4$, where $N$ is the number of basis functions used for the calculation. The origin of the $N^4$ scaling behavior is the calculation of the four-center two-electron integrals. In practice, scaling behavior is closer to $N^3$ as quantum chemistry programs for HF calculations can identify and neglect small two-electron integrals. In any case, a large scaling exponent imposes limitations on the system size that can be simulated using the HF method. Apart from that, the HF method is insufficient for the accurate quantitative description of the properties of many compounds due to the neglect of electron correlations.

The post-HF methods have similar limitations of poor scaling with system size. For instance, MP2, MP3 and MP4 methods scale as $N^5$, $N^6$ and $N^7$, respectively \cite{Cremer_2011_WIREsCMS.1.509}, whereas CCSD and CCSD(T) methods scale as $N^6$ and $N^7$, respectively \cite{Scuseria_1990_JCP.93.5851}. Such large scaling exponents impose strong limitations on the applicability of these methods for small-size molecular systems (typically containing up to several tens of atoms).

Another commonly discussed problem of \textit{ab initio} many-body methods is related to erratic or even divergent behavior of the MP$n$ series \cite{Cremer_2011_WIREsCMS.1.509, Leininger_2000_JCP.112.9213, Herman_2008_IJQC.109.210}. Systematic studies of MBPT have shown that it is not necessarily a convergent theory at high orders of perturbation and depends on the precise chemical system and basis set \cite{Leininger_2000_JCP.112.9213}. An oscillatory behavior of molecular energies and other properties with increasing the order of perturbation $n$ has been observed \cite{Cremer_2011_WIREsCMS.1.509, Cremer_1996_JPC.100.6173}, thus making it challenging to predict higher-order MP$n$ results and extrapolate them towards the ``exact'' results obtained using full CI.

One of the big challenges of the \textit{ab initio} many-body methods has been to reduce their scaling behavior with respect to the size of the basis set and develop efficient and low-scaling methods to compute large-size systems. This problem has been actively addressed since the 1990s \cite{Cremer_2011_WIREsCMS.1.509}, with the method development work being focused almost exclusively on MP2. Many computational techniques and approaches have been developed to convert MP2 from an $O(N^5)$ computational problem into a low-order or even a linear-scaling task \cite{Doser_2009_JCP.130.064107, Glasbrenner_2020_JCTC.16.6856} that can handle molecules containing $\sim$10$^3$ atoms. A review of these developments is given e.g. in ref~\citenum{Cremer_2011_WIREsCMS.1.509}.

One of the current challenges for the further development and utilization of \textit{ab initio} many-body methods would be the selection of the most efficient scaling methods and their widespread implementation in different existing quantum chemistry codes, which would enable a more straightforward application of \textit{ab initio} many-body methods for larger-size molecular systems containing $\sim 10^3 - 10^4$ atoms.

\subsubsection{Density Functional Theory}
\label{sec:Methods_DFT}

\textbf{Density Functional Theory (DFT)} is one of the basic tools for describing ground-state electronic properties of finite-size and periodic many-body systems and nanomaterials. The modern version in use today is Kohn-Sham (KS) DFT \cite{Kohn_Sham_1965}. It defines self-consistent equations solved for a set of electronic orbitals whose density, $\rho(r)$, is defined to be exactly that of the real system. In DFT the energy of a many-electron system includes the so-called exchange-correlation (XC) energy, defined in terms of $\rho(r)$.

In the DFT approach, one needs to define the XC potential to solve single-electron KS equations, although the form of this potential in the general case is unknown. Therefore, many different approximations have been introduced, allowing one to solve many-electron problems and describe the physical properties of many-electron systems \cite{Becke_2014_JCP.140.18A301}. The simplest XC approximation is the Local Density Approximation (LDA) \cite{Kohn_Sham_1965}, which became the popular standard in calculations of the properties of solids in the 1970s and 80s \cite{Parr_Yang_DFT_book}. In the late 1980s, Generalized Gradient Approximations (GGA) \cite{PBE_1996_PRL.77.3865, Scuseria_XC_funct_2005} reached a sufficient level of accuracy for chemical calculations. In the early 90s, hybrid XC functionals were introduced, where a fraction of GGA exchange interaction was replaced with the HF exchange contribution, leading to the ubiquitous B3LYP functional. Later on, more advanced long-range XC functionals, such as CAM-B3LYP \cite{CAM-B3LYP_Yanai_2004_CPL}, LC-$\omega$PBE \cite{Vydrov_2006_JCP.125.074106} and $\omega$B97X-D \cite{Chai_2008_PCCP.10.6615}, were developed, enabling a more accurate description of long-range electron-electron exchange interactions, which are essential for a correct description of charge-transfer and non-linear optical properties of molecular and condensed matter systems \cite{Tsuneda_2014_WIREsCMS.4.375, Li_2021_JCC.42.1486}.

\textbf{Software Tools}: Numerous quantum chemistry software tools nowadays enable calculations using DFT methods \cite{Wiki_QC_codes}. The most widely used tools include Gaussian \cite{GAUSSIAN_g16}, GAMESS \cite{GAMESS_Schmidt_1993_JCC.14.1347}, ORCA \cite{ORCA_Neese_2020_JCP.152.224108}, Dalton \cite{DALTON_Aidas_2014_WIREsCMS.4.269}, Octopus \cite{OCTOPUS_Castro_2006_PSSB.243.2465}, and QuantumEspresso \cite{QE_Giannozzi_2009_JPCM.21.395502}.

\textbf{Areas of Applications}: DFT has found numerous applications in chemistry and materials science by calculating the electronic ground-state properties of various systems \cite{Jones_2015_RMP.87.897, Makkar_2021_RSCAdv.11.27897}. In solid-state calculations, the LDA is still commonly used along with plane-wave basis sets, as an electron-gas approach is more appropriate for electrons delocalized through an infinite solid. However, in calculations involving molecular systems, more accurate long-range corrected XC functionals (e.g. the aforementioned CAM-B3LYP, LC-$\omega$PBE and $\omega$B97X-D) are typically required for a quantitative description of charge-transfer processes, e.g. in photochemistry and quantum biology studies.

\textbf{Limitations and Challenges}: DFT calculations employ the approximated XC interaction potentials, and the calculation results strongly depend on the approximation used. Many DFT approximations (particularly at the LDA and GGA levels) suffer from an incorrect asymptotic behavior of the XC potential. This issue is commonly associated in the literature with spurious self-interaction \cite{Perdew_Zunger_1981_PRB.23.5048}, arising from an approximate exchange functional. This self-interaction error is considered one of the major sources of error in most XC functionals for Kohn-Sham DFT. In order to address this problem, several self-interaction corrections have been developed, see e.g. a review \cite{Tsuneda_2014_JCP.140.18A513}.

The fact that long-range electron-electron exchange interactions are insufficiently incorporated in conventional exchange functionals has motivated the development of long-range correction (LC) of exchange functionals \cite{Tsuneda_2014_WIREsCMS.4.375}. This correction has been implemented in long-range XC functionals, e.g. the CAM-B3LYP, LC-$\omega$PBE and $\omega$B97X-D functionals mentioned above.

Conventional local and semi-local XC functionals do not describe the long-range dispersion (van der Waals) interaction, which plays an important role in the formation, stability, and functioning of molecules and materials. To address this problem, dispersion corrections to standard Kohn–Sham DFT have been developed, see e.g. reviews \cite{Grimme_2011_WIREsCMS.1.211, Grimme_2016_ChemRev.116.5105}.

The development of novel and the advancement of existing XC functionals is, therefore, one of the current challenges of DFT. This challenge can be addressed by validating the improved/developed functionals against larger datasets and making them more `flexible' and applicable to many periodic table elements.

Depending on the level of approximations in DFT, the maximum achieved system size in terms of the number of atoms ($N$) is typically on the order of $10^3$. A further increase in the size is often impractical, mainly due to the high cost associated with DFT \cite{Muller_2020_PRL.125.256402}, with the computational cost being proportional to $N^3$. Some recently developed DFT codes, such as CONQUEST \cite{CONQUEST_Nakata_2020_JCP.152.164112}, provide a linear scaling of the computer time with $N$, thus enabling large-scale electronic-structure calculations for systems containing $\sim 10^4-10^5$ atoms. A more widespread realization of efficient scaling algorithms in existing DFT codes is another challenge that could be addressed in the coming years.

\subsubsection{Time-Dependent Density Functional Theory}
\label{sec:Methods_TDDFT}

The ground-state DFT (Section~\ref{sec:Methods_DFT}) cannot describe electronic excited states and, therefore, many important irradiation-driven physical properties, such as optical absorption and emission, response to time-dependent fields, and the dynamical dielectric function. \textbf{Time-Dependent Density-Functional Theory (TDDFT)} extends the basic ideas of ground-state DFT to allow the treatment of electronic excitations or, more generally, time-dependent phenomena \cite{Runge_Gross_1984_PRL.52.997}.

TDDFT is an approximate method for solving the time-dependent Schr\"{o}dinger equation, which allows one to study the properties of many-electron systems as a function of time. The time-dependent Schrödinger equation is substituted in this approach with a set of time-dependent single-particle KS equations \cite{Ullrich_TDDFT_book_2012}. In an analogy with the KS method, which is the basic principle of the ground-state DFT, TDDFT considers an external time-dependent potential $v_{\rm ext}({\bf r},t)$ that describes the motion of a system of non-interacting particles and the time-dependent electron density $\rho({\bf r},t)$ of a real many-electron system.

TDDFT is usually realized either in the linear-response regime, where the electronic density of a system is considered in the frequency domain as a first-order response to an external perturbation potential, or directly in the time domain (so-called real-time TDDFT) by evolving the KS wavefunctions in time \cite{Maitra_2016_JCP.144.220901}. Real-time TDDFT allows studying both linear and non-linear regimes and the dynamics of electronic excitations in response to ultrafast laser pulses as observed in pump-probe spectroscopy.

\textbf{Software Tools}: The most widely used tools enabling calculations using TDDFT include Gaussian \cite{GAUSSIAN_g16}, Octopus \cite{OCTOPUS_Castro_2006_PSSB.243.2465}, QuantumEspresso \cite{QE_Giannozzi_2009_JPCM.21.395502}, ORCA \cite{ORCA_Neese_2020_JCP.152.224108}, NWChem \cite{NWChem_Valiev_2010_CPC.181.1477}, Dalton \cite{DALTON_Aidas_2014_WIREsCMS.4.269}, Molpro \cite{MOLPRO_Werner_2012_WIREsCMS.2.242}, and VASP \cite{VASP_Kresse_1996_CompMatSci.6.15}.

\textbf{Areas of Applications}: TDDFT has been used extensively to calculate excitation energies and optical absorption spectra of atoms, molecules, atomic clusters (containing up to $\sim$10$^2$ atoms) and solids \cite{Marques_TDDFT_book_2006, Adamo_2013_ChemSocRev.42.845, Lian_2018_AdvTheoSimul}. It has been widely used for photophysical and photochemical applications to study the excited-state dynamics (photoluminescence, electron transfer, etc.) in organic dye molecules and biomolecules \cite{Malcioglu_2011_JACS.133.15425, sjulstok2015quantifying, Shao_2020_JCTC.16.587}. It has also been used to calculate electron energy loss spectra for different solids at different values of the transferred momentum \cite{Timrov_2013_PRB.88.064301, Nicholls_2021_JPhysMater}.

The real-time TDDFT is particularly useful for studying the ultrafast dynamics of atoms, molecules, atomic clusters and solids in strong laser fields \cite{Moitra_2023_JPCL.14.1714, Ghosal_2022_CPL.796.139562}. With the advent of attosecond laser pulses, there is a great demand for efficient methods of studying the time-dependent evolution of systems beyond the linear response (see the case study in Section~\ref{sec:Case_study_Attosecond}).

\textbf{Limitations and Challenges}: As for the ground-state DFT, the accuracy of TDDFT calculations depends on the chosen XC functional. Due to the high computational costs of TDDFT calculations, a rigorous quantum-mechanical description of the irradiation-driven processes using TDDFT is only feasible for relatively small systems containing, at most, a few hundred atoms and is typically limited to femtosecond time scales \cite{Alberg-Flojborg_JPB_2020, Salo_2018_PRA.98.012702}.

TDDFT can be used for calculating probabilities of chemically-driven and irradiation-induced quantum processes, which serve as input for Reactive and Irradiation-Driven Molecular Dynamics simulations (see Sections~\ref{sec:Methods_RMD} and \ref{sec:Methods_IDMD}, respectively). However, determining such probabilities requires multiple real-time simulations of electron dynamics for statistical analysis of the calculated quantities. The development of novel computationally efficient approaches for performing multiple computationally expensive TDDFT calculations represents one of the current challenges for realizing the computational MM methodology illustrated by Figure~\ref{fig:MM_diagram}.

\subsubsection{\textit{Ab Initio} and Model Approaches for Calculating Scattering Cross Sections}
\label{sec:Methods_Cross_sections}

The scattering cross section is an important quantity to understand the physical interactions between radiation and a target. The quantum-based theory of scattering developed over the last century has been described in numerous textbooks on quantum mechanics, many-body quantum theory and quantum collision theory \cite{LL3, Newton_ScatTheory_book, Taylor_ScatTheory_book, Friedrich_ScatTheory_book, Dreizler_QuantCollTheory_book}. These books describe in great detail theoretical methods (such as the many-body scattering theory, Green's function, S-matrix formalism, partial-wave expansion, perturbation theory, the Born approximation, eikonal approximation, quasiclassical approximation, and others) used to calculate the scattering wave function and related physical quantities, such as scattering amplitudes, scattering phase shifts, and scattering cross sections. Apart from that, different theoretical methods have been developed to describe scattering processes involving relativistic particles \cite{Wachter_RelativQM_book}.

From a computational point of view, the scattering processes involving single atoms and simple few-atom molecules have been widely described through different \textit{ab initio} methods, including \textbf{many-body scattering theory} \cite{Hofierka_2022_Nature.66.688, Rawlins_2023_PRL.130.263001}, \textbf{configuration interaction} \cite{Samanta_2018_AdvQuantChem.77.317} and \textbf{R-matrix methods} \cite{Schneider_1975_CPL.31.237, Burke_1977_JPB.10.2497, Tennyson_2010_PhysRep.491.29}. Due to the computational complexity of the \textit{ab initio} methods, they have been utilized mainly to calculate the cross sections of low-energy electrons scattering from atoms and simple few-atom molecular targets while being impractical (or even unfeasible) for larger molecular systems and cross sections above the ionization energy of most molecular targets.

The need for photon-, electron- and ion-impact molecular ionization cross sections as input data for various radiation/particle transport modeling tools (see stage (ii) in Figure~\ref{fig:MM_diagram} and Section~\ref{sec:Methods_Particle_Transport} below) has stimulated the use of simpler (empirical and semiempirical) theoretical models for evaluating atomic and molecular ionization cross sections. The simplest approaches rely on the additivity rule concept \cite{Otvos_1956_JACS.78.546}, where the molecular ionization cross section is derived by adding the ionization cross sections, corresponding to the atomic constituents of a molecular system. In addition to these relatively simple approaches, several semiempirical approaches have been developed. The most commonly employed methods to compute ionization cross sections are the \textbf{Deutsch-M\"{a}rk (DM) formalism} \cite{Deutsch_Mark_2000_IJMS.197.37} and the more rigorous \textbf{Binary-Encounter Dipole (BED)} and \textbf{Binary-Encounter-Bethe (BEB) models} developed by Kim and Rudd \cite{BEB_Kim_Rudd_1994_PRA.50.3954}, which combine the additivity concept with molecular structure information calculated employing quantum mechanics.

Both the DM formalism and the BED/BEB formalism use an additivity concept so that the ionization cross section of a molecule is added from the contributions arising from the ejection of an electron from the different molecular orbitals. A detailed review of the BED, BEB and related models can be found in ref~\citenum{Tanaka_2016_RMP.88.025004}. The DM formalism was reviewed in ref~\citenum{Deutsch_Mark_2000_IJMS.197.37}. In brief, the BED model requires, as input, differential dipole oscillator strength (DOS) values of the target, which can be derived from theoretical or experimental photoionization cross sections. The BEB model is a simplification of the BED model in which the DOS term is approximated by a simple function of secondary electron energy \cite{BEB_Kim_Rudd_1994_PRA.50.3954}. Within the BEB framework, molecular ionization cross sections are evaluated using an analytic formula that requires only the incident particle energy and the binding and kinetic energy of molecular orbitals of the target. These energies can be obtained from electronic-structure calculations, e.g. using the HF method (see Section~\ref{sec:Methods_Many-body_theory}). The BEB method calculates the electron-impact molecular ionization cross section over the incident electron energies ranging from the ionization threshold to a few keV, or even up to thousands keV, using the relativistic version of BEB (RBEB) \cite{RBEB_Kim_2000_PRA.62.052710}.

Other analytical approaches exist for calculating cross sections of inelastic scattering (with respect to the energy transferred from the projectile to the medium) or ionization (in particular, as a function of the kinetic energy of emitted secondary electrons) and the related characteristics, such as the stopping power defined as the average projectile's energy loss per unit path length. The ionization cross section for different atomic and molecular targets irradiated with protons and heavier ions is commonly calculated using the semi-empirical Rudd's model \cite{Rudd_1992_RMP.64.441}, based on a combination of the experimental data and calculations within the plane-wave Born approximation. An alternative theoretical method \cite{deVera_2013_PRL.110.148104} for calculating ionization cross sections is based on the calculation of the energy-loss function of the target medium, ${\rm Im}(-1/\varepsilon(E, q))$, where $\varepsilon(E, q)$ is the complex dielectric function, and $\hbar q$ and $E$ are the momentum and energy transferred in the electronic excitation, respectively. This formalism allows obtaining the charged-particle-impact ionization cross sections for various condensed media, including liquid water and other biologically relevant media containing sugars, amino acids, etc., which are of great relevance for studying RADAM in biological systems (see an example in Section~\ref{sec:Intro_Ex_IBCT}).

\textbf{Software Tools}: The \textit{ab initio} methods for calculating scattering cross sections have been implemented in several computer codes, including:
\begin{enumerate}[label=(\roman*)]

\item
B-spline Atomic R-matrix code (BSR) \cite{BSR_Zatsarinny_2006_CPC.174.273}, which computes transition-matrix elements for electron collisions with many-electron atoms and ions as well as photoionization processes;

\item
The Convergent Close Coupling (CCC) computer code \cite{CCC_Bray_2002_JPB.35.R117} used for calculating cross sections for (anti)electron- (anti)proton- and photon-impact collision processes in one- and two-electron atomic and molecular systems;

\item
ePolyScat suite of codes \cite{ePolyScat_Gianturco_1994_JCP} for calculating electron-molecule scattering and molecular photoionization cross sections within the fixed-nuclei approximation;

\item
UK Molecular R-Matrix (UKRmol+) code \cite{UKRmol_Masin_2020_CPC}, which provides an implementation of the time-independent R-matrix method for molecules and permits calculating scattering cross sections for low-energy electrons and positrons as well as photons;

\item
XChem code for all-electron \textit{ab initio} calculations of ionization of atoms, small and medium-size molecules \cite{xCHEM_Borras_2021_JCTC}.

These and other \textit{ab initio} codes for calculating scattering cross sections are collected in the Atomic, Molecular, and Optical Science (AMOS) Gateway online platform \cite{AMOS_Gateway, AMOS_Gateway_paper2023}. Apart from these, one can also mention several other widely used codes:

\item
The ATOM computer program suite for studying the structure, transition probabilities and cross sections of various processes in multi-electron atoms \cite{Amusia_Photoeffect_book, ATOM_Chernysheva_VK_2022_Atoms.10.52}. This computational framework can calculate several key characteristics of the scattering process, including, in particular, amplitudes and cross sections of photoionization (including the oscillator strength for discrete electronic transitions) of atoms with filled and half-filled shells; characteristics of the angular distribution of photoelectrons and secondary electrons in the dipole approximation and beyond; cross sections of elastic and inelastic scattering of particles (electrons, positrons, and mesons) from atoms; cross sections of the ionization and excitation of atoms by electron impact; photoionization cross sections and angular anisotropy parameters of endohedral atoms, the decay of vacancies in such atoms, and many more. More recently, the ATOM program suite was extended for studying scattering processes involving few-atom molecules, as realized in the ATOM-M code \cite{ATOM_Chernysheva_VK_2022_Atoms.10.52}.

\item
The Dirac-Atomic R-matrix Codes (DARC) program suite \cite{DARC_McLaughlin} for relativistic scattering calculations.
\end{enumerate}

Scattering cross sections calculated using the empirical and semiempirical methods are implemented, to a large extent, into custom-made codes, which have been developed by many researchers over the decades (see numerous publications utilizing DM and BEB methods). The BEB model is also incorporated into the commercial Quantemol-Electron Collisions (Quantemol-EC) software \cite{Quantemol-EC_Cooper_2019} for calculating electron-molecule scattering cross sections.

\textbf{Areas of Applications}: As described above, the \textit{ab initio} scattering methods have been utilized mainly to calculate the cross sections of elastic and inelastic scattering from single atoms and few-atom molecular targets. The semiempirical models based on the DM and BED/BEB formalism have been widely used to evaluate electron-impact ionization cross sections for various atomic and molecular targets, including simpler molecules (e.g. H$_2$, CH$_4$ or CO$_2$) and more complex systems, such as e.g. DNA building blocks \cite{Bernhardt_2003_IJMS.223.599, Huber_2019_EPJD.73.137} and organometallic compounds \cite{Langer_2018_EPJD.72.112}, thus enabling to overcome the limitations of \textit{ab initio} scattering theory methods for calculating molecular ionization cross sections. The Rudd's model has been used, in particular, to calculate ionization cross sections of simple diatomic and few-atom molecular targets \cite{Rudd_1992_RMP.64.441} and DNA nucleobases \cite{Francis_2017_JAP.122.014701}. It has also been used to calculate the energy spectrum of secondary electrons produced in biologically relevant media \cite{Surdutovich_AVS_2014_EPJD.68.353, Surdutovich_2009_EPJD.51.63}. These characteristics are used as input for the track-structure MC simulations and analytical methods for modeling particle transport based on the solution of the diffusion equation (see Section~\ref{sec:Methods_Particle_Transport}).

\textbf{Limitations and Challenges}: The \textit{ab initio} scattering calculations are limited in application for single atoms and simple few-atom molecules due to their high computational costs, as described above. The main limitation of the semiempirical methods (DM, BEB, BED, etc.) for calculating scattering cross sections is their accuracy; these methods usually describe the general shape of the cross sections in a broad range of impact electron energies but cannot describe finer features of the cross sections, such as autoionization peaks. Moreover, the semiempirical methods often require the determination of a set of input parameters for each particular molecular target, and the existing parameters developed for particular systems might not be transferrable to other systems. For instance, parameters for calculating singly-differential ionization cross section using the Rudd's model \cite{Rudd_1992_RMP.64.441} have been developed for a limited number of atomic and molecular targets, namely noble gas atoms, simple diatomic molecules (e.g. H$_2$, N$_2$, O$_2$) and few-atom molecules (e.g. H$_2$O, CO$_2$, CH$_4$), but there are no parameters in the literature for more complex molecular systems (except from the recent study of DNA nucleobases \cite{Francis_2017_JAP.122.014701}). In such cases, the cross section can be estimated based on the available cross sections for smaller building blocks of similar targets, but the accuracy of this approach is not known \textit{a priori} and must be carefully validated against existing experimental data and more sophisticated calculations (see the discussion of the validation of MM methodologies in Section~\ref{sec:Validation}). Therefore, a significant current challenge is the systematization of the cross sections for different molecular targets in dedicated databases (see Section~\ref{sec:Validation}) and the development of universal approaches that can be applied directly to a large class of molecular and condensed matter systems.

\subsection{Particle Transport}
\label{sec:Methods_Particle_Transport}

\subsubsection{Monte Carlo-based Particle Transport Models}
\label{sec:Methods_MC_transport}

The \textbf{Monte Carlo (MC) method} is widely used for modeling stochastic processes of radiation transport, i.e. the propagation of particles such as photons, electrons, positrons, neutrons, protons, and heavier ions through various condensed media. In this approach, particle transport in a continuous medium is simulated stepwise, accounting for the stochastic nature of particles' interactions with atoms of molecules constituent the medium.

Particles propagating through a medium experience different quantum processes, such as elastic and inelastic scattering, electronic excitation, ionization, nuclear fragmentation, etc. The probabilities of such interactions are determined according to the cross sections of the corresponding quantum processes and the number density of target atoms/molecules in the medium. Secondary particles (e.g. secondary electrons or nuclear fragments) are created due to interactions of primary projectiles with a medium, and their subsequent transport can be simulated. The distance to the next step of the particle’s trajectory is determined by the particle’s mean free path, which depends on the cross sections of the considered quantum processes and the number density of constituent atoms or molecules.

The main methods for simulating radiation transport within the MC scheme are the \textbf{``condensed-history''} and the \textbf{``track-structure'' methods} \cite{Nikjoo_2006_RadiatMeas.41.1052, Kawrakow_1998_NIMB.142.253, Jenkins_MC_transport_book}.

The condensed-history approach is commonly utilized to evaluate the average energy loss due to the primary radiation calculated over a certain length of track traveled and the energy deposited along this track length. This method is helpful for computationally efficient calculations of many-particle interactions since it considers only the primary particles and disregards any interactions involving secondary particles. Condensed-history MC codes (some of which are listed below) only require the energy loss per track length, or stopping power, as input. Such data calculated using theoretical models (see Section~\ref{sec:Methods_Cross_sections}) and measured experimentally are tabulated for many atomic elements of the periodic table \cite{NIST_Stopping_Power}. Composite materials are modeled using weighted stopping power values based on their atomic composition and density scaling.

The ``track-structure'' (also known as ``event-by-event'') MC method simulates the trajectories of single particles in a medium, i.e. the complete track structure of the projectile and all the secondary particles generated in the medium \cite{Dingfelder_2012_HealthPhys.103.590}. Such simulations provide detailed information on the interactions, including spatial distributions of the deposited energy, different interaction types (ionization, excitation, elastic scattering, change of the charge state, etc.), and the radical species produced. Track-structure MC codes (listed below) use the total mean free path for determining the location of the next interaction, while the total cross section for all considered interactions and the medium density are used to model particle propagation in a medium. The corresponding cross sections are typically evaluated using the \textit{ab initio} and semiempirical methods described in Section~\ref{sec:Methods_Cross_sections} or taken from the experiment.

\textbf{Software Tools}: Some widely-known examples of condensed-history MC codes are FLUKA \cite{FLUKA_Bohlen_2014}, SRIM \cite{SRIM_Ziegler_2010_NIMB.268.1818}, PENELOPE \cite{PENELOPE_Salvat} and GEANT4 \cite{GEANT4_Agostinelli_2003_NIMA}. Several examples of track-structure codes are KURBUC \cite{Uehara_1992_PMB.37.1841}, PARTRAC \cite{PARTRAC_Friedland_2011_MutatRes}, NOTRE DAME \cite{Pimblott_1991_JPC.95.7291} and EPOTRAN \cite{EPOTRAN_Champion_2012_IJRB}.

One of the most widely known MC codes for particle transport is Geant4-DNA \cite{Incerti_2010_Geant4-DNA, Bernal_2015_PhysMed.31.861}, which is an extension of Geant4 for describing biological (mainly DNA) damage induced by ionizing radiation on the molecular scale. Geant4-DNA exploits cross sections for interactions of particles with various materials, particularly water \cite{Bernal_2015_PhysMed.31.861} and DNA nucleobases \cite{Zein_2021_NIMB.488.70}.

\textbf{Areas of Application}: Specific areas of application of the MC-based particle transport codes include (among others) radiation protection and dosimetry, modeling radiation effects in material (including radiation-induced material damage), radiation shielding, and medical physics. The condensed-history approach has been widely used to calculate macroscopic dose profiles. Track-structure MC methods employed in the codes mentioned above have been commonly used to simulate the first (physical) stage of the interaction of ionizing radiation with various condensed (including organic and biological) media, during which secondary electrons and ions are produced by the deposition of energy in matter. Some of the MC codes (e.g. Geant4-DNA) have been extended towards the simulation of later physicochemical and chemical stages involving the formation and transport of reactive species (mainly free radicals) and their interaction with macromolecules such as DNA. The outcomes of such MC simulations are used, for instance, in radiation biology to evaluate the effects of ionizing radiation on the biological response and to provide information on the initial patterns of radiation-induced damage to biological systems (see the case study in Section~\ref{sec:Case_study_RT_multiscale}).

\textbf{Limitations and Challenges}: A general limitation of the MC-based approach for modeling particle transport in condensed matter systems is that it can only simulate particle transport in a continuous static medium at equilibrium. Thus, the MC-based approach does not allow explicit simulations of the post-irradiation dynamics of molecular media and related physical and chemical phenomena occurring under non-equilibrium conditions (see stage~(iii) in Figure~\ref{fig:MM_diagram}) that, in some cases, might even be extreme \cite{surdutovich2010shock}.

A multiscale simulation of irradiation-driven chemistry processes and post-irradiation non-equilibrium medium dynamics is achieved \cite{DeVera2020} by combining the outputs of track-structure MC simulations of particle transport with the irradiation-driven molecular dynamics (IDMD) approach \cite{Sushko_IS_AS_FEBID_2016} described in Section~\ref{sec:Methods_IDMD}. The practical realization of such an interface is discussed in Section~\ref{sec:Interlinks}.

A current challenge for the widespread exploitation of this MM interface is the need for fast and efficient MC-based calculations of radiation fields produced by different primary and secondary particles propagating through various condensed matter systems. This challenge is related to the limited set of materials and processes for which interaction cross sections are implemented in the popular track-structure MC codes and the necessity of developing new sets of cross section data for many different combinations of a target medium and radiation modality (e.g. those related to the case studies presented in Section~\ref{sec:Case_studies}).

\subsubsection{Analytical Methods for Modeling Particle Transport}
\label{sec:Methods_Analytical_transport}

The propagation of primary particles, as well as the formation and transport of radiation-induced secondary particles, can also be studied using \textbf{analytical methods}.
For charged particles, one can determine the range of the particle's propagation in a uniform medium as a function of the particle’s initial energy and its type (mass and charge state). The range of the particle’s propagation depends on the density of the medium and the particle’s stopping cross section; the latter can be calculated using different theoretical approaches discussed above in Section~\ref{sec:Methods_Cross_sections}. The effect of energy straggling (and the related variation in the particle propagation range) due to multiple ion scattering can be taken into account using some phenomenological approaches \cite{Surdutovich_AVS_2014_EPJD.68.353, Kundrat_2007_PMB.52.6813}.

An alternative to the MC approach in simulating the formation and transport of radiation-induced secondary particles is based on continuous transport theories, e.g. the diffusion equation, the diffusion-reaction equation, kinetic equations, etc. The diffusion equation-based approach is well applicable for describing the transport of low-energy secondary electrons (with energies below $\sim$50~eV), mainly produced in the Bragg peak region of ions' trajectories. For other scenarios concerning more energetic ions out of the Bragg peak region, the diffusion equation-based analysis can be extended to account for the contribution of more energetic $\delta$-electrons \cite{deVera2017radial}. When justified, these analytical methods can provide faster and reliable solutions for the dynamics of propagating secondary and higher generations of particles. The temporal limit of the particle transport domain shown in Figure~\ref{fig:MM_diagram} has been established using the diffusion equation-based method \cite{ES_AVS_2015_EPJD.69.193}.

The analytical approach for modeling particle transport based on the solution of the diffusion equation complements the MC-based approach discussed in Section~\ref{sec:Methods_MC_transport}, particularly concerning the transport of low-energy electrons. Indeed, physical models implemented in most track-structure MC codes do not describe (or describe with limited accuracy) the interactions of low-energy secondary particles with organic and, particularly, metallic materials. At the same time, the analytical approach for modeling particle transport based on the solution of the diffusion equation is suitable for describing the transport of low-energy particles (with typical kinetic energy below $\sim$10$^2$~eV). The reliability of the diffusion equation-based approach for modeling the low-energy electron transport in liquid water was examined through the comparison with track-structure MC simulations and good agreement between the two approaches has been reported \cite{Surdutovich_AVS_2014_EPJD.68.353, AVS2017nanoscaleIBCT}.

\textbf{Software Tools}: The analytical methods for modeling particle transport in condensed media are based on the analytical (and sometimes also numerical) solution of equations of continuous transport theories, which are typically programmed in custom-made computer codes written in popular programming languages (e.g. Fortran or C++) or developed using the widely used programs like Wolfram Mathematica or Matlab.

\textbf{Areas of Applications}: Analytical approaches for modeling particle transport have been widely used to (i) determine the location of the Bragg peak for different ions of different energies and the corresponding range of their propagation in water and other biologically relevant media \cite{deVera2017radial, AVS2017nanoscaleIBCT} and (ii) simulate the transport of secondary electrons produced around the tracks of ions propagating through water \cite{Surdutovich_AVS_2014_EPJD.68.353, ES_AVS_2015_EPJD.69.193, Bug_2012_EPJD.66.291} in connection to the studies of radiation-induced damage of biomolecular systems (see a case study in Section~\ref{sec:Intro_Ex_IBCT}). These methods can also be applied to analyze radiation effects in non-biological condensed materials in connection to radiation-induced material damage and degradation.

\textbf{Limitations and Challenges}: Similar to the MC-based particle transport codes, the analytical approach can be used to provide input data for a MM interface with IDMD (see Section~\ref{sec:Interlinks}). Therefore, the associated challenge concerns a more widespread utilization of the analytical methods for calculating radiation fields created by primary and secondary particles for a broad range of systems and irradiation conditions, including those discussed in Section~\ref{sec:Case_studies}.

\subsubsection{Relativistic Molecular Dynamics}
\label{sec:Methods_RelativisticMD}

The \textbf{relativistic molecular dynamics (MD)} approach \cite{RelMD_2013_JCompPhys.252.404} enables simulations of the propagation of different energetic particles through various condensed matter systems, see refs~\citenum{Channeling_book, RelMD_2021_EPJD.75.107_review} and references therein. Using this approach, one can simulate the transport of negatively and positively charged, light and heavy projectiles propagating at relativistic and ultra-relativistic velocities, i.e. when a projectile's speed is comparable to the speed of light $c$. Due to the very high velocity of such projectiles, their dynamics takes place on typical time scales of $\sim 10^0 - 10^2$~fs, see the time domain of stage~(ii) in Figure~\ref{fig:MM_diagram} and the discussion in Section~\ref{sec:MM_key-definitions}.

Particles of such high energies move nearly classically and experience predominantly elastic collisions with atoms of the medium, which can be treated using classical relativistic theory. The relativistic MD is based on solving the equations of relativistic particle motion in a medium, which describe the classical motion of a particle in the electrostatic field of the medium atoms \cite{RelMD_2013_JCompPhys.252.404}.

The relativistic MD framework can also treat all the relevant quantum processes, including corrections due to the quantum scattering, ionization processes, photon emission, and recoil effect by emission of energetic photons. The current implementation of this methodology accounts for random events of inelastic scattering of a relativistic projectile from individual atoms of the medium that lead to quantum processes, such as atomic excitation or ionization and a random change in the direction of the particle's velocity \cite{MBNTutorials_2023_RelMD}. Considering that such events are random, fast and local, they are incorporated into the classical MD framework according to their probabilities. The implemented approach is similar to the one used in irradiation-driven molecular dynamics (see Section~\ref{sec:Methods_IDMD}). The probability of energy transfer due to ionizing collisions is calculated based on quantum mechanics \cite{Korol_2001_JPG.27.95}.

For ultra-relativistic projectiles, the radiative energy losses prevail over the losses due to the ionizing collisions \cite{LL4}. At energies above several tens of GeV, the radiation damping, i.e. the process of a gradual decrease in the particle's energy due to the emission of electromagnetic radiation, must be accounted for an accurate quantitative analysis of the projectile motion. Recent algorithmic implementations \cite{Sushko_2023_NIMB.535.117} have incorporated the radiative reaction force into the relativistic MD framework.

Due to the need to solve non-linear equations of motion with very high precision, the relativistic MD approach implies the use of a very small integration time step, which is typically several orders of magnitude smaller than the integration time steps used in ``conventional'' (non-relativistic) classical MD (see Section~\ref{sec:Methods_classicalMD}).

The dedicated computer algorithms enable simulations of particle dynamics on macroscopically large distances and radiation emission by propagating projectiles with atomistic accuracy for a wide range of condensed matter systems. This is achieved by choosing the interaction potential between the projectile particle and a target medium and utilizing the so-called ``dynamic simulation box'' \cite{RelMD_2013_JCompPhys.252.404}. Further details on the practical realization of the relativistic MD methodology for the atomistic simulation of particle propagation in macroscopically large media are given in Section~\ref{sec:Interlinks}.

\textbf{Software Tools}: The relativistic MD approach is a unique implementation in the MBN Explorer software package \cite{Solovyov_2012_JCC_MBNExplorer}.

\textbf{Areas of Application}: Relativistic MD can be utilized to study the dynamics of relativistic particles propagating through a broad range of molecular and condensed matter systems that can be simulated employing MBN Explorer, including crystals, amorphous bodies, nanostructured materials, and biological media. It can also be utilized to simulate the elastic collisions of energetic electrons with atoms of a deposit occurring in (scanning) transmission electron microscopy experiments.

Over the last years, the relativistic MD approach has been extensively applied to simulate the propagation of ultra-relativistic charged particles (within the sub-GeV up to ten GeV energy range) in oriented crystals accompanied by emission of intensive radiation. A comprehensive description of the case studies related to modeling the propagation of particles in straight, bent and periodically bent oriented crystals (including channeling phenomenon, multiple scattering, volume reflection, etc.) and photon emission are presented in a review article \cite{RelMD_2021_EPJD.75.107_review} and books \cite{Channeling_book, NovelLSs_Springer_book, DySoN_book_Springer_2022}.

\textbf{Limitations and Challenges}: The calculation of physically relevant characteristics of particle propagation through condensed media requires a rigorous statistical analysis of the particle trajectories and photon emission spectra, which implies carrying out a significant number of independent relativistic MD simulations, typically on the order of $\sim$10$^4$. The associated challenge is related to the development of efficient tools and computer scripts for the execution of such a large number of simulations and their statistical analysis.

One of the scientific challenges that can be addressed through the relativistic MD approach is the atomistic simulations of particle propagation in complex media (such as oriented crystals of different shapes, mosaic and granular crystals, or crystalline media under mechanical stress) and the analysis of physical processes occurring therein. Another challenge that can be addressed in future studies using relativistic MD is the analysis of secondary effects due to the formation of defects in irradiated materials on the characteristics of particle propagation and the emitted radiation.

\subsection{Molecular Transformations, Nonequilibrium Chemistry Processes, and Irradiated Medium Dynamics}
\label{sec:Methods_Nonequil_Chem}

\subsubsection{Classical Molecular Dynamics}
\label{sec:Methods_classicalMD}

The \textbf{classical molecular dynamics (MD)} approach occupies an essential niche on a MM time-space diagram between more accurate but computationally expensive \textit{ab initio} and DFT calculations (Section~\ref{sec:Methods_Quantum-proc}) and coarse-grained mesoscale models (Section~\ref{sec:Methods_large-scale}). It represents a powerful tool that can provide insights into system's nanoscale structural features and its thermal, mechanical and other properties employing advanced computer simulations \cite{MBNbook_Springer_2017}.

The concept of classical MD relies on the Born-Oppenheimer approximation \cite{Born_Oppenheimer_1927}, which justifies the separation of slow ionic and fast electronic degrees of freedom within a molecular system. Within the classical MD framework, the time evolution of a many-atom system is described through the integration of the classical coupled equations of motion, where interaction potentials or force fields acting between atoms are defined. Parameters of the force fields are usually derived from \textit{ab initio} studies of systems containing a much smaller number of atoms or fitted to experimental data.

Given the initial coordinates and velocities of the atoms in a system, the subsequent motion of individual atoms is described either by deterministic Newtonian dynamics or by Langevin-type stochastic dynamics. This dynamics corresponds to different types of thermodynamic ensembles characterized by the control of specific thermodynamic quantities. Most commonly used are (i) the microcanonical ($NVE$) ensemble, which implies that the number of atoms in the system ($N$), the system's volume ($V$), and the system's total energy ($E$) are constant during a simulation and (ii) a canonical ($NVT$) ensemble that describes the system’s dynamics in thermal equilibrium with a heat bath maintained at some fixed temperature $T$. The concepts of canonical and microcanonical ensembles are described in many textbooks on statistical mechanics \cite{LL5}, and thus we omit their further discussion here.

The molecular mechanics approach usually refers to running MD simulations with a specific force field, which describes the interatomic interactions in a molecular system through a parametric phenomenological potential that relies on the network of chemical bonds in the system. This network defines the so-called molecular topology, i.e., a set of rules that impose constraints on the system in order to maintain the topological structure of the system. The most widely known MM force fields are CHARMM \cite{CHARMM_MacKerell_1998_JPCB.102.3586, MacKerell_2004_JCC.25.1400} and AMBER \cite{Cornell_1995_JACS.117.5179}.

Several classical interatomic force fields have been developed over the past decades to model chemical transformations in molecular systems, particularly in carbon-based nanosystems such as graphene, fullerenes and carbon nanotubes. The \textbf{Reactive Empirical Bond-Order (REBO)} \cite{REBO_Tersoff_1988_PRL.61.2879, REBO_Tersoff_1988_PRB.37.6991, REBO_Brenner_1990_PRB.42.9458} and \textbf{Adaptive Intermolecular Reactive Empirical Bond-Order (AIREBO)} \cite{AIREBO_Stuart_2000_JCP.112.6472} potentials are based on the concept of bond order to represent the forces between interacting atoms in a system. \textbf{REBO-type potentials} permit calculation of the potential energy of covalent bonds and the associated interatomic forces. In this approach, the total potential energy of a system is modeled as a sum of nearest-neighbor pair interactions, which depend on the distance between atoms and their local atomic environment. A parametrized bond order function is used to describe chemical pair-bonded interactions.

The earliest formulation and parametrization of REBO for carbon systems \cite{REBO_Tersoff_1988_PRL.61.2879, REBO_Tersoff_1988_PRB.37.6991} enabled single-, double- and triple-bond energies in carbon structures such as hydrocarbons and diamond crystals to be described. In 1990, these Tersoff potential functions were extended for radical and conjugated hydrocarbon bonds by introducing two additional terms into the bond order function \cite{REBO_Brenner_1990_PRB.42.9458}. Compared to other classical force fields, REBO potentials are typically less time-consuming since only the 1$^{\textrm{st}}$- and 2$^{\textrm{nd}}$-nearest-neighbor interatomic interactions are considered. This computational efficiency is highly beneficial for large-scale atomic simulations (containing up to $\sim$10$^6$ atoms).

A second-generation REBO potential energy expression for solid carbon and hydrocarbon molecules was published in 2002 \cite{REBO2_Brenner_2002_JPCM.14.783}. This potential can describe covalent bond breaking and formation with associated changes in atomic hybridization within a classical potential, enabling the modeling of chemical transformations in large many-atom systems. This potential contains improved analytic functions to model interatomic interactions and includes an extended parameter database, as compared to the earlier version \cite{REBO_Brenner_1990_PRB.42.9458}. The improved REBO potential therefore permits a significantly more reliable description of bond energies, lengths, and force constants for hydrocarbon molecules, as well as elastic properties, interstitial defect energies, and surface energies for a diamond crystal.

In an extension of the Brenner potential, the so-called \textbf{AIREBO potential} \cite{AIREBO_Stuart_2000_JCP.112.6472}, the repulsive and attractive pair interaction functions of the original REBO potential are modified to fit bond properties, and the long-range atomic interactions and single bond torsional interactions are included.

\begin{sloppypar}
\textbf{Software Tools}: The classical MD and molecular mechanics approaches have been widely used throughout the past decades and have been implemented in a large number of well-established computational packages, including AMBER \cite{AMBER_program}, CHARMM \cite{CHARMM_program}, GROMACS \cite{GROMACS_program}, LAMMPS \cite{LAMMPS_program}, NAMD \cite{NAMD_program}, MBN Explorer \cite{Solovyov_2012_JCC_MBNExplorer}, MULTICOMP \cite{MULTICOMP_Akhukov_2023}, MMSTB Tool Set \cite{MMTSB_2004_Feig}, and Materials Studio \cite{MaterialsStudio_Shankar}.
\end{sloppypar}

\textbf{Areas of Application}: The classical MD technique is widely used in many research areas ranging from atomic cluster physics to materials science and biophysics \cite{Jellinek_At_Mol_Clusters_book, MBNbook_Springer_2017, Saito_CondensMatter_book, Mesirov_Biomol_Struct_Dyn_book}. The molecular mechanics approach is used for modeling biomolecular systems, such as DNA, proteins, or carbohydrates, and organic systems.

Classical MD has been used to simulate the dynamics of multimillion-atom systems \cite{Zhao_2013_Nature.497.643, Vashishta_2006_JPCB.110.3727} on a picosecond timescale or smaller-size systems (containing up to $\sim$10$^4$ atoms) on the time scale up to 1 millisecond \cite{Shaw_2010_Science.330.341, Pierce_2012_JCTC.8.2997}; see stage~(iii) in Figure~\ref{fig:MM_diagram}. Large-scale MD simulations of $\sim$100 million-atom systems on a nanosecond timescale have been reported recently \cite{Lu_2021_CPC.259.107624, Jia_100Mio_MD_arXiv} using advanced parallelization techniques with modern graphics processor units and advanced supercomputer facilities.

\textbf{Limitations and Challenges}: Despite its numerous advantages, standard classical MD cannot simulate irradiation-driven processes. It does not account for the coupling of the system to the incident radiation, nor does it describe quantum transformations in the molecular system induced by the irradiation. In particular, the widely used CHARMM \cite{CHARMM_MacKerell_1998_JPCB.102.3586, MacKerell_2004_JCC.25.1400} and AMBER \cite{Cornell_1995_JACS.117.5179} force fields employ harmonic approximations to describe the interatomic interactions, thereby limiting their applicability to small deformations of the molecular system close to the thermal equilibrium. Thus, this class of potentials can reproduce structural and conformational changes in the system but is usually unsuitable for modeling chemical reactions. REBO-type potentials can model the processes of bond rupture and formation in a particular class of systems, such as carbon-based materials and hydrocarbons, but significant efforts are required to develop parameters of these force fields for other systems.

In order to study chemical, irradiation- and collision-induced processes occurring in a broad range of molecular systems, where rupture and formation of chemical bonds play an essential role, it is essential to go beyond the harmonic approximation to describe the physics of molecular dissociation more accurately. This challenge has been addressed through the realization of reactive MD and irradiation-driven MD approaches, described in Sections~\ref{sec:Methods_Chem_equilibrium} and \ref{sec:Methods_large-scale}, respectively.

\subsubsection{Hybrid Quantum Mechanics/Molecular Mechanics (QM/MM) Methods}
\label{sec:Methods_QM-MM}

The importance of both QM-based calculations and MD simulations in the study of complex molecular systems was recognized by Warshel and Levitt \cite{Warshel_Levitt_1976}, who developed a new computational method that later became known as \textbf{Quantum Mechanics/Molecular Mechanics (QM/MM)} \cite{QM-MM_Karplus_1990_JCC.11.700}. Due to the accuracy of \textit{ab initio} QM calculations and the speed of classical MD simulations, the hybrid QM/MM approach enables studies of chemically- and irradiation-driven processes in molecular systems.

In the QM/MM approach, the simulated molecular system is split into two sub-systems, where one sub-system of particular interest is treated using the QM formalism, while the effects of the surrounding sub-system are included through classical molecular-mechanics simulations. An example could be a part of a protein in an aqueous solution. If we take the rest of the protein or the water out of the simulation, a crucial part might be missing, and a MD simulation can leave out specific effects only observed in QM-based calculations. This can be addressed by combining the QM and MD methods so that the system’s potential energy consists of a sum of potentials for the different methods \cite{Warshel_Levitt_1976}.

Several approaches have been developed to calculate the energy of the combined QM/MM system \cite{Senn_Thiel_2009_AngewChem}. In the so-called ``subtractive scheme'' \cite{Maseras_1995_JCC.16.1170, ONIOM_Svensson_1996_JPC}, the energy of the entire system is calculated using a classical molecular mechanics force field, adding the energy of the QM sub-system calculated using a QM method and subtracting the energy of this sub-system calculated using molecular mechanics. In this scheme, the interaction between the two sub-systems is treated only at the molecular-mechanics level of theory (see below). A more widely used approach is the ``additive scheme'' \cite{Senn_Thiel_2009_AngewChem}, where the energy of the entire system is determined through (i) a QM calculation for the QM sub-system and (ii) a molecular mechanics calculation for the ``classical'' sub-system and a QM/MM interface energy. The advantage of the additive QM/MM scheme is that no molecular mechanics parameters for atoms in the QM sub-region are needed because those energy terms are calculated only by QM.

The interaction between QM and molecular-mechanics sub-systems is typically dominated by electrostatics, but evaluating the Coulomb interaction between the QM and the molecular-mechanics sub-systems is known to be time-consuming. Therefore, several approaches of different sophistication levels have been developed to handle the electrostatic interaction between the charge density in the QM region and the charge model used in the molecular mechanics region. These approaches are characterized by the extent of mutual polarization and classified, accordingly, as mechanical, electrostatic, and polarized embedding \cite{Cao_2018_FrontChem.6.89, Senn_Thiel_2009_AngewChem}. In the ``mechanical embedding'' scheme, which is typically considered the least accurate method, the electrostatic QM-MM interaction is calculated at the MM level \cite{Maseras_1995_JCC.16.1170, ONIOM_Svensson_1996_JPC}. In the ``electrostatic embedding'', the electrostatic QM-MM interaction is treated at the QM level by including a point-charge model (i.e., atomic partial MM charges) of the molecular mechanics sub-system in the QM calculations \cite{QM-MM_Karplus_1990_JCC.11.700, Dapprich_1999_JMolStruct.461.1}. Hence, the molecular mechanics sub-system polarizes the QM sub-system but not vice versa. In the ``polarized embedding'' scheme, both QM and molecular mechanics sub-systems are mutually and self-consistently polarized in the QM calculations \cite{Soderhjelm_2009_JCTC.5.649, Olsen_2010_JCTC.6.3721}.

The QM/MM approach enables higher computational efficiency than pure QM-based calculations for a system of a specific size. In the case of classical simulations, the computational cost of simulations scales as $O(N^2)$ (where $N$ is the number of atoms in the system) due to electrostatic interactions between a given particle and all other particles in the system. Special computational algorithms, such as the particle mesh Ewald (PME) method \cite{Darden_1993_JCP.98.10089, Herce_2007_JCP.126.124106}, enable a reduction in this scaling dependence below $O(N^2)$, e.g. to $O(N \log{N})$ in the case of the PME \cite{Frenkel_Smit_MD_book}, while \textit{ab initio} calculations typically scale as $O(N^3)$ or even higher, as discussed in Section~\ref{sec:Methods_Quantum-proc}.

\begin{sloppypar}
\textbf{Software Tools}: The hybrid QM/MM approach is implemented in several computer codes and software packages, designed specifically for this purpose, e.g. pDynamo \cite{pDynamo_Field_2008_JCTC.4.1151}, JANUS \cite{JANUS_Zhang_2019_JCTC.15.4362} or QMMM \cite{QMMM_program}. Other software packages provide an interface between the existing QM-based and MD-based software tools, e.g. VIKING \cite{VIKING_2020_ACSOmega.5.1254} and ChemShell \cite{ChemShell_2014_WIREsCMS.4.101}, which can be interfaced to a large number of QM and MD codes, or INAQS \cite{INAQS_program}, which provides an interface between the widely used GROMACS \cite{GROMACS_program} and Q-CHEM \cite{Q-Chem_Kong_2000_JCC.21.1532, Q-Chem_Epifanovsky_2021_JCP.155.084801} software tools. The latest version of GROMACS also provides an interface to the CP2K package for QM-based electronic structure calculations \cite{GROMACS_CP2K_interface}. Recently, a new suite for QM/MM simulations has been developed \cite{NAMD_QMMM_2018_NatMeth.15.351} by combining the widely used MD and visualization programs NAMD \cite{NAMD_program} and VMD \cite{VMD_program_1996} with the quantum chemistry packages ORCA \cite{ORCA_Neese_2020_JCP.152.224108} and MOPAC \cite{MOPAC_Stewart_1990}.
\end{sloppypar}

\textbf{Areas of Application}: As mentioned above in this section, the QM/MM approach is applicable to study a broad range of chemically- and irradiation-driven processes in molecular systems. Particular examples include:
\begin{enumerate}[label=(\roman*)]

\item
\textit{Modeling Reactive Events}: QM/MM methods can provide insights into the electronic structure changes, charge transfer, and bond breaking/formation involved in photochemical reactions, ionization and excitation processes induced by irradiation. They allow one to study reactions that involve both quantum and classical effects simultaneously;

\item
\textit{Electronic Excitations}: Irradiation can excite electrons to higher energy levels. QM/MM methods can help in understanding the nature of these electronic excitations, such as their energies, transition probabilities, and effects on molecular properties;

\item
\textit{Radiation Damage}: Materials exposed to irradiation can undergo structural changes due to the displacement of atoms caused by energetic particles. QM/MM can be used to investigate the initial steps of this damage process, such as the interactions between the irradiating particles and the material;

\item
\textit{Solvation Effects}: Many irradiation-driven processes occur in a solvent environment. QM/MM methods can account for solvation effects, allowing one to study how the surrounding solvent influences the electronic and structural changes induced by irradiation;

\item
\textit{Non-Adiabatic Effects}: Irradiation processes often involve non-adiabatic transitions, where electronic and nuclear motions are coupled. QM/MM methods can help explore these complex dynamics.
\end{enumerate}

Particular application areas of the QM/MM approach include the simulations of large-scale biomolecular systems like enzymes, especially when chemical transformations are expected \cite{Shaik_2010_ChemRev.110.949}. Another application area concerns solid-state calculations of bulk defects and surface reactions using a finite cluster embedding model \cite{ChemShell_2014_WIREsCMS.4.101}, where a spherical or hemispherical cluster is cut from a periodic MM model of the material and simulated using the QM methods. A finite representation of the solid in QM/MM calculations avoids spurious interactions between periodic images of the defects.

\textbf{Limitations and Challenges}: Most QM/MM methods developed so far are used for structure optimization calculations of the whole large-scale molecular system  \cite{Senn_Thiel_2009_AngewChem, Gomes_2012_AnnRepSectC.108.222}, and detailed spectroscopic calculations are only performed on the isolated QM part. Therefore, a description of dynamical processes implying electron emission from the QM part and its propagation into an explicit molecular mechanics environment is challenging for existing QM/MM codes. Only limited attempts have been made to simulate the time propagation of the QM sub-system when coupled with a molecular mechanics environment and including electronic emission \cite{Dinh_2010_PhysRep.485.43}, but in general, this research area is currently in its infancy and has not been widely studied using the existing QM/MM codes.

Another computational challenge for QM/MM methods concerns the realization of the ``polarized embedding'' scheme (see above), which is the most accurate approach nowadays for the description of the interaction between the QM and molecular mechanics sub-systems. This scheme requires a polarizable molecular mechanics force field for the ``classical'' sub-system \cite{Lopes_2009_TheorChemAcc.124.11} and QM software that can treat polarizabilities.

\subsubsection{\textit{Ab Initio} Molecular Dynamics}
\label{sec:Methods_ab-initio_MD}

\textbf{\textit{Ab initio} Molecular Dynamics (AIMD)} is a computational method that uses first principles to simulate the dynamics of atoms in a system \cite{Marx_AIMD_book, Tuckerman_2002_JPCM.14.R1297, Iftimie_2005_PNAS.102.6654}. In contrast to classical MD (see Section~\ref{sec:Methods_classicalMD}), AIMD does not rely on empirical potentials or force fields to describe the interactions between atoms. Instead, forces acting on atoms are calculated ``on the fly'' from the electronic structure of the system using quantum mechanics, and the dynamics of atoms is treated classically by solving Newtonian equations of motion. Thus, AIMD permits chemical bond breakage and formation events to occur and accounts for electronic polarization effects.

There are three approaches for combining QM-based electronic structure calculations with classical MD for the ionic subsystem \cite{Marx_AIMD_book}: Born-Oppenheimer MD, Ehrenfest MD, and Car-Parinello MD.

In \textbf{Born-Oppenheimer MD}, the electronic wavefunction is considered the ground-state adiabatic wavefunction, and the dynamics of the nuclei is treated classically on the ground-state electronic potential energy surface (PES) \cite{Barnett_1993_PRB.48.2081}. The latter is obtained by solving the \textit{time-independent} Schr\"{o}dinger equation for the electronic subsystem for a fixed set of nuclear positions at a particular instant. Calculating the gradient of the PES gives the forces acting on the nuclei.

The time-independent Schr\"{o}dinger equation implies that there is no explicit time dependence of the electronic system in the Born-Oppenheimer MD, and the electronic subsystem adiabatically follows the motion of the nuclei. The electronic structure problem is solved self-consistently at each MD step at a particular nuclear configuration. Since the ground-state electronic problem cannot be solved exactly, approximate electronic structure methods are employed, the most common one being the KS formulation of DFT (see Section~\ref{sec:Methods_DFT}). Since there is no electron dynamics involved in solving the Born-Oppenheimer MD, the corresponding equations of motions can be integrated on the time scale given by nuclear motion (typically of $\sim$1~fs) \cite{Barnett_1993_PRB.48.2081, Niklasson_2023_JCP.158.154105}.

The Born-Oppenheimer approximation, which postulated that nuclear motion is much slower than electronic motion, might not hold in certain cases, for instance, during specific chemical reactions or in systems with ultrafast dynamics. In these cases, non-adiabatic effects (where electronic and nuclear motions are strongly coupled) can become significant and might require more advanced methods beyond the Born-Oppenheimer approximation \cite{Worth_Cederbaum_beyond_BOMD}.

The \textbf{Ehrenfest MD} intrinsically accounts for the time dependence of the electronic structure as a consequence of nuclear motion. In this approach, the nuclei are treated as classical point particles subjected to an effective force created by the electronic subsystem (similarly to the Born-Oppenheimer MD), but the time evolution of the electronic subsystem is treated explicitly by the \textit{time-dependent} electronic Schr\"{o}dinger equation \cite{Ehrenfest_Methods_chapter, Handbook_CompChem}. If the energy gap between the electronic ground-state and the excited states is large, Ehrenfest MD tends to the ground-state Born-Oppenheimer MD.

In Ehrenfest MD, the time scale and, thus, the time step to integrate, simultaneously, equations of motion for the electronic and nuclear subsystems are dictated by the intrinsic dynamics of the electrons. Since electronic motion is much faster than nuclear motion, the largest possible time step is that which allows the electronic equations of motion to be integrated.

The \textbf{Car-Parinello MD} \cite{Car_Parrinello_MD, Hutter_2012_WIREsCMS.2.604} is a computational method that employs fictitious dynamics for the electronic subsystem that mimics the Born-Oppenheimer MD and ensures that the electronic subsystem remains close to its ground state. Contrary to the Born-Oppenheimer MD that treats the electronic structure problem within the time-independent Schr\"{o}dinger equation, Car-Parinello MD explicitly includes the electrons as active degrees of freedom. In this approach, an extended Lagrangian for the system is written, leading to a system of coupled equations of motion for ions and electrons. In this way, a costly self-consistent iterative minimization at each time step (as done in the Born-Oppenheimer MD) is not needed: after an initial standard electronic minimization, the fictitious dynamics of the electrons keeps them on the electronic ground state corresponding to each new ionic configuration.

The fictitious dynamics relies on using fictitious electron mass of the electrons, chosen small enough to avoid a significant energy transfer from the ionic to the electronic degrees of freedom. This small fictitious mass requires that the equations of motion are integrated using a smaller time step ($\sim 0.01 - 0.1$~fs) than the one commonly used in Born-Oppenheimer molecular dynamics ($\sim 0.1 - 1$~fs). According to the Car-Parrinello equations of motion, the nuclei evolve in time at a particular physical temperature (determined naturally by the kinetic energy of the nuclear subsystem), whereas a ``fictitious temperature'' is associated with the electronic degrees of freedom. A condition of the smallness of the fictious electronic temperature implies that the electronic subsystem is close to its instantaneous minimum energy, i.e. close to the exact Born-Oppenheimer PES.

\textbf{Software Tools}: There are several software packages available for performing AIMD simulations. Some of the most widely used packages include: ABINIT \cite{ABINIT_Gonze_2002}, CASTEP \cite{CASTEP_Clark_2005}, CP2K \cite{CP2K_Kuhne_2020_JCP.152.194103}, CPMD \cite{CPMD_program_Kloeffel_CPC}, Gaussian \cite{GAUSSIAN_g16}, GPAW \cite{GPAW_Ojanpera_2012_JCP}, LAMMPS \cite{LAMMPS_program}, NWChem \cite{NWChem_Valiev_2010_CPC.181.1477}, OCTOPUS \cite{OCTOPUS_Castro_2006_PSSB.243.2465, OCTOPUS_Andrade_2012_JPCM}, Quantum Espresso \cite{QE_Giannozzi_2009_JPCM.21.395502}, SIESTA \cite{SIESTA_Soler_2002_JPCM}, and VASP \cite{VASP_Kresse_1996_CompMatSci.6.15}.

\textbf{Areas of Application}: Applications of AIMD are widespread in different areas of physics, chemistry, and life sciences \cite{Andreoni_2005_ChemPhysChem, Marx_AIMD_book, Boero_2015_CPMD}. Particular examples include:
\begin{enumerate}[label=(\roman*)]

\item
applications in solid-state physics and chemistry, such as analysis of the structure, pressure-induced structural transformations, and short-time dynamics of various crystal structures; the diffusion of atoms in solids; and structural and mechanical properties of polymers and macromolecular materials;

\item
the dynamics of molecules, small NPs and atoms adsorbed on surfaces, including surface chemistry reactions;

\item
mechanochemistry, that is the mechanical activation of covalent bonds by externally applied forces;

\item
studies of molecular liquids and aqueous solutions;

\item
the dynamics of atomic clusters, fullerenes, and nanotubes;

\item
chemical reactions and transformations (e.g. reactive scattering of small molecules in the gas phase or thermal decomposition of molecular systems);

\item
photo-induced physics and chemistry processes (e.g. isomerization, intramolecular proton transfer, etc.);

\item
structure and picosecond dynamics of proteins, and many more.
\end{enumerate}

An extensive list of original references that reported the applications of AIMD to the listed and other problems can be found in the reviews \cite{Andreoni_2005_ChemPhysChem, Marx_AIMD_book, Boero_2015_CPMD}.

\textbf{Limitations and Challenges}: Key challenges that need to be addressed in the field of AIMD are (i) the accuracy of the electronic structure methods and (ii) the high computational costs of the calculations.

The accuracy of AIMD calculations is limited by the level of theory used to obtain the electronic structure (see Section~\ref{sec:Methods_Quantum-proc}). Since using very accurate first principles methods implies very long computational times, the DFT is commonly used to describe the electronic subsystem in AIMD. Progress in this direction has been achieved by developing new DFT functionals and/or novel, computationally efficient algorithms that allow higher-level electronic structure methods to be used instead of KS DFT.

The inherent high computational cost associated with the electronic structure calculations has limited the affordable temporal scales and system sizes in AIMD. This particularly concerns the Ehrenfest MD, which explicitly treats electron dynamics and therefore requires a very short simulation time step, typically on the order of 1--10~as \cite{GPAW_Ojanpera_2012_JCP}. The time step is determined by the maximum electronic frequencies and is about three orders of magnitude less than the time step required to follow the nuclei in a Born-Oppenheimer MD ($\sim$1~fs). As a result, Ehrenfest MD is typically applied for the systems containing, at most, several tens of atoms and evolving on the 10--100~fs timescale.

Born-Oppenheimer and Car-Parinello MD are typically applied to larger systems evolving on longer time scales. Two decades ago, these methods were applied to systems consisting of a few tens or hundreds of atoms, accessing timescales on the order of tens of picoseconds \cite{Tuckerman_2002_JPCM.14.R1297}. With the development of advanced high-performance computer platforms, the system size has increased to $\sim$10$^3$ atoms, and simulation times have reached hundreds of picoseconds \cite{Boero_2015_CPMD}. Yet, many phenomena require the consideration of larger temporal scales, which can only be achieved by using other methods, such as the Irradiation-Driven MD described in Section~\ref{sec:Methods_IDMD}.

\subsubsection{Reactive Molecular Dynamics}
\label{sec:Methods_RMD}

The \textbf{Reactive Molecular Dynamics (RMD)} approach enables classical simulations of chemical reactions and chemistry-based nanoscale phenomena on temporal and spatial scales inaccessible by pure \textit{ab initio} methods (see stage~(iii) in Figure~\ref{fig:MM_diagram}). QM-based methods are computationally expensive and limited to systems with a size of a few nanometers, see Figure~\ref{fig:MM_diagram} and the previous subsections. On the other hand, conventional (non-reactive) classical MD provides a powerful tool to simulate larger-size systems on long timescales, but this approach cannot simulate chemical reactions or can only do it in a very limited way for selected systems using REBO-type potentials (see Section~\ref{sec:Methods_classicalMD}). In order to circumvent this problem, RMD methods have been developed which use interatomic force fields capable of locally mimicking the quantum effects due to chemical reactions. Typical examples of such force fields include the \textbf{Reactive Force Field (ReaxFF)} \cite{ReaxFF_vanDuin_2001_JPCA.105.9396, ReaxFF_Senftle_2016} and \textbf{reactive CHARMM (rCHARMM)} \cite{Sushko2016_rCHARMM}.

\textbf{ReaxFF} is a bond order-based force field that enables MD simulations of reactive (bond breakage and formation) and non-reactive interactions between atoms of a system. Bond order is calculated directly from interatomic distances using the empirical formula described elsewhere \cite{Russo_2011_NIMB.269.1549}.

The interatomic ReaxFF potential is constructed as a sum of several bond-order-dependent and independent energy contributions, which include: (i) the energy associated with forming bonds between atoms, (ii) the energies associated with the so-called three-body valence angle strain and four-body torsional angle strain, (iii) an energy penalty preventing the over-coordination of atoms which is based on atomic valence rules, for instance, a stiff energy penalty is applied if a carbon atom forms more than four bonds, (iv) electrostatic and (v) van der Waals interactions calculated between all atoms, regardless of their connectivity and bond order.

ReaxFF has been parameterized and tested for a large number of different systems and processes, such as reactions involving hydrocarbons, alkoxysilane gelation, transition-metal-catalyzed nanotube formation, and other material applications such as Li-ion batteries, TiO$_2$, polymers, and high-energy materials \cite{ReaxFF_Senftle_2016}. However, it should be stressed that although ReaxFF parameter sets exist for many elements of the periodic table \cite{ReaxFF_Senftle_2016}, they have limited transferability and cannot be used in any combination. There are two major ``branches'' of ReaxFF parameter sets that are intra-transferable with one another: (i) the ``combustion'' branch, which describes high-temperature chemical reactions between gaseous products, and (ii) the ``aqueous'' branch, which describes the interaction of liquid water with metal oxides, clays and biological molecules \cite{ReaxFF_Senftle_2016}.

In contrast to widely used non-reactive molecular mechanics force fields CHARMM \cite{CHARMM_MacKerell_1998_JPCB.102.3586, MacKerell_2004_JCC.25.1400} and AMBER \cite{Cornell_1995_JACS.117.5179} (see Section~\ref{sec:Methods_classicalMD}) and the reactive CHARMM \cite{Sushko2016_rCHARMM} (see below), ReaxFF does not employ different atom types for atoms of the same chemical element. In order to simulate bond-breaking and formation processes while having only one single atom type for each element, ReaxFF is constructed as a relatively complex force field with many parameters \cite{ReaxFF_Manual}. Therefore, an extensive training set is necessary to cover the relevant chemical phase space, including bond and angle stretches, activation and reaction energies, equations of state, surface energies, etc. \cite{Shchygol_2019_JCTC.15.6799}.

The reactive MD approach implemented in the MBN Explorer software package \cite{Solovyov_2012_JCC_MBNExplorer} is an alternative (and more versatile) approach that accounts for fast and local chemical transformations during the MD of molecular or condensed matter systems. Such transformations have a quantum nature and occur probabilistically. In the RMD methodology realized in MBN Explorer, chemical transformations are incorporated into MD on the basis of the MC approach (see Section~\ref{sec:Methods_MC_transport}) by coupling the chemical transformations having the quantum nature with the classical dynamics of a molecular medium. Such chemical transformations may involve:
\begin{enumerate}[label=(\roman*)]
\item the breaking and formation of covalent bonds in a system,

\item change of atomic types and valences,

\item change of bond multiplicities,

\item redistribution of atomic partial charges,

\item change of interatomic interaction potentials,

\item formation of specific chemical products associated with specific fragmentation channels,

\item changes in the molecular topology of the system, etc.
\end{enumerate}

The reactive MD realized in MBN Explorer operates with the \textbf{reactive CHARMM (rCHARMM) force field} \cite{Sushko2016_rCHARMM} that enables the description of bond rupture events and the formation of new covalent bonds by chemically active atoms in the system, monitoring its chemical composition and changes in the system's topology that occur during its transformations. The chemically active atoms carry information about their charges and valences, interactions with other atoms in the system, and multiplicities of the bonds that can be formed with other reactive atoms in the system.

Compared to the ``standard'' (non-reactive) CHARMM force field \cite{CHARMM_MacKerell_1998_JPCB.102.3586, MacKerell_2004_JCC.25.1400} rCHARMM requires the specification of two additional parameters for the bonded interactions, namely the dissociation energy of a covalent bond and the cutoff radius for bond breaking or formation \cite{Sushko2016_rCHARMM}. By specifying these parameters, all molecular mechanics interactions (i.e., bonded, angular and dihedral interactions) vanish as the distance between interacting atoms increases. To permit the rupture of covalent bonds in the molecular mechanics force field, rCHARMM employs a Morse potential instead of a harmonic potential. The rupture of covalent bonds in the simulation automatically employs a modification of the potential functions for valence and dihedral angles \cite{Sushko2016_rCHARMM}. The input parameters for RMD can be elaborated based on many-body theory, DFT and TDDFT (see Section~\ref{sec:Methods_Quantum-proc}), or can be taken from the experiment.

\textbf{Software Tools}: ReaxFF potentials can be utilized for reactive MD simulations using the LAMMPS software \cite{LAMMPS_program} and are also available in the Amsterdam Modeling Suite \cite{ADF_program_JCC_2021} and Materials Studio \cite{MaterialsStudio_Shankar}. The rCHARMM force field methodology is implemented in the MBN Explorer software \cite{Solovyov_2012_JCC_MBNExplorer}.

\textbf{Areas of Application}: As discussed above in this section, a large number of different ReaxFF potentials have been developed over the last two decades for studying two major classes of physicochemical problems, namely (i) high-temperature chemical reactions in the gas phase and (ii) chemical interactions between liquid water and different metal, inorganic and biological systems \cite{ReaxFF_Senftle_2016}.

As an extension of the non-reactive CHARMM force field, rCHARMM is directly applicable to organic and biomolecular systems \cite{Friis_2020_JCC, Friis_2021_PRE}. Its combination with many other pairwise and many-body force fields enables simulations of thermally driven and post-irradiation chemical transformations chemical transformations in various molecular and condensed matter systems while monitoring their molecular composition and topology changes \cite{MBNbook_Springer_2017, DySoN_book_Springer_2022}. Due to its versatility and universality, the rCHARMM force field has been applied to study a broad range of processes and phenomena, including collision-induced structural transformations and fragmentation (in the gas phase or after the collision with surfaces) \cite{Verkhovtsev_2017_EPJD.71.212}; radiation-induced fragmentation of molecular and cluster systems in the gas phase and placed in molecular environments \cite{Friis_2020_JCC, Friis_2021_PRE, deVera_2019_EPJD.73.215, Andreides_2023_JPCA.127.3757}; thermally-driven and radiation-induced chemistry of condensed systems, particularly water \cite{Sushko2016_rCHARMM, deVera_2018_EPJD.72.147}; and radiation-induced surface chemistry processes lying in the core of modern nanofabrication techniques \cite{Sushko_IS_AS_FEBID_2016, DeVera2020, Prosvetov2022_PCCP} (see a case study in Section~\ref{sec:Intro_Ex_FEBID}). An extended description of the applications of this methodology can be found in recent reviews \cite{DySoN_book_Springer_2022, verkhovtsev2021irradiation}.

\textbf{Limitations and Challenges}: From the computational point of view, reactive MD has similar computational limitations as the standard classical MD in terms of the time scales that can be simulated and the system sizes that could be modeled, see stage~(iii) in Fig.~\ref{fig:MM_diagram}. The accurate simulation of covalent bond breakage and formation events calls for shorter integration time steps than in non-reactive MD, typically on the order of $0.1 - 0.5$~fs. A challenge for this methodology is related to the development of more systematic approaches for determining and validating input parameters (such as, for instance, bond dissociation energies for different covalent bonds in a studied system or chemical reaction rates) using the QM-based methods (many-body theory, DFT, TDDFT, QM/MM, AIMD) described earlier in this section or experimental data available in the literature.

\subsubsection{Irradiation-Driven Molecular Dynamics}
\label{sec:Methods_IDMD}

A crucial part of the multistage scenario of irradiation-driven chemistry is the analysis of intermediate time and spatial domains where the irradiated molecular or condensed matter system is far from equilibrium, see stage~(iii) in Figure~\ref{fig:MM_diagram}. Atomistic simulations of irradiation-driven transformations in complex molecular and condensed matter systems can be performed using the \textbf{Irradiation-Driven Molecular Dynamics (IDMD)} methodology \cite{Sushko_IS_AS_FEBID_2016}. This methodology accounts, probabilistically, for fast and local radiation-induced transformations (listed in Section~\ref{sec:MM_key-definitions}) occurring during classical MD of systems. Such quantum processes happen on the sub-femto- to femtosecond time scales (i.e. over the intervals comparable or smaller than a typical time step of classical MD simulations) and typically involve a relatively small number of atoms.

The probability of each quantum process to happen is equal to the product of the process cross section and the flux density of incident particles \cite{LL3}. The cross sections of collision processes can be obtained from: (i) \textit{ab initio} calculations performed employing dedicated codes (e.g. those listed in Section~\ref{sec:Methods_Cross_sections}), (ii) analytical estimates and models (see Section~\ref{sec:Methods_Cross_sections}), (iii) experiments, or (iv) atomic and molecular databases. The flux densities of incident particles are usually specific to the problem and the system considered.

The IDMD methodology accounts for all the major dissociative transformations of irradiated molecular and condensed matter systems listed in Section~\ref{sec:Methods_RMD}. The properties of atoms or molecules (energy, momentum, charge, valence, interaction potentials with other atoms in the system, etc.) involved in such quantum transformations are changed according to their final quantum states in the corresponding quantum processes. These transformations are simulated using the rCHARMM force field \cite{Sushko2016_rCHARMM} (see Section~\ref{sec:Methods_RMD}) implemented in MBN Explorer \cite{Solovyov_2012_JCC_MBNExplorer}. The possibility of combining rCHARMM with many other pairwise and many-body force fields makes the IDMD approach universal and applicable to many different molecular and condensed matter systems.

IDMD enables the analysis of rapid energy transfer events into fragmenting covalent bonds caused by quantum processes and the analysis of post-irradiation energy relaxation processes, occurring typically on a picosecond timescale and leading to chemical transformations. The energy transferred to the system through irradiation is absorbed by the involved electronic and ionic degrees of freedom, and chemically reactive sites (atoms, molecules, molecular sites) in the irradiated system are created \cite{Sushko_IS_AS_FEBID_2016}. These events lead to changes in the system’s molecular topology, the number and type of atomic and molecular species present in the system, and other characteristics that may affect the dynamic behavior and chemical transformations of the molecular system on longer timescales. The subsequent dynamics of the reactive sites is determined by the classical MD and the thermodynamic state of the system until the system undergoes further irradiation-driven quantum transformations. The chemically reactive sites may also be involved in the chemical reactions leading to the change of their molecular and reactive properties and, ultimately, the formation of chemically stable atomic and molecular species \cite{Sushko_IS_AS_FEBID_2016, DeVera2020, Prosvetov2021_BJN, Prosvetov2022_PCCP, DySoN_book_Springer_2022, verkhovtsev2021irradiation}.

In the absence of irradiation, only reactive transformations become possible at larger time scales, which can be simulated using RMD (see Section~\ref{sec:Methods_RMD}). As such, IDMD together with RMD allow a computational analysis of physicochemical processes occurring in the system coupled to radiation and post-irradiation system's dynamics on time and spatial scales far beyond the limits of quantum mechanics-based computational schemes, such as TDDFT (Section~\ref{sec:Methods_TDDFT}) or \textit{ab initio} MD (Section~\ref{sec:Methods_ab-initio_MD}).

IDMD relies on several input parameters such as bond dissociation energies, molecular fragmentation cross sections, the amount of energy transferred to the system upon irradiation, energy relaxation rate, and spatial region in which the energy is relaxed. These characteristics can be elaborated based on many-body theory, DFT, TDDFT, and collision theory (see Section~\ref{sec:Methods_Quantum-proc}), or can be taken from the experiment. The probabilities for irradiation-induced quantum transformations are defined according to a specific irradiation field imposed on the system, which can be determined using various particle transport theories, such as the MC \cite{DeVera2020} or diffusion-equation based approaches (see Sections~\ref{sec:Methods_MC_transport} and \ref{sec:Methods_Analytical_transport}). Such an analysis provides the spatial distribution of the energy transferred to the medium through irradiation.

Due to the small number of physically meaningful parameters of IDMD and the reactive molecular force fields, together with the much larger number of output characteristics accessible for simulations and analysis, provide unique possibilities for modeling irradiation-driven modifications of complex molecular and condensed matter systems beyond the capabilities of pure quantum or pure classical MD.

\textbf{Software Tools}: IDMD is a unique implementation available in the MBN Explorer software package \cite{Solovyov_2012_JCC_MBNExplorer}.

\begin{sloppypar}
\textbf{Areas of Applications}: Due to its universality, IDMD can be utilized to study irradiation-induced processes in various molecular systems (in the gas phase or embedded into molecular environments) and condensed matter systems discussed throughout this paper. Particular areas for the application of this methodology include the analysis of nanoscopic mechanisms of radiation-induced damage of biomolecular systems (see Section~\ref{sec:Intro_Ex_IBCT} and case studies \ref{sec:Case_study_RADAM_biomol} and \ref{sec:Case_study_DNAorigami} in Section~\ref{sec:Case_studies}) and the nanoscopic mechanisms of the nanostructure formation and growth due to the irradiation with focused beams of charged particles \cite{Sushko_IS_AS_FEBID_2016, DeVera2020, Prosvetov2021_BJN, Prosvetov2022_PCCP} (see Section~\ref{sec:Intro_Ex_FEBID}). An extended description of the applications of this methodology can be found in recent reviews \cite{DySoN_book_Springer_2022, verkhovtsev2021irradiation}.
\end{sloppypar}

\textbf{Limitations and Challenges}: From the computational point of view, IDMD simulations have similar limitations as the standard classical and reactive MD in terms of the system sizes and time scales that can be simulated. The simulation of covalent bond breakage and formation events calls for typical integration time steps on the order of $0.1-0.5$~fs, similar to those used in reactive MD simulations (see Section~\ref{sec:Methods_RMD}).

A current challenge for the IDMD approach is the development of systematic methods for determining input parameters, particularly flux densities of primary and secondary irradiating particles provided by the MC-based particle transport codes and the cross sections for irradiation-induced quantum transformations. The first problem has been addressed by developing an interlink \cite{DeVera2020} between the particle transport methods and IDMD and it can be elaborated further by a widespread realization of the developed interlink for popular MC codes (see Section~\ref{sec:Methods_MC_transport}). The second problem calls for an interlink between the QM-based (TDDFT, QM/MM, AIMD, etc.) codes with IDMD. The problem of interlinking IDMD with particle transport and QM-based methods is discussed in greater detail in Section~\ref{sec:Interlinks}.

\subsection{Computational Methods for Chemical Equilibrium Calculations}
\label{sec:Methods_Chem_equilibrium}

The concept of chemical equilibrium is highly relevant to radiation research, especially in situations where radiation interacts with matter, leading to chemical changes. While chemical equilibrium is often associated with reactions occurring at ordinary temperature and pressure conditions, its principles can be extended to situations involving radiation-induced reactions and processes. Radiation interaction with matter can cause ionization, excitation, and other forms of energy deposition. These interactions can lead to chemical changes in molecules and materials. While the conditions might not be ordinary, the principles of chemical equilibrium can still apply to the balance between various chemical species formed due to radiation.

Computational methods for chemical equilibrium calculations involve determining the concentrations of reactants and products species in a chemical reaction mixture when the rates of forward and reverse reactions are equal. These methods are essential for understanding and predicting the composition of chemical systems at equilibrium, see stage~(iv) in Figure~\ref{fig:MM_diagram}. This information can serve as input for the analysis of large-scale effects at various irradiation or post-irradiation conditions, see stage~(v) in Figure~\ref{fig:MM_diagram}.

Several approaches can be used for equilibrium calculations.

\textbf{Gibbs Free Energy Minimization} involves minimizing the Gibbs free energy of the entire system with respect to the concentrations of the species involved \cite{Ebel_2000_JCC.21.247, deNevers_PhysChemEquil_book}. It is based on the principle that at equilibrium, the Gibbs free energy is minimized \cite{LL5}. The problem of Gibbs free energy minimization considers a closed chemical system (made of a specific number of chemical elements) at certain constant thermodynamic characteristics that could include temperature $T$, pressure $P$, volume $V$, number of particles (species) $N$, and energy $E$. The number of species $N$ can change due to chemical reactions.

The equilibrium problem can be reformulated as a mathematical problem in which the unknowns are the non-negative number of species $N$ that minimize the system's Gibbs energy $G$, at the defined temperature $T$, pressure $P$, and number of chemical elements. Furthermore, at the chemical equilibrium the reaction rates of all the forward and corresponding reverse reactions are equal; this condition could be used to probe if a studied system is at equilibrium. The occurrence of phases with multiple species and the complex dependencies of the species dynamics on their number make the Gibbs energy minimization problem non-linear and, thus, challenging to solve.

\textbf{Reaction Quotient Calculation} focuses on calculating the reaction quotient $Q$, which is the ratio of the concentrations of reaction products to reactants in a chemical reaction mixture at any given time instance. Comparing $Q$ to the equilibrium constant $K$ provides information about whether or not the system is at equilibrium.

\textbf{Iterative Methods} involve iteratively adjusting the concentrations of species based on their reactions and their deviation from equilibrium conditions until the system reaches equilibrium. This can be done using various algorithms, such as the Newton-Raphson method or other numerical optimization techniques.

Many equilibrium calculations are based on the parameters listed in various \textbf{thermodynamic databases} \cite{Thermocalc_DB, Thereda_DB, Blasco_2017_ProcEarthPlanetSci} that contain information about thermodynamic characteristics for various chemical substances, the most important ones being enthalpy, entropy, and Gibbs free energy. Numerical values of these thermodynamic characteristics are collected as tables or calculated, e.g. using the CALPHAD methodology \cite{Kaufman_Calphad_book_1970, Comput_Thermodyn_Mater_book} for determining thermodynamic, kinetic, and other properties of multicomponent materials using the corresponding properties of pure elements and binary and ternary systems. These databases are used to calculate the equilibrium constants and other thermodynamic quantities needed for equilibrium calculations.

The choice of the appropriate method should be dictated by the complexity of the reaction system, the accuracy required, and the available computational resources. Equilibrium calculations can be relatively simple for simple reactions with known thermodynamic data or involve more sophisticated techniques for complex systems.

While the principles of chemical equilibrium can be applied in radiation research, the conditions and kinetics might differ significantly from standard thermodynamic equilibrium. Radiation-induced processes often involve rapid and dynamic changes due to the transient nature of the species formed (see the case study in Section~\ref{sec:Case_study_UltrafastRadChem}). Therefore, specialized models and techniques are required to accurately describe the chemical and physical processes involved in radiation interactions.

\textbf{Software Tools}: There are several software tools that could be used for the prediction of concentrations of species in complex chemical systems at a thermodynamic equilibrium, for instance, ChemEQL \cite{ChemEQL_program}, MINEQL+ \cite{MINEQL_program}, or Visual MINTEQ \cite{Visual_MintEQ}.

\textbf{Areas of Application}: The concept of chemical equilibrium can be applied to different problems related to radiation research, for example:
\begin{enumerate}[label=(\roman*)]

\item
\textit{Equilibrium in Radiolysis}: Radiolysis refers to the process of breaking chemical bonds through the action of radiation. The resulting species can include radicals, ions, and excited molecules. Even though the radiation field might not be in thermodynamic equilibrium, the reactions and concentrations of these species can reach a dynamic equilibrium where the rates of formation and consumption are balanced.

\item
\textit{Equilibrium in Radioactive Decay}: In radiochemistry, the decay of radioactive isotopes follows first-order kinetics, where the rate of decay is proportional to the concentration of the radioactive species. When the decay product is itself radioactive, equilibrium can be established between the decay of the parent isotope and the subsequent decay of the daughter isotope.

\item
\textit{Equilibrium in Dosimetry}: In radiation dosimetry, which involves measuring the absorbed dose of radiation, equilibrium conditions play a role in determining the energy deposition within a material. Depending on the radiation type and energy, there can be different degrees of equilibrium or disequilibrium between the energy deposition and its distribution.

\item
\textit{Chemical Changes in Radiobiology}: In radiation biology, the effects of radiation on living organisms are often studied. Radiation can induce chemical changes in biological molecules, including DNA damage and free radical formation. Understanding the chemical equilibrium between various reactive species can provide insights into biological responses to radiation (see the case studies in Sections~\ref{sec:Case_study_RT_multiscale} and \ref{sec:Case_study_DNA_damage_repair}).

\item
\textit{Irradiation of Materials}: Computational methods for chemical equilibrium calculations in the context of irradiation of materials use thermodynamic principles, chemical reaction kinetics, and radiation transport modeling to predict how materials respond to various forms of radiation, including ionizing and non-ionizing radiation. These methods consider factors such as temperature, pressure, and radiation dose to calculate the equilibrium composition of materials and provide insights into radiation-induced changes, allowing researchers to assess material degradation, design radiation-resistant materials, and ensure the safety and reliability of technologies exposed to radiation.

\item
\textit{Radiation Protection}: Computational methods for chemical equilibrium calculations play a pivotal role in radiation protection by aiding in the design and evaluation of shielding materials and safety measures. These methods predict how materials interact with ionizing radiation, enabling the selection or design of optimal shielding materials, the estimation of radiation doses, and optimizing shielding configurations. They also assist in understanding material aging and durability under irradiation, ensuring the ongoing effectiveness of radiation protection measures. Whether in medical facilities, nuclear power plants, aerospace applications, or radiological emergency response, these computational tools contribute to safeguarding individuals and the environment from the harmful effects caused by ionizing radiation while optimizing cost and efficiency.

\item
\textit{Radiation Waste}: Computational methods for chemical equilibrium calculations are instrumental in managing radiation waste by predicting the behavior of radioactive materials and their interactions with surrounding environments. These methods help assess the solubility, stability, and speciation of radionuclides in waste materials. They can simulate how these materials might leach or react with geological formations and engineered barriers in disposal sites. By understanding the chemical processes governing radionuclide behavior, these computational tools aid in the safe disposal and long-term containment of radioactive waste, which is crucial for minimizing environmental risks and ensuring public safety in nuclear waste management.
\end{enumerate}

\textbf{Limitations and Challenges}: There are several limitations and challenges of computational methods used for chemical equilibrium calculations concerning radiation research. In particular, coupling chemical reactions with radiation transport or nuclear reactions can lead to highly complex and nonlinear mathematical equations. Modeling such reactions or related processes may be difficult due to the limited data for reaction rates, interaction cross sections, and thermodynamic properties of species under extreme conditions. The lack of such information can introduce uncertainty into computational predictions. The complexity and non-linearity of the equations underlying chemical equilibrium calculations often require intensive computational resources, especially if high accuracy results are expected, relying on few assumptions. On the other hand, model simplifications could lead to an insufficient description of the fundamental principles underlying radiation-chemical interactions. In this regard, it is unfortunate that experimental data for validating simplified computational models used to model scenarios involving both radiation and chemical reactions may be scarce, particularly for extreme conditions relevant to radiation research (see the case study in Section~\ref{sec:Case_study_MSA_SWs}).

\subsection{Large-Scale Processes}
\label{sec:Methods_large-scale}

\subsubsection{Stochastic Dynamics}
\label{sec:Methods_StochasticDyn}

As discussed in Sections~\ref{sec:Intro} and \ref{sec:MM_key-definitions}, the exposure of molecular and condensed matter systems to radiation may result in the manifestation of the processes and phenomena that span over much larger spatial and longer temporal scales as compared to those typical for the initial quantum processes; see stages~(iv) and (v) in Figure~\ref{fig:MM_diagram}. Examples of such large-scale processes include radiobiological phenomena (e.g. radiation-induced damage and repair, cell death, etc.), structure formation and evolution, material aging, morphological transitions, and others (see Section~\ref{sec:Intro}).

Although developments of MD-based accelerated dynamics \cite{Voter_Hyperdynamics_PRL_1997, Miron_2003_JCP.119.6210} have successfully extended the simulation time scales to micro- and even (sub-)milliseconds, it remains computationally inefficient to employ atomistic MD for the simulations of large-scale processes and phenomena occurring on time scales from milliseconds to hours and longer. \textbf{Stochastic Dynamics (SD)} \cite{MBNbook_Springer_2017, Stochastic_2022_JCC.43.1442} provides a valuable approach for modeling large-scale processes (including those induced by irradiation), especially when dealing with systems that involve complex interactions, multiple particles, and dynamic changes. Stochastic simulations consider the inherent randomness and discrete nature of molecular interactions, making them suitable for capturing the behavior of molecular systems subjected to radiation-induced effects.

In general, SD applies to the description of processes having a probabilistic nature that occur in very different complex systems on different temporal and spatial scales. SD does not describe the specific details of dynamical processes occurring in large-scale systems, but the major transitions of a system to new states are described by a certain number of kinetic rates that can be attributed to the main physical processes driving the system transformation.

In SD, a stochastically moving system of interest can be discretized into a set of constituent particles. The ``particles'' are defined as the building blocks of the system that carry specific properties sufficient for the description of the spatial and temporal evolution of the whole system with the desirable level of detail and accuracy. The different properties of particles are reflected in their types, which are assigned to each particle in the system. The time evolution of a many-particle system is then modeled stepwise in time. Instead of solving dynamical equations of motion, as done in MD, a SD approach assumes that the system undergoes a structural transformation at each step of evolution with a certain probability. The new configuration of the system is then used as the starting point for the next evolution step. The system’s transformation is governed by several kinetic rates for processes involved in the transformation chosen according to the model considered. The relevant values for the kinetic rates used to determine the probabilities in SD could often be established using MD simulations \cite{Stochastic_2022_JCC.43.1442, Dick_2011_PRB.84.115408, Panshenskov_2014_JCC.35.1317}.

The concept of SD in its most general form \cite{Stochastic_2022_JCC.43.1442} is implemented in the MBN Explorer software package \cite{Solovyov_2012_JCC_MBNExplorer}. The developed SD algorithms generalize the earlier methodologies based on kinetic MC algorithms \cite{Dick_2011_PRB.84.115408, Panshenskov_2014_JCC.35.1317, Solovyov_2012_JCC_MBNExplorer} by accounting for many additional channels of stochasticity in the system and by increasing the complexity of the system.

Currently, the SD approach permits the simulation of the following kinetic processes that may occur with the particles in a system \cite{Stochastic_2022_JCC.43.1442}: (i) free diffusion, (ii) diffusion along the periphery of a bound group of particles, (iii) detachment of particles from each other, (iv) uptake of particles from the system, (v) addition of particles to the system, and (vi) change of particle’s type. However, particles may also experience more complicated transformations, including (vii) dissociation reactions, (viii) fusion reactions, and (ix) substitution reactions. In dissociation reactions, one particle dissociates into two or more new particles of different types. This process is general and relevant, e.g. for describing chemical reactions (including irradiation-induced dissociation processes). Fusion of particles requires the participation of two or more particles of a specific type. Such particles may perform the fusion reaction with the given association rate and be merged into one new particle occupying the corresponding adjacent grid cells. Substitution and replacement of particles provide another important class of processes that are possible to simulate within the SD framework. In this case, the interaction between a pair of neighboring particles leads to the formation of several new product particles \cite{Stochastic_2022_JCC.43.1442}.

\textbf{Software Tools}: The SD methodology has been fully implemented recently in the MBN Explorer software package \cite{Solovyov_2012_JCC_MBNExplorer}. It can be used to study different systems with sizes ranging from subatomic to macroscopic. The realized computational approach describes the dynamics of systems in which all their constituent elements can move stochastically and may experience transformations and chemical (including irradiation-induced) reactions. These include different diffusion modes, dissociation and attachment (decay, fission, and fusion), uptake and injections (creation and annihilation) processes, reactive transformations and particle type alteration. The system’s constituent elements may have different nature, scale properties, and a set of interactions with other components within the system that affect their SD.

It is also worth emphasizing that since the SD module is fully integrated into MBN Explorer, it can be accessed via the multi-purpose toolkit MBN Studio \cite{Sushko_2019_MBNStudio}. MBN Studio permits an intuitive setup of SD simulations, visualization and the analysis of the corresponding results. This is especially important in the cases when simulations involve a variable number of particles (e.g. as a result of particle injection, removal or annihilation) as many other alternative visualization programs (e.g. VMD \cite{VMD_program_1996} or PyMol \cite{PyMOL}) do not support this feature.

\textbf{Areas of Application}: The SD computational framework is applicable for modeling many stochastically moving systems of different nature, different dynamical processes occurring and manifesting at different temporal and spatial scales. One can emphasize several avenues where SD can be used for multiscale modeling of irradiation-driven processes and phenomena:
\begin{enumerate}[label=(\roman*)]

\item
\textit{Radiation Transport}: Stochastic models simulate random interactions of radiation particles (such as photons, electrons, ions, or neutrons) with atoms or molecules of the medium as they traverse matter (see Section~\ref{sec:Methods_MC_transport}). MC simulations, a common stochastic approach, provide insights into radiation transport phenomena in diverse settings, from optimizing shielding materials for nuclear reactors to simulating the dose distribution in radiation therapy for cancer treatment.

\item
\textit{Nuclear Decay}: Radioactive decay follows a probabilistic pattern, and stochastic models are used to predict the decay rates of radioactive isotopes over time. These models are integral to nuclear physics and radiological safety, helping estimate the remaining radioactivity of materials and assess nuclear waste storage requirements.

\item
\textit{Nuclear Astrophysics}: Stochastic models play a role in studying nuclear reactions in astrophysical environments, where statistical laws govern reactions involving low-abundance nuclei. These models help scientists understand the probabilistic processes that govern element formation in stars, providing insights into the synthesis of heavy elements and the evolution of stellar systems.

\item
\textit{Radiation-Induced Damage}: Stochastic simulations model the random generation and distribution of defects and damage caused by radiation in materials and biological systems. For example, in materials science, these models help predict the probability of structural damage in materials exposed to radiation, impacting the safety of nuclear facilities. In radiobiology, they provide insights into the stochastic nature of DNA damage and mutations caused by ionizing radiation (see the case studies in Sections~\ref{sec:Case_study_RT_multiscale} and \ref{sec:Case_study_DNA_damage_repair}), which is vital in understanding cancer risk and radiation therapy planning (see the case study in Section~\ref{sec:Case_study_IBCT}).

\item
\textit{Radiation Effects in Biological Systems}: In radiobiology and radiation oncology, stochastic models are applied to study the probabilistic nature of biological responses to radiation. These models consider the inherent variability in cellular and molecular processes (see Sections~\ref{sec:Case_study_RT_multiscale} and \ref{sec:Case_study_DNA_damage_repair}) when assessing the risk of radiation-induced cancer or planning radiation treatments. They help optimize radiation therapy by predicting tumor control probability and typical tissue complications.

\item
\textit{Cellular Repair Mechanisms}: Applying stochastic dynamics involves modeling and understanding the probabilistic nature of biological responses to DNA damage at the molecular and cellular levels (see the case studies in Sections~\ref{sec:Case_study_RT_multiscale} and \ref{sec:Case_study_DNA_damage_repair}). Stochastic models elucidate the random encounters between repair enzymes and damaged DNA, repair pathway choices influenced by stochastic interactions, and the variable cellular responses to DNA damage, including activation of repair pathways and cell fate decisions. These models are critical in radiation biology, aging, disease, and drug development, shedding light on the stochastic processes underlying genomic instability, cancer development, and the effectiveness of therapeutic interventions targeting DNA repair pathways, ultimately enhancing our comprehension of cellular repair mechanisms and their implications in various biological contexts.

\item
\textit{Materials Ageing}: SD may also be applied to describe materials aging, which can help understand the probabilistic nature of material deterioration and degradation processes. Stochastic models consider the random environmental conditions, fluctuations in stress, temperature, and variability in material properties that influence aging mechanisms like corrosion, fatigue, and creep. These models are instrumental in assessing the reliability, durability, and safety of materials and structures by predicting the probability of failure, estimating remaining useful life, and optimizing maintenance schedules. They also aid in material selection and design, considering the impact of stochastic behavior on performance, making them valuable in fields ranging from infrastructure maintenance and aerospace engineering to nanomaterials development and composite materials design.

\item
\textit{Environmental Radiology}: SD is used to model the dispersion and behavior of radioactive contaminants in environmental systems. These models assess the random processes governing radionuclides' transport, deposition, and distribution in air, water, soil, and biota. They are crucial for assessing the environmental impact and human exposure risks following nuclear accidents or during environmental monitoring and remediation efforts.
\end{enumerate}

For these and many other application areas, the SD approach establishes the final links into the chain of theoretical and computational methods and algorithms, enabling computational MM of dynamics of systems and processes from atomic up to mesoscopic and even macroscopic scales with the temporal scales relevant to such modes of motion; see stages~(iv) and (v) in Figure~\ref{fig:MM_diagram}.

\textbf{Limitations and Challenges}: As described above, SD-based models require, as input, the probabilities for specific processes, which should be selected according to the model considered. The relevant values for the kinetic rates used to determine such probabilities can be determined through MD simulations, making setting up an SD model a non-trivial and multiscale task. Moreover, such an SD model might be very specific for a particular problem, and different SD models should be developed and validated in connection to the different application areas described above. Setting up comprehensive and complete SD models might be a big challenge due to the limited data for required kinetic rates and other relevant characteristics (e.g. interaction cross sections, activation energies, diffusion coefficients, etc.).

\subsubsection{Continuous Medium Models and Macroscopic Theories}
\label{sec:Methods_Macroscopic}

Stochastic dynamics (see Section~\ref{sec:Methods_StochasticDyn}), while being a powerful and universal approach, can impose significant computational demands, especially when modeling radiation-induced phenomena at the nanoscopic level, and may struggle to predict rare events accurately. The integration of SD with continuum medium computational models (e.g. those that employ the Finite Element Methods (FEM) \cite{Liu_FEM_book_2014, Whiteley_FEM_book_2017}) and infinite condensed matter theories (e.g. thermodynamics, hydrodynamics, theory of elasticity, and others) offers a valuable tool for the analysis of various macroscopic characteristics of condensed matter systems and large-scale phenomena. The development of such tools requires establishing the interlinks between these theories and methods, on the one hand, and MD and SD simulations, on the other hand, interlinking stages~(iii), (iv) and (v) of the MM diagram shown in Figure~\ref{fig:MM_diagram}. A practical realization of such an interlink is discussed in Section~\ref{sec:Interlinks}.

Continuum models and theories disregard a discrete and smaller-scale (atomistic or mesoscopic) structure of a material, which is assumed to be continuously distributed throughout its volume. Modeling at this scale can predict such large-scale irradiation-induced processes as material degradation and decomposition, defect formation, crack propagation, and phase transitions, which are highly relevant for industrial and technological applications.


\textbf{Software tools}: Among the most popular and widely used software tools for finite element analysis are Abaqus \cite{FEM_ABAQUS_book_2017, Boulbes_FEM_Abaqus_2020}, ANSYS, COMSOL Multiphysics, Nastran and others.

\textbf{Areas of application}: Macroscopic models and theories have been utilized to describe many macroscopic characteristics of materials, including their structural behavior, heat and mass transport, chemical reaction kinetics, electromagnetic properties, and others.

\textbf{Limitations and challenges}: In order to increase the predictive capability of continuum models and their applicability, these models need to include processes which originate at all the stages~(i)-(iv) of the time-space diagram shown in Fig.~\ref{fig:MM_diagram}. However, such effects are often neglected in the existing continuum-scale models. Therefore, developing accurate and predictive models accounting for the true multiscale nature of condensed matter systems and materials as well as radiation-induced phenomena therein is a big challenge.

\section{How to Approach Computationally Multiscale Problems? Practical Realization of a Multiscale Approach}
\label{sec:Interlinks}

This section discusses the practical realization of a MM approach through interlinking different theoretical and computational methodologies that are used to study different stages of the multiscale scenario of radiation-induced processes (see Fig.~\ref{fig:MM_diagram}). Each stage of the multiscale scenario is treated with the theoretical methods describing dynamics of systems at the corresponding temporal and spatial scales, see Section~\ref{sec:Methods}. The interlinking of different methods permits going beyond the limitations of each particular methodology and paves the way toward a coherent MM approach that unifies the description of radiation-induced processes.

The following interlinks should be emphasized in connection to Fig.~\ref{fig:MM_diagram}:

\begin{itemize}
\item
Methodologies for the description of radiation-induced quantum processes with those for particle transport simulations -- interlink between stages~(i) and (ii);

\item
QM-based methods with classical MD -- interlink between stages~(i) and (iii);

\item
Methodologies for particle transport with IDMD -- interlink between stages~(ii) and (iii);

\item
MD with SD -- interlinks between stages (iii) and (iv); (iii) and (v);

\item
MD and SD with thermodynamics and other macroscopic theories -- interlinks between stages (iii) and (iv); (iii) and (v); (iv) and (v).
\end{itemize}

Significant efforts have been made for the development of interlinks between the QM-based methods and particle transport codes (see Section~\ref{sec:Interlinks_QM-transport_MC} below) as well as between QM and classical MD methods, mainly through the realization of the QM/MM method (see Sections~\ref{sec:Methods_QM-MM} and \ref{sec:Interlinks_QM-MD_QM-MM}). Other aforementioned interlinks between the stages indicated in Fig.~\ref{fig:MM_diagram} have been established over the past decade mainly through the MBN Explorer software package \cite{Solovyov_2012_JCC_MBNExplorer}.

\subsection{Interfacing Methodologies for Studying Radiation-Induced Quantum Processes and Particle Transport}
\label{sec:Interlinks_QM-transport}

\subsubsection{Monte Carlo-based Codes}
\label{sec:Interlinks_QM-transport_MC}

\textbf{Nature of the Interface}: Particles propagating through a medium experience multiple collisions with atoms or molecules of the medium, which occur in a stochastic manner. Each collision event is a quantum process that occurs on the atto- to femtosecond time scales depending on the particle type, collision velocity and the type of the quantum process. High-energy particles can travel macroscopically large distances in condensed matter (see the spatial extent of stage (ii), illustrated in Fig.~\ref{fig:MM_diagram}) before these particles lose all their energy and stop. An interface that links the QM-based description of single collision events with the probabilistic simulation of the propagation of particles in a condensed medium is therefore needed.

\textbf{Practical Realization}: As explained in Section~\ref{sec:Methods_MC_transport}, the simulation of particle propagation over large distances is realized by employing MC methods in various track-structure codes such as Geant4-DNA \cite{Incerti_2010_Geant4-DNA, Bernal_2015_PhysMed.31.861}, PARTRAC \cite{PARTRAC_Friedland_2011_MutatRes}, and KURBUC \cite{Uehara_1992_PMB.37.1841}. The realization is based on the concept of ballistic (free) particle motion in a medium between the particle's random collisions with atoms or molecules of the medium. MC-based particle transport simulations require information on the average density of the medium and the cross sections of particle collision processes with the medium atoms or molecules. The cross sections are typically calculated using different \textit{ab initio} and semiempirical methods described in Section~\ref{sec:Methods_Cross_sections}.

The methodology for interfacing stages (i) and (ii) has been established long ago \cite{Practical_Manual_MC_1959}. In the last several decades, it has been advanced through modern computational methods and utilization of high-performance computers. It was widely utilized for simulations involving biologically relevant media, see e.g. refs \citenum{Garcia_Fuss_RADAM_BiomolSyst, AVS2017nanoscaleIBCT, Hahn_2023_JPhycCommun.7.042001} and references therein. Further advances included an accurate accounting for low-energy collision processes, e.g. down to $\sim$10~eV for projectile electrons \cite{Bernal_2015_PhysMed.31.861, Sakata_2016_JAP.120.244901}.

\textbf{Application Areas}: This interface of stages (i) and (ii) has been widely exploited in numerous case studies in dosimetry, medical physics, radiation protection, particle and accelerator physics, radiation-induced material damage, and many others, see Section~\ref{sec:Methods_MC_transport}.

\subsubsection{Relativistic Molecular Dynamics}
\label{sec:Interlinks_QM-transport_RelMD}

\textbf{Nature of the Interface}: Projectile particles of relativistic and ultra-relativistic energies (when a projectile's speed is comparable to the speed of light $c$) move nearly classically and experience predominantly elastic collisions with atoms of the medium. During their motion the propagating particles radiate photons and steadily lose their energy. Quantum processes and phenomena involving the particles (e.g. ionization, recoil by emission of energetic photons, electron-positron pair creation, etc.) may also take place. Such events occur during the particles' propagation in a stochastic manner. All these processes can be treated within the relativistic MD approach over the macroscopically large trajectories of the propagating particles, although the aforementioned quantum processes and phenomena occur on the temporal and spatial scales typical for them. Thus, an interface that links the quantum events and any atomistic level details with the relativistic MD simulation of the propagation of particles in a condensed medium over macroscopically large distances is needed.

\textbf{Practical Realization}: An interface of stages (i) and (ii) within the relativistic MD \cite{RelMD_2013_JCompPhys.252.404} framework is realized in MBN Explorer \cite{Solovyov_2012_JCC_MBNExplorer}. This implementation enables simulations of relativistic charged particles propagation in different media over macroscopically large distances with the atomistic level of details, see Section~\ref{sec:Methods_RelativisticMD}. The practical realization of this interface is based on using a dynamic simulation box \cite{RelMD_2013_JCompPhys.252.404}. In this approach, the simulation box `moves' along the simulated relativistic particle trajectory, thus allowing the ``on-the-fly'', accounting for the interactions of the particle with the medium atoms surrounding it at each instant while integrating the equations of the particle’s motion. The particle propagates within the simulation box, interacting with atoms defined through a cutoff distance. Once the projectile approaches the edge of the simulation box, a new box is generated, centered at the position of the projectile. To avoid spurious changes in the force acting on the projectile, the atoms located at the intersection of the old and the new simulation boxes are included in the latter, while atoms in the remaining part of the new simulation box are generated anew. The procedure is repeated as many times as necessary to propagate the projectile through a medium, which can be of macroscopically large sizes.

\textbf{Application Areas}: The interface of stages~(i) and (ii) realized in relativistic MD has been used in numerous case studies devoted to the analysis of propagation of different high-energy particles through various condensed matter systems, especially oriented crystals of different geometry (linear, bent and periodically bent), see refs \citenum{RelMD_2021_EPJD.75.107_review, Channeling_book} and references therein. Also, the photon emission processes (e.g. channeling radiation, crystalline undulator radiation) by ultra-relativistic particles have been in focus of these studies \cite{AVK_AVS_2020_EPJD.74.201_review, NovelLSs_Springer_book}. This research resulted in the practical realization of the novel gamma-ray crystal-based light sources within the currently running European project TECHNO-CLS \cite{TECHNO-CLS_website}, see the case study in Section~\ref{sec:Case_study_CLSs} for further details and references.

\subsection{Interfacing Quantum Mechanics (QM)-based Methods with Classical Molecular Dynamics (MD)}
\label{sec:Interlinks_QM-MD}

There are several theoretical and computational approaches for interlinking the QM based descriptions of quantum processes occurring in molecular and condensed matter systems [stage~(i)] with the classical description of the system's dynamics on the temporal and spatial scales beyond the capabilities of pure QM methods [stage~(iii)]. The interlink between the theoretical and computational approaches developed for the stages (i) and (iii) has been established via:
\begin{itemize}
\item
determining the parameters of the classical force field through the outcomes of QM-based calculations (Section~\ref{sec:Interlinks_QM-MD_ground-state}),

\item
the hybrid QM/MM approach (Section~\ref{sec:Interlinks_QM-MD_QM-MM}),

\item
reactive MD (Section~\ref{sec:Interlinks_QM-MD_RMD}), and

\item
irradiation-driven MD (Section~\ref{sec:Interlinks_QM-MD_IDMD}).
\end{itemize}

\subsubsection{Ground-state Dynamics}
\label{sec:Interlinks_QM-MD_ground-state}

\textbf{Nature of the Interface}: In the classical MD approach (see Section~\ref{sec:Methods_classicalMD}), one typically uses a force field or a classical potential energy function to approximate the ground-state potential energy of a quantum many-body system as a function of nuclear coordinates. Forces acting on atoms are derived in MD simulations as the negative gradient of the potential energy with respect to nuclear coordinates. The interlink between stages~(i) and (iii) is realized via determining the parameters of the force field through the outcomes of QM-based calculations.

\textbf{Practical Realization}: Parameters of pairwise and many-body interatomic potentials are commonly determined by fitting the characteristics obtained from \textit{ab initio} and DFT calculations, such as equilibrium bond lengths, lattice constants, equilibrium geometries of stable crystalline phases, cohesive energies, energies of defect formation, elastic moduli, vibrational frequencies, etc. QM methods are also commonly used to determine parameters of the bonded interactions in molecular mechanics force fields such as CHARMM \cite{CHARMM_MacKerell_1998_JPCB.102.3586, MacKerell_2004_JCC.25.1400}, AMBER \cite{Cornell_1995_JACS.117.5179}, and others.

Furthermore, different charge models (e.g. Mulliken charges, charges determined through the natural population analysis, or from the fitting of the electrostatic potential) available in many QM-based software tools can be used to parameterize the partial atomic charges in a molecular system to model electrostatic interactions in MD simulations \cite{Han_2021_JCTC.17.889, Bleiziffer_2018_JChemInfModel.58.579}.

A QM-MD interface has been successfully realized in a web platform VIKING \cite{VIKING_2020_ACSOmega.5.1254} that provides a convenient possibility for such multiscale computations on supercomputers. VIKING allows the integration of several popular quantum chemistry software packages (e.g. Gaussian \cite{GAUSSIAN_g16} and ORCA \cite{ORCA_Neese_2020_JCP.152.224108}) and software for classical MD simulations (NAMD \cite{NAMD_program} and MBN Explorer \cite{Solovyov_2012_JCC_MBNExplorer}) into a single platform that provides tools for setting up simulations, data analysis and visualization and can be used to model a broad range of molecular processes occurring at different scales. The computational tasks that can be solved through VIKING include, in particular, non-reactive MD simulations, reactive and irradiation-driven MD simulations using MBN Explorer, various types of quantum chemistry calculations (specifically, geometry optimization, electronic properties, NMR properties, as well as infrared, Raman, NMR and circular dichroism spectroscopy calculations), and spin chemistry.

\textbf{Application Areas}: This interlink between the QM-based and classical MD methods has been routinely realized over the decades to determine the parameters of the classical force fields for various types of molecular and condensed matter systems \cite{Mueser_2023_AdvPhysX.8.2093129, Solovyov_2012_JCC_MBNExplorer}. However, the overwhelming majority of classical interatomic potentials have been designed to reproduce only the ground-state equilibrium properties of a system. While matching the results of \textit{ab initio} calculations of ground-state parameters, these force fields often poorly describe highly excited vibrational states when the studied system is far from the potential energy minimum. MD simulations of non-equilibrium processes require other realizations of the interlink between the stages~(i) and (iii), described in the following sub-sections.

\subsubsection{Interface Between QM and MD Realized in Hybrid QM/MM}
\label{sec:Interlinks_QM-MD_QM-MM}

\textbf{Nature of the Interface}: The investigation of quantum processes within large-scale molecular and condensed matter systems necessitates the use of classical MD for treating the nuclear degrees of freedom (see Section~\ref{sec:Methods_classicalMD}). At the same time, a QM-based description is required for simulating the electron dynamics in those parts of the system where the quantum processes or chemical reactions take place. The classical molecular mechanics force field approach is used for the description of other parts of the molecular system that are not directly involved in the quantum processes or chemical reactions. The QM/MM approach thus provides an important interlink between the stages~(i) and (iii) in Fig.~\ref{fig:MM_diagram}.

\textbf{Practical Realization}: The interlink has been realized in different computer codes and software packages for QM/MM simulations, some of which are listed in Section~\ref{sec:Methods_QM-MM}. QM/MM calculations are often performed to optimize the ground-state geometry of a system.  The excited states in the QM region can then be treated with a broad set of QM methods, e.g. TDDFT or post-HF methods (see Section~\ref{sec:Methods_Quantum-proc}). A chemically important part of the system, where bond breaking/formation or electronic excitation occurs is treated using QM, while the remaining part of the system is modeled with a classical molecular mechanics force field.

\begin{sloppypar}
A recently developed QM/MM computational tool \cite{NAMD_QMMM_2018_NatMeth.15.351} combining the widely used NAMD \cite{NAMD_program} and VMD \cite{VMD_program_1996} programs with the quantum chemistry packages ORCA \cite{ORCA_Neese_2020_JCP.152.224108} and MOPAC \cite{MOPAC_Stewart_1990} allows multiple independent QM regions in the same molecular system to be set up. This approach enables the study of molecular complexes containing several active sites simultaneously calculated at a QM level. Each active site is calculated independently of others, keeping the low computational cost at an overall high efficiency of the simulation since all QM regions are calculated in parallel.
\end{sloppypar}

The spatial and temporal scales on which the QM/MM approach is applicable are defined by the size of the QM sub-system and the choice of the QM-based method (which depends on the problem of interest). While \textit{ab initio} or DFT methods are generally more reliable than semi-empirical QM methods (such as Density Functional based Tight Binding -- DFTB \cite{Elstner_1998_PRB.58.7260, Hourahine_2020_JCP.152.124101}), they are computationally more expensive. Therefore, currently, \textit{ab initio} or DFT-based QM/MM simulations are typically limited to tens to hundreds of picoseconds \cite{Vennelakanti_2022_CurrOpinStructBiol, Kubar_2023_AnnuRevBiophys}, which are usually too short for reliable computation of equilibrium (e.g. free energy) or dynamical properties of a complex many-body system. Another practical issue concerns the appropriate size of the QM region \cite{Kulik_2016_JPCB.120.11381, Jindal_2016_JPCB.120.9913, Das_2018_JCTC.13.1695}. It has become possible to conduct QM/MM simulations for QM sub-systems containing $10^2 - 10^3$ atoms \cite{Vennelakanti_2022_CurrOpinStructBiol}. However, the computational cost associated with such large QM regions limits the temporal scale for simulations of the system’s dynamics.

\textbf{Application Areas}: The interlink between stages~(i) and (iii) realized in the QM/MM approach has been used to simulate reactive events (such as charge transfer, bond breaking/formation involved in photochemical reactions, as well as ionization and excitation processes induced by irradiation) in various molecular systems, particularly large-scale biomolecular systems. More examples of the application of this interlink can be found in Section~\ref{sec:Methods_QM-MM}.

\subsubsection{Reactive Molecular Dynamics}
\label{sec:Interlinks_QM-MD_RMD}

\textbf{Nature of the Interface}: Chemical transformations involving atoms and molecules have the quantum nature. These are fast and local processes that occur probabilistically during characteristic times that are comparable with   the duration of a typical time step in classical MD simulations. Usually, only a few atoms are involved in such transformations that may lead to the creation of chemically active sites or species able to participate in further chemical reactions. The fast and local chemical transformations can be incorporated into the classical MD framework on the basis of the Monte Carlo approach.

\textbf{Practical Realization}: The interlink between stages~(i) and (iii) based on the reactive MD has been realized in MBN Explorer \cite{Solovyov_2012_JCC_MBNExplorer}, see Section~\ref{sec:Methods_RMD}. Chemical transformations in such MD simulations are described with the reactive rCHARMM force field \cite{Sushko2016_rCHARMM}, see Section~\ref{sec:Methods_RMD} for further details. The rCHARMM force field requires specification of the relevant input parameters, such as dissociation energies for different covalent bonds in a parent molecular system and molecular products, which can be routinely determined through potential energy scans carried out, for example, by means of DFT methods (see Section~\ref{sec:Interlinks_QM-MD_ground-state}). Formation of new covalent bonds is accompanied by partial charge redistribution in the system which can also be specified in the rCHARMM force field.

Simulations of chemical transformations in molecular and condensed matter systems may require the specification of several additional parameters, such as bond multiplicity, the valence of atoms, partial charges, etc., in reactants and possible reaction products. When these parameters are specified, MBN Explorer is instructed on how the existing covalent bonds can be broken and new ones can be formed. These capabilities make MBN Explorer \cite{Solovyov_2012_JCC_MBNExplorer} a unique software for simulating reactive molecular systems within the classical MD framework.

\textbf{Application Areas}: As discussed in Section~\ref{sec:Methods_RMD}, the interface between stages~(i) and (iii) realized in reactive MD has been employed for studying various chemical processes and phenomena \cite{MBNbook_Springer_2017, DySoN_book_Springer_2022, verkhovtsev2021irradiation}, including collision-induced structural transformations and fragmentation of molecular and cluster systems in the gas phase \cite{Verkhovtsev_2017_EPJD.71.212, deVera_2019_EPJD.73.215, Andreides_2023_JPCA.127.3757}, or placed in molecular environments \cite{Andreides_2023_JPCA.127.3757, Friis_2020_JCC, Friis_2021_PRE}, thermally-driven and collision-induced chemistry of condensed systems \cite{Sushko2016_rCHARMM, deVera_2018_EPJD.72.147}, including surface chemistry processes \cite{Sushko_IS_AS_FEBID_2016, DeVera2020, Prosvetov2022_PCCP}.

\subsubsection{Irradiation-Driven Molecular Dynamics}
\label{sec:Interlinks_QM-MD_IDMD}

\textbf{Nature of the Interface}: The nature of this interface is similar to the one described in Section~\ref{sec:Interlinks_QM-MD_RMD} for the reactive MD. As explained in Sections~\ref{sec:MM_key-definitions} and \ref{sec:Methods_IDMD}, irradiation-driven transformations in molecular and condensed matter systems are quantum and occur probabilistically. Due to the fast and localized nature of irradiation-induced transformations, they can be incorporated into the classical MD framework on the basis of the Monte Carlo approach.

\textbf{Practical Realization}: The interlink between the stages~(i) and (iii) [in the case of a single/isolated quantum processes like in molecules placed in a gas phase] and between the stages~(ii) and (iii) [in the case of multiple quantum events, e.g. induced by particle transport through a condensed matter system] based on IDMD \cite{Sushko_IS_AS_FEBID_2016} is realized in MBN Explorer \cite{Solovyov_2012_JCC_MBNExplorer}, see Section~\ref{sec:Methods_IDMD}. As described in Section~\ref{sec:Methods_IDMD}, IDMD accounts for local changes in the system's properties as a result of quantum transformations occurring due to irradiation-induced quantum processes. The probability of a quantum process to happen at a given space and time point is calculated as the product of the process cross section and the radiation flux density at this point \cite{LL3, Sushko_IS_AS_FEBID_2016}. The former can be obtained from \textit{ab initio} calculations, analytical estimates and models, experiments, or atomic and molecular databases. The latter can be determined by the chosen irradiation regime or using various particle transport theories, such as the MC or diffusion equation-based approaches (see Section~\ref{sec:Methods_Particle_Transport} and Section~\ref{sec:Interlinks_IDMD-transport} below).

Irradiation-induced transformations of molecular systems are simulated with MBN Explorer using the reactive rCHARMM force field \cite{Sushko2016_rCHARMM} described in Section~\ref{sec:Methods_RMD}. Similar to the reactive MD (see Section~\ref{sec:Interlinks_QM-MD_RMD}), simulations of irradiation-induced transformations in the system using IDMD requires the specification of several parameters in the rCHARMM force field, including equilibrium bonds distances and angles, force constants, bond dissociation energies, bond multiplicities, valences of atoms, and partial charges, in reactants and possible reaction products. Apart from that, one needs to define the rate (probability per unit time) of fragmentation of different covalent bonds and the amount of energy given to atoms of a particular bond upon its fragmentation. By specifying the parameters mentioned above, MBN Explorer is instructed on how the existing covalent bonds can be broken due to irradiation-induced quantum processes and how new covalent bonds in the system can be formed.

\textbf{Application Areas}: The interface between stages (i) and (iii) realized in IDMD can be exploited in connection to any molecular or condensed matter system exposed to radiation. Particular examples of the application of IDMD are discussed in Section~\ref{sec:Methods_IDMD}, Section~\ref{sec:Interlinks_IDMD-transport} and in the case study presented in Section~\ref{sec:Case_study_3D-nanoprinting}.

\subsection{Interfacing Irradiation-Driven Molecular Dynamics with Particle-Transport Methods}
\label{sec:Interlinks_IDMD-transport}

\textbf{Nature of the Interface}: As described in Section~\ref{sec:MM_key-definitions}, quantum processes occurring in molecular and condensed matter systems are often occur due to exposure of the system to external fields or irradiation by charged particles (electrons, protons, ions, etc.) or photons. Irradiation conditions for a system can be very different and depend on the radiation modality, duration of the system’s exposure to irradiation, and the system’s geometry. Irradiation can be homogeneous within a given volume or entirely inhomogeneous. The choice of irradiation conditions is determined by each particular case study. The quantum processes take place within stage~(i) of the multiscale scenario. They are coupled to the particle transport through the condensed matter systems represented by the stage~(ii), see Sections~\ref{sec:Methods_MC_transport} and \ref{sec:Methods_Analytical_transport} for particle transport techniques represented by track-structure MC codes or analytical methods. The interface between the two stages is discussed in Section~\ref{sec:Interlinks_QM-transport_MC}. The outcome of the stage~(ii) is typically represented by the spatial and temporal distributions of particles (primary, secondary, ternary etc.) created in the systems upon its irradiation at the relatively short time scales after the passage of projectiles through the systems, see Fig.~\ref{fig:MM_diagram}. These distributions can be utilized as the input for the IDMD simulations discussed in Section~\ref{sec:Methods_IDMD} and the corresponding analysis of the phenomena emerging at stage~(iv) of the multiscale scenario. Such possibility explains the nature of the interlink between the stages~(ii), (iii) and (iv).

\textbf{Practical Realization}: The interlink between the IDMD, discussed in Section~\ref{sec:Methods_IDMD}, and particle transport simulations (see Section~\ref{sec:Methods_Particle_Transport}) has been realized in MBN Explorer in ref~\citenum{DeVera2020}. The spatial and energetic distributions of primary particles/radiation and secondary, tertiary etc. particles have been obtained from particle-transport simulations and tabulated on a cubic grid consisting of voxels with pre-defined size and covering the whole IDMD simulation box. The convolution of the particle flux density with the cross section of the relevant quantum process (e.g. the cross section of molecular fragmentation, dissociative electron attachment, etc.) determines the process rate (i.e. the process probability per unit time) at any point within the system and at any given time.

In general, the yields and spatial distribution of secondary particles depend on the energy, shape and modality of the primary radiation beam and the irradiated material (geometry, composition and density). These characteristics can be obtained for different materials and irradiation conditions using the standard track-structure MC codes, such as Geant4-DNA, PARTRAC, KURBUC, SEED and other codes mentioned in Section~\ref{sec:Methods_MC_transport}. Similar methodology can be used to simulate various molecular systems placed into radiation fields of different modalities, geometries (e.g. a uniform irradiation field or a focused beam), and temporal profiles.

The first realization of this interlink in ref~\citenum{DeVera2020} was focused on atomistic simulations of irradiation-driven chemistry processes. In that case study the coupled MC-IDMD approach was employed to simulate irradiation-driven chemistry during the FEBID process of W(CO)$_6$ molecules deposited on a SiO$_2$ substrate using MBN Explorer. The yields and spatial distribution of secondary and backscattered electrons emitted from the substrate were obtained from the MC code SEED \cite{Azzolini_2018_JPCM.31.055901_SEED} at different irradiation conditions. This data was then used as an input for IDMD simulations on time scales up to several hundreds of nanoseconds, where a comparison with experimental data on the parameters of FEBID-grown nanostructures (structure height, lateral size and purity/elemental composition) can be performed.

\textbf{Application Areas}: By interlinking stages~(ii) and (iii) of the multiscale scenario depicted in Fig.~\ref{fig:MM_diagram}, one can explore novel features in the atomistic irradiation-driven molecular dynamics of molecular and condensed matter systems arising on the pico- and nanosecond time scales and achieve the multiscale description of irradiation-driven phenomena, chemistry and structure formation in many different systems, see refs~\citenum{MBNbook_Springer_2017, DySoN_book_Springer_2022, verkhovtsev2021irradiation} and references therein. Further examples of the exploitation of this interlink are given in Section~\ref{sec:Methods_IDMD} and in the case study presented in Section~\ref{sec:Case_study_3D-nanoprinting}.

\subsection{Interfacing Molecular Dynamics with Stochastic Dynamics}
\label{sec:Interlinks_MD-SD}

\textbf{Nature of the Interface}: As discussed in Sections~\ref{sec:Intro} and \ref{sec:MM_key-definitions}, the exposure of molecular and condensed matter systems to radiation may result in the manifestation of processes and phenomena that span over spatial and temporal scales, significantly exceeding the limits for the conventional atomistic MD, see stages~(iv) and (v) in Fig.~\ref{fig:MM_diagram}. Large-scale processes occurring in very different complex systems and having a probabilistic nature can be modeled using the Stochastic Dynamics (SD) approach \cite{Stochastic_2022_JCC.43.1442} (see Section~\ref{sec:Methods_StochasticDyn}), and the outcomes of atomistic (classical, reactive, irradiation-driven, relativistic) MD simulations can be used to construct such stochastic models.

\textbf{Practical Realization}: The interlink between stages~(iii), (iv), and (v) is realized in MBN Explorer \cite{Stochastic_2022_JCC.43.1442, SD_FEBID_abstract_Prague2023, Dick_2011_PRB.84.115408, Panshenskov_2014_JCC.35.1317, Moskovkin_2014_PSSB.251.1456, MBNbook_Springer_2017, DySoN_book_Springer_2022} via the combination of the MD and SD methods. This interlink significantly expands application areas of the software, allowing it to go beyond the limits of the pure MD codes that cannot support MM. The MD-SD interlink is realized by determining the probabilities for different stochastic processes utilized within the SD framework from atomistic MD simulations. For example, diffusion coefficients that characterize the kinetics of diffusing particles (molecules, atomic clusters, NPs, proteins, etc.) can be routinely obtained using MD and converted into a stochastic probability for the random translation of a particle into an adjacent position. Binding and activation energies can also be obtained directly from MD simulations and converted to a stochastic probability of particle detachment that governs the coalescence and fragmentation processes in complex large-scale systems.

\textbf{Application Areas}: For many application areas and case studies, the SD approach establishes the final links into the chain of theoretical and computational methods and algorithms, enabling computational MM of dynamics of systems and processes from atomic up to mesoscopic and even macroscopic scales with the temporal scales relevant to such modes of motion, see stages~(iv) and (v) in Fig.~\ref{fig:MM_diagram}.

\subsection{Interfacing Molecular Dynamics and Stochastic Dynamics with Thermodynamics and Other Macroscopic Theories}
\label{sec:Interlinks_MD-SD-macro_theories}

\textbf{Nature of the Interface}: Molecular-level processes and phenomena occurring in molecular and condensed matter systems exposed to radiation [stage (iii)] may affect the macroscopic characteristics of these systems [stages (iv) and (v)]. An interlink between stages (iii), (iv), and (v) can be established by combining the condensed-matter theories (such as hydrodynamics, acoustics, thermal conductivity, materials science, etc.) and MD and SD simulations, thus establishing a valuable tool for atomistic multiscale analysis of various micro- and macroscopic characteristics and phenomena.

\textbf{Practical Realization}: In general, the interlink between stages~(iii), (iv), and (v) is realized through comparison and coupling (where necessary) the outcomes of MD (including relativistic MD, RMD and IDMD) or SD simulations performed using MBN Explorer with the results of continuum methods-based simulations performed, for instance, using Abaqus \cite{FEM_ABAQUS_book_2017, Boulbes_FEM_Abaqus_2020} as well as with the analytical and/or numerical solutions of equations of continuum theories, which are typically programmed in custom-made computer codes or developed using the widely used programs like Wolfram Mathematica or Matlab.

The cross-comparison and interlink (where necessary) between MD and SD simulations, on the one hand, and the continuum theories and methods, on the other hand, depends on the particular system and process studied. One can highlight several examples of the realization of this interlink.
\begin{enumerate}[label=(\roman*)]

\item
\textit{Simulation of global conformational changes of biomacromolecules and structural transitions in proteins and bio-macromolecular complexes, including those induced by irradiation.} Statistical mechanics provides a practical theoretical framework for dealing with such processes. It defines the partition function, which is the sum over all possible system's states with the corresponding statistical weights \cite{LL5}. Knowing the partition function of the system, one can describe all its thermodynamic characteristics, e.g. evaluate its energy, pressure, or heat capacity at different temperatures. Establishing the fundamental connections of the statistical mechanics methods for calculating partition functions with the modern computational techniques for complex molecular and condensed matter systems based on MD is a promising research direction. Thus, as demonstrated in several case studies \cite{Yakubovich_2006_EPJD.40.363, Yakubovich_2008_EPJD.46.215, Solovyov_2008_EPJD.46.227, Yakubovich_2014_EPJD.68.145}, the combined statistical mechanics and MD methods are useful for the quantitative description of conformational changes and phase transitions in large biomacromolecules.

\item
\textit{The theoretical and computational manifestation of thermomechanical damage and related phenomena} (e.g. transport of reactive secondary species) caused by nanoscale shock waves that are created by heavy ions traversing biological medium \cite{surdutovich2010shock}, see also the case study in Section~\ref{sec:Case_study_MSA_SWs}. An ion-induced shock wave arises since ions can deposit a large amount of energy on the nanometer scale, resulting in the significant heating up of the medium in the localized vicinity of ion tracks. In a continuous medium this phenomenon is characterized by the so-called self-similar flow and the discontinuities of pressure and density of the medium at the wavefront as follows from the analytical solution of a set of corresponding hydrodynamic and thermodynamic equations \cite{surdutovich2010shock}. The solution of the hydrodynamic problem describing the strong explosion regime of the shock wave, as well as its mechanical features and limitations, are very well described in \cite{LL6, Zeldovich_ShockWaves}. The analytic solution for the case of the cylindrical shock wave \cite{surdutovich2010shock} has been well reproduced in MD simulations, analyzed and applied to the nanometer-scale dynamics of the DNA surroundings and the formation of complex lesions in the DNA molecule \cite{Surdutovich_AVS_2014_EPJD.68.353, surdutovich2013biodamage, Yakubovich_2012_NIMB.279.135, deVera2016molecular, deVera2017radial, Friis_2021_PRE}.

\item
\textit{The analysis of material damage as well as mechanical, thermal and transport properties of materials.} On a macroscopic level, these properties are commonly studied using the Finite Element Method (FEM), see Section~\ref{sec:Methods_Macroscopic}. Through the MM approach, the mechanical models from continuum-based elasticity theory can be examined by comparison with the outcomes of MD and SD simulations, and the limits of validity of the continuum-based elasticity theory can thus be established. In this way, the quantities of treated by continuum theories, such as the stress analysis employing the stress tensor, elastic constants, and elastic moduli, can be re-examined at the atomistic level.
\end{enumerate}

\textbf{Application Areas}: As follows from the examples given above, the interlinks between stages~(iii), (iv) and (v) realized through the combination of MD and SD methods and the continuum models and theories have been widely utilized in different research areas such as biophysics, radiation physics, chemistry, materials science, and technologies using plasma (see Section~\ref{sec:Case_study_Plasma}).

\section{Validation of Multiscale Modeling Methodologies}
\label{sec:Validation}

\subsection{Validation of Larger-Scale Theoretical and Computational Models on the Basis of Smaller-Scale Models}
\label{sec:Validation_TheoryComp}

The methodologies discussed in Section~\ref{sec:MM_key-definitions} in the context of the MM of irradiated condensed matter systems involve different model assumptions. This concerns practically all the theoretical approaches and methods presented in Section~\ref{sec:Methods}, including the most fundamental ones like HF, many-body and DFT theories. The assumptions utilized and their scope are different for different methods. There are more general assumptions, e.g. the fundamental postulates of HF or DFT theories, applicable to many (if not all) systems considered. However, as discussed in Section~\ref{sec:MM_key-definitions}, the numerical realization of such methodologies imposes significant limitations on their utilization. Hence, these methodologies are typically applied to relatively small and/or rather simple systems. The number of different modeling assumptions typically grows with increasing the system's complexity. The methods become more system-type and case-study-specific and often require multiscale approaches. However, the number of assumptions and/or the model parameters used to describe any complex system is usually relatively small compared with the number of systems and phenomena to which the corresponding model description can be applied.

For instance, the postulates of the HF, DFT, or TDDFT theories introduced in Section~\ref{sec:Methods} are applicable to the treatment of the electronic structure of atomic, molecular, and condensed matter systems. However, numerical practical solutions based on these methods are limited to relatively small system sizes and processes on relatively short time scales compared to the whole variety and complexity of condensed matter systems.

With the growth of size and complexity of a system, the utilized multiscale approaches typically involve models with additional parameters and more specific assumptions that are applicable to particular situations. The examples include transport theories, RMD and IDMD, SD, and models used in the specific case studies (FEBID, IBCT, nanofabrication) introduced in Section~\ref{sec:Intro} and further discussed in Section~\ref{sec:Case_studies}, devoted to the numerous concrete case studies.

As discussed in Section~\ref{sec:Interlinks}, most theoretical methods for MM of irradiated condensed matter systems are interconnected and interfaced. Therefore, numerous case studies permit defining a certain hierarchy of the relevant methods. The more fundamental techniques form the basis of such methodological chains. The upper levels of the methodological hierarchy are usually represented by approaches exploiting various additional assumptions, which may become more and more specific and empirical with the increase of the system's complexity. Typically, the methodological hierarchy levels considered in each case study correspond to different overlapping areas in the time-space diagram and related methodologies presented in Figure~\ref{fig:MM_diagram}.

An example of a methodological hierarchy is directly related to different types of MD (\textit{ab initio} MD based on DFT and TDDFT, classical MD, RMD, and IDMD) and interfaces between them with their further links to the upper-level theoretical methods, such as statistical mechanics and SD, see Section~\ref{sec:Interlinks}. Most of these methodologies and their mutual interlinks have been utilized for the MM of RADAM effects in materials and biological systems, including radiotherapy and space applications, radiation protection, nanofabrication technologies, and many more. Some of these applications have already been highlighted in Sections~\ref{sec:MM_key-definitions} and \ref{sec:Methods} and will be further discussed in Section~\ref{sec:Case_studies}, devoted to the representative case studies.

The hierarchal structure of the MM descriptions provides a possibility of their pure theoretical verification by testing the quality of the upper-level models and theoretical methods through their comparison with the lower-level ones based on the more fundamental and less empirical theoretical approaches. This comparison can be performed within the temporal and spatial scale ranges accessible for both types of descriptions (more fundamental and more empirical ones). Through such comparisons, the utilized model assumptions and parameters can be validated. Once validated, a model can be utilized for simulations of larger systems and longer processes that cannot be conducted using more fundamental theoretical methods. Typically, such situations arise at the interfaces between the different methods discussed in Section~\ref{sec:Interlinks}.

The MM of irradiated complex condensed matter systems and related processes may involve several validation procedures applied to different temporal and spatial scales presented in Figure~\ref{fig:MM_diagram}; see also the discussion of numerous related theoretical methods presented in Section~\ref{sec:Methods} and their interfaces in Section~\ref{sec:Interlinks}.

When a complete theoretical validation is impossible or too sophisticated, one can rely on the experimental validation of some parts of the MM schemes (see Section~\ref{sec:Validation_Experiment} below). Alternatively, one can make initial choices of some parameters of the models as they follow from analytical solutions, estimates, and comparative analysis with the already available knowledge, i.e. through the relevant educated guesses. The initial choices should then, in turn, be verified through the relevant experiments (Section~\ref{sec:Validation_Experiment}). In such situations, theoretical validation procedures can be utilized for a part of the entire MM scheme. The combined validation procedure of MM approaches becomes very useful when neither the entire theoretical nor entire experimental validation procedure is feasible due to various limitations.

The concrete examples of irradiated complex condensed matter systems and related processes that involve several validation procedures at different levels of theory and modeling have already been given in Section~\ref{sec:Intro}. Thus, the MM of RADAM of biological systems with ions involves analysis of the cross sections of elementary collision processes (quantum level), propagation of secondary particles including reactive species produced in the medium (interface of quantum theory and MC approach or analytical transport theory methods), their chemistry and interaction with the target DNA molecule resulting in its damage (IDMD, MC, or analytical approaches). These processes are followed by the relaxation of the energy deposited by ions into the medium with the follow-up dynamics of the medium, which at sufficiently large LET leads to further DNA damage, see Figure~\ref{fig:Intro_IBCT} and related references. The analysis of the DNA damage performed at the atomistic level can be linked to the cell survival rates through a model approach, which has been validated through many experiments with different cell lines, ions, and their energy ranges \cite{Surdutovich_AVS_2014_EPJD.68.353, verkhovtsev2016multiscale, AVS2017nanoscaleIBCT}. Finally, this MM approach has been linked to the extended biological systems \cite{ES_AVS_2017_EPJD.71.210, surdutovich2019multiscale}, providing a direct link and a guide to optimizing the existing treatment planning practices based on the MM approach. At present, MBN Explorer \cite{Solovyov_2012_JCC_MBNExplorer} and MBN Studio \cite{Sushko_2019_MBNStudio} provide the best platform for the computational MM of the aforementioned processes as these software tools are equipped with many unique implementations (including RMD, IDMD, SD) and support numerous interfaces between different methodologies, as discussed in Section~\ref{sec:Interlinks}.

Other examples of multiscale processes presented in Section~\ref{sec:Intro} concern the growth and self-organization processes in condensed matter systems, which can also be driven by irradiation of the system with focused electron or ion beams during the deposition of new precursor molecules on the surface of the growing system, as it occurs in the FEBID and FIBID processes. In these cases, the MM enables accounting for and interlinking all the stages of FEBID and FIBID, including all the phenomena relevant to each stage. The MM approach enables treating deposition, adsorption, desorption, diffusion, collision-induced excitation and fragmentation processes, chemical reactions, and structural transformations \cite{Sushko_IS_AS_FEBID_2016} while simulating the formation of growing FEBID/FIBID nanostructures with the atomistic level of details and determining their composition and morphology \cite{Sushko_IS_AS_FEBID_2016, DeVera2020, Prosvetov2021_BJN, Prosvetov2022_PCCP}. The MM approach for the FEBID/FIBID has been advanced further to the level of SD, enabling the modeling of these processes on the micro- and larger scales \cite{SD_FEBID_abstract_Prague2023}. The validation of such multistage, multilevel MM approaches also requires the multilevel validation procedure based on the abovementioned principles. Thus, one can validate the quantum inputs of the MM (cross sections of collision processes, parameters of the classical force fields), then perform analysis and validation of the results of simulations of various phenomena (adsorption, desorption, diffusion, etc.) on the molecular level and the level of the entire deposits (deposits composition, growth rate, morphology of the growing nanostructures and their dependence on the irradiation conditions). Finally, the results of MD-based atomistic modeling of FEBID/FIBID can be utilized to build FEBID/FIBID models based on the SD principles \cite{Stochastic_2022_JCC.43.1442}, significantly increasing the capacities of MM of such processes. These models can be validated by comparing MD-based modeling results with experiments. The advances in the MM of FEBID/FIBID make this computational approach very useful for various technological applications such as, for example, the FEBID/FIBID-based 3D nanofabrication, including 3D nanoprinting \cite{Winkler_2019_JAP_review, Plank_2020_Micromachines.11.48} (see Section~\ref{sec:Case_study_3D-nanoprinting}). The MM validation at this level of theory can only be achieved by comparing MM predictions with experiment results. Again, MBN Explorer and MBN Studio provide the best platforms for MM simulations of FEBID/FIBID processes and their technological applications. The MBN software has many unique implementations (including RMD, IDMD, SD) and supports numerous interfaces between different methodologies, as discussed in Section~\ref{sec:Interlinks}.

Once an MM methodology is fully validated, it can be applied to many relevant systems. For each particular system and the case study, one needs to determine a set of relevant model parameters that should be utilized for its simulations.

\subsection{Validation of Multiscale Models Through Experiment}
\label{sec:Validation_Experiment}

The major benefit of developing and utilizing multiscale models is that they can study a wide range of complex and intertwined processes. However, it is necessary to validate the `predictions' and simulations of the models with well-characterized data drawn from experiments and observations. Indeed, only once the model and methodologies have been `validated' can there be true confidence in the predictions and simulations arising from such models. This is no more than a revised statement of the well-known principles of the `scientific method', which has been defined for centuries as the process of objectively establishing facts through testing and experimentation. The basic process involves making an observation, forming a hypothesis, making a prediction, conducting an experiment, and, finally, analyzing the results.

Validation in a multiscale regime is complicated since the model/simulation typically involves scales ranging from the atomic level to a macroscale. Indeed, as already discussed above in Section~\ref{sec:Validation_TheoryComp}, MM typically involves models that use input parameters characterizing quantities or processes occurring on at least two different scales, but often on many more. Therefore, one can perform experimental validation of these parameters at each scale involved. Also, one can validate, through experiments, the outcomes/results of MM as a whole.

The validation of the whole MM approach should involve the validation at each of the five stages of the multiscale scenario shown in Figure~\ref{fig:MM_diagram} through dedicated experiments.
Let us consider how each stage of the multiscale scenario shown in Figure~\ref{fig:MM_diagram} can be `validated' using current experimental methodologies.


\subsubsection{Quantum Processes}
\label{sec:Validation_Exp_Quantum}

Radiation-induced elementary quantum processes are characterized by interactions at the atomic and molecular scale and include different types of spectroscopy, chemical reactions, and collisional phenomena, each of which has well-established experimental methodologies collating large amounts of data. This data is used as inputs to particle transport codes, RMD, IDMD and SD simulations and can be used to validate multiscale models. Such data is assembled in databases (Section~\ref{sec:Validation_Databases}), which, in turn, must be validated.

In most irradiation processes, energy is transferred from the incident radiation to the target atoms of molecules. Therefore, in any irradiation model, it is vital to have a good knowledge of the excited state spectroscopy of the target species. The spectroscopic properties of atoms and molecules are routinely derived using experimental techniques that are dependent upon the excitation energy of respective excited states, with UV or VUV spectroscopy being used to garner data on electronic states of atoms and molecules, IR or microwave-based spectroscopy to measure molecular vibration and rotation spectroscopy and x-rays to characterize inner-shell spectroscopy \cite{Fleming_Spectroscopy_OrgChem, Kolezynski_MolecSpectroscopy, Svanberg_AtMolSpectrosc}.

\subsubsubsection{Photoabsorption Spectroscopy}

Photoabsorption spectroscopy is the oldest and simplest form of spectroscopy. Light generated from an appropriate excitation source illuminates a sample of gas (or a transparent solid, e.g. condensed film). The absorption is characterized by the photoabsorption cross section $\sigma_{\rm abs}$, usually derived by applying the simple Beer-Lambert law, $I_{\rm out} = I_{\rm in} \exp{(- n \, x \, \sigma_{\rm abs})}$, where $I_{\rm in}$ is the intensity of the light incident on the sample, $I_{\rm out}$ the intensity of the light transmitted by the sample, $n$ the number density  (concentration) of the sample and $x$ the path length of the target.

The photoabsorption spectrum is obtained by measuring $\sigma_{\rm abs}$ as a function of incident wavelength/energy. Such spectra reveal the excited states of the target, and their excitation energies may be compared directly with derivations from many of the methods described in Section~\ref{sec:Methods_Quantum-proc} (DFT, TDDFT, etc.). Absorption cross sections (or oscillator strengths) measured by the Beer-Lambert law (often to an accuracy of a few percent) may also be directly compared with calculations.

In the case of UV spectroscopy, it must be noted that, below 200~nm, the sample must be placed within a vacuum since molecular oxygen strongly absorbs below 200~nm. Furthermore, while simple discharge lamps may produce light sources in visible and near UV regions, synchrotron light sources are required for light sources operating below 120~nm. Synchrotrons also have the advantage of producing continuous (or broadband) light, whereas lamp sources and sources based on high-harmonic generation often produce only discrete emission lines. While X-rays may be generated using metal cathodes, synchrotrons are now commonly used as bright, continuous X-ray sources \cite{Cramer_X-raySpectroscopy}.

\textbf{Areas of Application}:
Such photo-absorption methods are routinely used as analytical methods and are therefore found in most experimental laboratories, e.g. UV-vis spectrophotometers and Fourier-Transform Infra-Red (FTIR) spectrometers. Once $\sigma_{\rm abs}$ is known, the Beer-Lambert law may be used to determine the number density/concentration of species in the sample. Therefore, \textit{in situ} spectroscopy can be used to determine the number densities in any system that can be directly compared with the MM of that system. Such spectrometric techniques are widely used `in the field' to monitor concentrations of pollutants \cite{Spectrosc_Atmosphere} and are used \textit{in situ} in industrial plasma processing to determine purities and control processes \cite{Spectrosc_Semiconductor}. Using such spectroscopy, we have gathered most of our data on chemical inventories of the interstellar medium and planetary atmospheres by remote telescope operations \cite{Spectrosc_Astronomical}. These observations are the basic data used to benchmark and validate all of the models of these complex extraterrestrial systems, and the underlying spectral data is core to this work. Hence, large collections databases such as VAMDC \cite{VAMDC_portal, Albert_VAMDC_2020_Atoms.8.76} and VESPA \cite{VESPA_portal} have been developed (see Section~\ref{sec:Validation_Databases}).

\textbf{Limitations and Challenges}:
Photoabsorption spectroscopy has several inherent limitations. It is restricted to those states that obey the selection rules, and alternative methods, such as Electron Energy-Loss Spectroscopy (EELS), are needed to characterize `forbidden states'. The value of the $\sigma_{\rm abs}$ determines the experimental conditions and accuracy of the derived data (see below). A low $\sigma_{\rm abs}$ requires a long path length if an accurate measurement is to be made. While spectra may be dominated by a few large photoabsorption bands, the multitude of small, weaker absorption bands may provide a significant percentage (10 – 20\%) of the total absorbance. In order to measure low $\sigma_{\rm abs}$, cavity ring-down spectroscopy (CRDS) \cite{Berden_CavityRing_Spectrosc} has been developed to study weak absorption since it establishes a long path length by multiple reflections in a cavity. However, being a laser-based technique, CRDS is restricted to discrete absorption wavelengths.

Another common limitation in photoabsorption is the study of both short-lived (e.g. radical) species, such as the OH species, and long-lived (metastable) species. Short-lived reactive species may decay within the timescale of the measurement. Therefore, transient spectral techniques are required, for example, so-called `pump-probe' methods in which a `pump' (usually a laser) prepares the transient species and a `probe' (e.g. a short (nanosecond) laser pulse) is fired after the pump to measure the photoabsorption of the transient species \cite{Maiuri_2020_JACS.142.3}. Metastable species may be detected directly through their de-excitation of surfaces (liberating electrons), but since these arise from `forbidden' transitions, alternative methods such as energy-loss spectroscopy are more commonly used \cite{Fabrikant_1988_PhysRep.159.1}.

Despite many spectral measurements, the vast demand for spectral and $\sigma_{\rm abs}$ data required in models  to interpret observations cannot be satisfied. Therefore, we must rely on theoretical evaluations to provide the bulk of the data used for interpreting observations and input into models, with the experiments being used to benchmark and provide  confidence in the methods used. Where the atom or molecule is difficult to prepare for an experiment (e.g. radioactive atoms or biological molecules that cannot be prepared intact as in the gaseous phase), theory (e.g. DFT and TDDFT methods) is the only method to derive the required spectroscopic data, with these methods having been previously benchmarked against measurements of other atomic/molecular species.

\subsubsubsection{Energy-Loss Spectroscopy}

Electron Energy-Loss Spectroscopy (EELS) is another technique for studying the spectroscopy of atoms and molecules. In EELS, the energy of the scattered electron (prepared as a nearly monochromatic incident electron beam) is measured, often at a fixed scattering angle, and is equivalent to the energy transferred to the target atom/molecule. An `energy-loss spectrum' is a series of bands that record the excitation of different atomic/molecular states.

EELS has the advantage that it can observe both `allowed' and `forbidden' excited states, as electrons are not compelled to follow photon selection rules \cite{Brydson_EELS_book, Ibach_Mills_EELS}. EEL spectra recorded at low energy  ($<20$~eV) and small scattering energies ($<30^{\circ}$) typically reveal `allowed' transitions, while EEL spectra at high energies ($>100$~eV) and large (backward) scattering angles ($>90^{\circ}$) reveal `forbidden' transitions. EELS is also a standard method for measuring collision cross sections (see below).

\textbf{Areas of Application}:
EELS is a commonly used method to study molecules absorbed on a surface and surface reactions. However, it must be noted that the excited states of atoms and molecules in the condensed phase are shifted in energy due to their interactions with neighbors and/or the substrate. These shifts may be significant (e.g., there is a `blue shift' of 1~eV in the lowest excited state of water between the gas and ice phases) and may lead to quenching of some electronic states (e.g. Rydberg states), which can significantly change the excitation and dissociation of molecular species in different phases influencing the local chemistry \cite{Mason_2006_FaradayDisc.133.311}.

\textbf{Limitations and Challenges}:
Electron energy loss spectroscopy requires the samples to be placed within a high vacuum (typically $<10^{-5}$ torr). At a lower vacuum, the electrons will be scattered by the residual gas, and the electron detectors (channeltron and position-sensitive detectors) do not operate above $10^{-4}$ torr. The EEL spectrum may comprise many overlapping bands requiring careful deconvolution to resolve each individual atomic or molecular transition and high incident electron beam resolution, which requires considerable skill and experience by the operator.

\subsubsubsection{Photoelectron Spectroscopy}

The ionization potential (a particularly important parameter in plasma studies) and the ionic states of atoms and molecules may also be studied by EELS by detecting the ejected electron caused by the incident photon. This technique is commonly known as \textbf{Photoelectron Spectroscopy (PES)}, \textbf{Ultraviolet Photoelectron Spectroscopy (UPS)} if the incident photon is in the UV spectral range and liberates an outer electron \cite{Huefner_PES_book, Whitten_UV-PES_ApplSurfSci_2023}, or \textbf{X-ray Photoelectron Spectroscopy (XPS)} when an inner shell electron is released \cite{van_der_Heide_X-ray_PES}. Due to the characteristic energies of the ejected electrons, XPS enables one to measure the elemental composition and the electronic state of atoms in a material, making XPS a common analytical technique for determining the chemical composition of materials.

A UPS spectrum of a molecule contains a series of peaks. Each peak corresponds to a specific molecular-orbital energy level in the valence region. The high resolution enables the observation of fine structure due to vibrational levels of the molecular ion, which facilitates the assignment of particular peaks to bonding, nonbonding, or antibonding molecular orbitals. A valuable result of the characterization of solids using UPS is the determination of the work function of the material.

A typical XPS spectrum plots the number of electrons detected at a specific binding energy. Each chemical element produces a set of characteristic peaks in a spectrum, which correspond to the electron configuration of the atoms. The intensity of each peak is directly related to the amount of a particular element within the XPS sampling volume. In order to generate atomic percentage values, each raw XPS signal is corrected by dividing the intensity by a so-called relative sensitivity factor and normalized over all of the elements detected.

A UPS or XPS photoelectron spectrometer consists of a radiation source and an electron energy analyzer. The radiation source for UPS is usually a gas discharge lamp, with He discharge lamp operating at 58.4~nm (corresponding to 21.2~eV) being the most common. X-ray sources are typically either Mg or Al K$\alpha$ sources. However, today, synchrotron radiation sources are commonly used for XPS studies. Synchrotron radiation is especially useful as it provides continuous, polarized radiation.

\textbf{Areas of Application}:
XPS is routinely used as an analytical tool to measure the chemical content of substances, including inorganic compounds, metal alloys, polymers, elements, catalysts, glasses, ceramics, paints, papers, inks, woods, plant parts, make-up, teeth, bones, medical implants, bio-materials, coatings, electron and ion-modified materials and many others. Recently, it has been used as a tool for forensic science coupled with electron microscopy to analyze small samples, e.g. gunshot residue \cite{Blackledge_PES_Forensic}.

\textbf{Limitations and Challenges}:
The main limitation of both UPS and XPS is the finite resolution of the photoelectron spectrometer. The energy resolution may be increased, but the higher the resolution, the lower the sensitivity. Similarly, obtaining high spatial and energy resolutions comes at the expense of the signal intensity. The smallest analytical area that can be measured by XPS is $\sim$10~$\mu$m. XPS is also limited to measurements of elements with atomic numbers of 3 or greater, making it unable to detect hydrogen or helium.

UPS can only detect the ejected valence electrons, which limits the range and depth of surface experiments using UPS. Conventional UPS also has relatively poor resolution.
Once again, as with EELS, such methodology requires the samples to be placed within a  high vacuum (typically $<10^{-5}$ torr).

\subsubsubsection{Mass Spectrometry}

The fragmentation of target molecules after a collision or the products after a reaction is commonly measured using \textbf{mass spectrometry (MS)}. There are many different types of MS, but they are all comprised of at least these three components: (i) an ionization source, (ii) a mass analyzer, and (iii) an ion detection system \cite{deHoffmann_MassSpectrom_book, Gross_MS_textbook, McCullagh_Oldham_MassSpectrom}.

The ionization source (usually an electron beam) converts any neutral species into ions. Once ionized, the ions pass into the mass analyzer, where they are separated according to mass-to-charge ($m/z$) ratios using electric or magnetic fields. In a time-of-flight mass spectrometer, the mass of the ions is calculated by measuring the time the ions need to travel a fixed distance and their velocity. The signals from separated ions are then measured and sent to a data system, which stores information on the $m/z$ ratios and their relative abundance. A typical mass spectrum plots the intensities of different ions in a sample against their $m/z$ ratios. Each peak in a mass spectrum corresponds to a particular $m/z$ ratio, and the heights of the peaks denote the relative abundance of the various components in the sample. Many mass spectrometers operate with a 70-eV electron beam such that there are standard data on the expected mass of the ion fragments and relative height of the measured mass peaks for each molecule, which allow each species to be identified in any system.

In most conventional mass spectrometers, cations (positive ions) are detected. However, anion (negative ion) detection allows the study of the Dissociative Electron Attachment (DEA) process. This collisional process is a critical physical process in many natural and technological processes, including many examples discussed in this paper, such as FEBID and IBCT.

\textbf{Areas of Application}:
MS is one of the most widely used analytical methods applied in nearly every modern science and technology area. A few examples include \cite{ThermoFisher_MS}:
\begin{enumerate}[label=(\roman*)]

\item
\textit{Applications of MS in proteomics} -- Characterization of proteins and protein complexes, sequencing of peptides, and identification of posttranslational modifications.

\item
\textit{Applications of MS in metabolomics} -- Cancer screening and diagnosis, global metabolic fingerprinting analysis, biomarker discovery and profiling, biofuels generation and use, lipidomics studies, and metabolic disorder profiling.

\item
\textit{Applications of MS in environmental analysis} -- Drinking water testing, pesticide screening and quantitation, soil contamination assessment, carbon dioxide and pollution monitoring, and trace elemental analysis of heavy metals leaching.

\item
\textit{Applications of MS in pharmaceutical analysis} -- Drug discovery and absorption, distribution, metabolism, and elimination (ADME) studies, pharmacokinetic and pharmacodynamic analyses, metabolite screening, and preclinical development.

\item
\textit{Applications of MS in forensic analysis} -- Analysis of trace evidence (e.g. fibers in carpet, polymers in paint), arson investigation (e.g. fire accelerant), confirmation of drug abuse, and identification of explosive residues (bombing investigation).

\item
\textit{Clinical applications of MS} -- Clinical drug development, Phase 0 studies, clinical tests, disease screening, drug therapy monitoring, analysis of peptides used for diagnostic testing, and identification of infectious agents for targeted therapies.

\item
Mass spectrometers have also been miniaturized to operate onboard spacecraft to measure the atmospheric composition of planets and moons in the solar system and by vaporizing material from surfaces to study the composition of comets and asteroids \cite{Leseigneur_2022_AngewChemIntEd.61, Altwegg_2017_MNRAS.469.S130}.
\end{enumerate}

\textbf{Limitations and Challenges}:
One of the main disadvantages of MS is its limitations in identifying hydrocarbons that produce similar ions (as hydrocarbons have similar fragmentation patterns) and its inability to distinguish optical and geometrical isomers. In order to solve this issue, MS is often combined with other techniques, such as gas chromatography (GC-MS) \cite{Huebschmann_GC-MS_book}, to separate the different compounds.

GC–MS is composed of the gas chromatograph and the mass spectrometer. The gas chromatograph contains a capillary column whose properties regarding molecule separation depend on the column's length and diameter as well as on the chemical properties of the sample's constituents. Fragment molecules are separated as the sample travels through the column. The produced molecules are retained by the column and then eluted from it at different times (called the retention time). This allows the mass spectrometer downstream to detect the molecules separately.

Most MS techniques only provide qualitative data on the species produced, and it is difficult to derive absolute cross sections for the formation of such species. This is due to the inherent difficulty in detecting all of the product ions and energy discrimination effects in the mass analyzer, such that slow ions are detected in preference to fast ions that may escape the extraction field. This discrimination has been demonstrated to invalidate some of the reported dissociative ionization and DEA cross sections.

Neutral fragment species produced in dissociative collisions with molecules are harder to detect unless they are in an excited state (where fluorescence measurements may be used if the target is excited to a short-lived radiating excited state or direct detection if the state is ‘metastable’). However, laser-induced photoionization coupled with MS has been adopted to study neutrals in their ground state. This is important for determining the flux of such species in many plasmas, which, due to their reactivity, may dominate the local chemistry and be used to test predictions of MM.

\subsubsection{Particle Transport}
\label{sec:Validation_Exp_Particle-Transport}

In quantifying the transport of particles through any medium, the commonly used parameters are the measure of the energy deposited (dose deposition), the range or `penetration depth' of the incident radiation, and the production of secondary particles by the primary radiation.

\subsubsubsection{Measurement of Energy Deposited}

A commonly used measure of energy deposition is LET, which indicates the average energy lost per unit path length as a charged particle travels through a given material. The LET for electrons is traditionally expressed in units of MeV/cm or, when divided by the mass density, in units of MeV-cm$^2$/g. In radiobiology, the term RBE is used as a measure of the damage done by a given type of radiation per unit of energy deposited in biological tissues. Both LET and RBE are used as quantities in dosimetry in which the radiation dose is described by the term  ``absorbed dose'', which is the amount of radiation energy deposited in a target divided by the corresponding mass of the target. LET, RBE and absorbed dose may all be calculated in MM and are used in the clinic to derive radiotherapy treatment plans for each patient. Therefore, it is important to validate the models used for such treatment plans. RBE is a complex quantity dependent upon many different physical and biological parameters, so it is often hard to measure. However, a correlation between RBE and LET has been reported \cite{Sorensen_2011_ActaOncol.50.757}, which suggests LET is an acceptable physical surrogate for the purposes of treatment planning.


The most commonly used method for assessing dose deposition and LET employs \textbf{ionization chambers}, either an air-filled ionization chamber (IC) or a liquid-filled ionization chamber (LIC).

Ionization chambers have a simple design, being composed of two electrodes separated by an active medium consisting either of a gas or liquid. The electrodes may have the form of parallel plates or a cylinder with a coaxially located internal anode wire. As the ionizing beam passes through the chamber, it ionizes atoms and molecules in the medium, creating ion pairs with the resultant positive ions and electrons being attracted toward the electrodes of the opposite polarity. This generates an ionization current (in the range of fA to pA, depending on the chamber design) proportional to the radiation dose.

ICs have higher resolution than LICs in the beam direction, but LICs are preferable for dose distributions with narrow beam profiles. Since liquids have a higher mass density than gases, LICs may be significantly smaller than ICs. However, while LICs are advantageous due to their small diameter, the high density of liquids leads to recombination effects that suppress the detected signal. On the other hand, ICs, due to the low density of air, experience little recombination, so nearly all the electron-ion pairs created in ICs are captured by the collecting electrode. Recently, a methodology using IC and LIC in tandem was developed \cite{Tegami_2017_IJMPCERO} to derive the LET from the ratio between the IC and LIC signals.

\textbf{Areas of Application}:
Ionization chambers are used in many areas of science and technology. Small-size versions are routinely used to monitor beta and gamma radiation, particularly for high dose-rate measurements.
%
In medical physics and radiotherapy, ICs are the most widely used type of dosimeter, ensuring that the correct dose from a therapy unit or radiopharmaceutical is delivered. 
Each chamber has a calibration factor established by a national standards laboratory or has a factor determined by comparison against a standard chamber traceable to national standards at the user's site.

\textbf{Limitations and Challenges}:
ICs and LICs have restricted energy resolution. For air-filled chambers at standard temperature and pressure, the average energy required to eject an electron is 33.85~eV. Thus, they are unsuitable for detecting UV, low-energy X-rays, or low-energy ions.

The ion-collecting gas volume in the chamber must be precisely known if the IC/LIC is to be used as an absolute dosimeter. This is not usually practicable outside national standards laboratories, so most ICs and LICs are calibrated against such standards.

When using pulsed radiation, the pulses must be short compared to the transit time ($\sim$10$^{-3}$~s) and the repetition rate must be slow enough to remove all the ions from the chamber between pulses.

Moisture (humidity) is a major problem affecting the accuracy of ICs. The chamber's internal volume must be kept completely dry since they are sensitive to `leakage currents' due to the very low currents generated.
``Guard rings'' are typically used on higher voltage tubes to reduce leakage through or along the surface of tube connection insulators, which can require resistance in the order of $10^{13}~\Omega$.


\subsubsubsection{Measurement of Penetration Depth}

It is often important to know the range of the incident radiation; this can be characterized by the \textbf{penetration depth} commonly defined as the distance by which the incident radiation flux decreases by a factor $1/e$ with respect to its initial value. The penetration depth depends on the energy and type of the incident radiation and the atomic number, density, and thickness of the object. UV photons deposit most of their energy at shallow depths, so their penetration depth is low whereas high-energy charged particle radiation can penetrate deep into (or even through) materials. This is an important factor in designing spacecraft to protect internal facilities (electronics and/or astronauts) from RADAM.

A related term is the `stopping power' commonly used in nuclear and materials physics. Stopping powers are employed in a wide range of application areas, such as radiation protection, ion implantation, and nuclear medicine, where it is defined as the rate at which a material absorbs the kinetic energy of a charged particle. The stopping power of the material ($S$) is numerically equal to the loss of energy $E$ per unit path length $x$, $S = - dE/dx$. The stopping power is defined as a sum of two terms: nuclear and electronic. The nuclear term of the stopping power is the loss of energy per unit length from elastic Coulomb interactions between the particle and atomic nuclei, and the electronic term defines the energy loss per unit length from Coulomb interactions between the particle and electrons. The nuclear term dominates the stopping power at low energies (i.e. below 1~MeV per nucleon), $S$ may be routinely calculated and included in MM, e.g. using the MC code SRIM \cite{SRIM_Ziegler_2010_NIMB.268.1818}.

Traditionally, two methods have been used to measure stopping powers. The first relies on backscattering ions from materials \cite{Bauer_1987_NIMB.27.301, Mertens_1987_NIMB.27.315}, while in the second, ions are transmitted through a thin foil \cite{Raisanen_1996_NIMB.118.1}. Backscattering methods are ideally suited to measure the stopping powers of light ions in materials of high-Z elements, while the transmission method is the more versatile of the two because it can be used both for light and heavy ions, although it has been shown that the two approaches can be complementary \cite{Mertens_1986_NIMB.15.91}. More recently, these methods have been complemented and extended using time-of-flight (TOF) methods, which can reduce previous experimental uncertainties \cite{Fontana_2016_NIMB.366.104}.

\begin{figure}[t!]
\includegraphics[width=0.75\textwidth]{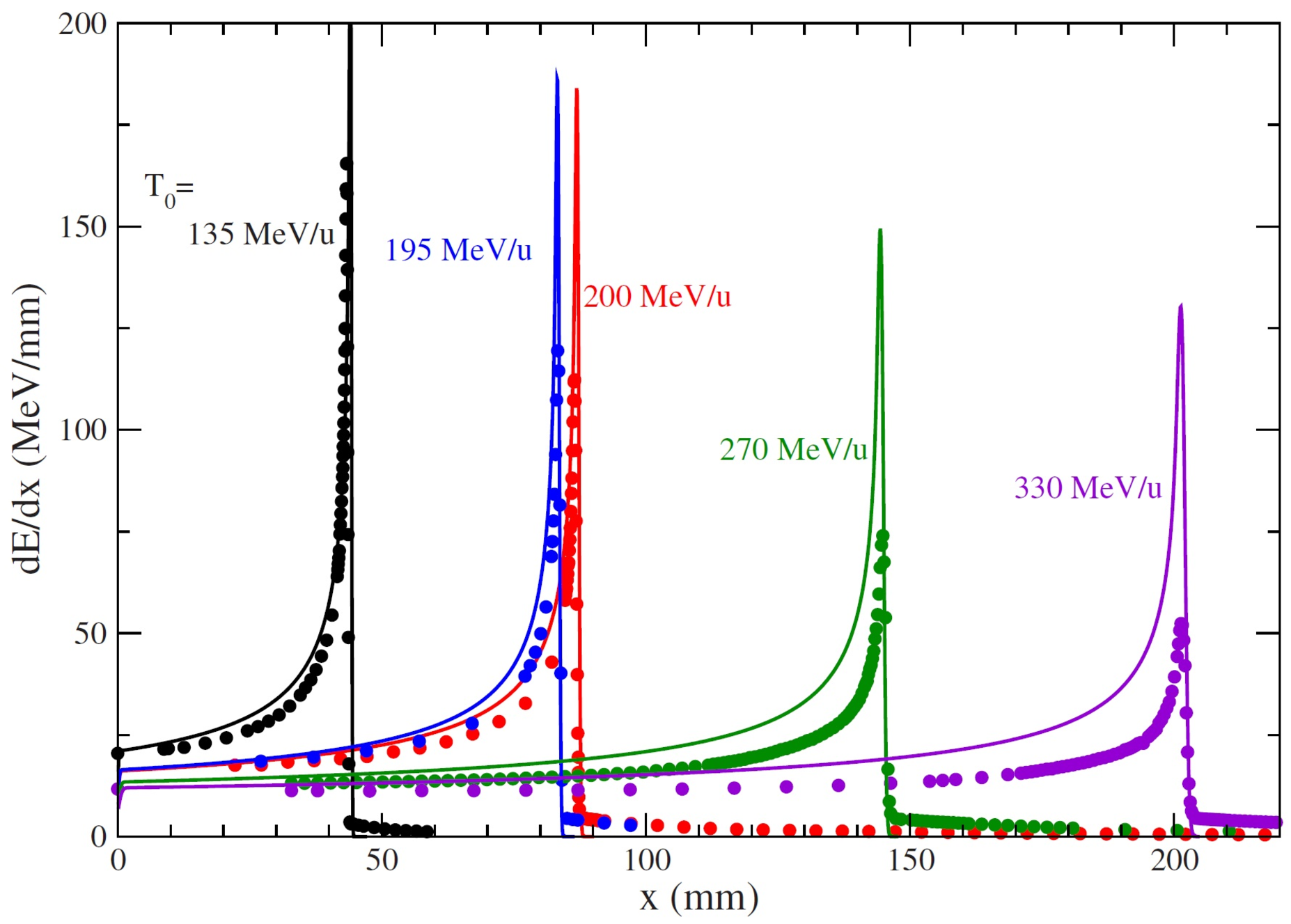}
\caption{Penetration depth and linear energy deposition for carbon ions in water, showing the Bragg peak for ions of different initial energies $T_0$. Model calculations \cite{Scifoni_2010_PRE.81.021903} performed within the MultiScale Approach to the physics of radiation damage with ions \cite{solov2009physics, Surdutovich_AVS_2014_EPJD.68.353} (MSA, see Section~\ref{sec:Intro_Ex_IBCT}) (lines) are compared to the results of experimental measurements \cite{Sihver_1998_JpnJMP.18.1, Haettner_2006_RPD.122.485} (symbols). Different labels and colors indicate curves at different initial ion energies. Figure is adapted from ref~\citenum{Scifoni_2010_PRE.81.021903}.}
\label{fig:Validation_Bragg-peak}
\end{figure}

The LET usually increases toward the end of the particle's range and reaches a maximum, \textbf{the Bragg peak}, shortly before the energy of the propagating particle drops to zero, see Figure~\ref{fig:Validation_Bragg-peak}. This is of great practical importance for radiation therapy, as discussed in Section~\ref{sec:Intro_Ex_IBCT} and case studies reported in Sections~\ref{sec:Case_study_MSA_SWs} and \ref{sec:Case_study_IBCT}.

Traditionally, the position of the Bragg peak can be measured using ionization chambers, with the energy deposited by radiation measured as a function of depth in the IC and LIC. Schauer \textit{et al.} \cite{Schauer_2022_FrontOncol.12.925542} recently developed an ionoacoustic method to measure the Bragg peak for pulsed proton beams with energies up to 220~MeV in water. The methodology of ref~\citenum{Schauer_2022_FrontOncol.12.925542} is based on the mechanism that when penetrating a medium, ions lose energy in electronic collisions, resulting in localized heating and a thermal expansion, which generates thermoacoustic emissions detectable with acoustic transducers. Since thermoacoustic emission is enhanced at the maximum of ion energy deposition (the Bragg peak), time-of-flight (TOF) measurements in combination with knowledge of the speed of sound in the traversed medium allow the ion range to be determined with (sub)millimeter accuracy providing a detailed validation of any particle transport code.

\textbf{Areas of Application}:
Penetration depth and stopping powers are routinely used when studying ionizing radiation. The Bragg peak is exploited in cancer particle therapy, specifically in proton and ion beam therapy, to concentrate the effect of ion beams on the irradiated tumor while minimizing the effect on the surrounding healthy tissue. A monoenergetic beam delivering a sharp Bragg peak may be widened by increasing the range of energies over which it is affecting so that a larger tumor volume can be treated. The plateau created by modifying the ion beam is called the spread-out Bragg Peak (SOBP), which allows the treatment to conform to larger tumors and more specific 3D shapes \cite{Jette_2011_PMB.56.N131} (see Section~\ref{sec:Case_study_IBCT}). An SOBP can be formed using variable thickness attenuators like spinning wheels \cite{Jia_2016_NIMA.806.101}.

\textbf{Limitations and Challenges}:
For LICs filled with dielectric liquids, the restricted stopping power relative to water shows minimal dependence on kinetic energy over clinical energy ranges. The higher density of liquids compared to gases also gives LICs greater sensitivity to ionizing radiation, allowing for constructing smaller sensitive volumes than ICs \cite{Stewart_2007_PMB.52.3089}. However, a disadvantage exists when using liquid-filled chambers to measure stopping powers since the distance between successive ionization events within a medium is inversely proportional to the density. As liquids are approximately three orders of magnitude denser than gases, successive ionization events are closer in LICs than in ICs, which increases the probability of electron--ion pairs recombination and may lead to net loss in recorded ion pairs.

In order to measure the track structure that can be derived in MM (for example, the spatial distribution of inelastic particle interactions along the track of the incident particle), detectors that are comparable in size to the radiation tracks are needed to investigate separately the structure of a particle track in the so-called track-core and penumbra regions. The latter region is exclusively formed by the interactions of so-called $\delta$-electrons, the secondary electrons created by ionization processes induced by the primary particles (see below). While ICs and LICs may provide submillimeter measurements along the depth dose curve, particularly focusing on the Bragg peak, they do not allow for studying radiation-induced phenomena on the nanoscale, which cannot be described by macroscopic quantities like absorbed dose or LET. Using the nanodosimetry approach, one can measure the radiation-induced frequency distribution of ionization clusters in liquid water (as a substitute for a sub-cellular material) in volumes comparable to those of the most probable radio-sensitive volumes of biological systems. This requires new types of detectors to be developed. One such detector is that at Legnaro National Laboratories of INFN \cite{Conte_2012_NJP.14.093010}. It simulates a target volume of about 20~nm in diameter, which can be moved with respect to a narrow primary particle beam. This allows measuring the ionization-cluster-size distribution, mainly representing the track-core region of a particle track but also describing the penumbra region as a function of the impact parameter with respect to the primary ion trajectory. The experimental data provides a direct validation of models and has proven to be the basis for modern nanodosimetry studies necessary for hadron therapy planning treatment models operating on sub-micron scales and authorized by the International Atomic Energy Agency (IAEA).

Another limitation and challenge for measurement (and definition) of the penetration depth of any radiation  is the production of secondary species (which, once created, have their own range or penetration depth) and the role of multiple scattering, which causes the diameter of the beam to expand with increasing depth. The beam profile spreads over the particle track due to the cumulative effect of small-angle scattering between the ionizing radiation and the medium through which it passes. As the particle’s energy decreases, the angular dispersion increases. This leads to the beam diameter increasing along the particle track due to small-angle scattering, leading to a beam spot larger than the initial diameter such that the ``field of view'' of the detector is required to take into account this effect or the percentage of the beam captured by the detector will decrease with increasing depth, this will be largest for light ions (protons) up to 15\% loss compared to $<5\%$ for carbon and heavier ions.

\subsubsubsection{Measurement of Secondary Species Produced by Incident Irradiation}

A major consequence of particle transport through the medium is the production of secondary particles (e.g. secondary electrons or nuclear fragments) created due to interactions of the primary projectiles within the medium, and their transport also needs to be simulated.

Electrons may be emitted from a surface by photons, the well-known photoelectron process, but secondary electrons may also be formed by electron or ion impact on surfaces \cite{Stein_1972_JAP.43.2617}. The energy of the secondary electrons ejected from the surface may be measured using EELS (discussed above). Surfaces often have adsorbates upon them, leading to ion yields that may be quantified by mass spectrometry with their energy spectra derived by TOF spectroscopy \cite{Lohmann_2020_NIMB.479.217}.

Another result of primary particle interactions is the production of chemical species. For instance, concentrations of the hydroxyl radicals (OH$\cdot$) and hydrogen peroxide (H$_2$O$_2$) produced by irradiation of water phantoms may be measured directly using \textit{in situ} IR spectroscopy (for H$_2$O$_2$) or by the addition of so-called scavengers such as methanol and tris(hydroxymethyl)aminomethane (for OH) with more recent refinements using ESR spin trapping \cite{Apak_2022_PureApplChem}.

\textbf{Areas of Application}:
The need for data on secondary particle production induced by primary radiations pervades all areas of radiation sciences.

In plasma sciences, where the plasma interacts with a surface (see Section~\ref{sec:Case_study_Plasma}), the constituent electrons and ions of the plasma release many secondary species that may dominate the subsequent behavior of the plasma itself and the chemistry it induces \cite{Zhang_2023_PSST.32.045015, Kondeti_2018_FreeRadicBiolMed}; this is often referred to as the `secondary plasma'.

The formation of new chemical species by irradiation underpins molecular formation in the interstellar medium, where reactions are induced by cosmic and UV irradiation of ice-covered dust grains \cite{Mason_2014_FaradayDisc.168.235}. Radiation-induced chemistry through the reaction of secondary species within the ices of planetary and lunar surfaces determines the chemical composition of the surface of these bodies, the measurement of which is a key objective of space missions to icy bodies in our solar system such as ESA JUICE mission to the icy moons on Jupiter which lie in the magnetosphere of the giant planet being constantly bombarded by ions and electrons \cite{Strazzulla_2023_EarthMoon}.

OH (and H$_2$O$_2$) production arising from irradiation of water in cells and the human body is believed to be the major route of radiation-induced damage in radiotherapy, with the primary radiation leading to a cascade of tens of thousands of secondary electrons leading to dissociation of \textit{in situ} water \cite{Halliwell_2021_ChemSocRev.50.8355}. Indeed, the role of radical species in DNA and ageing remains a hot topic in current biochemistry \cite{Schumacher_2021_Nature.592.695}.

\textbf{Limitations and Challenges}:
Measurements of secondary electron emission from surfaces are complicated by the charging of the surface, particularly in dielectrics and polymers. Recently, a new method utilizing a measurement of this charging to derive kinetic energy spectra of secondary electrons from common polymers (kapton, PTFE and ultem) has been reported \cite{Olano_2020_ResultsPhys.19.103456}.

Direct measurements of secondary yields in a liquid medium, needed to validate MM of biological systems, have proven equally challenging due to difficulties in performing scattering experiments with many liquids in a vacuum, so there has been only a very limited amount of experimental work on secondary electron emission from liquids \cite{Mehnaz_2020_MedPhys.47.759} using environmental SEMs \cite{Thiberge_2004_RevSciInstrum.75.2281, Joy_2006_JMiscosc.221.84}. However, these measurements are limited to high-energy electrons ($>5$~keV), and experiments with low-energy electrons are still needed. New experiments using liquid micro-jets \cite{Kaneda_2010_JCP.132.144502, Kitajima_2019_JCP.150.095102, Nag_2023_JPB.56.215201} are to be recommended.

\subsubsection{Irradiated Medium dynamics, Non-Equilibrium Chemistry and Molecular Transformations}
\label{sec:Validation_Exp_Non-equil_Chemistry}

At the end of the particle transport stage, a significant part of the energy of the primary radiation has been transferred into the system, resulting in the creation of molecular and ionic fragments and/or the formation of defects within the condensed matter. However, such an excited medium is created in a state far from its equilibrium, and it subsequently evolves towards the equilibrium through a cascade of processes, which can lead to the creation of new species through chemical reactions processes typically taking place on timescales of a picosecond or more. Additional processes operating on this time scale are diffusion through the medium, adsorption, and irreversible desorption from the surface.

Time-resolved studies of molecular transformations are challenging experimentally, with the main techniques being used to study systems once they have reached equilibrium (see Section~\ref{sec:Validation_Exp_Chem_Equil}), from which the intermediate processes leading to this equilibrium state can only be inferred. In contrast, such studies are inherent in MM, which can determine the intermediate states, energy transfer pathways, and dynamics involved in radiation-induced reactions and unravel the complex network of physical and chemical processes evolving in the system.

In order to explore the evolution of chemical and physical changes over time and study the transfer from a non-equilibrium to an equilibrium stage, it is therefore necessary to develop transient methods.


\textbf{Transient spectroscopy}, often called \textbf{time-resolved spectroscopy}, is used for the characterization of the electronic and structural properties of short-lived excited states (transient states) of molecules \cite{Knowles_2018_JMaterChemC.6.11853} and is an extension of the spectroscopic techniques described in Section~\ref{sec:Validation_Exp_Quantum}. It measures changes in the absorbance/transmittance in the sample after excitation of the molecule by a short burst of radiation. In a typical Transient Absorption (TA) experiment, both the excitation (`pump') and absorbance (`probe') are generated by a single pulsed laser. If the studied process is slow, the time resolution can be obtained with a continuous probe beam, and conventional spectroscopic techniques can be used. Using very short-pulsed lasers (femto- or now even attoseconds), it is possible to study processes that occur on time scales as short as $10^{-16}$ seconds (see Section~\ref{sec:Case_study_Attosecond}). The absorption of a probe pulse by the sample is recorded as a function of time at different wavelengths to study the dynamics of the excited state. The study of TA as a function of wavelength provides information regarding the evolution/decay of various intermediate species involved in a particular chemical reaction at different wavelengths. The transient absorption decay curve as a function of time contains information regarding the number of decay processes involved at a given wavelength and how fast or slow these decay processes occur.

Extensions of the TA method are time-emission or time-resolved fluorescence spectroscopy in which emission spectra are recorded post-pulsed excitation using Time-Correlated Single Photon Counting (TCSPC), using a streak camera or intensified CCD cameras (time resolution of picoseconds and slower). Time-resolved photoemission spectroscopy and two-photon photoelectron spectroscopy (2PPE) are important extensions to photoemission spectroscopy. The atom or molecule of interest is excited by the pump and then ionized by the probe. The electrons or positive ions resulting from the ionization event are then detected. By varying the time delay between the pump and the probe, the change in the energy of the photo-products is observed. EELS or TOF methods for detecting the products are commonly used.

Dromey and co-workers (their methodology is detailed in a case study in Section~\ref{sec:Case_study_UltrafastRadChem}) have recently shown that by using laser-driven ion accelerators, it has been possible to perform the first picosecond radiolysis studies for protons interacting with H$_2$O and transparent dielectrics \cite{Senje_2017_APL.110.104102, Taylor_2018_PlasmaPhys.60.054004}. In the future, we can expect the development of a suite of broadband probes that can interrogate the dynamics of a wide range of different chemical species in a single shot. Such advances in ultrafast radiation chemistry will aid our understanding of fundamental processes underpinning ionizing interactions in matter and, in turn, will feed back into the testing and benchmarking of MM.


Under certain conditions, the medium may see a significant increase in temperature and pressure within the relatively small volumes where the energy deposition takes place, and this may lead to new phenomena, such as the formation of nanoscopic shock waves \cite{surdutovich2010shock}. These shock waves may be sufficient to create an irreversible transformation of the medium, including the formation of defects, and may lead to bond breaks or even lethal damage in living cells \cite{Surdutovich_AVS_2014_EPJD.68.353, surdutovich2013biodamage, Friis_2021_PRE} (see Sections~\ref{sec:Intro_Ex_IBCT} and \ref{sec:Case_study_MSA_SWs}).

The effect of such shock waves on the final physical and chemical state of the system has been discovered using MM but to date, direct measurement of such shock waves has proven a significant challenge to experimental techniques. However, it has been shown recently that some thermoelectric sensors may have the response time and sensitivity to record the pulse in heat fluxes generated in shock-wave processes in short-duration high-speed gas dynamic experiments \cite{Kotov_2021_JPCS.2103.012218}.

\textbf{Areas of Application}: TA spectroscopy can be used to trace the intermediate states in a photo-chemical reaction; in a charge or electron transfer process; conformational changes, thermal relaxation, fluorescence or phosphorescence processes. With the increasing availability of ultrafast lasers and facilities such as the Extreme Light Infrastructure (ELI) \cite{ELI_website}, it is possible to excite a large molecule to desired excited states to study specific molecular dynamics.

TA microscopy enables measurement of excited state dynamics with sub-micron spatial resolution, which has been used to study the influence of film morphology on local excited state dynamics and image directly charge transport in solution-processed organic and hybrid organic--inorganic lead-halide perovskite semiconducting films \cite{Knowles_2018_JMaterChemC.6.11853}.

TA spectroscopy has become an important tool for characterizing various electronic states and energy transfer processes in NPs and locating trap states.

\textbf{Limitations and Challenges}: TA and time-emission measurements are highly sensitive to laser repetition rate, pulse duration, emission wavelength, polarization, intensity, sample chemistry, solvents, concentration and temperature. The excitation density (number of photons per unit area per second) must be kept low; otherwise, the sample may be saturated.

The novel method of DNA origami has been used to detect and quantify DNA damage \cite{Ebel_Bald_2022_JPCL.13.4871} (see also a case study in Section~\ref{sec:Case_study_DNAorigami}). DNA origami comprises long strands of DNA used to construct a variety of 3D nanostructures that allow for the arrangement of different functionalities such as specific DNA structures, nanoparticles, proteins and various chemical modifications with unprecedented precision. The arranged functional entities can be imaged using Atomic Force Microscopy (AFM) and spectroscopically characterized using Surface Enhanced Raman Scattering (SERS) and fluorescence spectroscopy.  In the future, DNA origami may provide a method for observing shock waves.

\subsubsection{Chemical (and Thermodynamic) Equilibrium}
\label{sec:Validation_Exp_Chem_Equil}

The fourth (and final if the system is closed) stage of MM is when the system has reached chemical and thermodynamic equilibrium, which follows the stage of irradiated medium dynamics, non-equilibrium chemistry, and molecular transformations. Experimental validation of this stage is the most common since the MM has determined the system's final state, which is to be compared with experiment or observation.

\subsubsubsection{Spectroscopic Analysis of Surfaces}

\begin{figure}[t!]
\includegraphics[width=1.0\textwidth]{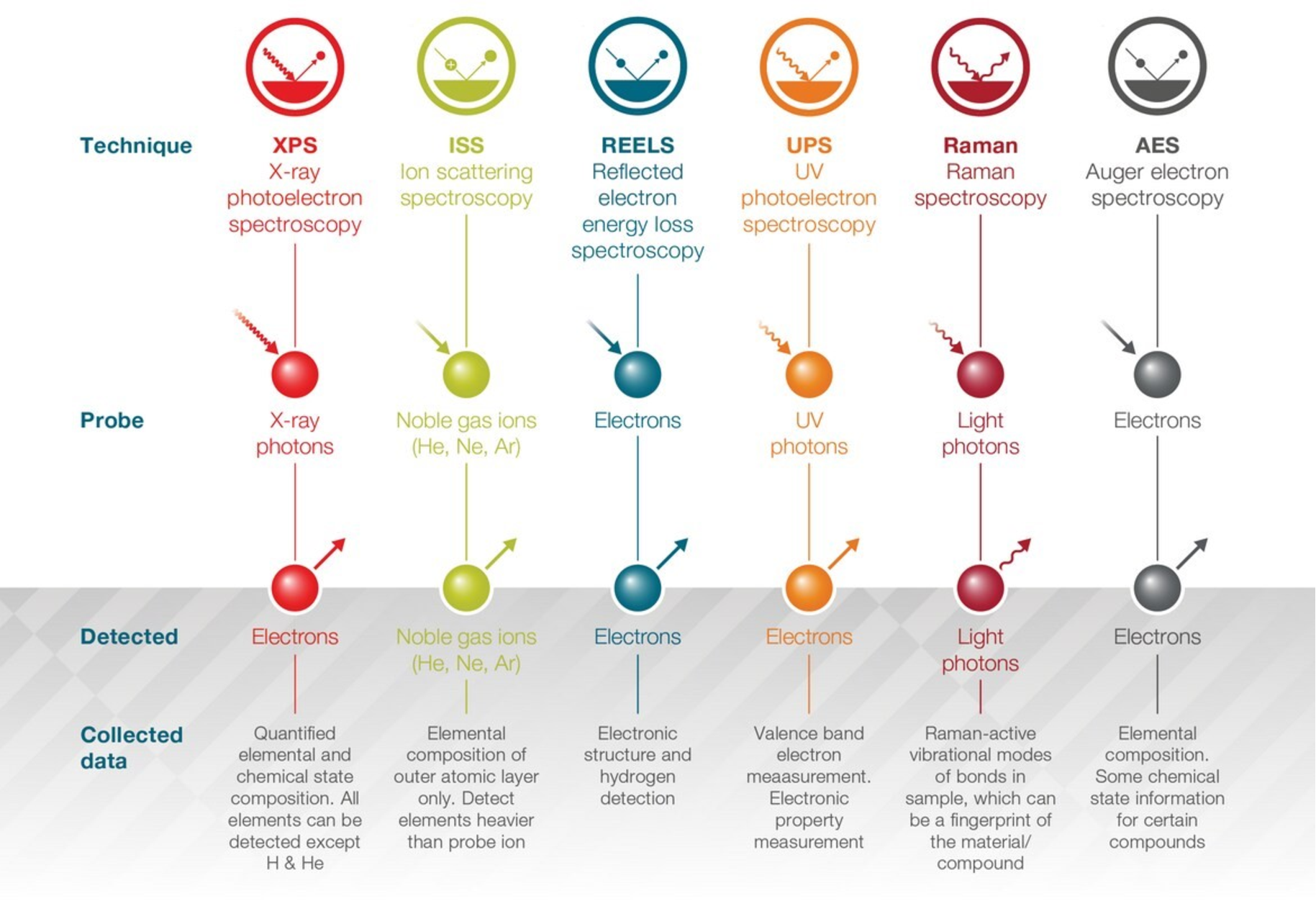}
\caption{Different analytical tools to study surface composition and reaction dynamics.}
\label{fig:Validation_Surface_Spectroscopy}
\end{figure}

Irradiation of surfaces, thin films and ices on the surface, as well as the formation of nanostructures through FEBID, have been discussed previously, and all have been subject to MM. Such surfaces and structures may be analyzed by a variety of what are now regarded as standard analytical techniques providing detailed chemical and morphological data, which may be directly compared with MM, and thus, these techniques may be used to validate such formalisms \cite{Sykes_SurfChemAnalysis, Erdogan_SurfCharacteriz}.
Figure~\ref{fig:Validation_Surface_Spectroscopy} summarizes different analytical tools to study surfaces.

\textbf{X-ray photoelectron spectroscopy (XPS)} has already been discussed above in relation to exploring the spectroscopy of atoms and molecules but is also commonly used to explore surface chemistry \cite{Krishna_2022_ApplSurfSciAdv}. The surface sensitivity of XPS reveals the surface chemistry of the first few nanometers of the surface at a level that other routinely used analytical techniques cannot. XPS analysis can be extended into a material through a process known as depth profiling, which slowly removes material from the surface using an ion beam, collecting XPS after each layer is removed. Depth profiling enables measuring a composition profile with a high-depth resolution. Using depth profiles, one can see how the composition changes from surface to bulk (e.g. due to corrosion or oxidation of the surface) or to understand the chemistry at interfaces between different materials.

\textbf{UV photoelectron spectroscopy (UPS)} is a technique similar to XPS, but in UPS, photoelectrons from the surface are excited by UV photons rather than X-ray photons \cite{Whitten_UV-PES_ApplSurfSci_2023}. As the kinetic energy of UV photons is lower than of X-rays, the detected photoelectrons are from the valence states, which provides a `fingerprint' of the surface species. Usually, XPS and UPS data are collected in tandem to allow more detailed chemical analysis. UPS can also be used to measure the work function of many surfaces.

\textbf{Auger electron spectroscopy (AES)} employs an electron beam to measure the surface composition \cite{Lannon_AugerSpectrosc_Surfaces}. The Auger emission process is caused by the relaxation of an atom after an electron has been emitted. The vacancy in an electron shell is filled by an electron from another orbital, and the extra energy released in this process causes the emission of another electron whose energy is detected. AES provides elemental and some chemical state information, complementing XPS.

\textbf{Ion scattering spectroscopy (ISS)} or \textbf{low-energy ion scattering (LEIS)} is a technique used to probe the elemental composition of the first layer of a surface \cite{Strehblow_2021_JElectrochemSoc}. As a probe, ISS uses a beam of noble gas ions scattered from the surface; their kinetic energy is measured similarly to EELS. As the energy of the incident beam, the mass of the ion, the scattering angle, and the energy of the scattered ion are known, conservation of momentum can be used to calculate the mass of the surface species. Since this interaction can occur only with the outermost surface layer, ISS is very effective and is used to investigate surface segregation and layer growth, complementing the composition information from XPS.

\textbf{Reflected electron energy-loss spectroscopy (REELS)} may be used to probe the electronic structure of the material at the surface \cite{Wang_ReflectEM_SurfAnal}. EELS as a spectroscopic tool has been discussed in Section~\ref{fig:Validation_Surface_Spectroscopy}. The incident electrons can lose energy by exciting electronic transitions in the sample, and these energy losses are measured in the REELS experiment, allowing properties such as electronic band gaps or the relative energy levels of unoccupied molecular orbitals to be measured.

\textbf{Reflection absorption infrared spectroscopy (RAIRS)}. In RAIRS, infrared radiation is directed onto the sample and reflected from an underlying metal surface \cite{Biliskov_2022_PCCP.24.19073}. The signal can be absorbed by bulk ice molecules, surface ice molecules, and the molecules adsorbed on the ice surface. The substrate is usually chemically inert, so it does not directly affect the behavior of the surface layer, but it will produce characteristic features in the IR spectrum. A photoabsorption spectrum is obtained by comparing the ice spectrum on the surface with a `background' spectrum of the substrate alone, similar to gas phase Beer-Lambert law measurements (see Section~\ref{sec:Validation_Exp_Quantum}).

\textbf{Raman spectroscopy} is highly sensitive to structural changes induced by incident irradiation and is commonly used to understand molecular bonding in materials \cite{Kudelski_2009_SurfSci.603.1328}. It is a scattering technique with photons (typically in the infrared to UV wavelengths) from a laser source being used. Photons undergo Raman scattering, losing energy through exciting vibrational modes of the molecules on the surface. These scattered photons are detected, and the chemical species from which they have scattered is derived through energy shift. Raman spectroscopy is a more sensitive technique than RAIRS.

Raman and RAIRS are complementary and have proven especially useful for studying polymers (where the bulk information complements the surface information) and nanomaterials, such as graphene and carbon nanotubes.

\textbf{Mass spectrometry}. When electron or ion beams are used to probe the surface, material may also be `sputtered' from the surface. When an ion is released from the surface, the method is commonly called Secondary Ion Mass Spectrometry (SIMS) \cite{Vickerman_ToF-SIMS_book}; when neutrals are released, the process is called Sputtered Neutral Mass Spectrometry (SNMS) \cite{Vad_NeutralMS_2009}, with the neutrals being subsequently ionized to be analyzed in a mass spectrometer (see Section~\ref{sec:Validation_Exp_Quantum}). SIMS and SNMS are used for depth analysis because of their high-depth resolution, achieving depth resolutions of less than 1~nm. This resolution allows studies of individual layers and can be used on various materials, including ceramics, metals, and semiconductors.

\textbf{Areas of Application}:
All these spectroscopic techniques are widely used for surface studies and form the basis of most surface science laboratories. Many different surfaces, from meteorites \cite{Ferus_2020_Icarus.341.113670} to dental materials \cite{Kaczmarek_2021_Materials.14.2624}, have been studied.

\textbf{Limitations and Challenges}:
The major limitation common to all of these spectroscopic methods is the spatial resolution such that often, analysis is limited to the whole surface rather than resolved parts of the surface. Also, as stated above, these methods cannot provide a temporal analysis of the changing surface, which requires many seconds for complete analysis. Therefore, alternative methods are needed for studies of transient and reactive species.

\subsubsubsection{Electron Microscopy}

In order to provide spatial information, the probe beam must be smaller than the region to be probed. Thus, the probe beams must be focused when using charged particle beams, which requires high energies, as at low energy, space charge leads to the divergence of the particles through mutual interactions. Instruments exploring smaller scales are categorized as `microscopes', but today they may provide chemical information as well as simple images of the surface.

The most common `microscopes' using electrons are the \textbf{Scanning Electron Microscope (SEM)} and \textbf{Transmission Electron Microscope (TEM)}, both of which can provide chemical data, while an \textbf{Atomic Force Microscope (AFM)} and a \textbf{Scanning Transmission Electron Microscope (STEM)} may provide nanoscale structural information \cite{Hunter_PracticalElMicroscopy}. Ion beams may also be focused and provide similar nanoscale and chemical data instruments commonly called ion (nano)probes. Indeed, ion beams may provide even higher resolution than electron beams because these heavier particles have more momentum, such that an ion beam has a smaller wavelength than an electron beam and suffers from less diffraction.

\begin{sloppypar}
Scanning electron microscopy combined with energy-dispersive X-ray spectroscopy (SEM/EDX) \cite{Goldstein_ScanningElMicroscopy} is a widely used surface analytical technique. High-resolution images of surface topography are produced using a highly focused, scanning primary electron (PE) beam. PEs in the beam enter a surface with an energy of $0.5 - 30$~keV, generating many low-energy secondary electrons (SEs). The intensity of SEs depends strongly on the surface topography of the sample. Therefore, an image of the sample surface can be constructed through the measurement of SE intensity at different positions of the PE beam. High spatial resolution is achievable by focusing the PE beam to a very small spot with a size of $<10$~nm. Details of the topographic features on the outermost surface ($<5$~nm) are achieved using a PE beam with an energy of $< 1$~keV.
\end{sloppypar}

In addition to low-energy SEs, backscattered electrons and X-rays are generated due to PE irradiation. The correlation of the backscattered electron intensity to the atomic number of the chemical element within the studied volume provides some qualitative elemental information. The characterization of X-rays emitted from the sample (by means of EDX, also known as Energy Dispersive Spectroscopy -- EDS) gives more quantitative elemental information. Such X-ray analysis probes analytical volumes as small as 1~$\mu$m$^3$.
SEM, accompanied by the X-ray analysis using EDX, is considered a relatively rapid, inexpensive, and non-destructive approach to the analysis of surfaces. It is often used to survey surface analytical problems before proceeding to more surface-sensitive and specialized techniques, such as the spectroscopic methods discussed above.

\textbf{Areas of Application}:
Electron microscopy is widely used, but in recent years, it has been deployed to study nanostructures fabricated by FEBID and FIBID \cite{Trummer_2019_ACSApplNanoMater} (see Section \ref{sec:Intro_Ex_FEBID} and case studies presented in Sections~\ref{sec:Case_study_3D-nanoprinting} and \ref{sec:Case_study_TESCAN}). In the case of FEBID or FIBID, the final nanostructures may be characterized in several ways, for example, by their physical size (shape, height, width) or their chemical composition (metallic purity of the structure). Such parameters may be derived from MM and thus compared directly with an experiment conducted with the same operational parameters (e.g. same organometallic precursor molecule, well-defined diameter, energy, and fluence of the electron/ion beam).

The metal content in fabricated nanostructures may be a good test of the underpinning model of the precursor electron/ion beam dissociation dynamics and the chemical reaction network used. By varying chemical parameters in the model, the sensitivity of the MM approach to atomic and molecular input data may be explored.
The chemical composition of nano-deposits can be determined experimentally by the analytical methods discussed above. XPS coupled with SEM is one of the most detailed methods where a focused electron beam is directed onto the deposited nanostructure to provide chemical information on the nanoscale. Focused ion beams may also be used to derive such data.

\begin{figure}[t!]
\includegraphics[width=0.9\textwidth]{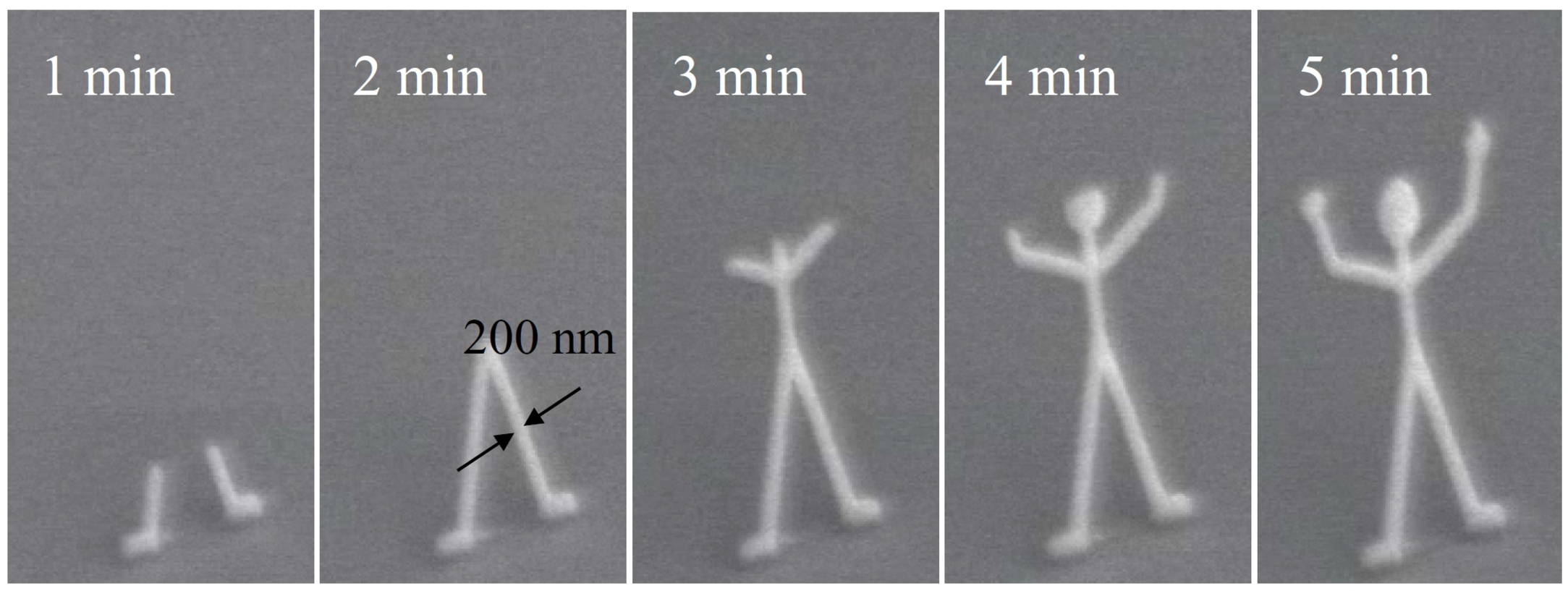}
\caption{Construction of a nano human using  FIBID providing a test of time evolution of the process; adapted from ref~\citenum{Kometani_2009_SciTechnolAdvMater}.}
\label{fig:Validation_FIBID_man}
\end{figure}

The diameter of structures fabricated by FEBID and FIBID is believed to be strongly dependent upon the SE flux. Thus, if the MM prediction matches the measured size of the product nanostructure and its growth rate (see Figure~\ref{fig:Validation_FIBID_man}), this may suggest that those parts of the MM that define the SE energy and flux are `correct'.
Experimentally, the shape of the fabricated nanostructure may be monitored \textit{in situ} with the same electron/ion beam used for the deposition since the electron source is commonly the same as used in a TEM. This approach also enables monitoring of the nanostructure growth rates, as illustrated in Figure~\ref{fig:Validation_FIBID_man}.

Comparison of such growth rates with MM provides another method for validating the MM of the whole nanofabrication process. Apart from the shape, surface-science imaging methods, such as AFM, SEM or TEM, also provide information about the internal structure of the deposits, e.g. a continuous or grain-like distribution of metal in the deposit. The morphology of the deposit is another characteristic, apart from the metal content, which can be used to validate the MM of the FEBID/FIBID process.

Apart from established experimental techniques for monitoring the FEBID/FIBID deposits, new approaches have been developed to bring insight into the deposition process. One example is the focused electron beam-induced mass spectrometry (FEBiMS) technique \cite{Jurczyk_2022_Nanomaterials.12.2710}, which enables the analysis of charged fragments generated on the substrate at real-time FEBID conditions without post-irradiation. Data obtained in such experiments can be used to validate the MM predictions for the irradiation phase of the deposition process.

\textbf{Limitations and Challenges}:
Since the penetration power of the electron beam is very low in any electron microscope, the studied object should be ultra-thin. An electron microscope should operate in a vacuum meaning living cells and liquid surfaces cannot be imaged. The images may suffer from some distortions due to surface charging and inherent aberrations in the electron optics (chromatic, diffraction and spherical).

\subsubsection{Larger-Scale Processes}
\label{sec:Validation_Exp_LargeScale}

The last (fifth) stage of the multiscale scenario depicted in Figure~\ref{fig:MM_diagram} corresponds to the large-scale processes. It arises in an irradiated system after it reaches the chemical equilibrium. Once the irradiated system has reached chemical (and thermodynamic) equilibrium, it may be characterized by its bulk phenomena. For example, FEBID and FIBID-derived nanostructures may have specific electrical and magnetic properties, while irradiated materials may have new thermal or structural characteristics, which may be a predicted outcome of the MM. The conductivity can be experimentally monitored, e.g. by real-time four-point resistance measurements, thus representing an additional way of validating MM. Similarly, the MM can predict the magnetization of structures, which can be validated experimentally.

The irradiation-induced phenomena may also manifest themselves on longer timescales, such as hours, weeks, or even years. For example, irradiation of materials may lead to the formation of defects leading to microscale cracks that can lead to longer-term material creep and fatigue. This is particularly important for studying materials for radioactive waste storage and in the choice of materials for use in space environments where high radiation doses can cause such effects, which are then further propagated by strong temperature variations, such as those encountered on the lunar surface.

The modeling of fatigue and creep failure scenarios necessitates a description of the relationship between the stress, strain, time, temperature, and damage of the material or structure. Several models have been developed, and some are available commercially. Fatigue testing is routine in materials engineering, and parameters such as yield may be compared with models \cite{Lepore_2021_EngFailAnal}. Coupling MM of the irradiation-induced phenomena in materials to bulk phenomena such as fatigue and creep remains a future challenge.

Another example of longer time scale processes is RADAM of cellular DNA and cellular structures such as the lipid membrane that may lead to functional deterioration on longer time scales. On the macroscale, MM of phenomena underlying radiotherapy has been validated through the calculation of cell survival curves \cite{surdutovich2019multiscale} that may be measured experimentally.
However, there are many parameters and effects that may influence cell inactivation, including intercellular signaling and the bystander effect \cite{Suzuki_2023_JRR.64.824} that may not be accounted for in the MM and may not be easily modeled.

MM may eventually successfully predict the damage to the tumor and its shrinking size, justifying the amount of radiation used. However, changes in the tumor may take many weeks to occur and may also be related to less mathematically controlled parameters, such as age, diet, and even social care \cite{Cheng_2020_Cancers.12.2835}.

\section{Case Studies for Multiscale Modeling of Irradiation-Driven Processes}
\label{sec:Case_studies}

\subsection{Light-Induced Electron Transfer Processes in Biological Systems}
\label{sec:Case_study_QuantumBiol}


\textbf{\textit{The problem.}}
Light-induced electron transfer (ET) processes play a crucial role in biological systems as these processes often provide the core mechanisms for sensory protein activation \cite{ding2021mapping, thamarath2010solid, losi2018blue, kottke2006blue, berlew2020optogenetic}, energy harvesting \cite{badura2006light, krassen2009photosynthetic}, and magnetic field sensing \cite{hore2016radical, ritz2004resonance, pedersen2016multiscale, gruning2022effects, gerhards2023modeling}. ET processes are also involved in detrimental phenomena, including the generation of hydrogen peroxide within cells \cite{husen2019molecular}. A precise understanding of the ET processes poses a significant challenge for modern biophysics. The spatial/temporal scales of these processes are typically at the nanometer/nanosecond scale, respectively, while the resulting impact is expected to be at the cellular scale. The involvement of ET reactions in diverse biological systems has been demonstrated \cite{sjulstok2015quantifying, moser1992nature,delaLande2012quantum, solovyov2012decrypting}, however, direct observation of these reactions under controlled experimental conditions remains difficult. In the last decades, it became clear that experimental studies alone are insufficient to elucidate the intricate details of ETs at an atomistic level, which is often necessary for a comprehensive description of the underlying biophysical mechanisms.

\begin{figure}[t!]
\includegraphics[width=0.9\textwidth]{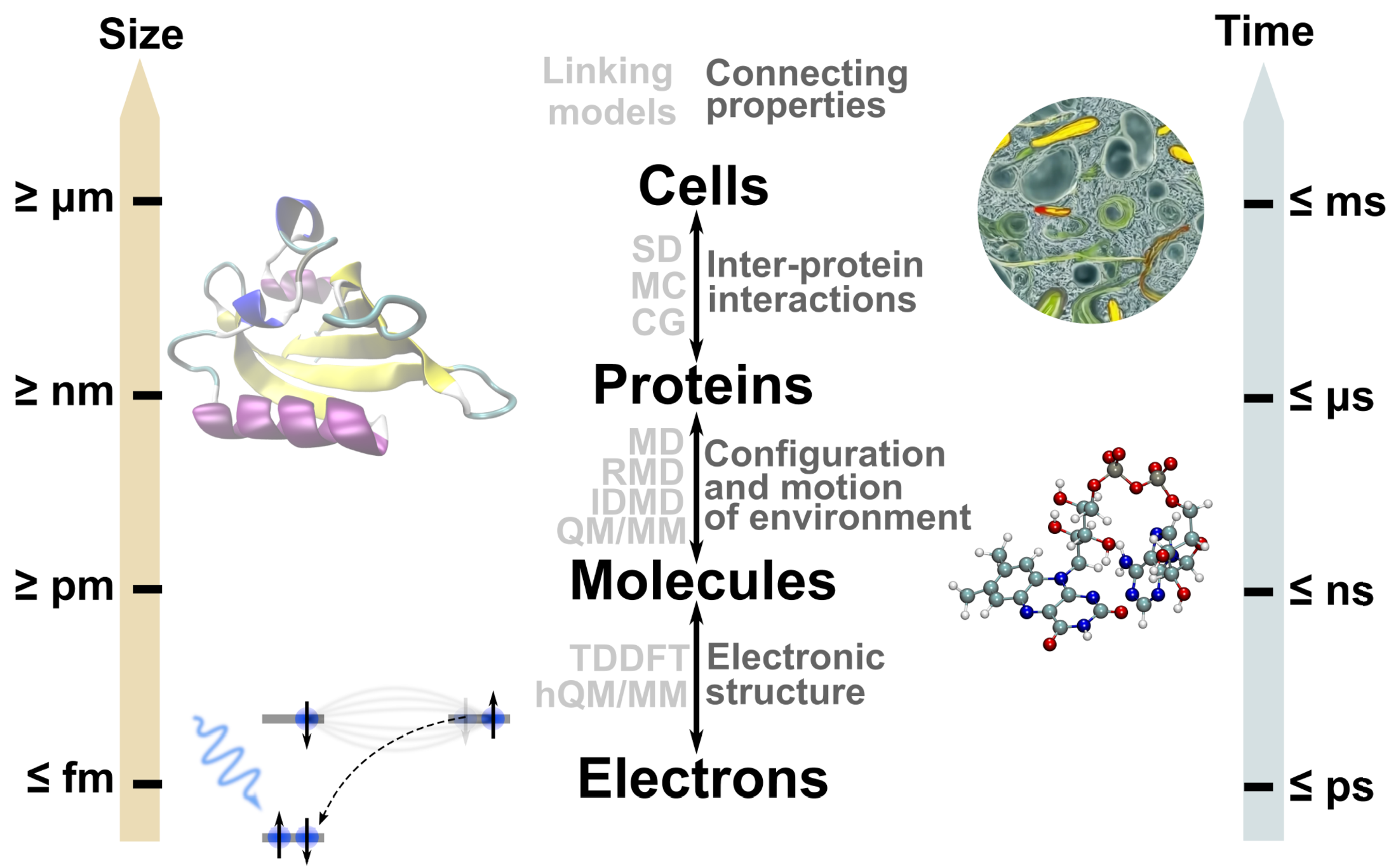}
\caption{Lengths (left) and time (right) scales of light-driven ET in biological systems. Light grey text shows the methods which link different scales, while dark grey text illustrates the connecting properties between the different scales. (SD = Stochastic dynamics, MC = Monte Carlo, CG = Coarse Graining, MD = Molecular Dynamics, RMD = Reactive MD, IDMD = Irradiation-driven MD, QM/MM = Quantum mechanics/Molecular mechanics, TDDFT = Time-dependent density functional theory, hQM/MM = Hybrid QM/MM).}
\label{fig:CaseStudy_ISolovyov_1}
\end{figure}

Light-induced ET processes are prominent in specific sensory proteins \cite{ding2021mapping,hore2016radical, solovyov2012decrypting, solovyov2014separation} and provide another example of a multiscale phenomenon as illustrated in Figure~\ref{fig:MM_diagram}. Indeed, a link between quantum mechanics (light absorption) and large-scale protein function spans many spatial and temporal scales, see Figure~\ref{fig:CaseStudy_ISolovyov_1}. Computational models provide robust approaches to characterize ET processes at these different scales \cite{moser1992nature, delaLande2012quantum, solovyov2012decrypting, solovyov2014separation}. An efficient interconnection of the approaches is essential for the description of the ET processes. Like for the solid-state systems \cite{gerhards2021quantum,gerhards2022theoretical}, it is crucial for light-induced ET processes in proteins to model the active site where light absorption and ET occurs quantum mechanically (quantum region). At the same time, the interaction of this quantum region with the rest of the molecular system is deterministic for accurately describing the underlying processes \cite{frederiksen2022quantum, nielsen2018absorption, timmer2023journal, guallar2008mapping, stevens2018exploring, ko2013charge}. Here, the interaction between the quantum region and the protein structure itself, the surrounding solvent and dissolved ions must be considered \cite{frederiksen2022quantum,nielsen2018absorption}. Furthermore, it is crucial to account for the continuous motion of the protein as an ET is a dynamic process and cannot be described statically.

Different time scales are involved in light-induced ET processes in biological systems. In proteins, light is typically absorbed by a specific chromophore \cite{zeugner2005light, losi2012evolution,stepanenko2011modern}, leading to electronic excitation. This process is often considered rapid and occurs nearly instantaneously, following the Franck-Condon principle \cite{lax1952franck}. Various examples of chromophores are found in nature. For instance, the light-receptor phototropin, present in plants like \emph{Avena sativa} (oat), contains the blue-light chromophore flavin mononucleotide (FMN) within the LOV-domain (light-oxygen-voltage). Phototropin regulates phototropism, participates in the phototaxis of chloroplasts, and contributes to the stomatal opening process \cite{ding2021mapping, christie2007phototropin, briggs2001phototropin}. Another example is the protein rhodopsin, which is found in rod cells of the retina and plays a vital role in regulating visual phototransduction by absorbing green-blue light ($\sim$500~nm) through the chromophore 11-cis-retinal \cite{mroginski2021frontiers}. Recently, flavoproteins called cryptochromes have received attention \cite{kottke2006blue,hore2016radical,ritz2004resonance,pedersen2016multiscale,gruning2022effects,gerhards2023modeling}. Cryptochromes are involved in several biological processes, including circadian rhythms \cite{matysik2023spin}, sensing of magnetic fields \cite{hore2016radical, ritz2004resonance, pedersen2016multiscale, gruning2022effects,gerhards2023modeling}, phototropism, and light capture in general, mediated by their chromophore flavin adenine dinucleotide (FAD) \cite{sjulstok2015quantifying}.

Light absorption in some cases may trigger electron transfer (for example, within the LOV-domain of phototropin \cite{matysik2023spin}), which may require timescales ranging from picoseconds \cite{timmer2023journal} to nanoseconds \cite{vanwonderen2021nanosecond}. ETs, in turn, trigger conformational changes required for protein activation. The latter process typically spans from nanoseconds to microseconds \cite{schuhmann2021exploring}. Therefore, the primary challenge for the computational modeling of ET in biological systems is to effectively connect these vastly different time scales in one multiscale description to guide the experiments where ET can only be observed through indirect measurements \cite{ludemann2015solvent}. Figure~\ref{fig:CaseStudy_ISolovyov_1} illustrates the multiscale nature of light-induced ET processes in biological systems and highlights the characteristic spatial and temporal scales of the processes involved.

\begin{figure}[t!]
\includegraphics[width=0.75\textwidth]{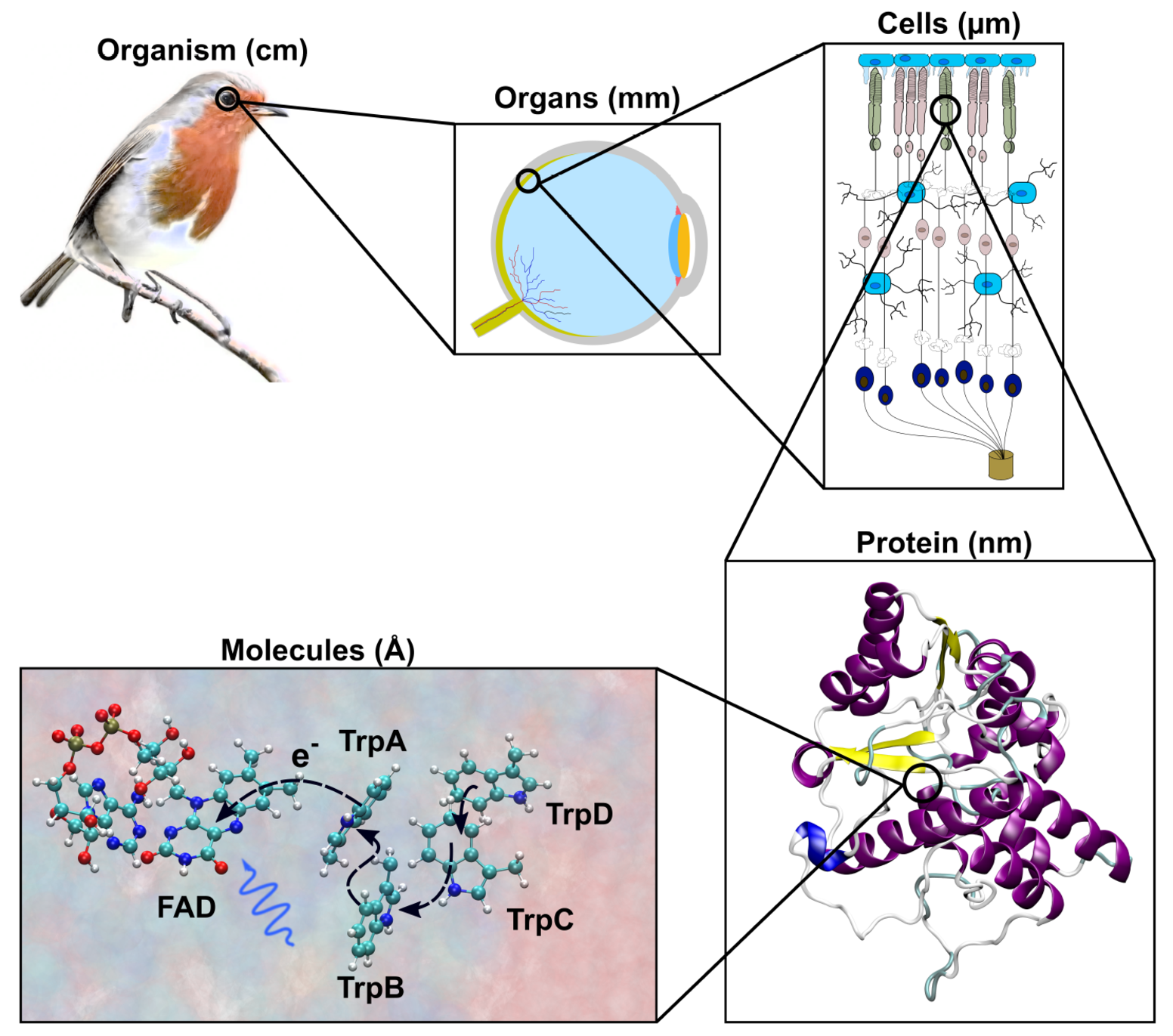}
\caption{Length scales and mechanism important for magnetoreception of migratory species. At the bottom length scale, the flavin adenine dinucleotide (FAD) is excited via blue light, triggering an ET transfer cascade from adjacent tryptophan residues (TrpX, X = A-D), leading to a correlated radical pair which interacts with the geomagnetic field. Spin-selective reactions eventually lead to conformational changes within the protein, which triggers a signal cascade over several length scales, leading to an action mechanism of the migratory species.}
\label{fig:CaseStudy_ISolovyov_2}
\end{figure}

A more specific example that demonstrates how MM can help advance a biological problem is related to a phenomenon called magnetoreception. In this example, one considers migratory species believed to sense the geomagnetic field through the light-induced formation of radical pairs (RPs) within a sensory cryptochrome (Cry) protein through efficient ET. The process is illustrated in Figure~\ref{fig:CaseStudy_ISolovyov_2}. Here, blue light is absorbed by the co-factor FAD inside Cry, which triggers a cascade of ET processes through adjacent tryptophan (Trp) residues. The resulting \(\left[\text{FAD}^{\bullet-} \ \text{TrpH}^{\bullet+} \right]\) RP exists in a non-equilibrium thermodynamic state where it can exhibit singlet or triplet spin character. The distinct spin states of Cry result in varied cellular behavior by enabling the formation of spin-selective reaction products. Interaction with the geomagnetic field affects the nonequilibrium RP dynamics which results in modulation of the spin-selective reaction products in Cry by magnetic field, which on a longer scale provides a sensory signal detected in the brain \cite{hore2016radical, ritz2004resonance, pedersen2016multiscale, gruning2022effects, gerhards2023modeling}.

Recently, cryptochrome 4a (ErCry4) from the night-migratory European robin (ErCry4a) was expressed and analyzed using transient absorption spectroscopy (TAS) \cite{timmer2023journal}. The study revealed the presence of light-induced radicals with lifetimes exceeding 100~ns \cite{timmer2023journal}. However, the sequential charge transfer steps could not be resolved experimentally due to limited time resolution \cite{timmer2023journal}. Here, computational modeling was the only way to investigate the possibility of a charge transfer.

However, achieving a precise description of the dynamics of the critical ET process within Cry necessitates combining different time and length scales (Figure~\ref{fig:CaseStudy_ISolovyov_1}). Only by incorporating these factors one can accurately interpret experimental techniques such as TAS or pump-probe experiments \cite{timmer2023journal}, enabling a comprehensive understanding of light-induced ET processes.

\textbf{\textit{How can MM address the problem.}}
The magnetoreception phenomenon and its association with cryptochrome (Cry) discussed below are illustrative examples for investigating light-induced ET processes in biological systems. However, the methodologies and techniques described below are universal and, therefore, can readily be applied to any biological system. MM offers an ideal framework for addressing the challenges associated with light-induced ET in biological systems, enabling the consideration of complex protein environments, such as Cry, and the configurational changes within the protein itself. This approach accurately describes the dynamics of ET in active sites, such as from Trp to FAD in Cry \cite{hore2016radical,ritz2004resonance,pedersen2016multiscale,gruning2022effects,gerhards2023modeling,ludemann2015solvent}, see Figure~\ref{fig:CaseStudy_ISolovyov_2}. Quantum mechanics and molecular mechanics combination, known as QM/MM, can be employed to incorporate protein motions, configurational changes, and solvent interactions using MD simulations. This hybrid approach allows calculating ET rate constants \cite{timmer2023journal,ludemann2015solvent} or absorption spectra \cite{nielsen2018absorption}, which can be compared to experimental data.

To achieve such a multiscale description, an embedding potential based on MD simulations with time trajectories spanning several hundred nanoseconds is required and usually relies on snapshots taken from a MD trajectory \cite{gruning2022effects,nielsen2018absorption,timmer2023journal}. These trajectories provide information about conformational changes in the protein and possible solvent effects that need to be considered in the QM calculations of interest \cite{nielsen2018absorption,timmer2023journal}. Additionally, the nuclear configurations of the active sites at different time instances can be used to account for vibronic effects in the QM description of the ET process.

Various methods have been developed in recent years to combine the different time and length scales inherent in MM and effectively model the environment's impact on the active sites in protein experiencing ET. For instance, environment can be represented using point charges \cite{gerhards2022theoretical,frederiksen2022quantum} or adjusted using polarizable embedding \cite{nielsen2018absorption,bondanza2020polarizable}, depending on the complexity of the surrounding.

For a quantum mechanical description of electron dynamics, TDDFT has emerged as a reliable technique for accurately capturing electron behavior \cite{lopata2011modeling,provorse2016electron,pedron2020electron}. This approach, for example, has been successfully used to compute absorption spectra for several selected Crys proteins and directly compares with experimental results \cite{nielsen2018absorption}. The TDDFT approach, however, only permits computations on time scales up to (sub)-picoseconds, which dictates that the method must be limited with some other approaches that take account for the slow processes in protein dynamics. Another method, the hybrid QM/MM density-functional tight-binding (hQM/MM DFTB) approach, introduced by L\"{u}demann \textit{et al.} \cite{ludemann2015solvent} in Cry of Arabidopsis thaliana (AtCry1), can be used to directly model light-induced electron propagation in proteins, permitting to evaluate the associated ET rate constants. This method is much more suitable to describe ET processes up to nanoseconds and was successfully used in several case studies \cite{ludemann2015solvent,nielsen2018absorption}.

For instance, Timmer et al. \cite{timmer2023journal} demonstrated the use of the hQM/MM DFTB approach in evaluating the ET cascade of the Trp chain in ErCry4 (see Figure~\ref{fig:CaseStudy_ISolovyov_2}), predicting the characteristic ET times to be between $60-960$~ps for the ETs between Trp sites. These computational predictions matched the results of the extensive TAS measurements \cite{timmer2023journal}.

Additionally, semiclassical Marcus-like theories can be employed to predict rate constants of light-induced ET processes by combining properties from MD simulations with quantum chemical calculations of the system's initial and final states \cite{timmer2023journal,xu2021magnetic,barragan2021theoretical}. A related Moser-Dutton theory has recently been used to evaluate several electron transfer rate constants in European robin Cry, ErCry4 \cite{xu2021magnetic}. Many advances of the theory exist, and notably, Marcus theory has also shown a remarkable potential in describing proton-coupled electron transfer reactions within the Cytochrome bc1 complex \cite{barragan2021theoretical}, which plays a critical role in ATP production.

The combination of mesoscale MD and microscale QM is furthermore crucial not only for accurately describing the interplay between the environment and the active sites but also for capturing the macroscopic statistics observed experimentally \cite{timmer2023journal, ludemann2015solvent}. Given the high complexity of the biological systems, several pathways and configurations may exist that ultimately lead to successful ETs. MD simulations may deliver sampling of the protein configurations over several hundreds of nanoseconds, which is often sufficient statistics of the ET rates to be established, and further justified through comparison with experimental observations \cite{timmer2023journal}. A highlighted example here is the ET in ErCry4a, where Timmer et al. demonstrated that various configurations of cryptochrome can inhibit the charge transfer cascade at the TrpB site (Figure~\ref{fig:CaseStudy_ISolovyov_2}) due to configurational stabilization, leading to an incomplete execution of the ET cascade \cite{timmer2023journal}.

Incorporating the microscopic details of ET dynamics and the mesoscale dynamics of the protein within a QM/MM approach is currently the most comprehensive way to understand the underlying principles of light-induced ET processes in biological systems.

\textbf{\textit{Future directions for 5-10 years period.}}
Significant progress has been made in comprehending ET processes in biological systems \cite{sjulstok2015quantifying, frederiksen2022quantum, timmer2023journal, stevens2018exploring, ludemann2015solvent}. However, several challenges persist, necessitating the application of MM.

One crucial aspect is the assessment of charge-separation lifetimes within proteins, as this feature is vital for subsequent reactions in activated proteins. For instance, within Cry, the lifetime of the radical pair plays a pivotal role in sensing the weak geomagnetic field and is estimated to be about a microsecond \cite{ritz2004resonance}. Achieving such timescales solely through all-atom MD simulations poses serious challenges. Modern techniques, such as coarse-grained (CG) methods \cite{joshi2021review} or SD \cite{Stochastic_2022_JCC.43.1442}, can be explored to overcome the existing problems (see Figure~\ref{fig:CaseStudy_ISolovyov_1}). In the SD approach, the ET transfer rates can be directly integrated into the framework, enabling the study of macroscopic dynamics influenced by the considered rates calculated on a microscale. In recent studies, it was demonstrated that this approach can be directly applied to macroscopic systems, including multiple reactions, diffusion, and other rates to align sufficiently with experimental observations \cite{Stochastic_2022_JCC.43.1442}.

Utilization of IDMD \cite{Sushko_IS_AS_FEBID_2016} can also be employed to incorporate ET rates directly into all-atom MD simulations. This allows for studying time-dependent charge transfers and their impact on intrinsic structural changes in proteins. For instance, it was shown that accurate modeling of electron irradiation-induced bond breaking and the formation of metal nanostructures on a surface could be made using the IDMD approach \cite{Sushko_IS_AS_FEBID_2016, DeVera2020, Prosvetov2021_BJN, Prosvetov2022_PCCP, verkhovtsev2021irradiation}. A similar ansatz may be helpful for studying the chemical dynamics of activated proteins, providing valuable insight into the dynamic behavior and functional mechanisms of ET processes in biological systems.

\textbf{\textit{Envisaged impact.}}
MM is an essential tool for elucidating the intricate dynamics of light-induced ET processes in biological systems \cite{sjulstok2015quantifying}. Integrating multiple size scales enables a comprehensive understanding of the complex cascades that occur in these systems (see Figure~\ref{fig:CaseStudy_ISolovyov_1}). The inherent complexity of the underlying processes presents challenges in quantifying experimental observations.

MM acts as a virtual microscope, facilitating the exploration and comprehension of mesoscopic phenomena arising from microscopic quantum effects that are challenging to observe and quantify experimentally. Through the synergistic combination of diverse computational techniques such as TDDFT \cite{nielsen2018absorption, lopata2011modeling, provorse2016electron, pedron2020electron}, MD \cite{timmer2023journal, ludemann2015solvent} and QM/MM \cite{nielsen2018absorption, timmer2023journal, ludemann2015solvent, bondanza2020polarizable}, MM provides valuable insights into the underlying mechanisms and dynamics driving light-induced ET in biological processes.

\subsection{Radiation Damage of Mass-Selected Biological Molecules}
\label{sec:Case_study_RADAM_biomol}


\textbf{\textit{The problem.}}
Radiation therapy using hard X-rays, MeV electrons or MeV protons and heavy ions is one of the most powerful tools in our battle against cancer. In particular, the initial stages of RADAM to biological systems, i.e. the primary excitation and/or ionization of biomolecular species such as DNA and the subsequent molecular dynamics still need to be understood better. The seminal work of Bouda\"{i}ffa et al.\cite{Boudaiffa_2000_Science.287.1658} showed that resonant attachment of low-energy electrons can lead to DNA single- and double-strand breaks. The notion that molecular mechanisms underlying biological RADAM can be investigated on the single-molecule level gave an intense impulse to the atomic and molecular collision community. The interaction of electrons, ions and photons with gas-phase DNA building blocks such as nucleobases could be studied with unprecedented accuracy using advanced experimental techniques from the atomic and molecular collision community \cite{Coupier_2002_EPJD.20.459, deVries_2003_PRL.91.053401, Hanel_2003_PRL.90.188104}, which led to the successful COST Action P9 ``Radiation Damage in Biomolecular Systems''.

The early studies on gas-phase DNA building blocks not only delivered a wealth of information but also proved the need for increasing complexity in order to be able to study biologically more realistic scenarios. Several groups have pioneered the use of electrospray ionization (ESI) for bringing complex molecular systems from solution into the gas phase \cite{Liu_2006_PRL.97.133401, Milosavljevic_2011_PCCP.13.15432, Gonzalez-Magana_2013_PRA}. The ESI approach has developed into a workhorse of the field, as it opens up a virtually unlimited repertoire of mass-selected gas-phase biomolecular targets for irradiation studies. Examples are short DNA strands \cite{Gonzalez-Magana_2013_PRA}, nanosolvated biomolecules \cite{Liu_2006_PRL.97.133401} and proteins \cite{Milosavljevic_2011_PCCP.13.15432, Lalande_2019_ChemBioChem}, while even more complex targets, such as large protein-DNA complexes, could also be studied easily.

\begin{figure}[t!]
\includegraphics[width=0.9\textwidth]{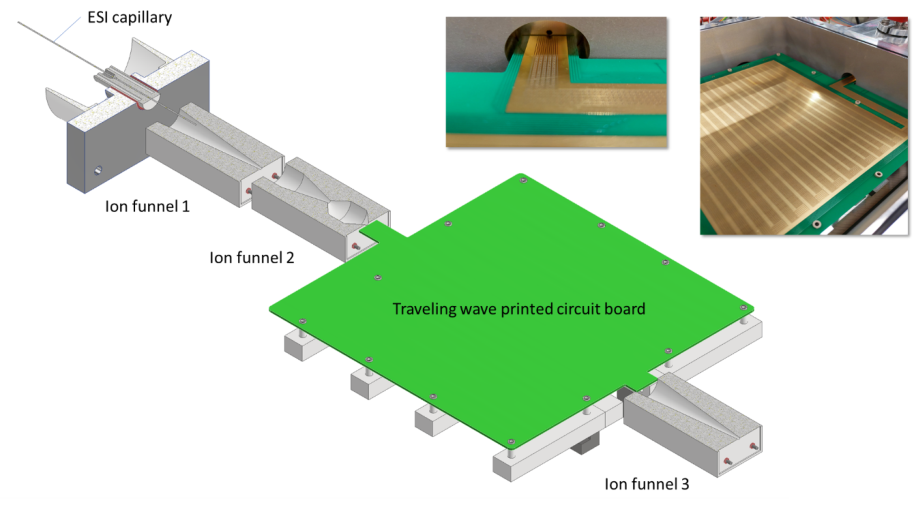}
\caption{Sketch of a traveling-wave ion mobility spectrometer for preparation of conformationally pure biomolecular ion targets, featuring an electrospray ion source, three radiofrequency ion funnels and two printed circuit boards (PCBs) with the traveling wave electrodes. The inset shows photographs of one of the printed circuit boards with pixelated electrodes.}
\label{fig:CaseStudy_Schlatholter}
\end{figure}

The current experimental challenge lies in the fact that most gas-phase biomolecular systems have potential energy surfaces that feature a multitude of local minima. Gas-phase targets therefore tend to contain various conformers, rendering the interpretation of experimental data and comparison to theoretical results very challenging. Several experimental groups are currently developing ESI systems with IMS stages (see Figure~\ref{fig:CaseStudy_Schlatholter}) that will soon be able to deliver mass-selected and conformationally pure biomolecular targets for irradiation studies.

Irradiation-driven dynamics in complex biomolecular systems pose a significant challenge to modelers. Even energetic interactions with gas-phase nucleobases are computationally challenging because of the large number of atoms that constitute such molecules. Therefore, accurate modeling provides a more straightforward way to interpret experimental data. For instance, Maclot et al. have studied the link between energy distribution and fragmentation for keV ion collisions with gas phase thymidine, by comparing experimental data with binding energies and potential energy surfaces from quantum chemical calculations \cite{Maclot_2016_PRL.117.073201}. Wang et al. have shown that high-resolution X-ray spectroscopy data from gas-phase oligonucleotides in combination with quantum chemical modeling allows to track X-ray absorption induced hydrogen transfer processes in DNA \cite{Wang_2022_PCCP.24.7815}.

To date, modeling typically focuses on particular aspects (or timescales) of the biomolecular response to the action of ionizing radiation shown in Figure~\ref{fig:MM_diagram}, such as the actual ionization/excitation process (sub-fs timescale \cite{Palacios_2019_CompMolSci}), hydrogen transfer processes, charge migration and direct bond cleavage (few-fs timescale \cite{Palacios_2019_CompMolSci}), internal conversion and intramolecular vibrational energy redistribution (typically, several-ps timescale \cite{West_2013_JPCA.117.5865}) and subsequent fragmentation processes which can span a wide range up to even longer timescales.

\textbf{\textit{How can MM address the problem.}}
Studies of RADAM to gas-phase biomolecules have benefitted from recent advances in experimental techniques. The response of mass-selected and conformationally pure gas-phase biomolecular targets to the action of ionizing radiation can be studied with an entire arsenal of molecular physics and collision physics techniques that give access to unprecedented details of the irradiation driven dynamics of molecules.

The MM approach is the perfect counterpart that allows one to model the entire cascade of molecular processes under a single umbrella. For the example of X-ray ionization of (nanosolvated) gas-phase DNA, this starts with a localized core ionization or excitation process that occurs on femtosecond timescales and that can, for instance, trigger intramolecular hydrogen transfer (e.g. between nucleobase and deoxyribose), hydrogen transfer to neighboring molecules and other fast rearrangement processes, both of the biomolecule itself or its surrounding molecular environment.  Subsequent internal conversion and intramolecular vibrational redistribution can ultimately lead to statistical processes such as the scission of the DNA backbone or glycosidic bond cleavage, occurring on longer timescales. In particular, entire cascades of chemical reactions can follow for very complex gas-phase systems, such as nanosolvated DNA, DNA-protein complexes, or DNA-based nanosystems.

\textbf{\textit{Future directions for 5-10 years period.}}
Combining the site selectivity of soft X-ray absorption with the recent availability of mass-selected and conformationally pure biomolecular targets allows for experimental studies of the molecular mechanisms of DNA RADAM that can be unambiguously compared to MM data.

Embedding nucleobase analogues containing heavier atoms, such as the F-containing fluorouracil, in a DNA strand makes it possible to selectively target the heavy atom site by excitation/ionization with X-ray at the respective absorption edge. Using synthetic oligonucleotides of different lengths and sequences, charge and energy transport, eventually  leading to direct DNA damage will be studied in unprecedented detail. In addition, the emission of secondary electrons from gas-phase oligonucleotides will be investigated. Fluorouracil and several other nucleobase analogues that can potentially serve as soft X-ray chromophores are also known as potent radiosensitizers. The described studies will, therefore, also help to improve our understanding of the molecular mechanisms of radiosensitization for this particular class of molecules.

Recent experimental data \cite{Li_2021_ChemSci.12.13177} suggests that inner-shell ionization and excitation of DNA results in the emission of more secondary electrons than previously thought, but a characterization and quantification of this effect is still lacking. An important additional aspect is the investigation (both by experiment and MM) of the influence of the DNA molecular environment (proteins, water molecules). Last but not least, the upper size limit of DNA-containing molecular systems that can be studied in the gas-phase has not at all been reached yet. ESI in principle allows to study systems with masses of thousands of atomic mass units. By close collaboration between experimentalists and modelers, new strategies will be developed to find experimental evidence for shock-wave induced ion damage to DNA in gas-phase systems.

\textbf{\textit{Envisaged impact.}}
Improved understanding of the fundamental molecular mechanisms that underly the effects of ionization radiation on DNA will have direct implications for radiation therapy. At present, there are fundamental gaps in our understanding of therapeutically relevant issues, for instance regarding the principles of action of radiosensitizers or of FLASH radiotherapy with protons or heavy ions. The combination of novel gas-phase collision studies on well-defined DNA containing nanosystems with multiscale modeling will help closing these gaps, which will be directly beneficial for further development of radiotherapy modalities.

\subsection{Irradiation-Induced Processes with DNA Origami}
\label{sec:Case_study_DNAorigami}


\textbf{\textit{The problem.}}
The DNA molecules in a human body may instantly become damaged by the direct and indirect effects of ionizing radiation from the surrounding environment while simultaneously undergoing efficient repair processes. Neutron showers caused by cosmic rays in the atmosphere and alpha particles from radon decay are primary sources of the current radiation pollution on Earth. Ionization-induced mutations in DNA influence all the stages of the cell cycle. Therefore, understanding the interaction of ionizing radiation with DNA is crucial for human radiation protection on Earth and during space missions. Furthermore, such an understanding would be instrumental for monitoring cancer evolution and various medical diagnostics and treatment applications. DNA is also an emerging material in nanotechnology \cite{Hu_2019_AdvMater.31.1806294}. Tailored ionizing radiation can pattern or modify DNA-based nanostructures, while natural radiation sources could also damage such nanostructures.

An in-depth understanding of RADAM to DNA is a rather nontrivial task due to the complexity of the DNA and its natural environment. Therefore, the research of DNA damage typically includes several levels of complexity ranging from the isolated DNA building blocks to a free plasmid DNA to post-irradiation analysis of DNA damage in cells and living tissue. The main challenge remains to interlink the results of such studies obtained at different spatial and temporal scales. A huge gap between the studies of DNA building blocks in the gas phase and plasmid DNA studies in solution is worth mentioning here. In order to fill this gap, precisely defined DNA sequences should be studied under controlled conditions. This can be achieved by experiments with DNA origami nanostructures \cite{Rajendran_2012_AngewChemIntEd.51.874}.

\begin{figure}[t!]
\includegraphics[width=1.0\textwidth]{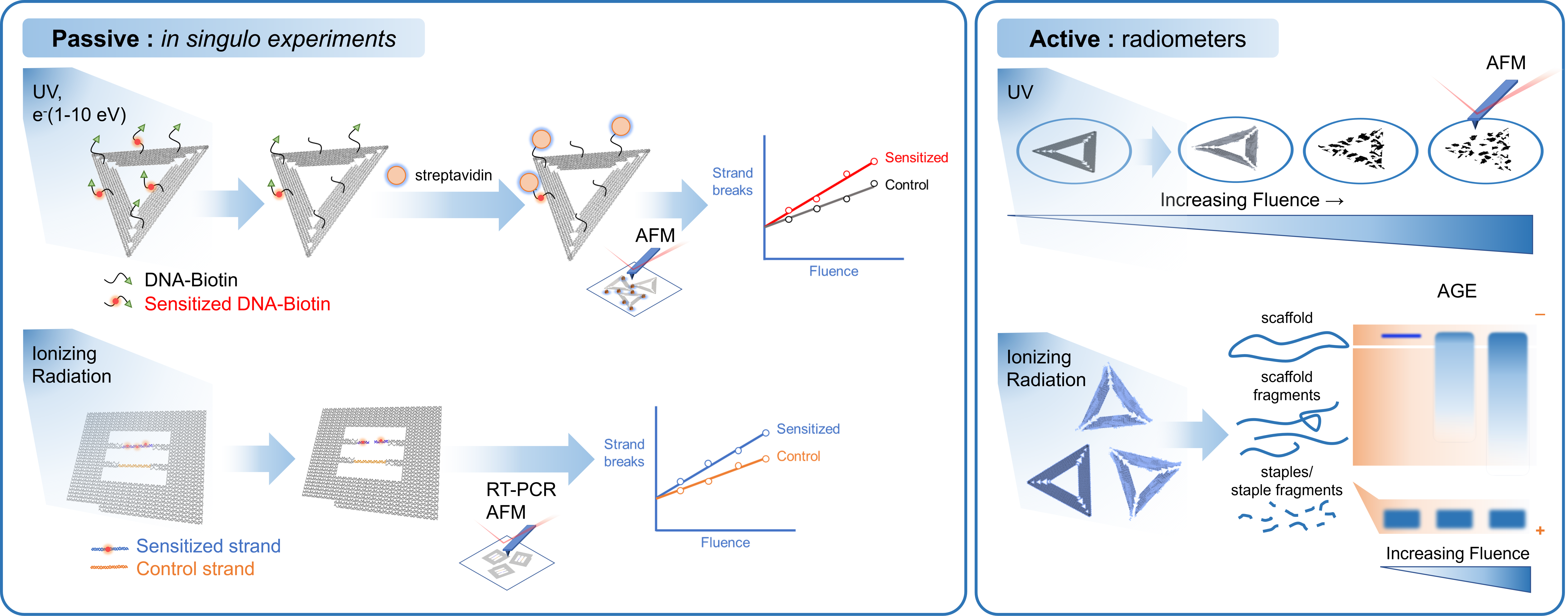}
\caption{Passive and active use of DNA origami nanostructures in fundamental studies of radiation damage to DNA. In the passive mode (left) DNA origami nanostructures serve as platforms to anchor DNA sequences of interest. Atomic force microscopy is widely used to extract information on cross-sections of DNA strand breaks with and without incorporated radiosensitizing molecules (e.g. ref~\citenum{Keller_2014_SciRep.4.7391}). RT-PCR can also be used for irradiation in solution for longer DNA strands such as DNA nanoframes (e.g. ref~\citenum{Sala_2022_JPCL.13.3922}). For higher dose regimes, dose-dependent damage can manifest in the nanostructures, and they can be used as ``nano-dosimeters'' (right); demonstrated for UV irradiation using AFM \cite{Fang_2020_JACS.142.8782} and for proton beam and gamma-ray irradiation \cite{Sala_2021_Nanoscale.13.11197} using agarose gel electrophoresis (AGE).}
\label{fig:CaseStudy_Kocisek}
\end{figure}

The first attempts to use DNA origami nanostructures for studying DNA RADAM could be attributed to Bald and coworkers \cite{Keller_2014_SciRep.4.7391}. More recently, the high stability of DNA origami nanostructures upon irradiation with various types of ionizing radiation was demonstrated \cite{Sala_2021_Nanoscale.13.11197}, paving the way for widening the area of their use in fundamental studies. Two kinds of irradiation experiments can be performed using DNA origami. The first is \textit{in singulo} experiments with precisely defined DNA sequences where the DNA origami is passively used as a substrate. The second includes experiments where the DNA origami nanostructures are used as active ``nano-dosimeters''. A sketch of these two approaches is shown in Figure~\ref{fig:CaseStudy_Kocisek}. In both cases, the interpretation of results critically depends on the outcomes of MM.

\textbf{\textit{How can MM address the problem.}}
Investigations of the interaction of ionizing radiation with DNA on a molecular level goes far beyond the present state-of-the-art. Excited or ionic state dynamics in simple DNA components can be studied by \textit{ab initio} methods, although neglecting the surrounding environment \cite{Matsika_2021_ChemRev.121.9407}. Models such as DNA base pairs or oligomers in realistic environments require a combination of QM and classical approaches \cite{Martinez-Fernandez_2023_DNAphotostability}. In some cases, dynamics on the nanosecond time scales could be probed by machine learning approaches \cite{Westermayr_2019_ChemSci.10.8100}. However, modeling long DNA strands is restricted to the coarse-grained approach \cite{Poppleton_2020_NuclAcidsRes.48.e72}. Coarse-grained modeling can be well used to describe processes of DNA folding in the environment and its mechanical deformation and give insights into environmental effects on DNA secondary structure. This approach, however, cannot be used to model the energy transfer and reactivity on the molecular level. These processes are crucial for describing DNA interaction with ionizing radiation and therefore call for MM, allowing to place DNA in a realistic environment mimicking its structure in biological systems and simultaneously include reactive and energy transfer processes occurring during the radiation event.

\textbf{\textit{Future directions for 5-10 years period.}}
MM has the necessary components to guide the experiments with DNA origami nanostructures and help to interpret the results, which will help better understand the fundamentals of DNA RADAM.

Precisely designed benchmark studies with DNA origami nanostructures should be performed to test the assumptions underlying the MM approach. Examples include irradiations using precisely defined ionizing radiation sources ranging from low-LET photons to high-LET ions, different ion charge states reproducing the effects such as electron capture during projectile penetration through matter, or dose dependencies to cover the whole range of interactions from under limit exposure to flash radiolysis. At the same time, DNA origami will need to be irradiated in different environments to allow the estimation of effects spanning over different scales of complexity, such as shock waves \cite{Surdutovich_AVS_2014_EPJD.68.353}. These benchmark studies will make MM a powerful tool that would go beyond providing an interpretation of experimental observations to a method with unique predictive power for studying DNA nanostructures.

The long-term vision here includes the possibilities of reactivity studies with large DNA systems and the development of methods to model DNA in its natural environment in chromosomes, paving the way for the incorporation of MM in software used in simulating radiation passage through matter, predicting material properties or creating inputs for state-of-the-art protocols in radiation oncology \cite{Huynh_2020_NatRevClinOncol.17.771} or nanotechnology \cite{Singh_2022_BiotechnolAdv.61.108052}.

\textbf{\textit{Envisaged impact.}}
In medicine, radiation therapy for cancer is booming due to the development of well-targeted ion-beam radiotherapies. So-called pencil ion beams have diameters of several millimeters. Transversal focusing is possible due to increased energy deposition in the Bragg peak of ion projectiles. Even though the technique is highly developed, modeling for highly targeted treatments is not ideal as the understanding of the underlying processes occurring during the physicochemical stage of radiation interaction with tissue needs to be included. Combining DNA origami experiments and MM may help to identify the most important of these processes.

Apart from the better technology of the ion beams, the targeting can be improved by combined chemo-radiation therapies. The advantage of such treatment is in the synergistic effect, which is the higher effect of chemotherapy combined with radiation compared to the individual treatments. It is well known that secondary DNA structure plays an important role in both sensitizing DNA towards radiation \cite{Rackwitz_2018_ChemEurJ.24.4680} and targeting \cite{Berardinelli_2017_MutatResRev.773.204}. Understanding the interplay between macroscopic properties of DNA concerning secondary structure, folding, interaction with proteins and molecular-level damage is crucial for the rational design of novel radio-theranostic agents.

Finally, the studies may impact the use of DNA origami as a material and a tool in bio-nanotechnology. The enhanced stability of origami against ionizing radiation compared to bare DNA predetermines the use of DNA origami in DNA storage or bio-nano-electronics. Their flexibility in design, already fully employed in biomedicine \cite{Keller_2020_AngewChemIntEd.59.15818}, can be used to develop novel chemo-radiation therapies. Nanostructure preparation techniques such as DNA origami-assisted lithography, nanopatterning of DNA origami by ion beams, ion implantation of DNA origami, or DNA origami-assisted deposition of metallic nanostructures could reach maternity thanks to the detailed modeling of the output nanostructures. With MM as a predictive tool, such preparation techniques could become a widely spread bio-nanotechnology routine.

\subsection{Attosecond and XFEL Science Applied to Complex Biomolecules and Clusters}
\label{sec:Case_study_Attosecond}


\textbf{\textit{The problem.}}
At the beginning of the 21st century, as it became possible to synthesize ultrashort light pulses with durations below one femtosecond \cite{Paul_2001_Science.292.1689}, photophysics and photochemistry entered an unexplored area in which the non-equilibrium properties of electrons can be tracked down to the angstrom length scale. This development was accompanied by the emergence of free electron lasers, which provide short and highly intense XUV X-ray pulses, offering the possibility to study non-linear interaction at extreme wavelengths, possibly down to the attosecond regime. The so-called ``attosecond science'' was born, and since then, it has attracted a lot of attention from the scientific community worldwide, essentially because it creates a new paradigm in which the properties of matter -- a molecular reaction, a phase transition or an electronic property -- can be manipulated by directly acting on electrons therefore modifying their localization and interaction with angstrom precision \cite{Drescher_2002_Nature.419.803}.

Attosecond pulses combine two properties that make them particularly suitable to study the interaction between matter and ionizing radiation. Firstly, attosecond pulses are so short that they can track the dynamics of charges (electron, hole, protons, nuclei) with ultrahigh time resolution using pump-probe spectroscopy. Secondly, attosecond pulses are generated in the XUV or X-ray domain, therefore they are by essence ionizing radiations with photon energy far above the ionization potential of molecules. As a consequence, attosecond science provides by definition direct insight into the first steps of the interaction between ionizing radiation and matter. Attosecond science can track the non-equilibrium properties induced by the oscillation of light electric field, ionization, electron scattering, coherent charge (electron/hole) dynamics, proton motion, etc.

Over the past 20 years, enormous efforts have been devoted to the development of highly performing attosecond light sources. In parallel, physicists have developed spectroscopic techniques that can benefit from these sources. Starting primarily with simple isolated atomic targets, experiments have shown the possibility of observing coherent ultrafast electronic wavepackets in atoms \cite{Goulielmakis_2010_Nature.466.739} and soon electron scattering at the atomic scale could be measured \cite{Schultze_2010_Science.328.1658}. While highly sophisticated experiments have been developed to address fundamental aspects of atomic physics, there also has been a constant race to investigate increasingly complex objects such as molecules or clusters \cite{Lepine_2014_NatPhot.8.195}. The first attosecond pump-probe experiment in molecules \cite{Sansone_2010_Nature.465.763} was performed in a simple molecule H$_2$ in which the electron localization in a dissociative H$_2^+$ cation was controlled. Soon after, the first attosecond experiment in more complex (polyatomic) molecules was developed \cite{Neidel_2013_PRL.111.033001}. By studying molecules such as N$_2$, C$_2$H$_4$ and CO$_2$, the instantaneous light-induced polarization of the electronic density could be observed. The first work on an isolated amino acid showed that an attosecond XUV pulse could create an ultrafast hole wavepacket in phenylalanine \cite{Calegari_2014_Science.346.336}.

\begin{figure}[t!]
\includegraphics[width=0.85\textwidth]{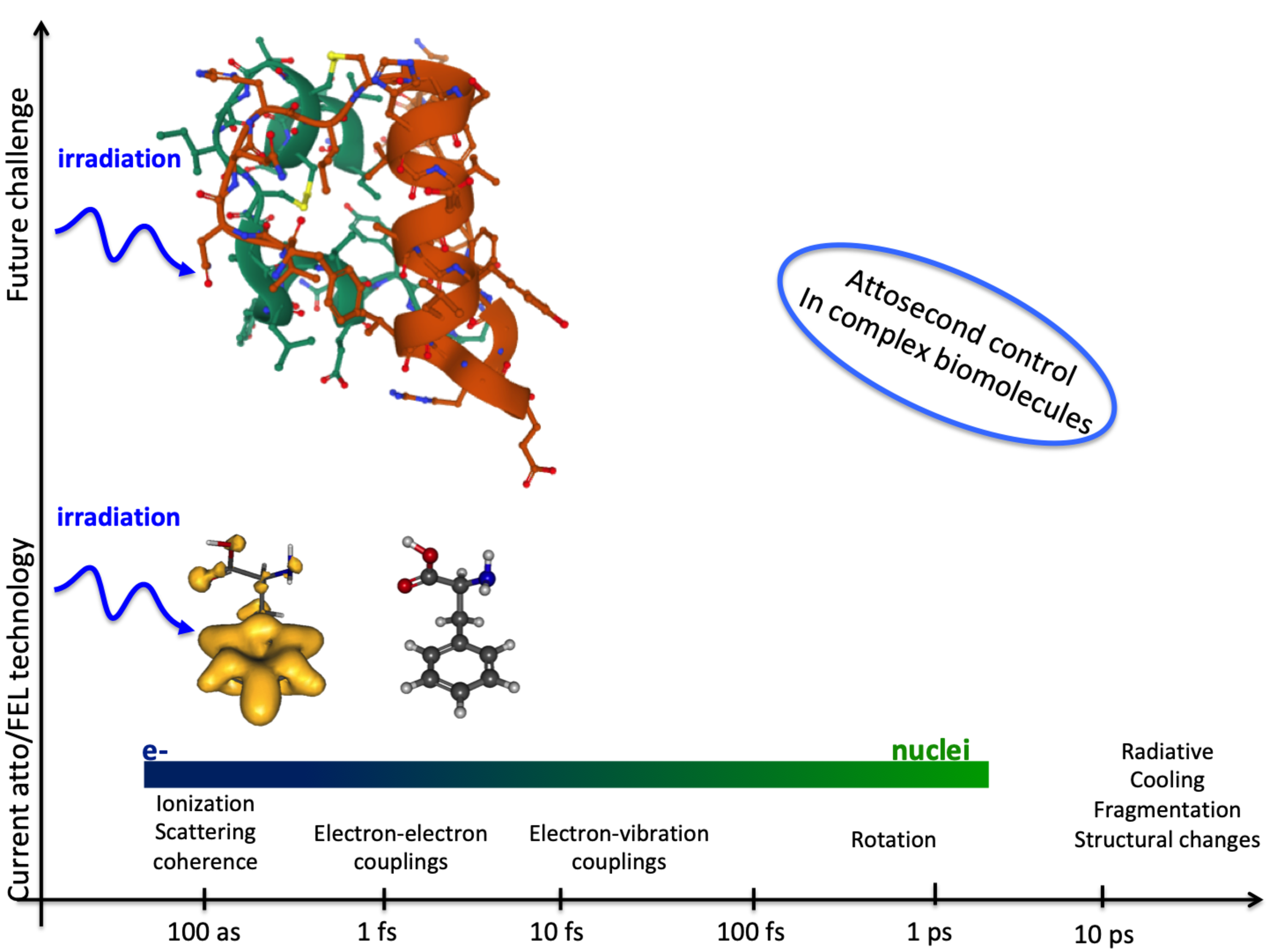}
\caption{Schematic representation of the multiscale character and challenges encountered in attosecond control in complex biomolecules.}
\label{fig:CaseStudy_Lepine}
\end{figure}

These proof-of-principle experiments have motivated further developments to study more complex objects, and future investigations will consider the control of molecular properties using attosecond pulses in increasingly complex systems such as a biomolecule or a cluster. However, attosecond technologies are still under development, and \textit{ad-hoc} spectroscopies are still limited. Therefore, the development of attosecond science in complex systems faces two challenges: First, the observables provided by experiments are not direct and do not provide a simple image of the studied dynamics. When dealing with increasingly complex systems, it is nowadays only possible to comprehend the dynamics with the support of sophisticated theories that connect the experimental observables measured on macroscopic timescales and the ultrafast microscopic dynamics. Second, manipulating electronic properties with angstrom precision is expected to significantly impact the macroscopic properties of matter. For instance, in the case of a photo-induced reaction, acting on electron localization will impact the way the chemical bond will rearrange and the molecule will transform. Understanding these processes requires to connect processes occurring at the attosecond timescale and angstrom length scale to processes occurring at femto- and picosecond timescales, where vibration and isomerization start, and macroscopic time scales where the chemical transformations occur. Consequently, attosecond control is by essence a multiscale problem (see Figure~\ref{fig:CaseStudy_Lepine}), especially when it deals with complex molecules.

\textbf{\textit{How can MM address the problem.}}
Because it is impossible to directly infer the physical process from the measurements, MM must provide the necessary translation of the experimental data into actual dynamics. MM is also crucial to learn how actions on an attosecond timescale can provide means to modify properties on a macroscopic timescale. For instance, MM could provide direct information on how charges are created in DNA upon irradiation after XUV or X-ray irradiation and how electron scattering, charge transfer and proton rearrangement occur. From these first steps, non-stationary properties will emerge, and structural rearrangement of the entire molecule will occur. How the first initial attosecond interaction will determine the final structure of the molecule could be addressed by MM.

\textbf{\textit{Future directions for 5-10 years period.}}
In order to develop attosecond and XFEL science in complex (bio)-molecules and clusters, it is crucial to connect ultrashort timescales where the interaction with light occurs and macroscopic observables where the consequence of the interaction is observed. From the experimental point of view, the interaction between attosecond pulses and complex biomolecules, such as an entire peptide or protein, is still in its infancy \cite{Herve_2022_AdvPhysX.7.2123283, Herve_2022_SciRep.12.13191}. Current theoretical developments deal with the description of the light--matter interaction where the time-dependent light electric field is explicitly considered. The light first interacts with electronic degrees of freedom and, depending on its frequency, outer valence, inner valence and inner shell electrons must be included in the description. At high photon energy, the accurate inclusion of electron correlation becomes an important issue. As a consequence of the interaction, quantum coherence and related charge/hole migration mechanisms need to be properly described, so it is for decoherence of the charge wavepacket \cite{Vacher_2017_PRL.118.083001, Despre_2018_PRL.121.203002}. Following this first excitation step, the inclusion of the nuclear degrees of freedom becomes essential, especially in terms of post Born-Oppenheimer treatment that includes a strong coupling between the electronic degrees of freedom and the vibrational ones \cite{Herve_2021_NatPhys.17.327}. For an increasingly large system, the problem becomes untreatable. In the coming years, there will be a need to develop approaches to identify the relevant degrees of freedom to decrease the dimensionality of the problem and to be able to include a limited number of particles in the MM simulation. The use of AI approaches is one possible way. An explicit treatment of time dependent light matter interaction, electron/hole dynamics at attosecond timescale and how it connects to nuclear dynamics and chemical transformation is therefore under development and applied to small systems. MM approaches will provide conceptually more potent and general tools to develop these approaches for biomolecules, clusters, and large molecular systems.

\textbf{\textit{Envisaged impact.}}
While structural biology has become a mature field of research, studying out-of-equilibrium properties of biomolecules is now a subject of high interest as it provides information on the inner functioning of biological processes. Time-resolved X-ray diffraction experiments provided by FEL installations offer the perfect tools to study these properties; however, light-induced ultrafast damages remain one of the major limitations to the development of these approaches \cite{Berrah_2019_NatPhys.15.1279}. From the combination of attosecond technology and MM, one will learn irradiation conditions to limit damages and predict highly non-linear effects encountered at FEL \cite{Wabnitz_2002_Nature.420.482}.

Like in femtosecond science before, where laser technologies offered new means to structure matter- providing tools for surgery and material machining- it is expected that attosecond science will provide similar breakthroughs in chemistry, biology and material science by allowing us to act directly on electrons, improving new means of control by reaching higher temporal and spatial resolution. In that context, MM is crucial to guide the experimental protocols and pave the way for developing attosecond technology applied to complex objects.

\subsection{Clusters in Molecular Beams}
\label{sec:Case_study_Cluster_Beams}


\textbf{\textit{The problem.}}
The ultimate goal of MM is to characterize macroscopic environments under irradiation. However, in specific cases, the bulk behavior is well approximated by aggregates of tens or hundreds of molecules, i.e. molecular clusters.
This is especially true if one is interested in chemical changes induced by irradiation. The interaction of radiation with one active molecule initiates a chain of subsequent events. For a cluster to represent a suitable mimic system, its size has to be sufficient to account for all the relevant relaxation events and chemical reactions. The second crucial factor is the composition of clusters. It should correspond as closely as possible to the macroscopic system it should represent, which is often a challenging task. However, when these two assumptions are met, the cluster-beam experiments provide an excellent way of benchmarking and validating the MM models since the experimental data (e.g. the fragmentation mass spectrum) are comparable with the results of the MM on a one-to-one basis.

Environments where a comparison of cluster-beam experiments with MM has been bringing an unprecedented level of insight into the irradiation-driven chemistry include (but are not limited to):
\begin{itemize}

\item \textit{Radiation damage.}
Biomolecular clusters have been used to bridge the experiments with isolated biomolecules in the gas phase and experiments studying the behavior of large biomolecular systems in macroscopic solutions and even biological tissue.\cite{baccarelli11, fabrikant17}. Most attention focused on the role of water molecules surrounding the biomolecules undergoing radiation-induced damage. An important effect revealed in this way is the suppression of molecular fragmentation by low-energy electrons \cite{kocisek16_uracil}. However, this relatively easy-to-understand phenomenon is not the only influence of clustering. Opening of new reaction pathways, such as the proton transfer upon electron attachment \cite{allan07_formic} or the dissociation of a glycosidic bond in nucleotides \cite{kocisek18}, have been observed experimentally. These experiments have pointed out the crucial role of the environment and the limited usability of data obtained at single-collision conditions.

\item \textit{FEBID and FIBID.}
Clusters of organometallic precursors have been used to reveal aggregation effects on the substrate during focused-beam nanofabrication \cite{postler15, lengyel17}. While the substrate is absent in this type of experiments, its elementary influence (presence of a heat bath) can be well approximated using large rare-gas clusters which serve as a nano-support for precursors or their clusters. The experimentally observed phenomena include, for example, the self-scavenging of electrons which changes the electron-energy range relevant for ligand dissociation \cite{lengyel16_scaveng}; suppression of the dissociative electron attachment (DEA) due to polarization screening \cite{lengyel17}; or the reactive role of water admixtures in organometallic precursors \cite{Lengyel2021}. The cluster-beam experiments can be directly compared with surface-based studies where thin condensed layers of precursor molecules are irradiated by electrons, and chemical changes in these layers are analyzed by various surface-science techniques \cite{Massey_Sanche2015, landheer11} (see Section~\ref{sec:Validation_Exp_Chem_Equil}).

\item \textit{Astrochemical synthesis on ice nanoparticles}.
Astrochemical ice and dust grains offer surfaces for chemical reactions in the interstellar medium, and the radiation serves as a trigger for such reactions (see Section~\ref{sec:Case_study_Space_Chemistry}). Irradiation-driven chemistry in space can be explored through experiments with clusters with a chemical composition similar to the ice mantles of the interstellar grains \cite{farnik21_persp}. Here, an additional advantage of clusters (created by a supersonic expansion) is their low internal temperature resulting from the evaporative cooling. Again, the comparison with astrochemically motivated surface-based experiments \cite{bohler13, Arumainayagam_2019_ChemSocRev} is straightforward.
\end{itemize}

\begin{figure}[t!]
\includegraphics[width=1.0\textwidth]{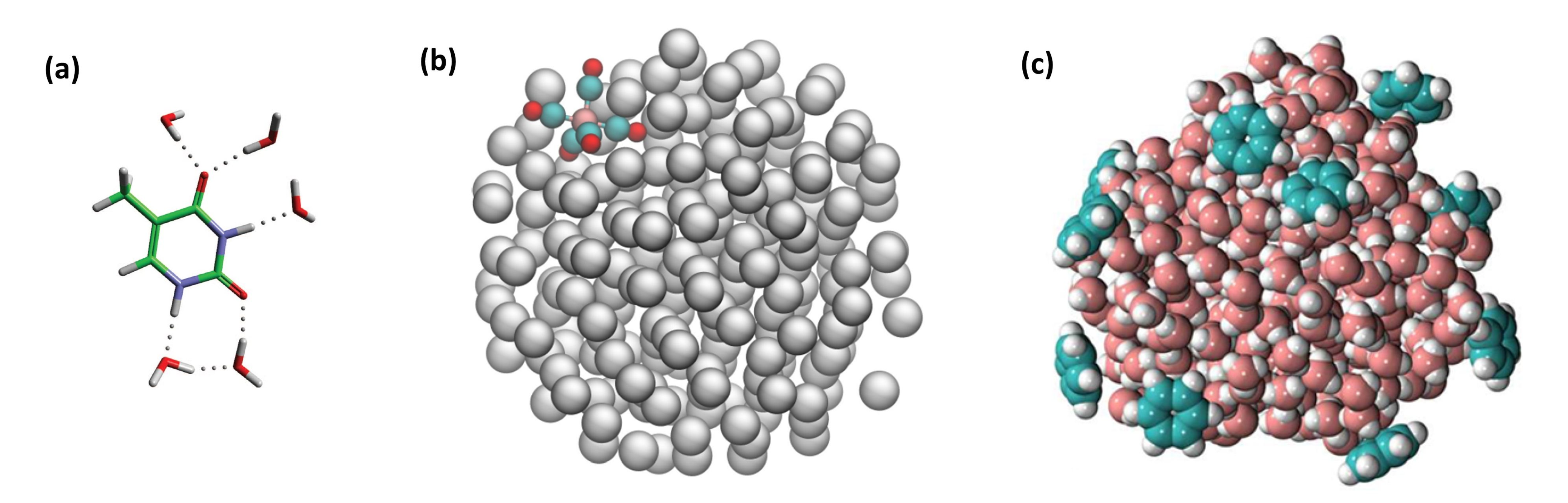}
\caption{Examples of clusters recently used for mimicking various environments. (a) Microhydrated thymine \cite{smyth14}, (b) Fe(CO)$_5$ embedded into a large argon cluster \cite{Andreides_2023_JPCA.127.3757}, (c) multiple benzene molecules adsorbed on a water cluster as a model system for interstellar ice nanoparticle \cite{pysanenko15}. The individual panels are scaled arbitrarily.}
\label{fig:CaseStudy_Fedor}
\end{figure}

From the point of view of cluster physics, these different problems share one aspect in common: a need for advanced experimental methods for the production of heteromolecular clusters (see Fig.~\ref{fig:CaseStudy_Fedor}). Such systems can be created either by a co-expansion of various samples or by first producing homomolecular clusters and then picking up guest molecules on them. Several pickup cells can be arranged in series, resulting in the formation of clusters, which are good proxies for chemically diverse environments. Such a cluster beam is then typically ionized by the required type of ionizing radiation, e.g by electron impact, ion impact or tunable X-rays. The primary experimental information is typically the fragmentation pattern (mass spectrum) (see Section~\ref{sec:Validation_Exp_Quantum}).

\textbf{\textit{How can MM address the problem.}}
It might be instructive first to define the timescales of interest. The typical mass-spectrometric detection time is units to tens of microseconds after the ionization. The detection times are prolonged to milliseconds (or longer) in the case of storage ring experiments. So, even if the MM approach is not necessary required for this range of  system sizes, it is needed because of the temporal scales involved. As outlined in Section~\ref{sec:Methods}, the quantum-mechanical treatment (e.g. by means of \textit{ab initio} MD) is typically feasible on the picosecond timescale, and the reactive and classical MD on a nanosecond timescale. For clusters, the longer timescales are typically treated by statistical methods, such as phase-space \cite{Chesnavich_1977_JACS.99.1705, Kato_1996_JCP.105.9502} or transition-state \cite{Garrett_1979_JPC.83.1052} theories.

A showcase example of how MM can address a cluster system relevant for RADAM is the suppression of electron-induced N-H bond cleavage in nucleobases. The isolated nucleobases are susceptible to DEA \cite{ptasinska05}; however, microhydration experiments have shown that the presence of just a few water molecules around the base prevents the fragmentation \cite{kocisek16_uracil}. The fixed-nuclei scattering calculations, in combination with a diatomic-like model for the dissociation, initially suggested \cite{smyth14} that the effect of water should be exactly the opposite (i.e. DEA fragmentation cross section should increase) since the water environment lowers the energy and the width of the electronic resonance responsible for this process. However, when the dissociation dynamics was simulated by the combination of DFT and MD \cite{mcallister19}, the caging effect due to water molecules was clearly revealed. The simulations thus explained the experimentally observed phenomenon. Furthermore, by evaluating the energy transfer from the base to the solvent (\textit{nanocalorimetry} via the combination of experiment and theory), it has been possible to estimate the enhancement of the LET to the environment originating from such a caging.

The multiple time scales are nicely demonstrated, e.g., in a computational study of electron-induced fragmentation of water clusters \cite{suchan22}. Here, the ionization events were identified by a kinetic MC procedure; subsequently, the fragmentation was modeled with classical MD simulations calibrated by non-adiabatic QM/MM simulations, and the fragmentation on microsecond timescale was modeled with a Rice–Ramsperger–Kassel (RRK) model. This combination of approaches yielded a good agreement with experimental mass spectra.

The combination of MM and cluster-beam experiments also brings insight into the problem of precursor dissociation in FEBID. For Fe(CO)$_5$, the most common precursor for the deposition of iron nanostructures, the gas-phase dissociative ionization is very fragmentative: the most abundant peak in the electron-impact mass spectra corresponds to a bare iron ion Fe$^+$. However, upon clustering, the fragmentation degree is considerably suppressed: the dominant channel is the removal of only two ligands from the Fe(CO)$_5^+$ molecule \cite{Lengyel2016_JPCC_2}. The reactive MD approach (see Section~\ref{sec:Methods_RMD}) has elucidated the mechanism of quenching the excess energy of the hot Fe(CO)$_5^+$ cation \cite{Andreides_2023_JPCA.127.3757} by the molecular environment. Such quenching alters the view of the elementary mechanisms that play a role in the FEBID process. Indeed, by comparing the number of cleaved ligands in the surface-based experiments with the gas-phase fragmentation spectra, it has been concluded that the fragmentation is driven by the DEA process \cite{thorman15}. The cluster-beam experiment and MM show that the same fragmentation degree as in bulk occurs in the dissociative ionization combined with the environmental energy quenching.

\textbf{\textit{Future directions for 5-10 years period.}}
In the combination of cluster-beam experiments with MM, there are clear challenges on both sides that should be addressed in the near future. For a direct comparison of the two approaches, it is essential to work with well-defined and controlled cluster targets. The main drawbacks of the present state-of-the-art experiments are: (i) all the experimental techniques can produce clusters with certain size distributions but not with a single size, (ii) the thermodynamic state of the clusters is often unknown, and (iii) it is often challenging to distinguish the pre- and post-interaction effects. All these problems have been receiving active attention in recent years. Better-defined clusters can be produced by an improved control of expansion conditions. A promising approach is the use of electrostatic deflectors to select the neutral cluster species according to their effective-dipole-moment-to-mass ratio \cite{KupperSelection}. Another approach for better defining the target state is the use of helium nanodroplets as the confining medium that cools the clusters down to 0.37~K \cite{mauracher18}.

Apart from the preparation of clusters with controlled composition, there are also experimental challenges on the detection side. A mass spectrum carries information about the fragmentation pattern, while the information about the bonding patterns is missing. Especially interesting is the question of which new covalent bonds are formed due to irradiation. Ion-trapping combined with action spectroscopy or collision-induced dissociation analysis of the fragmentation products can answer this question. A pertinent challenge, mainly in electron-induced reactions, is the detection of the neutral reaction byproducts. MM often predicts complex
rearrangement reactions in neutral dissociation products. While there have been initial attempts to characterize neutrals in electron interactions with gas-phase molecules, such techniques are yet to be implemented for clusters.

From the theoretical point of view, the main challenge is to include effects that have been neglected in the models so far. For reactive MD applied to FEBID precursors, this is, for example, an ``on-the-fly'' change of the bonding parameters. As the precursor molecule loses ligands successively, the bond dissociation energies and other force-field parameters may change, strongly influencing the subsequent fragmentation dynamics. Regarding the theoretical methodology, perhaps the least developed description is that of the systems where the electronic energy is in the continuum when resonances are formed in electron collisions with cluster constituents \cite{fabrikant17}. Here, the standard methods of quantum chemistry cannot be used, and the scattering calculations have to be utilized to parametrize the potentials for the dynamics of nuclei. At the same time, non-local and non-adiabatic effects may play an important role in the dynamics.
While huge advances have been made in this area in recent years in describing the dynamics of isolated molecules in the electronic continuum \cite{dvorak22_co2, kumar_pyrole22}, the application to clusters is yet to appear.

\textbf{\textit{Envisaged impact.}}
The main purpose of the cluster-beam experiments in combination with MM is twofold: (i) to provide insight into elementary irradiation-induced processes and (ii) to validate the MM methods on smaller-size systems. The impact of both of these directions is straightforward. The knowledge of elementary processes is essential for our ability to manipulate and control the outcome of irradiation-driven reactions. An example is the evaluation of the contribution of a specific bond-cleavage process to the LET in RADAM \cite{postulka17}. Validating the modeling methods with cluster experiments greatly enhances their credibility for simulating irradiation-driven macroscopic environments. This will bridge the fundamental gaps in our understanding of such systems and our abilities to utilize them for technological or biomedical applications.

\subsection{Time-Resolved Ultrafast Radiation Chemistry }
\label{sec:Case_study_UltrafastRadChem}


\textbf{\textit{The problem.}}
Linking the dynamic physics and chemistry generated in the immediate aftermath of ionising radiation interactions in matter to long-term chemical and/or biological changes in the medium is a grand challenge. It has the potential to unlock our understanding of some of the most fundamentally important processes in the universe. From the initial femtosecond-scale (fs, $10^{-15}$~s) reaction pathways to radiolytic yield over picosecond (ps, $10^{-12}$~s) and nanosecond (ns, $10^{-9}$~s) timeframes, quantitatively tracking radiation chemistry over multiple spatiotemporal resolutions (see Figure~\ref{fig:MM_diagram}) will allow for the identification and complete characterization of ionizing radiation induced chemical transformations in matter. Further still, such a capability would permit a detailed interrogation of the factors that influence these processes i.e. ionising species, material structure on the nanoscale, instantaneous dose etc.

Along with growing our understanding of ultrafast processes, new knowledge in this ultrafast regime can provide a platform for accessing and developing novel radiation-based technologies. By tracking the evolution of chemical species in real time, researchers can determine the intermediate states, energy transfer pathways, and dynamics involved in radiation seeded reactions. This information will, for example, help unravel the complex network of reactions that underpin water radiolysis and the yield of cytotoxic species relevant for improving different modalities of radiotherapy, including FLASH \cite{Binwei_2021_FrontOncol.11.644400} and hadrontherapy. Moreover, if the ultrafast radiation chemistry underpinning these applications can be tracked in real time it opens the possibility of controlling chemical reactions for applications in diverse fields such as catalysis (radiocatalytic reactions \cite{Zacheis_1999_JPCB.103.2142, Coekelbergs_1962_AdvCatal.13.55, Abedini_2013_NanoscaleResLett.8.474, Roy_2013_AnalChem.80.7504}) and drug discovery (radiation assisted strategies in nanotherapeutics \cite{Zhang_2023_AdvDrugDelivRev}). It is also important in the design of advanced materials where a better understanding of ionising species-dependent processes such as charge transfer, energy transfer, and structural change post irradiation is crucial for exploring enhanced properties, such as improved conductivity, efficient energy conversion, or tailored optical characteristics.

\textbf{\textit{How can MM address the problem.}}
To meet the challenges outlined in the above paragraphs, a considerable effort is now underway to reveal how the spatiotemporal evolution of the instantaneous dose distribution seeds long lasting, or even permanent, chemical and structural changes in matter. MM will be an essential part of the toolkit required to meet these challenges as it will provide a versatile and robust methodology to interrogate interactions as they evolve over multiple resolutions.

However, it is also crucial that experimental methodologies develop in lock step with advances on the modeling side. Firstly, providing raw data with minimum uncertainty as input for the physics packages that MM will be built around is essential. Assumptions about material structure (homogenous vs heterogeneous) and initial response that average over the epoch of ultrafast nanophysics can introduce significant divergence from the true evolution for the interaction as a whole. Secondly, benchmarking and testing the predictive power of MM will be an absolute necessity before high confidence can be established for sensitive applications such as modeling for radiotherapy and nuclear engineering.

Unfortunately, tracking radiation chemistry in real time experimentally is notoriously challenging. This is due to several reasons. Firstly, the spatiotemporal evolution that underpins the transition from initial ionization and excitation, to long-lived chemical species and permanent damage centre formation is inherently multiscale. The implication is that detectors with resolution that span many orders of magnitude i.e. fs to ns ($10^{-9}$~s) and $\mu$s ($10^{-6}$~s) are required. Next, as the initial dose evolves, a sequence of epochs emerges where different intermediate species and collective phenomena grow and dominate. These epochs play a crucial role in determining the final state of the material post-irradiation, but typically require different detection methodologies. One example would be the growth and decay of different photoabsorption bands corresponding to different reactants that partake in the sequence of water radiation chemistry. Overall, the implication is that it is challenging to get a single snapshot of the entire evolution with a high degree of absolute timing accuracy to investigate the interdependence of these species.

\begin{figure}[t!]
\includegraphics[width=1.0\textwidth]{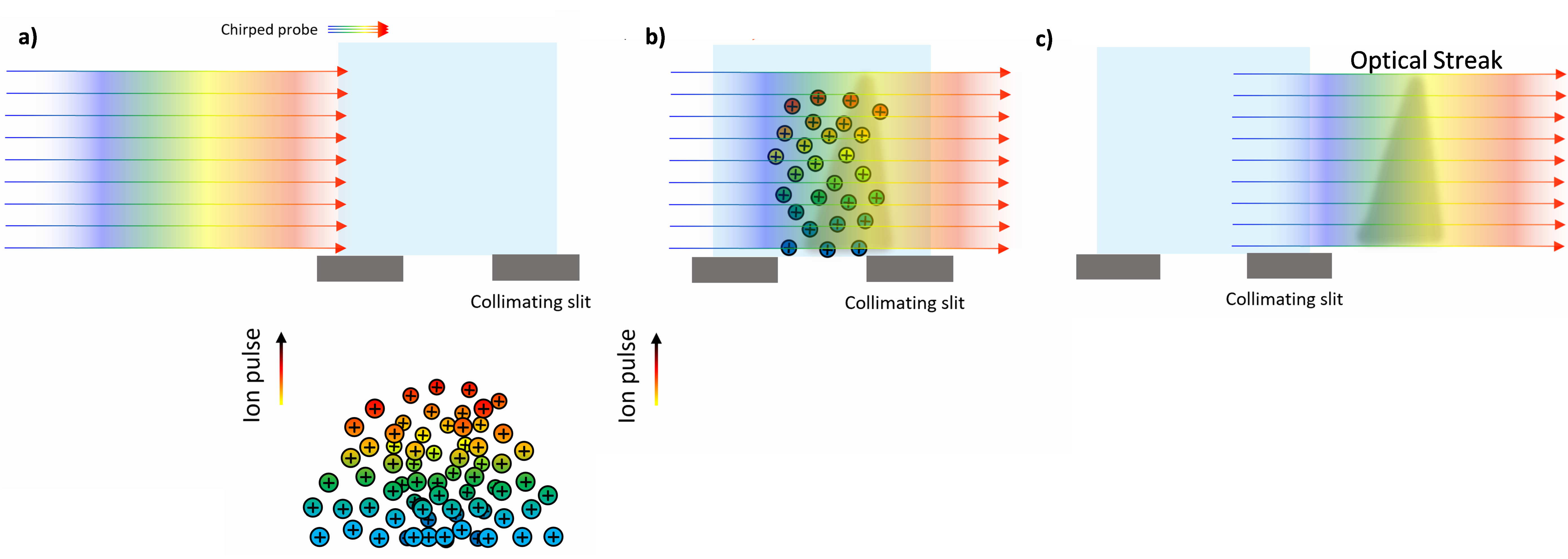}
\caption{ Schematic of the basic principle of chirped pulse optical streaking.}
\label{fig:CaseStudy_Dromey1}
\end{figure}

\begin{figure}[t!]
\includegraphics[width=1.0\textwidth]{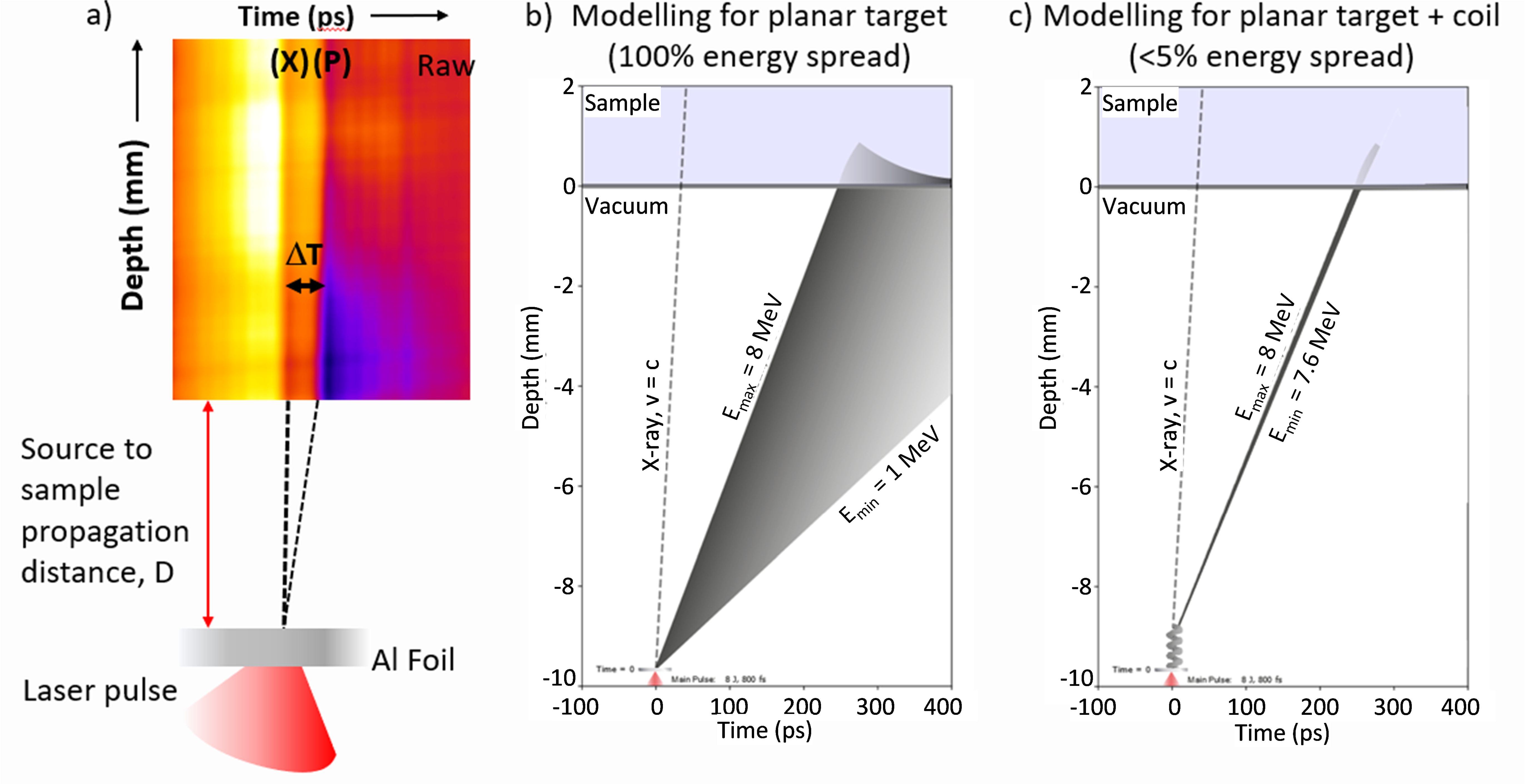}
\caption{Chirped pulse optical streak: current capabilities and upcoming improvements. The dotted line depicted in panels (b) and (c) shows the path of the X-rays in spacetime while the continuous shaded region shows the path in vacuum and stopping of the protons in H$_2$O modeled using the SRIM software \cite{Ziegler_1999_JAP.85.1249}.}
\label{fig:CaseStudy_Dromey2}
\end{figure}

To date, experimental work on ultrafast (fs- and ps-scale) radiation chemistry has focused on radiolysis using electrons \cite{Gauduel_2010_EPJD.60.121} and photolysis \cite{Svoboda_2020_SciAdv.6.3}. While there have been attempts to realize the same performance for ion interactions in matter using scavenging agents, this tends to result in large uncertainty in the resulting analysis as the scavenging agent itself must be considered in the radiolysis for the high concentrations required to access early time frames \cite{Baldacchino_2004_CPL.385.66}. Overall, this approach is required due to the lack of availability of ultrafast (picosecond timescale) sources of ions. In addition, achieving absolute timing in radiolysis using either electrons or ions is notoriously difficult as the electrical jitter associated with conventional radiofrequency accelerators is on the order of tens to hundreds of picoseconds, further adding to this uncertainty. Recently, however, novel experimental approaches using laser driven ion accelerators have opened the field of ultrafast radiation chemistry to proton interactions in matter \cite{Dromey_2016_NatCommun.7.10642}. By exploiting the ultrafast nature of Target Normal Sheath Acceleration (TNSA) and high synchronised probes from the driving laser, it has been possible to perform the first picosecond radiolysis studies for protons interacting pristine H$_2$O (solvated electron dynamics with no scavenging agents \cite{Senje_2017_APL.110.104102, Prasselsberger_2021_PRL.127.186001}) and transparent dielectrics \cite{Taylor_2018_PlasmaPhys.60.054004, Coughlan_2020_NJP.22.103023}. The underpinning methodology is outlined in Figures~\ref{fig:CaseStudy_Dromey1} and \ref{fig:CaseStudy_Dromey2}.

Figure~\ref{fig:CaseStudy_Dromey1} shows a schematic of the basic principle for interrogating ultrafast proton interactions in matter, chirped pulse optical streaking \cite{Dromey_2016_NatCommun.7.10642}. A chirped probe i.e. a linear variation of the instantaneous frequency in time, is incident on the sample (blue) travelling from left to right (transverse direction). At the same time, the ionising species (in this case ions) is incident from the bottom traveling towards the top (Figure~\ref{fig:CaseStudy_Dromey1}a). As the pulses overlap in the sample (see Fig.~\ref{fig:CaseStudy_Dromey1}b) the transient opacity (valence band electrons excited into the conduction band) or photoabsorption due to absorption bands associated with the dynamic yield of chemical species is encoded in the spectrum of the chirped probe as a temporally varying transmission (shaded area of spectrum in Figure~\ref{fig:CaseStudy_Dromey1}c). This spectrum is then analyzed using an imaging spectrometer to return a spatiotemporal image of the changing transmission with respect to depth in the sample. While the transform limited pulse duration of the probe sets the ultimate temporal resolution for this technique, in practice the spectrometer resolution, in combination with the width of the collimating slit in front of the sample, sets the experimental limit. At the same time, the spatial resolution is set by the imaging optics used to transport the beam to the entrance slit of the spectrometer.

Figure~\ref{fig:CaseStudy_Dromey2}a shows an example of raw data from a chirped probe optical streak taken in pristine H$_2$O and illustrates the onset of opacity due to both X-rays (X) and, a time $\Delta T$ later, a broadband pulse of TNSA protons (P). The darker color indicates lower transmission due to photoabsorption by solvated electrons forming post irradiation in the H$_2$O. Time increases left to right and depth into the target increases in the vertical direction. Below this is a schematic showing the laser-target interaction relative to the sample a defined distance, $D$, away. Figure~\ref{fig:CaseStudy_Dromey2}b shows the results of modeling for instantaneous interaction and demonstrates how the bremsstrahlung X-ray pulse provides a global timing fiducial for the interaction allowing absolute time of arrival of the proton beam to be ascertained. The broadband TNSA proton burst arrives $\sim$250~ps later. It is important to note that the sharp maximum energy cut-off in the TNSA spectrum \cite{Macchi_2013_RMP.85.751}, $E_{\rm max}$ (here modeled for 8~MeV), provides a well-defined onset time/leading edge for the opacity.

\textbf{\textit{Future directions for 5-10 years period.}}
Future improvements to the TNSA source are expected that will permit proton beams with less than 5\% energy bandwidth to be deployed for picosecond-scale pulsed radiolysis. This will be achieved using advanced coil targets that allow for the selection and compression of narrow bandwidths of the 100\% energy spread of the original TNSA pulse \cite{Kar_2016_NatCommun.7.10792}. Such an advance will permit the production of much clearer optical streaks, see Figure~\ref{fig:CaseStudy_Dromey2}c. The overarching aim here is to increase the precision achieved in experiments to better match the conditions that will be investigated using MM. The chirped pulse optical technique will be deployed for other laser-driven radiation sources such as heavier ions including carbon and aluminium, brilliant beams of alpha particles from proton-boron fusion, and narrow energy bandwidth beams of MeV electrons. Considering that these measurements will all be made with an absolute timing reference (prompt X-ray pulse), it will allow the first quantitative comparisons of the effects of different species on the subsequent ultrafast radiation chemistry. In the long term, another laser driven source, high harmonic generation from relativistically oscillating plasma mirrors \cite{Dromey_2006_NatPhys.2.456}, will be used to provide highly synchronised probes spanning from optical to X-ray wavelengths. This will provide experimentalists with a suite of broadband probes that can interrogate the dynamics of a wide range of different chemical species all on a single shot. This in turn will feed back into the testing and benchmarking of MM.

\textbf{\textit{Envisaged impact.}}
Advances in ultrafast radiation chemistry will support the overarching drive towards next generation radiation-based technologies by growing our understanding of fundamental processes underpinning ionising interactions in matter. From engineering carrier lifetimes in nanostructured electronics deployed in radiation harsh environments \cite{Ackland_2010_Science.327.1587} to investigating novel modalities for radiotherapy in healthcare \cite{surdutovich2019multiscale}, e.g. NP-enhanced deposition for highly-targeted dose delivery in addition to the development of patient-specific FLASH and hadrontherapy modalities, the ability to transform existing technologies will rely on predicting and controlling the evolution of dose on the shortest spatial and fastest temporal scales post-irradiation. Efficient and thoroughly benchmarked MM will allow for the rapid identification of optimal conditions required to realize these goals. This, in turn, will dramatically narrow the broad parameter space faced when developing their practical implementation, thereby enabling well-informed and cost-effective methodologies to be adopted.

\subsection{Multiscale Approach for the Radiation Damage of Biomolecular Systems by Ions}
\label{sec:Case_study_MSA_SWs}


\begin{sloppypar}
\textbf{\textit{The problem.}}
The elucidation of fundamental mechanisms of ion-induced RADAM of biomolecular and biological systems has attracted strong interest in the past several decades \cite{Garcia_Fuss_RADAM_BiomolSyst, AVS2017nanoscaleIBCT}, motivated by the development of radiotherapy with ion beams \cite{Amaldi_2005_RepProgPhys.68.1861, Schardt_2010_RMP.82.383, Surdutovich_AVS_2014_EPJD.68.353} and other applications of ions interacting with biological targets, e.g., radiation protection in space \cite{Durante_2011_RMP.83.1245, Kronenberg_2012_HealthPhys.103.556}.
\end{sloppypar}

\begin{figure}[t!]
\includegraphics[width=0.55\textwidth]{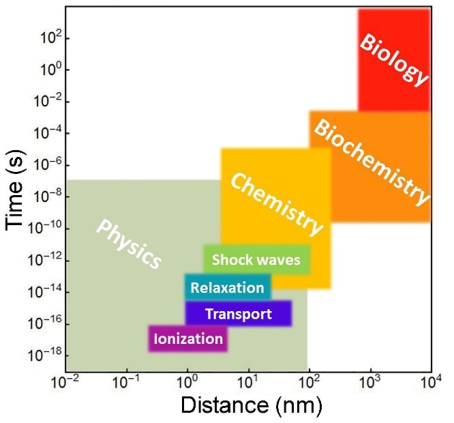}
\caption{A space-time diagram of features, processes and disciplines associated with hadron- / ion-beam therapy, indicating approximate scales of the key physical phenomena. Figure is adapted from ref~\citenum{Surdutovich_AVS_2014_EPJD.68.353}.}
\label{fig:CaseStudy_MSA_diagram}
\end{figure}

An understanding of the cascade of processes induced by irradiation of biomolecular and biological targets by ions and other radiation modalities requires a MM approach \cite{AVS2017nanoscaleIBCT, Surdutovich_AVS_2014_EPJD.68.353} that could bridge the (sub)-nanoscale atomic and molecular physics with the macroscale biophysics, biochemistry, and biology; see Figure~\ref{fig:CaseStudy_MSA_diagram} and Section~\ref{sec:Intro_Ex_IBCT}.

The multiscale approach (MSA) to the physics of RADAM with ions, discussed in Section~\ref{sec:Intro_Ex_IBCT}, has been developed to quantitatively describe the key physical, chemical and biological phenomena underlying molecular-level mechanisms of biological damage induced by ion-beam radiation \cite{AVS2017nanoscaleIBCT, Surdutovich_AVS_2014_EPJD.68.353, surdutovich2019multiscale, solov2009physics}. As described in Section~\ref{sec:Intro_Ex_IBCT}, the key phenomena and processes treated by the MSA are (i) propagation of charged particles in biological media, (ii) radiation-induced fast quantum processes within biomolecular environments, (iii) the time and spatial evolution of track structures and localized energy deposition into biological media, (iv) slower nanoscale post-irradiation relaxation, chemical and thermalization processes occurring in the irradiated biological media, and (v) evaluation of ion-induced biodamage and its link to larger-scale radiobiological phenomena, such as cell survival probabilities, RBE, etc.; see Figure~\ref{fig:Intro_IBCT} and refs~\citenum{AVS2017nanoscaleIBCT, Surdutovich_AVS_2014_EPJD.68.353, surdutovich2019multiscale, solov2009physics}.

An important physical effect predicted by the MSA concerns the significant RADAM arising due to the nanoscale shock waves (SWs) \cite{surdutovich2010shock} created by ions in a dense dielectric medium (such as liquid water or a biological system) at the high LET. As discussed in Section~\ref{sec:Intro_Ex_IBCT}, this phenomenon arises because ions can deposit a significant amount of energy on the nanometer scale, which leads to the strong heating of the medium in the vicinity of ion tracks. The energy lost by the ion is deposited into the medium due to (i) the production, transport, and stopping of secondary electrons and (ii) the relaxation of the electronic excitation energy of the medium into its vibrational degrees of freedom via the electron-phonon coupling mechanism \cite{surdutovich2010shock}.
The case study described in this section continues the case study on the multiscale scenario for the RADAM of biological systems with ions, described in Section~\ref{sec:Intro_Ex_IBCT}, with a particular focus on the ion-induced nanoscale SW phenomenon and the thermomechanical mechanism of RADAM by the ion-induced SWs.

It was demonstrated within the MSA that the ion-induced SWs play an essential role in the scenario of RADAM, see reviews~\citenum{AVS2017nanoscaleIBCT, Surdutovich_AVS_2014_EPJD.68.353, deVera_2019_CancerNano.10.5, DySoN_book_Springer_2022} and references therein. The two possible mechanisms of DNA damage by the ion-induced SW have been suggested \cite{Surdutovich_AVS_2014_EPJD.68.353, surdutovich2013biodamage}. First, the SW may inflict damage by the thermomechanical stress and induce covalent bond breakage in the DNA molecule \cite{surdutovich2013biodamage, Yakubovich_2012_NIMB.279.135, bottlander2015effect, deVera2016molecular, deVera_2018_EPJD.72.147, Friis_2020_JCC, Friis_2021_PRE}. As the strength of ion-induced SWs increases with LET, the SWs substantially contribute to radiation biodamage around the Bragg peak region of ion trajectories \cite{surdutovich2013biodamage}. The recent study \cite{Friis_2021_PRE} showed that thermomechanical stress of the DNA molecule caused by the ion-induced SW is the dominant mechanism of complex DNA damage for high-LET (e.g. iron) ion irradiation, resulting in cell inactivation. Apart from that, the radial collective motion of the medium induced by the SW helps to propagate highly reactive molecular species, such as OH radicals and solvated electrons, to distances up to tens of nanometers from the ion track, thus preventing their recombination \cite{AVS2017nanoscaleIBCT, Surdutovich_AVS_2014_EPJD.68.353}.

The transport of secondary electrons and radicals and RADAM induced by these particles have been commonly studied computationally using track-structure MC simulations \cite{Garcia_Fuss_RADAM_BiomolSyst, AVS2017nanoscaleIBCT} (see Section~\ref{sec:Methods_MC_transport}). However, such simulations consider the transport of particles in a static medium at equilibrium. This transport does not include the complete physical picture shown in Figures~\ref{fig:Intro_IBCT}, \ref{fig:MM_diagram} and \ref{fig:CaseStudy_MSA_diagram} because propagating secondary particles transfer the energy further, making the medium highly dynamic. As shown within the MSA, non-equilibrium dynamics of biomolecular systems in the environment plays an important role in ion irradiation-induced damage to biological systems \cite{Surdutovich_AVS_2014_EPJD.68.353}.

\textbf{\textit{How can MM address the problem.}}
The MSA-based description of radiation-driven biomolecular damage processes accounts for the system size, molecular interactions, radiation dynamics, post-irradiation chemistry involved, and their links to the large-scale biological effects, see Fig.~\ref{fig:CaseStudy_MSA_diagram}. A realistic approach to tackling a problem of such complexity must involve the multiscale theoretical and computational descriptions of the key phenomena (see Fig.~\ref{fig:Intro_IBCT} in Section~\ref{sec:Intro_Ex_IBCT}) and elaborate their major interlinks within a unifying MM framework (see Section~\ref{sec:Interlinks}). Several levels of interlinking should be emphasized: (i) an interlink between quantum-chemistry methods and MD through the development of reactive force fields for RMD simulations \cite{Sushko2016_rCHARMM} (Section~\ref{sec:Methods_RMD} and Section~\ref{sec:Interlinks_QM-MD_RMD}); (ii) interlink between track-structure MC methods and classical MD through the IDMD approach \cite{Sushko_IS_AS_FEBID_2016} (Sections~\ref{sec:Methods_IDMD} and \ref{sec:Interlinks_QM-MD_IDMD}), which permits the efficient simulations of irradiation-driven chemistry processes in complex molecular systems exposed to radiation; (iii) links of ``standard'' MD, RMD and IDMD with SD to simulate various large-scale dynamical processes on a probabilistic level (Section~\ref{sec:Interlinks_MD-SD}); (iv) links between the outcomes of RMD and SD with the statistical methods for the evaluation of probabilities of complex RADAM events, such as the formation of single- and double-strand breaks in DNA, complex DNA damages, etc. \cite{AVS2017nanoscaleIBCT, Surdutovich_AVS_2014_EPJD.68.353} (Section~\ref{sec:Interlinks_MD-SD-macro_theories}).

Further advancement of our understanding of the ion irradiation-induced damage mechanisms, including the thermomechanical mechanism caused by the ion-induced SWs, requires further elaboration and widening of the aforementioned interlinks. This includes:
\begin{enumerate}[label=(\roman*)]

\item
linking IDMD simulations (Section~\ref{sec:Methods_IDMD}) with track-structure MC simulations and analytical particle-transport models (Sections~\ref{sec:Methods_MC_transport} and \ref{sec:Methods_Analytical_transport}) for studying the transport of secondary electrons in the vicinity of an ion track propagating through biological media, such as water and DNA (see also a description of the corresponding interface in Section~\ref{sec:Interlinks_IDMD-transport}). This will enable the time and spatial evolution of ion track structures to be studied by accounting for the sub-picosecond and picosecond dynamics of the medium in which the track is created and the energy is deposited into the medium.

\item
studying the dynamics of biological media due to the ion-induced SW effect using the RMD approach (Section~\ref{sec:Methods_RMD}) and establishing the role of the thermomechanical mechanism in different scenarios of RADAM (for different ions, different LET values, different targets, etc.) \cite{surdutovich2013biodamage, deVera2016molecular, Friis_2021_PRE}.

\item
linking the outcomes of RMD simulations with SD simulations (see Section~\ref{sec:Interlinks_MD-SD}) to analyze the post-SW relaxation dynamics of biological media on the time scales from nanoseconds up to seconds, corresponding to the chemical stage of RADAM.
\end{enumerate}

\textbf{\textit{Future directions for 5-10 years period.}}
As described above in this section, a theoretical description of the production and propagation of reactive species in biologically relevant media should be extended beyond traditional MC approaches by accounting for the coupling of the reactive products with the thermally-driven movable biological medium in which they are produced. This can be done by elaborating on the interlinks between the theoretical and computational methods described in the previous paragraph. Triggered by multiscale simulations, new radiation-induced biodamage mechanisms, such as SW-induced damage, should also be explored experimentally. First experimental studies of the effects associated with the complex ultrafast dynamics of the medium in the vicinity of ion tracks were performed recently through the exploration of time-resolved picosecond dynamics of liquid water irradiated with laser-accelerated protons \cite{Senje_2017_APL.110.104102, Prasselsberger_2021_PRL.127.186001} (see the case study in Section~\ref{sec:Case_study_UltrafastRadChem}).

The computational MM approach for studying ion-induced RADAM in biomolecular and biological targets will provide further nanoscopic insights into the key ion-irradiation-induced phenomena, particularly the mechanisms of biodamage by the ion-induced SWs and the role of thermomechanical mechanisms in the overall scenario of biodamage \cite{AVS2017nanoscaleIBCT, Surdutovich_AVS_2014_EPJD.68.353, Friis_2021_PRE}. This knowledge can be integrated into current clinical radiotherapy planning protocols, which are based on the concept of ``macrodosimetry'', i.e. macroscale calculations of the radiation energy deposited per unit mass of patient. Advances in this direction may lead to the optimization of the currently utilized radiotherapy treatment models by accounting for the molecular-level phenomena in biodamage (see also the case study in Section~\ref{sec:Case_study_IBCT}).

\textbf{\textit{Envisaged impact.}}
Improved understanding of the fundamental molecular mechanisms that underlie the effects of ionization radiation on DNA, achieved through MM, will have direct implications for biotechnological and biomedical applications of ion-beam irradiation, especially hadron therapy. At present, there are fundamental gaps in our understanding of therapeutically relevant issues, for instance, regarding the functioning principles of radiosensitizing NPs (see Section~\ref{sec:Case_study_RadioNPs}) or the FLASH mechanism (Section~\ref{sec:Case_study_IBCT}), which uses ultra-high dose rate (UHDR) ion-beam irradiation \cite{Favaudon_2014_SciTranslMed} to spare healthy cells. Advanced atomistic and nanoscale understanding of the aforementioned phenomena will open new horizons for their efficient exploitation in biomedical applications.

\subsection{Innovative Radiation Therapy Strategies Based on Multiscale Processes Control}
\label{sec:Case_study_RT_multiscale}


\textbf{\textit{The problem.}}
With available experimental data, the fundamental basics of processes governing water radiation chemistry, radiation biology and radiation therapy to treat cancers have tremendously been enriched over the last 20~years. The advances were made possible with improved techniques giving the highest temporal (attoseconds) and spatial (nanometer) resolutions. Modern experimental techniques rely on probing the initial events by pulsed laser spectroscopies in microscopic systems after an energetic particle traverses the biological medium. The key objective of the success is the perfect control of the series of events from the beam interaction with the biological materials to the long-term health recovery of a patient.

Recently, the discovery of the electron FLASH radiation therapy has boosted the need of understanding and revisiting the radiobiological mechanisms, including the chemistry and the physical chemistry stages, under an increase of dose rate ($>40$~Gy/s) \cite{Favaudon_2014_SciTranslMed}. On the other hand, the progress of laser wakefield accelerators has activated the international race to the huge dose rate (currently up to $10^{12}$~Gy/s) promoted by the ultrashort duration of the electron bunch (a few tens of femtoseconds) they can provide \cite{Malka_2008_NatPhys.4.447}. Using these electron-bunches in radiation therapy, in an extreme-FLASH modality of very high energy electrons (VHEE, of few 100s of MeV) \cite{Fuchs_2009_PMB.54.3315}, is a real challenge. At the first stage, the effects on living cells must be validated: an improvement over the FLASH modality is expected. Secondly, a multi-timescale approach to the physical chemistry that governs the fate of the initial processes is necessary. The most sensitive step to depict and control is the initial energy deposition stage occurring in the first $10^{-15}$~s, because it influences the chemical reactions close to the ionization tracks and the long-term biology \cite{Malka_2010_MutatResRev.704.142}.

\begin{figure}[t!]
\includegraphics[width=0.8\textwidth]{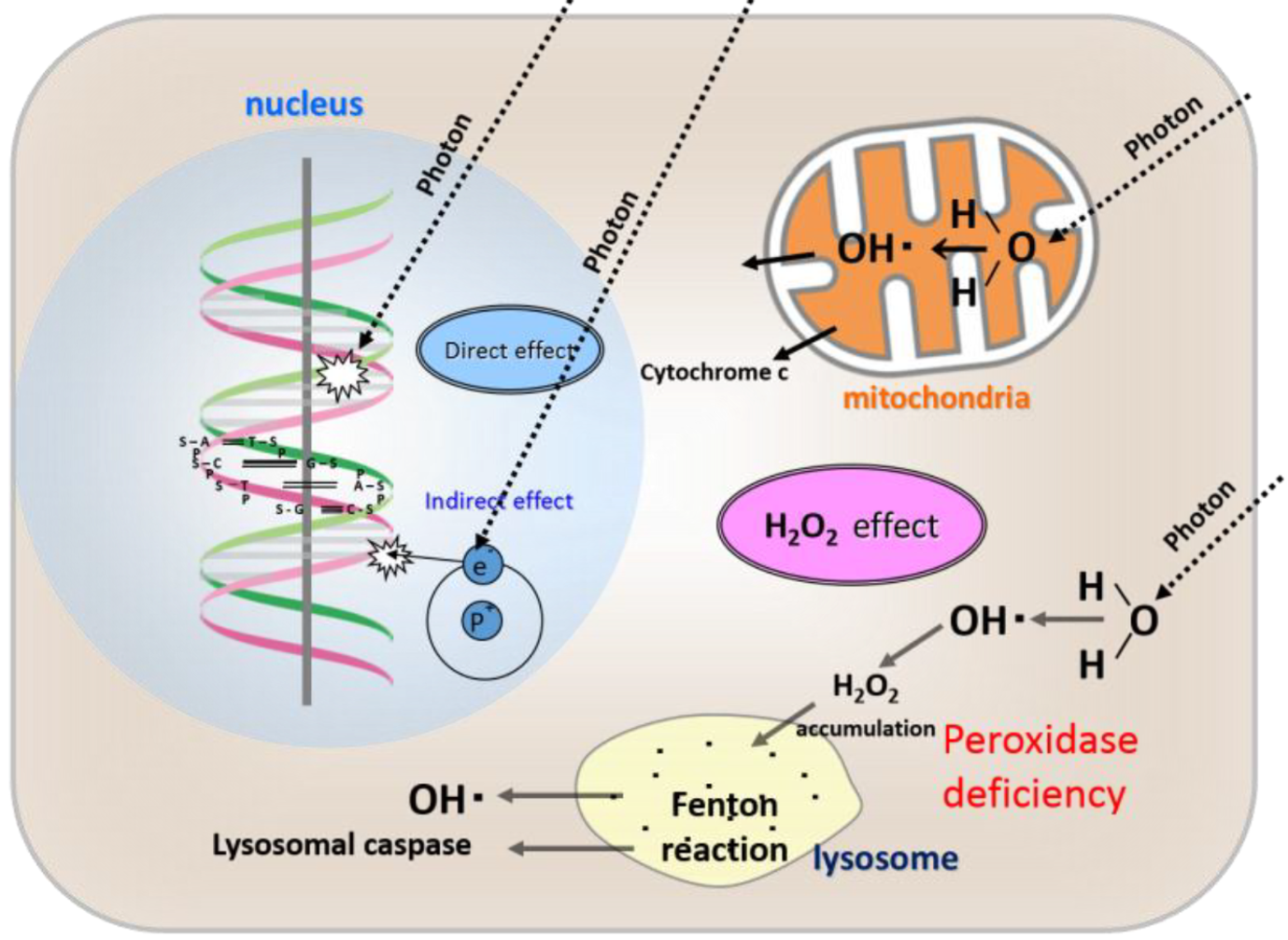}
\caption{H$_2$O$_2$ effect in radioresistant cell under X-rays (low-LET photons). Adapted from ref~\citenum{Ogawa_2016_Cancers.8.28}.}
\label{fig:CaseStudy_Baldacchino}
\end{figure}

Using proton or carbon ions in hadrontherapy is another radiation therapy modality that also evolves to improve the success of cancer treatments for defeating radiation resistant tumors, especially those having hypoxic environments or localized close to organs at risk \cite{Paganetti_2002_IJROBP.53.407}. Applying the FLASH delivery, combined with  spatial dose distribution modulation like it is used in Particle Minibeam Radiation Therapy (PMBRT) and carbon ions bring new hopes in hadrontherapy \cite{Mazal_2020_BJR.93.20190807}. The dose delivery all along the tracks until the ionizing particles stop in the Bragg peak and at the localization of the tumor is now well described. However, some issues, such as radio-induced tumors generation, give rise to several questions. This shows that something is not fully controlled in the overall multiscale processes. For example, the radiolysis processes in the Bragg peak is not well known because of a lack of experimental data concerning the chemical reactions occurring in this region \cite{Horendeck_2021_FrontOncol.11.690042}. Moreover, the complexity of physical chemistry  processes is addressed in every timescale and every compartment of the cell. For example, Figure~\ref{fig:CaseStudy_Baldacchino} shows the H$_2$O$_2$ effect and its implication in radioresistant tumor under X-ray treatment, among the radicals formed under indirect effect (such as OH), which react with biological materials (such as enzymes, DNA, mitochondria, etc) and the direct ionizations of DNA.

Multiscale simulations such as GEANT4-DNA have significantly progressed and nowadays provide useful data for treatment and for understanding some experimental outcomes \cite{Incerti_2010_Geant4-DNA}. Nevertheless, they rely on basic experimental data and regularly raise new questions such as ``what is the effect of oxygen depletion?'', or ``what is the effect of dose rate on reactive oxygen species (ROS) production''? This questioning involves a constant multidisciplinary research relying on a multiscale process: from the physical interaction of particles towards the tumor cells leading to death or healthy tissue repair.

\textbf{\textit{How can MM address the problem.}}
MM could be used to address the multidisciplinarity and the interactivity of models used in the different approaches shown in Figure~\ref{fig:MM_diagram} (Quantum Mechanics, Molecular dynamics, Stochastic Dynamics, Deterministic approaches, Biological models, Cell models, Organ models, etc.), that are involved in radiation therapy \cite{surdutovich2019multiscale}. The healing of a patient must be deterministic as far as possible and should be known before its treatment. Due to the complexity of the MM involved in radiation therapy, it is expected that Artificial Intelligence (AI) could help to make the procedures faster and consider all the parameters depending on the patient and the therapy method. In particular, AI will have to adapt radiotherapy procedures to the body's specificities, its oxidative stress, etc. Then, radiation therapy will have to rely on the data provided by an interface AI/MM by using a numerical twin of the patient. This should be considered the future of the ideal radiation therapy.

Coming back to the chemistry issues in Bragg peak of protons or heavy ions, MM could give an interpretation of new outcomes becoming more and more time and space resolved. In particular, it could give an interpretation of the chemical effect in the distal part of the proton Bragg peak. Similarly, the questions concerning FLASH effect should be addressed by MM: It is presently unknown why FLASH spares healthy tissues as it kills with the same efficiency tumor cells as conventional irradiation using X-rays. MM can address the complexity of the response at the cell level in the first approach.

\textbf{\textit{Future directions for 5-10 years period.}}
To correctly use MM in radiation chemistry for understanding and controlling the dose rate-like effect revealed by FLASH, it is essential to develop a methodology. This methodology must address the full particle beam, not a single particle track. It must also address the pulse duration. Every parameter that represents the beam reality has to be considered in MM, which is not the case in current simulations. It has been observed that the results of Geant4-DNA and continuous beam irradiations did not match \cite{Audouin_2023_SciRep}. The reason for the disagreement is that the chemistry is complex due to the overlapping of the tracks.

The huge dose rates delivered with the VHEE and the laser wakefield accelerator currently address a new domain of energies and very early time where experimental outcomes remain fragmented and unexplained even if the theory has been well established after World War II. An effort should be made to extend the frontiers in energies and earliest times to simulate the initial processes using MM.

A funadmental issue in radiation therapy is still to account for the continuity between the chemistry at the molecular level, the reactivity of radicals towards the biological materials and the biology that models the complex processes over another timescale, which is often involved in cycles. This should be a significant challenge for MM in the next 5-10 years.

There are also typically many experimental outcomes from DNA sequences interacting directly or indirectly with various particles. Other molecules like proteins, membranes, and organelles like mitochondria will reveal their reactivity under radiation. In the near future, the DNA model will evolve into a more complex but reliable molecular system to provide a numerical twin-cell to be implemented in MM.

These objectives must reasonably converge to an improved radiation therapy concomitantly to technical progress and novel modalities. These therapies should also be considered as improved if real time diagnostics could be implemented for each treatment and in the molecular system used under radiation: multifunction fluorescent systems can play this role as probes of sensitive changes of the dose, dose rate, pH or temperature and by giving signals of good health of the cell.

\textbf{\textit{Envisaged impact.}}
MM is an essential tool to go further in radiation therapies. By exploiting and comparing the new outcomes of the novel experiments to the models, the refinement of the MM will provide better and better radiation therapy results. Progress in the coupling of AI/MM is the key to success since a complete treatment needs to apply an incredible amount of knowledge, including the association of numerical twins of an entire body. Reaching this level of detail means that AI and MM will have made great progress.

\subsection{Radiation-Induced DNA Damage Repair and Response Mechanisms in Cellular Systems}
\label{sec:Case_study_DNA_damage_repair}


\textbf{\textit{The problem.}}
In plants, animals, humans and other eukaryotic organisms, the cell nucleus contains the whole program of species genesis, of all different cell types of an individual, of cellular functioning and replication, and of environmental stress response. This cell nucleus is a complex, self-organizing biological system \cite{Bizzarri_2020_Entropy.22.885} separated from the cytoplasm and enclosed by a membrane that allows the trafficking of molecular complexes in and out. Inside the cell nucleus, simultaneous reactions and functions take place to keep the cell as an individualized, specialized well-running system \cite{Erenpreisa_2023_IJMS.24.2658}. The cell nucleus contains chromatin consisting of DNA strands, histones and non-histone proteins, packed to various degrees of density. This organization ensures that different activities are separated in specific volumes of micro-scaled chromosome territories, sub-chromosomal domains and nano-scaled functional units \cite{Cremer_2015_FEBSLett.589.2931}. The base sequence of the DNA contains all the necessary information for the species-specific genome program. In addition to the sequence of the individual bases, information in the form of sequence motifs and motif arrangements plays a crucial role. In this context, very short, short, intermediate and long motifs are interspersed in the genome, which seem responsible for the appropriate spatial organization and folding of chromatin units \cite{Sievers_2021_Genes.12.1571, Sievers_2023_Genes.14.755}. This spatial organization can follow a specific dynamic regulation, opening and closing reactive chromatin units and thus controlling cell nuclear and cellular functioning \cite{Krigerts_2021_BiophysJ.120.711, Erenpreisa_2021_Cells.10.1582}.

Between the differently packed chromatin, there is enough ``free'' space for floating of differently-sized molecules, such as RNAs, proteins, enzymes, ATPs and water molecules, and atoms, differently charged ions or other entities, which are trafficking primarily by super-diffusion (i.e. anomalous diffusion supported and directed by active transport processes) or supra-diffusion (anomalous diffusion hindered e.g. by macro-molecular crowding) to the correct interaction points where they are required. Although sometimes ATP-supported, it seems that this trafficking works somehow self-propelled and drives and governs the system perfectly \cite{Erenpreisa_2023_IJMS.24.2658}. The molecular dynamics and macro-molecular organization must follow the chemical and physical laws of atomic and molecular binding, thermodynamics, electrodynamics and physical chemistry within a limited volume \cite{Hausmann2022networks}.

Exposure to UV or ionizing radiation (IR), such as X-rays, $\gamma$-rays, high-energy electrons or protons, high-energy particles or ions, etc. causes chromatin damage of different types, for instance, base modifications, single-strand breaks (SSBs) and especially double-strand breaks (DSBs) of different complexity \cite{Lee_Hausmann2021super-resolution, Jezkova_2018_Nanoscale.10.1162}. The character of DSBs (complexity, multiplicity) strongly depends on the dose and dose-rate of the exposing radiation, as well as on (cell type-specific) global and local chromatin organization \cite{MF_MH_2021_Cancers.13.18, Bobkova_2018_IJMS.19.3713}. From the perspective of (epi)genetics, the DNA damage induces an immediate response in terms of well-defined protein cascades that step by step repair the damage and eventually restore the system (functioning) to its original state \cite{Scully_2019_NatRevMolCellBiol}. These proteins sensor DSBs, cut the damaged ends of the DNA molecule, clear them to enable adding of new nucleotides, search for the correct nucleotides, exchange and incorporate new nucleotides, bind them entirely into the damaged strand and finally ligate both DNA strands and re-organize the folding of affected chromatin into the original status \cite{Lee_Hausmann2021super-resolution, Falk_2014_CritRev_PartA, Falk_2014_CritRev_PartB}. Such (epi)genetic pathways and control loops seem to follow a master plan, with no master telling them where and when to interact \cite{Lee_Hausmann2021super-resolution}.

The rationale described above suggests that the cell nucleus responds to environmental stress caused by irradiation as a whole (i.e., as a system) and triggers a damage-specific (epi)genetic response and molecular trafficking to the damaged sites in such a way that specific repair processes are initiated and continue at given sites until the end of restoring a fully functional and intact system. At the molecular level, many individual (epi)genetic pathways respond to SSBs and DSBs and allow repair of these types of DNA lesions \cite{Falk_2014_CritRev_PartA, Falk_2014_CritRev_PartB, Iliakis_2019_Cancers.11.1671}. However, how these responses are embedded into the coordinated response of the (chromatin) system is often neglected.

\textbf{\textit{How can MM address the problem.}}
The problem described above is a typical multi-scale problem that begins with the identification of specific linear DNA sequence motifs, proceeds to the analysis of their nuclear patterns and spatiotemporal organization of individual structurally and functionally distinct chromatin domains and analysis of DNA damage and repair mechanisms within these domains \cite{Falk_2007_BBA.1773.1534, Falk_2008_BBA.1783.2398, Falk_2008_JPCS.101.012018, Falk_2010_MutatRes.704.88}, and culminates in a complex response to irradiation at the level of the whole nucleus (chromatin network) as a system \cite{Erenpreisa_2023_IJMS.24.2658}. Chromatin folding and compaction ``coded'' by various DNA sequence motifs can be modeled using the laws of physics and chemistry \cite{Erenpreisa_2023_IJMS.24.2658, Krigerts_2021_BiophysJ.120.711, Erenpreisa_2021_Cells.10.1582, Hausmann2022networks}. This appears to be the starting point for simulating radiation-induced DNA damage in a more complex chromatin environment and how the exposure of DSB lesions to a specific surrounding environment (chromatin and molecules in inter-chromatin channels) can trigger and regulate a particular repair response \cite{MF_MH_2021_Cancers.13.18}. Finally, this approach needs to be extended in space and time to include larger parts of the cell nucleus or even the cell nucleus as a system as a whole. This may in the future (depending on computational capacity) help to answer how the chromatin network as a whole responds to the formation of different numbers of DNA breaks with different characteristics in different structural-functional chromatin domains and how specific repair mechanisms are activated at specific sites of damage, or how different repair pathways interact and coordinate with each other \cite{MF_MH_2021_Cancers.13.18}.

The MM approach, however, requires a serious basis of experimental investigations and results that can only be obtained by the novel but meanwhile established approaches, such as the alignment-free k-mer search \cite{Sievers_2018_Genes.9.482} and super-resolution single-molecule localization microscopy (SMLM) \cite{Lee_Hausmann2021super-resolution, MF_MH_2021_Cancers.13.18, Weidner2023advanced}, which are briefly described below.

The general concept of finding potentially interesting DNA sequence motifs and motif patterns is based on:
(1) Search = finding conserved DNA sequence patterns;
(2) Analysis = looking for associations with functional units;
(3) Interpretation = determining the relevance for chromatin organization and function.

\begin{figure}[t!]
\includegraphics[width=1.0\textwidth]{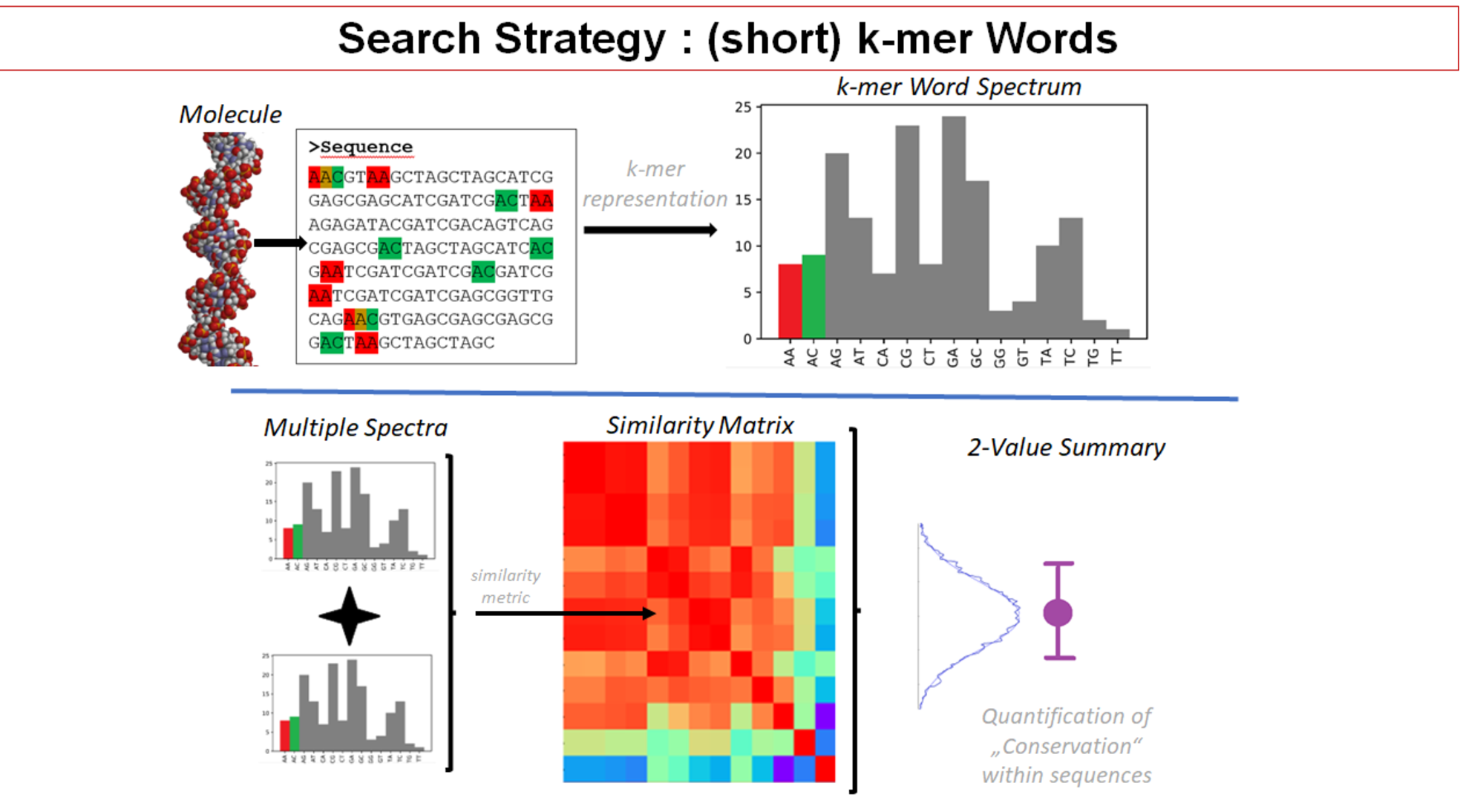}
\caption{Steps of the search strategy and analyses for k-mer words (= DNA sequence with a length of k bases). After sequencing the k-mers are searched for, their frequencies are counted, and the frequency spectra (histograms) of abundances are correlated pairwise. Finally, they may be presented as a heatmap with appropriate color-coding for the calculated correlation values. These values may be further condensed by averaging all the correlation values of a heatmap and giving the resulting distribution of the correlation values together with their mean and standard deviation. Adapted from ref~\citenum{Erenpreisa_2023_IJMS.24.2658}.}
\label{fig:CaseStudy_Hausmann1}
\end{figure}

The approach of searching for k-mers (all the possible DNA substrings of length k) and analyzing their distribution along the DNA sequence and their abundance is schematically shown in Figure~\ref{fig:CaseStudy_Hausmann1}. The results obtained by this approach show, for instance, that the abundance of k-mers appears to be highly conserved, especially in the intronic and intergenic regions of many eukaryotic genomes, indicating the general importance of particular k-mers for DNA and chromatin organization \cite{Cremer_2015_FEBSLett.589.2931, Erenpreisa_2021_Cells.10.1582}. This prompted a search for a specific subset of k-mers, called Super-Short-Tandem-Repeats (SSTR), with a precise number of base pairs, which revealed specific patterns in, for instance, centromeres and regions known as characteristic chromatin breakpoints after environmental stress.
%
%
Probes based on SSTRs may be applied to label these regions as indicators. The findings can be compared with known biophysical parameters of chromatin folding and arrangement \cite{Sievers_2023_Genes.14.755}.

\begin{figure}[t!]
\includegraphics[width=0.7\textwidth]{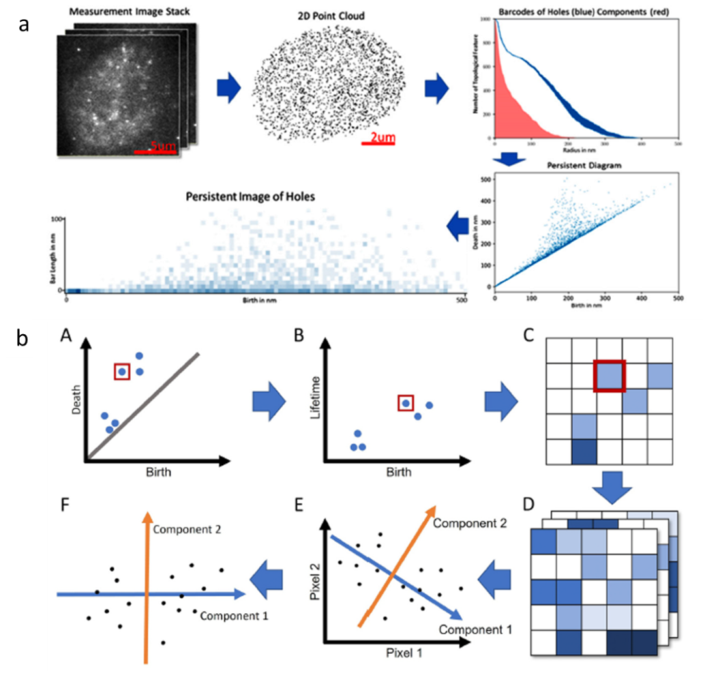}
\caption{Panel~(a): An example of persistent homology workflow. Persistent homology is applied after the \textit{Measured Image Stack} is converted into a \textit{2D Point Cloud}. The results are represented as \textit{Barcodes}, showing each component (red) and hole (blue) as one bar. The \textit{Persistent Diagram}, where each hole is shown as a point with birth and death as coordinates, is an equivalent representation. To vectorize the persistence diagram, it is converted into a \textit{Persistent Image} by laying a grid over it and counting the holes in each grid cell. Adapted from ref~\citenum{Erenpreisa_2023_IJMS.24.2658}. Panel~(b): (A) Generation of persistence image. In the first step, the persistence diagram is folded down by 45$^{\deg}$ (B). The y-axis thus shows the lifetime instead of the disappearance (death) of the hole. The diagram is converted to a grid (C) in the next step. The color intensity represents the number of points in each grid. The red box shows the path for one hole in the persistence diagram. Based on persistent images, principal component analysis (PCA) is applied. Multiple persistence images (D) are transferred into a vector space where each pixel is represented as a dimension. Values for pixels 1 and 2 are shown in the first plot (E). In the next step, the basis vectors are rotated. The first component (blue) points toward the largest variance. The next component (orange) must be perpendicular to all previous ones. Under this condition, it points in the direction of the largest variance. In the 2D case shown, only one possibility exist for the second component. Finally, the measurements are plotted with the new basis vectors (components 1 and 2) (F). Adapted from ref~\citenum{Weidner2023advanced}.}
\label{fig:CaseStudy_Hausmann2}
\end{figure}

Self-organization of chromatin in the cell nucleus \cite{Bizzarri_2020_Entropy.22.885} results in function-related networks \cite{Erenpreisa_2023_IJMS.24.2658} (e.g. heterochromatin or euchromatin network \cite{Erenpreisa_2021_Cells.10.1582}, network of ALU-repeats \cite{Erenpreisa_2023_IJMS.24.2658, Krufczik_2017_IJMS.18.1005}, network of L1-repeats \cite{Krufczik_2017_IJMS.18.1005}, etc.) and specific local topologies.
ALU repeats are short interspersed nuclear elements recognized by the Arthrobacter luteus (Alu) restriction endonuclease. The distribution of these elements in the genome forms a network contribution to the chromatin organization in the 3D genome architecture.
L1 (LINE-1) repeats are long interspersed nuclear elements. These elements play a role in the organization of heterochromatin and the re-arrangements of heterochromatin during genome functioning.
Such networks characteristically rearranged themselves depending on cell-fate \cite{Krigerts_2021_BiophysJ.120.711} and under environmental stress such as exposure to ionizing irradiation, thereby interacting with epigenetic pathways and response loops \cite{Hausmann2022networks}.

In order to obtain quantitative parameters for networks and network dynamics, super-resolution SMLM \cite{Lee_Hausmann2021super-resolution}  is used as a method for accurate (to within ten nanometers) registration of the coordinates of fluorescence labels of specific DNA sequences/chromatin domains and proteins. 
%
The major advantage of SMLM depends on this registration process that results in a map of molecule coordinates without primarily processing an image. Such nanoscaled coordinate matrices of points can be subjected to mathematical operations that allow for calculations of geometry and topology of molecular assemblies of interest (e.g. ionizing radiation-induced foci (IRIFs) and surrounding chromatin domains). The suitable mathematical approaches are well established and often applied beyond biology. They include, e.g., Ripley distance frequency statistics of pairwise molecular distances, persistence homology, persistence imaging, principal components analysis, etc. \cite{Weidner2023advanced} (see Figure~\ref{fig:CaseStudy_Hausmann2}).

Ripley statistics was developed by B.D. Ripley in Oxford as a form of spatial statistics that involves stochastic point processes, sampling, smoothing and interpolation of regional (areal unit) and lattice (gridded) point patterns, as well as the geometric interpretations of the statistical outcome.

Persistent homology is a well established mathematical method used in topological data analysis to study and compare qualitative features of data, for instance, network-like point patterns that persist across multiple scales. It is robust to perturbations like rotation or stretching of input data, independent of dimensions, and provides compact representations (e.g. by bar codes) of the qualitative features of the input data. Here these features are represented by births and deaths of bars of integrated components or holes of the network-like point patterns.
A representation of this bar code information is a persistence diagram in which the frequency of lifetimes of bars is summarized. The conversion of a persistence diagram into a finite-dimensional vector representation is called a persistence image.

Principal component analysis is a well-known mathematical procedure for dimensionality reduction. It is often used to reduce dimensionality by transforming a large set of multi-dimensional vector space into a smaller one that still contains the major information of the large set. The basis is thereby the original dimension with the largest variability. All further dimensions maintained are perpendicularly oriented to this basis. A trick in dimensionality reduction is to trade some accuracy or biological variability for simplicity.

Irradiation of cell nuclei induces DNA damage, e.g., DSBs, causing chromatin reorganization, which appears to govern the repair process at the given site of damage \cite{Hausmann2022networks, Scully_2019_NatRevMolCellBiol}. Broken ends in densely packed heterochromatin induce the relaxation of this part of the heterochromatin \cite{Falk_2007_BBA.1773.1534, Zhang_2015_PLoSONE.10.e0128555} so that the entropy-driven forces transfer the ends of the strands to the region between euchromatin and heterochromatin \cite{Falk_2007_BBA.1773.1534, Zhang_2014_Chromosoma}. After DSB induction, H2AX histones at DSB sites are phosphorylated at Ser139 in approximately 2~Mbp chromatin regions around the lesion. This molecular modification impacts the chromatin arrangement of the broken ends \cite{Hausmann_2018_Nanoscale.10.4320}. $\gamma$H2AX clusters are on average equally sized \cite{Hausmann_2021_IJMS.22.3636} and have a similar topology if they originate from heterochromatin \cite{Hofmann_2018_IJMS.19.2263}. This similarity of $\gamma$H2AX clusters is also higher early (about 30~min) after irradiation than in later periods of repair. However, if $\gamma$H2AX clusters persist a sufficiently long time (24~h) due to insufficient or impossible repair, they again maintain their topological similarity \cite{Hahn_2021_Cancers.13.5561}. Proteins associated with DSB repair (e.g. 53BP1, Mre11, pATM, Rad51, etc.) that attach the DSB sites in the frame of a given repair pathway also show a typical spatial arrangement \cite{Scherthan_2019_Cancers.11.1877}. This may indicate that a particular repair process and involved proteins require not only a special chromatin architecture allowing them to bind the damage site, but also that individual repair proteins in a timely- and spatially orchestrated way induce changes in a local environment that promote further steps of the particular repair pathway. In other words, the chromatin architecture at the sites of individual DSBs may (co)determine the activation of a particular repair mechanism by influencing the efficiency of transport and binding of specific proteins, whereas, the accumulation of these proteins may further stimulate the selected repair mechanism through dynamically inducing changes in chromatin architecture.

With the algorithms and approaches described above, local changes in chromatin architecture at damage sites were investigated. In addition, our analyses revealed that different cell types can be distinguished by the principal components of the topological characteristics of their heterochromatin and ALU-networks \cite{Erenpreisa_2023_IJMS.24.2658}. The ALU repeats form a network \cite{Erenpreisa_2023_IJMS.24.2658} complementary, for instance, to heterochromatin or euchromatin networks.
ALU repeats are involved in chromatin reorganization, especially after its exposure to ionizing radiation that causes DNA damage response and repair \cite{Morales_2015_PLoSGenet}. The number of ALU network signals decreases with dose in a linear-quadratic way \cite{Krufczik_2017_IJMS.18.1005, Hausmann_2017_IJMS.18.2066}.

Systematic changes in the chromatin architecture topologically expressed by mesh sizes of chromatin networks indicate rearrangements of chromatin architecture associated with repair activities. During the repair of DNA DSBs, the whole chromatin revealed a cyclic movement in the topological space of the two major principal components. 

\textbf{\textit{Future directions for 5-10 years period.}}
The recent advances in the application of DNA sequence pattern analysis by k-mer search and the application of Ripley distance frequency statistics of pairwise molecular distances, persistence homology, persistence imaging, and principal components analysis on SMLM data sets will offer novel perspectives for MM in order to investigate effects of ionizing radiation-induced chromatin damage response and repair processes of the cell nucleus as a complex system as a whole. Such a model is currently lacking but is essential if we want to understand these processes on different scales and obtain a predictive tool on how an irradiated cell nucleus would react.

It is envisaged that a complex MM approach for studying physical processes in the cell nucleus will be developed to create a new predictive model capable of interpreting investigations of DNA sequence motif patterns in relation to (topological and geometrical) SMLM data. Such a model can be first developed to assess chromatin damage and later extended to include DNA damage response with different repair mechanisms operating in different chromatin environments. Overall, this may provide a more accurate model for describing the individual sensitivity of cells to radiation and the risk of a cell becoming cancerous in the event of incomplete and/or incorrect chromatin repair.

\textbf{\textit{Envisaged impact.}}
MM can contribute significantly to unravelling the relationships between chromatin architecture from the micro- \cite{Vicar_2021_DeepFoci, Dobesova_2022_Pharmaceutics} to the nano-level \cite{MF_MH_2021_Cancers.13.18, Weidner2023advanced} and biophysical processes in cell nuclei. Thus, this approach can potentially bring about a tremendous shift in our understanding of the functional organization of chromatin as a system and the questions of how chromatin responds to and controls protein trafficking following exposure of cells to radiation (or environmental stress in general).

\subsection{Mechanisms of Nanoparticle Radiosensitization}
\label{sec:Case_study_RadioNPs}


\textbf{\textit{The problem.}}
Radiotherapy tumor dose is restricted by the highest tolerated dose to surrounding healthy tissue since both cancer and adjacent healthy cells have similar radiation interaction properties. Nanoparticles (NPs) can be designed to have higher radiation interaction properties (for example, high Z) and offer the potential for preferential uptake in tumor cells, owing to the enhanced permeability and retention (EPR) effect of nanoscale entities and their high surface area-to-volume ratio, offering a platform to conjugate cancer targeting moieties. Nanoparticle-enhanced radiotherapy (NERT) thereby offers potential to increase the therapeutic ratio of radiotherapy towards reduced side effects and enhanced tumor control \cite{Ahmad_2020_PartPartSystCharact}. Many preclinical studies of metallic NPs have demonstrated radiotherapy enhancement factors on the order of 10--100\% at clinically feasible concentrations \cite{Her_2017_AdvDrugDelivRev}. However, it is challenging to ascertain the relative advantages of the NERT strategies studied by the community owing to the high variation of NP and radiation characteristics, preclinical models and experimental read-outs reported throughout the literature \cite{Ricketts_2018_BJR.91.20180325}. To date, only two metal-based nanoformulations have translated to NERT clinical trials; gadolinium-based polysiloxanes theranostic particles (AGuIX, NH TherAguix SAS) and hafnium oxide particles (Nanobiotix SA).

The choice of experimental read-outs of NERT enhancement measurement is generally not driven by the latest knowledge of NP radiosensitization mechanisms, often focusing on physical dose enhancement as calculated through microscale MC simulations. MC simulations can be used to calculate the physical dose enhancement on the microscale stemming from photoelectrons and Auger electrons \cite{Douglass_2013_MedPhys.40.071710}. Such physical models underestimate the experimentally measured biological enhancement in cellular systems \cite{Butterworth_2012_Nanoscale.4.4830}. A potential reason is the lack of chemical interactions in MC simulations; chemical reactive oxygen species produced via water radiolysis have demonstrated a significant role in NP radiosensitation.

Additional mechanisms have been suggested including NP-induced cellular oxidative stress and modification of the cell cycle to radiosensitive phases \cite{Rosa_2017_CancerNano.8.2}. However, there is still no consensus nor significant evidence regarding the fundamental science governing these processes, and additional mechanisms may yet be at play. The layer surrounding the NP is known to impact NP radiosensitization \cite{Sicard-Roselli_2014_Small}, influencing reactive oxygen species production and the energy profile of emerging electrons. However, this factor is less widely studied than the NP core, and its incorporation into predictive modeling must be considered.

To fully exploit all the engineerability and tunability offered by NPs and their coatings, we need to understand the mechanisms of NP radiosensitization and the role played by the different components of the nanosystem (core, coating and environment), towards optimized NP design and improved experimental test platforms to benchmark developing nanoformulations.

\begin{figure}[t!]
\includegraphics[width=0.8\textwidth]{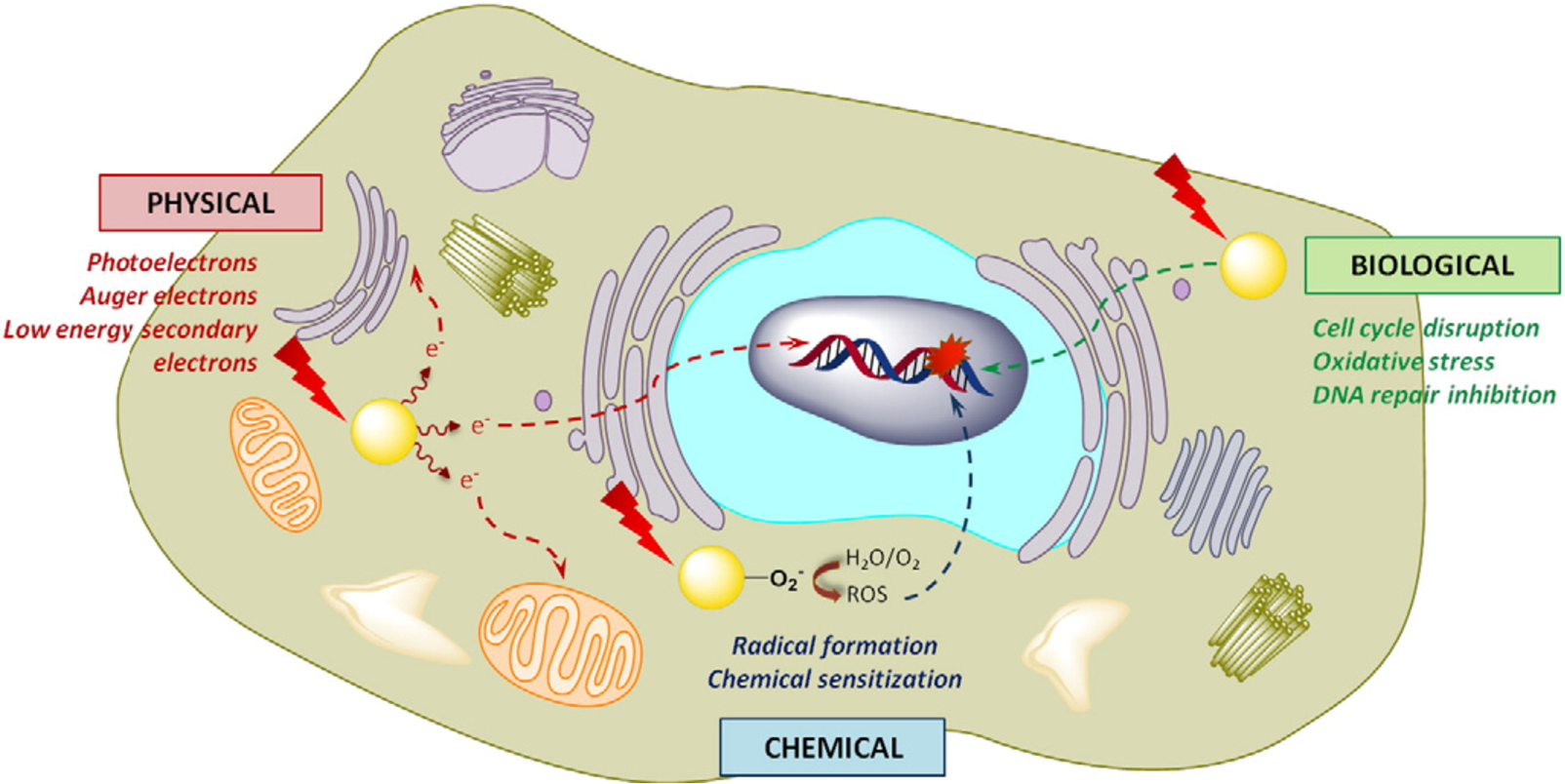}
\caption{The physical, chemical and biological mechanisms of gold NP radiosensitization. Adapted from ref~\citenum{Her_2017_AdvDrugDelivRev}.}
\label{fig:CaseStudy_Ricketts}
\end{figure}

\textbf{\textit{How can MM address the problem.}}
NP radiosensitization involves a range of initial fast interactions of radiation with NPs and biological tissue, slower post-irradiation relaxation, thermalization and track structure processes, and downstream biological processes, see Figure~\ref{fig:CaseStudy_Ricketts}. These processes occur over a range of time, space and energy scales \cite{AVS2017nanoscaleIBCT, Surdutovich_AVS_2014_EPJD.68.353}. No particular mechanism or process can fully model NP radiosensitization; instead a MM approach is essential. Pulling together the quantum physics and radiochemistry processes within a multiscale model offers the potential to determine the distribution of molecular damage (DNA single, double and complex strand breaks) to feed into biological models towards evaluating DNA repair mechanisms post irradiations and eventual modeling of cell survival.

A computational platform based on MD simulations has demonstrated the ability to model at the atomic level the components of the NP-biological environment system (NP core, NP coating and biomolecular environmental surround) \cite{Verkhovtsev_2022_JPCA.126.2170}. Understanding the impact of bioconjugation coatings on the resulting hydrodynamic and haemodynamic radii of NPs upon entering biological media can give information on radiosensitization capacity; forming inputs to models of energy spectra of low-energy electrons escaping the coating and resulting free radicals. Systematically modeling the structural components of the nanosystem will enable a more detailed description of radiochemistry effects in the vicinity of irradiated NPs, with the potential to inform future NP development.

\textbf{\textit{Future directions for 5-10 years period.}}
The multiscale model has shown good agreement with experimental data in predicting the survival probability of a broad range of cell lines under ion irradiation and demonstrated the capability to model hypoxic conditions and predict phenomena such as the oxygen enhancement ratio \cite{verkhovtsev2016multiscale}. Once the mechanisms of NP radiosensitization have been incorporated, this work can be extended to modeling cell survival in the presence of NP radiosensitisers for a range of cell lines (radiosensitive and radioresistant) and environmental conditions to make predictive effects of therapeutic ratio enhancement. This could inform treatment planning strategies for patient treatments.

MM should be developed in parallel with experimental efforts to determine appropriate preclinical models that represent the mechanisms, environments and readouts of the model towards benchmarking against future experimental work as it presents. Experimentalists are working towards introducing molecular analysis to this field including genomics and proteomics, which are a step change away from the fundamental physical / chemical / bio damage inputs contained with the multiscale model; however insights gleaned from such next generation molecular studies can be fed back into future versions of multiscale models towards a systems computational biology component to incorporate molecular mechanistic pathways.

\textbf{\textit{Envisaged impact.}}
An understanding and model of the fundamental mechanisms of NP radiosensitization and the impact of each component of the nanosystem (core, coating and environment) on downstream radiochemical effects has the potential to inform optimized NP design (size, shape, elemental composition, chemical coating). NP design is moving away from focusing on one mechanistic optimization (e.g. physical dose enhancement requiring high-Z elemental composition). In addition to modeled mechanistic design optimization, NP design must also be guided by factors that affect \textit{in vivo} NP microdistribution and radiation enhancement, including stability and aggregation, protein corona changes \textit{in vivo}, cellular and intracellular targeting moieties, circulation time, 3D penetration into tumor tissue, and toxicity profile and clearance pathways \cite{Ricketts_2018_BJR.91.20180325}. In addition, for translation, the design must also incorporate scalability of manufacture and potentially multifunctional components for imaging and drug delivery. A multidisciplinary team from the clinic, pharma, experimentalists and modelers should input into the MM process to determine the pertinent NP features to model with potential translational capability.

An improved understanding of mechanisms driving NERT will inform appropriate experimental read-outs, enabling study comparison throughout the community and satisfactory benchmarking of newly developed nanosolutions. A more in-depth understanding of the mechanisms of NP radiosensitization will uncover the spatial ranges of therapeutic action towards informing optimal microdistribution and NP uptake concentration levels. This has the potential to guide clinical infusion protocols, in addition to other methods of NP delivery, including implantation and direct injection. It will also inform the choice of targeting moieties for conjugation to the NP surface, with the potential to link to cell and cell organelle (e.g. nucleus, mitochondria) specific seekers.

\subsection{Medical Application of Radiation in Ion-Beam Cancer Therapy}
\label{sec:Case_study_IBCT}

\begin{figure}[t!]
\includegraphics[width=0.4\textwidth]{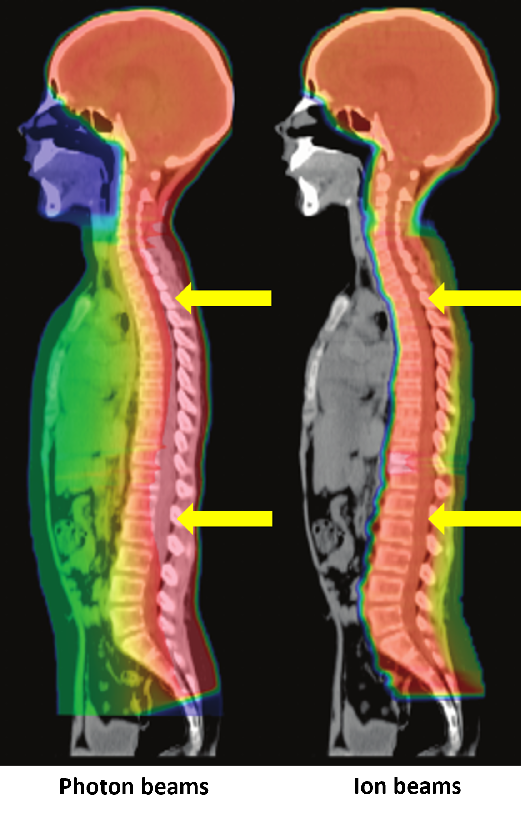}
\caption{Comparison of photon and ion dose distributions for cranio-spinal irradiation. Ion beams spare healthy tissue anterior to the target. Adapted from ref~\citenum{Gains_Amos_2019}.}
\label{fig:CaseStudy_Amos1}
\end{figure}

\textbf{\textit{The problem.}}
Radiotherapy, one of the three major components of trimodal cancer care, is well demonstrated to improve overall survival, spare healthy organs, and improve quality of life for select patients \cite{MC_Lecacy}. Indeed, the Royal College of Radiologists (RCR) in the UK has estimated that, of those cancer patients who are cured, 40\% are cured by radiotherapy. The significance of accurate and efficacious radiation-based techniques for treating cancer cannot be overstated. Radiotherapy aims to target the tumor within a patient with a sufficient dose of ionizing radiation for tumor control, while simultaneously reducing the risk of radiation-induced toxicities in surrounding healthy tissues. The use of proton and heavier-ion beams for radiotherapy continues to grow globally due to their favorable depth-dose distributions compared to traditional photon-based techniques. Depth-dose profiles for ion beams are characterized by a sharp increase in dose, the Bragg peak, near the end of the beam range, beyond which little-to-no dose is delivered, thus sparing healthy tissue distal to the tumor (Figure~\ref{fig:CaseStudy_Amos1}).

\begin{figure}[t!]
\includegraphics[width=1.0\textwidth]{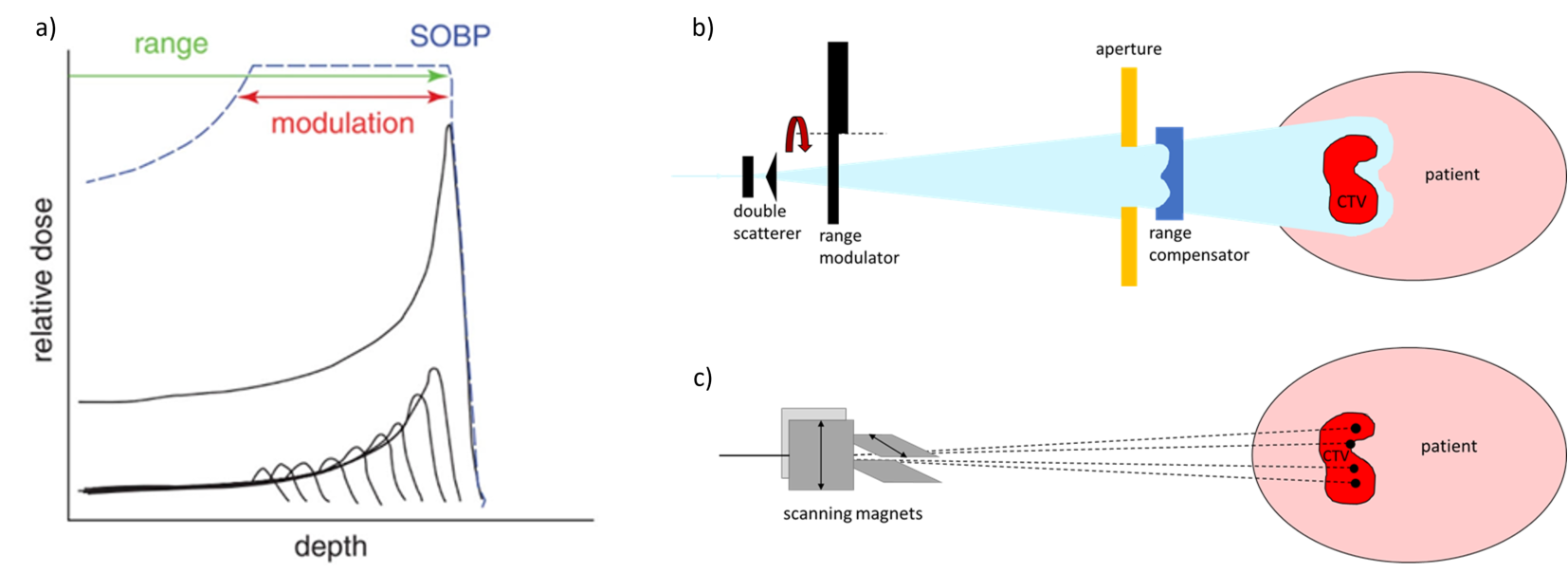}
\caption{Clinical beam delivery for IBCT: (a) superposition of modulated and weighted pristine Bragg peaks resulting in a spread-out Bragg peak (SOBP), (b) beam shaping and modulation using passive scattering (PS), (c) modulated beam delivery using pencil beam scanning (PBS). Adapted from ref~\citenum{Gains_Amos_2019}.}
\label{fig:CaseStudy_Amos2}
\end{figure}

The clinical practice of ion-beam cancer therapy (IBCT) involves the acquisition of an x-ray computed tomography (CT) image of the patient on which the \textit{gross tumor volume} (GTV) is delineated. The GTV is defined as the extent of the malignant disease that is palpable or visible on imaging. A margin extended beyond the GTV defines the \textit{clinical target volume} (CTV), which encompasses the GTV and sub-clinical microscopic disease, which may have to be eliminated. The treatment planning process involves the optimization of dose distributions from multiple ion-beams based on the CT dataset. Initial beam energies are selected such that beam ranges coincide with the target depth along each beam path and beams are modulated to create so-called spread-out Bragg peaks (SOBP) to cover the extent of the CTV with dose maxima (Figure~\ref{fig:CaseStudy_Amos2}a). Historically, clinical SOBP ion beams were delivered using a \textit{passive scattering} (PS) technique whereby the narrow ion beams transported from the accelerator to the delivery system are broadened and modulated by passing through scattering foils and a rotating range modulator wheel. The resulting broad-beam is shaped to conform to the CTV by the use of field-specific apertures, typically manufactured from brass, and wax compensators to shape the distal edge of the SOBP, see Figure~\ref{fig:CaseStudy_Amos2}b. Contemporary systems use a discrete \textit{pencil beam scanning} (PBS) technique to cover the target. With PBS beam delivery, narrow ion-beams from the accelerator system are magnetically scanned across the target in the plane positioned perpendicular to the beam direction. This is repeated layer-by-layer by changing the incident beam energy until the entire target volume is covered by discrete Bragg peaks, see Figure~\ref{fig:CaseStudy_Amos2}c. The clinical treatment planning process for IBCT is described by Zeng et al. \cite{Zeng_Amos_2018}. Due to a number of uncertainties, however, we are yet to fully exploit the physical characteristics of ions in the clinical practice of IBCT.

\begin{figure}[t!]
\includegraphics[width=0.8\textwidth]{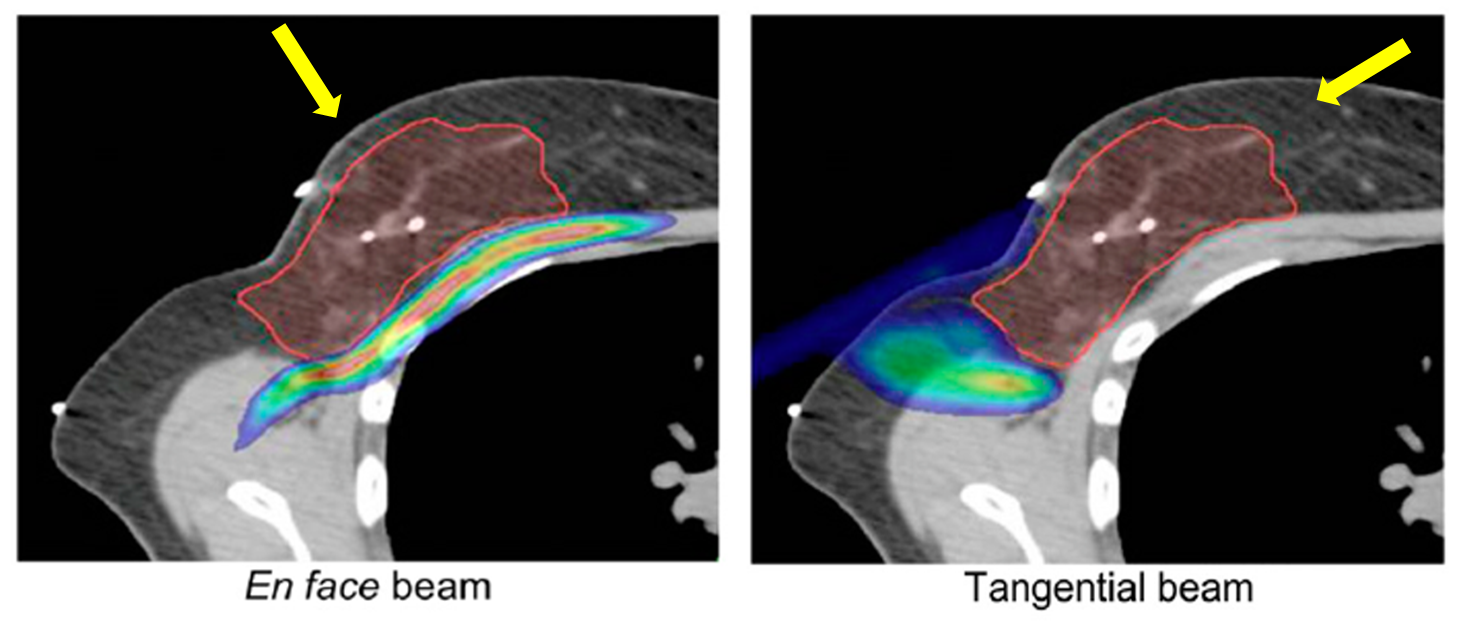}
\caption{Impact of ion beam range uncertainties on clinical treatment planning. A single \textit{en face} beam can cover the target in the shown breast cancer example while sparing healthy breast tissue. However, uncertainty in range (region of uncertainty shown in the color wash) risks high dose to ribs and lungs. This risk is mitigated by including a second tangential beam, but this reduces the optimal sparing of healthy breast tissue. Adapted from ref~\citenum{Wang_2013_BrJRadiol.86.20130176}.}
\label{fig:CaseStudy_Amos3}
\end{figure}

The calculation of ion range by the clinical treatment planning system (TPS) assumes a continuous slowing down approximation (CSDA), which integrates the total stopping power from zero energy to the initial energy of the ion-beam. This forms the basis of an analytical model of the depth-dose curve. However, uncertainties in converting from the Hounsfield units (HU) of the CT dataset to \textit{relative stopping powers} (RSP) for the ions lead to inaccuracies in the calculated beam range. HU, also known as CT numbers, are assigned to each voxel of a CT image and are defined as the fractional difference of the x-ray linear attenuation coefficient of the tissue in any voxel relative to water. RSP, however, is related to the rate of energy loss by charged particles that traverse the medium, dependent upon the beam energy and the composition of the medium. These uncertainties in the calculated ion-beam range often dictate clinical practice (Figure~\ref{fig:CaseStudy_Amos3}). Furthermore, lateral fluence distributions of the beams are essentially approximated analytically by a Gaussian function. The inaccuracies of these analytical models contribute to the margins required beyond the CTV for robust coverage, leading to the exposure of healthy tissue in the proximity of the tumor. Monte Carlo simulation potentially improves accuracy by considering the physics of the ion interactions, however this approach is yet to be adopted for routine clinical practice.

Another source of uncertainty in ion-beam treatment planning is related to increasing LET along the particle paths. As the ions lose energy along their path and slow down, their rate of energy loss as a function of distance increases, and thus the so-called LET increases. An increase in LET at the Bragg peak corresponds to an increase in RBE caused by clustering of ionization events leading to increased DNA damage. This is not accounted for in contemporary clinical proton beam therapy and there is a lack of consistency in biological models used for carbon-ion therapy \cite{Kim_Cho_2020_ProgMedPhys}.

In general, for clinical IBCT, only the location and quantity of energy deposition from the primary beam are modeled. MC simulations can help improve this accuracy, particularly by tracing the tracks of secondary particles \cite{Kelleter_2019_MedPhys.46.3734}. However, these are only the physical components of the ionizing interaction process, with detailed chemical, biological, and other modes of energy transfer, such as thermal and acoustic not accounted for. Consequently, the prescribed dose for a particular type of cancer is defined by population-based outcomes and is not personalized for any particular patient. Having the ability to deliver personalized IBCT based on individual patients' biochemical kinetics and also being able to adapt therapy as a function of response to treatment are highly desirable clinical goals \cite{Ree_2015_BrJRadiol.88.20150009}.

\textbf{\textit{How can MM address the problem.}}
With personalized precision IBCT as a clinical goal, accounting for all physical, chemical, biochemical, and biological processes associated with the interaction of ions at the sub-cellular, cellular, and micro-environment levels is required. The use of MM presents a methodology for achieving this \cite{Surdutovich_AVS_2014_EPJD.68.353, surdutovich2019multiscale, AVS2017nanoscaleIBCT}. The MM approach model could model these processes with the appropriate time scales and potentially predict radiation response as a function of individuals’ physiology, incorporating oxygenation and immune-response, for example. This approach could enable: \textit{in silico} design of pre-clinical experiments to verify simple end-points such as cell death \textit{in vivo}; development of personalized IBCT regimens based on appropriate biomarkers, and development of adaptive IBCT strategies as a function of tumor response.

\textbf{\textit{Future directions for 5-10 years period.}}
Advanced imaging techniques are enabling the application of imaging biomarkers to theragnostics in the field of radiotherapy. Such biomarkers, used in conjunction with MM, will contribute to the development of personalized IBCT over the next decade. Data generated by MM as a function of input biomarker data will contribute to the development of artificial intelligence (AI) aided predictive models of clinical outcomes for patients and guide clinicians in choosing personalized treatment strategies.

The MM approach will also contribute to the development of drugs for use in combination with ion-beam radiation. Modeling molecular pathways that may be pharmacologically targeted in combination with radiation will lead to the development of radiation-drug combination strategies including the use of: DNA damage response inhibitors; survival signaling pathway inhibitors; hypoxic cell sensitizers; immune modulators, and others \cite{Falls_2018_RadRes.190.350}.

The use of NPs to enhance radiation dose to cancer cells and improve the therapeutic ratio is of great interest to the radiotherapy community \cite{Cho_2017_CancerNanotechnology}. Development of NPs optimized for the ion-species used for irradiation and their targeted delivery agents will continue to be investigated over the next 5--10 years using the MM approach.

Boron neutron capture therapy (BNCT) is a radiotherapy modality currently experiencing a renaissance due to the development of accelerator-based BNCT sources for hospital use \cite{Suzuki_2020_IJClinOncol.25.43, Porra_2022_ActaOncol.61.269}. BNCT requires the delivery of $^{10}$B preferentially to cancer cells within the patient. External beams of thermal neutrons targeting the tumor volume are captured by $^{10}$B atoms resulting in a decay reaction yielding $^{4}$He- and $^{7}$Li-ions which deliver dose within $5-9$~$\mu$m. The challenge is the development of targeting agents to optimize $^{10}$B uptake in cancer cells. This modality offers great therapeutic potential for highly radioresistant tumors and MM can contribute significantly to its development in the coming years.

Novel delivery techniques for IBCT will be translated into clinical practice over the next 10 years. Of particular interest is the use of ultra-high dose rate (UHDR) ($> 40$~Gy/s) ion beams. Early evidence suggests that UHDR radiation has a sparing effect on healthy cells whilst maintaining tumor control \cite{Favaudon_2014_TransMed}. This effect has been termed the FLASH effect and is believed to represent a potential paradigm shift in radiotherapy practice. After having established accurate dosimetry for clinical trials, the world's first in-human trial for proton beam FLASH radiotherapy was completed in 2022 \cite{Lorenco_2023_SciRep.13.2054, Mascia_JAMAOncol.9.62}. However, the underlying mechanisms of the FLASH effect are not yet understood. Applying a MM approach to studying UHDR ion-beam interactions will contribute to understanding the FLASH mechanism, potentially expediting clinical translation from bench-to-bedside. The same arguments hold for electron beam UHDR treatments, which are also at an early stage of development.

Another emerging technique for IBCT delivery is the use of spatial fractionation. This involves the delivery of \textit{mini-beams} of ions in grid-like patterns, creating peaks and valleys in the dose profile across the target. Again, early evidence suggests that spatially-fractionated IBCT may enhance tumor control \cite{Prezado_2019_IJROBP.104.266}, but the underlying mechanisms for these observations are not understood. Using MM to study the spatially-fractionated ion-beam delivery processes will help develop that understanding.

\textbf{\textit{Envisaged impact.}}
The development of precise personalized treatment regimens is the Holy Grail across the medical field, including radiotherapy for localized cancers. Improving the targeting of disease with radiation-drug combinations; adopting radiotherapy protocols to adapt to tumor response; understanding the biological mechanisms underlying advanced IBCT techniques can all be realized by incorporating the MM approach to clinical treatment planning and pre-clinical investigations. The application of MM to develop precise personalized IBCT regimens will ultimately improve outcomes for cancer patients indicated for targeted radiotherapy. Not only will this save lives and/or improve quality of life for cancer patients, but reducing radiation-related toxicities among the cancer patient population also alleviates the burden on healthcare systems.

\subsection{Plasmon-Induced Chemistry}
\label{sec:Case_study_PlasmonChem}


\textbf{\textit{The problem.}}
Nowadays, humanity is facing the tremendous challenge of transitioning from energy and resource intensive processes into energy neutral and sustainable ones. This also concerns the production of chemicals, which gives about 2\% of the global greenhouse gas emissions. Two prominent examples of future challenges in this context are to find ways to produce green hydrogen and convert CO$_2$ into high-value chemicals. Photochemistry can help to convert molecules using the energy of light, but most of the photochemical transformations require UV light and they suffer from low specificity. In order to create new and valuable chemicals in an energy-neutral way, it is required to harvest the energy in the visible part of the solar spectrum, which corresponds to photon energies of about $1.6 - 3.0$~eV. In order to exploit the visible spectrum for photochemical transformations, substances are required that absorb photons in this range, such as dye molecules, transition metal complexes, semiconductors with a small band gap, or plasmonic NPs \cite{Strieth-Kalthoff_2018, Marzo_2018_AngewChem.57.10034, Zhan_2018_NatRevChem.2.216}. Plasmonic NPs such as gold NPs are particularly interesting in this context because they possess a very large absorption cross section that exceeds the extinction cross section of organic molecules by about five orders of magnitude \cite{Liu_2007_CollSurfB.58.3}. At the same time, it was demonstrated recently that plasmonic nanomaterials can mediate chemical processes upon excitation of their surface plasmon resonance (SPR) \cite{Gelle_2019_ChemRev.120.986}. The SPR represents the collective oscillation of free electrons in the metal, and for AuNPs the SPR lies around 520~nm for single spherical particles but can be shifted by size and shape \cite{Zhan_2018_NatRevChem.2.216}.

\begin{figure}[t!]
\includegraphics[width=0.6\textwidth]{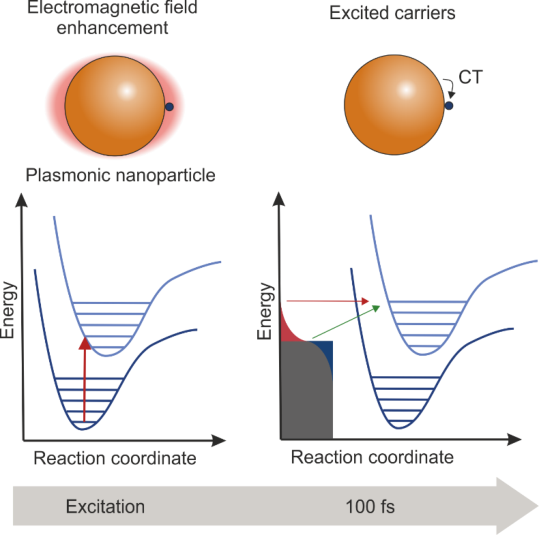}
\caption{Illustration of the interaction of a molecule with a plasmonic NP. Adapted from ref~\citenum{Zhan_2018_NatRevChem.2.216}. }
\label{fig:CaseStudy_Bald}
\end{figure}

The decay of SPRs typically takes place within $10 - 100$~fs and can result in the formation of hot electron-hole pairs (Figure~\ref{fig:CaseStudy_Bald}) \cite{Zhan_2018_NatRevChem.2.216}. These charge carriers can interact with molecules adsorbed on the NPs. A charge transfer from the NP to a molecule can result, for instance, in the formation of a transient molecular anion. This anion can release the extra electron but remain in an excited state, which then leads to further chemical transformations. A reported example is the oxidation of ethylene using photoexcited silver NPs, where the formation of a transient oxygen anion is assumed to be the critical intermediate \cite{Christopher_2011_NatChem.3.467}.  Alternatively, the transient ion itself is unstable towards dissociation, resulting in a bond breakage and further reactions. A prominent example is the hydrodehalogenation of brominated adenine on photoexcited gold and silver NPs \cite{Dutta_2021_ACSCatal.11.8370, Kogikoski_2021_ACSNano.15.20562, Schurmann_2017_Nanoscale.9.1951}.

The SPR can not only decay by charge transfer, but electron-electron scattering leads to the thermalization of excess energy in the picosecond timescale, and electron-phonon coupling then leads to a temperature increase of the NP lattice and eventually also a heating of the environment, including adsorbed molecules \cite{Zhan_2018_NatRevChem.2.216}. Such a temperature increase can drive chemical processes in the electronic ground state.

While there are many examples of plasmon-induced chemical transformations \cite{Gelle_2019_ChemRev.120.986}, it is still a matter of debate which mechanism prevails, i.e. whether charge transfer or simple heating leads to the observed reaction outcomes \cite{Zhou_2018_Science.362.69, Sivan_2019_Science.364, Robatjazi_2020_NatCatal.3.564, Dubi_2022_NatCatal.5.244, Swaminathan_221_AngewChem.60.12532}. This is, on the one hand, due to the complexity of the involved NP--molecular systems, and, on the other hand, due to the experimental and theoretical challenge to probe the different suggested mechanisms \cite{Baffou_2020_LightSciAppl}.

The chemical processes reported so far that can be initiated by plasmonic excitation include simple dissociation reactions, isomerizations, dimerizations and other organic coupling reactions as well as polymerizations \cite{Schurmann_2022_JCP.157.084708, Schurmann_2019_JPCL.10.3153, Sprague-Klein_2018_JACS.140.10583, Ding_2017_ACSPhot.4.1453}. Nevertheless, a systematic exploration is needed to identify which molecules can be plasmonically activated under which conditions (nanostructure, excitation conditions, environment) and how the further reaction pathways can be modified. An important aspect in this context is the interaction between a molecule and the nanoscale metal. It is known that small variations of chemical structure might have a notable impact on the reaction rate of plasmon-induced reactions \cite{Koopman_2021_AdvMaterInterf}. At the same time, it is important to exploit the respective reactions, whether molecules are strongly bound to NPs (e.g. via thiols) or unspecifically and rather loosely connected to the nanostructure. The latter is a prerequisite to use such NPs as catalysts.

Another important challenge concerns the exploration of new plasmonic materials that do not contain Au or Ag, but instead earth-abundant and low-cost elements and at the same time support SPRs and provide favorable photocatalytic properties \cite{Poulose_2022_NatNanotech.17.485}.

There is great potential that chemical pathways can be controlled using the above-listed parameters, however, it needs a holistic approach to assess the influence of each aspect on the reaction products.

\textbf{\textit{How can MM address the problem.}}
The processes described above cover many time and length scales, so MM is ideally suited to address these challenges. The experimental parameters that lead to a specific reaction outcome (defined by nanostructure, reactant molecule(s), excitation conditions and environment) are typically difficult to be described accurately by theory and computations, and simplifications are needed. MM provides the chance to identify the most relevant parameters and elementary steps determining the product formation. Particularly relevant are aggregates of NPs, which give rise to nanoscale interparticle gaps referred to as hot spots. The modeling of plasmonic properties needs to consider the size of nanostructures ranging from about 1~nm to about 100~nm, and the NP-molecular interface, especially within the interparticle gaps. Structural information such as NP facets and specific molecular adsorption geometries might strongly influence plasmonic reactivity and need to be considered by MM. The decay of SPR takes place within fs, but the rate-limiting steps might be the charge transfer between NPs and molecules or desorption and diffusion of molecules, which requires much longer timescales of up to seconds.

\textbf{\textit{Future directions for 5-10 years period.}}
A particular challenge in the theoretical description of plasmonic chemistry is the combination of the classical nature of the plasmonic excitation process (which can be described by classical electromagnetism) with the quantum mechanical nature of the molecules and specific charge transfer processes. Furthermore, the interaction of the molecular system with the nanostructured surface poses significant challenges for computations. While in recent years significant advancement was made in the description of the plasmonic excitation and calculation of e.g. energy distributions of charge carriers, more extensive models need to be developed that consider specific molecules and describe the elementary steps beyond the formation of hot charge carriers.

Plasmonic chemistry is expected to provide unprecedented control handles to tune and steer the energy flow from plasmonic excitation to a specific chemical transformation. However, convincing examples for controlling selective bond activation need to be identified and applied to interesting chemical problems. Plasmonic synthesis in the sense of the synthesis of chemical compounds with a high Free Gibbs energy driven by light needs to be developed, one important example being the synthesis of valuable organic chemicals out of CO$_2$ \cite{Ezendam_2022_ACSEnergyLett}.

\textbf{\textit{Envisaged impact.}}
Plasmonic chemistry has the potential to offer energy-neutral ways to produce value-added chemicals. Hydrogen production and CO$_2$ conversion are important prototypical reactions, which could be driven by sunlight and plasmonic NPs, thereby contributing to CO$_2$ reduction. However, other chemical reactions relevant for organic, pharmaceutical and polymer chemistry can be run under milder conditions using light-irradiated plasmonic NPs to reduce the energy consumption of the chemical industry. Apart from synthesizing relevant molecules, other emerging fields could be explored, such as plasmonic water remediation \cite{King_2022_ChemCatal.2.1880}. Many pharmaceutics present in water as micropollutants are prone to charge-induced degradation, and consequently plasmonic NPs represent an interesting strategy to get rid of these substances using visible light as a driver. Finally, plasmonic NPs are fascinating structures because they enable to concentrate light at a nanoscale, which can, for example, be exploited to functionalize NPs with nanoscale resolution and to create novel functional materials. All this requires a solid mechanistic understanding of the underlying multiscale phenomena, which can be provided by MM combined with a broad range of experiments that cover the different time and length scales.

\subsection{Self-Organization, Structure Formation, and Nanofractals}
\label{sec:Case_study_Self-organization}


\textbf{\textit{The problem.}}
Energy demand is constantly increasing, and the accompanying environmental penalties are intensifying. Proton exchange membrane fuel cells (PEMFCs) are among the most promising next-generation energy devices for clean power generation \cite{Takimoto_2023_NatCommun.14.19}. Over the past 30 years, PEMFC technology has rapidly developed, culminating in the first commercial sales of fuel-cell powered cars in 2015. Although a great success, mass market penetration by these zero-emission vehicles is currently hindered by a dependence on expensive platinum (Pt)-based catalysts, which are responsible for $\sim$46\% of the stack cost \cite{Kodama_2021_NatNanotechnol.16.140}. Of the platinum-group metals (PGMs), Pt has attracted particular attention because of its unique stability in acidic conditions, which makes it the best cathodic electrocatalyst candidate for the oxygen reduction reaction (ORR) in PEMFCs. However, despite this, the intrinsic cost considerations surrounding Pt are exacerbated by its large overpotential and correspondingly poor ORR kinetics, which necessitates a high metal loading to achieve a practical energy density \cite{Liu_2020_JACS.142.17182}. Furthermore, Pt-based cathode durability needs further improvement \cite{Liu_2020_JACS.142.17182}.

To address the issues mentioned above, several strategies present themselves. First, metal combinations can be employed \cite{Xie_2019_ChemRev.120.1184}. In the current case study, a catalytically active noble metal (such as a PGM) can be combined with a much cheaper, earth-abundant transition metal (TM). Creating intermixed PGM/TM systems offers to reduce the loading of noble metal, can promote activity and durability by modifying surface electronic structure \cite{Jenkinson_2020_AdvFunctMater.30.2002633}, and change reaction intermediates to prevent or limit poisoning \cite{Zhang_2021_ChemCommun.5.11}.

A second strategy for improving performance lies in controlling morphology. This offers a channel to develop more active and stable architectures by exposing more reaction sites and/or higher activity (111) crystal facets \cite{Jenkinson_2020_AdvFunctMater.30.2002633}. At the nanoscale, a range of morphologies have been reported. Particularly important are likely to be highly-branched 3D nanostructures such as pods \cite{Lei_2020_NanoRes.13.638} and fractals \cite{Mao_2016_ChemCommun.52.5985}. In particular, the last potentially display vast surface-to-volume ratios, with many more active sites and typically allowing superior performance and the use of smaller loadings than traditional or pseudospherical nanomaterials. Complex 3D structures like nanofractals have shown better resistance to corrosion \cite{Tian_2017_ACSEnergyLett.2.2035}.

\begin{figure}[t!]
\includegraphics[width=0.8\textwidth]{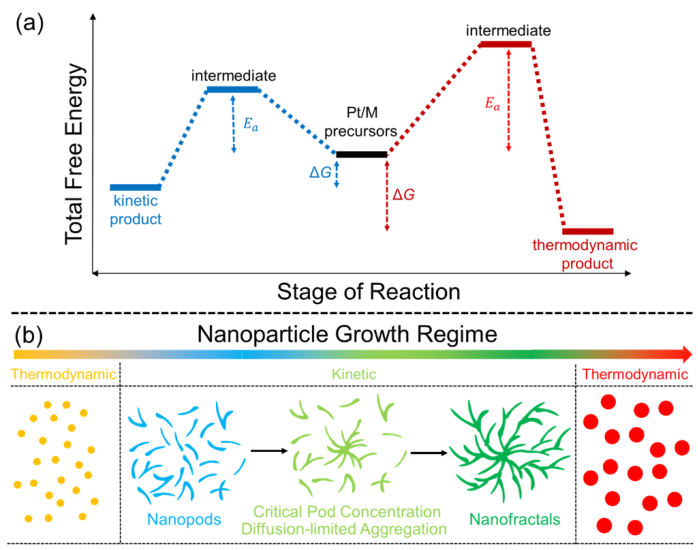}
\caption{Panel~(a): Scope for NP formation under kinetic or thermodynamic control -- the latter is vital for varying composition in heterometallics. Panel~(b): The regimes available as a function of feedstock supply and under increasingly forcing reaction conditions highlight the window of opportunity for nanofractal self-organization.}
\label{fig:CaseStudy_Wheatley}
\end{figure}

Combining the self-organization of highly-branched species such as nanofractals with control over composition promises a new generation of active materials with huge potential. However, this remains experimentally challenging; in practice, changing composition induces changes in morphology and vice versa \cite{Zhang_2018_AdvEnergyMater.8.1703597}. This problem has persisted because researchers have had an incomplete understanding of general growth mechanisms for complex nanoobjects and nanofractals. The corollary of this is that the synthesis of nanofractals has generally been empirical \cite{Chaudhari_2018_NanoRes.11.6111}. For straightforward geometries, temperature, chemical reductant and reaction time have all been harnessed to direct particle growth into the kinetic regime. However, the lack of a strategy for independently varying composition and shape has proved a major problem. That is, having manipulated one parameter e.g. to control morphology, varying a second parameter e.g. to change composition, has induced morphology to change further. What is only now emerging is the understanding that variables are synergetically related. This recognition is beginning to allow the design of compositionally controlled nanofractals by balancing the rate at which precursors react with the growth rate of the emerging particles. For nanofractals, nanopods must first be created under kinetic control, triggering fractal formation when a critical nanopod concentration is achieved, see Figure~\ref{fig:CaseStudy_Wheatley}. Meanwhile, compositional control requires that a thermodynamic reaction pathway be available. Understanding how and why to balance these factors is only now enabling the emergence of an embryonic strategy for nanofractal synthesis \cite{Ming_2023_Nanoscale.15.8814}.

\textbf{\textit{How can MM address the problem.}}
Only recently, a strategy has been suggested for non-empirical nanofractal preparation and the integration of this with multimetallic compositional control \cite{Ming_2023_Nanoscale.15.8814}. Despite this recent advance, the formation mechanism remains poorly understood. While time-resolved electron microscopy has offered insights into the agglomeration of NPs to give anisotropic architectures and the atomistic rearrangement of (hetero)junctions to explain oriented attachment \cite{Chen_2022_ChemNanoMater}, a complete understanding of such self-organization processes necessitates MM.

The importance of MM lies in its ability to probe nanofractal development at different scales. At the most basic level, how do adatom processes occur, and how do the relative rates of atomic deposition on individual crystal facets and adatom migration between facets vary with conditions? More complex, how do multiatom rearrangements of (hetero)junctions occur? In terms of nanofractals, what dictates the critical concentration of nanopods required for diffusion-controlled agglomeration? Furthermore, why is there an upper limit on this critical concentration? The particular appeal of MM lies in the use of stochastic dynamics to probe models of the random diffusion events that underpin the formation of a range of NPs. The ability to understand dynamical processes occurring on sufficiently large time scales so that fast microscopic degrees of freedom can be regarded as noise is a vital precursor to the predictive development of nanofractal chemistry.

\textbf{\textit{Future directions for 5-10 years period.}}
MM is now capable of explaining some hugely complex hierarchical nanochemistry. It can interrogate systems of very different levels of complexity and at different scales – enabling visualization of atomistic processes through to the collision/attachment/reorientation of pre-formed many-atom bodies. MM will allow us to understand the patterns of behavior emerging in PGM/TM systems and harness these for the wider non-empirical formation of complex anisotropic nanomaterials. As preliminary structural data from which information on forming anisotropic nanocatalysts can be extrapolated begins to emerge \cite{Ming_2023_Nanoscale.15.8814}, validation can be sought from ORR tests. Preliminary data suggest the significant outperformance of commercial Pt NPs by Pt/TM nanofractals in terms of mass activity (A/mg$_{\textrm{Pt}}$) and the onset and half-wave potentials (V). Establishing that these systems return impressive performance, it becomes logical to model their creation. This is likely to come to fruition over the next five years, providing a predictive tool for interpreting observations around catalyst creation. Closely linked to this, the evolution (reorganization) of anisotropic nanocatalysts during applications can be expected to form a major vector of study thereafter – a vital area as the principles of catalyst reuse and recycling grow in importance. In the current case, these ideas are beginning to be applied synthetically across the Pt/TM (TM = Fe, Co, Ni) series for PEMFC applications. More widely in the energy sector, new electrolysis systems will emerge for hydrogen production from water splitting, with diverse bimetallic and trimetallic combinations explored \cite{Gao_2023_NatCommun.14.2640}. There will also be implications for the synthetic chemical industry in catalysis and the production of the platform and fine chemicals \cite{Hughes_2021_Resources.10.93}. In the automobile industry, new generations of catalytic converters will become possible \cite{Sushma2018performance}. Electronics applications will likely focus on new electrode materials for devices like supercapacitors and sensors. Lastly, in environmental remediation, we will see the more effective degradation of volatile organic compounds (VOCs) and other pollutants \cite{Fujiwara_2017_EnvironSci}.

\textbf{\textit{Envisaged impact.}}
MM offers massive potential in augmenting our understanding of the hierarchical propagation of nanofractals. There is a vital need to understand processes from the atomistic level (conversion of reagents, irradiative degradation of reagents), combinatorial level (nucleation, interaction of intermediates, collisional particle growth), and structural level (surface morphology, composition, modification, self-organization, reorganization) in isolation, according to the diagrams shown in Figures~\ref{fig:MM_diagram} and \ref{fig:CaseStudy_Wheatley}. However, the interplay of processes at the same or different scales must be comprehended. Doing so will revolutionize the control we can exert at the boundaries of atomistic chemistry and materials science, impacting our ability to fabricate complex materials that offer currently unachievable levels of stability, efficiency, activity and durability in energy production and storage, but also the fields of chemical synthesis, transport, environmental remediation and sensing.

\subsection{3D Nanofabrication using Focused Beams of Charged Particles}
\label{sec:Case_study_3D-nanoprinting}


\textbf{\textit{The problem.}}
As already discussed in Section~\ref{sec:Intro}, the fabrication of nanometer-size devices is the major goal of the nanotechnology industry due to the unique electronic, magnetic, superconducting, mechanical and optical properties that emerge at the nanoscale \cite{Huth_2018_MicroelectronEng, DeTeresa_2016_JPD.49.243003}. However, the fabrication of three-dimensional (3D) nanostructures in a highly controllable manner (3D-nanoprinting) remains a considerable scientific and technological challenge. As the size of the structures falls below $\sim$10~nm, traditional fabrication methods (e.g. plasma etching or plasma-enhanced chemical vapor deposition) cannot control material properties and produce structures of desired size, shape, and chemical composition. Hence, there is currently a strong need to develop new nanofabrication methods based on `bottom-up' rather than traditional `top-down' etching processes, as discussed in Section~\ref{sec:Intro_Ex_FEBID}.

Focused Electron Beam-Induced Deposition (FEBID) \cite{Utke_book_2012, DeTeresa-book2020} introduced in Section~\ref{sec:Intro_Ex_FEBID}, is one of the promising technologies for 3D-nanoprinting since it enables the controlled direct-write fabrication of complex, free-standing 3D structures with feature sizes already produced down to $\sim$10~nm \cite{Winkler_2019_JAP_review, Huth_2021_JAP.130.170901}. The principle of FEBID is based upon a nanometer-sized focused electron beam impinging onto a surface exposed to a stream of precursor (typically organometallic) molecules \cite{Barth2020_JMaterChemC}; see also Section~\ref{sec:Intro_Ex_FEBID}. The decomposition of precursors is primarily induced by low-energy secondary electrons produced as the primary beam impinges on the substrate’s surface. The electron beam can be controlled in both position and pulse duration, with sub-nanometer and sub-microsecond precision, such that complex structures can be fabricated in a single step \cite{Winkler_2019_JAP_review}.

The capability to navigate the charged-particle beam in a well-defined manner is also attributable to the related technique of focused ion beam-induced deposition (FIBID) \cite{Utke_book_2012, DeTeresa-book2020, Reyntjens_FIBID_2000}, where the adsorbed precursor molecules are decomposed as a result of irradiation with a focused ion beam (typically, Ga$^+$ but also lighter ions such as He$^+$) \cite{Kometani_2009_SciTechnolAdvMater, Cui_2012_APL.100.143106, Nanda_2015_JVSTB.33.06F503, Cordoba_2018_NanoLett.18.1379}. Since FEBID and FIBID are based on the similar principle of charged-particle beam-induced deposition, they can complement each other, e.g., regarding the growth dynamics for 3D fabrication, achieving different material properties and structure resolution \cite{Winkler_2019_JAP_review}.

The reliable transfer from initial 3D design into the delivery of real nano-architectures in a routine way remains a significant challenge. The current major roadblock is a lack of molecular-level understanding of the Irradiation-Driven Chemistry (IDC) that governs nanostructure formation and growth during the FEBID and FIBID processes \cite{Sushko_IS_AS_FEBID_2016, Swiderek_2018_BJN.9.1317}. Of particular relevance is the incorporation of unwanted chemical elements (such as carbon) in the fabricated metal nanostructures \cite{Botman_2009_Nanotechnol.20.372001}, which can reduce or even mask the intended material properties. Most of the available post-processing purification procedures developed to alleviate this drawback are not readily applicable for 3D structures due to the severe structural deformations resulting from the high volume loss during impurity removal \cite{Geier_2014_JPCC.118.14009, Fowlkes_2015_PCCP.17.18294}.

The advancement of the methods for irradiation-driven 3D nanofabrication should be based on a deeper understanding of the molecular interactions and the key dynamical phenomena in nanosystems exposed to irradiation. This goal can be achieved by utilizing modern computational MM tools combined with experimental studies to validate such simulations.

\textbf{\textit{How can MM address the problem.}}
The study of the physicochemical phenomena that govern the formation and growth of nanostructures (both coupled to radiation and without irradiation) is a complex multi-parameter problem, as already highlighted in Section~\ref{sec:Intro}. Indeed, in the case of FEBID and other radiation-induced nanofabrication techniques such as FIBID, different precursor molecules, substrates, irradiation, replenishment and post-processing regimes, as well as additional molecular species facilitating the decomposition of precursors can be explored to improve the purity of fabricated deposits and increase their growth rate \cite{Utke_book_2012, DeTeresa-book2020}.

An understanding of the IDC in the FEBID process can be advanced using a computational MM approach that describes the whole set of FEBID-related processes occurring over different time and spatial scales and by establishing procedures for experimental validation of the MM results. The MM approach should embrace together different spatiotemporal stages and the corresponding processes and phenomena shown in Fig.~\ref{fig:MM_diagram}. Such an approach should combine: (i) \textit{ab initio} and DFT methods (see Sections~\ref{sec:Methods_Many-body_theory} and \ref{sec:Methods_DFT}) to evaluate parameters of irradiation- and chemically-induced quantum transformations occurring in the systems at the molecular level; (ii) classical and reactive MD (Sections~\ref{sec:Methods_classicalMD} and \ref{sec:Methods_RMD}) to study fragmentation of precursor molecules \cite{deVera_2019_EPJD.73.215, Andreides_2023_JPCA.127.3757}, and their interaction with the substrates on time and spatial scales accessible in classical MD; (iii) evaluation of cross sections of relevant collision-induced quantum processes (e.g. electronic excitation, ionization, dissociative ionization, and DEA) via \textit{ab initio} calculations and analytical estimates (Section~\ref{sec:Methods_Cross_sections}) or using data available in atomic and molecular databases; (iv) track-structure MC simulations (Section~\ref{sec:Methods_MC_transport}) to study the fluxes and fluences of primary and secondary electrons in order to evaluate the probabilities of the aforementioned quantum processes; (v) IDMD (Section~\ref{sec:Methods_IDMD}) to model random interactions of the electron beam and secondary electrons with the growing nanostructure taking into account possible chemical transformations therein; and (vi) the SD method (Section~\ref{sec:Methods_StochasticDyn}) to model the evolution of many-particle systems over the time scales significantly exceeding those accessible in MD simulations.

Different parts of the MM methodology have been successfully interlinked in MBN Explorer \cite{Solovyov_2012_JCC_MBNExplorer} to explore the mechanisms of formation and growth of metal-containing nanostructures under the action of focused electron beams \cite{Sushko_IS_AS_FEBID_2016, DeVera2020, Prosvetov2021_BJN, Prosvetov2022_PCCP, Prosvetov_2023_EPJD, SD_FEBID_abstract_Prague2023}.

\begin{figure}[t!]
\includegraphics[width=0.8\textwidth]{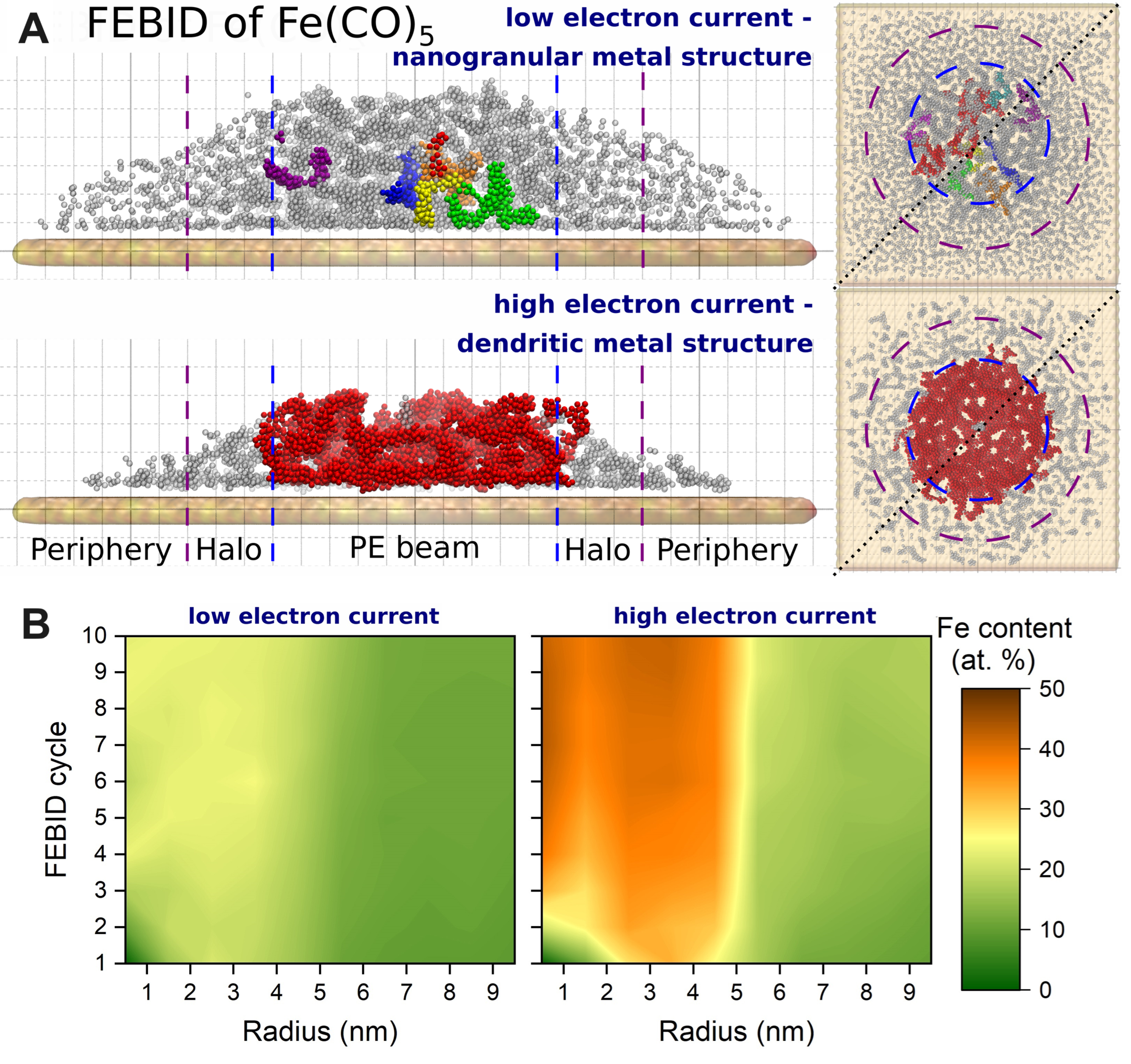}
\caption{Results of the IDMD simulations \cite{Prosvetov2022_PCCP} of the FEBID process for Fe(CO)$_5$. Panel (a) shows the snapshots of the simulated iron-containing nanostructures: side view on diagonal cross sections indicated by dotted lines (left) and top view (right). The top and bottom snapshots in panel (a) correspond to the electron current of 1 and 4~nA, respectively. Only iron atoms are shown for clarity. Topologically disconnected metal clusters containing more than 100 iron atoms are shown in different colors. Smaller clusters containing less than 100 iron atoms are shown in grey. Boundaries of the primary electron beam, halo and peripheral regions are indicated by dashed lines in the left column and by circles in the right column. Grid line spacing is equal to 1~nm in all dimensions. Panel (b) shows the atomic Fe content of the grown iron-containing structures as functions of the number of simulated irradiation-replenishment FEBID cycles for electron currents of 1 nA (left) and 4 nA (right). Figures are adapted from ref~\citenum{Prosvetov2022_PCCP}.}
\label{fig:CaseStudy_Nanofabrication}
\end{figure}

Figure \ref{fig:CaseStudy_Nanofabrication} illustrates the capabilities of the MM methodology for the atomistic level characterization of grown nanostructures, prediction of their morphology, growth rate, as well as geometrical (e.g. lateral size, height and volume) and chemical (metal content) characteristics. The morphology of deposits is an important characteristic that governs many physical properties, such as electrical and thermal conductivity and magnetic properties \cite{Huth_2009_NJP.11.033032, Huth_2020_Micromachines.11.28}. Figure~\ref{fig:CaseStudy_Nanofabrication}a shows snapshots of the IDMD simulations \cite{Prosvetov2022_PCCP} of the FEBID process for Fe(CO)$_5$, one of the most common FEBID precursors used to fabricate magnetic nanostructures. A variation of the electron current during the FEBID process significantly changes the deposit's morphology (Figure~\ref{fig:CaseStudy_Nanofabrication}a) and elemental composition (Figure~\ref{fig:CaseStudy_Nanofabrication}b). At a low beam current corresponding to a low degree of precursor fragmentation, the deposits consist of isolated small-size iron clusters surrounded by organic ligands, see the top panel in Figure~\ref{fig:CaseStudy_Nanofabrication}a. In this irradiation regime, metal clusters formed as a result of electron-irradiation-induced fragmentation of precursor molecules do not agglomerate with increasing electron fluence, resulting in relatively low metal content in the deposit. Higher beam current facilitates the precursor fragmentation, and the metal clusters coalesce into dendrite-like structures with the size corresponding to the PE beam, see the bottom panel of Fig.~\ref{fig:CaseStudy_Nanofabrication}a. In this regime, the deposit's metal content increases twofold compared to the case of low current, as shown in Figure~\ref{fig:CaseStudy_Nanofabrication}b.

The outcomes of atomistic simulations of the FEBID process using the IDMD approach can be used to construct stochastic models of FEBID using the SD methodology introduced in Section~\ref{sec:Methods_StochasticDyn}. An illustration of the application of SD for modeling the FEBID process was presented recently \cite{SD_FEBID_abstract_Prague2023}. It was demonstrated that the FEBID process can be described through step-by-step transformations occurring to particles of different types, representing intact and fragmented precursor molecules, ligands, isolated metal atoms, and the substrate. The probabilities of the underlying processes occurring in the system can be determined through atomistic MD simulations (see Section~\ref{sec:Methods_Nonequil_Chem}) and track-structure MC calculations (Section~\ref{sec:Methods_Particle_Transport}).

\textbf{\textit{Future directions for 5-10 years period.}}
A molecular-level understanding of the IDC in the FEBID process, including the mechanisms of electron-induced molecular fragmentation and the mechanisms of nanostructure formation and growth, can provide a deeper understanding of the relationship between deposition and irradiation conditions and their impact on the physicochemical characteristics of fabricated nanostructures (size, shape, purity, crystallinity, etc.). Developing such understanding is essential for broader exploitation of the FEBID 3D-nanoprinting technology.

From the MM side, future directions toward achieving this goal require the construction and validation of SD models, which enable simulations of the nanostructures' growth and the characterization of their properties on the time and spatial scales much larger than those accessible in pure atomistic MD simulations. Parameters of the SD models should be thoroughly validated through ``lower-scale'' modeling (e.g. through quantum-mechanical calculations or IDMD simulations) and experimental data (see Section~\ref{sec:Validation}).

\textbf{\textit{Envisaged impact.}}
As discussed above in this section and in Section~\ref{sec:Intro_Ex_FEBID}, FEBID is considered one of the most promising technologies for the controlled direct-write fabrication of complex nanostructures with nanometer resolution \cite{Winkler_2019_JAP_review}. Nanostructures created using this method can be used in electronic devices and other applications, including catalysis and nanoelectrochemistry \cite{Xu_2008_NatMater.7.992, Murray_2008_ChemRev.108.2688}; as sensors, nanoantennas and magnetic devices; surface coatings; and thin films with tailored properties.

An advantage of FEBID 3D-nanofabrication is that it can be performed using a conventional scanning electron microscope (SEM) with a mounted gas-injecting system to inject gaseous precursor molecules inside the SEM vacuum chamber. Considering the large number of SEMs installed at universities and research centers worldwide, the development of a reliable and easy-to-use methodology for 3D-nanofabrication using SEMs would open great opportunities for fundamental and applied research.

Further development and broader exploitation of a MM approach for molecular-level studies of the physicochemical processes behind FEBID would enable the prediction of elemental composition (particularly metal content) as well as nano- and microstructure of the growing nanostructures, e.g. their granularity properties and morphological transitions therein. Therefore, the MM approach for the description of FEBID 3D-nanofabrication can provide the necessary insights into the fundamental knowledge of the radiation chemistry required for optimizing the FEBID regimes and thus setting up a routine methodology for FEBID 3D-nanofabrication of new nanostructured systems with desired architecture and properties, e.g. electrical, magnetic, superconducting, plasmonic, mechanical, or thermomechanical ones.

Novel and more efficient methods of 3D-nanofabrication will allow for the miniaturization of the created electronic nanodevices and their cost-effective production. A better understanding of the mechanisms of radiation-induced formation, growth and modification of nanostructures will enable effective optimization of existing nanofabrication technologies, allowing more precise / better-controlled fabrication, and targeting specific compositions and morphologies of the fabricated nanostructures with tailored properties.

Finally, multiscale models similar to those applied for FEBID simulations can be utilized for simulations of other nanofabrication techniques using focused beams of charged particles, such as FIBID, or for other nano-processing methods exploiting ion beams, e.g. focused ion beam-induced etching \cite{Utke2008, Stanford_2016_ACSAMI.8.29155}. Further technological applications of the multiscale models include simulations of non-irradiative chemical synthesis \cite{Ming_2023_Nanoscale.15.8814} (see Section~\ref{sec:Case_study_Self-organization}) for the formation of nanofractals and other complex nanoobjects.

\subsection{Deposition and Quality Control During the Deposition Process of a 2.5D to 3D Structures as Close as Possible to the Design}
\label{sec:Case_study_TESCAN}


\textbf{\textit{The problem.}}
As the world of nanotechnology moves forward, more and more applications begin appearing where we explore the 3D dimension and the properties standing from this approach. So far, most 3D structures are created by masking the surface and using different techniques such as selective etching, deep reactive ion etching, and filling the space created by sputtering or electrochemistry. Nanowires can be grown by exploring the self-assembly properties of selected materials. All of these approaches are great at what they can achieve, but they still miss one fundamental property of the 3D structures, that is, complexity.

Novel nanostructures like double helix antennas in the field of plasmonics, tungsten nanowires for alternative electron source in the field of nanoelectronics, precise cobalt, iron or nickel-iron nanowires as tips for scanning probe microscopy in the field of magnetic are just few examples of complex structures which require not only precise control of chemical composition but also shape fidelity \cite{Seewald_2021_Micromachines.12.115, Utke_book_2012, Frabboni_2006_APL.88.213116, Winkler_2019_JAP_review, Keller_2018_SciRep.8.6160, Sanz-Hernandez_2020_ACSNano.14.8084, Passaseo_2017_AdvOptMater.5.1601079, Winkler_2017_ACSApplMatInt.9.8233, Esposito_2015_ACSPhoton.2.105, Plank_2020_Micromachines.11.48, Beard_2010_Nanotechnol.21.475702, Burbridge_2009_Nanotechnol.20.285308, Mutunga_2019_ACSNano.13.5198, Fowlkes_2020_Micromachines.11.8, Bret_2004_JVSTB.22.2504, Molhave_2004_Nanotechnol.15.1047, Fowlkes_2016_ACSNano.10.6163, Hirt_2017_AdvMater.29.1604211, Sanz-Hernandez_2017_BJN.8.2151, Plank_2008_Nanotechnol.19.485302}.

To retain a complex 3D shape using an electron or ion beam requires control of several parameters \cite{Kuhness_2021_ACSApplMatInt.13.1178, Pablo-Navarro_2019_Nanotechnol.30.505302, Winkler_2018_ACSApplNanoMater.1.1014, Fowlkes_2018_ACS_ApplNanoMater.1.1028, Toth_2015_BJN.6.1518, Guo_2012_JpnJAP.51.065001, Guo_2013_JVSTB.31.061601, Winkler_2014_ACS_ApplMatInt.6.2987, Olsen_2010_TheorChemAcc.125.207, Keller_2018_BJN.9.2581} and without modeling these complex interactions, it comes down to a long process of testing ideas and adjusting the processes based on the results.

\textbf{\textit{How can MM address the problem.}}
To shorten the time to result and improve the final quality of the devices, one could apply MM to control the changing parameters of the deposition to retain the desired shape. The main parameters involved here are the precursor chemical composition, the flow of the precursor, distance to the sample, surrounding pressure, the beam current, species, and the focus point of the beam itself.

Not all of these parameters can be changed in all the existing systems from different manufacturers, adding additional complexity to the MM, where some parameters might become easier to predict with the model but much more challenging to change in the real environment.

This property of the specific FIBSEM producers hardware limitation adds another level of complexity to the MM for the simulation and requires a large group of users to provide the initial data for the fine tuning of the process.

\begin{figure}[t!]
\includegraphics[width=1.0\textwidth]{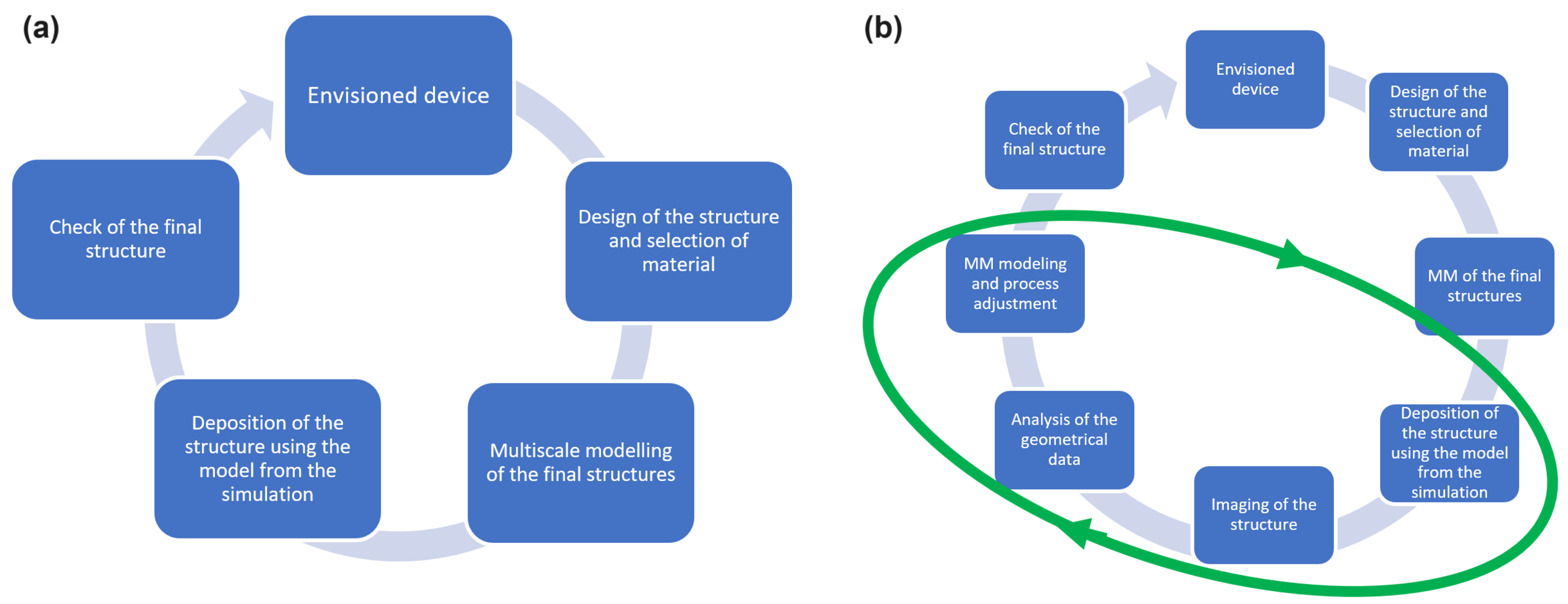}
\caption{Panel~(a): The standard process for MM usage in the deposition process. This process normally requires multiple repeated steps to produce the originally designed device. Panel~(b) shows a process where the MM is combined with imaging during the deposition process, feeding the geometrical information back to the MM and adjusting the model and parameters until the desired result is achieved.}
\label{fig:CaseStudy_Hrabovsky-1}
\end{figure}

\textbf{\textit{Future directions for 5-10 years period.}}
At the moment, the most used approach of 3D deposition would be, as shown in the Figure~\ref{fig:CaseStudy_Hrabovsky-1}a, to use MM simulation to calculate all the deposition parameters based on the provided design by creating a 3D patterning file with the beam position, used precursors, dwell time and focus length. This approach then creates the structure in one uninterrupted process, and only after this the design fidelity can be checked \cite{Schindelin_2012_NatMeth.9.676}, and if the structure does not have the desired shape, this information can be used to adjust the MM and try again.

A head of the field solution is shown in Figure~\ref{fig:CaseStudy_Hrabovsky-1}b to apply the same logic as in the previous approach but set a series of checks during the deposition process, analyze the geometrical properties of the structure \cite{Schindelin_2012_NatMeth.9.676}, use this data to adjust the MM, get updated deposition 3D file and repeat. This process should be repeated until the final desired shape is achieved.

\begin{figure}[t!]
\includegraphics[width=1.0\textwidth]{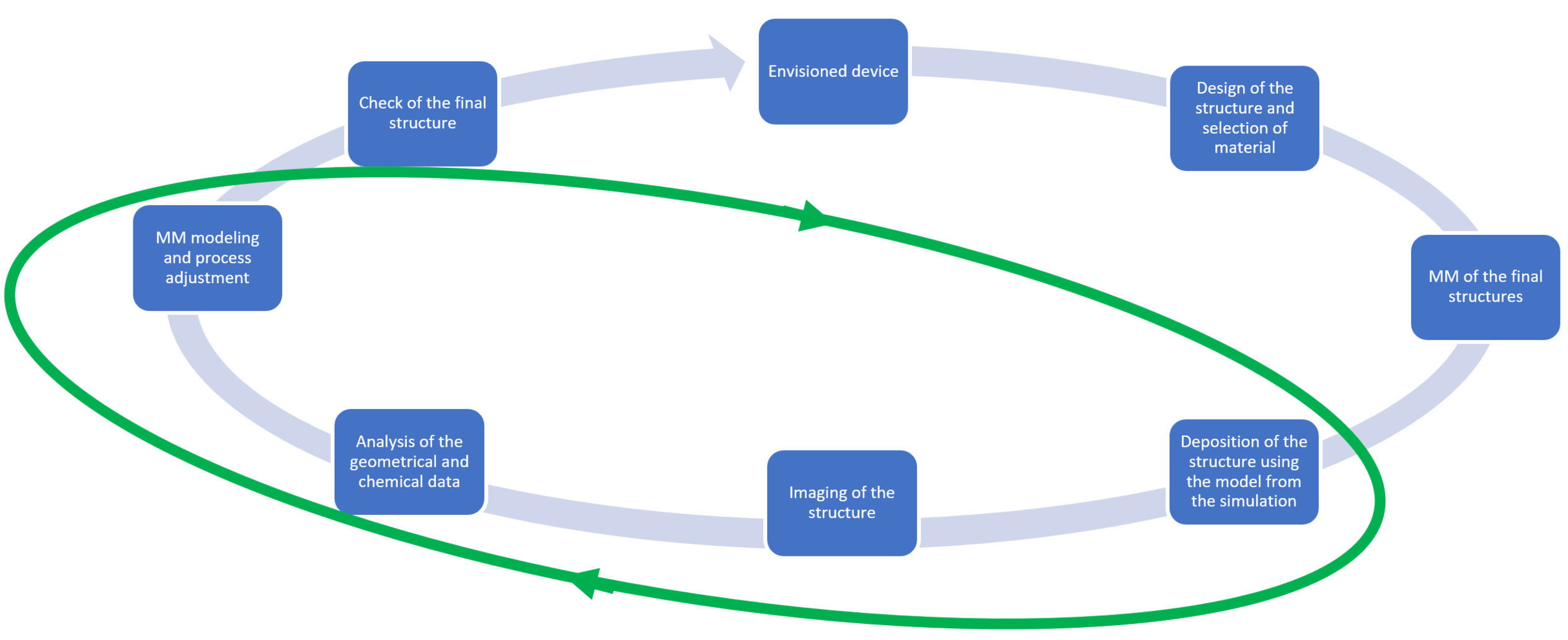}
\caption{Adjusted deposition process with the ability to collect both geometrical and chemical data from the process and use it for the MM and the process adjustments.}
\label{fig:CaseStudy_Hrabovsky-2}
\end{figure}

\textbf{\textit{Envisaged impact.}}
The envisioned impact of the MM on the community of 3D depositions driven by electron or ion beams is to use it in a process described in Figure~\ref{fig:CaseStudy_Hrabovsky-2}, where the data gathered from the process is not only geometrical but also chemical data. This can be achieved using TOF-SIMS analysis during the deposition to check for residues \cite{Jurczyk_2022_Nanomaterials.12.2710, Engmann_2012_PCCP.14.14611, Wnuk_2009_JPCC.113.2487} of the precursors and simulate the precursors molecules decomposition. The novel approach, coupled with cutting-edge technologies like gas-injection systems with controllable flow and a mixture of molecules, will be used to achieve more complex 3D structures for the next generation of nanodevices.

\subsection{Exploring New Frontiers -- Space Chemistry}
\label{sec:Case_study_Space_Chemistry}


\textbf{\textit{The problem.}}
Humanity is entering a new and exciting era of space exploration and exploitation (an era that has come to be known as Space~4.0) where an authoritative exploration programme of the planets and moons in and beyond our own Solar System will be conducted, and a permanent presence of humans in space such as on lunar bases or platforms in Earth and lunar orbit will be established \cite{Athanasopoulos2019_SpacePolicy}. With the commissioning of ground- (e.g. the Atacama Large Millimeter/submillimeter Array (ALMA), and the Extremely Large Telescope (ELT)) and space-based (e.g. GAIA, Eucid and  James Webb Space Telescope (JWST)) observatories and a suite of space missions (e.g. ExoMars (to study Mars), JUICE (to study moons of Jupiter) and the exoplanet missions Plato, and Ariel), the next two decades will provide a comprehensive view of the cosmos that may indeed allow us to address fundamental questions such as: (i) How do the stars and planets form? and (ii) How did life begin on Earth, and is it prevalent elsewhere?

Compared to the terrestrial surface, space represents a hostile environment characterised by a wide range of temperatures, extreme vacuum conditions, and an active radiation environment; all of which influence the physical (and chemical) processes that occur there \cite{vanDishoeck_2014_FadarayDisc, Arumainayagam_2019_ChemSocRev}. This requires a new understanding (and, often, re-interpretation) of many basic physical phenomena that, to date, have been assessed and defined purely in the terrestrial context. For example, there is a need to broaden the study of collision processes to include conditions relevant to space (e.g. low temperatures where tunnelling may be prevalent to overcome reaction barriers), as well as to study functional changes in space-borne materials (e.g. tensile strength, creep, fatigue) due to the altering of basic properties such as density, conductivity, and melting points caused through the broader accessibility of the phase diagram under space-relevant conditions (i.e., away from standard temperature and pressure). Indeed, new materials, such as novel alloys, are envisaged to be manufactured in space environments that could not be otherwise produced in terrestrial facilities \cite{Ghosh_2023_Materials.16.383}.

Replicating the conditions in space environments remains a significant challenge for the experimental community. Whilst modern cryogenic techniques allow the low temperatures of interstellar space and planetary/lunar surfaces to be replicated, the ultra-high vacuum conditions inherent in space are seldom met in any simulation facility (being $<10^{-12}$~torr) \cite{Mason_2006_FaradayDisc.133.311}. Thus, in the interstellar medium, the accretion of molecules on micron-sized dust surfaces is very slow, one molecule per month or longer. However, these ice-covered dust surfaces act as `chemical factories' for synthesizing the more complex molecules that are the prebiotic compounds leading to the evolution of life. Similarly, whilst all bodies (planets, moons or spacecraft) are subject to irradiation (from solar wind and cosmic rays) the flux at any one time is low, but the accumulated dose over many years is high. It is impossible to replicate the long time periods over which physical and chemical processes occur in any laboratory. Thus, a major challenge for astrochemistry and astrophysical studies is how and whether laboratory studies can accurately reflect the processes occurring in space.

\textbf{\textit{How can MM address the problem.}}
Faced with the inability to replicate space conditions accurately in the laboratory, it is necessary to develop detailed models of the space environment, which can then be used to test the accuracy (or otherwise) of the less-than-perfect laboratory simulations. For example, is the morphology of ices formed by very slow deposition in space replicated by deposition at higher and faster flux rates in an experimental simulation?  Are the chemical processes induced by low fluxes the same as high flux if the total dose received is the same? \cite{Dohnalek_2003_JCP.118.364, Kimmel_2001_JCP.114.5284, Stevenson_1999_Science.283.1505, Holtom_2006_PCCP.8.714}.

These questions can be answered using a MM approach. Indeed, MM is the perfect methodology for addressing the different time scales of experiments and space since, as demonstrated in Figure~\ref{fig:MM_diagram}, it can model molecule by molecule impact on a replicate of micron dust surface with each molecule being allowed to diffuse across the surface to create the `ice layer'.

\begin{figure}[t!]
\includegraphics[width=0.8\textwidth]{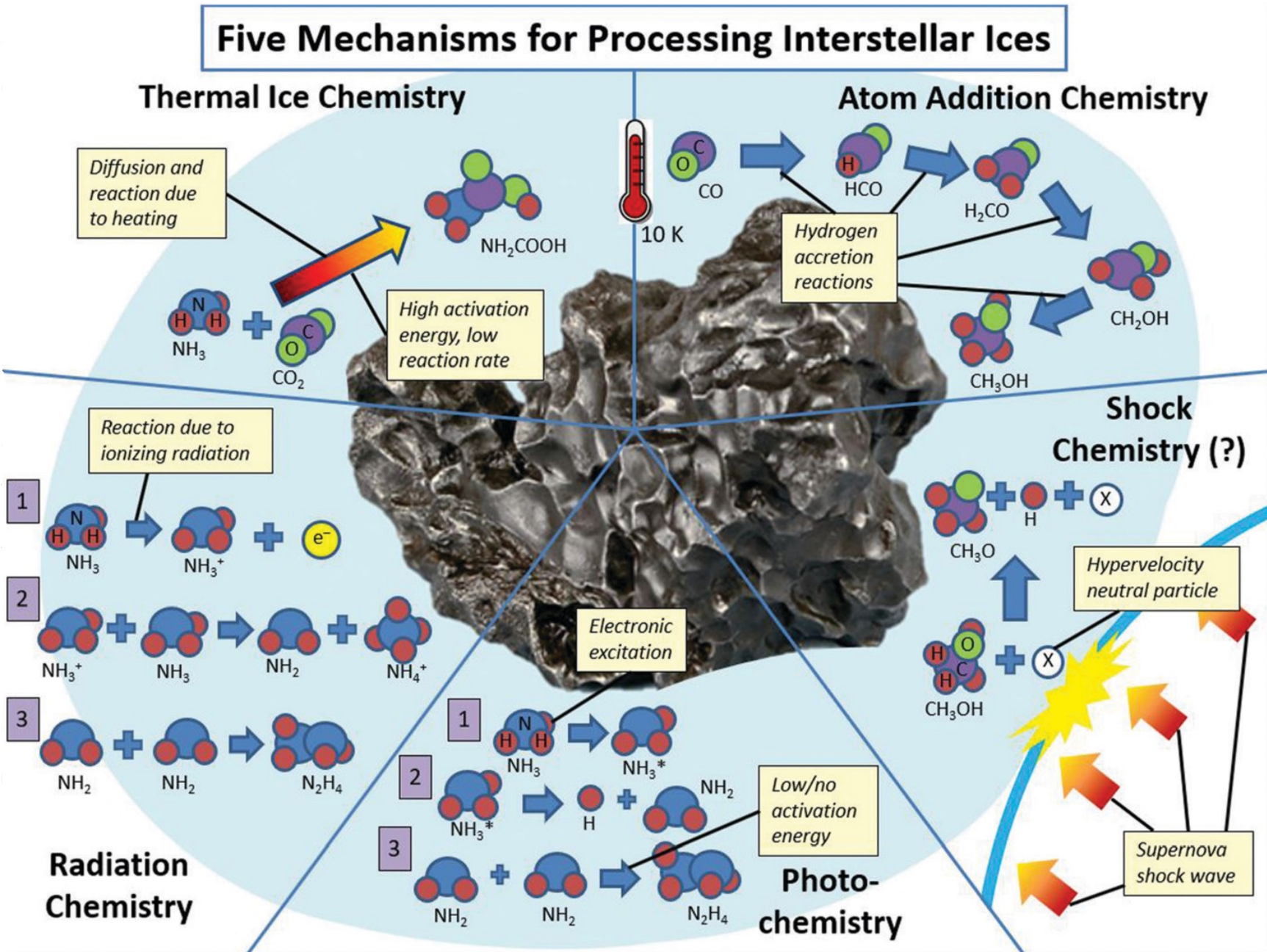}
\caption{Schematic diagram of five different mechanisms that can lead to physical and chemical alteration on an ice covered interstellar dust grain. Adapted from ref~\citenum{Arumainayagam_2019_ChemSocRev}.}
\label{fig:CaseStudy_Mason}
\end{figure}

The flux of incident absorbing species may be readily altered in the MM to determine how (or if) the final morphology of the ice is controlled by the flux rate. Indeed, a MM model is the only methodology by which we can explore the morphology of the ice under ISM conditions. By coupling particle transport over different length and time scales a MM can reveal, for the first time, the morphology of astrochemical ices on ISM dust grains.

Having established an ice layer on ISM dust, the synthesis of new species arising from various processes (as shown in Figure~\ref{fig:CaseStudy_Mason}) can be explored. Atom--atom (e.g. H + H to form H$_2$) or atom--molecule chemistry (e.g. H accretion reactions) driven by the Langmuir-Hinshelwood mechanism may create the simplest hydrocarbon species such as CH$_3$OH. Once again, MM is ideal for such studies through the coupling of radiation induced quantum processes (photon and radiation-induced chemistry in Figure~\ref{fig:CaseStudy_Mason}) to determine the non-equilibrium chemistry (e.g. creation of radicals (OH) and secondary electrons) that leads to the synthesis of complex molecular species which may be detected by remote observations either as spectral features within the ice or as gaseous species released by thermal processing of the grain or by shocks. MM allows the ice-covered dust grain to be exposed to various processes over different times, since sequential events are often separated by months or even years in the low pressure of space. Only by MM and its validation with experiments and observations can we `cycle' a representative ISM ice through its `life' in the ISM dust cloud (e.g. deposition-irradiation-thermal processing-fresh deposition-irradiation). The MM may then be used to understand whether a `Digital Twin' approach to space studies is feasible in which the MM model derives the chemical (and thermodynamic) equilibrium achieved in the observed ISM dust cloud. MM ISM ice-dust models may then be integrated into (form the basis of) star and planet formation models.

\textbf{\textit{Future directions for 5-10 years period.}}
The recent advances in replicating the chemistry in MM and the addition of stochastic dynamics are important in building a more detailed model of the effects of irradiation of materials (be they ice-covered surfaces or spacecraft materials) \cite{Bacon_2002_IntMaterRev, Shingledecker_2018_PCCP.20.5359, Shingledecker_2018_AstrophysJ.861.20}. Such a model is lacking at present. Therefore, a MM model of physico-chemistry occurring during irradiation of ice covered dust grains in the ISM is one of the most urgent challenges for MM.

Furthermore, the forthcoming establishment of permanent structure on the moon requires an assessment of how structural and operational materials (e.g. those used to extract water and oxygen from lunar regolith) are altered by radiation over periods of 10 to 20 years (a deliverable and commercially necessary lifetime for their use) \cite{Sherwood_2019_ActaAstron}. At present, we have little experience with how structural materials that may be used to construct a lunar base will be altered under lunar conditions where cosmic radiation is coupled to dramatic changes in temperature (the lunar surface varying from $+150^{\circ}$C to $-150^{\circ}$C) and the gamma ray flux inherent on the moon. MM of such materials is a feasible methodology for acquiring such data in the next decade, allowing materials to be preselected for deployment in any lunar base or be specifically designed for such a role. Given the high cost of transporting materials to the moon, the use of \textit{in situ} resources is recommended, requiring feasibility studies of using lunar regolith as a construction material. MM of irradiation of such material to explore its activation and physical properties when sintered into `lunar bricks' is a high priority,

It is therefore to be envisaged that a complete MM approach to the study of physical and chemical processes in space environments will be developed in the next five years with the objective of creating a new predictive model capable of both interpreting observations of chemical inventories on planetary and lunar surfaces (such and the Jovian moons to be explored by ESA and NASA missions in next decade) and interpreting the observations of chemical species found in regions of the ISM by the JWST. Such a model may also be developed for the assessment and selection of materials to be used in the construction of the next generation of space stations (in Earth and lunar orbit) and the first lunar bases with the future design of Martian missions, such a strategy being enshrined in ESA and NASA strategies.

A combination of MM of space environments and MM of radiation of DNA and cells (as discussed for ion-beam cancer therapy in Section~\ref{sec:Intro_Ex_IBCT}) may also provide a more accurate model of the risk to astronauts when in space and determine mitigation strategies for their exposure to space radiation during the longer term space missions now envisaged as humans establish a permanent presence in space \cite{Strigari_2021_FrontPublHealth, Mertens_2018_SpaceWeather}.

\textbf{\textit{Envisaged impact.}}
The exploitation of MM in revealing physical and chemical processes in space has the potential to have an enormous impact on our understanding of how stars and planets form and how life has evolved on Earth (and whether it is likely to have evolved elsewhere in the universe). MM will also provide the most extensive test of whether we can replicate structures and mechanisms in the ISM and on planetary/lunar bodies in terrestrial laboratories, determining the future direction of laboratory astrochemistry. MM of irradiated materials and development of a biological damage model may play a core role in determining how humanity establishes itself in space and the next stage of humanity's engagement in space exploration and exploitation -- `Space 5.0' humanity's settlement beyond the Earth.

\subsection{Construction of Novel Crystal-Based Light Sources}
\label{sec:Case_study_CLSs}


\textbf{\textit{The problem.}}
The development of Light Sources (LSs) operational at wavelengths $\lambda$ well below one angstrom (corresponding photon energies $\hbar \omega > 10$~keV) is a challenging goal of modern physics \cite{AVK_AVS_2020_EPJD.74.201_review, NovelLSs_Springer_book}. Sub-angstrom wavelength, ultrahigh brilliance, and tunable LSs will have a broad range of exciting potential cutting-edge applications. These applications include exploring elementary particles, probing nuclear structures and photonuclear physics, and examining quantum processes, which rely heavily on gamma-ray sources in the MeV to GeV range \cite{AVK_AVS_2020_EPJD.74.201_review, NovelLSs_Springer_book, Zhu_2020_SciAdv.6.eaaz7240, Howell_JPG_2022.49.010502}. Modern X-ray free-electron lasers (XFELs) can generate X-rays with wavelengths $\lambda \sim 1$~\AA \cite{Wu_2006_PRL.96.224801, Doerr_2018_NatMeth.15.33, Seddon_2017_RepProgPhys.80.115901, Milne_2017_ApplSci.7.720, Bostedt_2016_RMP.88.015007}. Existing synchrotron facilities provide radiation of shorter wavelengths but orders of magnitude less intensive \cite{Couprie_2014_JESRP.196.3, Tavares_2014_JSynchrRad.21.862, Yabashi_2017_NatPhoton.11.12}. Therefore, to create a powerful LS in the range $\lambda \ll 1$~\AA~new approaches and technologies are needed.

The practical realization of novel gamma-ray LSs that operate at photon energies from $\sim$100~keV up to the GeV range can be achieved by exposing oriented crystals (linear, bent and periodically bent) to the beams of ultra-relativistic charged particles \cite{AVK_AVS_2020_EPJD.74.201_review, NovelLSs_Springer_book}. In this way, novel gamma-ray Crystal-based LSs (CLSs) such as crystalline emitters of channeling radiation, crystalline emitters of synchrotron radiation, crystalline undulators (CUs) and others, can be constructed \cite{AVK_AVS_2020_EPJD.74.201_review, NovelLSs_Springer_book}. The practical realization of such crystal-based LSs is the main goal of the currently running European Project ``Emerging technologies for crystal-based gamma-ray light sources'' (TECHNO-CLS) \cite{TECHNO-CLS_website}. The realization of this goal is possible through (i) the development of breakthrough technologies needed for manufacturing the high-quality crystals of desired geometry and crystalline structure, (ii) creating the high-quality beams of ultra-relativistic electrons and positrons, and (iii) the apparatus design for generation, tuning and output of the beams of intensive gamma-rays with wavelengths significantly shorter than 1~Angstrom, i.e. within the range that cannot be reached in existing LSs based on magnetic undulators. Additionally, using a pre-bunched beam, a CU LS has the potential to generate coherent superradiant radiation of super-enhanced intensity \cite{AVK_AVS_2020_EPJD.74.201_review, NovelLSs_Springer_book}.

The motion of a projectile and the radiation emission in bent and periodically bent crystals are similar to those in magnet-based synchrotrons and undulators. The main difference is that in the latter, the particles and photons move in a vacuum, whereas in crystals, they propagate in a medium, thus leading to a number of limitations for the crystal length, bending curvature, and beam energy. However, the crystalline fields are so strong that they steer ultra-relativistic particles more effectively than the most advanced magnets. Strong fields bring the bending radius in bent crystals down to the cm range and the bending period in periodically bent crystals to the hundred or even ten microns range. These values are orders of magnitude smaller than those achievable with magnets \cite{Emma-EtAl_NaturePhotonics_v4_015006_2010}. As a result, the CLSs can be miniaturized, thus dramatically lowering their cost compared to conventional LSs.

The Horizon Europe EIC-Pathfinder Project TECHNO-CLS \cite{TECHNO-CLS_website} is a representative example of a high-risk/high-gain science-towards-technology breakthrough research program addressing the physics of the processes, which accompany the exposure of oriented crystals to irradiation by high-energy electron and positron beams, at the atomistic level of detail required for the realization of the aforementioned TECHNO-CLS goals. A broad interdisciplinary and international collaboration has been created previously in the frame of FP7 and H2020 projects, which performed initial experimental tests to demonstrate the CU idea \cite{AVK_AVS_WG_1998_JPG.24.L45}, production and characterization of periodically bent crystals, see refs~\citenum{AVK_AVS_2020_EPJD.74.201_review, NovelLSs_Springer_book} for references.

The aforementioned developments have been driven by the theory of CLSs and related phenomena \cite{AVK_AVS_2020_EPJD.74.201_review, NovelLSs_Springer_book}, as well as by the advanced computational methods and tools \cite{Solovyov_2012_JCC_MBNExplorer, RelMD_2013_JCompPhys.252.404, Sushko_2019_MBNStudio}. These theoretical and computational studies have ascertained the importance of the high quality of the CLS materials needed to achieve strong enhancement effects in the photon emission spectra \cite{AVK_AVS_2020_EPJD.74.201_review, NovelLSs_Springer_book}. By now, several methods for creating periodically bent crystalline structures have been proposed and/or realized; see refs~\citenum{AVK_AVS_2020_EPJD.74.201_review, NovelLSs_Springer_book} and references therein.

Advanced atomistic computational modeling of the channeling process, channeling and undulator radiation and other related phenomena beyond the simplistic continuous potential framework has been carried out using the multi-purpose computer package MBN Explorer \cite{Solovyov_2012_JCC_MBNExplorer, RelMD_2013_JCompPhys.252.404} and the special multi-task software toolkit with graphical user interface MBN Studio \cite{Sushko_2019_MBNStudio}. The MBN Explorer package is introduced in Sections~\ref{sec:Methods} and \ref{sec:Interlinks}. For simulations of the channeling and related phenomena, an additional module has been incorporated into MBN Explorer to compute the motion of relativistic projectiles along with dynamical simulations of their environments, including the crystalline structures. MBN Explorer enables such simulations for various materials, including biological ones \cite{MBNbook_Springer_2017, DySoN_book_Springer_2022}. The computation accounts for the interaction of projectiles with separate atoms of the environments, whereas many different interatomic potentials implemented in MBN Explorer support rigorous simulations of various media. This methodology, called relativistic MD, and its interfaces with other theoretical and computational methods are discussed in detail in Sections~\ref{sec:Methods} and \ref{sec:Interlinks}, demonstrating that MBN Explorer can be considered as a powerful tool to reveal the dynamics of relativistic projectiles in crystals and other materials including amorphous bodies, as well as in biological environments. Its efficiency and reliability have been benchmarked for the channeling of ultra-relativistic projectiles (within the sub-GeV to tens of GeV energy range) in straight, bent and periodically bent crystals \cite{AVK_AVS_2020_EPJD.74.201_review, NovelLSs_Springer_book, RelMD_2021_EPJD.75.107_review, Channeling_book}. In these papers, verification of the code against available experimental data and predictions of other theoretical models was carried out.

The radiometric unit frequently used to compare different LSs is brilliance, $B$. It is defined in terms of the number of photons $\Delta N_{\omega}$ of frequency $\omega$ within the interval $[\omega- \Delta\omega/2; \omega+ \Delta\omega/2]$ emitted in the cone $\Delta\Omega$ per unit time interval, unit source area, unit solid angle and per a bandwidth (BW) $\Delta \omega/\omega$, for details see refs~\citenum{AVK_AVS_2020_EPJD.74.201_review, NovelLSs_Springer_book}.

\begin{figure}[t!]
\includegraphics[width=0.85\textwidth]{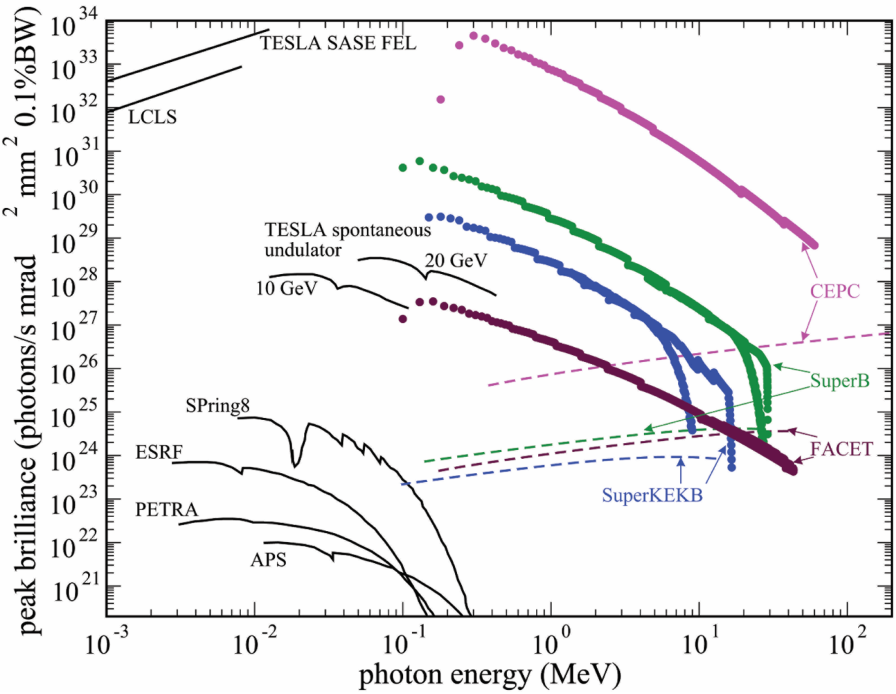}
\caption{Peak brilliance of spontaneous CU Radiation (CUR, dashed lines) and superradiant CUR (thick solid curves) from diamond(110) CUs calculated for the SuperKEKB, SuperB, FACET-II and CEPC positron beams versus modern synchrotrons, undulators and XFELs. The data on the latter are taken from ref~\citenum{Ayvazyan_2002_EPJD.20.149}. The figure is adapted from ref~\citenum{AVK_AVS_2020_EPJD.74.201_review}.}
\label{fig:CaseStudy_LightSources}
\end{figure}

This quantity is utilized to compare different light sources based on different physical principles and technologies. Figure \ref{fig:CaseStudy_LightSources} demonstrates such a comparison of the brilliance evaluated for CU LSs with that for modern synchrotron facilities and Free Electron Lasers (FELs) indicated by solid lines with various self-explanatory legends. Dashed lines present the peak brilliance calculated for positron-based diamond(110) CUs. The CU LSs curves refer to the optimal parameters of CUs, i.e. those which ensure the highest values of brilliance of the corresponding CU for each positron beam indicated; for further details, see Appendix B in ref~\citenum{AVK_AVS_2020_EPJD.74.201_review}.

The radiation emitted in an undulator is coherent (at the harmonics frequencies) with respect to the number of periods, $N_u$, but not with respect to the emitters since the positions of the beam particles are not correlated. As a result, the intensity of radiation emitted in a certain direction is proportional to $N_u^2$ and the number of particles, $I_{\rm inc} \sim N_p N_u^2$ (the subscript ``inc'' stands for ``incoherent''). In conventional undulators, $N_u$ is on the level of $10^3 - 10^4$ (ref~\citenum{Schmueser_FELs_book}); therefore, the enhancement due to the factor $N_u^2$ is large, making undulators a powerful source of spontaneous radiation. However, the incoherence with respect to the number of radiating particles causes a moderate (linear) increase in the radiated energy with the beam density.

More powerful and coherent radiation will be emitted by a beam in which the position of the particles is modulated in the longitudinal direction with the period equal to an integer multiple to the radiation wavelength $\lambda$. In this case, the electromagnetic waves emitted by different particles have approximately the same phase. Therefore, the amplitude of the emitted radiation is a coherent sum of the individual waves so that the intensity becomes proportional to the number of particles squared \cite{Bessonov_SovJQuantElectr_1986}, $I_{\rm coh} \sim N_p^^2 N_u^2$. Thus, the increase in the photon yield due to the beam pre-bunching (other terms used are ``bunching'' \cite{McNeil_2010_NatPhoton.4.814} or ``microbunching'' \cite{Bostedt_2016_RMP.88.015007}) can reach orders of magnitudes relative to radiation by a non-modulated beam of the same density (for more details see refs~\cite{AVK_AVS_2020_EPJD.74.201_review, NovelLSs_Springer_book}). Following ref~\citenum{Gover_2019_RMP.91.035003}, let us use the term ``superradiant'' to designate the coherent emission by a pre-bunched beam of particles. Methods preparing a pre-bunched beam with the parameters needed to amplify CU Radiation (CUR) are described in ref~\citenum{CU_LS_patent} and in Section~8.5 in ref~\citenum{Channeling_book}. The parasitic effect of the beam demodulation in a CU LS has been studied in ref~\citenum{Kostyuk_2010_JPB.43.151001}.

Figure \ref{fig:CaseStudy_LightSources} illustrates a boost in peak brilliance due to the beam modulation. Thick curves correspond to superradiant CUR calculated for fully modulated positron beams (as indicated) propagating in the channeling mode through diamond(110)-based CU. In the photon energy range $10^{-1} - 10^1$~MeV, the brilliance of superradiant CU LSs by orders of magnitudes (up to 8 orders in the case of CEPC) exceeds that of the spontaneous CU LS (dash-dotted curves) emitted by the random beams. A remarkable feature is that the superradiant CUR brilliance can not only be much higher than the spontaneous emission from the state-of-the-art magnetic undulator (see the curves for the TESLA undulator) but also be comparable with the values achievable at the XFEL facilities (LCLC (Stanford) and TESLA SASE FEL) which operate in much lower photon energy range.

The results shown in Figure~\ref{fig:CaseStudy_LightSources} demonstrate that the CU-based LSs possess exceptional characteristics capable of extending significantly the ranges of the currently operating LS. It is also important to emphasize that other suggested schemes for generating photons with comparable energies, e.g. based on the Compton backscattering of photons by the ultra-relativistic beams of electrons, result in the LS parameters that are much less favorable as compared to those achievable for the crystal-based gamma-ray LSs (see refs~\citenum{AVK_AVS_2020_EPJD.74.201_review, NovelLSs_Springer_book} and references therein).

\textbf{\textit{How can MM address the problem.}}
The practical realization of the novel gamma-ray CLSs is the goal of the large European project TECHNO-CLS \cite{TECHNO-CLS_website}. The experimental and technological developments toward this goal are mainly driven by theoretical predictions and computational modeling \cite{AVK_AVS_2020_EPJD.74.201_review, NovelLSs_Springer_book, Solovyov_2012_JCC_MBNExplorer, RelMD_2013_JCompPhys.252.404, Sushko_2019_MBNStudio, MBNbook_Springer_2017, DySoN_book_Springer_2022, RelMD_2021_EPJD.75.107_review, Channeling_book}. The utilization of MBN Explorer and MBN Studio play a crucial role in this process because these software packages have unique methodological and algorithmic implementations enabling relativistic MD of charged particles in all kinds of condensed matter systems, including oriented crystalline structures. The relativistic MD provides atomistic details for ultra-relativistic particles' propagation through condensed matter systems, including coherent effects in the particles' interaction with the medium, scattering phenomena, radiation, radiation-induced damage, etc. The relativistic MD has already been discussed above in this section and in more detail in Sections~\ref{sec:Methods} and \ref{sec:Interlinks}, where references to the original papers and recent reviews in this field have been made. Efficient algorithms and numerical solutions of the relativistic equations of motion of the propagating particles can provide such a description for the macroscopically large particle trajectories with all the necessary atomistic details. Such simulations contain all the necessary information needed to describe gamma-ray CLSs with desired properties and find the optimal technological solutions in their practical realization. Thus, one can describe both the photon emission process by propagating charged particles and study the effects of crystal geometry in crystals (i.e. macroscopic characteristics such as crystalline plane curvature radii, periods and amplitudes of periodical bindings) on this process with an atomistic level of detail. Important is that the predictions of theory and MM of CLSs can be verified and validated in various experiments on exposure of oriented crystalline structures to beams of electrons and positrons with the energies $\sim$1~GeV and above; see refs~\citenum{AVK_AVS_2020_EPJD.74.201_review, NovelLSs_Springer_book} and references therein. Therefore, such a MM approach possessing predictive power becomes an efficient and validated computational tool required for driving the science and technology of the gamma-ray CLSs toward their practical realization. This tool is only one component among many other components assembled within the universal and powerful software packages MBN Explorer and MBN Studio \cite{AVK_AVS_2020_EPJD.74.201_review, NovelLSs_Springer_book, Solovyov_2012_JCC_MBNExplorer, RelMD_2013_JCompPhys.252.404, Sushko_2019_MBNStudio, MBNbook_Springer_2017, DySoN_book_Springer_2022}.

The MM approach based on relativistic MD is also applicable in numerous research areas beyond the field of gamma-ray CLSs. Thus, the interface of the relativistic MD with IDMD, implemented in MBN Explorer and MBN Studio and discussed in detail in Section~\ref{sec:Interlinks}, provides unique possibilities for studying the effects of radiation-induced molecular transformations or damages in different condensed matter systems, including biological ones, within a multiscale approach linking the atomistic level descriptions with larger-scale phenomena (see Section~\ref{sec:MM_stage5}). Such an analysis is required in many applications in which condensed matter systems (including sensitive parts of electronic devices, samples analyzed by means of electron microscopy, materials in space or those utilized in reactors, biological systems) are exposed to radiation.

Radiation-induced phenomena in most systems mentioned above stretch across temporal and spatial scales and lead to large-scale phenomena discussed in Section~\ref{sec:MM_stage5}. The relativistic MD provides a possibility to develop MM descriptions not only for the propagating relativistic charged particles but also for the medium in which the propagation takes place. The possibility to link IDMD-based descriptions of the medium dynamics with the descriptions of large-scale processes based on the stochastic dynamics (as discussed in Section~\ref{sec:Methods_StochasticDyn}) opens many more possibilities for MM and its applications in various research fields and technological advances.

\textbf{\textit{Future directions for 5-10 years period.}}
The future developments in the gamma-ray CLS research area will lead to the practical realization of CLSs in a short-term perspective, at least on the prototype level. This goal is expected to be achieved within the TECHNO-CLS project within the next 2-3 years. It should open many new directions for further research and technological advances towards the optimization of the already developed technologies, the development of new ones that should enable the operation of gamma-ray CLSs in the superradiant regime, exploration of possibilities of gamma-ray CLSs with more energetic electron and positron beams (up to 100~GeV and above), manufacturing and characterization of crystals with desired properties for the gamma-ray CLS applications, construction of infrastructure/facilities for the exploitation of gamma-ray CLSs suitable for their end-users for both academic and industrial communities. The utilization of positron beams in gamma-ray CLSs provides advantages in their practical realization. Therefore, they should be preferably developed and utilized in the future for the construction of gamma-ray CLSs.

Sub-angstrom wavelength, ultrahigh brilliance, and tunable gamma-ray CLSs will have a broad range of exciting potential cutting-edge applications \cite{Zhu_2020_SciAdv.6.eaaz7240}. These applications include exploring elementary particles, probing nuclear structures and photonuclear physics, and examining quantum processes, which rely heavily on gamma-ray sources in the MeV to GeV range. Gamma rays induce \textit{nuclear reactions by photo-transmutationg}. For example \cite{Ledingham_2003_Science.300.1107}, a long-lived isotope can be converted into a short-lived one by irradiation with a gamma-ray bremsstrahlung pulse. However, the intensity of bremsstrahlung is orders of magnitudes less than CUR. Moreover, to increase the effectiveness of the photo-transmutation process, it is desirable to use photons whose energy is in resonance with the transition energies in the irradiated nucleus \cite{urRehman_2017_AnnNuclEn.105.150, urRehman_2018_IntJEnergyRes.42.236}. By tuning the energy of CUR, it is possible to induce the transmutation process in various isotopes. This opens the possibility for \textit{a novel technology for disposing of nuclear waste}. Another possible application of the CU-LSs concerns \textit{photo-induced nuclear fission} where a heavy nucleus is split into two or more fragments due to the irradiation with gamma-quanta, whose energy is tuned to match the transition energy between the nuclear states. This process can be used in a new type of nuclear reactor -- \textit{the photonuclear reactor} \cite{urRehman_2017_AnnNuclEn.105.150, urRehman_2018_IntJEnergyRes.42.236}. Photo-transmutation can also be used to produce much-needed \textit{medical isotopes}. Powerful monochromatic radiation within the MeV range can be used as an alternative source for producing beams of MeV protons by focusing a photon pulse onto a solid target \cite{Ledingham_2003_Science.300.1107}. Such protons can induce nuclear reactions in materials producing, in particular, light isotopes which serve as positron emitters to be used in Positron Emission Tomography (PET). The production of \textit{PET isotopes} using CUR exploiting the ($\gamma$; n) reaction in the region of the giant dipole resonance (typically $20-40$ MeV) is an important application of CLS since PET isotopes are used directly for medial PET and for Positron Emission Particle Tracking (PEPT) experiments.

Irradiation by hard X-ray strongly decreases the effects of natural surface tension of water \cite{Weon_2008_PRL.100.217403}. The possibility to tune the surface tension by CUR can be exploited to study the many phenomena affected by this parameter in physics, chemistry, and biology, such as, for example, the tendency of oil and water to segregate.

Last but not least, a micron-sized narrow CLS photon beam may be used in \textit{cancer therapy} \cite{AVS2017nanoscaleIBCT} to improve the precision and effectiveness of the therapy for the destruction of tumors by collimated radiation, allowing delicate operations to be performed in close vicinity of vital organs.

These developments will lead to establishing close links and cooperation between the TECHNO-CLS consortium and industrial companies that might be interested in these developments. The first steps in this direction have already been made at the recent highly successful TECHNO-CLS workshop held in October 2023 in Ferrara, Italy, with the participation of several leading companies representing some of the technological areas mentioned above.

The successful realization of the TECHNO-CLS project should be continued with the larger scale technological and industrial developments in the field along the aforementioned directions.

\textbf{\textit{Envisaged impact.}}
Development of gamma-ray CLSs will have enormous potential for both scientific and social-economic impact, providing European academic researchers and industry with internationally leading innovation capacity and unique possibilities for exploration of physical, chemical, and biological properties of condensed matter systems exposed to gamma-rays. As a result of the TECHNO-CLS project, the European Community will gain a core group of specialists who will pioneer the development of this novel and highly important field of research and technology with a wide range of applications. The radically new technology realized within TECHNO-CLS will ensure European R\&D is the first to create novel gamma-ray CLSs operating over a broad range of radiation wavelengths inaccessible by means of magnet-based synchrotrons and undulators. This will provide the European industry with the (once in a lifetime) opportunity to pioneer a new technology with all the commercial advantages such leadership provides. To quantify the scale of the impact within Europe and worldwide which the development of radically novel gamma-ray CLSs might have, let us draw historical parallels with synchrotrons, optical lasers and FELs. In each of these technologies, there was a time lag between the formulation of a pioneering idea, its practical realization, and follow-up industrial exploitation. However, each of these inventions has subsequently launched multi-billion dollar industries. Gamma-ray CLSs have the potential to become the new synchrotrons and lasers of the mid to late 21$^{\rm st}$ century, stimulating many applications in basic sciences, technology and medicine, opening a myriad of markets with their inherent employment opportunities and wealth creation.

\subsection{Application of Plasma-Driven Processes}
\label{sec:Case_study_Plasma}


\textbf{\textit{The problem.}}
Condensed matter systems in contact with plasma exhibit complex phenomena due to irradiation by photons, electrons, ions, and other particles from the plasma. These phenomena range from atomic-level physical and chemical interactions to macroscopic material responses. The multiscale nature of these processes has been appreciated for decades. However, the development of multiscale models poses challenges to the theoretical understanding of underlying processes and computational capabilities. The classical approach is often used to collect basic data on specific sub-models (e.g. sputtering yield, collision cross sections) through experiment or simulation and then the results as constants or functional data sets are used as input for a larger-scale model.

\begin{figure}[t!]
\includegraphics[width=0.85\textwidth]{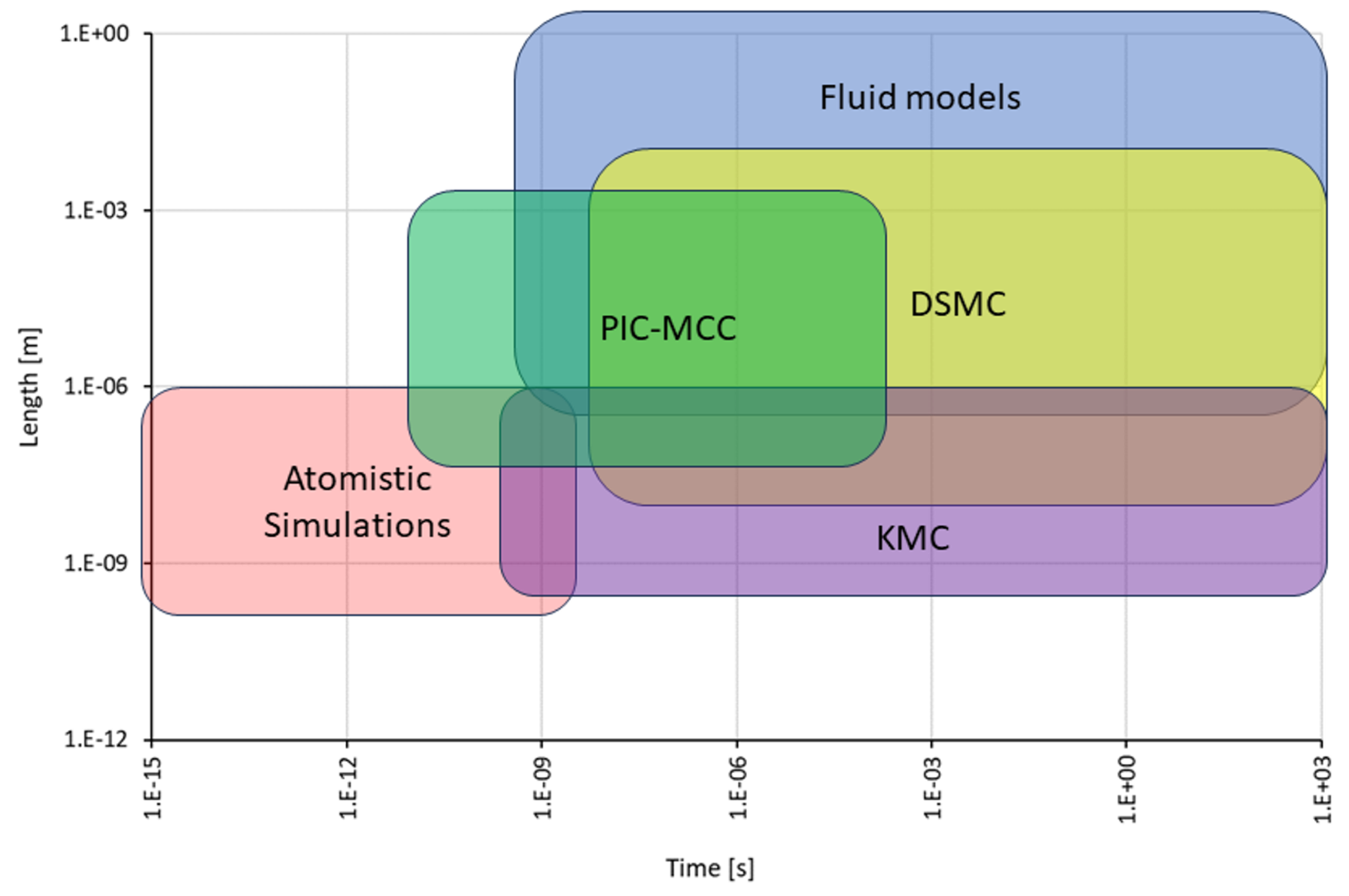}
\caption{Overview of approximate length and time scales of the most common methods for simulation of plasma--surface interactions. The abbreviations stand for Particle-in-Cell simulations with Monte Carlo collisions (PIC-MCC), kinetic Monte Carlo (KMC), and Direct-Simulation Monte Carlo (DSMC). Adapted from refs \citenum{Vanraes2021, Bonitz2019}. }
\label{fig:CaseStudy_Kadlec-1}
\end{figure}

Yet, multiscale models have been developed in different plasma physics and technology areas. Recent reviews summarize the progress and various approaches \cite{Vanraes2021, Ebert2006, Brault2018, Crose2018, Zhu2023, Adamovich2022, Bonitz2019, Dollet2004}. Figure~\ref{fig:CaseStudy_Kadlec-1} presents an overview of the approximate length and time scales of the most common methods for the simulation of plasma–surface interactions, see also Fig.~\ref{fig:MM_diagram} in Section~\ref{sec:MM_key-definitions}.
Atomistic simulations typically target quantum effects and individual particle–surface reactions. These simulations include the DFT (Section~\ref{sec:Methods_DFT}), TDDFT (Section~\ref{sec:Methods_TDDFT}), non-equilibrium Green's functions (NEGF), classical MD (Section~\ref{sec:Methods_classicalMD}), Born--Oppenheimer MD (Section~\ref{sec:Methods_ab-initio_MD}), quantum Boltzmann equation (QBE), and binary-collision approximation (BCA) \cite{Vanraes2021, Brault2018, Bonitz2019, Schleder2019, Neyts2017, Brault2023}.
The limitations regarding the system sizes and simulation times achievable in classical MD are indicated in Fig.~\ref{fig:MM_diagram} and described in Section~\ref{sec:Methods_classicalMD}.
Mesoscale models, such as kinetic Monte Carlo (KMC), use faster coarse-grained descriptions to overcome the computationally expensive atom-based simulations \cite{Bonitz2019, Yang2015}.
Finally, the macroscale models like fluid models based on computational fluid dynamics (CFD), Direct-Simulation Monte Carlo (DSMC), or Particle-in-Cell simulations with Monte Carlo collisions (PIC-MCC), abandon the fine details for a continuum representation of the entire system \cite{Crose2018, Bonitz2019, Dollet2004, Verboncoeur2005, Benilov2020, Murphy2017, Murphy2008, Trelles2018, Kadlec2007}.
%
Hybrid models combine macroscopic fluid and kinetic approaches on a similar spatial scale, using various time scales \cite{Kushner2009, Kim2005, Economou2017}.

The utilized models also vary in the nature of plasma and surface interactions. For example, thermal plasma applications (arc welding, plasma cutting, arcs in circuit breakers, etc.) mostly use CFD simulation tools, including data obtained from Boltzmann equation solvers and chemical kinetics models. The approaches are based either on local thermal equilibrium (LTE) or in some cases on non-equilibrium approaches \cite{Benilov2020, Murphy2017, Trelles2018}.

Atmospheric discharges, ranging from streamers to partial corona, breakdown, and dielectric barrier discharge, as well as lightning in the atmosphere, are important but rather complex and challenging phenomena \cite{Ebert2006, Adamovich2022, Nijdam2020}. Here, especially challenging multiscale problems are related to plasmas that shrink to filaments and exhibit large electric fields and density gradients of charged particles at the head (streamers, leaders, and sprites) \cite{Nijdam2020, Ebert2008, Ebert2010}. Surface flashover discharges on solid insulators are prone to branching and are sometimes described using fractal theory.

Plasma shrinking to narrow space is also typical for cathodic and anodic spots of an arc. The constriction to a cathodic spot and switching to an arc has been a challenge in glow discharge-type plasmas, especially in magnetron sputtering and high-power impulse magnetron sputtering (HiPIMS). This imposes challenges on power generators for these processes. However, other types of effects \cite{Jimenez2012, Gudmundsson2020, Anders2014, Brenning2013, Kadlec2017} and plasma instabilities in HiPIMS have also been studied in recent years, such as spokes \cite{Gudmundsson2020, Anders2014, Brenning2013}, presenting challenges in building multiscale models.

Plasmas with cathodic spots are directly used in vacuum (vacuum interrupters and vacuum circuit breakers), at low pressures (cathodic arc evaporation \cite{Anders2014, Anders2008}), and are also important in the arcs at atmospheric and higher pressure \cite{Benilov2020, Murphy2008, Trelles2018}.

The cathode spot has extremely high current and plasma density, with fractal features in time and space, resulting in high charge states of metal ions even at low discharge voltages. Quantum phenomena play a significant role there \cite{Anders2008}. There were multiple studies of cathode spots during the second half of the 20th century, both experimental and theoretical. However, this is one of the challenging topics for novel multiscale models with current computing and atomistic simulation capabilities.

Other issues related to arc plasma concern the plasma–wall interactions. The flux of ablated wall material (metal, plastic, etc.) alters the arc column, its chemistry, temperature, as well as fluid and radiation properties, which, in turn, affects the heat transfer to the walls \cite{Benilov2020, Trelles2018}. Proper modeling of the arc-wall interaction is another challenge for the MM.

\begin{sloppypar}
\textbf{\textit{How can MM address the problem.}}
Recent review and roadmap papers \cite{Brault2018, Crose2018, Zhu2023, Adamovich2022, Bonitz2019, Schneider2006, Cheimarios2021, Kambara2023} underlined the importance of MM combining plasma diagnostics, theory, modeling, and simulations in the plasma–surface interactions with data-driven approaches \cite{Jia_100Mio_MD_arXiv, Bleiziffer_2018_JChemInfModel.58.579, Westermayr_2019_ChemSci.10.8100, Kambara2023, Gunasegaram2021, Jetly_2021_MLST.2.035025, Nam_2021_PLA.387.127005}. MM can work hand in hand with the data-driven approaches that vary from using datasets from libraries (see also Section~\ref{sec:Validation_Databases}) to collecting and combining experimental data to the generation of data by modeling and machine learning (ML). In this section, we will focus more on these data-driven approaches based on ML.
\end{sloppypar}

If the MM needs input data like energy and angular distributions (EADs) of sputtered particles for a wide range of incident angles and energies, various approaches can be considered \cite{Vanraes2021, Krueger2019}. An example of using machine learning with artificial neural networks (ANN) for this goal is given in ref~\citenum{Krueger2019}. The paper deals with modeling the EADs of reflected and sputtered particles for Ar+ projectiles bombarding a Ti-Al composite. Interestingly, the ANN trained with reference distributions obtained by TRIDYN simulations using a limited sample of $10^4$ projectiles was shown to reliably generalize: the EADs predictions represented with good accuracy also large, smooth sample data obtained using $10^6$ projectiles \cite{Krueger2019}.

\begin{figure}[t!]
\includegraphics[width=0.85\textwidth]{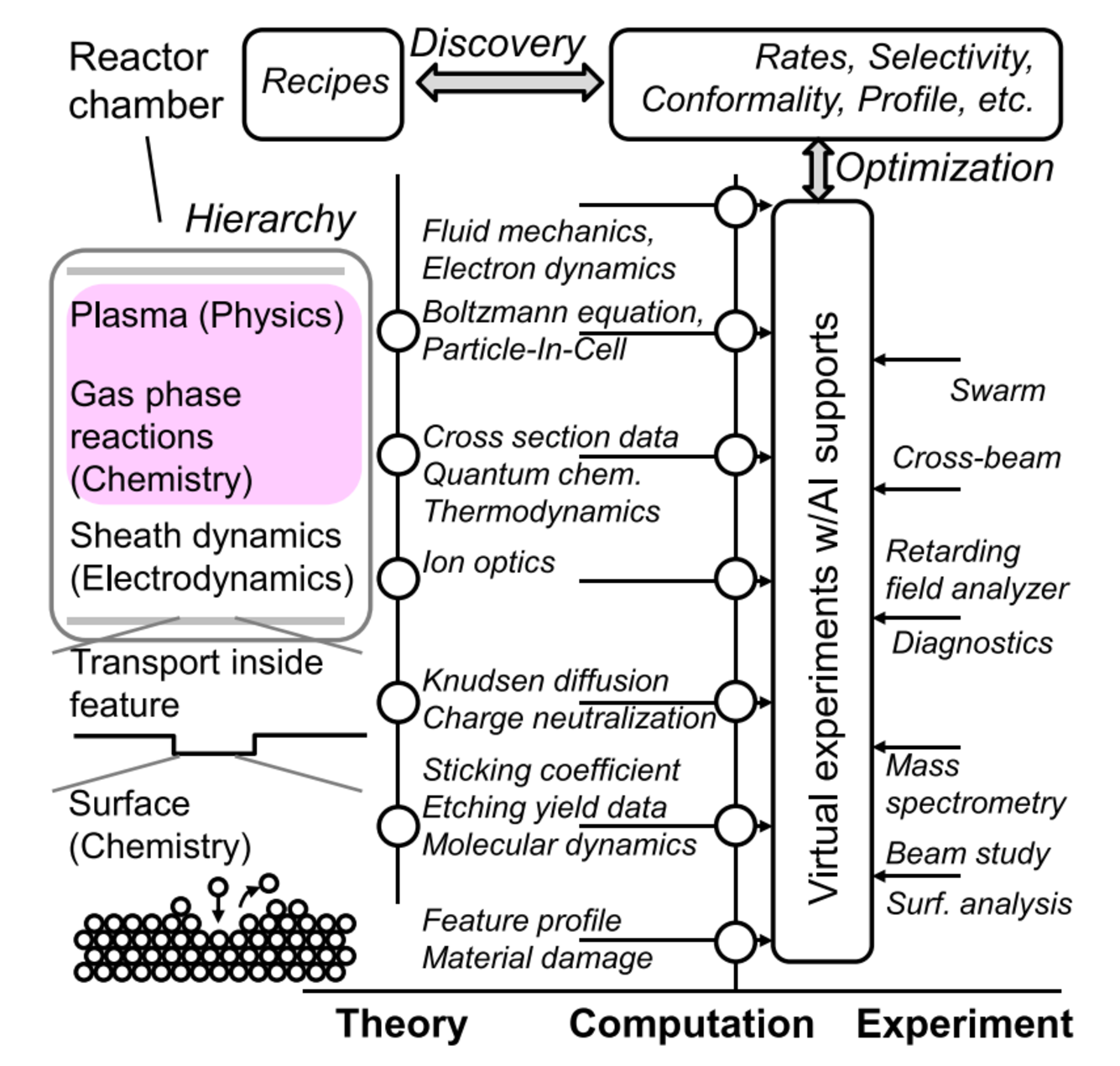}
\caption{Finding optimal process conditions using coordination between theory, computational methods, and experimental data aided with virtual experiments employing artificial intelligence (AI). Adapted from ref~\citenum{Kambara2023}.}
\label{fig:CaseStudy_Kadlec-2}
\end{figure}

An illustration of the suggested setup for a data-driven approach \cite{Kambara2023} for process optimization is depicted in Figure~\ref{fig:CaseStudy_Kadlec-2}. The approach is based on theoretical knowledge ranging from surface interaction in the features up to the bulk plasma. It is also supported by computation at all scales and together with experimental data, a Virtual Experiment (VE) is created with the support of Machine Learning (ML) or Artificial Intelligence (AI). Coupling MM with experimental techniques such as \textit{in situ} microscopy, spectroscopy, and diffraction enables model validation, parameterization, and optimization of the process.

Many papers appeared recently dealing with machine learning that includes equations of physics, referred to as Physics-Informed Machine Learning (PIML) or Physics-Informed Neural Networks (PINNs) \cite{Raissi2019, Karniadakis2021, Spears2018, Cuomo2022, Markidis2021}. This development has also been reflected in plasma physics and plasma-surface interactions \cite{Kambara2023, Park2020, Bonzanini2023, Kawaguchi2022}. The models have been used to solve the Boltzmann equation \cite{Kambara2023, Kawaguchi2022}, radiative transfer, heat transfer with Stefan problem \cite{Cai2021}, Poison equation with gas and a solid insulator \cite{Zeng2023}, and arc plasma \cite{Zhong2023}.

Based on Karniadakis \textit{et al.} \cite{Karniadakis2021}, there are three basic approaches to introducing physics into ML, namely observational, inductive, and learning biases.
The observational biases rely on large amounts of data used during the training phase of the ML model. These observational data should reflect the physics to be modeled. The inductive biases represent prior assumptions incorporated by direct interventions to an ML model architecture. It should guarantee the satisfaction of given physical laws, typically the conservation laws. The learning biases are introduced e.g. by choice of loss functions that softly modulate the training phase of an ML model to converge towards solutions that satisfy the laws underlying physics, e.g., the ordinary or partial differential equations.

Kawaguchi \textit{et al.} \cite{Kambara2023, Kawaguchi2022} explored the evaluation of an anisotropic electron velocity distribution function (EVDF) in SF$_6$ using PINNs. They used the PINNs (depth NL$ = 4$ and width NU$ = 100$) with 41700 parameters. The storage needed for calculation using the Direct numerical solution (DNS) -- 3D arrays with the size $10 000 \times 45 \times 720$. Note that the PINNs approach can represent the 3D EVDF with approximately 0.01\% of the memory capacity required in the DNS.

Other reports of implementing PINNs to plasma–related modeling are approaches to solving the Poisson equation \cite{Markidis2021}. One motivation for it is that PINNs methods are, in principle, grid-less: any point in the domain can be taken as input, and mesh definition is not required. Moreover, the trained PINNs model can be used for predicting the simulation values on different resolutions without retraining. Therefore, the computational cost in this case does not scale favorably with the number of grid points like many traditional computational methods.

The examples highlight the potential of machine learning, especially using PINNs for MM.

\textbf{\textit{Future directions for 5-10 years period.}}
The implementation and further development of multiscale methods can help advance the integration of different length and time scales to simulate the propagation plasmas like streamers. Similarly, modeling radiation-induced effects at solid surfaces will be possible by integrating models at different length scales from the atomic scale to the mesoscale and continuum.

\textbf{\textit{Envisaged impact.}}
The utilization of data-driven methods, including data mining and machine learning, is expected to expand, especially the PINNs embedding physical constraints in data analysis. However, these techniques must first prove their effective power in single-scale problems. The envisaged power is either increased computing power after the model has been trained or better memory usage, especially in multidimensional problems. The ability of PINNs to solve constrained optimizations and inverse problems has already been shown \cite{Cuomo2022}.

Another challenge lies in embedding the approach using PINNs in multiscale models. PINNs may fail in problems of a highly multiscale nature with large local gradients \cite{Cuomo2022}. Therefore, these methods may be first applied in sub-models that solve similar tasks often. Data obtained in these solutions by standard methods like finite elements could be used for training the neural networks, and then the prediction using PINNs can take over the rest of the calculation.

A limited list of the wide variety of multiscale problems encountered in plasma in contact with solid surfaces reviewed in this section indicates that the approaches will be highly diverse. Various levels of implementation of multiscale modeling and virtual experiments will improve predictive capabilities for material response under irradiation and especially accelerated development of new processes, nanostructures, and materials.

\section{Databases for Multiscale Modeling}
\label{sec:Validation_Databases}

A major input in all of the multiscale models discussed in this Roadmap paper are the datasets used to describe the nanoscale process. These include atomic and molecular data defining scattering cross sections, photoabsorption and dissociation (which drives much of the local chemistry) as well as chemical reaction rates, diffusion coefficients, absorption, and desorption times/probabilities, see e.g. the LXCat database \cite{Carbone2021}. Such data is often distributed across many reviews and papers such that the data used is often selected by each modeling team and hence subject to personal choice (and thence bias).

The outputs of the MM may be strongly dependent upon the selected data inputs, and it will often be difficult to compare the underlying physics and chemistry in the different multiscale models when the input data is so varied. Accordingly, in recent years, there has been a major effort to create databases where data is assembled and recommended datasets are provided to allow cross-comparison of multiscale models. This is particularly true in the atmospheric community where the HITRAN database is used as a standard. HITRAN \cite{HITRAN_database} is a compilation of spectroscopic parameters used by various computer codes to simulate the transmission and emission of light in the atmosphere and derive radiative forcing of aeronomic molecules \cite{Buehler_2022_JAMES.14}. HITRAN provides a self-consistent set of parameters (a mixture of calculated and experimental ones) that are widely adopted by the atmospheric community. Similarly, the fusion community, through IAEA, has developed the ALADDIN database \cite{ALADDIN_database}, providing numerical atomic, molecular and plasma-material interaction data of interest to fusion research. The data on ALADDIN are divided into (i) atomic and molecular data and (ii) particle--surface interactions. They are compiled mainly from commissioned IAEA data series, Atomic and Plasma-Material Interaction Data for Fusion.

Knowledge of nascent atomic and molecular data is fundamental to understanding and monitoring many plasmas. Therefore, the plasma industry has been particularly active in developing databases \cite{Celiberto_2016_PSST.25.033004, Industrial_Appl_Plasmas, Plasma_Roadmap_JPD_2012, Plasma_Roadmap_JPD_2017, Anirudh_PlasmaScience_2022} to support models and simulations, for example, to design the next generation of semiconductor chip manufacturing systems, the simulations being described as `virtual factories'.
A comprehensive list of databases providing atomic and molecular data can be found in ref~\cite{Weizmann_PlasmaLab_DBs}.


Most of our understanding of the universe is through observational astronomy. Therefore, that community has invested heavily in data compilation to interpret the huge sets of observational data being generated by the space telescopes. These include the Virtual Atomic and Molecular Data Centre (VAMDC) \cite{VAMDC_portal, Albert_VAMDC_2020_Atoms.8.76}, which provides access to 39 databases, and the Virtual European Solar and Planetary Access (VESPA) platform \cite{VESPA_portal} with access to 50 datasets. Both VAMDC and VESPA provide single-point portal access to a wide range of datasets primarily used for the interpretation of astronomical observations (identification of spectral lines) but also provide data necessary to model planetary atmospheres and star and planet formation in the interstellar medium. The growing study of exoplanets and the search for biosignatures requires enormous datasets of the infrared spectroscopy of atmospheric molecules, which can only be generated by theory. These are then compiled in databases, such as the ExoMol database \cite{ExoMol_database} of molecular line lists, which can be used for spectral characterization and as input to atmospheric models of exoplanets, brown dwarfs and cool stars, as well as other models, including those for combustion and sunspots.

The RADAM portal \cite{RADAM_portal, RADAM_DB_Denifl_JPCS} provides access to a network of RADAM (RAdiation DAMage) databases containing data on interactions of photons, electrons, positrons and ions with biomolecular systems, as well as on radiobiological effects and related phenomena occurring at different spatial, temporal and energy scales in irradiated targets during and after the irradiation.

Some databases may be more specific; for example, the ChannelingDB portal \cite{ChannelingDB_portal} is an interface to a database for collecting data on channeling and related phenomena. It provides data on beam deflection angles, channeling radiation, and characteristic channeling lengths of different projectiles in various crystalline media.

\textbf{Limitations and Challenges}:
While there is a high demand for databases, the compilation of datasets and support for establishing and maintaining such databases remains poor, with few funding opportunities. Thus, databases are often developed by small groups or even individuals. Many databases become obsolete as they are not updated with the latest results, and the supporting platform may become more challenging to integrate into new MM codes.

In order to meet this data need, some commercial software packages such as Quantemol \cite{Quantemol_website} and semi-empirical methods \cite{Joshipura_Mason_book} have been developed that may allow `users' to derive their own data sets. However, without a good knowledge of the underlying physics, there is a risk of producing erroneous data, which may percolate through the research community. Thus, there is a need for `database management' with recommendations for `approved' data and even `policing' of data.

Most recently, with the growth in machine learning and AI, there have been attempts to derive new data \cite{Jetly_2021_MLST.2.035025, Nam_2021_PLA.387.127005}, but, to date, these attempts have been only partially successful and require further evaluation before being widely adopted.

In inputting data into any multiscale model, it is essential to quantify any uncertainties in such values. While experimental data is standardly presented with uncertainties, this has only recently been the case for theoretically derived data \cite{DelZanna_2019_MNRAS.484.4754, Chung_2016_JPD.49.363002}. Such uncertainties are an important part of any sensitivity analysis of a multiscale model (see above) and may identify crucial parameters and data that need to be refined and quantified, providing inspiration for new experiments and calculations.

A key role in developing databases is to provide `recommended data sets', which should be self-consistent. For example, in defining a set of electron scattering cross sections (total, elastic, momentum transfer, excitation, and ionization), the summation of the individual cross sections should be consistent with the recommended total cross section. Similarly, sums of individual photoabsorption cross sections of atmospheric species should be consistent with a total atmospheric opacity. Reviewing data and selecting recommended datasets is challenging and requires a detailed knowledge of the field, the methods used to produce such data and often personal knowledge of the limitations and approximations used. Thus, the emphasis for such data compilation and recommendations must be placed upon the research community itself, even if financial support is often lacking.

\section{Expected Breakthrough in Multiscale Modeling of Irradiation-Driven Processes and Envisaged Impact Within the Next 5 to 10 Years}
\label{sec:Breakthrough}

\subsection{Breakthroughs in Fundamental Understanding of Multiscale Radiation-Induced Phenomena}
\label{sec:Breakthrough_Fundamental}

This paper has highlighted the advances made in the last ca. 15 years in the field of MM of condensed matter systems exposed to radiation and how the MM approach may address fundamental scientific and technological challenges. While there are many disparate systems (as illustrated by 18 case studies presented in Section~\ref{sec:Case_studies}), the fundamental phenomena, such as the quantum process initiated by the incident radiation and the propagation of radiation-induced damage, are common and can be understood based on the same fundamental theoretical principles and computational approaches (see Sections~\ref{sec:MM_key-definitions} and \ref{sec:Methods}). Thus, studying such processes requires a multiscale approach that incorporates different disciplines, such as physics, chemistry, biology, materials science, nanoscience, and biomedical research, allowing the interlinks between them to be accommodated in a single embracing model.

Computational MM methodologies, such as those espoused by MBN Explorer \cite{Solovyov_2012_JCC_MBNExplorer}, have provided a new and powerful tool for exploring many complex systems \cite{MBNbook_Springer_2017, DySoN_book_Springer_2022}, including irradiated condensed matter systems, through coupling five stages that define the processes and phenomena occurring over time scales from attoseconds to seconds and spatial scales from Angstroms to centimeters (see Figure~\ref{fig:MM_diagram}).

The potential of MM in the study of various irradiated condensed matter systems has been demonstrated through case studies discussed in Section~\ref{sec:Intro} and Section~\ref{sec:Case_studies}. The applicability of the MM approach to many other scientific and technological challenges is apparent. The implementation of MM is therefore expected to have a significant impact in the future decades, with MM being adopted by the broad research community as a critical tool.

Looking forward \textit{into the next decade, we foresee the full potential of MM to be realized}.

As discussed in Sections~\ref{sec:Methods_Macroscopic} and \ref{sec:Interlinks_MD-SD-macro_theories}, macroscopic theories describe the characteristics of molecular and condensed matter systems in the large scale limit of MM. For many systems, establishing a complete MM-based theory for condensed matter systems exposed to radiation, which goes across all the spatiotemporal scales depicted in Fig.~\ref{fig:MM_diagram} and links nano- and microscopic effects of radiation with macroscopic observables and large-scale phenomena, is a significant scientific challenge. It is a topic of intensive current investigations in many different areas of research discussed throughout this paper. Therefore, establishing such inclusive MM theories for each class of systems and processes will lead to a breakthrough in our fundamental understanding of multiscale radiation-driven phenomena occurring in such systems and the exploitation of this understanding in technological applications. Particular examples discussed throughout this paper include
\begin{enumerate}[label=(\roman*)]

\item
optimization of the existing treatment planning tools and protocols for radiotherapies (see Sections~\ref{sec:Case_study_MSA_SWs}--\ref{sec:Case_study_DNA_damage_repair} and \ref{sec:Case_study_IBCT});

\item
degradation of materials upon irradiation (as illustrated in Section~\ref{sec:Case_study_DNAorigami});

\item
nanotechnologies -- sustainability of functionality of nanosystems and nanodevices exposed to irradiation, e.g. in connection to 3D nanofabrication (Section~\ref{sec:Case_study_3D-nanoprinting}), plasmon-induced NP catalysts (Section~\ref{sec:Case_study_PlasmonChem}), or NP radiosensitization (Section~\ref{sec:Case_study_RadioNPs});

\item
the development of novel crystal-based light sources of intensive gamma-rays (Section~\ref{sec:Case_study_CLSs}), which is driven by the theoretical and computational modeling and, in particular, by the MM approach; and

\item
technological applications of plasma-driven processes (Section~\ref{sec:Case_study_Plasma}).
\end{enumerate}

As demonstrated through case studies presented in this paper, the MM approach has also enabled the \textit{discovery and interpretation of new phenomena} occurring in complex systems exposed to radiation, such as the formation of nanoscale shock waves induced by heavy ions passing through a biological medium and their role in the thermomechanical mechanism of ion-induced biological damage (Sections~\ref{sec:Intro_Ex_IBCT} and \ref{sec:Case_study_MSA_SWs}), magnetoreception phenomena driven by quantum processes occurring in biological systems (Section~\ref{sec:Case_study_QuantumBiol}), and nanoscopic mechanisms behind the NP radiosensitization (Section~\ref{sec:Case_study_RadioNPs}).

Moreover, the MM approach is also powerful in exploring systems that cannot be studied directly by experiment. For example, as discussed in Section~\ref{sec:Case_study_Space_Chemistry}, the time scales and physical conditions in which chemistry occurs in the regions of the ISM may not have a terrestrial analogue and cannot be genuinely replicated within the laboratory. By computationally exploring the inherently slow dynamical and irradiation-driven processes in space, MM may provide a unique digital twin of ISM molecular dynamics predicting the ISM ice morphologies that will influence star and planet formation.

The \textit{ultimate goal for the further development} of the area of MM-based research covered by this Roadmap is to ensure \textit{the adoption of the general MM methodology} (see Section~\ref{sec:MM_key-definitions} and Fig.~\ref{fig:MM_diagram}) in diverse scientific areas, e.g. radiation damage and protection research (Sections~\ref{sec:Case_study_RADAM_biomol}--\ref{sec:Case_study_RT_multiscale}), materials design (Sections~\ref{sec:Case_study_3D-nanoprinting}--\ref{sec:Case_study_TESCAN}), radiobiology (Sections~\ref{sec:Case_study_DNA_damage_repair}--\ref{sec:Case_study_IBCT}), astrochemistry (Section~\ref{sec:Case_study_Space_Chemistry}), quantum biology (Section~\ref{sec:Case_study_QuantumBiol}), processes involving plasmas (Section~\ref{sec:Case_study_Plasma}), etc., and the application of MM to a large number of case studies discussed throughout this paper.
Within the next 5 to 10 years, the MM approach should then become a general methodology with massive utilization. This goal can be achieved by \textit{broadening the interlinks between the different stages of the multiscale scenario} of the radiation-induced processes in condensed matter systems and related phenomena (see Section~\ref{sec:MM_key-definitions} and the following Section~\ref{sec:Breakthrough_TheoryCompExp}).

\subsection{Breakthroughs in Theoretical, Computational and Experimental Methodologies}
\label{sec:Breakthrough_TheoryCompExp}

MM of condensed matter systems exposed to radiation is achieved through interlinking different theoretical and computational methods for studying different stages of the multiscale scenario of radiation-induced processes (see Sections~\ref{sec:MM_key-definitions} and \ref{sec:Interlinks}). Further elaboration of the interlinks (through the development of novel computational algorithms) and their broader application to other systems and fields of research (including those represented by the case studies in Section~\ref{sec:Case_studies}) will allow an increase in the number of physical and chemical systems and radiation-induced processes therein that could be explored using MM. This will open many new possibilities for developing this whole field of research.

MM combining different theoretical and computational methods (Section~\ref{sec:Methods}) involves the transfer of large amounts of data generated by one group of computer programs to another group of programs where these data are usually used as input. Particular examples include the interlinks between the quantum chemistry and MD programs or between the MC-based particle transport codes and MD software, as described in Section~\ref{sec:Interlinks}. For most interlinks discussed in Section~\ref{sec:Interlinks}, there are no established standards for data transfer between the different classes of computer programs. The formulation of such standards and their utilization by the research community in the field of MM of condensed matter systems exposed to radiation will inevitably lead to a breakthrough in computational MM methodologies as it will enable more efficient use of different computer programs and tools.

The practical realization of the MM approach can be facilitated through the \textit{development of specialized computational instruments} enabling the efficient interlinking between particular computer programs (or groups of programs) for studying radiation-induced processes and phenomena at different spatiotemporal scales of the multiscale diagram shown in Fig.\ref{fig:MM_diagram}. One of such examples is the VIKING web-interface \cite{VIKING_2020_ACSOmega.5.1254}, introduced and discussed in Section~\ref{sec:Interlinks}.

As the amount of data generated at – and required for – different stages of the MM could be huge, it is essential to \textit{set up and maintain dedicated databases} (see Section~\ref{sec:Validation_Databases}), which store the key input parameters for multiscale models of a specific class of systems and physical and chemical processes. The use of specialized multi-task MM toolkits, such as MBN Studio \cite{Sushko_2019_MBNStudio} or VIKING \cite{VIKING_2020_ACSOmega.5.1254}, and their ability to retrieve the necessary information from different atomic and molecular physics and chemistry databases (see Section~\ref{sec:Validation_Databases}) can secure further advances in MM by enhancing the level of automatization of multistep MM simulation procedures.

Further advancement of \textit{computational algorithms for high-performance computing} (HPC) on HPC clusters, including computations using graphics processing units (GPUs), will open new, more efficient paths towards MM-based exploration of a larger number of scientific problems inaccessible with the conventional CPU-based computational technology. Cloud computing technology will enable more \textit{widespread utilization of MM through dedicated online services} supporting the various theoretical and computational methods described in Section~\ref{sec:Methods} and the corresponding interlinks (Section~\ref{sec:Interlinks}).

Many theoretical breakthroughs in the understanding of radiation-driven phenomena in different condensed matter systems are expected to occur through the \textit{formulation and further development of SD models} \cite{Stochastic_2022_JCC.43.1442},  which permit to explore novel and challenging physical, chemical, and biological phenomena, taking place in many different complex systems and involving a broad range of spatial and temporal scales (see Fig.~\ref{fig:MM_diagram} and Section~\ref{sec:Methods_StochasticDyn}).

Finally, theoretical and computational breakthroughs are expected within the next 5 to 10 years with further technical development of \textit{computational MM tools}, such as MBN Explorer \cite{Solovyov_2012_JCC_MBNExplorer} and MBN Studio \cite{Sushko_2019_MBNStudio}, and their application in various technological areas (e.g. bio-, nano-, material-based, and plasma technologies). In particular, such technical developments will aim to create new (or further elaborate the existing) modules dedicated to specific application areas of the software.


The aforementioned theoretical and computational advances will be necessarily accompanied by \textbf{experimental developments}, which will proceed in parallel and will be closely interconnected with each other.
Indeed, the development of MM has motivated and been influenced by experimental studies, as discussed in Section~\ref{sec:Validation_Experiment}. The ongoing challenging experimental work in different scientific disciplines -- physics, chemistry, biology, material science, and space research, has been presented in Section~\ref{sec:Case_studies}. These examples highlight \textit{similarities of multiscale phenomena that occur in very different systems} and involve very similar spatial and temporal scales. A significant part of this experimental work was centered around quantifying the predicted multiscale phenomena or providing the necessary experimental evidence to validate multiscale models.

In the next 5 to 10 years, many of the new concepts and methodologies outlined in this review in Sections~\ref{sec:Validation} and \ref{sec:Case_studies} are expected to come to fruition and offer new analytical tools that may be used to \textit{validate MMs and provide new data suitable for MM}.
However, as described in Section~\ref{sec:Validation}, existing experimental methods for studying different multiscale phenomena occurring in condensed matter systems exposed to radiation have certain limitations. Going beyond these limits will be a breakthrough in exploring the new dynamical regimes of condensed matter systems, achieving new spatial and temporal scales, and studying new types of condensed matter systems.
Particular examples include, e.g. \textit{the development of new analytical methodologies}, such as next generation of mass spectrometry, spectroscopy and microscopy techniques for structural and chemical analysis and radiation induced phenomena in biological systems, clusters and NPs, nanosystems and materials (see Section~\ref{sec:Validation_Experiment}). This includes the advancement of experimental techniques for producing complex (bio)-molecular systems with a specific conformational state (Section~\ref{sec:Case_study_RADAM_biomol}); molecular clusters of well-defined size in a given thermodynamic state (Section~\ref{sec:Case_study_Cluster_Beams}), as well as liquid jets and surfaces.

\textit{Absolute cross sections and rate constants of various radiation-induced quantum processes} play a key role in the fundamental understanding of the radiation phenomena in condensed matter systems. Their calculation and measurement in different environments are highly important for the field. Advances in this direction could be expected from the measurements of such quantities in the collision of particles, including photons, with well-defined and controlled systems, such as atomic and molecular clusters (Section~\ref{sec:Case_study_Cluster_Beams}), NPs (Section~\ref{sec:Case_study_RadioNPs}), complex (bio)-molecules (Section~\ref{sec:Case_study_RADAM_biomol}), DNA origami targets (Section~\ref{sec:Case_study_DNAorigami}), etc., being in the gas phase or on a surface.

Differences in cross sections in the gas and the condensed matter phases can be understood, for instance, because the excitation energy and propagation of electronic states are very different in these phases. However, the electronic state excitation and ionization energy in a liquid or on a liquid surface are largely unknown, nor is the ability of Rydberg states to be sustained in liquid media obvious.

Another obstacle concerns substantial challenges for the existing experimental techniques to \textit{detect neutral fragments} produced in collision processes involving the aforementioned molecular and condensed matter systems. Advances in the detection and analysis of neutral and reactive species will undoubtedly provide new insights to the physical and chemical processes discussed in this Roadmap.

The overview of the experimental techniques in Section~\ref{sec:Validation} elucidates that there are ranges of temporal and spatial scales at which experimental measurements of the structure and dynamics of irradiated condensed matter systems are currently not feasible. For example, direct measurements of the radiation-induced phenomena in condensed matter systems on the temporal scales from femtoseconds to hundreds of picoseconds are problematic with the available experimental techniques. Typically, such processes involve spatial scales up to tens of nanometers.

In such cases, the MM techniques described in Sections~\ref{sec:Methods} and \ref{sec:Interlinks} provide unique opportunities for predicting and quantifying the radiation-induced phenomena. Examples of such phenomena include ion-induced shock waves generated in the vicinity of the Bragg peak due to the deposition of large amounts of energy into the medium in the vicinity of the ion tracks (see Sections~\ref{sec:Intro_Ex_IBCT} and \ref{sec:Case_study_MSA_SWs}) or the atomistic characterization of the complex structure and physicochemical properties of radiosensitizing NPs (Section~\ref{sec:Case_study_RadioNPs}).

The number of novel radiation-induced phenomena in the entire field presented by this Roadmap that can be observed through the development of the existing and novel experimental techniques is very large. Let us mention several examples of such ongoing work.

\begin{itemize}

\item
\textit{Atomic and molecular clusters} have been used to study single-particle and collective irradiation-driven phenomena. These systems enable elucidation of the \textit{role of the environment} in the elementary quantum processes involved, the emergence of the collective response of the system upon its irradiation, and the evolution of the system properties from atomic toward bulk.

\item
Experimental studies are being conducted for isolated aerosols, liquid droplets and liquid jets \cite{Farnik_2023_JPCL.14.287, Kaneda_2010_JCP.132.144502, Kitajima_2019_JCP.150.095102, Nag_2023_JPB.56.215201} using acoustic \cite{Mason_2008_FaradayDisc.137.367, Dangi_2021_ACSOmega.6.10447} and optical levitation \cite{Rafferty_2023_PCCP.25.7066} systems. Within the condensed-matter field, \textit{the liquid phase} remains the least studied, and the transition of physical and chemical properties across the phases requires characterization. The development of new experimental systems to create liquid jets, in which one or more species may be present and can be probed by the spectroscopic methods described in Sections~\ref{sec:Validation_Exp_Quantum} and \ref{sec:Validation_Exp_Chem_Equil}, is expected to open a new era in the study of molecules in the liquid state, including probing electron and ion collisions in this phase \cite{Kitajima_2019_JCP.150.095102, Nag_2023_JPB.56.215201, Nomura_2017_JCP.147.225103}.

\item
New experimental methods for studying \textit{time-resolved radiation chemistry}, discussed in Section~\ref{sec:Validation_Exp_Non-equil_Chemistry}, are being developed to validate non-equilibrium chemistry and molecular transformations (see stage (iii) in Fig.~\ref{fig:MM_diagram}). Through such techniques, the spectroscopy, chemical reactivity, and the role of transient species in many irradiation processes may be quantified and MM models validated.

\item
A large area of research concerns studies of various \textit{processes in biological systems under irradiation conditions}. These studies are conducted with both animate and inanimate biological systems. The popular inanimate biological systems include DNA origami (Section~\ref{sec:Case_study_DNAorigami}), plasmid DNA (Section~\ref{sec:Case_study_RADAM_biomol}), proteins, lipid bilayers, etc. Animate biological systems vary from a single living cell (Section~\ref{sec:Case_study_DNA_damage_repair}) to a whole organism (Section~\ref{sec:Case_study_IBCT}).

The inanimate biological systems can be utilized for experimental measurements of the key processes that may take place in animate biological systems affecting the large-scale phenomena therein (see Fig.~\ref{fig:MM_diagram} and Sections~\ref{sec:MM_key-definitions} and \ref{sec:Methods_large-scale}). An example of such processes could be the formation of complex DNA damage in DNA origami or plasmid DNA systems and the impact of such events on the survival cells and larger biological systems upon their irradiation \cite{Surdutovich_AVS_2014_EPJD.68.353, verkhovtsev2016multiscale, Friis_2021_PRE}. Other examples may concern studies of the mechanisms of NP radiosensitization in biological systems (Section~\ref{sec:Case_study_RadioNPs}), dynamics of proteins, transport properties of cell membranes, etc.
\end{itemize}

Finally, the research area covered by this Roadmap could be extended further in its scope. Thus, the \textit{large-scale thermo-mechanical properties of materials, conductivity, fluidity}, and other classical phenomena in condensed matter systems exposed to radiation could be studied using relevant experimental and theoretical methods. It is also interesting to study the \textit{large-scale quantum phenomena (superconductivity, superfluidity, and magnetism)} in condensed matter systems in the presence of radiation. These and many other possible extensions of the presented studies will be conducted within the next decade.

\subsection{Envisaged Technological Advances}
\label{sec:Breakthrough_TechnolAdvances}

A fundamental understanding of radiation-induced multiscale processes and phenomena in condensed matter systems can facilitate technological advances in many different areas, including space research (e.g. the design and characterization of space-borne materials, radiation protection), renewable energy (capacitors, batteries, photovoltaics, solar panels, novel catalysis for green technologies), radiotherapy applications, nanomedicine, crystal-based light sources, fabrication of nanodevices (3D nanoprinting, (nano)-sensors), fabrication of new materials with tailored properties (including biomaterials, membranes, radiosensitizing NPs), plasma-driven technologies, and others. Each technological advance is necessarily linked to a specific multiscale scenario (see Fig.~\ref{fig:MM_diagram} and Sections~\ref{sec:Intro} and \ref{sec:MM_key-definitions}) because applications operate at the macroscale, but the fundamental physical and chemical processes behind them often occur on the molecular, atomistic or nanoscopic scales, as discussed in detail in Sections~\ref{sec:Intro} and \ref{sec:Case_studies}.
Several examples of technological advances envisaged through the exploitation of the MM approach have been discussed in greater detail in this Roadmap paper. Below, we highlight some of them in the context of future developments.

\textbf{(i)} MM has been remarkably successful in reproducing the results of FEBID \cite{Sushko_IS_AS_FEBID_2016, DeVera2020, Prosvetov2021_BJN, Prosvetov2022_PCCP}, one of the new `bottom-up' nanofabrication methods (see Sections~\ref{sec:Intro_Ex_FEBID} and \ref{sec:Case_study_3D-nanoprinting}). MM can guide the use of new precursor molecules and novel experimental methods for increasing the purity of the deposits (for example, it may explain why the addition of water to the precursor gas stream leads to purer deposits) and characterize the macroscale properties of the fabricated devices (e.g. magnetic, electrical and plasmonic properties). MM can also replicate 3D structures, which will be crucial to the economic viability of such methodologies.

The exploitation of the full potential of FEBID 3D nanofabrication for the vast diversity of materials and their desired properties requires significant advances in a molecular-level understanding of the Irradiation-Driven Chemistry (IDC) in the FEBID process (see Section~\ref{sec:Case_study_3D-nanoprinting}). Such knowledge is essential for transferring initial 3D designs into the fabrication of real nano-architectures with desired properties. An understanding of IDC will provide a deeper understanding of the relationship between deposition and irradiation parameters and their impact on the physical characteristics of fabricated nanostructures (size, shape, purity, crystallinity, etc.) and is an essential step towards commercial exploitation of FEBID 3D-nanofabrication.

\textbf{(ii)} Another new technology that has only become possible through the development of MM is the prospect of creating novel crystal-based sources of intensive gamma-ray radiation \cite{NovelLSs_Springer_book, AVK_AVS_2020_EPJD.74.201_review} through exposure of oriented crystals to the collimated beams of ultra-relativistic electrons and positrons. The design of such light sources similar to the UV and X-ray sources delivered by synchrotrons and free electron lasers but operating at much higher photon energies with comparable intensities (see Section~\ref{sec:Case_study_CLSs}) opens the possibilities for new imaging techniques for nanostructures, while establishing new technologies, including the role of intensive gamma-rays in biomedical applications and unique solutions for the nuclear waste problem, and many more \cite{NovelLSs_Springer_book, AVK_AVS_2020_EPJD.74.201_review, TECHNO-CLS_website}.

\textbf{(iii)} Through MM, we can develop the next generation of radiation treatment protocols to bring the full benefits of hadron therapies to the clinic. Developing a complete MM-based model of radiation-induced biological damage and related phenomena \cite{AVS2017nanoscaleIBCT, Surdutovich_AVS_2014_EPJD.68.353} may create the next generation of models for radiotherapy treatment planning based on nanoscale dosimetry rather than the macroscopic scale dosimetry used today. Accounting for the nanoscale (such as complex lesions of DNA \cite{Surdutovich_AVS_2014_EPJD.68.353}, DNA damage repair probabilities, ion-induced shock waves \cite{surdutovich2010shock} in biological media and the role of thermomechanical mechanisms in the overall scenario of biodamage \cite{AVS2017nanoscaleIBCT, Surdutovich_AVS_2014_EPJD.68.353, surdutovich2013biodamage, Friis_2021_PRE}) and larger-scale (such as NP radiosensitization and radioresistance \cite{AVS2017nanoscaleIBCT, Haume_2016_CNano.7.8}) phenomena should enable us to go beyond the macrodosimetry concept and establish more advanced, nanodosimetry-based protocols for the evaluation of radiobiological effects. This knowledge can then be used to optimize existing radiotherapy protocols and treatment planning tools. With the increasing construction of proton and ion-beam therapy centers worldwide, MM-based radiation treatment plans will become an urgent necessity. Similarly, new multiscale models of radiation-induced processes under space conditions, where low-flux, long-timescale irradiation induces chemical and physical changes in materials and affects human health on long space journeys, may be developed.

\textbf{(iv)} Current macrodosimetry significantly underestimates the radiosensitizing effects of metallic NPs under X-ray and particle-beam irradiation \cite{Ricketts_2018_BJR.91.20180325}. Hence, there is a need for a new theoretical background to determine the bio-physical-chemical mechanisms involved in radiation--NP interactions. Improved predictive models should incorporate the key processes known to impact cellular response to NP--radiation interaction, including nanodosimetry, electron transport, radiation chemistry and molecular processes in the vicinity of NPs \cite{AVS2017nanoscaleIBCT}, all of which can be incorporated within a MM scenario (see Section~\ref{sec:Case_study_RadioNPs}). MM methods combining nanoscale descriptions of radiation-driven molecular modifications/phenomena with larger-scale radiobiological effects are therefore central to developing such `next generation' radiotherapy treatments.

It is important to stress that MM enables the analysis of multidimensional parameter spaces relevant to the aforementioned and other technologically relevant case studies and finding the optimal parameters more efficiently than could be achieved through experiments. Indeed, in many case studies presented in this Roadmap, a systematic exploration of each of the parameters involved on the molecular/nanoscopic scale is a formidable and costly experimental task. However, it can be addressed using the MM approach, thus lowering the cost of such studies and facilitating technological advances.

\subsection{Community Building}
\label{sec:Breakthrough_CommunityBuilding}

In the next decade, MM has enormous potential to radically change how we treat models as a predictor and design tool for next-generation technologies while also becoming a core tool that addresses many of the most significant scientific challenges. However, to ensure the adoption of MM across these diverse communities, it will be necessary to raise the profile of current MM software and train new communities in its philosophy and use.

As demonstrated from a collection of case studies presented in this Roadmap (see Section~\ref{sec:Case_studies}), the community that can benefit from the broad utilization of the MM approach to achieve technological advances is extensive and comprises numerous interdisciplinary and intersectoral stakeholders. It is, therefore, of great importance to create a community-building EU platform to coordinate joint efforts towards a multiscale understanding of the fundamental processes arising due to the interaction of radiation with matter on the European level. Some efforts towards the creation of such a platform have been made through the ongoing European collaborative projects, such as the COST Action CA20129 ``MultIChem'' \cite{CA_MultIChem_website}, European projects supporting academic-industry interchange and direct applications of MM (e.g. RADON and N-LIGHT research and innovation staff exchange (RISE) projects \cite{H2020_RISE_RADON_website, H2020_RISE_N-LIGHT_website}), and the TECHNO-CLS Pathfinder project \cite{TECHNO-CLS_website}.

In Europe, we are fortunate to have the opportunity to build the MM user community and engage with stakeholders through national, bilateral, and pan-European-supported initiatives. Due to the importance of radiation-induced processes, we suggest a new EU scientific initiative – the RADIATION Flagship, which may be set up in the next Horizon Europe programme following the recently finished or ongoing large-scale initiatives, the Graphene Flagship \cite{Graphene_Flagship_website}, Human Brain Project \cite{Human-Brain-Project_website}, and the Quantum Technologies Flagship \cite{Quantum_Flagship_EU_website, Quantum_Flagship_website}. We plan new EU programs dedicated to training (Doctoral Networks) and staff exchange schemes (Marie Curie RISE programme and COST Actions). We envisage MM Clusters of Excellence (supported through Research Infrastructure and COFUND initiatives) and, as exemplified in the use of MM to develop crystal-based light sources \cite{TECHNO-CLS_website}, to exploit this new methodology to create new technology innovation programs through the EIC. Last, but not least, MM should also be adopted as a tool of choice in the larger European Science programs led by ESA and EUROATOM.

\subsection{Economic and Societal Impact}
\label{sec:Breakthrough_Impact}

The development of HPC in the latter part of the 20$^{\textrm{th}}$ century has led to the emergence of a third strand of science in between traditional theory and experiment: Numerical Modeling, which is now emerging as a dominant field of 21$^{\textrm{st}}$-century science and technology as it is evident from this Roadmap and similar recent works in other areas of research \cite{MBNbook_Springer_2017, DySoN_book_Springer_2022}. The traditional theoretical and experimental science must now be complemented by computational modeling science. The features of the computational modeling science are similar to those traditionally exploited in theory and experiment (logic, evidence-based, reproducibility of data) but also different, as modeling allows any question to be explored in multiple ways with the different `results' of the model often being equally valid within the model itself. Thus, the development of computational models, and in particular multiscale ones, for the most challenging scientific problems accomplished with their validation through experiment or comparison of model results with those derived from the already validated theories or models (see Section~\ref{sec:Validation}) will lead to the major breakthroughs in the related research and technological areas that necessarily will produce significant economic and societal impacts.

Many of today's most active research, like the one described in this Roadmap, requires a detailed understanding of complex multiscale phenomena linking the nanoscale with the macroscale while combining atomic and molecular processes with bulk phenomena (see Sections~\ref{sec:Intro} and \ref{sec:MM_key-definitions}). The MM provides the methodology by which the nano- and macroscale (quantum and classical worlds) can be intertwined, thus making the MM methodology a powerful instrument for modern scientific research and technology developments (see Section~\ref{sec:MM_key-definitions}), and the follow-up commercialization and exploitation of products in various branches of industry, and development of new products and markets \cite{AVS2017nanoscaleIBCT, Surdutovich_AVS_2014_EPJD.68.353, MBNbook_Springer_2017}.

These trends have been followed in studies of multiscale phenomena in condensed matter systems exposed to radiation that are now widely recognized as a rapidly emerging interdisciplinary research area developed through an ongoing COST Action CA20129 MultIChem \cite{CA_MultIChem_website}, two H2020 RISE projects -- RADON \cite{H2020_RISE_RADON_website} and N-LIGHT \cite{H2020_RISE_N-LIGHT_website}, and the Horizon Europe Pathfinder project TECHNO-CLS \cite{TECHNO-CLS_website}. To profile and highlight these achievements, the international community has worked together to produce this detailed Roadmap summarizing progress to date and outlining the potential for ground-breaking fundamental research and related innovation breakthroughs, economic and societal impacts for the next decade, should MM, its experimental verification and its links to technological applications be fully developed. This Roadmap provides several examples of technological developments with high economic and societal impacts (see Sections~\ref{sec:Intro} and \ref{sec:Case_studies}). Below, let us briefly evaluate the dimension of some of these impacts.

As explained in Sections~\ref{sec:Breakthrough_Fundamental} and \ref{sec:Breakthrough_TheoryCompExp}, using validated MM methods and advanced experimental techniques, one can achieve a breakthrough in understanding the key nano- and larger-scale phenomena underpinning radiation damage in general and radiation biodamage in particular \cite{AVS2017nanoscaleIBCT, Surdutovich_AVS_2014_EPJD.68.353}. This achievement has a tremendous societal impact because such knowledge is urgently required in many important application areas, such as radiotherapies, radiation protection, space missions, materials research, etc. Adopting nanodosimetry instead of macrodosimetry offers the potential to bridge the current disconnect between physics and biology in radiotherapy and will lead to improved prediction of cellular response to radiation \cite{verkhovtsev2016multiscale, Friis_2021_PRE}. Subsequently, it will be possible to optimize patient treatment plans based on predicted biological effects rather than simple deposition of radiation energy, minimizing radiation-induced toxicity while maximizing tumor degradation.

Further optimization of existing radiotherapy protocols on the MM basis and the development of `next generation' radiotherapy treatments will have enormous economic and technological impact. Indeed, there are over 100 operational proton therapy centers worldwide which delivered over 300,000 treatment cases by the end of 2022. A typical cost for such treatment is on the scale of 50--100 thousand Euros. Optimizing radiotherapy treatment protocols can significantly impact the efficiency of treatments, which will also have a strong societal impact, and the overall technological operation of these centers, thereby reducing the treatment costs and making them more economically feasible. Similar benefits can be presented for other radiation technology areas involving RADAM phenomena.

The successful realization of such a research program will have a strong societal impact in the medium to long term. As described above, each year, hundreds of thousands of patients undergo particle therapy to treat cancer. Although the treatment was successful in many cases, one has to accept that its efficiency could still be considerably improved. This issue can be addressed if a more fundamental understanding is available when designing the radiation treatment protocols. Such an understanding will be delivered by the research community represented by this Roadmap and is expected to have a considerable impact on society, as it would help save lives while improving the life quality of hundreds of thousands of patients globally.

Another example of important technological applications in which MM of irradiated condensed matter system provides important atomistic insights concerns the FEBID process. The details of this process, its MM, and its technological relevance have been discussed in Sections~\ref{sec:Intro_Ex_FEBID}, \ref{sec:Methods}, \ref{sec:Case_study_3D-nanoprinting}, and \ref{sec:Breakthrough_TechnolAdvances}. The further advancement of FEBID 3D-nanofabrication, achieving its higher technology readiness levels using MM methods and commercial exploitation, may have a technological impact on European and global scales. Beyond FEBID, the demonstration of the full potential of IDMD for MM of complex condensed matter systems coupled to radiation, their dynamics and IDC will open pathways toward broad exploitation of RMD, IDMD and SD methodologies by both academic and industrial communities (plasma research, radiation research, software development, etc).

Novel and more efficient methods of 3D-nanofabrication will allow for the miniaturization of the created electronic nanodevices and their cost-effective production. A better understanding of the mechanisms of radiation-induced formation, growth, and modification of nanostructures will enable effective optimization of existing nanofabrication technologies, allowing more precise / better-controlled fabrication and targeting specific compositions and morphologies of the fabricated nanostructures with tailored properties (see also Section~\ref{sec:Breakthrough_TechnolAdvances}). As a natural consequence, these technological developments will allow the next generation of nanoscale devices to be developed and produced, which will have a solid socioeconomic impact in the medium to long term. These processes will allow the industry to grow, thus creating more jobs and wealth.

The MM methodology, once approbated and validated, will also encompass and facilitate the near-future development of a wide range of societally significant end-products and applications in (i) virtual design and engineering of nanostructured materials; (ii) the electronic and chemical industry for constructing highly efficient batteries and catalyzers; (iii) avionics and automobile industry for designing nanostructured functionalized surface coatings, as well as the cosmic industry for radiation protection; (iv) radiotherapy and nanomedicine; (v) the pharmaceutical industry for drug design, etc. In most of these applications, it is necessary to identify and design specific systems' properties determined by their molecular structure on the nanoscale and to ensure their transfer to the macroscopic scale to make them functional and usable. Such a transition implies MM methods widely discussed in this Roadmap, which rely on combining different methodologies with interlinks relevant to different temporal and spatial scales (see Fig.~\ref{fig:MM_diagram} and Sections~\ref{sec:Methods} and \ref{sec:Interlinks}).

Similar evaluations of economic and societal impact can be made in connection with most of the case studies presented in Section~\ref{sec:Case_studies}, e.g. in Section~\ref{sec:Case_study_CLSs}, which discusses the practical realization of novel crystal-based light sources. Such challenging ideas, once they are realized, will open incredible opportunities for the commercialization of new products, the development of new markets, and the launching of multi-billion-euro industries. We can only overview some such opportunities in this paper. Instead, let us conclude this section by stating that the research area presented by this Roadmap, once developed within the next 10--15 years through the advancement of MM techniques, their validation, and exploitation, will produce enormous overall economic and societal impact.

\section{Conclusions}
\label{sec:Conclusions}

In conclusion to this Roadmap paper, let us emphasize that the multiscale methodology and its applications have developed rapidly over the last decade, providing new opportunities for many disciplines to advance their understanding of the fundamental processes and their applications. This paper has reviewed the state-of-the-art methodologies applicable to describing the behavior of irradiated condensed matter systems at different spatial and temporal scales (Section~\ref{sec:Methods}) and their interlinks (Section~\ref{sec:Interlinks}) to form a harmonized universal multiscale approach. Experimental techniques utilized in the field for measuring multiscale phenomena, providing the key input parameters of the multiscale models and their validation (Section~\ref{sec:Validation_Experiment}), have also been reviewed with an emphasis on the existing challenges and potential breakthroughs as well as their links to the novel and emerging technologies (Section~\ref{sec:Breakthrough}).

The Roadmap is addressed to: (i) the scientific community studying the behavior of condensed matter systems exposed to radiation, (ii) young researchers willing to advance their careers in the areas of modern interdisciplinary research, (iii) stakeholders such as funding agencies (including the European Commission who are already supporting several projects in this area and looking for the consolidation of research efforts of different, but relevant groups or even research communities), as well as (iv) broader publics. Therefore, the style and content of the paper in different sections differ by being adapted to these diverse 'audiences' of potential readers.

The paper provides a roadmap for the development of the field for the next decade (Section~\ref{sec:Breakthrough}) in terms of the fundamental tasks, computational methods and their practical realization in MM, its validation, experimental methods, technological and medical applications. It is worth to stress that the further development of the already identified applications in presented research area is coupled with enormous market place dealing with the large number of new products, services, technologies, as discussed in Section~\ref{sec:Breakthrough}.

The Roadmap presents a collection of case studies (Section~\ref{sec:Case_studies}), each can be seen as a possible direction for further development within the research area covered by this Roadmap. There are many more case studies in the area some of which have been mentioned but not presented. It is obvious that even more case studies will emerge in the area within the next 10--15 years.

Important is that most of them, if not all, have many similarities and common features. They can be understood on the basis of the methodologies reviewed in Section~\ref{sec:Methods} and all contribute to the fundamental understanding of condensed matter systems exposed to radiation through the multiscale approach (see Figure~\ref{fig:MM_diagram} and related discussion in Section~\ref{sec:MM_key-definitions}), and explore this knowledge in different technological applications or medicine.

\section*{Author Information}

\subsection*{Author Contributions}

The conceptualization of this Roadmap paper as well as writing most of its parts (except for most parts of Section~\ref{sec:Case_studies}), technical editing and the coordination of the whole project was done by Andrey V. Solov'yov, Alexey V. Verkhovtsev, Nigel J. Mason, and Ilia A. Solov'yov.

The following authors contributed to the case studies presented in Section~\ref{sec:Case_studies}:

\noindent
Section~\ref{sec:Case_study_QuantumBiol} -- Luca Gerhards and Ilia A. Solov'yov,

\noindent
Section~\ref{sec:Case_study_RADAM_biomol} -- Thomas Schlath\"{o}lter,

\noindent
Section~\ref{sec:Case_study_DNAorigami} -- Leo Sala and Jaroslav Ko\v{c}i\v{s}ek,

\noindent
Section~\ref{sec:Case_study_Attosecond} -- Franck L\'{e}pine,

\noindent
Section~\ref{sec:Case_study_Cluster_Beams} -- Juraj Fedor,

\noindent
Section~\ref{sec:Case_study_UltrafastRadChem} -- Brendan Dromey,

\noindent
Section~\ref{sec:Case_study_MSA_SWs} -- Alexey V. Verkhovtsev and Andrey V. Solov'yov,

\noindent
Section~\ref{sec:Case_study_RT_multiscale} -- G\'{e}rard Baldacchino,

\noindent
Section~\ref{sec:Case_study_DNA_damage_repair} -- Michael Hausmann, Martin Falk, and Georg Hildenbrand,

\noindent
Section~\ref{sec:Case_study_RadioNPs} -- Kate Ricketts,

\noindent
Section~\ref{sec:Case_study_IBCT} -- Richard A. Amos and Andrew Nisbet,

\noindent
Section~\ref{sec:Case_study_PlasmonChem} -- Ilko Bald,

\noindent
Section~\ref{sec:Case_study_Self-organization} -- Andrew Wheatley and Siyi Ming,

\noindent
Section~\ref{sec:Case_study_3D-nanoprinting} -- Alexey V. Verkhovtsev and Andrey V. Solov'yov,

\noindent
Section~\ref{sec:Case_study_TESCAN} -- Milo\v{s} Hrabovsk\'{y},

\noindent
Section~\ref{sec:Case_study_Space_Chemistry} -- Nigel J. Mason,

\noindent
Section~\ref{sec:Case_study_CLSs} -- Andrey V. Solov'yov,

\noindent
Section~\ref{sec:Case_study_Plasma} -- Stanislav Kadlec.

\subsection*{Biographies}

Professor Andrey V. Solov'yov graduated from the Peter the Great St. Petersburg Polytechnic University (St. Petersburg State Polytechnical University, Russia) and got Ph.D. in theoretical and mathematical physics at the A.F. Ioffe Physical-Technical Institute of the Russian Academy of Sciences in 1988. Since 2014, he is the Scientific and Executive Director of MBN Research Center (Frankfurt am Main, Germany), http://www.mbnresearch.com/. He holds the Scientific Degree of Doctor of Physical and Mathematical Sciences awarded at the A.F. Ioffe Physical-Technical Institute (1999) and the academic title of Professor of Theoretical Physics awarded by the Peter the Great St. Petersburg Polytechnic University (2016). In both institutions, he worked over the years at different positions ranging from young researcher to the leading scientist and the full professor. For many years (1996--2014), he also worked at the Goethe University Frankfurt am Main as a visiting professor at the Institute for Theoretical Physics and the fellow of the Frankfurt Institute for Advanced Studies. Over the years, he maintained close collaborations with the colleagues from the Imperial College of London, the Open University, the University of Kent in the UK and was a visiting professor in a number of other universities worldwide. He was the coordinator/leader of numerous research projects supported within the Framework Programmes of the European Commission including the most recent one -- Horizon Europe; by the Deutsche Forschungsgemeinschaft, Deutscher Academischer Austauschdienst, COST Association, Alexander von Humboldt Stiftung, and other funding agencies. He was the elected President of the Virtual Institute of Nano Films (2010--2021); the Editor-in-Chief of the European Physical Journal D: Atomic Molecular, Optical and Plasma Physics (2014--2021) and member of several other international journal editorial boards. He was the recipient of numerous scientific awards. He wrote over 430 journal articles and book chapters, 11 books devoted to different appealing topics in modern theoretical and computational physics, chemistry, biophysics, materials science and pioneered many interdisciplinary research directions at the interfaces of the aforementioned natural sciences including the one represented by this roadmap. He was an editor for 6 books and 15 journal topical issues devoted to various interdisciplinary topics.

Alexey V. Verkhovtsev is a Research Associate at the MBN Research Center gGmbH. He received his Master's degree in Condensed Matter Physics in 2010 from the St. Petersburg Polytechnic University (St. Petersburg, Russia) and a PhD degree in Physics in 2016 from Goethe University (Frankfurt am Main, Germany). He is the former Marie Skłodowska-Curie Early Stage Researcher (Spanish National Research Council -- CSIC, Madrid, 2015-2017) and Postdoctoral Fellow (MBN Research Center, Frankfurt am Main, 2019-2021) and the recipient of the International Postdoctoral Fellowship from the German Cancer Research Center (Heidelberg, 2018-2019). His research interests are linked to multiscale modeling of irradiation-driven processes involving nanoscopic and biomolecular systems. Particular research topics include the investigation of (i) nanoscale mechanisms of biological damage induced by ionizing radiation and nanoparticle-enhanced radiotherapies and (ii) fundamental processes and phenomena lying in the core of nanofabrication using charged particle beams.

Nigel John Mason is the Professor of Molecular Physics at the University of Kent and a scientific advisor to Atomki, Debrecen, Hungary. He graduated in Physics from University College London where he also obtained his PhD degree before receiving a Royal Society University Research Fellowship leading to a lectureship. In 2003 he moved to The Open University where he established the Atomic and Molecular Physics and Astrochemistry groups. In 2018 he moved to become Head of the School of Physical Sciences at the University of Kent. His research may be broadly classified as `Molecular Physics' which encompasses several interdisciplinary themes: astrochemistry and astrobiology, environmental and atmospheric physics, plasma physics, and next-generation radiotherapy. However, all of this research is centered upon fundamental studies exploring electron- and photon-induced fragmentation of molecules and the study of the subsequent reactivity that such processes may induce in local media. Since 2013 he has led the Europlanet Research Infrastructure, Europe's largest consortium in planetary sciences and he was the first President of the Europlanet Society (2019--2023). In 2007 he was awarded the Order of the British Empire (OBE) for his services to science. He has over 420 publications.

Richard A. Amos is Associate Professor of Proton Therapy in the Department of Medical Physics and Biomedical Engineering, University College London (UCL) and former Lead for Proton Therapy Physics at UCL Hospital. After graduating from the University of Birmingham he trained as a Clinical Scientist (Medical Physics) in London with the National Health Service (NHS) before working as a clinical radiotherapy physicist for a number of years in both the UK and Canada. He then spent three years involved in ion-microbeam research at the Gray Laboratory Cancer Research Trust in the UK. In 2002 he joined the faculty at Loma Linda University Medical Center in California, the world’s first hospital-based proton therapy facility, before moving to the renowned University of Texas MD Anderson Cancer Center in 2005 to implement and develop a proton therapy program, including the first clinical proton beam scanning system in the United States. He returned to London in 2013 to take leadership roles in both the development of a national proton therapy service for the NHS and in developing translational proton therapy physics research. He is a Chartered Physicist, a Chartered Scientist, and a Fellow of the Institute of Physics and Engineering in Medicine.

Ilko Bald graduated in chemistry from the Freie Universität Berlin (Germany) where he also obtained a PhD in Physical Chemistry. After a postdoc at the University of Iceland he moved to the Interdisciplinary Nanoscience Center (iNANO) at Aarhus University/DK working on scanning probe microscopy and DNA nanotechnology. In 2013 he established a junior research group ``Optical Spectroscopy and Chemical Imaging'' at the University of Potsdam and the Federal Institute for Materials Research and Testing (BAM, Berlin, Germany). Since 2019 he is Professor for Hybrid Nanostructures at the University of Potsdam investigating nanomaterials, their optical properties, and electron induced processes.

G\'{e}rard Baldacchino obtained the degree in Physical Chemistry Engineering and a Master's degree in Instrumentation from the University of Bordeaux in 1990, France. He obtained a PhD in Molecular Chemistry from the University of Orsay (now University Paris Saclay) in 1994 for a research work in the domain of femtochemistry, i.e. ultrafast charge transfer in excited states of dye molecules. He is working from 1994 as a researcher at the French Atomic Energy Commission (CEA) in the domain of Basic Research in Radiation Chemistry. After short periods as visiting scientist in Jay-Gerin Lab, Sherbrooke, Canada in 1996 and as visiting Professor in the Katsumura Lab in Tokyo, Japan in 2006. He has currently the leadership of the group working on ultrafast phenomena in condensed matter (atto- and femto-second time scale). As an expert senior fellow of CEA from 2023, his work is focused in the experimental observations of the effects of LET and dose rates on reactive oxygen species and hydrated electrons in water under extreme conditions ($T$, $P$, pH, etc.) in relation to nuclear applications and hadrontherapy. He has also some projects related to multiscale simulations of radiation--matter interactions using MC codes such as Geant4-DNA.

Brendan Dromey completed his Undergraduate and Masters studies in University College Dublin before going on to complete his Ph.D. in Queen's University Belfast studying high harmonic generation from relativistic laser plasmas for which he received the European Physical Society Plasma Physics Division Thesis prize in 2007. Following a stint on a research fellowship in the Max Planck Institute for Quantum Optics in 2007-08, Brendan returned to Queen’s University Belfast on a Career Acceleration Fellowship to study intense attosecond pulse generation from relativistic plasmas through the  award of an EPSRC Career Acceleration Fellowship. After being appointed Reader in the Centre for Plasma Physics, Brendan was involved in a series of pioneering experiments on both the TARANIS laser at Queen’s and the Gemini laser at the Central Laser Facility that have provided the first insights in to proton interactions in scavenger free matter on picosecond timeframes. Now a Professor in the Centre for Light Matter Interactions in Queen's, Brendan's primary research focus is on developing the novel discipline of Ultrafast Nanodosimetry which aims to unlock precisely how ionising radiation, with an particular emphasis on ions, results in macroscopic, long term radiation damage in matter.

Martin Falk and his team investigate the biological effects of different types of ionizing radiation, in particular radiation damage to DNA and DNA repair in the context of chromatin architecture at micro- and nanoscale. Martin Falk is the head of the Department of Cell Biology and Radiobiology and the Laboratory of Chromatin Function, Damage and Repair at the Institute of Biophysics, Academy of Sciences of the Czech Republic in Brno. Currently, he has also been selected as a visiting professor at the University of Heidelberg, Germany, and teaches regular courses on radiation biophysics and related topics at Masaryk University in Brno. Martin Falk was recently elected as the Chairman of the Radiobiological Society for Crisis Planning of the Czech Medical Society J.E. Purkyn\v{e}, is a member of the Council of the European Radiation Research Society (ERRS) and an ERRS councilor at the International Association for Radiation Research (IARR). He is a recipient of the Otto Wichterle Prize for outstanding young scientists (from the Academy of Sciences of the Czech Republic) and several other international and national awards.

Juraj Fedor received his Ph.D. from the University of Innsbruck in 2006. He has been a postdoctoral researcher at the University of Fribourg, Switzerland, Marie-Curie Fellow in Prague, Czech Republic and SNSF Ambizione Fellow in Fribourg, Switzerland. Since 2015, he has been a senior scientist at the J. Heyrovsk\'{y} Institute of Physical Chemistry in Prague. Since 2020, he has been a head of the Department of Molecular and Cluster Dynamics. His research expertise lies mainly in electron collisions with molecules and clusters, experiments with liquid microjets, and cold ion physics.

Luca Gerhards graduated in chemistry in 2015 and obtained his Master's degree in chemistry in 2017. In 2021, he received his PhD in chemistry from the Carl-Von-Ossietzky University Oldenburg. Currently, he is a postdoctoral researcher at the Institute of Physics of the University of Oldenburg. His research focus is on the development of spin dynamics and calculation of electronic structure for the investigation of magnetic fields effects in biological systems.

Michael Hausmann received his diploma in Physics in 1984, the PhD degree in natural sciences in 1988, and his habilitation degree (venia legend in physics) in 1996 from Heidelberg University. In 2004, he got his appointment as adjunct professor and became leader of the Experimental Biophysics group at the Kirchhoff-Institute for Physics, Heidelberg University. His research is focused on experimental and radiation biophysics, especially on chromatin organization and re-arrangements on the micro- and nano-scale during tumor genesis and after ionizing radiation exposure, protein arrangements on the nano-scale in cell membranes and the endoplasmatic reticulum, technical developments and applications of super- resolution fluorescence far-field microscopy, comparative sequence pattern analysis in genome data bases in relation to 3D genome architecture, specific DNA labelling by computer designed nano-probes (COMBO-FISH), and astrobiophysics (development of life and radiation energy models). In 2014, he was awarded by the Faculty of Physics and Astronomy for excellent teaching. In 2021, he obtained the Ulrich-Hagen Award of the German Society for Biological Radiation Research (DeGBS) in honor for his life work in science and his engagement for the German radiation research. In 2022, he was nominated for the professor of the year. He is author and co-author of more than 250 research articles in renowned journals and books, and about 10 patents.

Georg Hildenbrand obtained a master's degree in biology and a PhD degree in physics from Heidelberg University, Germany, in 1999 and 2003, respectively. In 2011 he also obtained a licence to practice medicine from Mainz University, Germany. Since 2007 he has taught in several disciplines in the area of biophysics at Heidelberg University. In 2022 he became a professor for medical engineering at Aschaffenburg University of Applied Sciences. His research focuses in the field of biophysics on bioinformatics and radiation biology.

Milo\v{s} Hrabovsk\'{y} received his maters degree in applied physics in 2014, during his PhD studies he joined Tescan as an application developer for electron and ion beam patterning. In 2017, he left his PhD in applied physics to focus solely on work in the field of electron and ion beam microscopy. Currently, Milo\v{s} represents TESCAN Group as a Product Marketing Manager for Nanoprototyping, focusing on electron- and ion-beam patterning, depositions and enhanced etching, and automation of microscopes with the use of scripting.

Stanislav Kadlec received his master's degree in 1983 from the Charles University in Prague, and the PhD in Plasma Physics from the Institute of Physics, Czechoslovak Academy of Sciences, in 1990. Most of his scientific life in academia and industry he was devoted to physical vapor deposition, especially magnetron sputtering including high power pulsing, and reactive deposition of thin films. In 2017, he joined Eaton corporation, dealing with dielectric breakdown in gases, switching arcs, and other atmospheric plasmas.

Jaroslav Ko\v{c}i\v{s}ek received PhDs in physics from Comenius University in Bratislava (2010) and Charles University in Prague (2013). After postdocs studying electron scattering at the University of Fribourg and collisions of multiply charged ions at GANIL, he started a junior research group at J. Heyrovsk\'{y} Institute of Physical Chemistry in 2016, where he is now holding senior scientist position. His research is focused on environmental effects on reaction dynamics and elementary processes in radiation interaction with living tissue, studied experimentally on model systems ranging from isolated molecules and clusters to self-assembled nanostructures.

Franck L\'{e}pine is CNRS Research Director and group leader at Institut Lumi\`{e}re Mati\`{e}re in Lyon, France. He graduated in Physics at the University Lyon 1 (France), and obtained his Ph.D. degree at Laboratoire de spectrom\'{e}trie ionique et mol\'{e}culaire, Lyon, in 2003. From 2003-2005, he worked as a post-doc fellow at FOM-AMOLF institut, Amsterdam, Netherlands. In 2005 he was recruited at CNRS as a permanent researcher. Since 2016, he is leading the French National Network on ultrafast phenomenon (GDR UP). His work is dedicated to the study of ultrafast dynamics, especially attosecond processes. He developed new experimental approaches to study complex systems with attosecond precision and to understand ultrafast processes using ultrashort light pulses in wavelength range from THz to soft X-ray.

Siyi Ming earned his BSc in Light Chemical Engineering from South China University of Technology in 2017, dedicating three years to researching versatile nanofibrillated cellulose-clay materials aimed at addressing the issue of "white pollution." He subsequently embarked on an MPhil in Chemistry at the University of Cambridge in 2020, where he focused on the development of edible, additive-free, color-retentive microparticles using hydroxypropyl cellulose. Currently, Siyi is in the final year of his PhD program in Chemistry at Cambridge, working under the guidance of Professor Andrew Wheatley. During this time, he made a significant breakthrough in understanding the non-empirical growth mechanism of nanodendrites. His research is centered on the production of nanopods, nanodendrites, single atom alloys, as well as broadening the scope of compositional variations in bimetallics and trimetallics. He has also delved into comprehending particle formation and catalyst surface tailoring to facilitate the design of cost-effective nanostructures for oxygen reduction reaction (ORR) catalysts, which are essential for large-scale hydrogen fuel cell applications. Siyi's scientific interests are primarily focused on exploring the next generation of cost-effective nanomaterials with the potential to significantly enhance the quality of life for people.

Andrew Nisbet graduated in Physics in 1988 from the University of Edinburgh, UK and obtained his Master’s degree in Medical Physics in 1989 and his Ph.D. degree in Medical Physics in 1994, both from the University of Aberdeen, UK. He is now Head of the Department of Medical Physics and Biomedical Engineering at University College London, UK. His current research interests are focussed on radiotherapy physics, particularly on the development of radiation dosimeters for use in micro-dosimetry, preclinical and in vivo applications and how information derived from radiobiological measurements employing 3D scaffolds may impact radiotherapy treatment planning models.

Kate Ricketts is Associate Professor of Cancer Physics at the Division of Surgery and Interventional Science, University College London, and NHS clinically accredited Radiotherapy Physicist. She obtained a BA in Physical Natural Sciences from Cambridge University in 2004, IPEM NHS Part 1 Training in Medical Physics (Diagnostic Radiology, Nuclear Medicine and Radiotherapy) at the Royal Berkshire Hospital and MSc in Radiation Physics from UCL in 2008, continuing with a PhD in Nanoparticles for Tumour Diagnostics from UCL awarded in 2011. After a postdoc in Big Data at UCLH Radiotherapy Department, she completed Part 2 NHS Training in Radiotherapy Physics, clinically accredited in 2014. She commenced a lectureship at UCL Division of Surgery and Interventional Science in 2013, making Associate Professor in 2016. She currently works at the interface of radiation physics, AI and cancer biology towards improving patient response to radiotherapy treatment.

Leo Sala graduated with a degree in chemistry and materials science and engineering in 2011 at the Ateneo de Manila University. He finished his master’s degree in chemistry at the University of Porto and the University of Paris-Saclay in 2015. At the latter, he completed his PhD in Chemistry in 2018. In 2019, he started working as a postdoctoral researcher at the J. Heyrovsk\'{y} Institute of Physical Chemistry (Prague, Czech Republic) where he is currently continuing as an associate researcher. His work mostly focuses on investigating radiation damage to DNA origami nanostructures and their preparation for in vitro and in vivo applications.

Thomas Schlathölter is Associate Professor at the Zernike Institute for Advanced Materials at the University of Groningen, the Netherlands. He received his diploma in Physics in 1993 from the University of Osnabrück, where he also obtained his PhD in 1996. He joined the University of Groningen as a Marie-Curie Fellow in 1997. From 2001-2006, Schlathölter started his work on the interactions of ionizing radiation with gas-phase biomolecules as a fellow of the Royal Dutch Academy of Arts and Sciences. In 2011, he pioneered the combination of electrospray ionization and radiofrequency ion trapping with synchrotron and keV ion beamlines for studying radiation damage in large biomolecular ions. Since 2013, Schlathölter is associate professor at the Zernike Institute for Advanced Materials at the University of Groningen.

Andrew Wheatley received his BSc from the University of Kent at Canterbury in 1995. He then worked at Cambridge under the guidance of Dr. Ron Snaith, receiving his PhD in 1999. After his PhD studies he spent time as a Junior Research Fellow at Gonville \& Caius College before becoming a University Lecturer at Cambridge (2000) and a Fellow of Fitzwilliam College. He became a University Senior Lecturer (2010) and a University Reader in Materials Chemistry in 2018. He is now a University Professor of Materials Chemistry. He has received the Harrison Memorial Medal from the Royal Society of Chemistry and he has been a Visiting Professor at Tohoku University, Japan. In early 2024, he will be a Visiting Professor at the Shanghai Institute of Optics and Fine Mechanics. His research interests are in the areas of organometallic synthesis, synergistic effects in organometallics and nanomaterials, and novel encapsulation strategies for active species. He is interested in new solutions to photo- and electrocatalytic problems and his work has applications in pollution abatement, sensing, gas storage, fuel production, and electrocatalysis. He recently co-edited a book on Polar Organometallic Reagents, is the author of 6 book chapters, 4 patents and more than 175 communications, full papers and reviews on organometallics, nanoscience and composite materials.

Ilia A. Solov'yov, currently serving as a Professor of Theoretical Molecular Physics at the Carl von Ossietzky University in Oldenburg, Germany, since 2019, previously established the Quantum Biology and Computational Physics Group in 2013 at the University of Southern Denmark in Odense. His academic journey includes graduating from Goethe University in Frankfurt am Main, Germany, in 2008 and receiving a Candidate of Science degree in theoretical physics from the Ioffe Physical-Technical Institute in St. Petersburg, Russia, a year later. His research interests encompass a diverse spectrum of topics within biomolecules and smart inorganic materials, focusing on biological processes that facilitate energy conversion into forms suitable for chemical transformations underpinned by quantum mechanical principles. To explore these phenomena, he adeptly employs and advances classical and quantum molecular dynamics, Monte Carlo simulations, and multiscale techniques, which are pivotal in his investigations of biophysical processes, spanning chemical reactions, light absorption, the formation of excited electronic states, as well as the transfer of excitation energy, electrons, and protons.

\begin{acknowledgement}

The authors acknowledge the COST Action CA20129 MultIChem, supported by COST (European Cooperation in Science and Technology), as well as the European Commission through the RADON (GA 872494) and N-LIGHT (GA 872196) projects within the H2020-MSCA-RISE-2019 call, and the Horizon Europe EIC Pathfinder Project TECHNO-CLS (Project No. 101046458).
The authors would like to thank the Volkswagen Foundation (Lichtenberg professorship awarded to I.A.S.), the Deutsche Forschungsgemeinschaft (GRK1885 Molecular Basis of Sensory Biology; SFB 1372 Magnetoreception and Navigation in Vertebrates, no. 395940726; TRR386/1-2023 HYP*MOL Hyperpolarization in molecular systems, no 514664767 to I.A.S.), the Ministry for Science and Culture of Lower Saxony Simulations Meet Experiments on the Nanoscale: Opening up the Quantum World to Artificial Intelligence (SMART) to I.A.S.; Dynamik auf der Nanoskala: Von koh\"{a}renten Elementarprozessen zur Funktionalit\"{a}t (DyNano) to I.A.S. The authors also gratefully acknowledge the computing time granted by the Resource Allocation Board and provided on the supercomputer Lise and Emmy at NHR@ZIB and NHR@ G\"{o}ttingen as part of the NHR infrastructure. The calculations for this research were conducted with computing resources under the project nip00058.
N.J.M. recognises support from Europlanet 2024 RI funded from the European Union's Horizon 2020 research and innovation programme under grant agreement No 871149.
I.B. acknowledges funding by the German Research Foundation (DFG, project number 450169704).
Michael Hausmann and G.H. deeply thank Aaron Sievers and Jonas Weidner as well as many bachelor students at the Kirchhoff-Institute for Physics, Heidelberg, for their work, which contributed to the scientific outcome summarized in Section~\ref{sec:Case_study_DNA_damage_repair}. In addition, the financial support of DFG (H1601/16-1)/GACR (project GACR 20-04109J), BMBF (FKZ 02NUK058A), and DAAD to Michael Hausmann and M.F. in the context of the referred results in that section is gratefully acknowledged.
A.W. thanks CSC Cambridge for funding.
The project was also partly supported by the Technological Agency of the Czech Republic, project no. TK04020069 (S.K., J.K. and J.F.). J.F. acknowledges the Czech Science Foundation project 21-26601X and OP RDE project ``QUEENTEC'' CZ.02.01.01/00/22\_008/0004649.

\end{acknowledgement}

\section*{Abbreviations}

\noindent
2PPE -- two-photon photoelectron spectroscopy

\noindent
AES -- Auger electron spectroscopy

\noindent
AGE -- agarose gel electrophoresis

\noindent
AI -- artificial intelligence

\noindent
AIMD -- \textit{ab initio} molecular dynamics

\noindent
AIREBO -- adaptive intermolecular reactive empirical bond-order

\noindent
AFM -- atomic force microscope

\noindent
ANN -- artificial neural network

\noindent
BCA -- binary-collision approximation

\noindent
BEB -- binary-encounter-Bethe

\noindent
BED -- binary-encounter dipole

\noindent
BNCT -- boron neutron capture therapy

\noindent
CC -- coupled cluster

\noindent
CFD -- computational fluid dynamics

\noindent
CG -- coarse graining

\noindent
CI -- configuration interaction

\noindent
CLS -- crystal-based light source

\noindent
CRDS -- cavity ring-down spectroscopy

\noindent
CS -- cross section

\noindent
CSDA -- continuous slowing-down approximation

\noindent
CT -- computed tomography

\noindent
CTV -- clinical target volume

\noindent
CU -- crystalline undulator

\noindent
CUR -- crystalline undulator radiation

\noindent
DEA -- dissociative electron attachment

\noindent
DFT -- density functional theory

\noindent
DFTB -- density functional-based tight binding

\noindent
DHF -- Dirac-Hartree-Fock

\noindent
DM -- Deutsch-M\"{a}rk

\noindent
DNS -- direct numerical solution

\noindent
DOS -- dipole oscillator strength

\noindent
DSB -- double-strand break (in DNA)

\noindent
EAD -- energy and angular distribution

\noindent
EDS -- energy-dispersive spectroscopy

\noindent
EDX -- energy-dispersive X-ray spectroscopy

\noindent
EELS -- electron energy-loss spectroscopy

\noindent
EPR -- enhanced permeability and retention

\noindent
ESI -- electrospray ionization

\noindent
ET -- electron transfer

\noindent
EVDF -- electron velocity distribution function

\noindent
FAD -- flavin adenine dinucleotide

\noindent
FEBID -- focused electron beam-induced deposition

\noindent
FEBiMS -- focused electron beam-induced mass spectrometry

\noindent
FEL -- free electron laser

\noindent
FEM -- finiteelement method

\noindent
FIBID -- focused ion beam-induced deposition

\noindent
FTIR -- Fourier-transform infra-red

\noindent
GC-MS -- gas chromatography mass spectrometry

\noindent
GGA -- generalized gradient approximations

\noindent
GPU -- graphics processing units

\noindent
GTV -- gross tumor volume

\noindent
HF -- Hartree-Fock

\noindent
HiPIMS -- high-power impulse magnetron sputtering

\noindent
HPC -- high-performance computing

\noindent
IAEA -- International Atomic Energy Agency

\noindent
IBCT -- ion-beam cancer therapy

\noindent
IC -- (air-filled) ionization chamber

\noindent
IDC -- irradiation-driven chemistry

\noindent
IDMD -- irradiation-driven molecular dynamics

\noindent
IMS -- ion mobility spectrometry

\noindent
IR -- ionizing radiation

\noindent
IRIF -- ionizing radiation-induced focus

\noindent
ISM -- interstellar medium

\noindent
ISS -- ion scattering spectroscopy 

\noindent
JWST -- James Webb space telescope

\noindent
KMC -- kinetic Monte Carlo

\noindent
KS -- Kohn-Sham

\noindent
LC -- long-range correction

\noindent
LDA -- local density approximation

\noindent
LEIS -- low-energy ion scattering

\noindent
LET -- linear energy transfer

\noindent
LIC -- liquid-filled ionization chamber

\noindent
LS -- light source

\noindent
LTE -- local thermal equilibrium

\noindent
MBPT -- many-body perturbation theory

\noindent
MC -- Monte Carlo

\noindent
MD -- molecular dynamics

\noindent
ML -- machine learning

\noindent
MM -- multiscale modeling

\noindent
MP -- M{\o}ller-Plesset

\noindent
MS -- mass spectrometry

\noindent
MSA -- multiscale approach to the physics of radiation damage with ions

\noindent
NEGF -- non-equilibrium Green's function

\noindent
NERT -- nanoparticle-enhanced radiotherapy

\noindent
NP -- nanoparticle

\noindent
ORR -- oxygen reduction reaction

\noindent
PBS -- pencil beam scanning

\noindent
PCB -- printed circuit boards

\noindent
PE -- primary electron

\noindent
PEMFC -- proton exchange membrane fuel cell

\noindent
PEPT -- positron emission particle tracking

\noindent
PES -- potential energy surface

\noindent
PES -- photoelectron spectroscopy

\noindent
PET -- positron emission tomography

\noindent
PGM -- platinum-group metal

\noindent
PIC-MCC -- particle-in-cell simulations with Monte Carlo collisions

\noindent
PIML -- physics-informed machine learning

\noindent
PINN -- physics-informed neural network

\noindent
PMBRT -- particle minibeam radiation therapy

\noindent
PME -- particle mesh Ewald

\noindent
PS -- passive scattering

\noindent
QBE -- quantum Boltzmann equation
 
\noindent
QM/MM -- quantum mechanics / molecular mechanics

\noindent
RADAM -- radiation damage

\noindent
RAIRS -- reflection absorption infrared spectroscopy

\noindent
RBE -- relative biological effectiveness

\noindent
REBO -- reactive empirical bond-order

\noindent
REELS -- reflected electron energy-loss spectroscopy

\noindent
RMD -- reactive molecular dynamics

\noindent
ROS -- reactive oxygen species

\noindent
RP -- radical pair

\noindent
RSP -- relative stopping power

\noindent
SD -- stochastic dynamics

\noindent
SDCS -- singly-differentiated cross section

\noindent
SE -- secondary electron

\noindent
SEM -- scanning electron microscope

\noindent
SERS -- surface-enhanced Raman scattering

\noindent
SIMS -- secondary ion mass spectrometry

\noindent
SMLM -- single-molecule localization microscopy

\noindent
SNMS -- sputtered neutral mass spectrometry

\noindent
SOBP -- spread-out Bragg peak

\noindent
SPR -- surface plasmon resonance

\noindent
SSB -- single-strand break (in DNA)

\noindent
SSTR -- super-short-tandem-repeats

\noindent
STEM -- scanning transmission electron microscope

\noindent
SW -- shock wave

\noindent
TA -- transient absorption

\noindent
TAS -- transient absorption spectroscopy

\noindent
TDDFT -- time-dependent density functional theory

\noindent
TEM -- transmission electron microscope

\noindent
TM -- transition metal

\noindent
TNSA -- target normal sheath acceleration

\noindent
TOF -- time-of-flight

\noindent
TPS -- treatment planning system

\noindent
UHDR -- ultra-high dose rate

\noindent
UPS -- ultraviolet photoelectron spectroscopy

\noindent
UV -- ultraviolet

\noindent
VAMDC -- Virtual Atomic and Molecular Data Centre

\noindent
VESPA -- Virtual European Solar and Planetary Access

\noindent
VOC -- volatile organic compound

\noindent
VUV -- vacuum ultraviolet

\noindent
XFEL -- X-ray free-electron laser

\noindent
XPS -- X-ray photoelectron spectroscopy

\noindent
XUV -- extreme ultraviolet





\bibliography{Roadmap_paper_bibliography}

\providecommand{\latin}[1]{#1}
\makeatletter
\providecommand{\doi}
  {\begingroup\let\do\@makeother\dospecials
  \catcode`\{=1 \catcode`\}=2 \doi@aux}
\providecommand{\doi@aux}[1]{\endgroup\texttt{#1}}
\makeatother
\providecommand*\mcitethebibliography{\thebibliography}
\csname @ifundefined\endcsname{endmcitethebibliography}
  {\let\endmcitethebibliography\endthebibliography}{}
\begin{mcitethebibliography}{845}
\providecommand*\natexlab[1]{#1}
\providecommand*\mciteSetBstSublistMode[1]{}
\providecommand*\mciteSetBstMaxWidthForm[2]{}
\providecommand*\mciteBstWouldAddEndPuncttrue
  {\def\EndOfBibitem{\unskip.}}
\providecommand*\mciteBstWouldAddEndPunctfalse
  {\let\EndOfBibitem\relax}
\providecommand*\mciteSetBstMidEndSepPunct[3]{}
\providecommand*\mciteSetBstSublistLabelBeginEnd[3]{}
\providecommand*\EndOfBibitem{}
\mciteSetBstSublistMode{f}
\mciteSetBstMaxWidthForm{subitem}{(\alph{mcitesubitemcount})}
\mciteSetBstSublistLabelBeginEnd
  {\mcitemaxwidthsubitemform\space}
  {\relax}
  {\relax}

\bibitem[Landau and Lifshitz(1981)Landau, and Lifshitz]{LL3}
Landau,~L.~D.; Lifshitz,~E.~M. \emph{{Quantum Mechanics: Non-Relativistic
  Theory}}, 3rd ed.; Butterworth-Heinemann, Oxford, 1981\relax
\mciteBstWouldAddEndPuncttrue
\mciteSetBstMidEndSepPunct{\mcitedefaultmidpunct}
{\mcitedefaultendpunct}{\mcitedefaultseppunct}\relax
\EndOfBibitem
\bibitem[Landau and Lifshitz(1980)Landau, and Lifshitz]{LL5}
Landau,~L.~D.; Lifshitz,~E.~M. \emph{{Statistical Physics}}, 3rd ed.; Pergamon
  Press, Oxford, 1980\relax
\mciteBstWouldAddEndPuncttrue
\mciteSetBstMidEndSepPunct{\mcitedefaultmidpunct}
{\mcitedefaultendpunct}{\mcitedefaultseppunct}\relax
\EndOfBibitem
\bibitem[Landau and Lifshitz(1980)Landau, and Lifshitz]{LL9}
Landau,~L.~D.; Lifshitz,~E.~M. \emph{{Statistical Physics: Theory of the
  Condensed State}}; Butterworth-Heinemann, Oxford, 1980\relax
\mciteBstWouldAddEndPuncttrue
\mciteSetBstMidEndSepPunct{\mcitedefaultmidpunct}
{\mcitedefaultendpunct}{\mcitedefaultseppunct}\relax
\EndOfBibitem
\bibitem[Lifshitz and Pitaevskii(1981)Lifshitz, and Pitaevskii]{LL10}
Lifshitz,~E.~M.; Pitaevskii,~L.~P. \emph{{Physical Kinetics}};
  Butterworth-Heinemann, Oxford, 1981\relax
\mciteBstWouldAddEndPuncttrue
\mciteSetBstMidEndSepPunct{\mcitedefaultmidpunct}
{\mcitedefaultendpunct}{\mcitedefaultseppunct}\relax
\EndOfBibitem
\bibitem[Solov'yov \latin{et~al.}(2017)Solov'yov, Korol, and
  Solov'yov]{MBNbook_Springer_2017}
Solov'yov,~I.~A.; Korol,~A.~V.; Solov'yov,~A.~V. \emph{{Multiscale Modeling of
  Complex Molecular Structure and Dynamics with MBN Explorer}}; Springer
  International Publishing: Cham, Switzerland, 2017\relax
\mciteBstWouldAddEndPuncttrue
\mciteSetBstMidEndSepPunct{\mcitedefaultmidpunct}
{\mcitedefaultendpunct}{\mcitedefaultseppunct}\relax
\EndOfBibitem
\bibitem[Solov'yov \latin{et~al.}(2022)Solov'yov, Verkhovtsev, Korol, and
  Solov'yov]{DySoN_book_Springer_2022}
Solov'yov,~I.~A., Verkhovtsev,~A.~V., Korol,~A.~V., Solov'yov,~A.~V., Eds.
  \emph{{Dynamics of Systems on the Nanoscale}}; Springer Nature Switzerland:
  Cham, Switzerland, 2022\relax
\mciteBstWouldAddEndPuncttrue
\mciteSetBstMidEndSepPunct{\mcitedefaultmidpunct}
{\mcitedefaultendpunct}{\mcitedefaultseppunct}\relax
\EndOfBibitem
\bibitem[Workman \latin{et~al.}(2022)Workman, Burkert, Crede, Klempt, Thoma,
  Tiator, Agashe, Aielli, Allanach, Amsler, and \textit{et al.} (Particle
  Data~Group).]{Workman_PDG_2022_review}
Workman,~R.~L.; Burkert,~V.~D.; Crede,~V.; Klempt,~E.; Thoma,~U.; Tiator,~L.;
  Agashe,~K.; Aielli,~G.; Allanach,~B.~C.; Amsler,~C.; \textit{et al.}
  (Particle Data~Group). {Review of Particle Physics}. \emph{Prog. Theor. Exp.
  Phys.} \textbf{2022}, \emph{2022}, 083C01\relax
\mciteBstWouldAddEndPuncttrue
\mciteSetBstMidEndSepPunct{\mcitedefaultmidpunct}
{\mcitedefaultendpunct}{\mcitedefaultseppunct}\relax
\EndOfBibitem
\bibitem[Landau \latin{et~al.}(1984)Landau, Lifshitz, and Pitaevskii]{LL8}
Landau,~L.~D.; Lifshitz,~E.~M.; Pitaevskii,~L.~P. \emph{{Electrodynamics of
  Continuous Media}}, 2nd ed.; Elsevier Butterworth-Heinemann, Oxford,
  1984\relax
\mciteBstWouldAddEndPuncttrue
\mciteSetBstMidEndSepPunct{\mcitedefaultmidpunct}
{\mcitedefaultendpunct}{\mcitedefaultseppunct}\relax
\EndOfBibitem
\bibitem[Berestetskii \latin{et~al.}(1982)Berestetskii, Lifshitz, and
  Pitaevskii]{LL4}
Berestetskii,~V.~B.; Lifshitz,~E.~M.; Pitaevskii,~L.~P. \emph{{Quantum
  Electrodynamics}}, 2nd ed.; Elsevier Butterworth-Heinemann, Oxford,
  1982\relax
\mciteBstWouldAddEndPuncttrue
\mciteSetBstMidEndSepPunct{\mcitedefaultmidpunct}
{\mcitedefaultendpunct}{\mcitedefaultseppunct}\relax
\EndOfBibitem
\bibitem[Solov'yov(2017)]{AVS2017nanoscaleIBCT}
Solov'yov,~A.~V., Ed. \emph{{Nanoscale Insights into Ion-Beam Cancer Therapy}};
  Springer International Publishing: Cham, Switzerland, 2017\relax
\mciteBstWouldAddEndPuncttrue
\mciteSetBstMidEndSepPunct{\mcitedefaultmidpunct}
{\mcitedefaultendpunct}{\mcitedefaultseppunct}\relax
\EndOfBibitem
\bibitem[Surdutovich and Solov'yov(2014)Surdutovich, and
  Solov'yov]{Surdutovich_AVS_2014_EPJD.68.353}
Surdutovich,~E.; Solov'yov,~A.~V. {Multiscale Approach to the Physics of
  Radiation Damage with Ions}. \emph{Eur. Phys. J. D} \textbf{2014}, \emph{68},
  353\relax
\mciteBstWouldAddEndPuncttrue
\mciteSetBstMidEndSepPunct{\mcitedefaultmidpunct}
{\mcitedefaultendpunct}{\mcitedefaultseppunct}\relax
\EndOfBibitem
\bibitem[Surdutovich and Solov'yov(2019)Surdutovich, and
  Solov'yov]{surdutovich2019multiscale}
Surdutovich,~E.; Solov'yov,~A.~V. Multiscale Modeling for Cancer
  Radiotherapies. \emph{Cancer Nanotechnol.} \textbf{2019}, \emph{10}, 6\relax
\mciteBstWouldAddEndPuncttrue
\mciteSetBstMidEndSepPunct{\mcitedefaultmidpunct}
{\mcitedefaultendpunct}{\mcitedefaultseppunct}\relax
\EndOfBibitem
\bibitem[Sushko \latin{et~al.}(2016)Sushko, Solov'yov, and
  Solov'yov]{Sushko_IS_AS_FEBID_2016}
Sushko,~G.~B.; Solov'yov,~I.~A.; Solov'yov,~A.~V. {Molecular Dynamics for
  Irradiation Driven Chemistry: Application to the {FEBID} Process}. \emph{Eur.
  Phys. J. D} \textbf{2016}, \emph{70}, 217\relax
\mciteBstWouldAddEndPuncttrue
\mciteSetBstMidEndSepPunct{\mcitedefaultmidpunct}
{\mcitedefaultendpunct}{\mcitedefaultseppunct}\relax
\EndOfBibitem
\bibitem[de~Vera \latin{et~al.}(2020)de~Vera, Azzolini, Sushko, Abril,
  Garcia-Molina, Dapor, Solov'yov, and Solov'yov]{DeVera2020}
de~Vera,~P.; Azzolini,~M.; Sushko,~G.; Abril,~I.; Garcia-Molina,~R.; Dapor,~M.;
  Solov'yov,~I.~A.; Solov'yov,~A.~V. {Multiscale Simulation of the Focused
  Electron Beam Induced Deposition Process}. \emph{Sci. Rep.} \textbf{2020},
  \emph{10}, 20827\relax
\mciteBstWouldAddEndPuncttrue
\mciteSetBstMidEndSepPunct{\mcitedefaultmidpunct}
{\mcitedefaultendpunct}{\mcitedefaultseppunct}\relax
\EndOfBibitem
\bibitem[Solov'yov \latin{et~al.}(2014)Solov'yov, Solov'yov, K\'{e}baili,
  Masson, and Br\'{e}chignac]{Solovyov_2014_PhysStatSolB.251.609}
Solov'yov,~I.~A.; Solov'yov,~A.~V.; K\'{e}baili,~N.; Masson,~A.;
  Br\'{e}chignac,~C. {Thermally Induced Morphological Transition of Silver
  Fractals}. \emph{Phys. Stat. Sol. B} \textbf{2014}, \emph{251},
  609--622\relax
\mciteBstWouldAddEndPuncttrue
\mciteSetBstMidEndSepPunct{\mcitedefaultmidpunct}
{\mcitedefaultendpunct}{\mcitedefaultseppunct}\relax
\EndOfBibitem
\bibitem[Solov'yov \latin{et~al.}(2022)Solov'yov, Sushko, Friis, and
  Solov'yov]{Stochastic_2022_JCC.43.1442}
Solov'yov,~I.~A.; Sushko,~G.; Friis,~I.; Solov'yov,~A.~V. {Multiscale Modeling
  of Stochastic Dynamics Processes With MBN Explorer}. \emph{J. Comput. Chem.}
  \textbf{2022}, \emph{43}, 1442--1458\relax
\mciteBstWouldAddEndPuncttrue
\mciteSetBstMidEndSepPunct{\mcitedefaultmidpunct}
{\mcitedefaultendpunct}{\mcitedefaultseppunct}\relax
\EndOfBibitem
\bibitem[Korol and Solov'yov(2022)Korol, and Solov'yov]{NovelLSs_Springer_book}
Korol,~A.~V.; Solov'yov,~A.~V. \emph{{Novel Lights Sources Beyond Free Electron
  Lasers}}; Particle Acceleration and Detection series, Springer Nature
  Switzerland: Cham, Switzerland, 2022\relax
\mciteBstWouldAddEndPuncttrue
\mciteSetBstMidEndSepPunct{\mcitedefaultmidpunct}
{\mcitedefaultendpunct}{\mcitedefaultseppunct}\relax
\EndOfBibitem
\bibitem[Korol and Solov'yov(2020)Korol, and
  Solov'yov]{AVK_AVS_2020_EPJD.74.201_review}
Korol,~A.~V.; Solov'yov,~A.~V. {Crystal-Based Intensive Gamma-Ray Light
  Sources}. \emph{Eur. Phys. J. D} \textbf{2020}, \emph{74}, 201\relax
\mciteBstWouldAddEndPuncttrue
\mciteSetBstMidEndSepPunct{\mcitedefaultmidpunct}
{\mcitedefaultendpunct}{\mcitedefaultseppunct}\relax
\EndOfBibitem
\bibitem[Korol \latin{et~al.}(2014)Korol, Solov'yov, and
  Greiner]{Channeling_book}
Korol,~A.~V.; Solov'yov,~A.~V.; Greiner,~W. \emph{{Channeling and Radiation in
  Periodically Bent Crystals}}, 2nd ed.; Springer Series on Atomic, Optical,
  and Plasma Physics, vol. 69. Springer-Verlag: Heidelberg, New York,
  Dordrecht, London, 2014\relax
\mciteBstWouldAddEndPuncttrue
\mciteSetBstMidEndSepPunct{\mcitedefaultmidpunct}
{\mcitedefaultendpunct}{\mcitedefaultseppunct}\relax
\EndOfBibitem
\bibitem[Connerade \latin{et~al.}(2002)Connerade, Solov'yov, and
  Greiner]{JPC_AVS_WG_EurophysNews.33.200}
Connerade,~J.-P.; Solov'yov,~A.~V.; Greiner,~W. {The Science of Clusters: An
  Emerging Field}. \emph{Europhys. News} \textbf{2002}, \emph{33},
  200--202\relax
\mciteBstWouldAddEndPuncttrue
\mciteSetBstMidEndSepPunct{\mcitedefaultmidpunct}
{\mcitedefaultendpunct}{\mcitedefaultseppunct}\relax
\EndOfBibitem
\bibitem[Kim \latin{et~al.}(2001)Kim, Tripp, and Wei]{Kim_2001_JACS.123.7955}
Kim,~B.; Tripp,~S.~L.; Wei,~A. {Self-Organization of Large Gold Nanoparticle
  Arrays}. \emph{J. Am. Chem. Soc.} \textbf{2001}, \emph{123}, 7955--7956\relax
\mciteBstWouldAddEndPuncttrue
\mciteSetBstMidEndSepPunct{\mcitedefaultmidpunct}
{\mcitedefaultendpunct}{\mcitedefaultseppunct}\relax
\EndOfBibitem
\bibitem[Eichhorn and Yu(2015)Eichhorn, and Yu]{Eichhorn_SelfOrg_MetalNPs}
Eichhorn,~S.~H.; Yu,~J.~K. In \emph{Anisotropic Nanomaterials: NanoScience and
  Technology}; Li,~Q., Ed.; Springer, Cham, 2015; pp 289--336\relax
\mciteBstWouldAddEndPuncttrue
\mciteSetBstMidEndSepPunct{\mcitedefaultmidpunct}
{\mcitedefaultendpunct}{\mcitedefaultseppunct}\relax
\EndOfBibitem
\bibitem[Kim \latin{et~al.}(2007)Kim, Kwon, Kim, Kwon, Ihm, and
  Min]{Kim_2007_JPCC.111.11252}
Kim,~J.-Y.; Kwon,~M.-H.; Kim,~J.-T.; Kwon,~S.; Ihm,~D.-W.; Min,~Y.-K.
  {Crystallization Growth and Micropatterning on Self-Assembled Conductive
  Polymer Nanofilms}. \emph{J. Phys. Chem. C} \textbf{2007}, \emph{111},
  11252--11258\relax
\mciteBstWouldAddEndPuncttrue
\mciteSetBstMidEndSepPunct{\mcitedefaultmidpunct}
{\mcitedefaultendpunct}{\mcitedefaultseppunct}\relax
\EndOfBibitem
\bibitem[Zhang \latin{et~al.}(2018)Zhang, Xing, and
  Li]{Zhang_2018_IJMS.19.1641}
Zhang,~S.; Xing,~M.; Li,~B. {Biomimetic Layer-by-Layer Self-Assembly of
  Nanofilms, Nanocoatings, and 3D Scaffolds for Tissue Engineering}. \emph{Int.
  J. Mol. Sci.} \textbf{2018}, \emph{19}, 1641\relax
\mciteBstWouldAddEndPuncttrue
\mciteSetBstMidEndSepPunct{\mcitedefaultmidpunct}
{\mcitedefaultendpunct}{\mcitedefaultseppunct}\relax
\EndOfBibitem
\bibitem[Mae \latin{et~al.}(2017)Mae, Toyama, Nawa-Okita, Yamamoto, Chen,
  Yoshikawa, Toshimitsu, Nakashima, Matsuda, and Shioi]{Mae_2017_SciRep.7.5267}
Mae,~K.; Toyama,~H.; Nawa-Okita,~E.; Yamamoto,~D.; Chen,~Y.-J.; Yoshikawa,~K.;
  Toshimitsu,~F.; Nakashima,~N.; Matsuda,~K.; Shioi,~A. {Self-Organized
  Micro-Spiral of Single-Walled Carbon Nanotubes}. \emph{Sci. Rep.}
  \textbf{2017}, \emph{7}, 5267\relax
\mciteBstWouldAddEndPuncttrue
\mciteSetBstMidEndSepPunct{\mcitedefaultmidpunct}
{\mcitedefaultendpunct}{\mcitedefaultseppunct}\relax
\EndOfBibitem
\bibitem[Macak \latin{et~al.}(2007)Macak, Tsuchiya, Ghicov, Yasuda, Hahn,
  Bauer, and Schmuki]{Macak_2007_CurrOpinSolidState.11.3}
Macak,~J.~M.; Tsuchiya,~H.; Ghicov,~A.; Yasuda,~K.; Hahn,~R.; Bauer,~S.;
  Schmuki,~P. {TiO$_2$ Nanotubes: Self-Organized Electrochemical Formation,
  Properties and Applications}. \emph{Curr. Opin. Solid State Mater. Sci.}
  \textbf{2007}, \emph{11}, 3--18\relax
\mciteBstWouldAddEndPuncttrue
\mciteSetBstMidEndSepPunct{\mcitedefaultmidpunct}
{\mcitedefaultendpunct}{\mcitedefaultseppunct}\relax
\EndOfBibitem
\bibitem[Shimizu \latin{et~al.}(2014)Shimizu, Minamikawa, Kogiso, Aoyagi,
  Kameta, Ding, and Masuda]{Shimizu_2014_PolymJ.46.831}
Shimizu,~T.; Minamikawa,~H.; Kogiso,~M.; Aoyagi,~M.; Kameta,~N.; Ding,~W.;
  Masuda,~M. {Self-Organized Nanotube Materials and Their Application in
  Bioengineering}. \emph{Polym. J.} \textbf{2014}, \emph{46}, 831--858\relax
\mciteBstWouldAddEndPuncttrue
\mciteSetBstMidEndSepPunct{\mcitedefaultmidpunct}
{\mcitedefaultendpunct}{\mcitedefaultseppunct}\relax
\EndOfBibitem
\bibitem[Fan \latin{et~al.}(2006)Fan, Werner, and
  Zacharias]{Fan_2006_Small.2.700}
Fan,~H.~J.; Werner,~P.; Zacharias,~M. {Semiconductor Nanowires: From
  Self-Organization to Patterned Growth}. \emph{Small} \textbf{2006}, \emph{2},
  700--717\relax
\mciteBstWouldAddEndPuncttrue
\mciteSetBstMidEndSepPunct{\mcitedefaultmidpunct}
{\mcitedefaultendpunct}{\mcitedefaultseppunct}\relax
\EndOfBibitem
\bibitem[Yi \latin{et~al.}(223)Yi, Peres, Pierrot, Cayez, Cours, Warot-Fonrose,
  Marcelot, Roblin, Soulantica, and Blon]{Yi_2023_NanoRes.16.1606}
Yi,~D.; Peres,~L.; Pierrot,~A.; Cayez,~S.; Cours,~R.; Warot-Fonrose,~B.;
  Marcelot,~C.; Roblin,~P.; Soulantica,~K.; Blon,~T. {Self-Organization and
  Tunable Characteristic Lengths of Two-Dimensional Hexagonal Superlattices of
  Nanowires Directly Grown on Substrates}. \emph{Nano Res.} \textbf{223},
  \emph{16}, 1606--1613\relax
\mciteBstWouldAddEndPuncttrue
\mciteSetBstMidEndSepPunct{\mcitedefaultmidpunct}
{\mcitedefaultendpunct}{\mcitedefaultseppunct}\relax
\EndOfBibitem
\bibitem[Zhang \latin{et~al.}(2023)Zhang, Park, Jia, Chang, Ng, and
  Ooi]{Zhang_2023_CrystGrowthDes}
Zhang,~X.; Park,~T.-Y.; Jia,~Y.; Chang,~H.; Ng,~T.~K.; Ooi,~B.~S.
  {Self-Organized Growth of Nanowires on a Graphene Film}. \emph{Cryst. Growth
  Des.} \textbf{2023}, \emph{23}, 3813--3819\relax
\mciteBstWouldAddEndPuncttrue
\mciteSetBstMidEndSepPunct{\mcitedefaultmidpunct}
{\mcitedefaultendpunct}{\mcitedefaultseppunct}\relax
\EndOfBibitem
\bibitem[Jensen(1999)]{Jensen_1999_RMP.71.1695}
Jensen,~P. {Growth of Nanostructures by Cluster Deposition: Experiments and
  Simple Models}. \emph{Rev. Mod. Phys.} \textbf{1999}, \emph{71},
  1695--1735\relax
\mciteBstWouldAddEndPuncttrue
\mciteSetBstMidEndSepPunct{\mcitedefaultmidpunct}
{\mcitedefaultendpunct}{\mcitedefaultseppunct}\relax
\EndOfBibitem
\bibitem[Lando \latin{et~al.}(2006)Lando, K\'{e}ba\"{i}li, Cahuzac, Masson, and
  Br\'{e}chignac]{Lando_2006_PRL.97.133402}
Lando,~A.; K\'{e}ba\"{i}li,~N.; Cahuzac,~P.; Masson,~A.; Br\'{e}chignac,~C.
  {Coarsening and Pearling Instabilities in Silver Nanofractal Aggregates}.
  \emph{Phys. Rev. Lett.} \textbf{2006}, \emph{97}, 133402\relax
\mciteBstWouldAddEndPuncttrue
\mciteSetBstMidEndSepPunct{\mcitedefaultmidpunct}
{\mcitedefaultendpunct}{\mcitedefaultseppunct}\relax
\EndOfBibitem
\bibitem[Dick \latin{et~al.}(2010)Dick, Solov'yov, and
  Solov'yov]{Dick_2010_JPCS.248.012025}
Dick,~V.~V.; Solov'yov,~I.~A.; Solov'yov,~A.~V. {Nanoparticles Dynamics on a
  Surface: Fractal Pattern Formation and Fragmentation}. \emph{J. Phys.: Conf.
  Ser.} \textbf{2010}, \emph{248}, 012025\relax
\mciteBstWouldAddEndPuncttrue
\mciteSetBstMidEndSepPunct{\mcitedefaultmidpunct}
{\mcitedefaultendpunct}{\mcitedefaultseppunct}\relax
\EndOfBibitem
\bibitem[Dick \latin{et~al.}(2011)Dick, Solov'yov, and
  Solov'yov]{Dick_2011_PRB.84.115408}
Dick,~V.~V.; Solov'yov,~I.~A.; Solov'yov,~A.~V. {Fragmentation Pathways of
  Nanofractal Structures on Surface}. \emph{Phys. Rev. B} \textbf{2011},
  \emph{84}, 115408\relax
\mciteBstWouldAddEndPuncttrue
\mciteSetBstMidEndSepPunct{\mcitedefaultmidpunct}
{\mcitedefaultendpunct}{\mcitedefaultseppunct}\relax
\EndOfBibitem
\bibitem[Panshenskov \latin{et~al.}(2014)Panshenskov, Solov'yov, and
  Solov'yov]{Panshenskov_2014_JCC.35.1317}
Panshenskov,~M.; Solov'yov,~I.~A.; Solov'yov,~A.~V. {Efficient 3D Kinetic Monte
  Carlo Method for Modeling of Molecular Structure and Dynamics}. \emph{J.
  Comput. Chem.} \textbf{2014}, \emph{35}, 1317--1329\relax
\mciteBstWouldAddEndPuncttrue
\mciteSetBstMidEndSepPunct{\mcitedefaultmidpunct}
{\mcitedefaultendpunct}{\mcitedefaultseppunct}\relax
\EndOfBibitem
\bibitem[Br\'{e}chignac \latin{et~al.}(2007)Br\'{e}chignac, Houdy, and
  Lahmani]{Brechignac_2007_NanomaterNanochem}
Br\'{e}chignac,~C., Houdy,~P., Lahmani,~M., Eds. \emph{{Nanometerials and
  Nanochemistry}}; Cambridge University Press, Cambridge, 2007\relax
\mciteBstWouldAddEndPuncttrue
\mciteSetBstMidEndSepPunct{\mcitedefaultmidpunct}
{\mcitedefaultendpunct}{\mcitedefaultseppunct}\relax
\EndOfBibitem
\bibitem[Yu \latin{et~al.}(2013)Yu, Regulacio, Yea, and
  Han]{Yu_2013_ChemSocRev.42.6006}
Yu,~H.-D.; Regulacio,~M.~D.; Yea,~E.; Han,~M.-Y. {Chemical Routes to Top-Down
  Nanofabrication}. \emph{Chem. Soc. Rev.} \textbf{2013}, \emph{42},
  6006--6018\relax
\mciteBstWouldAddEndPuncttrue
\mciteSetBstMidEndSepPunct{\mcitedefaultmidpunct}
{\mcitedefaultendpunct}{\mcitedefaultseppunct}\relax
\EndOfBibitem
\bibitem[Biswas \latin{et~al.}(2012)Biswas, Bayer, Biris, Wang, Dervishi, and
  Faupel]{Biswas_2012_AdvColloidInterfaceSci}
Biswas,~A.; Bayer,~I.~S.; Biris,~A.~S.; Wang,~T.; Dervishi,~E.; Faupel,~F.
  {Advances in Top--Down and Bottom--Up Surface Nanofabrication: Techniques,
  Applications \& Future Prospects}. \emph{Adv. Colloid Interface Sci.}
  \textbf{2012}, \emph{170}, 2--27\relax
\mciteBstWouldAddEndPuncttrue
\mciteSetBstMidEndSepPunct{\mcitedefaultmidpunct}
{\mcitedefaultendpunct}{\mcitedefaultseppunct}\relax
\EndOfBibitem
\bibitem[Shimomura and Sawadaishi(2001)Shimomura, and
  Sawadaishi]{Shimomura_2001_BottomUp}
Shimomura,~M.; Sawadaishi,~T. {Bottom-Up Strategy of Materials Fabrication: A
  New Trend in Nanotechnology of Soft Materials}. \emph{Curr. Opin. Colloid
  Interface Sci.} \textbf{2001}, \emph{6}, 11--16\relax
\mciteBstWouldAddEndPuncttrue
\mciteSetBstMidEndSepPunct{\mcitedefaultmidpunct}
{\mcitedefaultendpunct}{\mcitedefaultseppunct}\relax
\EndOfBibitem
\bibitem[Mei(2000)]{Meiwes-Broer_ClustersSurfaces}
\emph{{Metal Clusters at Surfaces: Structure, Quantum Properties, Physical
  Chemistry}}; Springer-Verlag, Berlin Heidelberg, 2000\relax
\mciteBstWouldAddEndPuncttrue
\mciteSetBstMidEndSepPunct{\mcitedefaultmidpunct}
{\mcitedefaultendpunct}{\mcitedefaultseppunct}\relax
\EndOfBibitem
\bibitem[Lando \latin{et~al.}(2007)Lando, K\'{e}ba\"{i}li, Cahuzac, Colliex,
  Couillard, Masson, Schmidt, and Br\'{e}chignac]{Lando_2007_EPJD.43.151}
Lando,~A.; K\'{e}ba\"{i}li,~N.; Cahuzac,~P.; Colliex,~C.; Couillard,~M.;
  Masson,~A.; Schmidt,~M.; Br\'{e}chignac,~C. {Chemically Induced Morphology
  Change in Cluster-Based Nanostructures}. \emph{Eur. Phys. J. D}
  \textbf{2007}, \emph{43}, 151--154\relax
\mciteBstWouldAddEndPuncttrue
\mciteSetBstMidEndSepPunct{\mcitedefaultmidpunct}
{\mcitedefaultendpunct}{\mcitedefaultseppunct}\relax
\EndOfBibitem
\bibitem[Br\'{e}chignac \latin{et~al.}(2003)Br\'{e}chignac, Cahuzac, Carlier,
  Colliex, {de Frutos}, K\'{e}ba\"{i}li, Roux, Masson, and
  Yoon]{Brechignac_2003_EPJD.24.265}
Br\'{e}chignac,~C.; Cahuzac,~P.; Carlier,~F.; Colliex,~C.; {de Frutos},~M.;
  K\'{e}ba\"{i}li,~N.; Roux,~J.~L.; Masson,~A.; Yoon,~B. {Thermal and Chemical
  Nanofractal Relaxation}. \emph{Eur. Phys. J. D} \textbf{2003}, \emph{24},
  265--268\relax
\mciteBstWouldAddEndPuncttrue
\mciteSetBstMidEndSepPunct{\mcitedefaultmidpunct}
{\mcitedefaultendpunct}{\mcitedefaultseppunct}\relax
\EndOfBibitem
\bibitem[Liu and Reinke(2006)Liu, and Reinke]{Liu_2006_JCP.124.164707}
Liu,~H.; Reinke,~P. {C$_{60}$ Thin Film Growth on Graphite: Coexistence of
  Spherical and Fractal-Dendritic Islands}. \emph{J. Chem. Phys.}
  \textbf{2006}, \emph{124}, 164707\relax
\mciteBstWouldAddEndPuncttrue
\mciteSetBstMidEndSepPunct{\mcitedefaultmidpunct}
{\mcitedefaultendpunct}{\mcitedefaultseppunct}\relax
\EndOfBibitem
\bibitem[{De Teresa}(2020)]{DeTeresa-book2020}
{De Teresa},~J.~M., Ed. \emph{{Nanofabrication: Nanolithography Techniques and
  Their Applications}}; IOP Publishing Ltd: Bristol, 2020\relax
\mciteBstWouldAddEndPuncttrue
\mciteSetBstMidEndSepPunct{\mcitedefaultmidpunct}
{\mcitedefaultendpunct}{\mcitedefaultseppunct}\relax
\EndOfBibitem
\bibitem[Utke \latin{et~al.}(2008)Utke, Hoffmann, and Melngailis]{Utke2008}
Utke,~I.; Hoffmann,~P.; Melngailis,~J. {Gas-Assisted Focused Electron Beam and
  Ion Beam Processing and Fabrication}. \emph{J. Vac. Sci. Technol. B}
  \textbf{2008}, \emph{26}, 1197--1276\relax
\mciteBstWouldAddEndPuncttrue
\mciteSetBstMidEndSepPunct{\mcitedefaultmidpunct}
{\mcitedefaultendpunct}{\mcitedefaultseppunct}\relax
\EndOfBibitem
\bibitem[Plant \latin{et~al.}(2014)Plant, Cao, and
  Palmer]{Plant_2014_JACS.136.7559}
Plant,~S.~R.; Cao,~L.; Palmer,~R.~E. {Atomic Structure Control of Size-Selected
  Gold Nanoclusters During Formation}. \emph{J. Am. Chem. Soc.} \textbf{2014},
  \emph{136}, 7559--7562\relax
\mciteBstWouldAddEndPuncttrue
\mciteSetBstMidEndSepPunct{\mcitedefaultmidpunct}
{\mcitedefaultendpunct}{\mcitedefaultseppunct}\relax
\EndOfBibitem
\bibitem[Huth \latin{et~al.}(2012)Huth, Porrati, Schwalb, Winhold, Sachser,
  Dukic, Adams, and Fantner]{Huth_2012_BJN.3.597}
Huth,~M.; Porrati,~F.; Schwalb,~C.; Winhold,~M.; Sachser,~R.; Dukic,~M.;
  Adams,~J.; Fantner,~G. {Focused Electron Beam Induced Deposition: A
  Perspective}. \emph{Beilstein J. Nanotechnol.} \textbf{2012}, \emph{3},
  597--619\relax
\mciteBstWouldAddEndPuncttrue
\mciteSetBstMidEndSepPunct{\mcitedefaultmidpunct}
{\mcitedefaultendpunct}{\mcitedefaultseppunct}\relax
\EndOfBibitem
\bibitem[Utke \latin{et~al.}(2012)Utke, Moshkalev, and Russell]{Utke_book_2012}
Utke,~I., Moshkalev,~S., Russell,~P., Eds. \emph{{Nanofabrication Using Focused
  Ion and Electron Beams}}; Oxford University Press: New York, 2012\relax
\mciteBstWouldAddEndPuncttrue
\mciteSetBstMidEndSepPunct{\mcitedefaultmidpunct}
{\mcitedefaultendpunct}{\mcitedefaultseppunct}\relax
\EndOfBibitem
\bibitem[Cui(2017)]{Cui_Nanofabrication_book}
Cui,~Z. \emph{{Nanofabrication. Principles, Capabilities and Limits}}; Springer
  International Publishing: Cham, Switzerland, 2017\relax
\mciteBstWouldAddEndPuncttrue
\mciteSetBstMidEndSepPunct{\mcitedefaultmidpunct}
{\mcitedefaultendpunct}{\mcitedefaultseppunct}\relax
\EndOfBibitem
\bibitem[Xu \latin{et~al.}(2008)Xu, Kong, Yeh, and
  Chen]{Xu_2008_NatMater.7.992}
Xu,~W.; Kong,~J.~S.; Yeh,~Y. T.~E.; Chen,~P. {Single-Molecule Nanocatalysis
  Reveals Heterogeneous Reaction Pathways and Catalytic Dynamics}. \emph{Nat.
  Mater.} \textbf{2008}, \emph{7}, 992--996\relax
\mciteBstWouldAddEndPuncttrue
\mciteSetBstMidEndSepPunct{\mcitedefaultmidpunct}
{\mcitedefaultendpunct}{\mcitedefaultseppunct}\relax
\EndOfBibitem
\bibitem[Murray(2008)]{Murray_2008_ChemRev.108.2688}
Murray,~R. {Nanoelectrochemistry: Metal Nanoparticles, Nanoelectrodes, and
  Nanopores}. \emph{Chem. Rev.} \textbf{2008}, \emph{108}, 2688--2720\relax
\mciteBstWouldAddEndPuncttrue
\mciteSetBstMidEndSepPunct{\mcitedefaultmidpunct}
{\mcitedefaultendpunct}{\mcitedefaultseppunct}\relax
\EndOfBibitem
\bibitem[Barth \latin{et~al.}(2020)Barth, Huth, and
  Jungwirth]{Barth2020_JMaterChemC}
Barth,~S.; Huth,~M.; Jungwirth,~F. {Precursors for Direct-Write Nanofabrication
  with Electrons}. \emph{J. Mater. Chem. C} \textbf{2020}, \emph{8},
  15884--15919\relax
\mciteBstWouldAddEndPuncttrue
\mciteSetBstMidEndSepPunct{\mcitedefaultmidpunct}
{\mcitedefaultendpunct}{\mcitedefaultseppunct}\relax
\EndOfBibitem
\bibitem[Prosvetov \latin{et~al.}(2021)Prosvetov, Verkhovtsev, Sushko, and
  Solov'yov]{Prosvetov2021_BJN}
Prosvetov,~A.; Verkhovtsev,~A.~V.; Sushko,~G.; Solov'yov,~A.~V.
  {Irradiation-Driven Molecular Dynamics Simulation of the FEBID Process for
  Pt(PF$_3$)$_4$}. \emph{Beilstein J. Nanotechnol.} \textbf{2021}, \emph{12},
  1151--1172\relax
\mciteBstWouldAddEndPuncttrue
\mciteSetBstMidEndSepPunct{\mcitedefaultmidpunct}
{\mcitedefaultendpunct}{\mcitedefaultseppunct}\relax
\EndOfBibitem
\bibitem[Prosvetov \latin{et~al.}(2022)Prosvetov, Verkhovtsev, Sushko, and
  Solov'yov]{Prosvetov2022_PCCP}
Prosvetov,~A.; Verkhovtsev,~A.~V.; Sushko,~G.; Solov'yov,~A.~V. {Atomistic
  Simulation of the FEBID-Driven Growth of Iron-Based Nanostructures}.
  \emph{Phys. Chem. Chem. Phys.} \textbf{2022}, \emph{24}, 10807--10819\relax
\mciteBstWouldAddEndPuncttrue
\mciteSetBstMidEndSepPunct{\mcitedefaultmidpunct}
{\mcitedefaultendpunct}{\mcitedefaultseppunct}\relax
\EndOfBibitem
\bibitem[Ragesh~Kumar \latin{et~al.}(2018)Ragesh~Kumar, Weirich, Hrachowina,
  Hanefeld, Bjornsson, Hrodmarsson, Barth, Fairbrother, Huth, and
  Ing\'{o}lfsson]{Kumar2018}
Ragesh~Kumar,~T.~P.; Weirich,~P.; Hrachowina,~L.; Hanefeld,~M.; Bjornsson,~R.;
  Hrodmarsson,~H.~R.; Barth,~S.; Fairbrother,~D.~H.; Huth,~M.;
  Ing\'{o}lfsson,~O. {Electron Interactions with the Heteronuclear Carbonyl
  Precursor H$_2$FeRu$_3$(CO)$_{13}$ and Comparison with HFeCo$_3$(CO)$_{12}$:
  From Fundamental Gas Phase and Surface Science Studies to Focused Electron
  Beam Induced Deposition}. \emph{Beilstein J. Nanotechnol.} \textbf{2018},
  \emph{9}, 555--579\relax
\mciteBstWouldAddEndPuncttrue
\mciteSetBstMidEndSepPunct{\mcitedefaultmidpunct}
{\mcitedefaultendpunct}{\mcitedefaultseppunct}\relax
\EndOfBibitem
\bibitem[Wysocki \latin{et~al.}(1987)Wysocki, Kentt\"amaa, and
  Cooks]{Wysocki_1987_IJMSIP.75.181}
Wysocki,~V.~H.; Kentt\"amaa,~H.~I.; Cooks,~R.~G. {Internal Energy Distributions
  of Isolated Ions After Activation by Various Methods}. \emph{Int. J. Mass
  Spectrom. Ion Proc.} \textbf{1987}, \emph{75}, 181--208\relax
\mciteBstWouldAddEndPuncttrue
\mciteSetBstMidEndSepPunct{\mcitedefaultmidpunct}
{\mcitedefaultendpunct}{\mcitedefaultseppunct}\relax
\EndOfBibitem
\bibitem[Beranov\'{a} and Wesdemiotis(1994)Beranov\'{a}, and
  Wesdemiotis]{Beranova_1994_JAMS.5.1093}
Beranov\'{a},~S.; Wesdemiotis,~C. {Internal Energy Distributions of Tungsten
  Hexacarbonyl Ions After Neutralization--Reionization}. \emph{J. Am. Soc. Mass
  Spectrom.} \textbf{1994}, \emph{5}, 1093--1101\relax
\mciteBstWouldAddEndPuncttrue
\mciteSetBstMidEndSepPunct{\mcitedefaultmidpunct}
{\mcitedefaultendpunct}{\mcitedefaultseppunct}\relax
\EndOfBibitem
\bibitem[Cooks \latin{et~al.}(1990)Cooks, Ast, Kralj, Kramer, and
  \v{Z}igon]{Cooks_1990_JASMS.1.16}
Cooks,~R.~G.; Ast,~T.; Kralj,~B.; Kramer,~V.; \v{Z}igon,~D. {Internal energy
  distributions deposited in doubly and singly charged tungsten hexacarbonyl
  ions generated by charge stripping, electron impact, and charge exchange}.
  \emph{J. Am. Soc. Mass Spectrom.} \textbf{1990}, \emph{1}, 16--27\relax
\mciteBstWouldAddEndPuncttrue
\mciteSetBstMidEndSepPunct{\mcitedefaultmidpunct}
{\mcitedefaultendpunct}{\mcitedefaultseppunct}\relax
\EndOfBibitem
\bibitem[Wnorowski \latin{et~al.}(2012)Wnorowski, Stano, Barszczewska,
  J\'{o}wko, and Matej\v{c}\'{i}k]{Wnorowski_2012_IJMS.314.42}
Wnorowski,~K.; Stano,~M.; Barszczewska,~W.; J\'{o}wko,~A.;
  Matej\v{c}\'{i}k,~{\v{S}}. {Electron Ionization of W(CO)$_6$: Appearance
  Energies}. \emph{Int. J. Mass Spectrom.} \textbf{2012}, \emph{314},
  42--48\relax
\mciteBstWouldAddEndPuncttrue
\mciteSetBstMidEndSepPunct{\mcitedefaultmidpunct}
{\mcitedefaultendpunct}{\mcitedefaultseppunct}\relax
\EndOfBibitem
\bibitem[Wnorowski \latin{et~al.}(2012)Wnorowski, Stano, Matias, Denifl,
  Barszczewska, and Matej\v{c}\'{i}k]{Wnorowski_2012_RCMS.26.2093}
Wnorowski,~K.; Stano,~M.; Matias,~C.; Denifl,~S.; Barszczewska,~W.;
  Matej\v{c}\'{i}k,~{\v{S}}. {Low-Energy Electron Interactions with Tungsten
  Hexacarbonyl -- W(CO)$_6$}. \emph{Rapid Commun. Mass Spectrom.}
  \textbf{2012}, \emph{26}, 2093--2098\relax
\mciteBstWouldAddEndPuncttrue
\mciteSetBstMidEndSepPunct{\mcitedefaultmidpunct}
{\mcitedefaultendpunct}{\mcitedefaultseppunct}\relax
\EndOfBibitem
\bibitem[Neustetter \latin{et~al.}(2016)Neustetter, Jabbour Al~Maalouf,
  Lim\~{a}o Vieira, and Denifl]{Neustetter_2016_JCP.145.054301}
Neustetter,~M.; Jabbour Al~Maalouf,~E.; Lim\~{a}o Vieira,~P.; Denifl,~S.
  {Fragmentation Pathways of Tungsten Hexacarbonyl Clusters Upon Electron
  Ionization}. \emph{J. Chem. Phys.} \textbf{2016}, \emph{145}, 054301\relax
\mciteBstWouldAddEndPuncttrue
\mciteSetBstMidEndSepPunct{\mcitedefaultmidpunct}
{\mcitedefaultendpunct}{\mcitedefaultseppunct}\relax
\EndOfBibitem
\bibitem[Lacko \latin{et~al.}(2015)Lacko, Papp, Wnorowski, and
  Matej\v{c}\'{i}k]{Lacko_2015_EPJD.69.84}
Lacko,~M.; Papp,~P.; Wnorowski,~K.; Matej\v{c}\'{i}k,~{\v{S}}.
  {Electron-Induced Ionization and Dissociative Ionization of Iron
  Pentacarbonyl Molecules}. \emph{Eur. Phys. J. D} \textbf{2015}, \emph{69},
  84\relax
\mciteBstWouldAddEndPuncttrue
\mciteSetBstMidEndSepPunct{\mcitedefaultmidpunct}
{\mcitedefaultendpunct}{\mcitedefaultseppunct}\relax
\EndOfBibitem
\bibitem[Lengyel \latin{et~al.}(2016)Lengyel, Fedor, and
  F{\'{a}}rn\'{i}k]{Lengyel2016_JPCC_2}
Lengyel,~J.; Fedor,~J.; F{\'{a}}rn\'{i}k,~M. {Ligand Stabilization and Charge
  Transfer in Dissociative Ionization of Fe(CO)$_5$ Aggregates}. \emph{J. Phys.
  Chem. C} \textbf{2016}, \emph{120}, 17810--17816\relax
\mciteBstWouldAddEndPuncttrue
\mciteSetBstMidEndSepPunct{\mcitedefaultmidpunct}
{\mcitedefaultendpunct}{\mcitedefaultseppunct}\relax
\EndOfBibitem
\bibitem[Lengyel \latin{et~al.}(2021)Lengyel, Pysanenko, Swiderek, Heiz,
  F{\'{a}}rn\'{i}k, and Fedor]{Lengyel2021}
Lengyel,~J.; Pysanenko,~A.; Swiderek,~P.; Heiz,~U.; F{\'{a}}rn\'{i}k,~M.;
  Fedor,~J. {Water-Assisted Electron-Induced Chemistry of the Nanofabrication
  Precursor Iron Pentacarbonyl}. \emph{J. Phys. Chem. A} \textbf{2021},
  \emph{125}, 1919--1926\relax
\mciteBstWouldAddEndPuncttrue
\mciteSetBstMidEndSepPunct{\mcitedefaultmidpunct}
{\mcitedefaultendpunct}{\mcitedefaultseppunct}\relax
\EndOfBibitem
\bibitem[Massey \latin{et~al.}(2015)Massey, Bass, and
  Sanche]{Massey_Sanche2015}
Massey,~S.; Bass,~A.~D.; Sanche,~L. {Role of Low-Energy Electrons ($<35$~eV) in
  the Degradation of Fe(CO)$_5$ for Focused Electron Beam Induced Deposition
  Applications: Study by Electron Stimulated Desorption of Negative and
  Positive Ions}. \emph{J. Phys. Chem. C} \textbf{2015}, \emph{119},
  12708--12719\relax
\mciteBstWouldAddEndPuncttrue
\mciteSetBstMidEndSepPunct{\mcitedefaultmidpunct}
{\mcitedefaultendpunct}{\mcitedefaultseppunct}\relax
\EndOfBibitem
\bibitem[Bilgilisoy \latin{et~al.}(2021)Bilgilisoy, Thorman, Barclay, Marbach,
  and Fairbrother]{Bilgilisoy_Fairbrother2021}
Bilgilisoy,~E.; Thorman,~R.~M.; Barclay,~M.~S.; Marbach,~H.; Fairbrother,~D.~H.
  {Low energy electron- and ion-induced surface reactions of Fe(CO)$_5$ thin
  films}. \emph{J. Phys. Chem. C} \textbf{2021}, \emph{125}, 17749--17760\relax
\mciteBstWouldAddEndPuncttrue
\mciteSetBstMidEndSepPunct{\mcitedefaultmidpunct}
{\mcitedefaultendpunct}{\mcitedefaultseppunct}\relax
\EndOfBibitem
\bibitem[Sushko \latin{et~al.}(2016)Sushko, Solov'yov, Verkhovtsev, Volkov, and
  Solov'yov]{Sushko2016_rCHARMM}
Sushko,~G.~B.; Solov'yov,~I.~A.; Verkhovtsev,~A.~V.; Volkov,~S.~N.;
  Solov'yov,~A.~V. {Studying Chemical Reactions in Biological Systems with MBN
  Explorer: Implementation of Molecular Mechanics with Dynamical Topology}.
  \emph{Eur. Phys. J. D} \textbf{2016}, \emph{70}, 12\relax
\mciteBstWouldAddEndPuncttrue
\mciteSetBstMidEndSepPunct{\mcitedefaultmidpunct}
{\mcitedefaultendpunct}{\mcitedefaultseppunct}\relax
\EndOfBibitem
\bibitem[Solov'yov \latin{et~al.}(2012)Solov'yov, Yakubovich, Nikolaev,
  Volkovets, and Solov'yov]{Solovyov_2012_JCC_MBNExplorer}
Solov'yov,~I.~A.; Yakubovich,~A.~V.; Nikolaev,~P.~V.; Volkovets,~I.;
  Solov'yov,~A.~V. {MesoBioNano Explorer -- A Universal Program for Multiscale
  Computer Simulations of Complex Molecular Structure and Dynamics}. \emph{J.
  Comput. Chem.} \textbf{2012}, \emph{33}, 2412--2439\relax
\mciteBstWouldAddEndPuncttrue
\mciteSetBstMidEndSepPunct{\mcitedefaultmidpunct}
{\mcitedefaultendpunct}{\mcitedefaultseppunct}\relax
\EndOfBibitem
\bibitem[Fowlkes and Rack(2010)Fowlkes, and Rack]{Fowlkes2010}
Fowlkes,~J.~D.; Rack,~P.~D. {Fundamental Electron-Precursor-Solid Interactions
  Derived From Time-Dependent Electron-Beam-Induced Deposition Simulations and
  Experiments}. \emph{ACS Nano} \textbf{2010}, \emph{4}, 1619--1629\relax
\mciteBstWouldAddEndPuncttrue
\mciteSetBstMidEndSepPunct{\mcitedefaultmidpunct}
{\mcitedefaultendpunct}{\mcitedefaultseppunct}\relax
\EndOfBibitem
\bibitem[Solov'yov \latin{et~al.}(2009)Solov'yov, Surdutovich, Scifoni,
  Mishustin, and Greiner]{solov2009physics}
Solov'yov,~A.~V.; Surdutovich,~E.; Scifoni,~E.; Mishustin,~I.; Greiner,~W.
  {Physics of Ion Beam Cancer Therapy: A Multiscale Approach}. \emph{Phys. Rev.
  E} \textbf{2009}, \emph{79}, 011909\relax
\mciteBstWouldAddEndPuncttrue
\mciteSetBstMidEndSepPunct{\mcitedefaultmidpunct}
{\mcitedefaultendpunct}{\mcitedefaultseppunct}\relax
\EndOfBibitem
\bibitem[Surdutovich \latin{et~al.}(2009)Surdutovich, Obolensky, Scifoni,
  Pshenichnov, Mishustin, Solov'yov, and Greiner]{Surdutovich_2009_EPJD.51.63}
Surdutovich,~E.; Obolensky,~O.~I.; Scifoni,~E.; Pshenichnov,~I.; Mishustin,~I.;
  Solov'yov,~A.~V.; Greiner,~W. {Ion-Induced Electron Production in Tissue-Like
  Media and DNA Damage Mechanisms}. \emph{Eur. Phys. J. D} \textbf{2009},
  \emph{51}, 63--71\relax
\mciteBstWouldAddEndPuncttrue
\mciteSetBstMidEndSepPunct{\mcitedefaultmidpunct}
{\mcitedefaultendpunct}{\mcitedefaultseppunct}\relax
\EndOfBibitem
\bibitem[Scifoni \latin{et~al.}(2010)Scifoni, Surdutovich, and
  Solov'yov]{Scifoni_2010_PRE.81.021903}
Scifoni,~E.; Surdutovich,~E.; Solov'yov,~A.~V. {Spectra of Secondary Electrons
  Generated in Water by Energetic Ions}. \emph{Phys. Rev. E} \textbf{2010},
  \emph{81}, 021903\relax
\mciteBstWouldAddEndPuncttrue
\mciteSetBstMidEndSepPunct{\mcitedefaultmidpunct}
{\mcitedefaultendpunct}{\mcitedefaultseppunct}\relax
\EndOfBibitem
\bibitem[Pshenichnov \latin{et~al.}(2008)Pshenichnov, Mishustin, and
  Greiner]{Pshenichnov_2008_NIMB.266.1094}
Pshenichnov,~I.; Mishustin,~I.; Greiner,~W. {Comparative Study of Depth--Dose
  Distributions for Beams of Light and Heavy Nuclei in Tissue-Like Media}.
  \emph{Nucl. Instrum. Meth. B} \textbf{2008}, \emph{266}, 1094--1098\relax
\mciteBstWouldAddEndPuncttrue
\mciteSetBstMidEndSepPunct{\mcitedefaultmidpunct}
{\mcitedefaultendpunct}{\mcitedefaultseppunct}\relax
\EndOfBibitem
\bibitem[de~Vera \latin{et~al.}(2013)de~Vera, Garcia-Molina, Abril, and
  Solov'yov]{deVera_2013_PRL.110.148104}
de~Vera,~P.; Garcia-Molina,~R.; Abril,~I.; Solov'yov,~A.~V. {Semiempirical
  Model for the Ion Impact Ionization of Complex Biological Media}. \emph{Phys.
  Rev. Lett.} \textbf{2013}, \emph{110}, 148104\relax
\mciteBstWouldAddEndPuncttrue
\mciteSetBstMidEndSepPunct{\mcitedefaultmidpunct}
{\mcitedefaultendpunct}{\mcitedefaultseppunct}\relax
\EndOfBibitem
\bibitem[Nikjoo \latin{et~al.}(2006)Nikjoo, Uehara, Emfietzoglou, and
  Cucinotta]{Nikjoo_2006_RadiatMeas.41.1052}
Nikjoo,~H.; Uehara,~S.; Emfietzoglou,~D.; Cucinotta,~F.~A. {Track-Structure
  Codes in Radiation Research}. \emph{Radiat. Meas.} \textbf{2006}, \emph{41},
  1052--1074\relax
\mciteBstWouldAddEndPuncttrue
\mciteSetBstMidEndSepPunct{\mcitedefaultmidpunct}
{\mcitedefaultendpunct}{\mcitedefaultseppunct}\relax
\EndOfBibitem
\bibitem[Surdutovich and Solov'yov(2012)Surdutovich, and
  Solov'yov]{ES_AVS_2012_EPJD.66.206}
Surdutovich,~E.; Solov'yov,~A.~V. {Double Strand Breaks in DNA Resulting from
  Double Ionization Events}. \emph{Eur. Phys. J. D} \textbf{2012}, \emph{66},
  206\relax
\mciteBstWouldAddEndPuncttrue
\mciteSetBstMidEndSepPunct{\mcitedefaultmidpunct}
{\mcitedefaultendpunct}{\mcitedefaultseppunct}\relax
\EndOfBibitem
\bibitem[Surdutovich and Solov'yov(2015)Surdutovich, and
  Solov'yov]{ES_AVS_2015_EPJD.69.193}
Surdutovich,~E.; Solov'yov,~A.~V. {Transport of Secondary Electrons and
  Reactive Species in Ion Tracks}. \emph{Eur. Phys. J. D} \textbf{2015},
  \emph{69}, 193\relax
\mciteBstWouldAddEndPuncttrue
\mciteSetBstMidEndSepPunct{\mcitedefaultmidpunct}
{\mcitedefaultendpunct}{\mcitedefaultseppunct}\relax
\EndOfBibitem
\bibitem[Bug \latin{et~al.}(2012)Bug, Surdutovich, Rabus, Rosenfeld, and
  Solov'yov]{Bug_2012_EPJD.66.291}
Bug,~M.~U.; Surdutovich,~E.; Rabus,~H.; Rosenfeld,~A.~B.; Solov'yov,~A.~V.
  {Nanoscale Characterization of Ion Tracks: MC Simulations Versus Analytical
  Approach}. \emph{Eur. Phys. J. D} \textbf{2012}, \emph{66}, 291\relax
\mciteBstWouldAddEndPuncttrue
\mciteSetBstMidEndSepPunct{\mcitedefaultmidpunct}
{\mcitedefaultendpunct}{\mcitedefaultseppunct}\relax
\EndOfBibitem
\bibitem[Surdutovich and Solov'yov(2010)Surdutovich, and
  Solov'yov]{surdutovich2010shock}
Surdutovich,~E.; Solov'yov,~A.~V. {Shock Wave Initiated by an Ion Passing
  Through Liquid Water}. \emph{Phys. Rev. E} \textbf{2010}, \emph{82},
  051915\relax
\mciteBstWouldAddEndPuncttrue
\mciteSetBstMidEndSepPunct{\mcitedefaultmidpunct}
{\mcitedefaultendpunct}{\mcitedefaultseppunct}\relax
\EndOfBibitem
\bibitem[Surdutovich \latin{et~al.}(2013)Surdutovich, Yakubovich, and
  Solov'yov]{surdutovich2013biodamage}
Surdutovich,~E.; Yakubovich,~A.~V.; Solov'yov,~A.~V. {Biodamage via Shock Waves
  Initiated by Irradiation with Ions}. \emph{Sci. Rep.} \textbf{2013},
  \emph{3}, 1289\relax
\mciteBstWouldAddEndPuncttrue
\mciteSetBstMidEndSepPunct{\mcitedefaultmidpunct}
{\mcitedefaultendpunct}{\mcitedefaultseppunct}\relax
\EndOfBibitem
\bibitem[Yakubovich \latin{et~al.}(2012)Yakubovich, Surdutovich, and
  Solov'yov]{Yakubovich_2012_NIMB.279.135}
Yakubovich,~A.~V.; Surdutovich,~E.; Solov'yov,~A.~V. {Thermomechanical Damage
  of Nucleosome by the Shock Wave Initiated by Ion Passing Through Liquid
  Water}. \emph{Nucl. Instrum. Meth. B} \textbf{2012}, \emph{279},
  135--139\relax
\mciteBstWouldAddEndPuncttrue
\mciteSetBstMidEndSepPunct{\mcitedefaultmidpunct}
{\mcitedefaultendpunct}{\mcitedefaultseppunct}\relax
\EndOfBibitem
\bibitem[Yakubovich \latin{et~al.}(2011)Yakubovich, Surdutovich, and
  Solov'yov]{Yakubovich_2011_AIP.1344.230}
Yakubovich,~A.~V.; Surdutovich,~E.; Solov'yov,~A.~V. {Atomic and Molecular Data
  Needs for Radiation Damage Modeling: Multiscale Approach}. \emph{AIP Conf.
  Proc.} \textbf{2011}, \emph{1344}, 230--238\relax
\mciteBstWouldAddEndPuncttrue
\mciteSetBstMidEndSepPunct{\mcitedefaultmidpunct}
{\mcitedefaultendpunct}{\mcitedefaultseppunct}\relax
\EndOfBibitem
\bibitem[de~Vera \latin{et~al.}(2016)de~Vera, Mason, Currell, and
  Solov'yov]{deVera2016molecular}
de~Vera,~P.; Mason,~N.~J.; Currell,~F.~J.; Solov'yov,~A.~V. {Molecular Dynamics
  Study of Accelerated Ion-Induced Shock Waves in Biological Media}. \emph{Eur.
  Phys. J. D} \textbf{2016}, \emph{70}, 183\relax
\mciteBstWouldAddEndPuncttrue
\mciteSetBstMidEndSepPunct{\mcitedefaultmidpunct}
{\mcitedefaultendpunct}{\mcitedefaultseppunct}\relax
\EndOfBibitem
\bibitem[de~Vera \latin{et~al.}(2017)de~Vera, Surdutovich, Mason, and
  Solov'yov]{deVera2017radial}
de~Vera,~P.; Surdutovich,~E.; Mason,~N.~J.; Solov'yov,~A.~V. {Radial Doses
  Around Energetic Ion Tracks and the Onset of Shock Waves on the Nanoscale}.
  \emph{Eur. Phys. J. D} \textbf{2017}, \emph{71}, 281\relax
\mciteBstWouldAddEndPuncttrue
\mciteSetBstMidEndSepPunct{\mcitedefaultmidpunct}
{\mcitedefaultendpunct}{\mcitedefaultseppunct}\relax
\EndOfBibitem
\bibitem[de~Vera \latin{et~al.}(2018)de~Vera, Surdutovich, Mason, Currell, and
  Solov'yov]{deVera_2018_EPJD.72.147}
de~Vera,~P.; Surdutovich,~E.; Mason,~N.~J.; Currell,~F.~J.; Solov'yov,~A.~V.
  {Simulation of the Ion-Induced Shock Waves Effects on the Transport of
  Chemically Reactive Species in Ion Tracks}. \emph{Eur. Phys. J. D}
  \textbf{2018}, \emph{72}, 147\relax
\mciteBstWouldAddEndPuncttrue
\mciteSetBstMidEndSepPunct{\mcitedefaultmidpunct}
{\mcitedefaultendpunct}{\mcitedefaultseppunct}\relax
\EndOfBibitem
\bibitem[Friis \latin{et~al.}(2020)Friis, Verkhovtsev, Solov'yov, and
  Solov'yov]{Friis_2020_JCC}
Friis,~I.; Verkhovtsev,~A.; Solov'yov,~I.~A.; Solov'yov,~A.~V. {Modeling the
  Effect of Ion-Induced Shock Waves and DNA Breakage with the Reactive CHARMM
  Force Field}. \emph{J. Comput. Chem.} \textbf{2020}, \emph{41},
  2429--2439\relax
\mciteBstWouldAddEndPuncttrue
\mciteSetBstMidEndSepPunct{\mcitedefaultmidpunct}
{\mcitedefaultendpunct}{\mcitedefaultseppunct}\relax
\EndOfBibitem
\bibitem[Friis \latin{et~al.}(2021)Friis, Verkhovtsev, Solov'yov, and
  Solov'yov]{Friis_2021_PRE}
Friis,~I.; Verkhovtsev,~A.; Solov'yov,~I.~A.; Solov'yov,~A.~V. {Lethal DNA
  Damage Caused by Ion-Induced Shock Waves in Cells}. \emph{Phys. Rev. E}
  \textbf{2021}, \emph{104}, 054408\relax
\mciteBstWouldAddEndPuncttrue
\mciteSetBstMidEndSepPunct{\mcitedefaultmidpunct}
{\mcitedefaultendpunct}{\mcitedefaultseppunct}\relax
\EndOfBibitem
\bibitem[Ward(1988)]{Ward_1988_ProgNuclAcid.35.95}
Ward,~J.~F. {DNA Damage Produced by Ionizing Radiation in Mammalian Cells:
  Identities, Mechanisms of Formation and Repairability}. \emph{Prog. Nucleic
  Acid Res. Mol. Biol.} \textbf{1988}, \emph{35}, 95--125\relax
\mciteBstWouldAddEndPuncttrue
\mciteSetBstMidEndSepPunct{\mcitedefaultmidpunct}
{\mcitedefaultendpunct}{\mcitedefaultseppunct}\relax
\EndOfBibitem
\bibitem[Ward(1995)]{Ward_1995_RadiatRes.142.362}
Ward,~J.~F. {Radiation Mutagenesis: The Initial DNA Lesions Responsible}.
  \emph{Radiat. Res.} \textbf{1995}, \emph{142}, 362--368\relax
\mciteBstWouldAddEndPuncttrue
\mciteSetBstMidEndSepPunct{\mcitedefaultmidpunct}
{\mcitedefaultendpunct}{\mcitedefaultseppunct}\relax
\EndOfBibitem
\bibitem[Malyarchuk \latin{et~al.}(2008)Malyarchuk, Castore, and
  Harrison]{Malyarchuk_2008_NuclAcidsRes.36.4872}
Malyarchuk,~S.; Castore,~R.; Harrison,~L. {DNA Repair of Clustered Lesions in
  Mammalian Cells: Involvement of Non-Homologous End-Joining}. \emph{Nucleic
  Acids Res.} \textbf{2008}, \emph{36}, 4872--4882\relax
\mciteBstWouldAddEndPuncttrue
\mciteSetBstMidEndSepPunct{\mcitedefaultmidpunct}
{\mcitedefaultendpunct}{\mcitedefaultseppunct}\relax
\EndOfBibitem
\bibitem[Malyarchuk \latin{et~al.}(2009)Malyarchuk, Castore, and
  Harrison]{Malyarchuk_2009_DNARepair.8.1343}
Malyarchuk,~S.; Castore,~R.; Harrison,~L. {Apex1 Can Cleave Complex Clustered
  DNA Lesions in Cells}. \emph{DNA Repair} \textbf{2009}, \emph{8},
  1343--1354\relax
\mciteBstWouldAddEndPuncttrue
\mciteSetBstMidEndSepPunct{\mcitedefaultmidpunct}
{\mcitedefaultendpunct}{\mcitedefaultseppunct}\relax
\EndOfBibitem
\bibitem[Sage and Harrison(2011)Sage, and Harrison]{Sage_2011_MutatRes.711.123}
Sage,~E.; Harrison,~L. {Clustered DNA Lesion Repair in Eukaryotes: Relevance to
  Mutagenesis and Cell Survival}. \emph{Mutat. Res.} \textbf{2011}, \emph{711},
  123--133\relax
\mciteBstWouldAddEndPuncttrue
\mciteSetBstMidEndSepPunct{\mcitedefaultmidpunct}
{\mcitedefaultendpunct}{\mcitedefaultseppunct}\relax
\EndOfBibitem
\bibitem[Verkhovtsev \latin{et~al.}(2016)Verkhovtsev, Surdutovich, and
  Solov'yov]{verkhovtsev2016multiscale}
Verkhovtsev,~A.; Surdutovich,~E.; Solov'yov,~A.~V. {Multiscale Approach
  Predictions for Biological Outcomes in Ion-Beam Cancer Therapy}. \emph{Sci.
  Rep.} \textbf{2016}, \emph{6}, 27654\relax
\mciteBstWouldAddEndPuncttrue
\mciteSetBstMidEndSepPunct{\mcitedefaultmidpunct}
{\mcitedefaultendpunct}{\mcitedefaultseppunct}\relax
\EndOfBibitem
\bibitem[Linz(2012)]{IBT_book_Linz}
Linz,~U., Ed. \emph{{Ion Beam Therapy: Fundamentals, Technology, Clinical
  Applications}}; Springer-Verlag: Berlin-Heidelberg, 2012\relax
\mciteBstWouldAddEndPuncttrue
\mciteSetBstMidEndSepPunct{\mcitedefaultmidpunct}
{\mcitedefaultendpunct}{\mcitedefaultseppunct}\relax
\EndOfBibitem
\bibitem[Taylor(2006)]{Taylor_ScatTheory_book}
Taylor,~J.~R. \emph{{Scattering Theory: The Quantum Theory of Nonrelativistic
  Collisions}}; Dover Publications Inc., 2006\relax
\mciteBstWouldAddEndPuncttrue
\mciteSetBstMidEndSepPunct{\mcitedefaultmidpunct}
{\mcitedefaultendpunct}{\mcitedefaultseppunct}\relax
\EndOfBibitem
\bibitem[Newton(1982)]{Newton_ScatTheory_book}
Newton,~R.~G. \emph{{Scattering Theory of Waves and Particles}}, 2nd ed.;
  Springer Berlin, Heidelberg, 1982\relax
\mciteBstWouldAddEndPuncttrue
\mciteSetBstMidEndSepPunct{\mcitedefaultmidpunct}
{\mcitedefaultendpunct}{\mcitedefaultseppunct}\relax
\EndOfBibitem
\bibitem[Sushko \latin{et~al.}(2013)Sushko, Bezchastnov, Solov'yov, Korol,
  Greiner, and Solov'yov]{RelMD_2013_JCompPhys.252.404}
Sushko,~G.~B.; Bezchastnov,~V.~G.; Solov'yov,~I.~A.; Korol,~A.~V.; Greiner,~W.;
  Solov'yov,~A.~V. {Simulation of Ultra-Relativistic Electrons and Positrons
  Channeling in Crystals with MBN Explorer}. \emph{J. Comput. Phys.}
  \textbf{2013}, \emph{252}, 404--418\relax
\mciteBstWouldAddEndPuncttrue
\mciteSetBstMidEndSepPunct{\mcitedefaultmidpunct}
{\mcitedefaultendpunct}{\mcitedefaultseppunct}\relax
\EndOfBibitem
\bibitem[Froese~Fischer(1977)]{FroeseFischer_HF_book}
Froese~Fischer,~C. \emph{{The Hartree-Fock Method for Atoms: A Numerical
  Approach}}; John Wiley \& Sons Inc., 1977\relax
\mciteBstWouldAddEndPuncttrue
\mciteSetBstMidEndSepPunct{\mcitedefaultmidpunct}
{\mcitedefaultendpunct}{\mcitedefaultseppunct}\relax
\EndOfBibitem
\bibitem[Kohn and Sham(1965)Kohn, and Sham]{Kohn_Sham_1965}
Kohn,~W.; Sham,~L.~J. {Self-Consistent Equations Including Exchange and
  Correlation Effects}. \emph{Phys. Rev.} \textbf{1965}, \emph{140},
  A1133--A1138\relax
\mciteBstWouldAddEndPuncttrue
\mciteSetBstMidEndSepPunct{\mcitedefaultmidpunct}
{\mcitedefaultendpunct}{\mcitedefaultseppunct}\relax
\EndOfBibitem
\bibitem[Parr and Yang(1989)Parr, and Yang]{Parr_Yang_DFT_book}
Parr,~R.~G.; Yang,~W. \emph{{Density-Functional Theory of Atoms and
  Molecules}}; Oxford University Press, New York, 1989\relax
\mciteBstWouldAddEndPuncttrue
\mciteSetBstMidEndSepPunct{\mcitedefaultmidpunct}
{\mcitedefaultendpunct}{\mcitedefaultseppunct}\relax
\EndOfBibitem
\bibitem[Ekardt(1999)]{Ekardt_MetalClusters}
Ekardt,~W., Ed. \emph{{Metal Clusters}}; Wiley, 1999\relax
\mciteBstWouldAddEndPuncttrue
\mciteSetBstMidEndSepPunct{\mcitedefaultmidpunct}
{\mcitedefaultendpunct}{\mcitedefaultseppunct}\relax
\EndOfBibitem
\bibitem[Connerade and Solov'yov(2004)Connerade, and
  Solov'yov]{LatestAdv_ISAAC_2004}
Connerade,~J.-P., Solov'yov,~A.~V., Eds. \emph{{Latest Advances in Atomic
  Clusters Collision: Fission, Fusion, Electron, Ion and Photon Impact}};
  Imperial College Press, London, 2004\relax
\mciteBstWouldAddEndPuncttrue
\mciteSetBstMidEndSepPunct{\mcitedefaultmidpunct}
{\mcitedefaultendpunct}{\mcitedefaultseppunct}\relax
\EndOfBibitem
\bibitem[Mariscal \latin{et~al.}(2013)Mariscal, Oviedo, and
  Leiva]{Mariscal_2013_book}
Mariscal,~M.~M., Oviedo,~O.~A., Leiva,~E. P.~M., Eds. \emph{{Metal Clusters and
  Nanoalloys: From Modeling to Applications}}; Springer Science+Business Media,
  New York, 2013\relax
\mciteBstWouldAddEndPuncttrue
\mciteSetBstMidEndSepPunct{\mcitedefaultmidpunct}
{\mcitedefaultendpunct}{\mcitedefaultseppunct}\relax
\EndOfBibitem
\bibitem[Jalkanen \latin{et~al.}(2009)Jalkanen, Suhai, and
  Bohr]{Jalkanen_2009_DFT_molbiol}
Jalkanen,~K.~J.; Suhai,~S.; Bohr,~H. In \emph{Handbook of Molecular Biophysics.
  Methods and Applications}; Bohr,~H.~G., Ed.; Wiley-VCH Verlag, Weinheim,
  2009; pp 7--66\relax
\mciteBstWouldAddEndPuncttrue
\mciteSetBstMidEndSepPunct{\mcitedefaultmidpunct}
{\mcitedefaultendpunct}{\mcitedefaultseppunct}\relax
\EndOfBibitem
\bibitem[Becke(2014)]{Becke_2014_JCP.140.18A301}
Becke,~A.~D. {Fifty Years of Density-Functional Theory in Chemical Physics}.
  \emph{J. Chem. Phys.} \textbf{2014}, \emph{140}, 18A301\relax
\mciteBstWouldAddEndPuncttrue
\mciteSetBstMidEndSepPunct{\mcitedefaultmidpunct}
{\mcitedefaultendpunct}{\mcitedefaultseppunct}\relax
\EndOfBibitem
\bibitem[Scuseria and Staroverov(2005)Scuseria, and
  Staroverov]{Scuseria_XC_funct_2005}
Scuseria,~G.~E.; Staroverov,~V.~N. In \emph{Theory and Applications of
  Computational Chemistry: The First Forty Years}; Dykstra,~C., Frenking,~G.,
  Kim,~K., Scuseria,~G., Eds.; Elsevier, Amsterdam, 2005; pp 669--724\relax
\mciteBstWouldAddEndPuncttrue
\mciteSetBstMidEndSepPunct{\mcitedefaultmidpunct}
{\mcitedefaultendpunct}{\mcitedefaultseppunct}\relax
\EndOfBibitem
\bibitem[Jones(2015)]{Jones_2015_RMP.87.897}
Jones,~R.~O. {Density Functional Theory: Its Origins, Rise to Prominence, and
  Future}. \emph{Rev. Mod. Phys.} \textbf{2015}, \emph{87}, 897--923\relax
\mciteBstWouldAddEndPuncttrue
\mciteSetBstMidEndSepPunct{\mcitedefaultmidpunct}
{\mcitedefaultendpunct}{\mcitedefaultseppunct}\relax
\EndOfBibitem
\bibitem[Leszczynski \latin{et~al.}(2017)Leszczynski, Kaczmarek-Kedziera,
  Puzyn, Papadopoulos, Reis, and Shukla]{Handbook_CompChem}
Leszczynski,~J., Kaczmarek-Kedziera,~A., Puzyn,~T., Papadopoulos,~M.~G.,
  Reis,~H., Shukla,~M.~K., Eds. \emph{{Handbook of Computational Chemistry}},
  2nd ed.; Springer International Publishing, Cham, 2017\relax
\mciteBstWouldAddEndPuncttrue
\mciteSetBstMidEndSepPunct{\mcitedefaultmidpunct}
{\mcitedefaultendpunct}{\mcitedefaultseppunct}\relax
\EndOfBibitem
\bibitem[Kaplan \latin{et~al.}(2023)Kaplan, Levy, and
  Perdew]{Kaplan_2023_AnnuRevPhysChem}
Kaplan,~A.~D.; Levy,~M.; Perdew,~J.~P. {The Predictive Power of Exact
  Constraints and Appropriate Norms in Density Functional Theory}. \emph{Annu.
  Rev. Phys. Chem.} \textbf{2023}, \emph{74}, 193--218\relax
\mciteBstWouldAddEndPuncttrue
\mciteSetBstMidEndSepPunct{\mcitedefaultmidpunct}
{\mcitedefaultendpunct}{\mcitedefaultseppunct}\relax
\EndOfBibitem
\bibitem[Canc\`{e}s and Friesecke(2023)Canc\`{e}s, and
  Friesecke]{Cances_DFT_book_2023}
Canc\`{e}s,~E., Friesecke,~G., Eds. \emph{{Density Functional Theory. Modeling,
  Mathematical Analysis, Computational Methods, and Applications}}; Springer
  Nature Switzerland AG, Cham, 2023\relax
\mciteBstWouldAddEndPuncttrue
\mciteSetBstMidEndSepPunct{\mcitedefaultmidpunct}
{\mcitedefaultendpunct}{\mcitedefaultseppunct}\relax
\EndOfBibitem
\bibitem[Grant(1970)]{Grant_AdvPhys1970}
Grant,~I.~P. {Relativistic Calculation of Atomic Structures}. \emph{Adv. Phys.}
  \textbf{1970}, \emph{19}, 747--811\relax
\mciteBstWouldAddEndPuncttrue
\mciteSetBstMidEndSepPunct{\mcitedefaultmidpunct}
{\mcitedefaultendpunct}{\mcitedefaultseppunct}\relax
\EndOfBibitem
\bibitem[Mohanty and Clementi(1991)Mohanty, and
  Clementi]{Mohanty_1991_IJQC.39.487}
Mohanty,~A.; Clementi,~E. {Dirac-Fock Self-Consistent Field Method for
  Closed-Shell Molecules with Kinetic Balance and Finite Nuclear Size}.
  \emph{Int. J. Quantum Chem.} \textbf{1991}, \emph{39}, 487--517\relax
\mciteBstWouldAddEndPuncttrue
\mciteSetBstMidEndSepPunct{\mcitedefaultmidpunct}
{\mcitedefaultendpunct}{\mcitedefaultseppunct}\relax
\EndOfBibitem
\bibitem[Pyykk\"{o}(1988)]{Pyykko_1988_ChemRev.88.563}
Pyykk\"{o},~P. {Relativistic Effects in Structural Chemistry}. \emph{Chem.
  Rev.} \textbf{1988}, \emph{88}, 563--594\relax
\mciteBstWouldAddEndPuncttrue
\mciteSetBstMidEndSepPunct{\mcitedefaultmidpunct}
{\mcitedefaultendpunct}{\mcitedefaultseppunct}\relax
\EndOfBibitem
\bibitem[Shavitt and Bartlett(2009)Shavitt, and
  Bartlett]{MBPT_methods_book_Shavitt}
Shavitt,~I.; Bartlett,~R.-J. \emph{{Many-Body Methods in Chemistry and Physics:
  MBPT and Coupled-Cluster Theory}}; Cambridge University Press, Cambridge,
  2009\relax
\mciteBstWouldAddEndPuncttrue
\mciteSetBstMidEndSepPunct{\mcitedefaultmidpunct}
{\mcitedefaultendpunct}{\mcitedefaultseppunct}\relax
\EndOfBibitem
\bibitem[Cizek and Paldus(1980)Cizek, and Paldus]{Cizek_1980_PhysScr.21.251}
Cizek,~J.; Paldus,~J. {Coupled Cluster Approach}. \emph{Phys. Scr.}
  \textbf{1980}, \emph{21}, 251--254\relax
\mciteBstWouldAddEndPuncttrue
\mciteSetBstMidEndSepPunct{\mcitedefaultmidpunct}
{\mcitedefaultendpunct}{\mcitedefaultseppunct}\relax
\EndOfBibitem
\bibitem[Bartlett and Musia{\l}(2007)Bartlett, and
  Musia{\l}]{Bartlett_2007_RMP.79.291}
Bartlett,~R.~J.; Musia{\l},~M. {Coupled-Cluster Theory in Quantum Chemistry}.
  \emph{Rev. Mod. Phys.} \textbf{2007}, \emph{79}, 291--352\relax
\mciteBstWouldAddEndPuncttrue
\mciteSetBstMidEndSepPunct{\mcitedefaultmidpunct}
{\mcitedefaultendpunct}{\mcitedefaultseppunct}\relax
\EndOfBibitem
\bibitem[Zhang and Gr\"{u}neis(2019)Zhang, and
  Gr\"{u}neis]{Zhang_2019_FrontMater.6.123}
Zhang,~I.~Y.; Gr\"{u}neis,~A. {Coupled Cluster Theory in Materials Science}.
  \emph{Front. Mater.} \textbf{2019}, \emph{6}, 123\relax
\mciteBstWouldAddEndPuncttrue
\mciteSetBstMidEndSepPunct{\mcitedefaultmidpunct}
{\mcitedefaultendpunct}{\mcitedefaultseppunct}\relax
\EndOfBibitem
\bibitem[Szalay \latin{et~al.}(2012)Szalay, M\"{u}ller, Gidofalvi, Lischka, and
  Shepard]{Szalay_2012_ChemRev.112.108}
Szalay,~P.~G.; M\"{u}ller,~T.; Gidofalvi,~G.; Lischka,~H.; Shepard,~R.
  {Multiconfiguration Self-Consistent Field and Multireference Configuration
  Interaction Methods and Applications}. \emph{Chem. Rev.} \textbf{2012},
  \emph{112}, 108--181\relax
\mciteBstWouldAddEndPuncttrue
\mciteSetBstMidEndSepPunct{\mcitedefaultmidpunct}
{\mcitedefaultendpunct}{\mcitedefaultseppunct}\relax
\EndOfBibitem
\bibitem[Knowles and Handy(1989)Knowles, and Handy]{Knowles_1989_CPC.54.75}
Knowles,~P.~J.; Handy,~N.~C. {A Determinant Based Full Configuration
  Interaction Program}. \emph{Comput. Phys. Commun.} \textbf{1989}, \emph{54},
  75--83\relax
\mciteBstWouldAddEndPuncttrue
\mciteSetBstMidEndSepPunct{\mcitedefaultmidpunct}
{\mcitedefaultendpunct}{\mcitedefaultseppunct}\relax
\EndOfBibitem
\bibitem[Rontani \latin{et~al.}(2006)Rontani, Cavazzoni, Bellucci, and
  Goldoni]{Rontani_2006_JCP.124.124102}
Rontani,~M.; Cavazzoni,~C.; Bellucci,~D.; Goldoni,~G. {Full Configuration
  Interaction Approach to the Few-Electron Problem in Artificial Atoms}.
  \emph{J. Chem. Phys.} \textbf{2006}, \emph{124}, 124102\relax
\mciteBstWouldAddEndPuncttrue
\mciteSetBstMidEndSepPunct{\mcitedefaultmidpunct}
{\mcitedefaultendpunct}{\mcitedefaultseppunct}\relax
\EndOfBibitem
\bibitem[Joecker \latin{et~al.}(2021)Joecker, Baczewski, Gamble, Pla, Saraiva,
  and Morello]{Joecker_2021_NJP.23.073007}
Joecker,~B.; Baczewski,~A.~D.; Gamble,~J.~K.; Pla,~J.~J.; Saraiva,~A.;
  Morello,~A. {Full Configuration Interaction Simulations of Exchange-Coupled
  Donors in Silicon Using Multi-Valley Effective Mass Theory}. \emph{New J.
  Phys.} \textbf{2021}, \emph{23}, 073007\relax
\mciteBstWouldAddEndPuncttrue
\mciteSetBstMidEndSepPunct{\mcitedefaultmidpunct}
{\mcitedefaultendpunct}{\mcitedefaultseppunct}\relax
\EndOfBibitem
\bibitem[Xu \latin{et~al.}(2018)Xu, Uejima, and
  Ten-no]{Xu_Uejima_2018_PRL.121.113001}
Xu,~E.; Uejima,~M.; Ten-no,~S.~L. {Full Coupled-Cluster Reduction for Accurate
  Description of Strong Electron Correlation}. \emph{Phys. Rev. Lett.}
  \textbf{2018}, \emph{121}, 113001\relax
\mciteBstWouldAddEndPuncttrue
\mciteSetBstMidEndSepPunct{\mcitedefaultmidpunct}
{\mcitedefaultendpunct}{\mcitedefaultseppunct}\relax
\EndOfBibitem
\bibitem[Purvis~III and Bartlett(1982)Purvis~III, and
  Bartlett]{Purvis_1982_JCP.76.1910}
Purvis~III,~G.~D.; Bartlett,~R.~J. {A Full Coupled-Cluster Singles and Doubles
  Model: The Inclusion of Disconnected Triples}. \emph{J. Chem. Phys.}
  \textbf{1982}, \emph{76}, 1910--1918\relax
\mciteBstWouldAddEndPuncttrue
\mciteSetBstMidEndSepPunct{\mcitedefaultmidpunct}
{\mcitedefaultendpunct}{\mcitedefaultseppunct}\relax
\EndOfBibitem
\bibitem[Cullen and Zerner(1982)Cullen, and Zerner]{Cullen_1982_JCP.77.4088}
Cullen,~J.~M.; Zerner,~M.~C. {The Linked Singles and Doubles Model: An
  Approximate Theory of Electron Correlation Based on the Coupled-Cluster
  Ansatz}. \emph{J. Chem. Phys.} \textbf{1982}, \emph{77}, 4088--4109\relax
\mciteBstWouldAddEndPuncttrue
\mciteSetBstMidEndSepPunct{\mcitedefaultmidpunct}
{\mcitedefaultendpunct}{\mcitedefaultseppunct}\relax
\EndOfBibitem
\bibitem[Raghavachari \latin{et~al.}(1989)Raghavachari, Trucks, Pople, and
  Head-Gordon]{Raghavachari_1989_CPL.157.479}
Raghavachari,~K.; Trucks,~G.~W.; Pople,~J.~A.; Head-Gordon,~M. {A Fifth-Order
  Perturbation Comparison of Electron Correlation Theories}. \emph{Chem. Phys.
  Lett.} \textbf{1989}, \emph{157}, 479--483\relax
\mciteBstWouldAddEndPuncttrue
\mciteSetBstMidEndSepPunct{\mcitedefaultmidpunct}
{\mcitedefaultendpunct}{\mcitedefaultseppunct}\relax
\EndOfBibitem
\bibitem[M{\o}ller and Plesset(1934)M{\o}ller, and
  Plesset]{Moller_Plesset_1934}
M{\o}ller,~C.; Plesset,~M.~S. {Note on an Approximation Treatment for
  Many-Electron Systems}. \emph{Phys. Rev.} \textbf{1934}, \emph{46},
  618--622\relax
\mciteBstWouldAddEndPuncttrue
\mciteSetBstMidEndSepPunct{\mcitedefaultmidpunct}
{\mcitedefaultendpunct}{\mcitedefaultseppunct}\relax
\EndOfBibitem
\bibitem[Cremer(2011)]{Cremer_2011_WIREsCMS.1.509}
Cremer,~D. {M{\o}ller--Plesset Perturbation Theory: From Small Molecule Methods
  to Methods for Thousands of Atoms}. \emph{WIREs Comput. Molec. Sci.}
  \textbf{2011}, \emph{1}, 509--530\relax
\mciteBstWouldAddEndPuncttrue
\mciteSetBstMidEndSepPunct{\mcitedefaultmidpunct}
{\mcitedefaultendpunct}{\mcitedefaultseppunct}\relax
\EndOfBibitem
\bibitem[Wik((accessed 2023-11-15))]{Wiki_QC_codes}
List of Quantum Chemistry and Solid-State Physics Software.
  \url{https://en.wikipedia.org/wiki/List_of_quantum_chemistry_and_solid-state_physics_software},
  (accessed 2023-11-15)\relax
\mciteBstWouldAddEndPuncttrue
\mciteSetBstMidEndSepPunct{\mcitedefaultmidpunct}
{\mcitedefaultendpunct}{\mcitedefaultseppunct}\relax
\EndOfBibitem
\bibitem[K\"{u}hne \latin{et~al.}(2020)K\"{u}hne, Iannuzzi, Del~Ben, Rybkin,
  Seewald, Stein, Laino, Khaliullin, Sch\"{u}tt, Schiffmann, and \textit{et
  al.}]{CP2K_Kuhne_2020_JCP.152.194103}
K\"{u}hne,~T.~D.; Iannuzzi,~M.; Del~Ben,~M.; Rybkin,~V.~V.; Seewald,~P.;
  Stein,~F.; Laino,~T.; Khaliullin,~R.~Z.; Sch\"{u}tt,~O.; Schiffmann,~F.;
  \textit{et al.} {CP2K: An Electronic Structure and Molecular Dynamics
  Software Package -- Quickstep: Efficient and Accurate Electronic Structure
  Calculations}. \emph{J. Chem. Phys.} \textbf{2020}, \emph{152}, 194103\relax
\mciteBstWouldAddEndPuncttrue
\mciteSetBstMidEndSepPunct{\mcitedefaultmidpunct}
{\mcitedefaultendpunct}{\mcitedefaultseppunct}\relax
\EndOfBibitem
\bibitem[Aidas \latin{et~al.}(2014)Aidas, Angeli, Bak, Bakken, Bast, Boman,
  Christiansen, Cimiraglia, Coriani, Dahle, and \textit{et
  al.}]{DALTON_Aidas_2014_WIREsCMS.4.269}
Aidas,~K.; Angeli,~C.; Bak,~K.~L.; Bakken,~V.; Bast,~R.; Boman,~L.;
  Christiansen,~O.; Cimiraglia,~R.; Coriani,~S.; Dahle,~P.; \textit{et al.}
  {The Dalton Quantum Chemistry Program System}. \emph{WIREs Comput. Mol. Sci.}
  \textbf{2014}, \emph{4}, 269--284\relax
\mciteBstWouldAddEndPuncttrue
\mciteSetBstMidEndSepPunct{\mcitedefaultmidpunct}
{\mcitedefaultendpunct}{\mcitedefaultseppunct}\relax
\EndOfBibitem
\bibitem[Schmidt \latin{et~al.}(1993)Schmidt, Baldridge, Boatz, Elbert, Gordon,
  Jensen, Koseki, Matsunaga, Nguyen, Su, and \textit{et
  al.}]{GAMESS_Schmidt_1993_JCC.14.1347}
Schmidt,~M.~W.; Baldridge,~K.~K.; Boatz,~J.~A.; Elbert,~S.~T.; Gordon,~M.~S.;
  Jensen,~J.~H.; Koseki,~S.; Matsunaga,~N.; Nguyen,~K.~A.; Su,~S.; \textit{et
  al.} {General Atomic and Molecular Electronic Structure System}. \emph{J.
  Comput. Chem.} \textbf{1993}, \emph{14}, 1347--1363\relax
\mciteBstWouldAddEndPuncttrue
\mciteSetBstMidEndSepPunct{\mcitedefaultmidpunct}
{\mcitedefaultendpunct}{\mcitedefaultseppunct}\relax
\EndOfBibitem
\bibitem[Frisch \latin{et~al.}(2016)Frisch, Trucks, Schlegel, Scuseria, Robb,
  Cheeseman, Scalmani, Barone, Petersson, Nakatsuji, and \textit{et
  al.}]{GAUSSIAN_g16}
Frisch,~M.~J.; Trucks,~G.~W.; Schlegel,~H.~B.; Scuseria,~G.~E.; Robb,~M.~A.;
  Cheeseman,~J.~R.; Scalmani,~G.; Barone,~V.; Petersson,~G.~A.; Nakatsuji,~H.;
  \textit{et al.} Gaussian 16 {R}evision {C}.01. 2016; Gaussian Inc.
  Wallingford CT\relax
\mciteBstWouldAddEndPuncttrue
\mciteSetBstMidEndSepPunct{\mcitedefaultmidpunct}
{\mcitedefaultendpunct}{\mcitedefaultseppunct}\relax
\EndOfBibitem
\bibitem[Valiev \latin{et~al.}(2010)Valiev, Bylaska, Govind, Kowalski,
  Straatsma, Dam, Wang, Nieplocha, Apra, Windus, and
  Jong]{NWChem_Valiev_2010_CPC.181.1477}
Valiev,~M.; Bylaska,~E.~J.; Govind,~N.; Kowalski,~K.; Straatsma,~T.~P.; Dam,~H.
  J. J.~V.; Wang,~D.; Nieplocha,~J.; Apra,~E.; Windus,~T.~L.; Jong,~W. A.~D.
  {NWChem: A Comprehensive and Scalable Open-Source Solution for Large Scale
  Molecular Simulations}. \emph{Comput. Phys. Commun.} \textbf{2010},
  \emph{181}, 1477--1489\relax
\mciteBstWouldAddEndPuncttrue
\mciteSetBstMidEndSepPunct{\mcitedefaultmidpunct}
{\mcitedefaultendpunct}{\mcitedefaultseppunct}\relax
\EndOfBibitem
\bibitem[Neese \latin{et~al.}(2020)Neese, Wennmohs, Becker, and
  Riplinger]{ORCA_Neese_2020_JCP.152.224108}
Neese,~F.; Wennmohs,~F.; Becker,~U.; Riplinger,~C. {The ORCA Quantum Chemistry
  Program Package}. \emph{J. Chem. Phys.} \textbf{2020}, \emph{152},
  224108\relax
\mciteBstWouldAddEndPuncttrue
\mciteSetBstMidEndSepPunct{\mcitedefaultmidpunct}
{\mcitedefaultendpunct}{\mcitedefaultseppunct}\relax
\EndOfBibitem
\bibitem[Grant and Quiney(2022)Grant, and
  Quiney]{GRASP_Grant_2022_Atoms.10.108}
Grant,~I.; Quiney,~H. {GRASP: The Future?} \emph{Atoms} \textbf{2022},
  \emph{10}, 108\relax
\mciteBstWouldAddEndPuncttrue
\mciteSetBstMidEndSepPunct{\mcitedefaultmidpunct}
{\mcitedefaultendpunct}{\mcitedefaultseppunct}\relax
\EndOfBibitem
\bibitem[Visscher \latin{et~al.}(1994)Visscher, Visser, Aerts, Merenga, and
  Nieuwpoort]{MOLFDIR_Visscher_1994_CPC.81.120}
Visscher,~L.; Visser,~O.; Aerts,~P.; Merenga,~H.; Nieuwpoort,~W. {Relativistic
  Quantum Chemistry: The MOLFDIR Program Package}. \emph{Comput. Phys. Commun.}
  \textbf{1994}, \emph{81}, 120--144\relax
\mciteBstWouldAddEndPuncttrue
\mciteSetBstMidEndSepPunct{\mcitedefaultmidpunct}
{\mcitedefaultendpunct}{\mcitedefaultseppunct}\relax
\EndOfBibitem
\bibitem[Saue \latin{et~al.}(1997)Saue, F{\ae}gri, Helgaker, and
  Gropen]{DIRAC_Saue_1997_MolPhys.91.937}
Saue,~T.; F{\ae}gri,~K.; Helgaker,~T.; Gropen,~O. {Principles of Direct
  4-Component Relativistic SCF: Application to Caesium Auride}. \emph{Mol.
  Phys.} \textbf{1997}, \emph{91}, 937--950\relax
\mciteBstWouldAddEndPuncttrue
\mciteSetBstMidEndSepPunct{\mcitedefaultmidpunct}
{\mcitedefaultendpunct}{\mcitedefaultseppunct}\relax
\EndOfBibitem
\bibitem[Grant and Quiney(2000)Grant, and
  Quiney]{BERTHA_Grant_2000_IJQC.80.283}
Grant,~I.~P.; Quiney,~H.~M. {Application of Relativistic Theories and Quantum
  Electrodynamics to Chemical Problems}. \emph{Int. J. Quantum Chem.}
  \textbf{2000}, \emph{80}, 283--297\relax
\mciteBstWouldAddEndPuncttrue
\mciteSetBstMidEndSepPunct{\mcitedefaultmidpunct}
{\mcitedefaultendpunct}{\mcitedefaultseppunct}\relax
\EndOfBibitem
\bibitem[Belpassi \latin{et~al.}(2020)Belpassi, de~Santis, Quiney, Tarantelli,
  and Storchi]{BERTHA_Belpassi_2020_JCP.152.164118}
Belpassi,~L.; de~Santis,~M.; Quiney,~H.~M.; Tarantelli,~F.; Storchi,~L.
  {BERTHA: Implementation of a Four-Component Dirac-Kohn-Sham Relativistic
  Framework}. \emph{J. Chem. Phys.} \textbf{2020}, \emph{152}, 164118\relax
\mciteBstWouldAddEndPuncttrue
\mciteSetBstMidEndSepPunct{\mcitedefaultmidpunct}
{\mcitedefaultendpunct}{\mcitedefaultseppunct}\relax
\EndOfBibitem
\bibitem[Grant and Quiney(2006)Grant, and Quiney]{Grant_BERTHA_chapter_2006}
Grant,~I.~P.; Quiney,~H.~M. In \emph{Recent Advances in the Theory of Chemical
  and Physical Systems}; Julien,~J.-P., Maruani,~J., Mayou,~D., Wilson,~S.,
  Delgado-Barrio,~G., Eds.; Springer, Dordrecht, 2006; pp 199--215\relax
\mciteBstWouldAddEndPuncttrue
\mciteSetBstMidEndSepPunct{\mcitedefaultmidpunct}
{\mcitedefaultendpunct}{\mcitedefaultseppunct}\relax
\EndOfBibitem
\bibitem[Chernysheva and Ivanov(2022)Chernysheva, and
  Ivanov]{ATOM_Chernysheva_VK_2022_Atoms.10.52}
Chernysheva,~L.~V.; Ivanov,~V.~K. {ATOM Program System and Computational
  Experiment}. \emph{Atoms} \textbf{2022}, \emph{10}, 52\relax
\mciteBstWouldAddEndPuncttrue
\mciteSetBstMidEndSepPunct{\mcitedefaultmidpunct}
{\mcitedefaultendpunct}{\mcitedefaultseppunct}\relax
\EndOfBibitem
\bibitem[March(2003)]{March_2003_HF_beyond}
March,~N.~H. In \emph{Advanced Topics in Theoretical Chemical Physics};
  Maruani,~J., Lefebvre,~R., Br\"{a}ndas,~E.~J., Eds.; Springer, Dordrecht,
  2003; pp 53--70\relax
\mciteBstWouldAddEndPuncttrue
\mciteSetBstMidEndSepPunct{\mcitedefaultmidpunct}
{\mcitedefaultendpunct}{\mcitedefaultseppunct}\relax
\EndOfBibitem
\bibitem[Evarestov(2012)]{Evarestov_QC_Solids}
Evarestov,~R.~A. \emph{{Quantum Chemistry of Solids: LCAO Treatment of Crystals
  and Nanostructures}}, 2nd ed.; Springer-Verlag Berlin Heidelberg, 2012\relax
\mciteBstWouldAddEndPuncttrue
\mciteSetBstMidEndSepPunct{\mcitedefaultmidpunct}
{\mcitedefaultendpunct}{\mcitedefaultseppunct}\relax
\EndOfBibitem
\bibitem[Ren \latin{et~al.}(2012)Ren, Rinke, Joas, and
  Scheffler]{Ren_2012_JMaterSci.47.7447}
Ren,~X.; Rinke,~P.; Joas,~C.; Scheffler,~M. {Random-Phase Approximation and Its
  Applications in Computational Chemistry and Materials Science}. \emph{J.
  Mater. Sci.} \textbf{2012}, \emph{47}, 7447--7471\relax
\mciteBstWouldAddEndPuncttrue
\mciteSetBstMidEndSepPunct{\mcitedefaultmidpunct}
{\mcitedefaultendpunct}{\mcitedefaultseppunct}\relax
\EndOfBibitem
\bibitem[Starace(2006)]{Starace_PI_atoms_2006}
Starace,~A. In \emph{Springer Handbook of Atomic, Molecular, and Optical
  Physics}; Drake,~G., Ed.; Springer, New York, NY, 2006; pp 379--390\relax
\mciteBstWouldAddEndPuncttrue
\mciteSetBstMidEndSepPunct{\mcitedefaultmidpunct}
{\mcitedefaultendpunct}{\mcitedefaultseppunct}\relax
\EndOfBibitem
\bibitem[Solov'yov(2005)]{Solovyov_2005_IJMPB.19.4143}
Solov'yov,~A.~V. {Plasmon Excitations in Metal Clusters and Fullerenes}.
  \emph{Int. J. Mod. Phys. B} \textbf{2005}, \emph{19}, 4143--4184\relax
\mciteBstWouldAddEndPuncttrue
\mciteSetBstMidEndSepPunct{\mcitedefaultmidpunct}
{\mcitedefaultendpunct}{\mcitedefaultseppunct}\relax
\EndOfBibitem
\bibitem[Strout and Scuseria(1995)Strout, and
  Scuseria]{Strout_1995_JCP.102.8448}
Strout,~D.~L.; Scuseria,~G.~E. {A Quantitative Study of the Scaling Properties
  of the Hartree--Fock Method}. \emph{J. Chem. Phys.} \textbf{1995},
  \emph{102}, 8448--8452\relax
\mciteBstWouldAddEndPuncttrue
\mciteSetBstMidEndSepPunct{\mcitedefaultmidpunct}
{\mcitedefaultendpunct}{\mcitedefaultseppunct}\relax
\EndOfBibitem
\bibitem[Scuseria and Lee(1990)Scuseria, and Lee]{Scuseria_1990_JCP.93.5851}
Scuseria,~G.~E.; Lee,~T.~J. {Comparison of Coupled-Cluster Methods Which
  Include the Effects of Connected Triple Excitations}. \emph{J. Chem. Phys.}
  \textbf{1990}, \emph{93}, 5851--5855\relax
\mciteBstWouldAddEndPuncttrue
\mciteSetBstMidEndSepPunct{\mcitedefaultmidpunct}
{\mcitedefaultendpunct}{\mcitedefaultseppunct}\relax
\EndOfBibitem
\bibitem[Leininger \latin{et~al.}(2000)Leininger, Allen, Schaefer~III, and
  Sherrill]{Leininger_2000_JCP.112.9213}
Leininger,~M.~L.; Allen,~W.~D.; Schaefer~III,~H.~F.; Sherrill,~C.~D. {Is
  M{\o}ller--Plesset Perturbation Theory a Convergent Ab Initio Method?}
  \emph{J. Chem. Phys.} \textbf{2000}, \emph{112}, 9213--9222\relax
\mciteBstWouldAddEndPuncttrue
\mciteSetBstMidEndSepPunct{\mcitedefaultmidpunct}
{\mcitedefaultendpunct}{\mcitedefaultseppunct}\relax
\EndOfBibitem
\bibitem[Herman and Hagedorn(2008)Herman, and
  Hagedorn]{Herman_2008_IJQC.109.210}
Herman,~M.~S.; Hagedorn,~G.~A. {Does M{\o}ller--Plesset Perturbation Theory
  Converge? A Look at Two-Electron Systems}. \emph{Int. J. Quantum Chem.}
  \textbf{2008}, \emph{109}, 210--225\relax
\mciteBstWouldAddEndPuncttrue
\mciteSetBstMidEndSepPunct{\mcitedefaultmidpunct}
{\mcitedefaultendpunct}{\mcitedefaultseppunct}\relax
\EndOfBibitem
\bibitem[Cremer and He(1996)Cremer, and He]{Cremer_1996_JPC.100.6173}
Cremer,~D.; He,~Z. {Sixth-Order M{\o}ller--Plesset Perturbation Theory -- On
  the Convergence of the MP$n$ Series}. \emph{J. Phys. Chem.} \textbf{1996},
  \emph{100}, 6173--6188\relax
\mciteBstWouldAddEndPuncttrue
\mciteSetBstMidEndSepPunct{\mcitedefaultmidpunct}
{\mcitedefaultendpunct}{\mcitedefaultseppunct}\relax
\EndOfBibitem
\bibitem[Doser \latin{et~al.}(2009)Doser, Lambrecht, Kussmann, and
  Ochsenfeld]{Doser_2009_JCP.130.064107}
Doser,~B.; Lambrecht,~D.~S.; Kussmann,~J.; Ochsenfeld,~C. {Linear-Scaling
  Atomic Orbital-Based Second-Order M{\o}ller--Plesset Perturbation Theory by
  Rigorous Integral Screening Criteria}. \emph{J. Chem. Phys.} \textbf{2009},
  \emph{130}, 064107\relax
\mciteBstWouldAddEndPuncttrue
\mciteSetBstMidEndSepPunct{\mcitedefaultmidpunct}
{\mcitedefaultendpunct}{\mcitedefaultseppunct}\relax
\EndOfBibitem
\bibitem[Glasbrenner \latin{et~al.}(2020)Glasbrenner, Graf, and
  Ochsenfeld]{Glasbrenner_2020_JCTC.16.6856}
Glasbrenner,~M.; Graf,~D.; Ochsenfeld,~C. {Efficient Reduced-Scaling
  Second-Order M{\o}ller--Plesset Perturbation Theory with Cholesky-Decomposed
  Densities and an Attenuated Coulomb Metric}. \emph{J. Chem. Theory Comput.}
  \textbf{2020}, \emph{16}, 6856--6868\relax
\mciteBstWouldAddEndPuncttrue
\mciteSetBstMidEndSepPunct{\mcitedefaultmidpunct}
{\mcitedefaultendpunct}{\mcitedefaultseppunct}\relax
\EndOfBibitem
\bibitem[Perdew \latin{et~al.}(1996)Perdew, Burke, and
  Ernzerhof]{PBE_1996_PRL.77.3865}
Perdew,~J.~P.; Burke,~K.; Ernzerhof,~M. {Generalized Gradient Approximation
  Made Simple}. \emph{Phys. Rev. Lett.} \textbf{1996}, \emph{77},
  3865--3868\relax
\mciteBstWouldAddEndPuncttrue
\mciteSetBstMidEndSepPunct{\mcitedefaultmidpunct}
{\mcitedefaultendpunct}{\mcitedefaultseppunct}\relax
\EndOfBibitem
\bibitem[Yanai \latin{et~al.}(2004)Yanai, Tew, and
  Handy]{CAM-B3LYP_Yanai_2004_CPL}
Yanai,~T.; Tew,~D.~P.; Handy,~N.~C. {A New Hybrid Exchange--Correlation
  Functional Using the Coulomb-Attenuating Method (CAM-B3LYP)}. \emph{Chem.
  Phys. Lett.} \textbf{2004}, \emph{393}, 51--57\relax
\mciteBstWouldAddEndPuncttrue
\mciteSetBstMidEndSepPunct{\mcitedefaultmidpunct}
{\mcitedefaultendpunct}{\mcitedefaultseppunct}\relax
\EndOfBibitem
\bibitem[Vydrov \latin{et~al.}(2006)Vydrov, Heyd, Krukau, and
  Scuseria]{Vydrov_2006_JCP.125.074106}
Vydrov,~O.~A.; Heyd,~J.; Krukau,~A.; Scuseria,~G.~E. {Importance of Short-Range
  Versus Long-Range Hartree-Fock Exchange for the Performance of Hybrid Density
  Functionals}. \emph{J. Chem. Phys.} \textbf{2006}, \emph{125}, 074106\relax
\mciteBstWouldAddEndPuncttrue
\mciteSetBstMidEndSepPunct{\mcitedefaultmidpunct}
{\mcitedefaultendpunct}{\mcitedefaultseppunct}\relax
\EndOfBibitem
\bibitem[Chai and Head-Gordon(2008)Chai, and
  Head-Gordon]{Chai_2008_PCCP.10.6615}
Chai,~J.-D.; Head-Gordon,~M. {Long-Range Corrected Hybrid Density Functionals
  with Damped Atom-Atom Dispersion Corrections}. \emph{Phys. Chem. Chem. Phys.}
  \textbf{2008}, \emph{10}, 6615--6620\relax
\mciteBstWouldAddEndPuncttrue
\mciteSetBstMidEndSepPunct{\mcitedefaultmidpunct}
{\mcitedefaultendpunct}{\mcitedefaultseppunct}\relax
\EndOfBibitem
\bibitem[Tsuneda and Hirao(2014)Tsuneda, and
  Hirao]{Tsuneda_2014_WIREsCMS.4.375}
Tsuneda,~T.; Hirao,~K. {Long-Range Correction for Density Functional Theory}.
  \emph{WIREs Comput. Mol. Sci.} \textbf{2014}, \emph{4}, 375--390\relax
\mciteBstWouldAddEndPuncttrue
\mciteSetBstMidEndSepPunct{\mcitedefaultmidpunct}
{\mcitedefaultendpunct}{\mcitedefaultseppunct}\relax
\EndOfBibitem
\bibitem[Li \latin{et~al.}(2021)Li, Reimers, Ford, Kobayashi, and
  Amos]{Li_2021_JCC.42.1486}
Li,~M.; Reimers,~J.~R.; Ford,~M.~J.; Kobayashi,~R.; Amos,~R.~D. {Accurate
  Prediction of the Properties of Materials Using the CAM-B3LYP Density
  Functional}. \emph{J. Comput. Chem.} \textbf{2021}, \emph{42},
  1486--1497\relax
\mciteBstWouldAddEndPuncttrue
\mciteSetBstMidEndSepPunct{\mcitedefaultmidpunct}
{\mcitedefaultendpunct}{\mcitedefaultseppunct}\relax
\EndOfBibitem
\bibitem[Castro \latin{et~al.}(2006)Castro, Appel, Oliveira, Rozzi, Andrade,
  Lorenzen, Marques, Gross, and Rubio]{OCTOPUS_Castro_2006_PSSB.243.2465}
Castro,~A.; Appel,~H.; Oliveira,~M.; Rozzi,~C.~A.; Andrade,~X.; Lorenzen,~F.;
  Marques,~M. A.~L.; Gross,~E. K.~U.; Rubio,~A. {Octopus: A Tool for the
  Application of Time-Dependent Density Functional Theory}. \emph{Phys. Stat.
  Sol. B} \textbf{2006}, \emph{243}, 2465--2488\relax
\mciteBstWouldAddEndPuncttrue
\mciteSetBstMidEndSepPunct{\mcitedefaultmidpunct}
{\mcitedefaultendpunct}{\mcitedefaultseppunct}\relax
\EndOfBibitem
\bibitem[Giannozzi \latin{et~al.}(2009)Giannozzi, Baroni, Bonini, Calandra,
  Car, Cavazzoni, Ceresoli, Chiarotti, Cococcioni, Dabo, and \textit{et
  al.}]{QE_Giannozzi_2009_JPCM.21.395502}
Giannozzi,~P.; Baroni,~S.; Bonini,~N.; Calandra,~M.; Car,~R.; Cavazzoni,~C.;
  Ceresoli,~D.; Chiarotti,~G.~L.; Cococcioni,~M.; Dabo,~I.; \textit{et al.}
  {QUANTUM ESPRESSO: A Modular and Open-Source Software Project for Quantum
  Simulations of Materials}. \emph{J. Phys.: Condens. Matter} \textbf{2009},
  \emph{21}, 395502\relax
\mciteBstWouldAddEndPuncttrue
\mciteSetBstMidEndSepPunct{\mcitedefaultmidpunct}
{\mcitedefaultendpunct}{\mcitedefaultseppunct}\relax
\EndOfBibitem
\bibitem[Makkar and Ghosh(2021)Makkar, and Ghosh]{Makkar_2021_RSCAdv.11.27897}
Makkar,~P.; Ghosh,~N.~N. {A Review on the Use of DFT for the Prediction of the
  Properties of Nanomaterials}. \emph{RSC Adv.} \textbf{2021}, \emph{11},
  27897--27924\relax
\mciteBstWouldAddEndPuncttrue
\mciteSetBstMidEndSepPunct{\mcitedefaultmidpunct}
{\mcitedefaultendpunct}{\mcitedefaultseppunct}\relax
\EndOfBibitem
\bibitem[Perdew and Zunger(1981)Perdew, and
  Zunger]{Perdew_Zunger_1981_PRB.23.5048}
Perdew,~J.~P.; Zunger,~A. {Self-Interaction Correction to Density-Functional
  Approximations for Many-Electron Systems}. \emph{Phys. Rev. B} \textbf{1981},
  \emph{23}, 5048--5079\relax
\mciteBstWouldAddEndPuncttrue
\mciteSetBstMidEndSepPunct{\mcitedefaultmidpunct}
{\mcitedefaultendpunct}{\mcitedefaultseppunct}\relax
\EndOfBibitem
\bibitem[Tsuneda and Hirao(2014)Tsuneda, and
  Hirao]{Tsuneda_2014_JCP.140.18A513}
Tsuneda,~T.; Hirao,~K. {Self-Interaction Corrections in Density Functional
  Theory}. \emph{J. Chem. Phys.} \textbf{2014}, \emph{140}, 18A513\relax
\mciteBstWouldAddEndPuncttrue
\mciteSetBstMidEndSepPunct{\mcitedefaultmidpunct}
{\mcitedefaultendpunct}{\mcitedefaultseppunct}\relax
\EndOfBibitem
\bibitem[Grimme(2011)]{Grimme_2011_WIREsCMS.1.211}
Grimme,~S. {Density Functional Theory with London Dispersion Corrections}.
  \emph{WIREs Comput. Mol. Sci.} \textbf{2011}, \emph{1}, 211--228\relax
\mciteBstWouldAddEndPuncttrue
\mciteSetBstMidEndSepPunct{\mcitedefaultmidpunct}
{\mcitedefaultendpunct}{\mcitedefaultseppunct}\relax
\EndOfBibitem
\bibitem[Grimme \latin{et~al.}(2016)Grimme, Hansen, Brandenburg, and
  Bannwarth]{Grimme_2016_ChemRev.116.5105}
Grimme,~S.; Hansen,~A.; Brandenburg,~J.~G.; Bannwarth,~C. {Dispersion-Corrected
  Mean-Field Electronic Structure Methods}. \emph{Chem. Rev.} \textbf{2016},
  \emph{116}, 5105--5154\relax
\mciteBstWouldAddEndPuncttrue
\mciteSetBstMidEndSepPunct{\mcitedefaultmidpunct}
{\mcitedefaultendpunct}{\mcitedefaultseppunct}\relax
\EndOfBibitem
\bibitem[M\"uller \latin{et~al.}(2020)M\"uller, an~E.~K. U.~Gross, and
  Dewhurst]{Muller_2020_PRL.125.256402}
M\"uller,~T.; an~E.~K. U.~Gross,~S.~S.; Dewhurst,~J.~K. {Extending Solid-State
  Calculations to Ultra-Long-Range Length Scales}. \emph{Phys. Rev. Lett.}
  \textbf{2020}, \emph{125}, 256402\relax
\mciteBstWouldAddEndPuncttrue
\mciteSetBstMidEndSepPunct{\mcitedefaultmidpunct}
{\mcitedefaultendpunct}{\mcitedefaultseppunct}\relax
\EndOfBibitem
\bibitem[Nakata \latin{et~al.}(2020)Nakata, Baker, Mujahed, Poulton, Arapan,
  Lin, Raza, Yadav, Truflandier, Miyazaki, and
  Bowler]{CONQUEST_Nakata_2020_JCP.152.164112}
Nakata,~A.; Baker,~J.~S.; Mujahed,~S.~Y.; Poulton,~J. T.~L.; Arapan,~S.;
  Lin,~J.; Raza,~Z.; Yadav,~S.; Truflandier,~L.; Miyazaki,~T.; Bowler,~D.~R.
  {Large Scale and Linear Scaling DFT With the CONQUEST Code}. \emph{J. Chem.
  Phys.} \textbf{2020}, \emph{152}, 164112\relax
\mciteBstWouldAddEndPuncttrue
\mciteSetBstMidEndSepPunct{\mcitedefaultmidpunct}
{\mcitedefaultendpunct}{\mcitedefaultseppunct}\relax
\EndOfBibitem
\bibitem[Runge and Gross(1984)Runge, and Gross]{Runge_Gross_1984_PRL.52.997}
Runge,~E.; Gross,~E. K.~U. {Density-Functional Theory for Time-Dependent
  Systems}. \emph{Phys. Rev. Lett.} \textbf{1984}, \emph{52}, 997--1000\relax
\mciteBstWouldAddEndPuncttrue
\mciteSetBstMidEndSepPunct{\mcitedefaultmidpunct}
{\mcitedefaultendpunct}{\mcitedefaultseppunct}\relax
\EndOfBibitem
\bibitem[Ullrich(2012)]{Ullrich_TDDFT_book_2012}
Ullrich,~C.~A. \emph{{Time-Dependent Density-Functional Theory: Concepts and
  Applications}}; Oxford University Press, Oxford, 2012\relax
\mciteBstWouldAddEndPuncttrue
\mciteSetBstMidEndSepPunct{\mcitedefaultmidpunct}
{\mcitedefaultendpunct}{\mcitedefaultseppunct}\relax
\EndOfBibitem
\bibitem[Maitra(2016)]{Maitra_2016_JCP.144.220901}
Maitra,~N.~T. {Fundamental Aspects of Time-Dependent Density Functional
  Theory}. \emph{J. Chem. Phys.} \textbf{2016}, \emph{144}, 220901\relax
\mciteBstWouldAddEndPuncttrue
\mciteSetBstMidEndSepPunct{\mcitedefaultmidpunct}
{\mcitedefaultendpunct}{\mcitedefaultseppunct}\relax
\EndOfBibitem
\bibitem[Werner \latin{et~al.}(2012)Werner, Knowles, Knizia, Manby, and
  Sch\"{u}tz]{MOLPRO_Werner_2012_WIREsCMS.2.242}
Werner,~H.-J.; Knowles,~P.~J.; Knizia,~G.; Manby,~F.~R.; Sch\"{u}tz,~M.
  {Molpro: A General-Purpose Quantum Chemistry Program Package}. \emph{WIREs
  Comput. Mol. Sci.} \textbf{2012}, \emph{2}, 242--253\relax
\mciteBstWouldAddEndPuncttrue
\mciteSetBstMidEndSepPunct{\mcitedefaultmidpunct}
{\mcitedefaultendpunct}{\mcitedefaultseppunct}\relax
\EndOfBibitem
\bibitem[Kresse and Furthm\"{u}ller(1996)Kresse, and
  Furthm\"{u}ller]{VASP_Kresse_1996_CompMatSci.6.15}
Kresse,~G.; Furthm\"{u}ller,~J. {Efficiency of Ab-Initio Total Energy
  Calculations for Metals and Semiconductors Using a Plane-Wave Basis Set}.
  \emph{Comput. Mater. Sci.} \textbf{1996}, \emph{6}, 15--50\relax
\mciteBstWouldAddEndPuncttrue
\mciteSetBstMidEndSepPunct{\mcitedefaultmidpunct}
{\mcitedefaultendpunct}{\mcitedefaultseppunct}\relax
\EndOfBibitem
\bibitem[Marques \latin{et~al.}(2006)Marques, Ullrich, Nogueira, Rubio, Burke,
  and Gross]{Marques_TDDFT_book_2006}
Marques,~M. A.~L., Ullrich,~C.~A., Nogueira,~F., Rubio,~A., Burke,~K.,
  Gross,~E. K.~U., Eds. \emph{{Time-Dependent Density Functional Theory
  (Lecture Notes in Physics, vol. 706)}}; Springer, Berlin Heidelberg,
  2006\relax
\mciteBstWouldAddEndPuncttrue
\mciteSetBstMidEndSepPunct{\mcitedefaultmidpunct}
{\mcitedefaultendpunct}{\mcitedefaultseppunct}\relax
\EndOfBibitem
\bibitem[Adamo and Jacquemin(2013)Adamo, and
  Jacquemin]{Adamo_2013_ChemSocRev.42.845}
Adamo,~C.; Jacquemin,~D. {The Calculations of Excited-State Properties with
  Time-Dependent Density Functional Theory}. \emph{Chem. Soc. Rev.}
  \textbf{2013}, \emph{42}, 845--856\relax
\mciteBstWouldAddEndPuncttrue
\mciteSetBstMidEndSepPunct{\mcitedefaultmidpunct}
{\mcitedefaultendpunct}{\mcitedefaultseppunct}\relax
\EndOfBibitem
\bibitem[Lian \latin{et~al.}(218)Lian, Guan, Hu, Zhang, and
  Meng]{Lian_2018_AdvTheoSimul}
Lian,~C.; Guan,~M.; Hu,~S.; Zhang,~J.; Meng,~S. {Photoexcitation in Solids:
  First-Principles Quantum Simulations by Real-Time TDDFT}. \emph{Adv. Theory
  Simul.} \textbf{218}, \emph{1}, 1800055\relax
\mciteBstWouldAddEndPuncttrue
\mciteSetBstMidEndSepPunct{\mcitedefaultmidpunct}
{\mcitedefaultendpunct}{\mcitedefaultseppunct}\relax
\EndOfBibitem
\bibitem[Malcio\u{g}lu \latin{et~al.}(2011)Malcio\u{g}lu, Calzolari, Gebauer,
  Varsano, and Baroni]{Malcioglu_2011_JACS.133.15425}
Malcio\u{g}lu,~O.~B.; Calzolari,~A.; Gebauer,~R.; Varsano,~D.; Baroni,~S.
  {Dielectric and Thermal Effects on the Optical Properties of Natural Dyes: A
  Case Study on Solvated Cyanin}. \emph{J. Am. Chem. Soc.} \textbf{2011},
  \emph{133}, 15425--15433\relax
\mciteBstWouldAddEndPuncttrue
\mciteSetBstMidEndSepPunct{\mcitedefaultmidpunct}
{\mcitedefaultendpunct}{\mcitedefaultseppunct}\relax
\EndOfBibitem
\bibitem[Sjulstok \latin{et~al.}(2015)Sjulstok, Olsen, and
  Solov'yov]{sjulstok2015quantifying}
Sjulstok,~E.; Olsen,~J. M.~H.; Solov'yov,~I.~A. {Quantifying Electron Transfer
  Reactions in Biological Systems: What Interactions Play the Major Role?}
  \emph{Sci. Rep.} \textbf{2015}, \emph{5}, 18446\relax
\mciteBstWouldAddEndPuncttrue
\mciteSetBstMidEndSepPunct{\mcitedefaultmidpunct}
{\mcitedefaultendpunct}{\mcitedefaultseppunct}\relax
\EndOfBibitem
\bibitem[Shao \latin{et~al.}(2020)Shao, Mei, Sundholm, and
  Kaila]{Shao_2020_JCTC.16.587}
Shao,~Y.; Mei,~Y.; Sundholm,~D.; Kaila,~V. R.~I. {Benchmarking the Performance
  of Time-Dependent Density Functional Theory Methods on Biochromophores}.
  \emph{J. Chem. Theory Comput.} \textbf{2020}, \emph{16}, 587--600\relax
\mciteBstWouldAddEndPuncttrue
\mciteSetBstMidEndSepPunct{\mcitedefaultmidpunct}
{\mcitedefaultendpunct}{\mcitedefaultseppunct}\relax
\EndOfBibitem
\bibitem[Timrov \latin{et~al.}(2013)Timrov, Vast, Gebauer, and
  Baroni]{Timrov_2013_PRB.88.064301}
Timrov,~I.; Vast,~N.; Gebauer,~R.; Baroni,~S. {Electron Energy Loss and
  Inelastic X-Ray Scattering Cross Sections From Time-Dependent
  Density-Functional Perturbation Theory}. \emph{Phys. Rev. B} \textbf{2013},
  \emph{88}, 064301\relax
\mciteBstWouldAddEndPuncttrue
\mciteSetBstMidEndSepPunct{\mcitedefaultmidpunct}
{\mcitedefaultendpunct}{\mcitedefaultseppunct}\relax
\EndOfBibitem
\bibitem[Nicholls(2021)]{Nicholls_2021_JPhysMater}
Nicholls,~R.~J. {Advances in Modelling Electron Energy Loss Spectra From First
  Principles}. \emph{J. Phys. Mater.} \textbf{2021}, \emph{4}, 024008\relax
\mciteBstWouldAddEndPuncttrue
\mciteSetBstMidEndSepPunct{\mcitedefaultmidpunct}
{\mcitedefaultendpunct}{\mcitedefaultseppunct}\relax
\EndOfBibitem
\bibitem[Moitra \latin{et~al.}(2023)Moitra, Konecny, Kadek, Rubio, and
  Repisky]{Moitra_2023_JPCL.14.1714}
Moitra,~T.; Konecny,~L.; Kadek,~M.; Rubio,~A.; Repisky,~M. {Accurate
  Relativistic Real-Time Time-Dependent Density Functional Theory for Valence
  and Core Attosecond Transient Absorption Spectroscopy}. \emph{J. Phys. Chem.
  Lett.} \textbf{2023}, \emph{14}, 1714--1724\relax
\mciteBstWouldAddEndPuncttrue
\mciteSetBstMidEndSepPunct{\mcitedefaultmidpunct}
{\mcitedefaultendpunct}{\mcitedefaultseppunct}\relax
\EndOfBibitem
\bibitem[Ghosal and Roy(2022)Ghosal, and Roy]{Ghosal_2022_CPL.796.139562}
Ghosal,~A.; Roy,~A.~K. {A Real-Time TDDFT Scheme for Strong-Field Interaction
  in Cartesian Coordinate Grid}. \emph{Chem. Phys. Lett.} \textbf{2022},
  \emph{796}, 139562\relax
\mciteBstWouldAddEndPuncttrue
\mciteSetBstMidEndSepPunct{\mcitedefaultmidpunct}
{\mcitedefaultendpunct}{\mcitedefaultseppunct}\relax
\EndOfBibitem
\bibitem[Alberg-Fl{\o}jborg \latin{et~al.}(2020)Alberg-Fl{\o}jborg, Salo, and
  Solov'yov]{Alberg-Flojborg_JPB_2020}
Alberg-Fl{\o}jborg,~A.; Salo,~A.~B.; Solov'yov,~I.~A. {Quantum Mechanical
  Simulations of a Carbon Ion Colliding With a Biological Target}. \emph{J.
  Phys. B: At. Mol. Opt. Phys.} \textbf{2020}, \emph{53}, 145202\relax
\mciteBstWouldAddEndPuncttrue
\mciteSetBstMidEndSepPunct{\mcitedefaultmidpunct}
{\mcitedefaultendpunct}{\mcitedefaultseppunct}\relax
\EndOfBibitem
\bibitem[Salo \latin{et~al.}(2018)Salo, Alberg-Fl{\o}jborg, and
  Solov'yov]{Salo_2018_PRA.98.012702}
Salo,~A.~B.; Alberg-Fl{\o}jborg,~A.; Solov'yov,~I.~A. {Free-Electron Production
  From Nucleotides Upon Collision With Charged Carbon Ions}. \emph{Phys. Rev.
  A} \textbf{2018}, \emph{98}, 012702\relax
\mciteBstWouldAddEndPuncttrue
\mciteSetBstMidEndSepPunct{\mcitedefaultmidpunct}
{\mcitedefaultendpunct}{\mcitedefaultseppunct}\relax
\EndOfBibitem
\bibitem[Friedrich(2013)]{Friedrich_ScatTheory_book}
Friedrich,~H. \emph{{Scattering Theory (Lecture Notes in Physics, vol. 872)}};
  Springer Berlin, Heidelberg, 2013\relax
\mciteBstWouldAddEndPuncttrue
\mciteSetBstMidEndSepPunct{\mcitedefaultmidpunct}
{\mcitedefaultendpunct}{\mcitedefaultseppunct}\relax
\EndOfBibitem
\bibitem[Dreizler \latin{et~al.}(2022)Dreizler, Kirchner, and
  L\"{u}dde]{Dreizler_QuantCollTheory_book}
Dreizler,~R.~M.; Kirchner,~T.; L\"{u}dde,~C. \emph{{Quantum Collision Theory of
  Nonrelativistic Particles}}; Springer-Verlag GmbH Germany, 2022\relax
\mciteBstWouldAddEndPuncttrue
\mciteSetBstMidEndSepPunct{\mcitedefaultmidpunct}
{\mcitedefaultendpunct}{\mcitedefaultseppunct}\relax
\EndOfBibitem
\bibitem[Wachter(2010)]{Wachter_RelativQM_book}
Wachter,~A. \emph{{Relativistic Quantum Mechanics}}; Springer Dordrecht,
  2010\relax
\mciteBstWouldAddEndPuncttrue
\mciteSetBstMidEndSepPunct{\mcitedefaultmidpunct}
{\mcitedefaultendpunct}{\mcitedefaultseppunct}\relax
\EndOfBibitem
\bibitem[Hofierka \latin{et~al.}(2022)Hofierka, Cunningham, Rawlins, Patterson,
  and Green]{Hofierka_2022_Nature.66.688}
Hofierka,~J.; Cunningham,~B.; Rawlins,~C.~M.; Patterson,~C.~H.; Green,~D.
  {Many-Body Theory of Positron Binding to Polyatomic Molecules}. \emph{Nature}
  \textbf{2022}, \emph{606}, 688--693\relax
\mciteBstWouldAddEndPuncttrue
\mciteSetBstMidEndSepPunct{\mcitedefaultmidpunct}
{\mcitedefaultendpunct}{\mcitedefaultseppunct}\relax
\EndOfBibitem
\bibitem[Rawlins \latin{et~al.}(2023)Rawlins, Hofierka, Cunningham, Patterson,
  and Green]{Rawlins_2023_PRL.130.263001}
Rawlins,~C.~M.; Hofierka,~J.; Cunningham,~B.; Patterson,~C.~H.; Green,~D.~G.
  {Many-Body Theory Calculations of Positron Scattering and Annihilation in
  H$_2$, N$_2$, and CH$_4$}. \emph{Phys. Rev. Lett.} \textbf{2023}, \emph{130},
  263001\relax
\mciteBstWouldAddEndPuncttrue
\mciteSetBstMidEndSepPunct{\mcitedefaultmidpunct}
{\mcitedefaultendpunct}{\mcitedefaultseppunct}\relax
\EndOfBibitem
\bibitem[Samanta \latin{et~al.}(2018)Samanta, Tsogbayar, Zhang, and
  Yeager]{Samanta_2018_AdvQuantChem.77.317}
Samanta,~K.; Tsogbayar,~T.; Zhang,~S.~B.; Yeager,~D.~L. {Electron--Atom and
  Electron--Molecule Resonances: Some Theoretical Approaches Using Complex
  Scaled Multiconfigurational Methods}. \emph{Adv. Quantum Chem.}
  \textbf{2018}, \emph{77}, 317--390\relax
\mciteBstWouldAddEndPuncttrue
\mciteSetBstMidEndSepPunct{\mcitedefaultmidpunct}
{\mcitedefaultendpunct}{\mcitedefaultseppunct}\relax
\EndOfBibitem
\bibitem[Schneider(1975)]{Schneider_1975_CPL.31.237}
Schneider,~B. {R-matrix Theory for Electron-Atom and Electron-Molecule
  Collisions Using Analytic Basis Set Expansions}. \emph{Chem. Phys. Lett.}
  \textbf{1975}, \emph{31}, 237--241\relax
\mciteBstWouldAddEndPuncttrue
\mciteSetBstMidEndSepPunct{\mcitedefaultmidpunct}
{\mcitedefaultendpunct}{\mcitedefaultseppunct}\relax
\EndOfBibitem
\bibitem[Burke \latin{et~al.}(1977)Burke, Mackey, and
  Shimamura]{Burke_1977_JPB.10.2497}
Burke,~P.~G.; Mackey,~I.; Shimamura,~I. {R-Matrix Theory of Electron--Molecule
  Scattering}. \emph{J. Phys. B: Atom. Mol. Phys.} \textbf{1977}, \emph{10},
  2497--2512\relax
\mciteBstWouldAddEndPuncttrue
\mciteSetBstMidEndSepPunct{\mcitedefaultmidpunct}
{\mcitedefaultendpunct}{\mcitedefaultseppunct}\relax
\EndOfBibitem
\bibitem[Tennyson(2010)]{Tennyson_2010_PhysRep.491.29}
Tennyson,~J. {Electron--Molecule Collision Calculations Using the R-matrix
  Method}. \emph{Phys. Rep.} \textbf{2010}, \emph{491}, 29--76\relax
\mciteBstWouldAddEndPuncttrue
\mciteSetBstMidEndSepPunct{\mcitedefaultmidpunct}
{\mcitedefaultendpunct}{\mcitedefaultseppunct}\relax
\EndOfBibitem
\bibitem[Otvos and Stevenson(1956)Otvos, and Stevenson]{Otvos_1956_JACS.78.546}
Otvos,~J.~W.; Stevenson,~D.~P. {Cross-Sections of Molecules for Ionization by
  Electrons}. \emph{J. Am. Chem. Soc.} \textbf{1956}, \emph{78}, 546--551\relax
\mciteBstWouldAddEndPuncttrue
\mciteSetBstMidEndSepPunct{\mcitedefaultmidpunct}
{\mcitedefaultendpunct}{\mcitedefaultseppunct}\relax
\EndOfBibitem
\bibitem[Deutsch \latin{et~al.}(2000)Deutsch, Becker, Matt, and
  M\"{a}rk]{Deutsch_Mark_2000_IJMS.197.37}
Deutsch,~H.; Becker,~K.; Matt,~S.; M\"{a}rk,~T.~D. {Theoretical Determination
  of Absolute Electron-Impact Ionization Cross Sections of Molecules}.
  \emph{Int. J. Mass Spectrom.} \textbf{2000}, \emph{197}, 37--69\relax
\mciteBstWouldAddEndPuncttrue
\mciteSetBstMidEndSepPunct{\mcitedefaultmidpunct}
{\mcitedefaultendpunct}{\mcitedefaultseppunct}\relax
\EndOfBibitem
\bibitem[Kim and Rudd(1994)Kim, and Rudd]{BEB_Kim_Rudd_1994_PRA.50.3954}
Kim,~Y.-K.; Rudd,~M.~E. {Binary-Encounter-Dipole Model for Electron-Impact
  Ionization}. \emph{Phys. Rev. A} \textbf{1994}, \emph{50}, 3954--3967\relax
\mciteBstWouldAddEndPuncttrue
\mciteSetBstMidEndSepPunct{\mcitedefaultmidpunct}
{\mcitedefaultendpunct}{\mcitedefaultseppunct}\relax
\EndOfBibitem
\bibitem[Tanaka \latin{et~al.}(2016)Tanaka, Brunger, Campbell, Kato, Hoshino,
  and Rau]{Tanaka_2016_RMP.88.025004}
Tanaka,~H.; Brunger,~M.~J.; Campbell,~L.; Kato,~H.; Hoshino,~M.; Rau,~A. R.~P.
  {Scaled Plane-Wave Born Cross Sections for Atoms and Molecules}. \emph{Rev.
  Mod. Phys.} \textbf{2016}, \emph{88}, 025004\relax
\mciteBstWouldAddEndPuncttrue
\mciteSetBstMidEndSepPunct{\mcitedefaultmidpunct}
{\mcitedefaultendpunct}{\mcitedefaultseppunct}\relax
\EndOfBibitem
\bibitem[Kim \latin{et~al.}(2000)Kim, Santos, and
  Parente]{RBEB_Kim_2000_PRA.62.052710}
Kim,~Y.-K.; Santos,~J.~P.; Parente,~F. {Extension of the
  Binary-Encounter-Dipole Model to Relativistic Incident Electrons}.
  \emph{Phys. Rev. A} \textbf{2000}, \emph{62}, 052710\relax
\mciteBstWouldAddEndPuncttrue
\mciteSetBstMidEndSepPunct{\mcitedefaultmidpunct}
{\mcitedefaultendpunct}{\mcitedefaultseppunct}\relax
\EndOfBibitem
\bibitem[Rudd \latin{et~al.}(1992)Rudd, Kim, Madison, and
  Gay]{Rudd_1992_RMP.64.441}
Rudd,~M.~E.; Kim,~Y.~K.; Madison,~D.~H.; Gay,~T. {Electron Production in Proton
  Collisions with Atoms and Molecules: Energy Distributions}. \emph{Rev. Mod.
  Phys.} \textbf{1992}, \emph{64}, 441--490\relax
\mciteBstWouldAddEndPuncttrue
\mciteSetBstMidEndSepPunct{\mcitedefaultmidpunct}
{\mcitedefaultendpunct}{\mcitedefaultseppunct}\relax
\EndOfBibitem
\bibitem[Zatsarinny(2006)]{BSR_Zatsarinny_2006_CPC.174.273}
Zatsarinny,~O. {BSR: B-spline Atomic R-matrix Codes}. \emph{Comput. Phys.
  Commun.} \textbf{2006}, \emph{174}, 273--356\relax
\mciteBstWouldAddEndPuncttrue
\mciteSetBstMidEndSepPunct{\mcitedefaultmidpunct}
{\mcitedefaultendpunct}{\mcitedefaultseppunct}\relax
\EndOfBibitem
\bibitem[Bray \latin{et~al.}(2002)Bray, Fursa, Kheifets, and
  Stelbovics]{CCC_Bray_2002_JPB.35.R117}
Bray,~I.; Fursa,~D.~V.; Kheifets,~A.~S.; Stelbovics,~A.~T. {Electrons and
  Photons Colliding with Atoms: Development and Application of the Convergent
  Close-Coupling Method}. \emph{J. Phys. B: At. Mol. Opt. Phys.} \textbf{2002},
  \emph{35}, R117--R146\relax
\mciteBstWouldAddEndPuncttrue
\mciteSetBstMidEndSepPunct{\mcitedefaultmidpunct}
{\mcitedefaultendpunct}{\mcitedefaultseppunct}\relax
\EndOfBibitem
\bibitem[Gianturco \latin{et~al.}(1994)Gianturco, Lucchese, and
  Sanna]{ePolyScat_Gianturco_1994_JCP}
Gianturco,~F.~A.; Lucchese,~R.~R.; Sanna,~N. {Calculation of Low-Energy Elastic
  Cross Sections for Electron--CF$_4$ Scattering}. \emph{J. Chem. Phys.}
  \textbf{1994}, \emph{100}, 6464--6471\relax
\mciteBstWouldAddEndPuncttrue
\mciteSetBstMidEndSepPunct{\mcitedefaultmidpunct}
{\mcitedefaultendpunct}{\mcitedefaultseppunct}\relax
\EndOfBibitem
\bibitem[Ma\v{s}\'{i}n \latin{et~al.}(2020)Ma\v{s}\'{i}n, Benda, Gorfinkiel,
  Harvey, and Tennyson]{UKRmol_Masin_2020_CPC}
Ma\v{s}\'{i}n,~Z.; Benda,~J.; Gorfinkiel,~J.~D.; Harvey,~A.~G.; Tennyson,~J.
  {UKRmol+: A Suite for Modelling Electronic Processes in Molecules Interacting
  With Electrons, Positrons and Photons Using the R-matrix Method}.
  \emph{Comput. Phys. Commun.} \textbf{2020}, \emph{249}, 107092\relax
\mciteBstWouldAddEndPuncttrue
\mciteSetBstMidEndSepPunct{\mcitedefaultmidpunct}
{\mcitedefaultendpunct}{\mcitedefaultseppunct}\relax
\EndOfBibitem
\bibitem[Borr\`{a}s \latin{et~al.}(2021)Borr\`{a}s, Gonz\'{a}lez-V\'{a}zquez,
  Argenti, and Mart\'{i}n]{xCHEM_Borras_2021_JCTC}
Borr\`{a}s,~V.~J.; Gonz\'{a}lez-V\'{a}zquez,~J.; Argenti,~L.; Mart\'{i}n,~F.
  {Molecular-Frame Photoelectron Angular Distributions of CO in the Vicinity of
  Feshbach Resonances: An XCHEM Approach}. \emph{J. Chem. Theory Comput.}
  \textbf{2021}, \emph{17}, 6330--6339\relax
\mciteBstWouldAddEndPuncttrue
\mciteSetBstMidEndSepPunct{\mcitedefaultmidpunct}
{\mcitedefaultendpunct}{\mcitedefaultseppunct}\relax
\EndOfBibitem
\bibitem[AMO((accessed 2023-11-15))]{AMOS_Gateway}
AMOS Gateway. A Portal for Research and Education in Atomic, Molecular, and
  Optical Science. (accessed 2023-11-15); \url{https://amosgateway.org/}\relax
\mciteBstWouldAddEndPuncttrue
\mciteSetBstMidEndSepPunct{\mcitedefaultmidpunct}
{\mcitedefaultendpunct}{\mcitedefaultseppunct}\relax
\EndOfBibitem
\bibitem[Hamilton \latin{et~al.}(2023)Hamilton, Bartschat, Douguet,
  Pamidighantam, and Schneider]{AMOS_Gateway_paper2023}
Hamilton,~K.~R.; Bartschat,~K.; Douguet,~N.; Pamidighantam,~S.~V.;
  Schneider,~B.~I. {Simulation for All: The Atomic, Molecular, and Optical
  Science Gateway}. \emph{Comput. Sci. Eng.} \textbf{2023}, \emph{25},
  68--72\relax
\mciteBstWouldAddEndPuncttrue
\mciteSetBstMidEndSepPunct{\mcitedefaultmidpunct}
{\mcitedefaultendpunct}{\mcitedefaultseppunct}\relax
\EndOfBibitem
\bibitem[Amusia(1990)]{Amusia_Photoeffect_book}
Amusia,~M.~Y. \emph{{Atomic Photoeffect}}; Plenum Press, New York, 1990\relax
\mciteBstWouldAddEndPuncttrue
\mciteSetBstMidEndSepPunct{\mcitedefaultmidpunct}
{\mcitedefaultendpunct}{\mcitedefaultseppunct}\relax
\EndOfBibitem
\bibitem[McLaughlin and Balance(2014)McLaughlin, and Balance]{DARC_McLaughlin}
McLaughlin,~B.; Balance,~C.~P. In \emph{Sustained Simulation Performance 2014};
  Resch,~M.~M., Bez,~W., Focht,~E., Kobayashi,~H., Patel,~N., Eds.; Springer,
  Cham, 2014; pp 173--185\relax
\mciteBstWouldAddEndPuncttrue
\mciteSetBstMidEndSepPunct{\mcitedefaultmidpunct}
{\mcitedefaultendpunct}{\mcitedefaultseppunct}\relax
\EndOfBibitem
\bibitem[Cooper \latin{et~al.}(2019)Cooper, Tudorovskaya, Mohr, O'Hare,
  Hanicinec, Dzarasova, Gorfinkiel, Benda, Ma\v{s}\'{i}n, Al-Refaie, Knowles,
  and Tennyson]{Quantemol-EC_Cooper_2019}
Cooper,~B.; Tudorovskaya,~M.; Mohr,~S.; O'Hare,~A.; Hanicinec,~M.;
  Dzarasova,~A.; Gorfinkiel,~J.~D.; Benda,~J.; Ma\v{s}\'{i}n,~Z.;
  Al-Refaie,~A.~F.; Knowles,~P.~J.; Tennyson,~J. {Quantemol Electron Collisions
  (QEC): An Enhanced Expert System for Performing Electron Molecule Collision
  Calculations Using the R-matrix Method}. \emph{Atoms} \textbf{2019},
  \emph{7}, 97\relax
\mciteBstWouldAddEndPuncttrue
\mciteSetBstMidEndSepPunct{\mcitedefaultmidpunct}
{\mcitedefaultendpunct}{\mcitedefaultseppunct}\relax
\EndOfBibitem
\bibitem[Bernhardt and Paretzke(2003)Bernhardt, and
  Paretzke]{Bernhardt_2003_IJMS.223.599}
Bernhardt,~P.; Paretzke,~H.~G. {Calculation of Electron Impact Ionization Cross
  Sections of DNA Using the Deutsch--M\"{a}rk and Binary--Encounter--Bethe
  Formalisms}. \emph{Int. J. Mass Spectrom.} \textbf{2003}, \emph{223-224},
  599--611\relax
\mciteBstWouldAddEndPuncttrue
\mciteSetBstMidEndSepPunct{\mcitedefaultmidpunct}
{\mcitedefaultendpunct}{\mcitedefaultseppunct}\relax
\EndOfBibitem
\bibitem[Huber and Mauracher(2019)Huber, and Mauracher]{Huber_2019_EPJD.73.137}
Huber,~S.~E.; Mauracher,~A. {Electron Impact Ionisation Cross Sections of
  Fluoro-Substituted Nucleosides}. \emph{Eur. Phys. J. D} \textbf{2019},
  \emph{73}, 137\relax
\mciteBstWouldAddEndPuncttrue
\mciteSetBstMidEndSepPunct{\mcitedefaultmidpunct}
{\mcitedefaultendpunct}{\mcitedefaultseppunct}\relax
\EndOfBibitem
\bibitem[Langer \latin{et~al.}(2018)Langer, Zawadzki, F\'{a}rn\'{i}k, Pinkas,
  Fedor, and Ko\v{c}i\v{s}ek]{Langer_2018_EPJD.72.112}
Langer,~J.; Zawadzki,~M.; F\'{a}rn\'{i}k,~M.; Pinkas,~J.; Fedor,~J.;
  Ko\v{c}i\v{s}ek,~J. {Electron Interactions with
  Bis(pentamethylcyclopentadienyl) Titanium(IV) Dichloride and Difluoride}.
  \emph{Eur. Phys. J. D} \textbf{2018}, \emph{72}, 112\relax
\mciteBstWouldAddEndPuncttrue
\mciteSetBstMidEndSepPunct{\mcitedefaultmidpunct}
{\mcitedefaultendpunct}{\mcitedefaultseppunct}\relax
\EndOfBibitem
\bibitem[Francis \latin{et~al.}(2017)Francis, Bitar, Incerti, Bernal,
  Karamitros, and Tran]{Francis_2017_JAP.122.014701}
Francis,~Z.; Bitar,~Z.~E.; Incerti,~S.; Bernal,~M.~A.; Karamitros,~M.;
  Tran,~H.~N. {Calculation of Lineal Energies for Water and DNA Bases Using the
  Rudd Model Cross Sections Integrated Within the Geant4-DNA Processes}.
  \emph{J. Appl. Phys.} \textbf{2017}, \emph{122}, 014701\relax
\mciteBstWouldAddEndPuncttrue
\mciteSetBstMidEndSepPunct{\mcitedefaultmidpunct}
{\mcitedefaultendpunct}{\mcitedefaultseppunct}\relax
\EndOfBibitem
\bibitem[Kawrakow and Bielajew(1998)Kawrakow, and
  Bielajew]{Kawrakow_1998_NIMB.142.253}
Kawrakow,~I.; Bielajew,~A.~F. {On the Condensed History Technique for Electron
  Transport}. \emph{Nucl. Instrum. Meth. B} \textbf{1998}, \emph{142},
  253--280\relax
\mciteBstWouldAddEndPuncttrue
\mciteSetBstMidEndSepPunct{\mcitedefaultmidpunct}
{\mcitedefaultendpunct}{\mcitedefaultseppunct}\relax
\EndOfBibitem
\bibitem[Jenkins \latin{et~al.}(1988)Jenkins, Nelson, and
  Rindi]{Jenkins_MC_transport_book}
Jenkins,~T.~M., Nelson,~W.~R., Rindi,~A., Eds. \emph{{Monte Carlo Transport of
  Electrons and Photons}}; Springer New York, NY, 1988\relax
\mciteBstWouldAddEndPuncttrue
\mciteSetBstMidEndSepPunct{\mcitedefaultmidpunct}
{\mcitedefaultendpunct}{\mcitedefaultseppunct}\relax
\EndOfBibitem
\bibitem[Berger \latin{et~al.}((accessed 2023-09-08))Berger, Coursey, Zucker,
  and Chang]{NIST_Stopping_Power}
Berger,~M.~J.; Coursey,~J.~S.; Zucker,~M.~A.; Chang,~J. Stopping-Power \& Range
  Tables for Electrons, Protons, and Helium Ions (NIST Standard Reference
  Database 124). (accessed 2023-09-08);
  \url{https://www.nist.gov/pml/stopping-power-range-tables-electrons-protons-and-helium-ions}\relax
\mciteBstWouldAddEndPuncttrue
\mciteSetBstMidEndSepPunct{\mcitedefaultmidpunct}
{\mcitedefaultendpunct}{\mcitedefaultseppunct}\relax
\EndOfBibitem
\bibitem[Dingfelder(2012)]{Dingfelder_2012_HealthPhys.103.590}
Dingfelder,~M. {Track-structure Simulations for Charged Particles}.
  \emph{Health Phys.} \textbf{2012}, \emph{103}, 590--595\relax
\mciteBstWouldAddEndPuncttrue
\mciteSetBstMidEndSepPunct{\mcitedefaultmidpunct}
{\mcitedefaultendpunct}{\mcitedefaultseppunct}\relax
\EndOfBibitem
\bibitem[B\"{o}hlen \latin{et~al.}(2014)B\"{o}hlen, Cerutti, Chin, Fass\`{o},
  Ferrari, Ortega, Mairani, Sala, Smirnov, and Vlachoudis]{FLUKA_Bohlen_2014}
B\"{o}hlen,~T.~T.; Cerutti,~F.; Chin,~M. P.~W.; Fass\`{o},~A.; Ferrari,~A.;
  Ortega,~P.~G.; Mairani,~A.; Sala,~P.~R.; Smirnov,~G.; Vlachoudis,~V. {The
  FLUKA Code: Developments and Challenges for High Energy and Medical
  Applications}. \emph{Nucl. Data Sheets} \textbf{2014}, \emph{120},
  211--214\relax
\mciteBstWouldAddEndPuncttrue
\mciteSetBstMidEndSepPunct{\mcitedefaultmidpunct}
{\mcitedefaultendpunct}{\mcitedefaultseppunct}\relax
\EndOfBibitem
\bibitem[Ziegler \latin{et~al.}(2010)Ziegler, Ziegler, and
  Biersack]{SRIM_Ziegler_2010_NIMB.268.1818}
Ziegler,~J.~F.; Ziegler,~M.~D.; Biersack,~J.~P. {SRIM -- The Stopping and Range
  of Ions in Matter (2010)}. \emph{Nucl. Instrum. Meth. B} \textbf{2010},
  \emph{268}, 1818--1823\relax
\mciteBstWouldAddEndPuncttrue
\mciteSetBstMidEndSepPunct{\mcitedefaultmidpunct}
{\mcitedefaultendpunct}{\mcitedefaultseppunct}\relax
\EndOfBibitem
\bibitem[Salvat \latin{et~al.}((accessed on 2023-06-13))Salvat,
  Fern\'{a}ndez-Varea, and Sempau]{PENELOPE_Salvat}
Salvat,~F.; Fern\'{a}ndez-Varea,~J.~M.; Sempau,~J. PENELOPE-2011: A Code System
  for Monte Carlo Simulation of Electron and Photon Transport. (accessed on
  2023-06-13);
  \url{https://www.oecd-nea.org/science/docs/2011/nsc-doc2011-5}\relax
\mciteBstWouldAddEndPuncttrue
\mciteSetBstMidEndSepPunct{\mcitedefaultmidpunct}
{\mcitedefaultendpunct}{\mcitedefaultseppunct}\relax
\EndOfBibitem
\bibitem[Agostinelli \latin{et~al.}(2003)Agostinelli, Allison, Amako,
  Apostolakis, Araujo, Arce, Asai, Axen, Banerjee, Barrand, and \textit{et
  al.}]{GEANT4_Agostinelli_2003_NIMA}
Agostinelli,~S.; Allison,~J.; Amako,~K.; Apostolakis,~J.; Araujo,~H.; Arce,~P.;
  Asai,~M.; Axen,~D.; Banerjee,~S.; Barrand,~G.; \textit{et al.} {GEANT4 -- A
  Simulation Toolkit}. \emph{Nucl. Instrum. Meth. A} \textbf{2003}, \emph{506},
  250--303\relax
\mciteBstWouldAddEndPuncttrue
\mciteSetBstMidEndSepPunct{\mcitedefaultmidpunct}
{\mcitedefaultendpunct}{\mcitedefaultseppunct}\relax
\EndOfBibitem
\bibitem[Uehara \latin{et~al.}(1992)Uehara, Nikjoo, and
  Goodhead]{Uehara_1992_PMB.37.1841}
Uehara,~S.; Nikjoo,~H.; Goodhead,~D.~T. {Cross-Sections for Water Vapour for
  the Monte Carlo Electron Track Structure Code From 10 eV to the MeV Region}.
  \emph{Phys. Med. Biol.} \textbf{1992}, \emph{37}, 1841--1858\relax
\mciteBstWouldAddEndPuncttrue
\mciteSetBstMidEndSepPunct{\mcitedefaultmidpunct}
{\mcitedefaultendpunct}{\mcitedefaultseppunct}\relax
\EndOfBibitem
\bibitem[Friedland \latin{et~al.}(2011)Friedland, Dingfelder, Kundr\'{a}t, and
  Jacob]{PARTRAC_Friedland_2011_MutatRes}
Friedland,~W.; Dingfelder,~M.; Kundr\'{a}t,~P.; Jacob,~P. {Track Structures,
  DNA Targets and Radiation Effects in the Biophysical Monte Carlo Simulation
  Code PARTRAC}. \emph{Mutat. Res.} \textbf{2011}, \emph{711}, 28--40\relax
\mciteBstWouldAddEndPuncttrue
\mciteSetBstMidEndSepPunct{\mcitedefaultmidpunct}
{\mcitedefaultendpunct}{\mcitedefaultseppunct}\relax
\EndOfBibitem
\bibitem[Pimblott and Mozumder(1991)Pimblott, and
  Mozumder]{Pimblott_1991_JPC.95.7291}
Pimblott,~S.~M.; Mozumder,~A. {Structure of Electron Tracks in Water. 2.
  Distribution of Primary Ionizations and Excitations in Water Radiolysis}.
  \emph{J. Phys. Chem.} \textbf{1991}, \emph{95}, 7291--7300\relax
\mciteBstWouldAddEndPuncttrue
\mciteSetBstMidEndSepPunct{\mcitedefaultmidpunct}
{\mcitedefaultendpunct}{\mcitedefaultseppunct}\relax
\EndOfBibitem
\bibitem[Champion \latin{et~al.}(2012)Champion, Loirec, and
  Stosic]{EPOTRAN_Champion_2012_IJRB}
Champion,~C.; Loirec,~C.~L.; Stosic,~B. {EPOTRAN: A Full-differential Monte
  Carlo Code for Electron and Positron Transport in Liquid and Gaseous Water}.
  \emph{Int. J. Radiat. Biol.} \textbf{2012}, \emph{88}, 54--61\relax
\mciteBstWouldAddEndPuncttrue
\mciteSetBstMidEndSepPunct{\mcitedefaultmidpunct}
{\mcitedefaultendpunct}{\mcitedefaultseppunct}\relax
\EndOfBibitem
\bibitem[Incerti \latin{et~al.}(2010)Incerti, Baldacchino, Bernal, Capra,
  Champion, Francis, Gu\`{e}ye, Mantero, Mascialino, Moretto, and
  et~al.]{Incerti_2010_Geant4-DNA}
Incerti,~S.; Baldacchino,~G.; Bernal,~M.; Capra,~R.; Champion,~C.; Francis,~Z.;
  Gu\`{e}ye,~P.; Mantero,~A.; Mascialino,~B.; Moretto,~P.; et~al. {The
  Geant4-DNA Project}. \emph{Int. J. Model. Simul. Sci. Comput.} \textbf{2010},
  \emph{1}, 157--178\relax
\mciteBstWouldAddEndPuncttrue
\mciteSetBstMidEndSepPunct{\mcitedefaultmidpunct}
{\mcitedefaultendpunct}{\mcitedefaultseppunct}\relax
\EndOfBibitem
\bibitem[Bernal \latin{et~al.}(2015)Bernal, Bordage, Brown, Dav\'{i}dkov\'{a},
  Delage, Bitar, Enger, Francis, Guatelli, Ivanchenko, and \textit{et
  al.}]{Bernal_2015_PhysMed.31.861}
Bernal,~M.~A.; Bordage,~M.~C.; Brown,~J. M.~C.; Dav\'{i}dkov\'{a},~M.;
  Delage,~E.; Bitar,~Z.~E.; Enger,~S.~A.; Francis,~Z.; Guatelli,~S.;
  Ivanchenko,~V.~N.; \textit{et al.} {Track Structure Modeling in Liquid Water:
  A Review of the Geant4-DNA Very Low Energy Extension of the Geant4 Monte
  Carlo Simulation Toolkit}. \emph{Phys. Med.} \textbf{2015}, \emph{31},
  861--874\relax
\mciteBstWouldAddEndPuncttrue
\mciteSetBstMidEndSepPunct{\mcitedefaultmidpunct}
{\mcitedefaultendpunct}{\mcitedefaultseppunct}\relax
\EndOfBibitem
\bibitem[Zein \latin{et~al.}(2021)Zein, Bordage, Francis, Macetti, Genoni,
  Cappello, Shin, and Incerti]{Zein_2021_NIMB.488.70}
Zein,~S.~A.; Bordage,~M.-C.; Francis,~Z.; Macetti,~G.; Genoni,~A.;
  Cappello,~C.~D.; Shin,~W.-G.; Incerti,~S. {Electron Transport in DNA Bases:
  An Extension of the Geant4-DNA Monte Carlo Toolkit}. \emph{Nucl. Instrum.
  Meth. B} \textbf{2021}, \emph{488}, 70--82\relax
\mciteBstWouldAddEndPuncttrue
\mciteSetBstMidEndSepPunct{\mcitedefaultmidpunct}
{\mcitedefaultendpunct}{\mcitedefaultseppunct}\relax
\EndOfBibitem
\bibitem[Kundr\'{a}t(2007)]{Kundrat_2007_PMB.52.6813}
Kundr\'{a}t,~P. {A Semi-Analytical Radiobiological Model May Assist Treatment
  Planning in Light Ion Radiotherapy}. \emph{Phys. Med. Biol.} \textbf{2007},
  \emph{52}, 6813--6830\relax
\mciteBstWouldAddEndPuncttrue
\mciteSetBstMidEndSepPunct{\mcitedefaultmidpunct}
{\mcitedefaultendpunct}{\mcitedefaultseppunct}\relax
\EndOfBibitem
\bibitem[Korol \latin{et~al.}(2021)Korol, Sushko, and
  Solov'yov]{RelMD_2021_EPJD.75.107_review}
Korol,~A.~V.; Sushko,~G.~B.; Solov'yov,~A.~V. {All-Atom Relativistic Molecular
  Dynamics Simulations of Channeling and Radiation Processes in Oriented
  Crystals}. \emph{Eur. Phys. J. D} \textbf{2021}, \emph{75}, 107\relax
\mciteBstWouldAddEndPuncttrue
\mciteSetBstMidEndSepPunct{\mcitedefaultmidpunct}
{\mcitedefaultendpunct}{\mcitedefaultseppunct}\relax
\EndOfBibitem
\bibitem[Solov'yov \latin{et~al.}(2023)Solov'yov, Sushko, Verkhovtsev, and
  an~A.~V.~Solov'yov]{MBNTutorials_2023_RelMD}
Solov'yov,~I.~A.; Sushko,~G.~B.; Verkhovtsev,~A.~V.; an~A.~V.~Solov'yov,~A.
  V.~K. \emph{{MBN Explorer and MBN Studio Tutorials: Version 5.0, Chapter 10
  ``Relativistic molecular dynamics''}}; 2023\relax
\mciteBstWouldAddEndPuncttrue
\mciteSetBstMidEndSepPunct{\mcitedefaultmidpunct}
{\mcitedefaultendpunct}{\mcitedefaultseppunct}\relax
\EndOfBibitem
\bibitem[Korol \latin{et~al.}(2001)Korol, Solov'yov, and
  Greiner]{Korol_2001_JPG.27.95}
Korol,~A.~V.; Solov'yov,~A.~V.; Greiner,~W. {The Influence of the Dechannelling
  Process on the Photon Emission by an Ultra-Relativistic Positron Channelling
  in a Periodically Bent Crystal}. \emph{J. Phys. G: Nucl. Part. Phys.}
  \textbf{2001}, \emph{27}, 95--125\relax
\mciteBstWouldAddEndPuncttrue
\mciteSetBstMidEndSepPunct{\mcitedefaultmidpunct}
{\mcitedefaultendpunct}{\mcitedefaultseppunct}\relax
\EndOfBibitem
\bibitem[Sushko \latin{et~al.}(2023)Sushko, Korol, and
  Solov'yov]{Sushko_2023_NIMB.535.117}
Sushko,~G.~B.; Korol,~A.~V.; Solov'yov,~A.~V. {Atomistic Modeling of the
  Channeling Process with Radiation Reaction Force Included}. \emph{Nucl.
  Instrum. Meth. B} \textbf{2023}, \emph{535}, 117--125\relax
\mciteBstWouldAddEndPuncttrue
\mciteSetBstMidEndSepPunct{\mcitedefaultmidpunct}
{\mcitedefaultendpunct}{\mcitedefaultseppunct}\relax
\EndOfBibitem
\bibitem[Born and Oppenheimer(1927)Born, and
  Oppenheimer]{Born_Oppenheimer_1927}
Born,~M.; Oppenheimer,~J.~R. {Zur Quantentheorie der Molekeln}. \emph{Ann.
  Phys.} \textbf{1927}, \emph{389}, 457--484\relax
\mciteBstWouldAddEndPuncttrue
\mciteSetBstMidEndSepPunct{\mcitedefaultmidpunct}
{\mcitedefaultendpunct}{\mcitedefaultseppunct}\relax
\EndOfBibitem
\bibitem[{MacKerell, Jr.} \latin{et~al.}(1998){MacKerell, Jr.}, Bashford,
  Bellott, {Dunbrack, Jr.}, Evanseck, Field, Fischer, Gao, Guo, Ha, and
  \textit{et al.}]{CHARMM_MacKerell_1998_JPCB.102.3586}
{MacKerell, Jr.},~A.~D.; Bashford,~D.; Bellott,~M.; {Dunbrack, Jr.},~R.~L.;
  Evanseck,~J.~D.; Field,~M.~J.; Fischer,~S.; Gao,~J.; Guo,~H.; Ha,~S.;
  \textit{et al.} {All-Atom Empirical Potential for Molecular Modeling and
  Dynamics Studies of Proteins}. \emph{J. Phys. Chem. B} \textbf{1998},
  \emph{102}, 3586--3616\relax
\mciteBstWouldAddEndPuncttrue
\mciteSetBstMidEndSepPunct{\mcitedefaultmidpunct}
{\mcitedefaultendpunct}{\mcitedefaultseppunct}\relax
\EndOfBibitem
\bibitem[{MacKerell Jr.} \latin{et~al.}(2004){MacKerell Jr.}, Feig, and {Brooks
  III}]{MacKerell_2004_JCC.25.1400}
{MacKerell Jr.},~A.~D.; Feig,~M.; {Brooks III},~C.~L. {Extending the Treatment
  of Backbone Energetics in Protein Force Fields: Limitations of Gas-Phase
  Quantum Mechanics in Reproducing Protein Conformational Distributions in
  Molecular Dynamics Simulations}. \emph{J. Comput. Chem.} \textbf{2004},
  \emph{25}, 1400--1415\relax
\mciteBstWouldAddEndPuncttrue
\mciteSetBstMidEndSepPunct{\mcitedefaultmidpunct}
{\mcitedefaultendpunct}{\mcitedefaultseppunct}\relax
\EndOfBibitem
\bibitem[Cornell \latin{et~al.}(1995)Cornell, Cieplak, Bayly, Gould, {Merz Jr},
  Ferguson, Spellmeyer, Fox, Caldwell, and Kollman]{Cornell_1995_JACS.117.5179}
Cornell,~W.~D.; Cieplak,~P.; Bayly,~C.~I.; Gould,~I.~R.; {Merz Jr},~K.~M.;
  Ferguson,~D.~M.; Spellmeyer,~D.~C.; Fox,~T.; Caldwell,~J.~W.; Kollman,~P.~A.
  {A Second Generation Force Field for the Simulation of Proteins, Nucleic
  Acids, and Organic Molecules}. \emph{J. Am. Chem. Soc.} \textbf{1995},
  \emph{117}, 5179--5197\relax
\mciteBstWouldAddEndPuncttrue
\mciteSetBstMidEndSepPunct{\mcitedefaultmidpunct}
{\mcitedefaultendpunct}{\mcitedefaultseppunct}\relax
\EndOfBibitem
\bibitem[Tersoff(1988)]{REBO_Tersoff_1988_PRL.61.2879}
Tersoff,~J. {Empirical Interatomic Potential for Carbon, with Applications to
  Amorphous Carbon}. \emph{Phys. Rev. Lett.} \textbf{1988}, \emph{61},
  2879--2882\relax
\mciteBstWouldAddEndPuncttrue
\mciteSetBstMidEndSepPunct{\mcitedefaultmidpunct}
{\mcitedefaultendpunct}{\mcitedefaultseppunct}\relax
\EndOfBibitem
\bibitem[Tersoff(1988)]{REBO_Tersoff_1988_PRB.37.6991}
Tersoff,~J. {New Empirical Approach for the Structure and Energy of Covalent
  Systems}. \emph{Phys. Rev. B} \textbf{1988}, \emph{37}, 6991--7000\relax
\mciteBstWouldAddEndPuncttrue
\mciteSetBstMidEndSepPunct{\mcitedefaultmidpunct}
{\mcitedefaultendpunct}{\mcitedefaultseppunct}\relax
\EndOfBibitem
\bibitem[Brenner(1990)]{REBO_Brenner_1990_PRB.42.9458}
Brenner,~D.~W. {Empirical Potential for Hydrocarbons for Use in Simulating the
  Chemical Vapor Deposition of Diamond Films}. \emph{Phys. Rev. B}
  \textbf{1990}, \emph{42}, 9458--9471\relax
\mciteBstWouldAddEndPuncttrue
\mciteSetBstMidEndSepPunct{\mcitedefaultmidpunct}
{\mcitedefaultendpunct}{\mcitedefaultseppunct}\relax
\EndOfBibitem
\bibitem[Stuart \latin{et~al.}(2000)Stuart, Tutein, and
  Harrison]{AIREBO_Stuart_2000_JCP.112.6472}
Stuart,~S.~J.; Tutein,~A.~B.; Harrison,~J.~A. {A Reactive Potential for
  Hydrocarbons with Intermolecular Interactions}. \emph{J. Chem. Phys.}
  \textbf{2000}, \emph{112}, 6472--6486\relax
\mciteBstWouldAddEndPuncttrue
\mciteSetBstMidEndSepPunct{\mcitedefaultmidpunct}
{\mcitedefaultendpunct}{\mcitedefaultseppunct}\relax
\EndOfBibitem
\bibitem[Brenner \latin{et~al.}(2002)Brenner, Shenderova, Harrison, Stuart, Ni,
  and Sinnott]{REBO2_Brenner_2002_JPCM.14.783}
Brenner,~D.~W.; Shenderova,~O.~A.; Harrison,~J.~A.; Stuart,~S.~J.; Ni,~B.;
  Sinnott,~S.~B. {A Second-Generation Reactive Empirical Bond Order (REBO)
  Potential Energy Expression for Hydrocarbons}. \emph{J. Phys.: Condens.
  Matter} \textbf{2002}, \emph{14}, 783--802\relax
\mciteBstWouldAddEndPuncttrue
\mciteSetBstMidEndSepPunct{\mcitedefaultmidpunct}
{\mcitedefaultendpunct}{\mcitedefaultseppunct}\relax
\EndOfBibitem
\bibitem[Case \latin{et~al.}(2005)Case, {Cheatham III}, Darden, Gohlke, Luo,
  {Merz Jr.}, Onufriev, Simmerling, Wang, and Woods]{AMBER_program}
Case,~D.~A.; {Cheatham III},~T.~E.; Darden,~T.; Gohlke,~H.; Luo,~R.; {Merz
  Jr.},~K.~M.; Onufriev,~A.; Simmerling,~C.; Wang,~B.; Woods,~R.~J. {The Amber
  Biomolecular Simulation Programs}. \emph{J. Comput. Chem.} \textbf{2005},
  \emph{26}, 1668--1688\relax
\mciteBstWouldAddEndPuncttrue
\mciteSetBstMidEndSepPunct{\mcitedefaultmidpunct}
{\mcitedefaultendpunct}{\mcitedefaultseppunct}\relax
\EndOfBibitem
\bibitem[Brooks \latin{et~al.}(1983)Brooks, Bruccoleri, Olafson, States,
  Swaminathan, and Karplus]{CHARMM_program}
Brooks,~B.~R.; Bruccoleri,~R.~E.; Olafson,~B.~D.; States,~D.~J.;
  Swaminathan,~S.; Karplus,~M. {CHARMM: A Program for Macromolecular Energy,
  Minimization, and Dynamics Calculations}. \emph{J. Comput. Chem.}
  \textbf{1983}, \emph{4}, 187--217\relax
\mciteBstWouldAddEndPuncttrue
\mciteSetBstMidEndSepPunct{\mcitedefaultmidpunct}
{\mcitedefaultendpunct}{\mcitedefaultseppunct}\relax
\EndOfBibitem
\bibitem[Spoel \latin{et~al.}(2005)Spoel, Lindahl, Hess, Groenhof, Mark, and
  Berendsen]{GROMACS_program}
Spoel,~D. V.~D.; Lindahl,~E.; Hess,~B.; Groenhof,~G.; Mark,~A.~E.;
  Berendsen,~H.~J. {GROMACS: Fast, Flexible, and Free}. \emph{J. Comput. Chem.}
  \textbf{2005}, \emph{26}, 1701--1718\relax
\mciteBstWouldAddEndPuncttrue
\mciteSetBstMidEndSepPunct{\mcitedefaultmidpunct}
{\mcitedefaultendpunct}{\mcitedefaultseppunct}\relax
\EndOfBibitem
\bibitem[Plimpton(1995)]{LAMMPS_program}
Plimpton,~S. Fast Parallel Algorithms for Short-Range Molecular Dynamics.
  \emph{J. Comput. Phys.} \textbf{1995}, \emph{117}, 1--19\relax
\mciteBstWouldAddEndPuncttrue
\mciteSetBstMidEndSepPunct{\mcitedefaultmidpunct}
{\mcitedefaultendpunct}{\mcitedefaultseppunct}\relax
\EndOfBibitem
\bibitem[Phillips \latin{et~al.}(2005)Phillips, Braun, Wang, Gumbart,
  Tajkhorshid, Villa, Chipot, Skeel, Kal\'{e}, and Schulten]{NAMD_program}
Phillips,~J.~C.; Braun,~R.; Wang,~W.; Gumbart,~J.; Tajkhorshid,~E.; Villa,~E.;
  Chipot,~C.; Skeel,~R.~D.; Kal\'{e},~L.; Schulten,~K. {Scalable Molecular
  Dynamics with NAMD}. \emph{J. Comput. Chem.} \textbf{2005}, \emph{26},
  1781--1802\relax
\mciteBstWouldAddEndPuncttrue
\mciteSetBstMidEndSepPunct{\mcitedefaultmidpunct}
{\mcitedefaultendpunct}{\mcitedefaultseppunct}\relax
\EndOfBibitem
\bibitem[Akhukov \latin{et~al.}(2023)Akhukov, Chorkov, Gavrilov, Guseva,
  Khalatur, Khokhlov, Kniznik, Komarov, Okun, Potapkin, Rudyak, Shirabaykin,
  Skomorokhov, and Trepalin]{MULTICOMP_Akhukov_2023}
Akhukov,~M.~A.; Chorkov,~V.~A.; Gavrilov,~A.~A.; Guseva,~D.~V.;
  Khalatur,~P.~G.; Khokhlov,~A.~R.; Kniznik,~A.~A.; Komarov,~P.~V.;
  Okun,~M.~V.; Potapkin,~B.~V.; Rudyak,~V.~Y.; Shirabaykin,~D.~B.;
  Skomorokhov,~A.~S.; Trepalin,~S.~V. {MULTICOMP Package for Multilevel
  Simulation of Polymer Nanocomposites}. \emph{Comput. Mater. Sci.}
  \textbf{2023}, \emph{216}, 111832\relax
\mciteBstWouldAddEndPuncttrue
\mciteSetBstMidEndSepPunct{\mcitedefaultmidpunct}
{\mcitedefaultendpunct}{\mcitedefaultseppunct}\relax
\EndOfBibitem
\bibitem[Feig \latin{et~al.}(2004)Feig, Karanicolas, and III]{MMTSB_2004_Feig}
Feig,~M.; Karanicolas,~J.; III,~C. L.~B. {MMTSB Tool Set: Enhanced Sampling and
  Multiscale Modeling Methods for Applications in Structural Biology}. \emph{J.
  Molec. Graph. Model.} \textbf{2004}, \emph{22}, 377--395\relax
\mciteBstWouldAddEndPuncttrue
\mciteSetBstMidEndSepPunct{\mcitedefaultmidpunct}
{\mcitedefaultendpunct}{\mcitedefaultseppunct}\relax
\EndOfBibitem
\bibitem[Shankar \latin{et~al.}(2022)Shankar, Gogoi, Sethi, and
  Verma]{MaterialsStudio_Shankar}
Shankar,~U.; Gogoi,~R.; Sethi,~S.~K.; Verma,~A. In \emph{Forcefields for
  Atomistic-Scale Simulations: Materials and Applications}; Verma,~A.,
  Rangappa,~S.~M., Ogata,~S., Siengchin,~S., Eds.; Springer Singapore, 2022; pp
  299--313\relax
\mciteBstWouldAddEndPuncttrue
\mciteSetBstMidEndSepPunct{\mcitedefaultmidpunct}
{\mcitedefaultendpunct}{\mcitedefaultseppunct}\relax
\EndOfBibitem
\bibitem[Jellinek(1999)]{Jellinek_At_Mol_Clusters_book}
Jellinek,~J., Ed. \emph{{Theory of Atomic and Molecular Clusters}};
  Springer-Verlag Berlin Heidelberg, 1999\relax
\mciteBstWouldAddEndPuncttrue
\mciteSetBstMidEndSepPunct{\mcitedefaultmidpunct}
{\mcitedefaultendpunct}{\mcitedefaultseppunct}\relax
\EndOfBibitem
\bibitem[Saito(2020)]{Saito_CondensMatter_book}
Saito,~K. \emph{{Chemical Physics of Molecular Condensed Matter}}; Springer
  Singapore, 2020\relax
\mciteBstWouldAddEndPuncttrue
\mciteSetBstMidEndSepPunct{\mcitedefaultmidpunct}
{\mcitedefaultendpunct}{\mcitedefaultseppunct}\relax
\EndOfBibitem
\bibitem[Mesirov \latin{et~al.}(1996)Mesirov, Schulten, and
  Sumners]{Mesirov_Biomol_Struct_Dyn_book}
Mesirov,~J.~P., Schulten,~K., Sumners,~D.~W., Eds. \emph{{Mathematical
  Approaches to Biomolecular Structure and Dynamics}}; Springer New York,
  1996\relax
\mciteBstWouldAddEndPuncttrue
\mciteSetBstMidEndSepPunct{\mcitedefaultmidpunct}
{\mcitedefaultendpunct}{\mcitedefaultseppunct}\relax
\EndOfBibitem
\bibitem[Zhao \latin{et~al.}(2013)Zhao, Perilla, Yufenyuy, Meng, Chen, Ning,
  Ahn, Gronenborn, Schulten, Aiken, and Zhang]{Zhao_2013_Nature.497.643}
Zhao,~G.; Perilla,~J.~R.; Yufenyuy,~E.~L.; Meng,~X.; Chen,~B.; Ning,~J.;
  Ahn,~J.; Gronenborn,~A.~M.; Schulten,~K.; Aiken,~C.; Zhang,~P. {Mature HIV-1
  Capsid Structure by Cryo-Electron Microscopy and All-Atom Molecular
  Dynamics}. \emph{Nature} \textbf{2013}, \emph{497}, 643--646\relax
\mciteBstWouldAddEndPuncttrue
\mciteSetBstMidEndSepPunct{\mcitedefaultmidpunct}
{\mcitedefaultendpunct}{\mcitedefaultseppunct}\relax
\EndOfBibitem
\bibitem[Vashishta \latin{et~al.}(2006)Vashishta, Kalia, and
  Nakano]{Vashishta_2006_JPCB.110.3727}
Vashishta,~P.; Kalia,~R.~K.; Nakano,~A. {Multimillion Atom Simulations of
  Dynamics of Oxidation of an Aluminum Nanoparticle and Nanoindentation on
  Ceramics}. \emph{J. Phys. Chem. B} \textbf{2006}, \emph{110},
  3727--3733\relax
\mciteBstWouldAddEndPuncttrue
\mciteSetBstMidEndSepPunct{\mcitedefaultmidpunct}
{\mcitedefaultendpunct}{\mcitedefaultseppunct}\relax
\EndOfBibitem
\bibitem[Shaw \latin{et~al.}(2010)Shaw, Maragakis, and
  Wriggers]{Shaw_2010_Science.330.341}
Shaw,~D.~E.; Maragakis,~P.; Wriggers,~W. {Atomic-Level Characterization of the
  Structural Dynamics of Proteins}. \emph{Science} \textbf{2010}, \emph{330},
  341--346\relax
\mciteBstWouldAddEndPuncttrue
\mciteSetBstMidEndSepPunct{\mcitedefaultmidpunct}
{\mcitedefaultendpunct}{\mcitedefaultseppunct}\relax
\EndOfBibitem
\bibitem[Pierce \latin{et~al.}(2012)Pierce, Salomon-Ferrer, de~Oliveira,
  McCammon, and Walker]{Pierce_2012_JCTC.8.2997}
Pierce,~L. C.~T.; Salomon-Ferrer,~R.; de~Oliveira,~C. A.~F.; McCammon,~J.~A.;
  Walker,~R.~C. {Routine Access to Millisecond Time Scale Events with
  Accelerated Molecular Dynamics}. \emph{J. Chem. Theory Comput.}
  \textbf{2012}, \emph{8}, 2997--3002\relax
\mciteBstWouldAddEndPuncttrue
\mciteSetBstMidEndSepPunct{\mcitedefaultmidpunct}
{\mcitedefaultendpunct}{\mcitedefaultseppunct}\relax
\EndOfBibitem
\bibitem[Lu \latin{et~al.}(2021)Lu, Wang, Chen, Lin, Car, E, Jia, and
  Zhang]{Lu_2021_CPC.259.107624}
Lu,~D.; Wang,~H.; Chen,~M.; Lin,~L.; Car,~R.; E,~W.; Jia,~W.; Zhang,~L. {86
  PFLOPS Deep Potential Molecular Dynamics Simulation of 100 Million Atoms With
  Ab Initio Accuracy}. \emph{Comput. Phys. Commun.} \textbf{2021}, \emph{259},
  107624\relax
\mciteBstWouldAddEndPuncttrue
\mciteSetBstMidEndSepPunct{\mcitedefaultmidpunct}
{\mcitedefaultendpunct}{\mcitedefaultseppunct}\relax
\EndOfBibitem
\bibitem[Jia \latin{et~al.}((accessed 2023-09-08))Jia, Wang, Chen, Lu, Lin,
  Car, E, and Zhang]{Jia_100Mio_MD_arXiv}
Jia,~W.; Wang,~H.; Chen,~M.; Lu,~D.; Lin,~L.; Car,~R.; E,~W.; Zhang,~L. Pushing
  the Limit of Molecular Dynamics With Ab Initio Accuracy to 100 Million Atoms
  with Machine Learning. (accessed 2023-09-08);
  \url{https://arxiv.org/abs/2005.00223}\relax
\mciteBstWouldAddEndPuncttrue
\mciteSetBstMidEndSepPunct{\mcitedefaultmidpunct}
{\mcitedefaultendpunct}{\mcitedefaultseppunct}\relax
\EndOfBibitem
\bibitem[Warshel and Levitt(1976)Warshel, and Levitt]{Warshel_Levitt_1976}
Warshel,~A.; Levitt,~M. {Theoretical Studies of Enzymic Reactions: Dielectric,
  Electrostatic and Steric Stabilization of the Carbonium Ion in the Reaction
  of Lysozyme}. \emph{J. Molec. Biol.} \textbf{1976}, \emph{103},
  227--249\relax
\mciteBstWouldAddEndPuncttrue
\mciteSetBstMidEndSepPunct{\mcitedefaultmidpunct}
{\mcitedefaultendpunct}{\mcitedefaultseppunct}\relax
\EndOfBibitem
\bibitem[Field \latin{et~al.}(1990)Field, Bash, and
  Karplus]{QM-MM_Karplus_1990_JCC.11.700}
Field,~M.~J.; Bash,~P.~A.; Karplus,~M. {A Combined Quantum Mechanical and
  Molecular Mechanical Potential for Molecular Dynamics Simulations}. \emph{J.
  Comput. Chem.} \textbf{1990}, \emph{11}, 700--733\relax
\mciteBstWouldAddEndPuncttrue
\mciteSetBstMidEndSepPunct{\mcitedefaultmidpunct}
{\mcitedefaultendpunct}{\mcitedefaultseppunct}\relax
\EndOfBibitem
\bibitem[Senn and Thiel(2009)Senn, and Thiel]{Senn_Thiel_2009_AngewChem}
Senn,~H.~M.; Thiel,~W. {QM/MM Methods for Biomolecular Systems}. \emph{Angew.
  Chem. Int. Ed.} \textbf{2009}, \emph{48}, 1198--1229\relax
\mciteBstWouldAddEndPuncttrue
\mciteSetBstMidEndSepPunct{\mcitedefaultmidpunct}
{\mcitedefaultendpunct}{\mcitedefaultseppunct}\relax
\EndOfBibitem
\bibitem[Maseras and Morokuma(1995)Maseras, and
  Morokuma]{Maseras_1995_JCC.16.1170}
Maseras,~F.; Morokuma,~K. {IMOMM: A New Integrated \textit{Ab Initio} +
  Molecular Mechanics Geometry Optimization Scheme of Equilibrium Structures
  and Transition States}. \emph{J. Comput. Chem.} \textbf{1995}, \emph{16},
  1170--1179\relax
\mciteBstWouldAddEndPuncttrue
\mciteSetBstMidEndSepPunct{\mcitedefaultmidpunct}
{\mcitedefaultendpunct}{\mcitedefaultseppunct}\relax
\EndOfBibitem
\bibitem[Svensson \latin{et~al.}(1996)Svensson, Humbel, Froese, Matsubara,
  Sieber, and Morokuma]{ONIOM_Svensson_1996_JPC}
Svensson,~M.; Humbel,~S.; Froese,~R. D.~J.; Matsubara,~T.; Sieber,~S.;
  Morokuma,~K. {ONIOM: A Multilayered Integrated MO + MM Method for Geometry
  Optimizations and Single Point Energy Predictions. A Test for Diels-Alder
  Reactions and Pt(P(t-Bu)$_3$)$_2$ + H$_2$ Oxidative Addition}. \emph{J. Phys.
  Chem.} \textbf{1996}, \emph{100}, 19357--19363\relax
\mciteBstWouldAddEndPuncttrue
\mciteSetBstMidEndSepPunct{\mcitedefaultmidpunct}
{\mcitedefaultendpunct}{\mcitedefaultseppunct}\relax
\EndOfBibitem
\bibitem[Cao and Ryde(2018)Cao, and Ryde]{Cao_2018_FrontChem.6.89}
Cao,~L.; Ryde,~U. {On the Difference Between Additive and Subtractive QM/MM
  Calculations}. \emph{Front. Chem.} \textbf{2018}, \emph{6}, 89\relax
\mciteBstWouldAddEndPuncttrue
\mciteSetBstMidEndSepPunct{\mcitedefaultmidpunct}
{\mcitedefaultendpunct}{\mcitedefaultseppunct}\relax
\EndOfBibitem
\bibitem[Dapprich \latin{et~al.}(1999)Dapprich, Kom\'{a}romi, Byun, Morokuma,
  and Frisch]{Dapprich_1999_JMolStruct.461.1}
Dapprich,~S.; Kom\'{a}romi,~I.; Byun,~K.~S.; Morokuma,~K.; Frisch,~M.~J. {A New
  ONIOM Implementation in Gaussian98. Part I. The Calculation of Energies,
  Gradients, Vibrational Frequencies and Electric Field Derivatives}. \emph{J.
  Molec. Struct.} \textbf{1999}, \emph{461--462}, 1--21\relax
\mciteBstWouldAddEndPuncttrue
\mciteSetBstMidEndSepPunct{\mcitedefaultmidpunct}
{\mcitedefaultendpunct}{\mcitedefaultseppunct}\relax
\EndOfBibitem
\bibitem[S\"{o}derhjelm \latin{et~al.}(2009)S\"{o}derhjelm, Husberg, Strambi,
  Olivucci, and Ryde]{Soderhjelm_2009_JCTC.5.649}
S\"{o}derhjelm,~P.; Husberg,~C.; Strambi,~A.; Olivucci,~M.; Ryde,~U. {Protein
  Influence on Electronic Spectra Modeled by Multipoles and Polarizabilities}.
  \emph{J. Chem. Theory Comput.} \textbf{2009}, \emph{5}, 649--658\relax
\mciteBstWouldAddEndPuncttrue
\mciteSetBstMidEndSepPunct{\mcitedefaultmidpunct}
{\mcitedefaultendpunct}{\mcitedefaultseppunct}\relax
\EndOfBibitem
\bibitem[Olsen \latin{et~al.}(2010)Olsen, Aidas, and
  Kongsted]{Olsen_2010_JCTC.6.3721}
Olsen,~J.~M.; Aidas,~K.; Kongsted,~J. {Excited States in Solution Through
  Polarizable Embedding}. \emph{J. Chem. Theory Comput.} \textbf{2010},
  \emph{6}, 3721--3734\relax
\mciteBstWouldAddEndPuncttrue
\mciteSetBstMidEndSepPunct{\mcitedefaultmidpunct}
{\mcitedefaultendpunct}{\mcitedefaultseppunct}\relax
\EndOfBibitem
\bibitem[Darden \latin{et~al.}(1993)Darden, York, and
  Pedersen]{Darden_1993_JCP.98.10089}
Darden,~T.; York,~D.; Pedersen,~L. {Particle Mesh Ewald: An $N \cdot \log{(N)}$
  Method for Ewald Sums in Large Systems}. \emph{J. Chem. Phys.} \textbf{1993},
  \emph{98}, 10089--10092\relax
\mciteBstWouldAddEndPuncttrue
\mciteSetBstMidEndSepPunct{\mcitedefaultmidpunct}
{\mcitedefaultendpunct}{\mcitedefaultseppunct}\relax
\EndOfBibitem
\bibitem[Herce \latin{et~al.}(2007)Herce, Garcia, and
  Darden]{Herce_2007_JCP.126.124106}
Herce,~H.~D.; Garcia,~A.~E.; Darden,~T. {The Electrostatic Surface Term: (I)
  Periodic Systems}. \emph{J. Chem. Phys.} \textbf{2007}, \emph{126},
  124106\relax
\mciteBstWouldAddEndPuncttrue
\mciteSetBstMidEndSepPunct{\mcitedefaultmidpunct}
{\mcitedefaultendpunct}{\mcitedefaultseppunct}\relax
\EndOfBibitem
\bibitem[Frenkel and Smit(2002)Frenkel, and Smit]{Frenkel_Smit_MD_book}
Frenkel,~D.; Smit,~B. \emph{{Understanding Molecular Simulation: From
  Algorithms to Applications}}, 2nd ed.; Academic Press, San Diego, 2002\relax
\mciteBstWouldAddEndPuncttrue
\mciteSetBstMidEndSepPunct{\mcitedefaultmidpunct}
{\mcitedefaultendpunct}{\mcitedefaultseppunct}\relax
\EndOfBibitem
\bibitem[Field(2008)]{pDynamo_Field_2008_JCTC.4.1151}
Field,~M.~J. {The pDynamo Program for Molecular Simulations Using Hybrid
  Quantum Chemical and Molecular Mechanical Potentials}. \emph{J. Chem. Theory
  Comput.} \textbf{2008}, \emph{4}, 1151--1161\relax
\mciteBstWouldAddEndPuncttrue
\mciteSetBstMidEndSepPunct{\mcitedefaultmidpunct}
{\mcitedefaultendpunct}{\mcitedefaultseppunct}\relax
\EndOfBibitem
\bibitem[Zhang \latin{et~al.}(2019)Zhang, Altarawy, Barnes, Turney, and
  {Schaefer III}]{JANUS_Zhang_2019_JCTC.15.4362}
Zhang,~B.; Altarawy,~D.; Barnes,~T.; Turney,~J.~M.; {Schaefer III},~H.~F.
  {Janus: An Extensible Open-Source Software Package for Adaptive QM/MM
  Methods}. \emph{J. Chem. Theory Comput.} \textbf{2019}, \emph{15},
  4362--4373\relax
\mciteBstWouldAddEndPuncttrue
\mciteSetBstMidEndSepPunct{\mcitedefaultmidpunct}
{\mcitedefaultendpunct}{\mcitedefaultseppunct}\relax
\EndOfBibitem
\bibitem[Lin \latin{et~al.}((accessed on 2023-09-13))Lin, Zhang, Pezeshki,
  Wang, Wu, Gagliardi, and Truhlar]{QMMM_program}
Lin,~H.; Zhang,~Y.; Pezeshki,~S.; Wang,~B.; Wu,~X.-P.; Gagliardi,~L.;
  Truhlar,~D.~G. QMMM 2018. (accessed on 2023-09-13);
  \url{http://comp.chem.umn.edu/qmmm}, University of Minnesota, Minneapolis,
  2018\relax
\mciteBstWouldAddEndPuncttrue
\mciteSetBstMidEndSepPunct{\mcitedefaultmidpunct}
{\mcitedefaultendpunct}{\mcitedefaultseppunct}\relax
\EndOfBibitem
\bibitem[Korol \latin{et~al.}(2020)Korol, Husen, Sjulstok, Nielsen, Friis,
  Frederiksen, Salo, and Solov'yov]{VIKING_2020_ACSOmega.5.1254}
Korol,~V.; Husen,~P.; Sjulstok,~E.; Nielsen,~C.; Friis,~I.; Frederiksen,~A.;
  Salo,~A.~B.; Solov'yov,~I.~A. {Introducing VIKING: A Novel Online Platform
  for Multiscale Modeling}. \emph{ACS Omega} \textbf{2020}, \emph{5},
  1254--1260\relax
\mciteBstWouldAddEndPuncttrue
\mciteSetBstMidEndSepPunct{\mcitedefaultmidpunct}
{\mcitedefaultendpunct}{\mcitedefaultseppunct}\relax
\EndOfBibitem
\bibitem[Metz \latin{et~al.}(2014)Metz, K\"{a}stner, Sokol, Keal, and
  Sherwood]{ChemShell_2014_WIREsCMS.4.101}
Metz,~S.; K\"{a}stner,~J.; Sokol,~A.~A.; Keal,~T.~W.; Sherwood,~P.
  {ChemShell--A Modular Software Package for QM/MM Simulations}. \emph{WIREs
  Comput. Molec. Sci.} \textbf{2014}, \emph{4}, 101--110\relax
\mciteBstWouldAddEndPuncttrue
\mciteSetBstMidEndSepPunct{\mcitedefaultmidpunct}
{\mcitedefaultendpunct}{\mcitedefaultseppunct}\relax
\EndOfBibitem
\bibitem[Cofer-Shabica \latin{et~al.}(2022)Cofer-Shabica, Menger, Ou, Shao,
  Subotnik, and Faraji]{INAQS_program}
Cofer-Shabica,~D.~V.; Menger,~M. F. S.~J.; Ou,~Q.; Shao,~Y.; Subotnik,~J.~E.;
  Faraji,~S. {INAQS, A Generic Interface for Nonadiabatic QM/MM Dynamics:
  Design, Implementation, and Validation for GROMACS/Q-CHEM Simulations}.
  \emph{J. Chem. Theory Comput.} \textbf{2022}, \emph{18}, 4601--4614\relax
\mciteBstWouldAddEndPuncttrue
\mciteSetBstMidEndSepPunct{\mcitedefaultmidpunct}
{\mcitedefaultendpunct}{\mcitedefaultseppunct}\relax
\EndOfBibitem
\bibitem[Kong \latin{et~al.}(2000)Kong, White, Krylov, Sherrill, Adamson,
  Furlani, Lee, Lee, Gwaltney, Adams, and \textit{et
  al.}]{Q-Chem_Kong_2000_JCC.21.1532}
Kong,~J.; White,~C.~A.; Krylov,~A.~I.; Sherrill,~D.; Adamson,~R.~D.;
  Furlani,~T.~R.; Lee,~M.~S.; Lee,~A.~M.; Gwaltney,~S.~R.; Adams,~T.~R.;
  \textit{et al.} {Q-Chem 2.0: A High-Performance \textit{Ab Initio} Electronic
  Structure Program Package}. \emph{J. Comput. Chem.} \textbf{2000}, \emph{21},
  1532--1548\relax
\mciteBstWouldAddEndPuncttrue
\mciteSetBstMidEndSepPunct{\mcitedefaultmidpunct}
{\mcitedefaultendpunct}{\mcitedefaultseppunct}\relax
\EndOfBibitem
\bibitem[Epifanovsky \latin{et~al.}(2021)Epifanovsky, Gilbert, Feng, Lee, Mao,
  Mardirossian, Pokhilko, White, Coons, Dempwolff, and \textit{et
  al.}]{Q-Chem_Epifanovsky_2021_JCP.155.084801}
Epifanovsky,~E.; Gilbert,~A. T.~B.; Feng,~X.; Lee,~J.; Mao,~Y.;
  Mardirossian,~N.; Pokhilko,~P.; White,~A.~F.; Coons,~M.~P.; Dempwolff,~A.~L.;
  \textit{et al.} {Software for the Frontiers of Quantum Chemistry: An Overview
  of Developments in the Q-Chem 5 Package}. \emph{J. Chem. Phys.}
  \textbf{2021}, \emph{155}, 084801\relax
\mciteBstWouldAddEndPuncttrue
\mciteSetBstMidEndSepPunct{\mcitedefaultmidpunct}
{\mcitedefaultendpunct}{\mcitedefaultseppunct}\relax
\EndOfBibitem
\bibitem[GRO((accessed on 2023-09-13))]{GROMACS_CP2K_interface}
Hybrid Quantum-Classical simulations (QM/MM) with CP2K interface. (accessed on
  2023-09-13);
  \url{https://manual.gromacs.org/current/reference-manual/special/qmmm.html}\relax
\mciteBstWouldAddEndPuncttrue
\mciteSetBstMidEndSepPunct{\mcitedefaultmidpunct}
{\mcitedefaultendpunct}{\mcitedefaultseppunct}\relax
\EndOfBibitem
\bibitem[Melo \latin{et~al.}(2018)Melo, Bernardi, Rudack, Scheurer, Riplinger,
  Phillips, Maia, Rocha, Ribeiro, Stone, Neese, Schulten, and
  Luthey-Schulten]{NAMD_QMMM_2018_NatMeth.15.351}
Melo,~M. C.~R.; Bernardi,~R.~C.; Rudack,~T.; Scheurer,~M.; Riplinger,~C.;
  Phillips,~J.~C.; Maia,~J. D.~C.; Rocha,~G.~B.; Ribeiro,~J.~V.; Stone,~J.~E.;
  Neese,~F.; Schulten,~K.; Luthey-Schulten,~Z. {NAMD Goes Quantum: An
  Integrative Suite for Hybrid Simulations}. \emph{Nat. Meth.} \textbf{2018},
  \emph{15}, 351--354\relax
\mciteBstWouldAddEndPuncttrue
\mciteSetBstMidEndSepPunct{\mcitedefaultmidpunct}
{\mcitedefaultendpunct}{\mcitedefaultseppunct}\relax
\EndOfBibitem
\bibitem[Humphrey \latin{et~al.}(1996)Humphrey, Dalke, and
  Schulten]{VMD_program_1996}
Humphrey,~W.; Dalke,~A.; Schulten,~K. {VMD: Visual Molecular Dynamics}.
  \emph{J. Mol. Graph.} \textbf{1996}, \emph{14}, 33--38\relax
\mciteBstWouldAddEndPuncttrue
\mciteSetBstMidEndSepPunct{\mcitedefaultmidpunct}
{\mcitedefaultendpunct}{\mcitedefaultseppunct}\relax
\EndOfBibitem
\bibitem[Stewart(1990)]{MOPAC_Stewart_1990}
Stewart,~J.~J. {MOPAC: A Semiempirical Molecular Orbital Program}. \emph{J.
  Comp.-Aided Mol. Design} \textbf{1990}, \emph{4}, 1--103\relax
\mciteBstWouldAddEndPuncttrue
\mciteSetBstMidEndSepPunct{\mcitedefaultmidpunct}
{\mcitedefaultendpunct}{\mcitedefaultseppunct}\relax
\EndOfBibitem
\bibitem[Shaik \latin{et~al.}(2010)Shaik, Cohen, Wang, Chen, Kumar, and
  Thiel]{Shaik_2010_ChemRev.110.949}
Shaik,~S.; Cohen,~S.; Wang,~Y.; Chen,~H.; Kumar,~D.; Thiel,~W. {P450 Enzymes:
  Their Structure, Reactivity, and Selectivity--Modeled by QM/MM Calculations}.
  \emph{Chem. Rev.} \textbf{2010}, \emph{110}, 949--1017\relax
\mciteBstWouldAddEndPuncttrue
\mciteSetBstMidEndSepPunct{\mcitedefaultmidpunct}
{\mcitedefaultendpunct}{\mcitedefaultseppunct}\relax
\EndOfBibitem
\bibitem[Gomes and Jacob(2012)Gomes, and Jacob]{Gomes_2012_AnnRepSectC.108.222}
Gomes,~A. S.~P.; Jacob,~C.~R. {Quantum-Chemical Embedding Methods for Treating
  Local Electronic Excitations in Complex Chemical Systems}. \emph{Ann. Rep.
  Sect. C (Phys. Chem.)} \textbf{2012}, \emph{108}, 222--277\relax
\mciteBstWouldAddEndPuncttrue
\mciteSetBstMidEndSepPunct{\mcitedefaultmidpunct}
{\mcitedefaultendpunct}{\mcitedefaultseppunct}\relax
\EndOfBibitem
\bibitem[Dinh \latin{et~al.}(2010)Dinh, Reinhard, and
  Suraud]{Dinh_2010_PhysRep.485.43}
Dinh,~P.~M.; Reinhard,~P.~G.; Suraud,~E. {Dynamics of Clusters and Molecules in
  Contact with an Environment}. \emph{Phys. Rep.} \textbf{2010}, \emph{485},
  43--107\relax
\mciteBstWouldAddEndPuncttrue
\mciteSetBstMidEndSepPunct{\mcitedefaultmidpunct}
{\mcitedefaultendpunct}{\mcitedefaultseppunct}\relax
\EndOfBibitem
\bibitem[Lopes \latin{et~al.}(2009)Lopes, Roux, and {MacKerell
  Jr.}]{Lopes_2009_TheorChemAcc.124.11}
Lopes,~P. E.~M.; Roux,~B.; {MacKerell Jr.},~A.~D. {Molecular Modeling and
  Dynamics Studies with Explicit Inclusion of Electronic Polarizability: Theory
  and Applications}. \emph{Theor. Chem. Acc.} \textbf{2009}, \emph{124},
  11--28\relax
\mciteBstWouldAddEndPuncttrue
\mciteSetBstMidEndSepPunct{\mcitedefaultmidpunct}
{\mcitedefaultendpunct}{\mcitedefaultseppunct}\relax
\EndOfBibitem
\bibitem[Marx and Hutter(2009)Marx, and Hutter]{Marx_AIMD_book}
Marx,~D.; Hutter,~J. \emph{{Ab Initio Molecular Dynamics. Basic Theory and
  Advanced Methods}}; Cambridge University Press, Cambridge, 2009\relax
\mciteBstWouldAddEndPuncttrue
\mciteSetBstMidEndSepPunct{\mcitedefaultmidpunct}
{\mcitedefaultendpunct}{\mcitedefaultseppunct}\relax
\EndOfBibitem
\bibitem[Tuckerman(2002)]{Tuckerman_2002_JPCM.14.R1297}
Tuckerman,~M.~E. {\textit{Ab initio} Molecular Dynamics: Basic Concepts,
  Current Trends and Novel Applications}. \emph{J. Phys.: Condens. Matter}
  \textbf{2002}, \emph{14}, R1297--R1355\relax
\mciteBstWouldAddEndPuncttrue
\mciteSetBstMidEndSepPunct{\mcitedefaultmidpunct}
{\mcitedefaultendpunct}{\mcitedefaultseppunct}\relax
\EndOfBibitem
\bibitem[Iftimie \latin{et~al.}(2005)Iftimie, Minary, and
  Tuckerman]{Iftimie_2005_PNAS.102.6654}
Iftimie,~R.; Minary,~P.; Tuckerman,~M.~E. {\textit{Ab initio} Molecular
  Dynamics: Concepts, Recent Developments, and Future Trends}. \emph{Proc.
  Natl. Acad. Sci. U.S.A.} \textbf{2005}, \emph{102}, 6654--6659\relax
\mciteBstWouldAddEndPuncttrue
\mciteSetBstMidEndSepPunct{\mcitedefaultmidpunct}
{\mcitedefaultendpunct}{\mcitedefaultseppunct}\relax
\EndOfBibitem
\bibitem[Barnett and Landman(1993)Barnett, and
  Landman]{Barnett_1993_PRB.48.2081}
Barnett,~R.~N.; Landman,~U. {Born-Oppenheimer Molecular-Dynamics Simulations of
  Finite Systems: Structure and Dynamics of (H$_2$O)$_2$}. \emph{Phys. Rev. B}
  \textbf{1993}, \emph{48}, 2081--2097\relax
\mciteBstWouldAddEndPuncttrue
\mciteSetBstMidEndSepPunct{\mcitedefaultmidpunct}
{\mcitedefaultendpunct}{\mcitedefaultseppunct}\relax
\EndOfBibitem
\bibitem[Niklasson and Negre(2023)Niklasson, and
  Negre]{Niklasson_2023_JCP.158.154105}
Niklasson,~A. M.~N.; Negre,~C. F.~A. {Shadow Energy Functionals and Potentials
  in Born--Oppenheimer Molecular Dynamics}. \emph{J. Chem. Phys.}
  \textbf{2023}, \emph{158}, 154105\relax
\mciteBstWouldAddEndPuncttrue
\mciteSetBstMidEndSepPunct{\mcitedefaultmidpunct}
{\mcitedefaultendpunct}{\mcitedefaultseppunct}\relax
\EndOfBibitem
\bibitem[Worth and Cederbaum(2004)Worth, and
  Cederbaum]{Worth_Cederbaum_beyond_BOMD}
Worth,~G.~A.; Cederbaum,~L.~S. {Beyond Born-Oppenheimer: Molecular Dynamics
  Through a Conical Intersection}. \emph{Annu. Rev. Phys. Chem.} \textbf{2004},
  \emph{55}, 127--158\relax
\mciteBstWouldAddEndPuncttrue
\mciteSetBstMidEndSepPunct{\mcitedefaultmidpunct}
{\mcitedefaultendpunct}{\mcitedefaultseppunct}\relax
\EndOfBibitem
\bibitem[Kirrander and Vacher(2020)Kirrander, and
  Vacher]{Ehrenfest_Methods_chapter}
Kirrander,~A.; Vacher,~M. In \emph{Quantum Chemistry and Dynamics of Excited
  States: Methods and Applications}; Gonz\'{a}lez,~L., Lindh,~R., Eds.; John
  Wiley \& Sons Ltd., 2020; Chapter 15, pp 469--497\relax
\mciteBstWouldAddEndPuncttrue
\mciteSetBstMidEndSepPunct{\mcitedefaultmidpunct}
{\mcitedefaultendpunct}{\mcitedefaultseppunct}\relax
\EndOfBibitem
\bibitem[Car and Parrinello(1985)Car, and Parrinello]{Car_Parrinello_MD}
Car,~R.; Parrinello,~M. {Unified Approach for Molecular Dynamics and
  Density-Functional Theory}. \emph{Phys. Rev. Lett.} \textbf{1985}, \emph{55},
  2471--2474\relax
\mciteBstWouldAddEndPuncttrue
\mciteSetBstMidEndSepPunct{\mcitedefaultmidpunct}
{\mcitedefaultendpunct}{\mcitedefaultseppunct}\relax
\EndOfBibitem
\bibitem[Hutter(2012)]{Hutter_2012_WIREsCMS.2.604}
Hutter,~J. {Car-Parrinello Molecular Dynamics}. \emph{WIREs Comput. Molec.
  Sci.} \textbf{2012}, \emph{2}, 604--612\relax
\mciteBstWouldAddEndPuncttrue
\mciteSetBstMidEndSepPunct{\mcitedefaultmidpunct}
{\mcitedefaultendpunct}{\mcitedefaultseppunct}\relax
\EndOfBibitem
\bibitem[Gonze \latin{et~al.}(2002)Gonze, Beuken, Caracas, Detraux, Fuchs,
  Rignanese, Sindic, Verstraete, Zerah, Jollet, and \textit{et
  al.}]{ABINIT_Gonze_2002}
Gonze,~X.; Beuken,~J.-M.; Caracas,~R.; Detraux,~F.; Fuchs,~M.;
  Rignanese,~G.-M.; Sindic,~L.; Verstraete,~M.; Zerah,~G.; Jollet,~F.;
  \textit{et al.} {First-Principles Computation of Material Properties: The
  ABINIT Software Project}. \emph{Comput. Mater. Sci.} \textbf{2002},
  \emph{25}, 478--492\relax
\mciteBstWouldAddEndPuncttrue
\mciteSetBstMidEndSepPunct{\mcitedefaultmidpunct}
{\mcitedefaultendpunct}{\mcitedefaultseppunct}\relax
\EndOfBibitem
\bibitem[Clark \latin{et~al.}(2005)Clark, Segall, Pickard, Hasnip, Probert,
  Refson, and Payne]{CASTEP_Clark_2005}
Clark,~S.~J.; Segall,~M.~D.; Pickard,~C.~J.; Hasnip,~P.~J.; Probert,~M.~J.;
  Refson,~K.; Payne,~M.~C. {First Principles Methods Using CASTEP}. \emph{Z.
  Kristallogr. -- Cryst. Mater.} \textbf{2005}, \emph{220}, 567--570\relax
\mciteBstWouldAddEndPuncttrue
\mciteSetBstMidEndSepPunct{\mcitedefaultmidpunct}
{\mcitedefaultendpunct}{\mcitedefaultseppunct}\relax
\EndOfBibitem
\bibitem[Kl\"{o}ffel \latin{et~al.}(2021)Kl\"{o}ffel, Mathias, and
  Meyer]{CPMD_program_Kloeffel_CPC}
Kl\"{o}ffel,~T.; Mathias,~G.; Meyer,~B. {Integrating State of the Art Compute,
  Communication, and Autotuning Strategies to Multiply the Performance of
  \textit{Ab Initio} Molecular Dynamics on Massively Parallel Multi-Core
  Supercomputers}. \emph{Comput. Phys. Commun.} \textbf{2021}, \emph{260},
  107745\relax
\mciteBstWouldAddEndPuncttrue
\mciteSetBstMidEndSepPunct{\mcitedefaultmidpunct}
{\mcitedefaultendpunct}{\mcitedefaultseppunct}\relax
\EndOfBibitem
\bibitem[Ojanper\"{a} \latin{et~al.}(2012)Ojanper\"{a}, Havu, Lehtovaara, and
  Puska]{GPAW_Ojanpera_2012_JCP}
Ojanper\"{a},~A.; Havu,~V.; Lehtovaara,~L.; Puska,~M. {Nonadiabatic Ehrenfest
  Molecular Dynamics Within the Projector Augmented-Wave Method}. \emph{J.
  Chem. Phys.} \textbf{2012}, \emph{136}, 144103\relax
\mciteBstWouldAddEndPuncttrue
\mciteSetBstMidEndSepPunct{\mcitedefaultmidpunct}
{\mcitedefaultendpunct}{\mcitedefaultseppunct}\relax
\EndOfBibitem
\bibitem[Andrade \latin{et~al.}(2012)Andrade, Alberdi-Rodriguez, Strubbe,
  Oliveira, Nogueira, Castro, Muguerza, Arruabarrena, Louie, Aspuru-Guzik,
  Rubio, and Marques]{OCTOPUS_Andrade_2012_JPCM}
Andrade,~X.; Alberdi-Rodriguez,~J.; Strubbe,~D.~A.; Oliveira,~M.~J.;
  Nogueira,~F.; Castro,~A.; Muguerza,~J.; Arruabarrena,~A.; Louie,~S.~G.;
  Aspuru-Guzik,~A.; Rubio,~A.; Marques,~M.~A. {Time-Dependent
  Density-Functional Theory in Massively Parallel Computer Architectures: The
  OCTOPUS Project}. \emph{J. Phys.: Condens. Matter} \textbf{2012}, \emph{24},
  233202\relax
\mciteBstWouldAddEndPuncttrue
\mciteSetBstMidEndSepPunct{\mcitedefaultmidpunct}
{\mcitedefaultendpunct}{\mcitedefaultseppunct}\relax
\EndOfBibitem
\bibitem[Soler \latin{et~al.}(2002)Soler, Artacho, Gale, Garc\'{i}a, Junquera,
  Ordej\'{o}n, and S\'{a}nchez-Portal]{SIESTA_Soler_2002_JPCM}
Soler,~J.~M.; Artacho,~E.; Gale,~J.~D.; Garc\'{i}a,~A.; Junquera,~J.;
  Ordej\'{o}n,~P.; S\'{a}nchez-Portal,~D. {The SIESTA Method for \textit{Ab
  Initio} Order-$N$ Materials Simulation}. \emph{J. Phys.: Condens. Matter}
  \textbf{2002}, \emph{14}, 2745--2779\relax
\mciteBstWouldAddEndPuncttrue
\mciteSetBstMidEndSepPunct{\mcitedefaultmidpunct}
{\mcitedefaultendpunct}{\mcitedefaultseppunct}\relax
\EndOfBibitem
\bibitem[Andreoni \latin{et~al.}(2005)Andreoni, Marx, and
  Sprik]{Andreoni_2005_ChemPhysChem}
Andreoni,~W.; Marx,~D.; Sprik,~M. {Editorial: A Tribute to Michele Parrinello:
  From Physics via Chemistry to Biology}. \emph{ChemPhysChem} \textbf{2005},
  \emph{6}, 1671--1676\relax
\mciteBstWouldAddEndPuncttrue
\mciteSetBstMidEndSepPunct{\mcitedefaultmidpunct}
{\mcitedefaultendpunct}{\mcitedefaultseppunct}\relax
\EndOfBibitem
\bibitem[Boero and Oshiyama(2015)Boero, and Oshiyama]{Boero_2015_CPMD}
Boero,~M.; Oshiyama,~A. In \emph{Encyclopedia of Nanotechnology}; Bhushan,~B.,
  Ed.; Springer, Dordrecht, 2015; pp 1--10\relax
\mciteBstWouldAddEndPuncttrue
\mciteSetBstMidEndSepPunct{\mcitedefaultmidpunct}
{\mcitedefaultendpunct}{\mcitedefaultseppunct}\relax
\EndOfBibitem
\bibitem[{van Duin} \latin{et~al.}(2001){van Duin}, Dasgupta, Lorant, and
  Goddard]{ReaxFF_vanDuin_2001_JPCA.105.9396}
{van Duin},~A. C.~T.; Dasgupta,~S.; Lorant,~F.; Goddard,~W.~A. {ReaxFF: A
  Reactive Force Field for Hydrocarbons}. \emph{J. Phys. Chem. A}
  \textbf{2001}, \emph{105}, 9396--9409\relax
\mciteBstWouldAddEndPuncttrue
\mciteSetBstMidEndSepPunct{\mcitedefaultmidpunct}
{\mcitedefaultendpunct}{\mcitedefaultseppunct}\relax
\EndOfBibitem
\bibitem[Senftle \latin{et~al.}(2016)Senftle, Hong, Islam, Kylasa, Zheng, Shin,
  Junkermeier, Engel-Herbert, Janik, Aktulga, Verstraelen, Grama, and {van
  Duin}]{ReaxFF_Senftle_2016}
Senftle,~T.~P.; Hong,~S.; Islam,~M.~M.; Kylasa,~S.~B.; Zheng,~Y.; Shin,~Y.~K.;
  Junkermeier,~C.; Engel-Herbert,~R.; Janik,~M.~J.; Aktulga,~H.~M.;
  Verstraelen,~T.; Grama,~A.; {van Duin},~A. C.~T. {The ReaxFF Reactive
  Force-Field: Development, Applications and Future Directions}. \emph{npj
  Comput. Mater.} \textbf{2016}, \emph{2}, 15011\relax
\mciteBstWouldAddEndPuncttrue
\mciteSetBstMidEndSepPunct{\mcitedefaultmidpunct}
{\mcitedefaultendpunct}{\mcitedefaultseppunct}\relax
\EndOfBibitem
\bibitem[{Russo Jr.} and {van Duin}(2011){Russo Jr.}, and {van
  Duin}]{Russo_2011_NIMB.269.1549}
{Russo Jr.},~M.~F.; {van Duin},~A. C.~T. {Atomistic-Scale Simulations of
  Chemical Reactions: Bridging From Quantum Chemistry to Engineering}.
  \emph{Nucl. Instrum. Meth. B} \textbf{2011}, \emph{269}, 1549--1554\relax
\mciteBstWouldAddEndPuncttrue
\mciteSetBstMidEndSepPunct{\mcitedefaultmidpunct}
{\mcitedefaultendpunct}{\mcitedefaultseppunct}\relax
\EndOfBibitem
\bibitem[Rea((accessed 2023-09-14))]{ReaxFF_Manual}
ReaxxFF Manual, Amsterdam Modeling Suite 2023.1. (accessed 2023-09-14);
  \url{https://www.scm.com/doc/ReaxFF/_downloads/af5ba007160596ded1785a11e54b6c8b/ReaxFF.pdf}\relax
\mciteBstWouldAddEndPuncttrue
\mciteSetBstMidEndSepPunct{\mcitedefaultmidpunct}
{\mcitedefaultendpunct}{\mcitedefaultseppunct}\relax
\EndOfBibitem
\bibitem[Shchygol \latin{et~al.}(2019)Shchygol, Yakovlev, Trnka, {van Duin},
  and Verstraelen]{Shchygol_2019_JCTC.15.6799}
Shchygol,~G.; Yakovlev,~A.; Trnka,~T.; {van Duin},~A. C.~T.; Verstraelen,~T.
  {ReaxFF Parameter Optimization with Monte-Carlo and Evolutionary Algorithms:
  Guidelines and Insights}. \emph{J. Chem. Theory Comput.} \textbf{2019},
  \emph{15}, 6799--6812\relax
\mciteBstWouldAddEndPuncttrue
\mciteSetBstMidEndSepPunct{\mcitedefaultmidpunct}
{\mcitedefaultendpunct}{\mcitedefaultseppunct}\relax
\EndOfBibitem
\bibitem[{te Velde} \latin{et~al.}(2001){te Velde}, Bickelhaupt, Baerends,
  {Fonseca Guerra}, {van Gisbergen}, Snijders, and
  Ziegler]{ADF_program_JCC_2021}
{te Velde},~G.; Bickelhaupt,~F.~M.; Baerends,~E.~J.; {Fonseca Guerra},~C.; {van
  Gisbergen},~S. J.~A.; Snijders,~J.~G.; Ziegler,~T. {Chemistry with ADF}.
  \emph{J. Comput. Chem.} \textbf{2001}, \emph{22}, 931--967\relax
\mciteBstWouldAddEndPuncttrue
\mciteSetBstMidEndSepPunct{\mcitedefaultmidpunct}
{\mcitedefaultendpunct}{\mcitedefaultseppunct}\relax
\EndOfBibitem
\bibitem[Verkhovtsev \latin{et~al.}(2017)Verkhovtsev, Korol, and
  Solov'yov]{Verkhovtsev_2017_EPJD.71.212}
Verkhovtsev,~A.~V.; Korol,~A.~V.; Solov'yov,~A.~V. {Classical Molecular
  Dynamics Simulations of Fusion and Fragmentation in Fullerene--Fullerene
  Collisions}. \emph{Eur. Phys. J. D} \textbf{2017}, \emph{71}, 212\relax
\mciteBstWouldAddEndPuncttrue
\mciteSetBstMidEndSepPunct{\mcitedefaultmidpunct}
{\mcitedefaultendpunct}{\mcitedefaultseppunct}\relax
\EndOfBibitem
\bibitem[{de Vera} \latin{et~al.}(2019){de Vera}, Verkhovtsev, Sushko, and
  Solov'yov]{deVera_2019_EPJD.73.215}
{de Vera},~P.; Verkhovtsev,~A.; Sushko,~G.; Solov'yov,~A.~V. {Reactive
  Molecular Dynamics Simulations of Organometallic Compound W(CO)$_6$
  Fragmentation}. \emph{Eur. Phys. J. D} \textbf{2019}, \emph{73}, 215\relax
\mciteBstWouldAddEndPuncttrue
\mciteSetBstMidEndSepPunct{\mcitedefaultmidpunct}
{\mcitedefaultendpunct}{\mcitedefaultseppunct}\relax
\EndOfBibitem
\bibitem[Andreides \latin{et~al.}(2023)Andreides, Verkhovtsev, Fedor, and
  Solov'yov]{Andreides_2023_JPCA.127.3757}
Andreides,~B.; Verkhovtsev,~A.~V.; Fedor,~J.; Solov'yov,~A.~V. {Role of the
  Molecular Environment in Quenching the Irradiation-Driven Fragmentation of
  Fe(CO)$_5$: A Reactive Molecular Dynamics Study}. \emph{J. Phys. Chem. A}
  \textbf{2023}, \emph{127}, 3757--3767\relax
\mciteBstWouldAddEndPuncttrue
\mciteSetBstMidEndSepPunct{\mcitedefaultmidpunct}
{\mcitedefaultendpunct}{\mcitedefaultseppunct}\relax
\EndOfBibitem
\bibitem[Verkhovtsev \latin{et~al.}(2021)Verkhovtsev, Solov'yov, and
  Solov'yov]{verkhovtsev2021irradiation}
Verkhovtsev,~A.~V.; Solov'yov,~I.~A.; Solov'yov,~A.~V. {Irradiation-Driven
  Molecular Dynamics: A Review}. \emph{Eur. Phys. J. D} \textbf{2021},
  \emph{75}, 213\relax
\mciteBstWouldAddEndPuncttrue
\mciteSetBstMidEndSepPunct{\mcitedefaultmidpunct}
{\mcitedefaultendpunct}{\mcitedefaultseppunct}\relax
\EndOfBibitem
\bibitem[Ebel \latin{et~al.}(2000)Ebel, Ghiorso, Sack, and
  Grossman]{Ebel_2000_JCC.21.247}
Ebel,~D.~S.; Ghiorso,~M.~S.; Sack,~R.~O.; Grossman,~L. {Gibbs Energy
  Minimization in Gas + Liquid + Solid Systems}. \emph{J. Comput. Chem.}
  \textbf{2000}, \emph{21}, 247--256\relax
\mciteBstWouldAddEndPuncttrue
\mciteSetBstMidEndSepPunct{\mcitedefaultmidpunct}
{\mcitedefaultendpunct}{\mcitedefaultseppunct}\relax
\EndOfBibitem
\bibitem[de~Nevers(2012)]{deNevers_PhysChemEquil_book}
de~Nevers,~N. \emph{{Physical and Chemical Equilibrium for Chemical
  Engineers}}; John Wiley \& Sons, Inc., 2012\relax
\mciteBstWouldAddEndPuncttrue
\mciteSetBstMidEndSepPunct{\mcitedefaultmidpunct}
{\mcitedefaultendpunct}{\mcitedefaultseppunct}\relax
\EndOfBibitem
\bibitem[The((accessed 2023-09-15))]{Thermocalc_DB}
Thermodynamic Databases. (accessed 2023-09-15);
  \url{https://thermocalc.com/products/databases/}\relax
\mciteBstWouldAddEndPuncttrue
\mciteSetBstMidEndSepPunct{\mcitedefaultmidpunct}
{\mcitedefaultendpunct}{\mcitedefaultseppunct}\relax
\EndOfBibitem
\bibitem[The((accessed 2023-09-15))]{Thereda_DB}
The Thermodynamic Reference Database (THEREDA). (accessed 2023-09-15);
  \url{https://www.thereda.de/en}\relax
\mciteBstWouldAddEndPuncttrue
\mciteSetBstMidEndSepPunct{\mcitedefaultmidpunct}
{\mcitedefaultendpunct}{\mcitedefaultseppunct}\relax
\EndOfBibitem
\bibitem[Blasco \latin{et~al.}(2017)Blasco, Gimeno, and
  Auqu\'{e}]{Blasco_2017_ProcEarthPlanetSci}
Blasco,~M.; Gimeno,~M.~J.; Auqu\'{e},~L.~F. {Comparison of Different
  Thermodynamic Databases used in a Geothermometrical Modelling Calculation}.
  \emph{Procedia Earth Planet. Sci.} \textbf{2017}, \emph{17}, 120--123\relax
\mciteBstWouldAddEndPuncttrue
\mciteSetBstMidEndSepPunct{\mcitedefaultmidpunct}
{\mcitedefaultendpunct}{\mcitedefaultseppunct}\relax
\EndOfBibitem
\bibitem[L. and H.(1970)L., and H.]{Kaufman_Calphad_book_1970}
L.,~K.; H.,~B. \emph{{Computer Calculation of Phase Diagrams}}; Academic Press,
  NY, 1970\relax
\mciteBstWouldAddEndPuncttrue
\mciteSetBstMidEndSepPunct{\mcitedefaultmidpunct}
{\mcitedefaultendpunct}{\mcitedefaultseppunct}\relax
\EndOfBibitem
\bibitem[Liu and Yi(2016)Liu, and Yi]{Comput_Thermodyn_Mater_book}
Liu,~Z.-K.; Yi,~W. \emph{{Computational Thermodynamics of Materials}};
  Cambridge University Press, 2016\relax
\mciteBstWouldAddEndPuncttrue
\mciteSetBstMidEndSepPunct{\mcitedefaultmidpunct}
{\mcitedefaultendpunct}{\mcitedefaultseppunct}\relax
\EndOfBibitem
\bibitem[Che((accessed 2023-09-15))]{ChemEQL_program}
ChemEQL -- a software for the calculation of chemical equilibria. (accessed
  2023-09-15);
  \url{https://www.eawag.ch/en/department/surf/projects/chemeql/}\relax
\mciteBstWouldAddEndPuncttrue
\mciteSetBstMidEndSepPunct{\mcitedefaultmidpunct}
{\mcitedefaultendpunct}{\mcitedefaultseppunct}\relax
\EndOfBibitem
\bibitem[MIN((accessed 2023-09-15))]{MINEQL_program}
MINEQL+ Version 5.0. (accessed 2023-09-15); \url{https://www.mineql.com/}\relax
\mciteBstWouldAddEndPuncttrue
\mciteSetBstMidEndSepPunct{\mcitedefaultmidpunct}
{\mcitedefaultendpunct}{\mcitedefaultseppunct}\relax
\EndOfBibitem
\bibitem[Vis((accessed 2023-09-15))]{Visual_MintEQ}
Visual MINTEQ -- a chemical equilibrium model. (accessed 2023-09-15);
  \url{https://vminteq.lwr.kth.se/}\relax
\mciteBstWouldAddEndPuncttrue
\mciteSetBstMidEndSepPunct{\mcitedefaultmidpunct}
{\mcitedefaultendpunct}{\mcitedefaultseppunct}\relax
\EndOfBibitem
\bibitem[Voter(1997)]{Voter_Hyperdynamics_PRL_1997}
Voter,~A.~F. {Hyperdynamics: Accelerated Molecular Dynamics of Infrequent
  Events}. \emph{Phys. Rev. Lett.} \textbf{1997}, \emph{78}, 3908--3911\relax
\mciteBstWouldAddEndPuncttrue
\mciteSetBstMidEndSepPunct{\mcitedefaultmidpunct}
{\mcitedefaultendpunct}{\mcitedefaultseppunct}\relax
\EndOfBibitem
\bibitem[Miron and Fichthorn(2003)Miron, and
  Fichthorn]{Miron_2003_JCP.119.6210}
Miron,~R.~A.; Fichthorn,~K.~A. {Accelerated Molecular Dynamics with the
  Bond-Boost Method}. \emph{J. Chem. Phys.} \textbf{2003}, \emph{119},
  6210--6216\relax
\mciteBstWouldAddEndPuncttrue
\mciteSetBstMidEndSepPunct{\mcitedefaultmidpunct}
{\mcitedefaultendpunct}{\mcitedefaultseppunct}\relax
\EndOfBibitem
\bibitem[Sushko \latin{et~al.}(2019)Sushko, Solov'yov, and
  Solov'yov]{Sushko_2019_MBNStudio}
Sushko,~G.~B.; Solov'yov,~I.~A.; Solov'yov,~A.~V. {Modeling MesoBioNano Systems
  with MBN Studio Made Easy}. \emph{J. Mol. Graph. Model.} \textbf{2019},
  \emph{88}, 247--260\relax
\mciteBstWouldAddEndPuncttrue
\mciteSetBstMidEndSepPunct{\mcitedefaultmidpunct}
{\mcitedefaultendpunct}{\mcitedefaultseppunct}\relax
\EndOfBibitem
\bibitem[PyM((accessed 2023-09-16))]{PyMOL}
PyMOL, a molecular visualization system. (accessed 2023-09-16);
  \url{https://pymol.org/2/}\relax
\mciteBstWouldAddEndPuncttrue
\mciteSetBstMidEndSepPunct{\mcitedefaultmidpunct}
{\mcitedefaultendpunct}{\mcitedefaultseppunct}\relax
\EndOfBibitem
\bibitem[Liu and Quek(2014)Liu, and Quek]{Liu_FEM_book_2014}
Liu,~G.~R.; Quek,~S.~S. \emph{{The Finite Element Method. A Practical Course}},
  2nd ed.; Elsevier, 2014\relax
\mciteBstWouldAddEndPuncttrue
\mciteSetBstMidEndSepPunct{\mcitedefaultmidpunct}
{\mcitedefaultendpunct}{\mcitedefaultseppunct}\relax
\EndOfBibitem
\bibitem[Whiteley(2017)]{Whiteley_FEM_book_2017}
Whiteley,~J. \emph{{Finite Element Methods. A Practical Guide}}; Springer
  International Publishing, Cham, 2017\relax
\mciteBstWouldAddEndPuncttrue
\mciteSetBstMidEndSepPunct{\mcitedefaultmidpunct}
{\mcitedefaultendpunct}{\mcitedefaultseppunct}\relax
\EndOfBibitem
\bibitem[Ataei and Mamaghani(2017)Ataei, and Mamaghani]{FEM_ABAQUS_book_2017}
Ataei,~H.; Mamaghani,~M. \emph{{Finite Element Analysis Applications and Solved
  Problems using ABAQUS}}; CreateSpace Independent Publishing Platform,
  2017\relax
\mciteBstWouldAddEndPuncttrue
\mciteSetBstMidEndSepPunct{\mcitedefaultmidpunct}
{\mcitedefaultendpunct}{\mcitedefaultseppunct}\relax
\EndOfBibitem
\bibitem[Boulbes(2020)]{Boulbes_FEM_Abaqus_2020}
Boulbes,~R.~J. \emph{{Troubleshooting Finite-Element Modeling with Abaqus}};
  Springer, Cham, 2020\relax
\mciteBstWouldAddEndPuncttrue
\mciteSetBstMidEndSepPunct{\mcitedefaultmidpunct}
{\mcitedefaultendpunct}{\mcitedefaultseppunct}\relax
\EndOfBibitem
\bibitem[Cashwell and Everett(1959)Cashwell, and
  Everett]{Practical_Manual_MC_1959}
Cashwell,~E.~D.; Everett,~C.~J. \emph{{A Practical Manual on the Monte Carlo
  Method for Random Walk Problems}}; Pergamon Press, London, 1959\relax
\mciteBstWouldAddEndPuncttrue
\mciteSetBstMidEndSepPunct{\mcitedefaultmidpunct}
{\mcitedefaultendpunct}{\mcitedefaultseppunct}\relax
\EndOfBibitem
\bibitem[{Garc\'{i}a G\'{o}mez-Tejedor} and Fuss(2012){Garc\'{i}a
  G\'{o}mez-Tejedor}, and Fuss]{Garcia_Fuss_RADAM_BiomolSyst}
{Garc\'{i}a G\'{o}mez-Tejedor},~G., Fuss,~M.~C., Eds. \emph{{Radiation Damage
  in Biomolecular Systems}}; Springer Dordrecht, 2012\relax
\mciteBstWouldAddEndPuncttrue
\mciteSetBstMidEndSepPunct{\mcitedefaultmidpunct}
{\mcitedefaultendpunct}{\mcitedefaultseppunct}\relax
\EndOfBibitem
\bibitem[Hahn(2023)]{Hahn_2023_JPhycCommun.7.042001}
Hahn,~M.~B. {Accessing Radiation Damage to Biomolecules on the Nanoscale by
  Particle-Scattering Simulations}. \emph{J. Phys. Commun.} \textbf{2023},
  \emph{7}, 042001\relax
\mciteBstWouldAddEndPuncttrue
\mciteSetBstMidEndSepPunct{\mcitedefaultmidpunct}
{\mcitedefaultendpunct}{\mcitedefaultseppunct}\relax
\EndOfBibitem
\bibitem[Sakata \latin{et~al.}(2016)Sakata, Incerti, Bordage, Lampe, Okada,
  Emfietzoglou, Kyriakou, Murakami, Sasaki, Tran, Guatelli, and
  Ivantchenko]{Sakata_2016_JAP.120.244901}
Sakata,~D.; Incerti,~S.; Bordage,~M.~C.; Lampe,~N.; Okada,~S.;
  Emfietzoglou,~D.; Kyriakou,~I.; Murakami,~K.; Sasaki,~T.; Tran,~H.;
  Guatelli,~S.; Ivantchenko,~V.~N. {An Implementation of Discrete Electron
  Transport Models for Gold in the Geant4 Simulation Toolkit}. \emph{J. Appl.
  Phys.} \textbf{2016}, \emph{120}, 244901\relax
\mciteBstWouldAddEndPuncttrue
\mciteSetBstMidEndSepPunct{\mcitedefaultmidpunct}
{\mcitedefaultendpunct}{\mcitedefaultseppunct}\relax
\EndOfBibitem
\bibitem[TEC((accessed 2023-10-26))]{TECHNO-CLS_website}
EIC-Pathfinder Project ``Emerging technologies for crystal-based gamma-ray
  light sources'' (TECHNO-CLS). (accessed 2023-10-26);
  \url{https://www.mbnresearch.com/TECHNO-CLS/Main}\relax
\mciteBstWouldAddEndPuncttrue
\mciteSetBstMidEndSepPunct{\mcitedefaultmidpunct}
{\mcitedefaultendpunct}{\mcitedefaultseppunct}\relax
\EndOfBibitem
\bibitem[Han \latin{et~al.}(2021)Han, Isborn, and Shi]{Han_2021_JCTC.17.889}
Han,~B.; Isborn,~C.~M.; Shi,~L. {Determining Partial Atomic Charges for Liquid
  Water: Assessing Electronic Structure and Charge Models}. \emph{J. Chem.
  Theory Comput.} \textbf{2021}, \emph{17}, 889--901\relax
\mciteBstWouldAddEndPuncttrue
\mciteSetBstMidEndSepPunct{\mcitedefaultmidpunct}
{\mcitedefaultendpunct}{\mcitedefaultseppunct}\relax
\EndOfBibitem
\bibitem[Bleiziffer \latin{et~al.}(2018)Bleiziffer, Schaller, and
  Riniker]{Bleiziffer_2018_JChemInfModel.58.579}
Bleiziffer,~P.; Schaller,~K.; Riniker,~S. {Machine Learning of Partial Charges
  Derived From High-Quality Quantum-Mechanical Calculations}. \emph{J. Chem.
  Inf. Model.} \textbf{2018}, \emph{58}, 579--590\relax
\mciteBstWouldAddEndPuncttrue
\mciteSetBstMidEndSepPunct{\mcitedefaultmidpunct}
{\mcitedefaultendpunct}{\mcitedefaultseppunct}\relax
\EndOfBibitem
\bibitem[M\"{u}ser \latin{et~al.}(2023)M\"{u}ser, Sukhomlinov, and
  Pastewka]{Mueser_2023_AdvPhysX.8.2093129}
M\"{u}ser,~M.~H.; Sukhomlinov,~S.~V.; Pastewka,~L. {Interatomic Potentials:
  Achievements and Challenges}. \emph{Adv. Phys. X} \textbf{2023}, \emph{8},
  2093129\relax
\mciteBstWouldAddEndPuncttrue
\mciteSetBstMidEndSepPunct{\mcitedefaultmidpunct}
{\mcitedefaultendpunct}{\mcitedefaultseppunct}\relax
\EndOfBibitem
\bibitem[Elstner \latin{et~al.}(1998)Elstner, Porezaga, Jungnickel, Elsner,
  Haugk, Frauenheim, Suhai, and Seifert]{Elstner_1998_PRB.58.7260}
Elstner,~M.; Porezaga,~D.; Jungnickel,~G.; Elsner,~J.; Haugk,~M.;
  Frauenheim,~T.; Suhai,~S.; Seifert,~G. {Self-Consistent-Charge
  Density-Functional Tight-Binding Method for Simulations of Complex Materials
  Properties}. \emph{Phys. Rev. B} \textbf{1998}, \emph{58}, 7260--7268\relax
\mciteBstWouldAddEndPuncttrue
\mciteSetBstMidEndSepPunct{\mcitedefaultmidpunct}
{\mcitedefaultendpunct}{\mcitedefaultseppunct}\relax
\EndOfBibitem
\bibitem[Hourahine \latin{et~al.}(2020)Hourahine, Aradi, Blum, Bonaf\'{e},
  Buccheri, Camacho, Cevallos, Deshaye, Dumitric\v{a}, Dominguez, and
  \textit{et al.}]{Hourahine_2020_JCP.152.124101}
Hourahine,~B.; Aradi,~B.; Blum,~V.; Bonaf\'{e},~F.; Buccheri,~A.; Camacho,~C.;
  Cevallos,~C.; Deshaye,~M.~Y.; Dumitric\v{a},~T.; Dominguez,~A.; \textit{et
  al.} {DFTB+, A Software Package for Efficient Approximate Density Functional
  Theory Based Atomistic Simulations}. \emph{J. Chem. Phys.} \textbf{2020},
  \emph{152}, 124101\relax
\mciteBstWouldAddEndPuncttrue
\mciteSetBstMidEndSepPunct{\mcitedefaultmidpunct}
{\mcitedefaultendpunct}{\mcitedefaultseppunct}\relax
\EndOfBibitem
\bibitem[Vennelakanti \latin{et~al.}(2022)Vennelakanti, Nazemi, Mehmood,
  Steeves, and Kulik]{Vennelakanti_2022_CurrOpinStructBiol}
Vennelakanti,~V.; Nazemi,~A.; Mehmood,~R.; Steeves,~A.~H.; Kulik,~H.~J.
  {Harder, Better, Faster, Stronger: Large-scale QM and QM/MM for Predictive
  Modeling in Enzymes and Proteins}. \emph{Curr. Opin. Struct. Biol.}
  \textbf{2022}, \emph{72}, 9--17\relax
\mciteBstWouldAddEndPuncttrue
\mciteSetBstMidEndSepPunct{\mcitedefaultmidpunct}
{\mcitedefaultendpunct}{\mcitedefaultseppunct}\relax
\EndOfBibitem
\bibitem[Kuba\v{r} \latin{et~al.}(2023)Kuba\v{r}, Elstner, and
  Cui]{Kubar_2023_AnnuRevBiophys}
Kuba\v{r},~T.; Elstner,~M.; Cui,~Q. {Hybrid Quantum Mechanical/Molecular
  Mechanical Methods For Studying Energy Transduction in Biomolecular
  Machines}. \emph{Annu. Rev. Biophys.} \textbf{2023}, \emph{52},
  525--551\relax
\mciteBstWouldAddEndPuncttrue
\mciteSetBstMidEndSepPunct{\mcitedefaultmidpunct}
{\mcitedefaultendpunct}{\mcitedefaultseppunct}\relax
\EndOfBibitem
\bibitem[Kulik \latin{et~al.}(2016)Kulik, Zhang, Klinman, and
  Martinez]{Kulik_2016_JPCB.120.11381}
Kulik,~H.~J.; Zhang,~J.; Klinman,~J.~P.; Martinez,~T.~J. {How Large Should the
  QM Region Be in QM/MM Calculations? The Case of Catechol
  $O$-Methyltransferase}. \emph{J. Phys. Chem. B} \textbf{2016}, \emph{120},
  11381--11394\relax
\mciteBstWouldAddEndPuncttrue
\mciteSetBstMidEndSepPunct{\mcitedefaultmidpunct}
{\mcitedefaultendpunct}{\mcitedefaultseppunct}\relax
\EndOfBibitem
\bibitem[Jindal and Warshel(2016)Jindal, and
  Warshel]{Jindal_2016_JPCB.120.9913}
Jindal,~G.; Warshel,~A. {Exploring the Dependence of QM/MM Calculations of
  Enzyme Catalysis on the Size of the QM Region}. \emph{J. Phys. Chem. B}
  \textbf{2016}, \emph{120}, 9913--9921\relax
\mciteBstWouldAddEndPuncttrue
\mciteSetBstMidEndSepPunct{\mcitedefaultmidpunct}
{\mcitedefaultendpunct}{\mcitedefaultseppunct}\relax
\EndOfBibitem
\bibitem[Das \latin{et~al.}(2018)Das, Nam, and Major]{Das_2018_JCTC.13.1695}
Das,~S.; Nam,~K.; Major,~D.~T. {Rapid Convergence of Energy and Free Energy
  Profiles with Quantum Mechanical Size in Quantum Mechanical--Molecular
  Mechanical Simulations of Proton Transfer in DNA}. \emph{J. Chem. Theory
  Comput.} \textbf{2018}, \emph{14}, 1695--1705\relax
\mciteBstWouldAddEndPuncttrue
\mciteSetBstMidEndSepPunct{\mcitedefaultmidpunct}
{\mcitedefaultendpunct}{\mcitedefaultseppunct}\relax
\EndOfBibitem
\bibitem[Azzolini \latin{et~al.}(2018)Azzolini, Angelucci, Cimino, Larciprete,
  Pugno, Taioli, and Dapor]{Azzolini_2018_JPCM.31.055901_SEED}
Azzolini,~M.; Angelucci,~M.; Cimino,~R.; Larciprete,~R.; Pugno,~N.~M.;
  Taioli,~S.; Dapor,~M. {Secondary Electron Emission and Yield Spectra of
  Metals from Monte Carlo Simulations and Experiments}. \emph{J. Phys. Condens.
  Matter} \textbf{2018}, \emph{31}, 055901\relax
\mciteBstWouldAddEndPuncttrue
\mciteSetBstMidEndSepPunct{\mcitedefaultmidpunct}
{\mcitedefaultendpunct}{\mcitedefaultseppunct}\relax
\EndOfBibitem
\bibitem[Solov'yov \latin{et~al.}(2023)Solov'yov, Prosvetov, Sushko, and
  Solov'yov]{SD_FEBID_abstract_Prague2023}
Solov'yov,~I.~A.; Prosvetov,~A.; Sushko,~G.; Solov'yov,~A.~V. \emph{The Second
  Conference ``Multiscale Irradiation and Chemistry Driven Processes and
  Related Technologies'' (Prague, Czech Republic, April 26-28, 2023). Book of
  Abstracts}; 2023; p~26\relax
\mciteBstWouldAddEndPuncttrue
\mciteSetBstMidEndSepPunct{\mcitedefaultmidpunct}
{\mcitedefaultendpunct}{\mcitedefaultseppunct}\relax
\EndOfBibitem
\bibitem[Moskovkin \latin{et~al.}(2014)Moskovkin, Panshenskov, Lucas, and
  Solov'yov]{Moskovkin_2014_PSSB.251.1456}
Moskovkin,~P.; Panshenskov,~M.; Lucas,~S.; Solov'yov,~A.~V. {Simulation of
  Nanowire Fragmentation By Means of Kinetic Monte Carlo Approach: 2D Case}.
  \emph{Phys. Stat. Sol. B} \textbf{2014}, \emph{251}, 1456--1462\relax
\mciteBstWouldAddEndPuncttrue
\mciteSetBstMidEndSepPunct{\mcitedefaultmidpunct}
{\mcitedefaultendpunct}{\mcitedefaultseppunct}\relax
\EndOfBibitem
\bibitem[Yakubovich \latin{et~al.}(2006)Yakubovich, Solov'yov, Solov'yov, and
  Greiner]{Yakubovich_2006_EPJD.40.363}
Yakubovich,~A.~V.; Solov'yov,~I.~A.; Solov'yov,~A.~V.; Greiner,~W. {Phase
  Transition in Polypeptides: A Step Towards the Understanding of Protein
  Folding}. \emph{Eur. Phys. J. D} \textbf{2006}, \emph{40}, 363--367\relax
\mciteBstWouldAddEndPuncttrue
\mciteSetBstMidEndSepPunct{\mcitedefaultmidpunct}
{\mcitedefaultendpunct}{\mcitedefaultseppunct}\relax
\EndOfBibitem
\bibitem[Yakubovich \latin{et~al.}(2008)Yakubovich, Solov'yov, Solov'yov, and
  Greiner]{Yakubovich_2008_EPJD.46.215}
Yakubovich,~A.~V.; Solov'yov,~I.~A.; Solov'yov,~A.~V.; Greiner,~W. {\textit{Ab
  initio} Theory of Helix$\leftrightarrow$Coil Phase Transition}. \emph{Eur.
  Phys. J. D} \textbf{2008}, \emph{46}, 215--225\relax
\mciteBstWouldAddEndPuncttrue
\mciteSetBstMidEndSepPunct{\mcitedefaultmidpunct}
{\mcitedefaultendpunct}{\mcitedefaultseppunct}\relax
\EndOfBibitem
\bibitem[Solov'yov \latin{et~al.}(2008)Solov'yov, Yakubovich, Solov'yov, and
  Greiner]{Solovyov_2008_EPJD.46.227}
Solov'yov,~I.~A.; Yakubovich,~A.~V.; Solov'yov,~A.~V.; Greiner,~W.
  {$\alpha$-Helix$\leftrightarrow$Random Coil Phase Transition: Analysis of
  \textit{Ab Initio} Theory Predictions}. \emph{Eur. Phys. J. D} \textbf{2008},
  \emph{46}, 227--240\relax
\mciteBstWouldAddEndPuncttrue
\mciteSetBstMidEndSepPunct{\mcitedefaultmidpunct}
{\mcitedefaultendpunct}{\mcitedefaultseppunct}\relax
\EndOfBibitem
\bibitem[Yakubovich and Solov'yov(2014)Yakubovich, and
  Solov'yov]{Yakubovich_2014_EPJD.68.145}
Yakubovich,~A.~V.; Solov'yov,~A.~V. {Quantitative Thermodynamic Model for
  Globular Protein Folding}. \emph{Eur. Phys. J. D} \textbf{2014}, \emph{68},
  145\relax
\mciteBstWouldAddEndPuncttrue
\mciteSetBstMidEndSepPunct{\mcitedefaultmidpunct}
{\mcitedefaultendpunct}{\mcitedefaultseppunct}\relax
\EndOfBibitem
\bibitem[Landau and Lifshitz(1987)Landau, and Lifshitz]{LL6}
Landau,~L.~D.; Lifshitz,~E.~M. \emph{{Fluid Mechanics}}, 2nd ed.;
  Butterworth-Heinemann: Oxford, 1987\relax
\mciteBstWouldAddEndPuncttrue
\mciteSetBstMidEndSepPunct{\mcitedefaultmidpunct}
{\mcitedefaultendpunct}{\mcitedefaultseppunct}\relax
\EndOfBibitem
\bibitem[Zel'dovich and Raiser(1966)Zel'dovich, and
  Raiser]{Zeldovich_ShockWaves}
Zel'dovich,~Y.~B.; Raiser,~Y.~P. \emph{{Physics of Shock Waves and
  High-Temperature Hydrodynamic Phenomena}}; Academic Press: New York,
  1966\relax
\mciteBstWouldAddEndPuncttrue
\mciteSetBstMidEndSepPunct{\mcitedefaultmidpunct}
{\mcitedefaultendpunct}{\mcitedefaultseppunct}\relax
\EndOfBibitem
\bibitem[Surdutovich and Solov'yov(2017)Surdutovich, and
  Solov'yov]{ES_AVS_2017_EPJD.71.210}
Surdutovich,~E.; Solov'yov,~A.~V. {Cell Survival Probability in a Spread-Out
  Bragg Peak for Novel Treatment Planning}. \emph{Eur. Phys. J. D}
  \textbf{2017}, \emph{71}, 210\relax
\mciteBstWouldAddEndPuncttrue
\mciteSetBstMidEndSepPunct{\mcitedefaultmidpunct}
{\mcitedefaultendpunct}{\mcitedefaultseppunct}\relax
\EndOfBibitem
\bibitem[Winkler \latin{et~al.}(2019)Winkler, Fowlkes, Rack, and
  Plank]{Winkler_2019_JAP_review}
Winkler,~R.; Fowlkes,~J.~D.; Rack,~P.~D.; Plank,~H. {3D Nanoprinting via
  Focused Electron Beams}. \emph{J. Appl. Phys.} \textbf{2019}, \emph{125},
  210901\relax
\mciteBstWouldAddEndPuncttrue
\mciteSetBstMidEndSepPunct{\mcitedefaultmidpunct}
{\mcitedefaultendpunct}{\mcitedefaultseppunct}\relax
\EndOfBibitem
\bibitem[Plank \latin{et~al.}(2020)Plank, Winkler, Schwalb, H{\"{u}}tner,
  Fowlkes, Rack, Utke, and Huth]{Plank_2020_Micromachines.11.48}
Plank,~H.; Winkler,~R.; Schwalb,~C.~H.; H{\"{u}}tner,~J.; Fowlkes,~J.~D.;
  Rack,~P.~D.; Utke,~I.; Huth,~M. {Focused Electron Beam-Based 3D Nanoprinting
  for Scanning Probe Microscopy: A Review}. \emph{Micromachines} \textbf{2020},
  \emph{11}, 48\relax
\mciteBstWouldAddEndPuncttrue
\mciteSetBstMidEndSepPunct{\mcitedefaultmidpunct}
{\mcitedefaultendpunct}{\mcitedefaultseppunct}\relax
\EndOfBibitem
\bibitem[Fleming and Williams(2019)Fleming, and
  Williams]{Fleming_Spectroscopy_OrgChem}
Fleming,~I.; Williams,~D. \emph{{Spectroscopic Methods in Organic Chemistry}},
  7th ed.; Springer: Cham, 2019\relax
\mciteBstWouldAddEndPuncttrue
\mciteSetBstMidEndSepPunct{\mcitedefaultmidpunct}
{\mcitedefaultendpunct}{\mcitedefaultseppunct}\relax
\EndOfBibitem
\bibitem[Kole\.{z}y\'{n}ski and Kr\'{o}l(219)Kole\.{z}y\'{n}ski, and
  Kr\'{o}l]{Kolezynski_MolecSpectroscopy}
Kole\.{z}y\'{n}ski,~A., Kr\'{o}l,~M., Eds. \emph{{Molecular Spectroscopy --
  Experiment and Theory: From Molecules to Functional Materials}}; Springer:
  Cham, 219\relax
\mciteBstWouldAddEndPuncttrue
\mciteSetBstMidEndSepPunct{\mcitedefaultmidpunct}
{\mcitedefaultendpunct}{\mcitedefaultseppunct}\relax
\EndOfBibitem
\bibitem[Svanberg(2022)]{Svanberg_AtMolSpectrosc}
Svanberg,~S. \emph{{Atomic and Molecular Spectroscopy: Basic Aspects and
  Practical Applications}}, 5th ed.; Springer: Cham, 2022\relax
\mciteBstWouldAddEndPuncttrue
\mciteSetBstMidEndSepPunct{\mcitedefaultmidpunct}
{\mcitedefaultendpunct}{\mcitedefaultseppunct}\relax
\EndOfBibitem
\bibitem[Cramer(2020)]{Cramer_X-raySpectroscopy}
Cramer,~S.~P. \emph{{X-Ray Spectroscopy with Synchrotron Radiation:
  Fundamentals and Applications}}; Springer: Cham, 2020\relax
\mciteBstWouldAddEndPuncttrue
\mciteSetBstMidEndSepPunct{\mcitedefaultmidpunct}
{\mcitedefaultendpunct}{\mcitedefaultseppunct}\relax
\EndOfBibitem
\bibitem[Chen \latin{et~al.}(2021)Chen, Venables, and
  Sigrist]{Spectrosc_Atmosphere}
Chen,~W., Venables,~D.~S., Sigrist,~M.~W., Eds. \emph{{Advances in
  Spectroscopic Monitoring of the Atmosphere}}; Elsevier: Amsterdam, 2021\relax
\mciteBstWouldAddEndPuncttrue
\mciteSetBstMidEndSepPunct{\mcitedefaultmidpunct}
{\mcitedefaultendpunct}{\mcitedefaultseppunct}\relax
\EndOfBibitem
\bibitem[Protopopov(2022)]{Spectrosc_Semiconductor}
Protopopov,~V. \emph{{Spectroscopic Techniques For Semiconductor Industry}};
  World Scientific Publishing: Singapore, 2022\relax
\mciteBstWouldAddEndPuncttrue
\mciteSetBstMidEndSepPunct{\mcitedefaultmidpunct}
{\mcitedefaultendpunct}{\mcitedefaultseppunct}\relax
\EndOfBibitem
\bibitem[Tennyson(2019)]{Spectrosc_Astronomical}
Tennyson,~J. \emph{{Astronomical Spectroscopy: An Introduction to the Atomic
  and Molecular Physics of Astronomical Spectroscopy}}, 3rd ed.; World
  Scientific Publishing Europe: London, 2019\relax
\mciteBstWouldAddEndPuncttrue
\mciteSetBstMidEndSepPunct{\mcitedefaultmidpunct}
{\mcitedefaultendpunct}{\mcitedefaultseppunct}\relax
\EndOfBibitem
\bibitem[VAM((accessed 2023-10-18))]{VAMDC_portal}
VAMDC (Virtual Atomic and Molecular Database Consortium) Portal. (accessed
  2023-10-18); \url{https://portal.vamdc.org/vamdc_portal/home.seam}\relax
\mciteBstWouldAddEndPuncttrue
\mciteSetBstMidEndSepPunct{\mcitedefaultmidpunct}
{\mcitedefaultendpunct}{\mcitedefaultseppunct}\relax
\EndOfBibitem
\bibitem[Albert \latin{et~al.}(2020)Albert, Antony, Ba, Babikov, Bollard,
  Boudon, Delahaye, {Del Zanna}, Dimitrijevi\'{c}, Drouin, and \textit{et
  al.}]{Albert_VAMDC_2020_Atoms.8.76}
Albert,~D.; Antony,~B.~K.; Ba,~Y.~A.; Babikov,~Y.~L.; Bollard,~P.; Boudon,~V.;
  Delahaye,~F.; {Del Zanna},~G.; Dimitrijevi\'{c},~M.~S.; Drouin,~B.~J.;
  \textit{et al.} {A Decade with VAMDC: Results and Ambitions}. \emph{Atoms}
  \textbf{2020}, \emph{8}, 76\relax
\mciteBstWouldAddEndPuncttrue
\mciteSetBstMidEndSepPunct{\mcitedefaultmidpunct}
{\mcitedefaultendpunct}{\mcitedefaultseppunct}\relax
\EndOfBibitem
\bibitem[VES((accessed 2023-10-18))]{VESPA_portal}
VESPA (Virtual European Solar and Planetary Access) Portal. (accessed
  2023-10-18); \url{https://vespa.obspm.fr/planetary/data/}\relax
\mciteBstWouldAddEndPuncttrue
\mciteSetBstMidEndSepPunct{\mcitedefaultmidpunct}
{\mcitedefaultendpunct}{\mcitedefaultseppunct}\relax
\EndOfBibitem
\bibitem[Giel~Berden(2009)]{Berden_CavityRing_Spectrosc}
Giel~Berden,~R.~E., Ed. \emph{{Cavity Ring-Down Spectroscopy: Techniques and
  Applications}}; Blackwell Publishing: Chichester, UK, 2009\relax
\mciteBstWouldAddEndPuncttrue
\mciteSetBstMidEndSepPunct{\mcitedefaultmidpunct}
{\mcitedefaultendpunct}{\mcitedefaultseppunct}\relax
\EndOfBibitem
\bibitem[Maiuri \latin{et~al.}(2020)Maiuri, Garavelli, and
  Cerullo]{Maiuri_2020_JACS.142.3}
Maiuri,~M.; Garavelli,~M.; Cerullo,~G. {Ultrafast Spectroscopy: State of the
  Art and Open Challenges}. \emph{J. Am. Chem. Soc.} \textbf{2020}, \emph{142},
  3--15\relax
\mciteBstWouldAddEndPuncttrue
\mciteSetBstMidEndSepPunct{\mcitedefaultmidpunct}
{\mcitedefaultendpunct}{\mcitedefaultseppunct}\relax
\EndOfBibitem
\bibitem[Fabrikant \latin{et~al.}(1988)Fabrikant, Shpenik, Snegursky, and
  Zavilopulo]{Fabrikant_1988_PhysRep.159.1}
Fabrikant,~I.~I.; Shpenik,~O.~B.; Snegursky,~A.~V.; Zavilopulo,~A.~N. {Electron
  Impact Formation of Metastable Atoms}. \emph{Phys. Rep.} \textbf{1988},
  \emph{159}, 1--97\relax
\mciteBstWouldAddEndPuncttrue
\mciteSetBstMidEndSepPunct{\mcitedefaultmidpunct}
{\mcitedefaultendpunct}{\mcitedefaultseppunct}\relax
\EndOfBibitem
\bibitem[Brydson(2001)]{Brydson_EELS_book}
Brydson,~R. \emph{{Electron Energy Loss Spectroscopy}}, 1st ed.; Garland
  Science, London, 2001\relax
\mciteBstWouldAddEndPuncttrue
\mciteSetBstMidEndSepPunct{\mcitedefaultmidpunct}
{\mcitedefaultendpunct}{\mcitedefaultseppunct}\relax
\EndOfBibitem
\bibitem[Ibach and Mills(1982)Ibach, and Mills]{Ibach_Mills_EELS}
Ibach,~H.; Mills,~D.~L. \emph{{Electron Energy Loss Spectroscopy and Surface
  Vibration}}; Academic Press, New York, 1982\relax
\mciteBstWouldAddEndPuncttrue
\mciteSetBstMidEndSepPunct{\mcitedefaultmidpunct}
{\mcitedefaultendpunct}{\mcitedefaultseppunct}\relax
\EndOfBibitem
\bibitem[Mason \latin{et~al.}(2006)Mason, Dawes, Holtom, Mukerji, Davis,
  Sivaraman, Kaiser, Hoffmann, and Shaw]{Mason_2006_FaradayDisc.133.311}
Mason,~N.~J.; Dawes,~A.; Holtom,~P.~D.; Mukerji,~R.~J.; Davis,~M.~P.;
  Sivaraman,~B.; Kaiser,~R.~I.; Hoffmann,~S.~V.; Shaw,~D.~A. {VUV Spectroscopy
  and Photo-Processing of Astrochemical Ices: An Experimental Study}.
  \emph{Faraday Discus.} \textbf{2006}, \emph{133}, 311--329\relax
\mciteBstWouldAddEndPuncttrue
\mciteSetBstMidEndSepPunct{\mcitedefaultmidpunct}
{\mcitedefaultendpunct}{\mcitedefaultseppunct}\relax
\EndOfBibitem
\bibitem[H\"{u}fner(2003)]{Huefner_PES_book}
H\"{u}fner,~S. \emph{{Photoelectron Spectroscopy: Principles and
  Applications}}, 3rd ed.; Springer Berlin, Heidelberg, 2003\relax
\mciteBstWouldAddEndPuncttrue
\mciteSetBstMidEndSepPunct{\mcitedefaultmidpunct}
{\mcitedefaultendpunct}{\mcitedefaultseppunct}\relax
\EndOfBibitem
\bibitem[Whitten(2023)]{Whitten_UV-PES_ApplSurfSci_2023}
Whitten,~J.~E. {Ultraviolet Photoelectron Spectroscopy: Practical Aspects and
  Best Practices}. \emph{Appl. Surf. Sci. Adv.} \textbf{2023}, \emph{13},
  100384\relax
\mciteBstWouldAddEndPuncttrue
\mciteSetBstMidEndSepPunct{\mcitedefaultmidpunct}
{\mcitedefaultendpunct}{\mcitedefaultseppunct}\relax
\EndOfBibitem
\bibitem[{van der Heide}(2012)]{van_der_Heide_X-ray_PES}
{van der Heide},~P. \emph{{X-ray Photoelectron Spectroscopy: An introduction to
  Principles and Practices}}; John Wiley \& Sons, Hoboken, NJ, 2012\relax
\mciteBstWouldAddEndPuncttrue
\mciteSetBstMidEndSepPunct{\mcitedefaultmidpunct}
{\mcitedefaultendpunct}{\mcitedefaultseppunct}\relax
\EndOfBibitem
\bibitem[Blackledge(2023)]{Blackledge_PES_Forensic}
Blackledge,~R.~D., Ed. \emph{{Leading Edge Techniques in Forensic Trace
  Evidence Analysis: More New Trace Analysis Methods}}; John Wiley \& Sons,
  Hoboken, NJ, 2023\relax
\mciteBstWouldAddEndPuncttrue
\mciteSetBstMidEndSepPunct{\mcitedefaultmidpunct}
{\mcitedefaultendpunct}{\mcitedefaultseppunct}\relax
\EndOfBibitem
\bibitem[{de Hoffmann} and Stroobant(2007){de Hoffmann}, and
  Stroobant]{deHoffmann_MassSpectrom_book}
{de Hoffmann},~E.; Stroobant,~V. \emph{{Mass Spectrometry: Principles and
  Applications}}, 3rd ed.; John Wiley \& Sons, Chichester, UK, 2007\relax
\mciteBstWouldAddEndPuncttrue
\mciteSetBstMidEndSepPunct{\mcitedefaultmidpunct}
{\mcitedefaultendpunct}{\mcitedefaultseppunct}\relax
\EndOfBibitem
\bibitem[Gross(2017)]{Gross_MS_textbook}
Gross,~J.~H. \emph{{Mass Spectrometry: A Textbook}}, 3rd ed.; Springer
  International Publishing, Cham, 2017\relax
\mciteBstWouldAddEndPuncttrue
\mciteSetBstMidEndSepPunct{\mcitedefaultmidpunct}
{\mcitedefaultendpunct}{\mcitedefaultseppunct}\relax
\EndOfBibitem
\bibitem[McCullagh and Oldham(2019)McCullagh, and
  Oldham]{McCullagh_Oldham_MassSpectrom}
McCullagh,~J.; Oldham,~N. \emph{{Mass Spectrometry}}; Oxford University Press,
  Oxford, 2019\relax
\mciteBstWouldAddEndPuncttrue
\mciteSetBstMidEndSepPunct{\mcitedefaultmidpunct}
{\mcitedefaultendpunct}{\mcitedefaultseppunct}\relax
\EndOfBibitem
\bibitem[The((accessed 2023-10-18))]{ThermoFisher_MS}
Mass Spectrometry Applications Areas. (accessed 2023-10-18);
  \url{https://www.thermofisher.com/de/de/home/industrial/mass-spectrometry/mass-spectrometry-learning-center/mass-spectrometry-applications-area.html}\relax
\mciteBstWouldAddEndPuncttrue
\mciteSetBstMidEndSepPunct{\mcitedefaultmidpunct}
{\mcitedefaultendpunct}{\mcitedefaultseppunct}\relax
\EndOfBibitem
\bibitem[Leseigneur \latin{et~al.}(2022)Leseigneur, Bredeh\"{o}ft, Gautier,
  Giri, Kr\"{u}ger, MacDermott, Meierhenrich, {Mu\~{n}oz Caro}, Raulin, Steele,
  and \textit{et al.}]{Leseigneur_2022_AngewChemIntEd.61}
Leseigneur,~G.; Bredeh\"{o}ft,~J.~H.; Gautier,~T.; Giri,~C.; Kr\"{u}ger,~H.;
  MacDermott,~A.~J.; Meierhenrich,~U.~J.; {Mu\~{n}oz Caro},~G.~M.; Raulin,~F.;
  Steele,~A.; \textit{et al.} {ESA's Cometary Mission Rosetta --
  Re-Characterization of the COSAC Mass Spectrometry Results}. \emph{Angew.
  Chem. Int. Ed.} \textbf{2022}, \emph{61}, e202201925\relax
\mciteBstWouldAddEndPuncttrue
\mciteSetBstMidEndSepPunct{\mcitedefaultmidpunct}
{\mcitedefaultendpunct}{\mcitedefaultseppunct}\relax
\EndOfBibitem
\bibitem[Altwegg \latin{et~al.}(2017)Altwegg, Balsiger, Berthelier, Bieler,
  Calmonte, Fuselier, Goesmann, Gasc, Gombosi, {Le Roy}, and \textit{et
  al.}]{Altwegg_2017_MNRAS.469.S130}
Altwegg,~K.; Balsiger,~H.; Berthelier,~J.~J.; Bieler,~A.; Calmonte,~U.;
  Fuselier,~S.~A.; Goesmann,~F.; Gasc,~S.; Gombosi,~T.~I.; {Le Roy},~L.;
  \textit{et al.} {Organics in Comet 67P -- A First Comparative Analysis of
  Mass Spectra From ROSINA--DFMS, COSAC and Ptolemy}. \emph{Mon. Notices Royal
  Astron. Soc.} \textbf{2017}, \emph{469}, S130--S141\relax
\mciteBstWouldAddEndPuncttrue
\mciteSetBstMidEndSepPunct{\mcitedefaultmidpunct}
{\mcitedefaultendpunct}{\mcitedefaultseppunct}\relax
\EndOfBibitem
\bibitem[H\"{u}bschmann(2015)]{Huebschmann_GC-MS_book}
H\"{u}bschmann,~H.-J. \emph{{Handbook of GC-MS: Fundamentals and
  Applications}}; Wiley-VCH Verlag, Weinheim, 2015\relax
\mciteBstWouldAddEndPuncttrue
\mciteSetBstMidEndSepPunct{\mcitedefaultmidpunct}
{\mcitedefaultendpunct}{\mcitedefaultseppunct}\relax
\EndOfBibitem
\bibitem[S{\o}rensen \latin{et~al.}(2011)S{\o}rensen, Overgaard, and
  Bassler]{Sorensen_2011_ActaOncol.50.757}
S{\o}rensen,~B.~S.; Overgaard,~J.; Bassler,~N. {In Vitro RBE--LET Dependence
  for Multiple Particle Types}. \emph{Acta Oncol.} \textbf{2011}, \emph{50},
  757--762\relax
\mciteBstWouldAddEndPuncttrue
\mciteSetBstMidEndSepPunct{\mcitedefaultmidpunct}
{\mcitedefaultendpunct}{\mcitedefaultseppunct}\relax
\EndOfBibitem
\bibitem[Tegami \latin{et~al.}(2017)Tegami, Bello, Luan, Mairani, Parodi, and
  Holzscheiter]{Tegami_2017_IJMPCERO}
Tegami,~S.; Bello,~S.~D.; Luan,~S.; Mairani,~A.; Parodi,~K.;
  Holzscheiter,~M.~H. {LET Monitoring Using Liquid Ionization Chambers}.
  \emph{Int. J. Med. Phys. Clin. Eng. Radiat. Oncol.} \textbf{2017}, \emph{6},
  197--207\relax
\mciteBstWouldAddEndPuncttrue
\mciteSetBstMidEndSepPunct{\mcitedefaultmidpunct}
{\mcitedefaultendpunct}{\mcitedefaultseppunct}\relax
\EndOfBibitem
\bibitem[Bauer(1987)]{Bauer_1987_NIMB.27.301}
Bauer,~P. {How to Measure Absolute Stopping Cross Sections by Backscattering
  and by Transmission Methods: Part I. Backscattering}. \emph{Nucl. Instrum.
  Meth. B} \textbf{1987}, \emph{27}, 301--314\relax
\mciteBstWouldAddEndPuncttrue
\mciteSetBstMidEndSepPunct{\mcitedefaultmidpunct}
{\mcitedefaultendpunct}{\mcitedefaultseppunct}\relax
\EndOfBibitem
\bibitem[Mertens(1987)]{Mertens_1987_NIMB.27.315}
Mertens,~P. {How to Measure Absolute Stopping Cross Sections by Backscattering
  and by Transmission Methods: Part II. Transmission}. \emph{Nucl. Instrum.
  Meth. B} \textbf{1987}, \emph{27}, 315--322\relax
\mciteBstWouldAddEndPuncttrue
\mciteSetBstMidEndSepPunct{\mcitedefaultmidpunct}
{\mcitedefaultendpunct}{\mcitedefaultseppunct}\relax
\EndOfBibitem
\bibitem[R\"{a}is\"{a}nen \latin{et~al.}(1996)R\"{a}is\"{a}nen, W\"{a}tjen,
  Plompen, and Munnik]{Raisanen_1996_NIMB.118.1}
R\"{a}is\"{a}nen,~J.; W\"{a}tjen,~U.; Plompen,~A. J.~M.; Munnik,~F. {Stopping
  Power Determinations by the Transmission Technique}. \emph{Nucl. Instrum.
  Meth. B} \textbf{1996}, \emph{118}, 1--6\relax
\mciteBstWouldAddEndPuncttrue
\mciteSetBstMidEndSepPunct{\mcitedefaultmidpunct}
{\mcitedefaultendpunct}{\mcitedefaultseppunct}\relax
\EndOfBibitem
\bibitem[Mertens \latin{et~al.}(1986)Mertens, Bauer, and
  Semrad]{Mertens_1986_NIMB.15.91}
Mertens,~P.; Bauer,~P.; Semrad,~D. {Proton Stopping Powers in Al, Ni, Cu, Ag
  and Au Measured Comparatively on Identical Targets in Backscattering and
  Transmission Geometry}. \emph{Nucl. Instrum. Meth. B} \textbf{1986},
  \emph{15}, 91--95\relax
\mciteBstWouldAddEndPuncttrue
\mciteSetBstMidEndSepPunct{\mcitedefaultmidpunct}
{\mcitedefaultendpunct}{\mcitedefaultseppunct}\relax
\EndOfBibitem
\bibitem[Fontana \latin{et~al.}(2016)Fontana, Chen, Crespillo, Graham, Xue,
  Zhang, and Weber]{Fontana_2016_NIMB.366.104}
Fontana,~C.~L.; Chen,~C.-H.; Crespillo,~M.~L.; Graham,~J.~T.; Xue,~H.;
  Zhang,~Y.; Weber,~W.~J. {Stopping power measurements with the Time-of-Flight
  (ToF) technique}. \emph{Nucl. Instrum. Meth. B} \textbf{2016}, \emph{366},
  104--116\relax
\mciteBstWouldAddEndPuncttrue
\mciteSetBstMidEndSepPunct{\mcitedefaultmidpunct}
{\mcitedefaultendpunct}{\mcitedefaultseppunct}\relax
\EndOfBibitem
\bibitem[Sihver \latin{et~al.}(1998)Sihver, Schardt, and
  Kanai]{Sihver_1998_JpnJMP.18.1}
Sihver,~L.; Schardt,~D.; Kanai,~T. {Depth-Dose Distributions of High-Energy
  Carbon, Oxygen and Neon Beams in Water}. \emph{Jpn. J. Med. Phys.}
  \textbf{1998}, \emph{18}, 1\relax
\mciteBstWouldAddEndPuncttrue
\mciteSetBstMidEndSepPunct{\mcitedefaultmidpunct}
{\mcitedefaultendpunct}{\mcitedefaultseppunct}\relax
\EndOfBibitem
\bibitem[E.~Haettner(2006)]{Haettner_2006_RPD.122.485}
E.~Haettner,~D.~S.,~H.~Iwase {Experimental Fragmentation Studies with $^{12}$C
  Therapy Beams}. \emph{Radiat. Prot. Dosim.} \textbf{2006}, \emph{122},
  485--487\relax
\mciteBstWouldAddEndPuncttrue
\mciteSetBstMidEndSepPunct{\mcitedefaultmidpunct}
{\mcitedefaultendpunct}{\mcitedefaultseppunct}\relax
\EndOfBibitem
\bibitem[Schauer \latin{et~al.}(2022)Schauer, Wieser, Huang, Ruser, Lascaud,
  W\"{u}rl, Chmyrov, Vidal, Herault, Ntziachristos, Assmann, Parodi, and
  Dollinger]{Schauer_2022_FrontOncol.12.925542}
Schauer,~J.; Wieser,~H.~P.; Huang,~Y.; Ruser,~H.; Lascaud,~J.; W\"{u}rl,~M.;
  Chmyrov,~A.; Vidal,~M.; Herault,~J.; Ntziachristos,~V.; Assmann,~W.;
  Parodi,~K.; Dollinger,~G. {Proton Beam Range Verification by Means of
  Ionoacoustic Measurements at Clinically Relevant Doses Using a
  Correlation-Based Evaluation}. \emph{Front. Oncol.} \textbf{2022}, \emph{12},
  925542\relax
\mciteBstWouldAddEndPuncttrue
\mciteSetBstMidEndSepPunct{\mcitedefaultmidpunct}
{\mcitedefaultendpunct}{\mcitedefaultseppunct}\relax
\EndOfBibitem
\bibitem[Jette and Chen(2011)Jette, and Chen]{Jette_2011_PMB.56.N131}
Jette,~D.; Chen,~W. {Creating a Spread-Out Bragg Peak in Proton Beams}.
  \emph{Phys. Med. Biol.} \textbf{2011}, \emph{56}, N131--N138\relax
\mciteBstWouldAddEndPuncttrue
\mciteSetBstMidEndSepPunct{\mcitedefaultmidpunct}
{\mcitedefaultendpunct}{\mcitedefaultseppunct}\relax
\EndOfBibitem
\bibitem[Jia \latin{et~al.}(2016)Jia, Romano, Cirrone, Cuttone, Hadizadeh,
  Mowlavi, and Raffaele]{Jia_2016_NIMA.806.101}
Jia,~S.~B.; Romano,~F.; Cirrone,~G. A.~P.; Cuttone,~G.; Hadizadeh,~M.~H.;
  Mowlavi,~A.~A.; Raffaele,~L. {Designing a range modulator wheel to spread-out
  the Bragg peak for a passive proton therapy facility}. \emph{Nucl. Instrum.
  Meth. A} \textbf{2016}, \emph{806}, 101--108\relax
\mciteBstWouldAddEndPuncttrue
\mciteSetBstMidEndSepPunct{\mcitedefaultmidpunct}
{\mcitedefaultendpunct}{\mcitedefaultseppunct}\relax
\EndOfBibitem
\bibitem[Stewart \latin{et~al.}(2007)Stewart, Elliott, and
  Seuntjens]{Stewart_2007_PMB.52.3089}
Stewart,~K.~J.; Elliott,~A.; Seuntjens,~J.~P. {Development of a Guarded Liquid
  Ionization Chamber for Clinical Dosimetry}. \emph{Phys. Med. Biol.}
  \textbf{2007}, \emph{52}, 3089--3104\relax
\mciteBstWouldAddEndPuncttrue
\mciteSetBstMidEndSepPunct{\mcitedefaultmidpunct}
{\mcitedefaultendpunct}{\mcitedefaultseppunct}\relax
\EndOfBibitem
\bibitem[Conte \latin{et~al.}(2012)Conte, Colautti, Grosswendt, Moro, and {De
  Nardo}]{Conte_2012_NJP.14.093010}
Conte,~V.; Colautti,~P.; Grosswendt,~B.; Moro,~D.; {De Nardo},~L. {Track
  Structure of Light Ions: Experiments and Simulations}. \emph{New J. Phys.}
  \textbf{2012}, \emph{14}, 093010\relax
\mciteBstWouldAddEndPuncttrue
\mciteSetBstMidEndSepPunct{\mcitedefaultmidpunct}
{\mcitedefaultendpunct}{\mcitedefaultseppunct}\relax
\EndOfBibitem
\bibitem[Stein and White(1972)Stein, and White]{Stein_1972_JAP.43.2617}
Stein,~J.~D.; White,~F.~A. {New Method for the Measurement of Electron Yield
  from Ion Bombardment}. \emph{J. Appl. Phys.} \textbf{1972}, \emph{43},
  2617--2620\relax
\mciteBstWouldAddEndPuncttrue
\mciteSetBstMidEndSepPunct{\mcitedefaultmidpunct}
{\mcitedefaultendpunct}{\mcitedefaultseppunct}\relax
\EndOfBibitem
\bibitem[Lohmann \latin{et~al.}(2020)Lohmann, Niggas, Charnay, Hole\v{n}\'{a}k,
  and Primetzhofer]{Lohmann_2020_NIMB.479.217}
Lohmann,~S.; Niggas,~A.; Charnay,~V.; Hole\v{n}\'{a}k,~R.; Primetzhofer,~D.
  {Assessing Electron Emission Induced by Pulsed Ion Beams: A Time-of-Flight
  Approach}. \emph{Nucl. Instrum. Meth. B} \textbf{2020}, \emph{479},
  217--221\relax
\mciteBstWouldAddEndPuncttrue
\mciteSetBstMidEndSepPunct{\mcitedefaultmidpunct}
{\mcitedefaultendpunct}{\mcitedefaultseppunct}\relax
\EndOfBibitem
\bibitem[Apak \latin{et~al.}(2022)Apak, Calokerinos, Gorinstein, {Alves
  Segundo}, Hibbert, G\"{u}l\c{c}in, \c{C}eki\c{c}, G\"{u}\c{c}l\"{u},
  \"{O}zy\"{u}rek, \c{C}elik, and \textit{et al.}]{Apak_2022_PureApplChem}
Apak,~R.; Calokerinos,~A.; Gorinstein,~S.; {Alves Segundo},~M.; Hibbert,~D.~B.;
  G\"{u}l\c{c}in,~I.; \c{C}eki\c{c},~S.~D.; G\"{u}\c{c}l\"{u},~K.;
  \"{O}zy\"{u}rek,~M.; \c{C}elik,~S.~E.; \textit{et al.} {Methods to Evaluate
  the Scavenging Activity of Antioxidants Toward Reactive Oxygen and Nitrogen
  Species (IUPAC Technical Report)}. \emph{Pure Appl. Chem.} \textbf{2022},
  \emph{94}, 87--144\relax
\mciteBstWouldAddEndPuncttrue
\mciteSetBstMidEndSepPunct{\mcitedefaultmidpunct}
{\mcitedefaultendpunct}{\mcitedefaultseppunct}\relax
\EndOfBibitem
\bibitem[Zhang \latin{et~al.}(2023)Zhang, Ming, Sun, and
  Tiwari]{Zhang_2023_PSST.32.045015}
Zhang,~K.; Ming,~Z.; Sun,~D.-W.; Tiwari,~B.~K. {Correlation of Plasma Generated
  Long-Lived Reactive Species in Aqueous and Gas Phases with Different Feeding
  Gases}. \emph{Plasma Sources Sci. Technol.} \textbf{2023}, \emph{32},
  045015\relax
\mciteBstWouldAddEndPuncttrue
\mciteSetBstMidEndSepPunct{\mcitedefaultmidpunct}
{\mcitedefaultendpunct}{\mcitedefaultseppunct}\relax
\EndOfBibitem
\bibitem[Kondeti \latin{et~al.}(2018)Kondeti, an~K.~Wende, Jablonowski, Gangal,
  Granick, Hunter, and Bruggeman]{Kondeti_2018_FreeRadicBiolMed}
Kondeti,~V. S. S.~K.; an~K.~Wende,~C. Q.~P.; Jablonowski,~H.; Gangal,~U.;
  Granick,~J.~L.; Hunter,~R.~C.; Bruggeman,~P.~J. {Long-Lived and Short-Lived
  Reactive Species Produced by a Cold Atmospheric Pressure Plasma Jet for the
  Inactivation of \textit{Pseudomonas Aeruginosa} and \textit{Staphylococcus
  Aureus}}. \emph{Free Radic. Biol. Med.} \textbf{2018}, \emph{124},
  275--287\relax
\mciteBstWouldAddEndPuncttrue
\mciteSetBstMidEndSepPunct{\mcitedefaultmidpunct}
{\mcitedefaultendpunct}{\mcitedefaultseppunct}\relax
\EndOfBibitem
\bibitem[Mason \latin{et~al.}(2014)Mason, Nair, Jheeta, and
  Szyma\'{n}ska]{Mason_2014_FaradayDisc.168.235}
Mason,~N.~J.; Nair,~B.; Jheeta,~S.; Szyma\'{n}ska,~E. {Electron Induced
  Chemistry: A New Frontier in Astrochemistry}. \emph{Faraday Discuss.}
  \textbf{2014}, \emph{168}, 235--247\relax
\mciteBstWouldAddEndPuncttrue
\mciteSetBstMidEndSepPunct{\mcitedefaultmidpunct}
{\mcitedefaultendpunct}{\mcitedefaultseppunct}\relax
\EndOfBibitem
\bibitem[Strazzulla \latin{et~al.}(2023)Strazzulla, Palumbo, Boduch, and
  Rothard]{Strazzulla_2023_EarthMoon}
Strazzulla,~G.; Palumbo,~M.~E.; Boduch,~P.; Rothard,~H. {Ion Implantation and
  Chemical Cycles in the Icy Galilean Satellites}. \emph{Earth Moon Planets}
  \textbf{2023}, \emph{2}, 127\relax
\mciteBstWouldAddEndPuncttrue
\mciteSetBstMidEndSepPunct{\mcitedefaultmidpunct}
{\mcitedefaultendpunct}{\mcitedefaultseppunct}\relax
\EndOfBibitem
\bibitem[Halliwell \latin{et~al.}(2021)Halliwell, Adhikary, Dingfelder, and
  Dizdaroglu]{Halliwell_2021_ChemSocRev.50.8355}
Halliwell,~B.; Adhikary,~A.; Dingfelder,~M.; Dizdaroglu,~M. {Hydroxyl Radical
  is a Significant Player in Oxidative DNA Damage \textit{In Vivo}}.
  \emph{Chem. Soc. Rev.} \textbf{2021}, \emph{50}, 8355--8360\relax
\mciteBstWouldAddEndPuncttrue
\mciteSetBstMidEndSepPunct{\mcitedefaultmidpunct}
{\mcitedefaultendpunct}{\mcitedefaultseppunct}\relax
\EndOfBibitem
\bibitem[Schumacher \latin{et~al.}(2021)Schumacher, Pothof, Vijg, and
  Hoeijmakers]{Schumacher_2021_Nature.592.695}
Schumacher,~B.; Pothof,~J.; Vijg,~J.; Hoeijmakers,~J. H.~J. {The Central Role
  of DNA Damage in the Ageing Process}. \emph{Nature} \textbf{2021},
  \emph{592}, 695--703\relax
\mciteBstWouldAddEndPuncttrue
\mciteSetBstMidEndSepPunct{\mcitedefaultmidpunct}
{\mcitedefaultendpunct}{\mcitedefaultseppunct}\relax
\EndOfBibitem
\bibitem[Olano and Montero(2020)Olano, and
  Montero]{Olano_2020_ResultsPhys.19.103456}
Olano,~L.; Montero,~I. {Energy Spectra of Secondary Electrons in Dielectric
  Materials by Charging Analysis}. \emph{Results Phys.} \textbf{2020},
  \emph{19}, 103456\relax
\mciteBstWouldAddEndPuncttrue
\mciteSetBstMidEndSepPunct{\mcitedefaultmidpunct}
{\mcitedefaultendpunct}{\mcitedefaultseppunct}\relax
\EndOfBibitem
\bibitem[Mehnaz \latin{et~al.}(2020)Mehnaz, Yang, Zou, Da, Mao, Li, Zhao, and
  Ding]{Mehnaz_2020_MedPhys.47.759}
Mehnaz; Yang,~L.~H.; Zou,~Y.~B.; Da,~B.; Mao,~S.~F.; Li,~H.~M.; Zhao,~Y.~F.;
  Ding,~Z.~J. {A Comparative Study on Monte Carlo Simulations of Electron
  Emission From Liquid Water}. \emph{Med. Phys.} \textbf{2020}, \emph{47},
  759--771\relax
\mciteBstWouldAddEndPuncttrue
\mciteSetBstMidEndSepPunct{\mcitedefaultmidpunct}
{\mcitedefaultendpunct}{\mcitedefaultseppunct}\relax
\EndOfBibitem
\bibitem[Thiberge \latin{et~al.}(2004)Thiberge, Zik, and
  Moses]{Thiberge_2004_RevSciInstrum.75.2281}
Thiberge,~S.; Zik,~O.; Moses,~E. {An Apparatus for Imaging Liquids, Cells, and
  Other Wet Samples in the Scanning Electron Microscopy}. \emph{Rev. Sci.
  Instrum.} \textbf{2004}, \emph{75}, 2281--2289\relax
\mciteBstWouldAddEndPuncttrue
\mciteSetBstMidEndSepPunct{\mcitedefaultmidpunct}
{\mcitedefaultendpunct}{\mcitedefaultseppunct}\relax
\EndOfBibitem
\bibitem[Joy and Joy(2006)Joy, and Joy]{Joy_2006_JMiscosc.221.84}
Joy,~D.~C.; Joy,~C.~S. {Scanning Electron Microscope Imaging in Liquids -- Some
  Data on Electron Interactions in Water}. \emph{J. Microsc.} \textbf{2006},
  \emph{221}, 84--88\relax
\mciteBstWouldAddEndPuncttrue
\mciteSetBstMidEndSepPunct{\mcitedefaultmidpunct}
{\mcitedefaultendpunct}{\mcitedefaultseppunct}\relax
\EndOfBibitem
\bibitem[Kaneda \latin{et~al.}(2010)Kaneda, Shimizu, Hayakawa, Iriki, Tsuchida,
  and Itoh]{Kaneda_2010_JCP.132.144502}
Kaneda,~M.; Shimizu,~M.; Hayakawa,~T.; Iriki,~Y.; Tsuchida,~H.; Itoh,~A.
  {Positive and Negative Cluster Ions From Liquid Ethanol by Fast Ion
  Bombardment}. \emph{J. Chem. Phys.} \textbf{2010}, \emph{132}, 144502\relax
\mciteBstWouldAddEndPuncttrue
\mciteSetBstMidEndSepPunct{\mcitedefaultmidpunct}
{\mcitedefaultendpunct}{\mcitedefaultseppunct}\relax
\EndOfBibitem
\bibitem[Kitajima \latin{et~al.}(2019)Kitajima, Tsuchida, Majima, and
  Saito]{Kitajima_2019_JCP.150.095102}
Kitajima,~K.; Tsuchida,~H.; Majima,~T.; Saito,~M. {Secondary Electron-Induced
  Biomolecular Fragmentation in Fast Heavy-Ion Irradiation of Microdroplets of
  Glycine Solution}. \emph{J. Chem. Phys.} \textbf{2019}, \emph{150},
  095102\relax
\mciteBstWouldAddEndPuncttrue
\mciteSetBstMidEndSepPunct{\mcitedefaultmidpunct}
{\mcitedefaultendpunct}{\mcitedefaultseppunct}\relax
\EndOfBibitem
\bibitem[Nag \latin{et~al.}(2023)Nag, Rankovi\'{c}, Schewe, Rakovsk\'{y}, Sala,
  Ko\v{c}i\v{s}ek, and Fedor]{Nag_2023_JPB.56.215201}
Nag,~P.; Rankovi\'{c},~M.; Schewe,~H.~C.; Rakovsk\'{y},~J.; Sala,~L.;
  Ko\v{c}i\v{s}ek,~J.; Fedor,~J. {Experimental Setup for Probing
  Electron-Induced Chemistry in Liquid Micro-Jets}. \emph{J. Phys. B: At. Mol.
  Opt. Phys.} \textbf{2023}, \emph{56}, 215201\relax
\mciteBstWouldAddEndPuncttrue
\mciteSetBstMidEndSepPunct{\mcitedefaultmidpunct}
{\mcitedefaultendpunct}{\mcitedefaultseppunct}\relax
\EndOfBibitem
\bibitem[Knowles \latin{et~al.}(2018)Knowles, Koch, and
  Shelton]{Knowles_2018_JMaterChemC.6.11853}
Knowles,~K.~E.; Koch,~M.~D.; Shelton,~J.~L. {Three Applications of Ultrafast
  Transient Absorption Spectroscopy of Semiconductor Thin Films:
  Spectroelectrochemistry, Microscopy, and Identification of Thermal
  Contributions}. \emph{J. Mater. Chem. C} \textbf{2018}, \emph{6},
  11853--11867\relax
\mciteBstWouldAddEndPuncttrue
\mciteSetBstMidEndSepPunct{\mcitedefaultmidpunct}
{\mcitedefaultendpunct}{\mcitedefaultseppunct}\relax
\EndOfBibitem
\bibitem[Senje \latin{et~al.}(2017)Senje, Coughlan, Jung, Taylor, Nersisyan,
  Riley, Lewis, Lundh, Wahlstr\"{o}m, Zepf, and
  Dromey]{Senje_2017_APL.110.104102}
Senje,~L.; Coughlan,~M.; Jung,~D.; Taylor,~M.; Nersisyan,~G.; Riley,~D.;
  Lewis,~C. L.~S.; Lundh,~O.; Wahlstr\"{o}m,~C.-G.; Zepf,~M.; Dromey,~B.
  {Experimental Investigation of Picosecond Dynamics Following Interactions
  Between Laser Accelerated Protons and Water}. \emph{Appl. Phys. Lett.}
  \textbf{2017}, \emph{110}, 104102\relax
\mciteBstWouldAddEndPuncttrue
\mciteSetBstMidEndSepPunct{\mcitedefaultmidpunct}
{\mcitedefaultendpunct}{\mcitedefaultseppunct}\relax
\EndOfBibitem
\bibitem[Taylor \latin{et~al.}(2018)Taylor, Coughlan, Nersisyan, Senje, Jung,
  Currell, Riley, Lewis, Zepf, and Dromey]{Taylor_2018_PlasmaPhys.60.054004}
Taylor,~M.; Coughlan,~M.; Nersisyan,~G.; Senje,~L.; Jung,~D.; Currell,~F.;
  Riley,~D.; Lewis,~C. L.~S.; Zepf,~M.; Dromey,~B. {Probing Ultrafast Proton
  Induced Dynamics in Transparent Dielectrics}. \emph{Plasma Phys. Control.
  Fusion} \textbf{2018}, \emph{60}, 054004\relax
\mciteBstWouldAddEndPuncttrue
\mciteSetBstMidEndSepPunct{\mcitedefaultmidpunct}
{\mcitedefaultendpunct}{\mcitedefaultseppunct}\relax
\EndOfBibitem
\bibitem[Kotov \latin{et~al.}(2021)Kotov, Shemyakin, Solovyov, Yakimov, Glebov,
  Dubrova, Malyutin, Popov, Poniaev, Lapushkina, Monakhov, and
  Sakharov]{Kotov_2021_JPCS.2103.012218}
Kotov,~M.~A.; Shemyakin,~A.~N.; Solovyov,~N.~G.; Yakimov,~M.~Y.; Glebov,~V.~N.;
  Dubrova,~G.~A.; Malyutin,~A.~M.; Popov,~P.~A.; Poniaev,~S.~A.;
  Lapushkina,~T.~A.; Monakhov,~N.~A.; Sakharov,~V.~A. {The Analysis of the
  Applicability of Thermoelectric Radiation Detectors for Heat Flux
  Measurements Behind a Reflected Shock Wave}. \emph{J. Phys.: Conf. Ser.}
  \textbf{2021}, \emph{2103}, 012218\relax
\mciteBstWouldAddEndPuncttrue
\mciteSetBstMidEndSepPunct{\mcitedefaultmidpunct}
{\mcitedefaultendpunct}{\mcitedefaultseppunct}\relax
\EndOfBibitem
\bibitem[ELI((accessed 2023-11-15))]{ELI_website}
Extreme Light Infrastructure -- A European project for investigating
  light-matter interactions at highest intensities and shortest time scales.
  (accessed 2023-11-15); \url{https://eli-laser.eu/}\relax
\mciteBstWouldAddEndPuncttrue
\mciteSetBstMidEndSepPunct{\mcitedefaultmidpunct}
{\mcitedefaultendpunct}{\mcitedefaultseppunct}\relax
\EndOfBibitem
\bibitem[Ebel and Bald(2022)Ebel, and Bald]{Ebel_Bald_2022_JPCL.13.4871}
Ebel,~K.; Bald,~I. {Low-Energy ($5-20$ eV) Electron-Induced Single and Double
  Strand Breaks in Well-Defined DNA Sequences}. \emph{J. Phys. Chem. Lett.}
  \textbf{2022}, \emph{13}, 4871--4876\relax
\mciteBstWouldAddEndPuncttrue
\mciteSetBstMidEndSepPunct{\mcitedefaultmidpunct}
{\mcitedefaultendpunct}{\mcitedefaultseppunct}\relax
\EndOfBibitem
\bibitem[Sykes(2017)]{Sykes_SurfChemAnalysis}
Sykes,~D. In \emph{Springer Handbook of Electronic and Photonic Materials}, 2nd
  ed.; Kasap,~S., Capper,~P., Eds.; Springer, Cham, 2017; pp 413--423\relax
\mciteBstWouldAddEndPuncttrue
\mciteSetBstMidEndSepPunct{\mcitedefaultmidpunct}
{\mcitedefaultendpunct}{\mcitedefaultseppunct}\relax
\EndOfBibitem
\bibitem[Erdo\u{g}an \latin{et~al.}(2017)Erdo\u{g}an, G\"{u}ler, Kili\c{c},
  Kili\c{c}, Erdo\u{g}an, Tosun, Kivrak, T\"{u}rkan, \"{O}zcan, G\"{u}rsoy, and
  Karaman]{Erdogan_SurfCharacteriz}
Erdo\u{g}an,~G.; G\"{u}ler,~G.; Kili\c{c},~T.; Kili\c{c},~D.~O.;
  Erdo\u{g}an,~B.; Tosun,~Z.; Kivrak,~H.~D.; T\"{u}rkan,~U.; \"{O}zcan,~F.;
  G\"{u}rsoy,~M.; Karaman,~M. In \emph{Surface Treatments for Biological,
  Chemical, and Physical Applications}; G\"{u}rsoy,~M., Karaman,~M., Eds.;
  WILEY-VCH Verlag, Weinheim, 2017; pp 67--114\relax
\mciteBstWouldAddEndPuncttrue
\mciteSetBstMidEndSepPunct{\mcitedefaultmidpunct}
{\mcitedefaultendpunct}{\mcitedefaultseppunct}\relax
\EndOfBibitem
\bibitem[Krishna and Philip(2022)Krishna, and
  Philip]{Krishna_2022_ApplSurfSciAdv}
Krishna,~D. N.~G.; Philip,~J. {Review on Surface-Characterization Applications
  of X-Ray Photoelectron Spectroscopy (XPS): Recent Developments and
  Challenges}. \emph{Appl. Surf. Sci. Adv.} \textbf{2022}, \emph{12},
  100332\relax
\mciteBstWouldAddEndPuncttrue
\mciteSetBstMidEndSepPunct{\mcitedefaultmidpunct}
{\mcitedefaultendpunct}{\mcitedefaultseppunct}\relax
\EndOfBibitem
\bibitem[{Lannon Jr.} and Stinespring(2006){Lannon Jr.}, and
  Stinespring]{Lannon_AugerSpectrosc_Surfaces}
{Lannon Jr.},~J.~M.; Stinespring,~C.~D. In \emph{Encyclopedia of Analytical
  Chemistry: Applications, Theory, and Instrumentation}; Meyers,~R., Ed.; John
  Wiley \& Sons Ltd, Chichester, 2006; pp 1--15\relax
\mciteBstWouldAddEndPuncttrue
\mciteSetBstMidEndSepPunct{\mcitedefaultmidpunct}
{\mcitedefaultendpunct}{\mcitedefaultseppunct}\relax
\EndOfBibitem
\bibitem[Strehblow(2021)]{Strehblow_2021_JElectrochemSoc}
Strehblow,~H.-H. {Review -- Ion Scattering as a Surface Analytical Tool for the
  Study of Passive Layers}. \emph{J. Electrochem. Soc.} \textbf{2021},
  \emph{168}, 021510\relax
\mciteBstWouldAddEndPuncttrue
\mciteSetBstMidEndSepPunct{\mcitedefaultmidpunct}
{\mcitedefaultendpunct}{\mcitedefaultseppunct}\relax
\EndOfBibitem
\bibitem[Wang(1996)]{Wang_ReflectEM_SurfAnal}
Wang,~Z.~L. \emph{{Reflection Electron Microscopy and Spectroscopy for Surface
  Analysis}}; Cambridge University Press, Cambridge, 1996\relax
\mciteBstWouldAddEndPuncttrue
\mciteSetBstMidEndSepPunct{\mcitedefaultmidpunct}
{\mcitedefaultendpunct}{\mcitedefaultseppunct}\relax
\EndOfBibitem
\bibitem[Bili\v{s}kov(2022)]{Biliskov_2022_PCCP.24.19073}
Bili\v{s}kov,~N. {Infrared Spectroscopic Monitoring of Solid-State Processes}.
  \emph{Phys. Chem. Chem. Phys.} \textbf{2022}, \emph{24}, 19073--19120\relax
\mciteBstWouldAddEndPuncttrue
\mciteSetBstMidEndSepPunct{\mcitedefaultmidpunct}
{\mcitedefaultendpunct}{\mcitedefaultseppunct}\relax
\EndOfBibitem
\bibitem[Kudelski(2009)]{Kudelski_2009_SurfSci.603.1328}
Kudelski,~A. {Raman Spectroscopy of Surfaces}. \emph{Surf. Sci.} \textbf{2009},
  \emph{603}, 1328--1334\relax
\mciteBstWouldAddEndPuncttrue
\mciteSetBstMidEndSepPunct{\mcitedefaultmidpunct}
{\mcitedefaultendpunct}{\mcitedefaultseppunct}\relax
\EndOfBibitem
\bibitem[Vickerman and Briggs(2013)Vickerman, and
  Briggs]{Vickerman_ToF-SIMS_book}
Vickerman,~J.~C.; Briggs,~D. \emph{{ToF-SIMS: Materials Analysis by Mass
  Spectrometry}}, 2nd ed.; IM Publications, Chichester, 2013\relax
\mciteBstWouldAddEndPuncttrue
\mciteSetBstMidEndSepPunct{\mcitedefaultmidpunct}
{\mcitedefaultendpunct}{\mcitedefaultseppunct}\relax
\EndOfBibitem
\bibitem[Vad \latin{et~al.}(2009)Vad, Csik, and Langer]{Vad_NeutralMS_2009}
Vad,~K.; Csik,~A.; Langer,~G.~A. {Secondary Neutral Mass Spectrometry -- A
  Powerful Technique for Quantitative Elemental and Depth Profiling Analyses of
  Nanostructures}. \emph{Spectrosc. Eur.} \textbf{2009}, \emph{21},
  13--16\relax
\mciteBstWouldAddEndPuncttrue
\mciteSetBstMidEndSepPunct{\mcitedefaultmidpunct}
{\mcitedefaultendpunct}{\mcitedefaultseppunct}\relax
\EndOfBibitem
\bibitem[Ferus \latin{et~al.}(2020)Ferus, Petera, Koukal, Len\v{z}a,
  Drtinov\'{a}, Haloda, Mat\'{y}sek, Pastorek, Laitl, Poltronieri, and
  \textit{et al.}]{Ferus_2020_Icarus.341.113670}
Ferus,~M.; Petera,~L.; Koukal,~J.; Len\v{z}a,~L.; Drtinov\'{a},~B.; Haloda,~J.;
  Mat\'{y}sek,~D.; Pastorek,~A.; Laitl,~V.; Poltronieri,~R.~C.; \textit{et al.}
  {Elemental Composition, Mineralogy and Orbital Parameters of the Porangaba
  Meteorite}. \emph{Icarus} \textbf{2020}, \emph{341}, 113670\relax
\mciteBstWouldAddEndPuncttrue
\mciteSetBstMidEndSepPunct{\mcitedefaultmidpunct}
{\mcitedefaultendpunct}{\mcitedefaultseppunct}\relax
\EndOfBibitem
\bibitem[Kaczmarek \latin{et~al.}(2021)Kaczmarek, Leniart, Lapinsk, Skrzypek,
  and Lukomska-Szymanska]{Kaczmarek_2021_Materials.14.2624}
Kaczmarek,~K.; Leniart,~A.; Lapinsk,~B.; Skrzypek,~S.; Lukomska-Szymanska,~M.
  {Selected Spectroscopic Techniques for Surface Analysis of Dental Materials:
  A Narrative Review}. \emph{Materials} \textbf{2021}, \emph{14}, 2624\relax
\mciteBstWouldAddEndPuncttrue
\mciteSetBstMidEndSepPunct{\mcitedefaultmidpunct}
{\mcitedefaultendpunct}{\mcitedefaultseppunct}\relax
\EndOfBibitem
\bibitem[Hunter(1993)]{Hunter_PracticalElMicroscopy}
Hunter,~E.~E. \emph{{Practical Electron Microscopy: A Beginner's Illustrated
  Guide}}, 2nd ed.; Cambridge University Press, Cambridge, 1993\relax
\mciteBstWouldAddEndPuncttrue
\mciteSetBstMidEndSepPunct{\mcitedefaultmidpunct}
{\mcitedefaultendpunct}{\mcitedefaultseppunct}\relax
\EndOfBibitem
\bibitem[Goldstein \latin{et~al.}(2017)Goldstein, Newbury, Michael, Ritchie,
  Scott, and Joy]{Goldstein_ScanningElMicroscopy}
Goldstein,~J.~I.; Newbury,~D.~E.; Michael,~J.~R.; Ritchie,~N. W.~M.; Scott,~J.
  H.~J.; Joy,~D.~C. \emph{{Scanning Electron Microscopy and X-Ray
  Microanalysis}}, 4th ed.; Springer New York, NY, 2017\relax
\mciteBstWouldAddEndPuncttrue
\mciteSetBstMidEndSepPunct{\mcitedefaultmidpunct}
{\mcitedefaultendpunct}{\mcitedefaultseppunct}\relax
\EndOfBibitem
\bibitem[Trummer \latin{et~al.}(2019)Trummer, Winkler, Plank, Kothleitner, and
  Haberfehlner]{Trummer_2019_ACSApplNanoMater}
Trummer,~C.; Winkler,~R.; Plank,~H.; Kothleitner,~G.; Haberfehlner,~G.
  {Analyzing the Nanogranularity of Focused-Electron-Beam-Induced-Deposited
  Materials by Electron Tomography}. \emph{ACS Appl. Nano Mater.}
  \textbf{2019}, \emph{2}, 5356--5359\relax
\mciteBstWouldAddEndPuncttrue
\mciteSetBstMidEndSepPunct{\mcitedefaultmidpunct}
{\mcitedefaultendpunct}{\mcitedefaultseppunct}\relax
\EndOfBibitem
\bibitem[Kometani and Ishihara(2009)Kometani, and
  Ishihara]{Kometani_2009_SciTechnolAdvMater}
Kometani,~R.; Ishihara,~S. {Nanoelectromechanical Device Fabrications by 3-D
  Nanotechnology Using Focused-Ion Beams}. \emph{Sci. Technol. Adv. Mater.}
  \textbf{2009}, \emph{10}, 034501\relax
\mciteBstWouldAddEndPuncttrue
\mciteSetBstMidEndSepPunct{\mcitedefaultmidpunct}
{\mcitedefaultendpunct}{\mcitedefaultseppunct}\relax
\EndOfBibitem
\bibitem[Jurczyk \latin{et~al.}(2022)Jurczyk, Pillatsch, Berger, Priebe,
  Madajska, Kapusta, Szyma\'{n}ska, Michler, and
  Utke]{Jurczyk_2022_Nanomaterials.12.2710}
Jurczyk,~J.; Pillatsch,~L.; Berger,~L.; Priebe,~A.; Madajska,~K.; Kapusta,~C.;
  Szyma\'{n}ska,~I.~B.; Michler,~J.; Utke,~I. {In Situ Time-of-Flight Mass
  Spectrometry of Ionic Fragments Induced by Focused Electron Beam Irradiation:
  Investigation of Electron Driven Surface Chemistry inside an SEM under High
  Vacuum}. \emph{Nanomaterials} \textbf{2022}, \emph{12}, 2710\relax
\mciteBstWouldAddEndPuncttrue
\mciteSetBstMidEndSepPunct{\mcitedefaultmidpunct}
{\mcitedefaultendpunct}{\mcitedefaultseppunct}\relax
\EndOfBibitem
\bibitem[Lepore \latin{et~al.}(2021)Lepore, Maligno, and
  Berto]{Lepore_2021_EngFailAnal}
Lepore,~M.~A.; Maligno,~A.~R.; Berto,~F. {A Unified Approach to Simulate the
  Creep-Fatigue Crack Growth in P91 Steel at Elevated Temperature Under SSY and
  SSC Conditions}. \emph{Eng. Fail. Anal.} \textbf{2021}, \emph{127},
  105569\relax
\mciteBstWouldAddEndPuncttrue
\mciteSetBstMidEndSepPunct{\mcitedefaultmidpunct}
{\mcitedefaultendpunct}{\mcitedefaultseppunct}\relax
\EndOfBibitem
\bibitem[Suzuki \latin{et~al.}(2023)Suzuki, Funayama, Suzuki, and
  Kobayashi]{Suzuki_2023_JRR.64.824}
Suzuki,~M.; Funayama,~T.; Suzuki,~M.; Kobayashi,~Y.
  {Radiation-Quality-Dependent Bystander Cellular Effects Induced by Heavy-Ion
  Microbeams Through Different Pathways}. \emph{J. Radiat. Res.} \textbf{2023},
  \emph{64}, 824--832\relax
\mciteBstWouldAddEndPuncttrue
\mciteSetBstMidEndSepPunct{\mcitedefaultmidpunct}
{\mcitedefaultendpunct}{\mcitedefaultseppunct}\relax
\EndOfBibitem
\bibitem[Cheng \latin{et~al.}(2020)Cheng, Cheadle, and
  Illidge]{Cheng_2020_Cancers.12.2835}
Cheng,~S.; Cheadle,~E.~J.; Illidge,~T.~M. {Understanding the Effects of
  Radiotherapy on the Tumour Immune Microenvironment to Identify Potential
  Prognostic and Predictive Biomarkers of Radiotherapy Response }.
  \emph{Cancers} \textbf{2020}, \emph{12}, 2835\relax
\mciteBstWouldAddEndPuncttrue
\mciteSetBstMidEndSepPunct{\mcitedefaultmidpunct}
{\mcitedefaultendpunct}{\mcitedefaultseppunct}\relax
\EndOfBibitem
\bibitem[Ding \latin{et~al.}(2021)Ding, Zhao, Matysik, G\"{a}rtner, and
  Losi]{ding2021mapping}
Ding,~Y.; Zhao,~Z.; Matysik,~J.; G\"{a}rtner,~W.; Losi,~A. {Mapping the Role of
  Aromatic Amino Acids Within a Blue-Light Sensing LOV Domain}. \emph{Phys.
  Chem. Chem. Phys.} \textbf{2021}, \emph{23}, 16767--16775\relax
\mciteBstWouldAddEndPuncttrue
\mciteSetBstMidEndSepPunct{\mcitedefaultmidpunct}
{\mcitedefaultendpunct}{\mcitedefaultseppunct}\relax
\EndOfBibitem
\bibitem[Thamarath \latin{et~al.}(2010)Thamarath, Heberle, Hore, Kottke, and
  Matysik]{thamarath2010solid}
Thamarath,~S.~S.; Heberle,~J.; Hore,~P.~J.; Kottke,~T.; Matysik,~J.
  {Solid-State Photo-CIDNP Effect Observed in Phototropin LOV1-C57S by $^{13}$C
  Magic-Angle Spinning NMR Spectroscopy}. \emph{J. Am. Chem. Soc.}
  \textbf{2010}, \emph{132}, 15542--15543\relax
\mciteBstWouldAddEndPuncttrue
\mciteSetBstMidEndSepPunct{\mcitedefaultmidpunct}
{\mcitedefaultendpunct}{\mcitedefaultseppunct}\relax
\EndOfBibitem
\bibitem[Losi \latin{et~al.}(2018)Losi, Gardner, and M\"{o}glich]{losi2018blue}
Losi,~A.; Gardner,~K.~H.; M\"{o}glich,~A. {Blue-Light Receptors for
  Optogenetics}. \emph{Chem. Rev.} \textbf{2018}, \emph{118},
  10659--10709\relax
\mciteBstWouldAddEndPuncttrue
\mciteSetBstMidEndSepPunct{\mcitedefaultmidpunct}
{\mcitedefaultendpunct}{\mcitedefaultseppunct}\relax
\EndOfBibitem
\bibitem[Kottke \latin{et~al.}(2006)Kottke, Batschauer, Ahmad, and
  Heberle]{kottke2006blue}
Kottke,~T.; Batschauer,~A.; Ahmad,~M.; Heberle,~J. {Blue-Light-Induced Changes
  in Arabidopsis Cryptochrome 1 Probed by FTIR Difference Spectroscopy}.
  \emph{Biochemistry} \textbf{2006}, \emph{45}, 2472--2479\relax
\mciteBstWouldAddEndPuncttrue
\mciteSetBstMidEndSepPunct{\mcitedefaultmidpunct}
{\mcitedefaultendpunct}{\mcitedefaultseppunct}\relax
\EndOfBibitem
\bibitem[Berlew \latin{et~al.}(2020)Berlew, Kuznetsov, Yamada, Bugaj, and
  Chow]{berlew2020optogenetic}
Berlew,~E.~E.; Kuznetsov,~I.~A.; Yamada,~K.; Bugaj,~L.~J.; Chow,~B.~Y.
  {Optogenetic Rac1 Engineered from Membrane Lipid-Binding RGS-LOV for
  Inducible Lamellipodia Formation}. \emph{Photochem. Photobiol. Sci.}
  \textbf{2020}, \emph{19}, 353--361\relax
\mciteBstWouldAddEndPuncttrue
\mciteSetBstMidEndSepPunct{\mcitedefaultmidpunct}
{\mcitedefaultendpunct}{\mcitedefaultseppunct}\relax
\EndOfBibitem
\bibitem[Badura \latin{et~al.}(2006)Badura, Esper, Ataka, Grunwald, W\"{o}ll,
  Kuhlmann, Heberle, and R\"{o}gner]{badura2006light}
Badura,~A.; Esper,~B.; Ataka,~K.; Grunwald,~C.; W\"{o}ll,~C.; Kuhlmann,~J.;
  Heberle,~J.; R\"{o}gner,~M. {Light-Driven Water Splitting for (Bio-)Hydrogen
  Production: Photosystem 2 as the Central Part of a Bioelectrochemical
  Device}. \emph{Photochem. Photobiol.} \textbf{2006}, \emph{82},
  1385--1390\relax
\mciteBstWouldAddEndPuncttrue
\mciteSetBstMidEndSepPunct{\mcitedefaultmidpunct}
{\mcitedefaultendpunct}{\mcitedefaultseppunct}\relax
\EndOfBibitem
\bibitem[Krassen \latin{et~al.}(2009)Krassen, Schwarze, Friedrich, Ataka, Lenz,
  and Heberle]{krassen2009photosynthetic}
Krassen,~H.; Schwarze,~A.; Friedrich,~B.; Ataka,~K.; Lenz,~O.; Heberle,~J.
  {Photosynthetic Hydrogen Production by a Hybrid Complex of Photosystem I and
  [NiFe]-Hydrogenase}. \emph{ACS Nano} \textbf{2009}, \emph{3},
  4055--4061\relax
\mciteBstWouldAddEndPuncttrue
\mciteSetBstMidEndSepPunct{\mcitedefaultmidpunct}
{\mcitedefaultendpunct}{\mcitedefaultseppunct}\relax
\EndOfBibitem
\bibitem[Hore and Mouritsen(2016)Hore, and Mouritsen]{hore2016radical}
Hore,~P.; Mouritsen,~H. {The Radical-Pair Mechanism of Magnetoreception}.
  \emph{Annu. Rev. Biophys.} \textbf{2016}, \emph{45}, 299--344\relax
\mciteBstWouldAddEndPuncttrue
\mciteSetBstMidEndSepPunct{\mcitedefaultmidpunct}
{\mcitedefaultendpunct}{\mcitedefaultseppunct}\relax
\EndOfBibitem
\bibitem[Ritz \latin{et~al.}(2004)Ritz, Thalau, Phillips, Wiltschko, and
  Wiltschko]{ritz2004resonance}
Ritz,~T.; Thalau,~P.; Phillips,~J.~B.; Wiltschko,~R.; Wiltschko,~W. {Resonance
  Effects Indicate a Radical-Pair Mechanism for Avian Magnetic Compass}.
  \emph{Nature} \textbf{2004}, \emph{429}, 177--180\relax
\mciteBstWouldAddEndPuncttrue
\mciteSetBstMidEndSepPunct{\mcitedefaultmidpunct}
{\mcitedefaultendpunct}{\mcitedefaultseppunct}\relax
\EndOfBibitem
\bibitem[Pedersen \latin{et~al.}(2016)Pedersen, Nielsen, and
  Solov'yov]{pedersen2016multiscale}
Pedersen,~J.~B.; Nielsen,~C.; Solov'yov,~I.~A. {Multiscale Description of Avian
  Migration: From Chemical Compass to Behaviour Modeling}. \emph{Sci. Rep.}
  \textbf{2016}, \emph{6}, 36709\relax
\mciteBstWouldAddEndPuncttrue
\mciteSetBstMidEndSepPunct{\mcitedefaultmidpunct}
{\mcitedefaultendpunct}{\mcitedefaultseppunct}\relax
\EndOfBibitem
\bibitem[Gr\"{u}ning \latin{et~al.}(2022)Gr\"{u}ning, Wong, Gerhards,
  Schuhmann, Kattnig, Hore, and Solov'yov]{gruning2022effects}
Gr\"{u}ning,~G.; Wong,~S.~Y.; Gerhards,~L.; Schuhmann,~F.; Kattnig,~D.~R.;
  Hore,~P.~J.; Solov'yov,~I.~A. {Effects of Dynamical Degrees of Freedom on
  Magnetic Compass Sensitivity: A Comparison of Plant and Avian Cryptochromes}.
  \emph{J. Am. Chem. Soc.} \textbf{2022}, \emph{144}, 22902--22914\relax
\mciteBstWouldAddEndPuncttrue
\mciteSetBstMidEndSepPunct{\mcitedefaultmidpunct}
{\mcitedefaultendpunct}{\mcitedefaultseppunct}\relax
\EndOfBibitem
\bibitem[Gerhards \latin{et~al.}(2023)Gerhards, Nielsen, Kattnig, Hore, and
  Solov'yov]{gerhards2023modeling}
Gerhards,~L.; Nielsen,~C.; Kattnig,~D.; Hore,~P.~J.; Solov'yov,~I.~A. {Modeling
  Spin Relaxation in Complex Radical Systems Using MolSpin}. \emph{J. Computat.
  Chem.} \textbf{2023}, \emph{44}, 1704--1714\relax
\mciteBstWouldAddEndPuncttrue
\mciteSetBstMidEndSepPunct{\mcitedefaultmidpunct}
{\mcitedefaultendpunct}{\mcitedefaultseppunct}\relax
\EndOfBibitem
\bibitem[Husen \latin{et~al.}(2019)Husen, Nielsen, Martino, and
  Solov'yov]{husen2019molecular}
Husen,~P.; Nielsen,~C.; Martino,~C.~F.; Solov'yov,~I.~A. {Molecular Oxygen
  Binding in the Mitochondrial Electron Transfer Flavoprotein}. \emph{J. Chem.
  Inf. Model.} \textbf{2019}, \emph{59}, 4868--4879\relax
\mciteBstWouldAddEndPuncttrue
\mciteSetBstMidEndSepPunct{\mcitedefaultmidpunct}
{\mcitedefaultendpunct}{\mcitedefaultseppunct}\relax
\EndOfBibitem
\bibitem[Moser \latin{et~al.}(1992)Moser, Keske, Warncke, Farid, and
  Dutton]{moser1992nature}
Moser,~C.~C.; Keske,~J.~M.; Warncke,~K.; Farid,~R.~S.; Dutton,~P.~L. {Nature of
  Biological Electron Transfer}. \emph{Nature} \textbf{1992}, \emph{355},
  796--802\relax
\mciteBstWouldAddEndPuncttrue
\mciteSetBstMidEndSepPunct{\mcitedefaultmidpunct}
{\mcitedefaultendpunct}{\mcitedefaultseppunct}\relax
\EndOfBibitem
\bibitem[de~la Lande \latin{et~al.}(2012)de~la Lande, Babcock, Rez\'{a}c,
  L\'{e}vy, Sanders, and Salahub]{delaLande2012quantum}
de~la Lande,~A.; Babcock,~N.~S.; Rez\'{a}c,~J.; L\'{e}vy,~B.; Sanders,~B.~C.;
  Salahub,~D.~R. {Quantum Effects in Biological Electron Transfer}. \emph{Phys.
  Chem. Chem. Phys.} \textbf{2012}, \emph{14}, 5902--5918\relax
\mciteBstWouldAddEndPuncttrue
\mciteSetBstMidEndSepPunct{\mcitedefaultmidpunct}
{\mcitedefaultendpunct}{\mcitedefaultseppunct}\relax
\EndOfBibitem
\bibitem[Solov'yov \latin{et~al.}(2012)Solov'yov, Domratcheva, Moughal~Shahi,
  and Schulten]{solovyov2012decrypting}
Solov'yov,~I.~A.; Domratcheva,~T.; Moughal~Shahi,~A.~R.; Schulten,~K.
  {Decrypting Cryptochrome: Revealing the Molecular Identity of the
  Photoactivation Reaction}. \emph{J. Am. Chem. Soc.} \textbf{2012},
  \emph{134}, 18046--18052\relax
\mciteBstWouldAddEndPuncttrue
\mciteSetBstMidEndSepPunct{\mcitedefaultmidpunct}
{\mcitedefaultendpunct}{\mcitedefaultseppunct}\relax
\EndOfBibitem
\bibitem[Solov'yov \latin{et~al.}(2014)Solov'yov, Domratcheva, and
  Schulten]{solovyov2014separation}
Solov'yov,~I.~A.; Domratcheva,~T.; Schulten,~K. {Separation of Photo-Induced
  Radical Pair in Cryptochrome to a Functionally Critical Distance}. \emph{Sci.
  Rep.} \textbf{2014}, \emph{4}, 3845\relax
\mciteBstWouldAddEndPuncttrue
\mciteSetBstMidEndSepPunct{\mcitedefaultmidpunct}
{\mcitedefaultendpunct}{\mcitedefaultseppunct}\relax
\EndOfBibitem
\bibitem[Gerhards and Kl\"{u}ner(2021)Gerhards, and
  Kl\"{u}ner]{gerhards2021quantum}
Gerhards,~L.; Kl\"{u}ner,~T. {Quantum Chemical Investigation of
  Photocatalytical Sulfoxidation of Hydrocarbons on TiO$_2$}. \emph{J. Phys.
  Chem. C} \textbf{2021}, \emph{125}, 13313--13323\relax
\mciteBstWouldAddEndPuncttrue
\mciteSetBstMidEndSepPunct{\mcitedefaultmidpunct}
{\mcitedefaultendpunct}{\mcitedefaultseppunct}\relax
\EndOfBibitem
\bibitem[Gerhards and Kl\"{u}ner(2022)Gerhards, and
  Kl\"{u}ner]{gerhards2022theoretical}
Gerhards,~L.; Kl\"{u}ner,~T. {Theoretical Investigation of CH-Bond Activation
  by Photocatalytic Excited SO$_2$ and the Effects of C-, N-, S-, and Se-doped
  TiO$_2$}. \emph{Phys. Chem. Chem. Phys.} \textbf{2022}, \emph{24},
  2051--2069\relax
\mciteBstWouldAddEndPuncttrue
\mciteSetBstMidEndSepPunct{\mcitedefaultmidpunct}
{\mcitedefaultendpunct}{\mcitedefaultseppunct}\relax
\EndOfBibitem
\bibitem[Frederiksen \latin{et~al.}(2022)Frederiksen, Teusch, and
  Solov'yov]{frederiksen2022quantum}
Frederiksen,~A.; Teusch,~T.; Solov'yov,~I.~A. In \emph{Dynamics of Systems on
  the Nanoscale}; Solov'yov,~I.~A., Verkhovtsev,~A.~V., Korol,~A.~V.,
  Solov'yov,~A.~V., Eds.; Springer Nature Switzerland, 2022; pp 201--247\relax
\mciteBstWouldAddEndPuncttrue
\mciteSetBstMidEndSepPunct{\mcitedefaultmidpunct}
{\mcitedefaultendpunct}{\mcitedefaultseppunct}\relax
\EndOfBibitem
\bibitem[Nielsen \latin{et~al.}(2018)Nielsen, N{\o}rby, Kongsted, and
  Solov'yov]{nielsen2018absorption}
Nielsen,~C.; N{\o}rby,~M.~S.; Kongsted,~J.; Solov'yov,~I.~A. {Absorption
  Spectra of FAD Embedded in Cryptochromes}. \emph{J. Phys. Chem. Lett.}
  \textbf{2018}, \emph{9}, 3618--3623\relax
\mciteBstWouldAddEndPuncttrue
\mciteSetBstMidEndSepPunct{\mcitedefaultmidpunct}
{\mcitedefaultendpunct}{\mcitedefaultseppunct}\relax
\EndOfBibitem
\bibitem[Timmer \latin{et~al.}(2023)Timmer, Frederiksen, L\"{u}nemann, Thomas,
  Xu, Bart\"{o}lke, Schmidt, Kuba\v{r}, De~Sio, Solov'yov, Mouritsen, and
  Lienau]{timmer2023journal}
Timmer,~D.; Frederiksen,~A.; L\"{u}nemann,~D.~C.; Thomas,~A.~R.; Xu,~J.;
  Bart\"{o}lke,~R.; Schmidt,~J.; Kuba\v{r},~T.; De~Sio,~A.; Solov'yov,~I.~A.;
  Mouritsen,~H.; Lienau,~C. {Tracking the Electron Transfer Cascade in European
  Robin Cryptochrome 4 Mutants}. \emph{J. Am. Chem. Soc.} \textbf{2023},
  \emph{145}, 11566--11578\relax
\mciteBstWouldAddEndPuncttrue
\mciteSetBstMidEndSepPunct{\mcitedefaultmidpunct}
{\mcitedefaultendpunct}{\mcitedefaultseppunct}\relax
\EndOfBibitem
\bibitem[Guallar and Frank(2008)Guallar, and Frank]{guallar2008mapping}
Guallar,~V.; Frank,~W. {Mapping Protein Electron Transfer Pathways with QM/MM
  Methods}. \emph{J. R. Soc. Interface} \textbf{2008}, \emph{5},
  S233--S239\relax
\mciteBstWouldAddEndPuncttrue
\mciteSetBstMidEndSepPunct{\mcitedefaultmidpunct}
{\mcitedefaultendpunct}{\mcitedefaultseppunct}\relax
\EndOfBibitem
\bibitem[Stevens and Hammes-Schiffer(2018)Stevens, and
  Hammes-Schiffer]{stevens2018exploring}
Stevens,~D.~R.; Hammes-Schiffer,~S. {Exploring the Role of the Third Active
  Site Metal Ion in DNA Polymerase with QM/MM Free Energy Simulations}.
  \emph{J. Am. Chem. Soc.} \textbf{2018}, \emph{140}, 8965--8969\relax
\mciteBstWouldAddEndPuncttrue
\mciteSetBstMidEndSepPunct{\mcitedefaultmidpunct}
{\mcitedefaultendpunct}{\mcitedefaultseppunct}\relax
\EndOfBibitem
\bibitem[Ko and Hammes-Schiffer(2013)Ko, and Hammes-Schiffer]{ko2013charge}
Ko,~C.; Hammes-Schiffer,~S. {Charge-Transfer Excited States and Proton Transfer
  in Model Guanine--Cytosine DNA Duplexes in Water}. \emph{J. Phys. Chem.
  Lett.} \textbf{2013}, \emph{4}, 2540--2545\relax
\mciteBstWouldAddEndPuncttrue
\mciteSetBstMidEndSepPunct{\mcitedefaultmidpunct}
{\mcitedefaultendpunct}{\mcitedefaultseppunct}\relax
\EndOfBibitem
\bibitem[Zeugner \latin{et~al.}(2005)Zeugner, Byrdin, Bouly, Bakrim, Giovani,
  Brettel, and Ahmad]{zeugner2005light}
Zeugner,~A.; Byrdin,~M.; Bouly,~J.~P.; Bakrim,~N.; Giovani,~B.; Brettel,~K.;
  Ahmad,~M. {Light-Induced Electron Transfer in Arabidopsis Cryptochrome-1
  Correlates with In Vivo Function}. \emph{J. Biol. Chem.} \textbf{2005},
  \emph{280}, 19437--19440\relax
\mciteBstWouldAddEndPuncttrue
\mciteSetBstMidEndSepPunct{\mcitedefaultmidpunct}
{\mcitedefaultendpunct}{\mcitedefaultseppunct}\relax
\EndOfBibitem
\bibitem[Losi and G\"{a}rtner(2012)Losi, and G\"{a}rtner]{losi2012evolution}
Losi,~A.; G\"{a}rtner,~W. {The Evolution of Flavin-Binding Photoreceptors: An
  Ancient Chromophore Serving Trendy Blue-Light Sensors}. \emph{Annu. Rev.
  Plant Biol.} \textbf{2012}, \emph{63}, 49--72\relax
\mciteBstWouldAddEndPuncttrue
\mciteSetBstMidEndSepPunct{\mcitedefaultmidpunct}
{\mcitedefaultendpunct}{\mcitedefaultseppunct}\relax
\EndOfBibitem
\bibitem[Stepanenko \latin{et~al.}(2011)Stepanenko, Stepanenko, Shcherbakova,
  Kuznetsova, Turoverov, and Verkhusha]{stepanenko2011modern}
Stepanenko,~O.~V.; Stepanenko,~O.~V.; Shcherbakova,~D.~M.; Kuznetsova,~I.~M.;
  Turoverov,~K.~K.; Verkhusha,~V.~V. {Modern Fluorescent Proteins: From
  Chromophore Formation to Novel Intracellular Applications}.
  \emph{Biotechniques} \textbf{2011}, \emph{51}, 313--327\relax
\mciteBstWouldAddEndPuncttrue
\mciteSetBstMidEndSepPunct{\mcitedefaultmidpunct}
{\mcitedefaultendpunct}{\mcitedefaultseppunct}\relax
\EndOfBibitem
\bibitem[Lax(1952)]{lax1952franck}
Lax,~M. {The Franck-Condon Principle and Its Application to Crystals}. \emph{J.
  Chem. Phys.} \textbf{1952}, \emph{20}, 1752--1760\relax
\mciteBstWouldAddEndPuncttrue
\mciteSetBstMidEndSepPunct{\mcitedefaultmidpunct}
{\mcitedefaultendpunct}{\mcitedefaultseppunct}\relax
\EndOfBibitem
\bibitem[Christie(2007)]{christie2007phototropin}
Christie,~J.~M. {Phototropin Blue-Light Receptors}. \emph{Annu. Rev. Plant
  Biol.} \textbf{2007}, \emph{58}, 21--45\relax
\mciteBstWouldAddEndPuncttrue
\mciteSetBstMidEndSepPunct{\mcitedefaultmidpunct}
{\mcitedefaultendpunct}{\mcitedefaultseppunct}\relax
\EndOfBibitem
\bibitem[Briggs \latin{et~al.}(2001)Briggs, Beck, Cashmore, Christie, Hughes,
  Jarillo, Kagawa, Kanegae, Liscum, Nagatani, Okada, Salomon, R\"{u}diger,
  Sakai, Takano, Wada, and Watson]{briggs2001phototropin}
Briggs,~W.~R. \latin{et~al.}  {The Phototropin Family of Photoreceptors}.
  \emph{Plant Cell} \textbf{2001}, \emph{13}, 993--997\relax
\mciteBstWouldAddEndPuncttrue
\mciteSetBstMidEndSepPunct{\mcitedefaultmidpunct}
{\mcitedefaultendpunct}{\mcitedefaultseppunct}\relax
\EndOfBibitem
\bibitem[Mroginski \latin{et~al.}(2021)Mroginski, Adam, Amoyal, Barnoy, Bondar,
  Borin, Church, Domratcheva, Ensing, Fanelli, and
  et~al.]{mroginski2021frontiers}
Mroginski,~M.-A.; Adam,~S.; Amoyal,~G.~S.; Barnoy,~A.; Bondar,~A.-N.;
  Borin,~V.~A.; Church,~J.~R.; Domratcheva,~T.; Ensing,~B.; Fanelli,~F.; et~al.
  {Frontiers in Multiscale Modeling of Photoreceptor Proteins}.
  \emph{Photochem. Photobiol.} \textbf{2021}, \emph{97}, 243--269\relax
\mciteBstWouldAddEndPuncttrue
\mciteSetBstMidEndSepPunct{\mcitedefaultmidpunct}
{\mcitedefaultendpunct}{\mcitedefaultseppunct}\relax
\EndOfBibitem
\bibitem[Matysik \latin{et~al.}(2023)Matysik, Gerhards, Theiss, Timmermann,
  Kurle-Tucholski, Musabirova, Qin, Ortmann, Solov'yov, and
  Gulder]{matysik2023spin}
Matysik,~J.; Gerhards,~L.; Theiss,~T.; Timmermann,~L.; Kurle-Tucholski,~P.;
  Musabirova,~G.; Qin,~R.; Ortmann,~F.; Solov'yov,~I.~A.; Gulder,~T. {Spin
  Dynamics of Flavoproteins}. \emph{Int. J. Mol. Sci.} \textbf{2023},
  \emph{24}, 8218\relax
\mciteBstWouldAddEndPuncttrue
\mciteSetBstMidEndSepPunct{\mcitedefaultmidpunct}
{\mcitedefaultendpunct}{\mcitedefaultseppunct}\relax
\EndOfBibitem
\bibitem[van Wonderen \latin{et~al.}(2021)van Wonderen, Adamczyk, Wu, Jiang,
  Piper, Hall, Edwards, Clarke, Zhang, Jeuken, Sazanovich, Towrie, Blumberger,
  Meech, and Butt]{vanwonderen2021nanosecond}
van Wonderen,~J.~H.; Adamczyk,~K.; Wu,~X.; Jiang,~X.; Piper,~S. E.~H.;
  Hall,~C.~R.; Edwards,~M.~J.; Clarke,~T.~A.; Zhang,~H.; Jeuken,~L. J.~C.;
  Sazanovich,~I.~V.; Towrie,~M.; Blumberger,~J.; Meech,~S.~R.; Butt,~J.~N.
  {Nanosecond Heme-to-Heme Electron Transfer Rates in a Multiheme Cytochrome
  Nanowire Reported by a Spectrally Unique His/Met-Ligated Heme}. \emph{Proc.
  Natl. Acad. Sci. U.S.A.} \textbf{2021}, \emph{118}, e2107939118\relax
\mciteBstWouldAddEndPuncttrue
\mciteSetBstMidEndSepPunct{\mcitedefaultmidpunct}
{\mcitedefaultendpunct}{\mcitedefaultseppunct}\relax
\EndOfBibitem
\bibitem[Schuhmann \latin{et~al.}(2021)Schuhmann, Kattnig, and
  Solov'yov]{schuhmann2021exploring}
Schuhmann,~F.; Kattnig,~D.~R.; Solov'yov,~I.~A. {Exploring Post-Activation
  Conformational Changes in Pigeon Cryptochrome 4}. \emph{J. Phys. Chem. B}
  \textbf{2021}, \emph{125}, 9652--9659\relax
\mciteBstWouldAddEndPuncttrue
\mciteSetBstMidEndSepPunct{\mcitedefaultmidpunct}
{\mcitedefaultendpunct}{\mcitedefaultseppunct}\relax
\EndOfBibitem
\bibitem[L\"{u}demann \latin{et~al.}(2015)L\"{u}demann, Solov'yov, Kubar, and
  Elstner]{ludemann2015solvent}
L\"{u}demann,~G.; Solov'yov,~I.~A.; Kubar,~T.; Elstner,~M. {Solvent Driving
  Force Ensures Fast Formation of a Persistent and Well-Separated Radical Pair
  in Plant Cryptochrome}. \emph{J. Am. Chem. Soc.} \textbf{2015}, \emph{137},
  1147--1156\relax
\mciteBstWouldAddEndPuncttrue
\mciteSetBstMidEndSepPunct{\mcitedefaultmidpunct}
{\mcitedefaultendpunct}{\mcitedefaultseppunct}\relax
\EndOfBibitem
\bibitem[Bondanza \latin{et~al.}(2020)Bondanza, Nottoli, Cupellini, Lipparini,
  and Mennucci]{bondanza2020polarizable}
Bondanza,~M.; Nottoli,~M.; Cupellini,~L.; Lipparini,~F.; Mennucci,~B.
  {Polarizable Embedding QM/MM: The Future Gold Standard for Complex
  (Bio)systems?} \emph{Phys. Chem. Chem. Phys.} \textbf{2020}, \emph{22},
  14433--14448\relax
\mciteBstWouldAddEndPuncttrue
\mciteSetBstMidEndSepPunct{\mcitedefaultmidpunct}
{\mcitedefaultendpunct}{\mcitedefaultseppunct}\relax
\EndOfBibitem
\bibitem[Lopata and Govind(2011)Lopata, and Govind]{lopata2011modeling}
Lopata,~K.; Govind,~N. {Modeling Fast Electron Dynamics with Real-Time
  Time-Dependent Density Functional Theory: Application to Small Molecules and
  Chromophores}. \emph{J. Chem. Theory Comput.} \textbf{2011}, \emph{7},
  1344--1355\relax
\mciteBstWouldAddEndPuncttrue
\mciteSetBstMidEndSepPunct{\mcitedefaultmidpunct}
{\mcitedefaultendpunct}{\mcitedefaultseppunct}\relax
\EndOfBibitem
\bibitem[Provorse and Isborn(2016)Provorse, and Isborn]{provorse2016electron}
Provorse,~M.~R.; Isborn,~C.~M. {Electron Dynamics with Real-Time Time-Dependent
  Density Functional Theory}. \emph{Int. J. Quantum Chem.} \textbf{2016},
  \emph{116}, 739--749\relax
\mciteBstWouldAddEndPuncttrue
\mciteSetBstMidEndSepPunct{\mcitedefaultmidpunct}
{\mcitedefaultendpunct}{\mcitedefaultseppunct}\relax
\EndOfBibitem
\bibitem[Pedron \latin{et~al.}(2020)Pedron, Issoglio, Estrin, and
  Scherlis]{pedron2020electron}
Pedron,~F.~N.; Issoglio,~F.; Estrin,~D.~A.; Scherlis,~D.~A. {Electron Transfer
  Pathways from Quantum Dynamics Simulations}. \emph{J. Chem. Phys.}
  \textbf{2020}, \emph{153}, 225102\relax
\mciteBstWouldAddEndPuncttrue
\mciteSetBstMidEndSepPunct{\mcitedefaultmidpunct}
{\mcitedefaultendpunct}{\mcitedefaultseppunct}\relax
\EndOfBibitem
\bibitem[Xu \latin{et~al.}(2021)Xu, Jarocha, Zollitsch, Konowalczyk, Henbest,
  Richert, Golesworthy, Schmidt, D\'{e}jean, Sowood, and
  et~al.]{xu2021magnetic}
Xu,~J.; Jarocha,~L.~E.; Zollitsch,~T.; Konowalczyk,~M.; Henbest,~K.~B.;
  Richert,~S.; Golesworthy,~M.~J.; Schmidt,~J.; D\'{e}jean,~V.; Sowood,~D.
  J.~C.; et~al. {Magnetic Sensitivity of Cryptochrome 4 from a Migratory
  Songbird}. \emph{Nature} \textbf{2021}, \emph{594}, 535--540\relax
\mciteBstWouldAddEndPuncttrue
\mciteSetBstMidEndSepPunct{\mcitedefaultmidpunct}
{\mcitedefaultendpunct}{\mcitedefaultseppunct}\relax
\EndOfBibitem
\bibitem[Barragan \latin{et~al.}(2021)Barragan, Soudackov, Luthey-Schulten,
  Hammes-Schiffer, Schulten, and Solov'yov]{barragan2021theoretical}
Barragan,~A.~M.; Soudackov,~A.~V.; Luthey-Schulten,~Z.; Hammes-Schiffer,~S.;
  Schulten,~K.; Solov'yov,~I.~A. {Theoretical Description of the Primary
  Proton-Coupled Electron Transfer Reaction in the Cytochrome bc 1 Complex}.
  \emph{J. Am. Chem. Soc.} \textbf{2021}, \emph{143}, 715--723\relax
\mciteBstWouldAddEndPuncttrue
\mciteSetBstMidEndSepPunct{\mcitedefaultmidpunct}
{\mcitedefaultendpunct}{\mcitedefaultseppunct}\relax
\EndOfBibitem
\bibitem[Joshi and Deshmukh()Joshi, and Deshmukh]{joshi2021review}
Joshi,~S.~Y.; Deshmukh,~S.~A. A Review of Advancements in Coarse-Grained
  Molecular Dynamics Simulations. \relax
\mciteBstWouldAddEndPunctfalse
\mciteSetBstMidEndSepPunct{\mcitedefaultmidpunct}
{}{\mcitedefaultseppunct}\relax
\EndOfBibitem
\bibitem[Bouda\"{i}ffa \latin{et~al.}(2000)Bouda\"{i}ffa, Cloutier, Hunting,
  Huels, and Sanche]{Boudaiffa_2000_Science.287.1658}
Bouda\"{i}ffa,~B.; Cloutier,~P.; Hunting,~D.; Huels,~M.~A.; Sanche,~L.
  {Resonant Formation of DNA Strand Breaks by Low-Energy (3 to 20 eV)
  Electrons}. \emph{Science} \textbf{2000}, \emph{287}, 1658--1660\relax
\mciteBstWouldAddEndPuncttrue
\mciteSetBstMidEndSepPunct{\mcitedefaultmidpunct}
{\mcitedefaultendpunct}{\mcitedefaultseppunct}\relax
\EndOfBibitem
\bibitem[Coupier \latin{et~al.}(2002)Coupier, Farizon, Farizon, Gaillard,
  Gobet, de~Castro~Faria, Jalbert, Ouaskit, Carr\'{e}, Gstir, Hanel, Denifl,
  Feketeova, Scheier, and M\"{a}rk]{Coupier_2002_EPJD.20.459}
Coupier,~B.; Farizon,~B.; Farizon,~M.; Gaillard,~M.~J.; Gobet,~F.;
  de~Castro~Faria,~N.~V.; Jalbert,~G.; Ouaskit,~S.; Carr\'{e},~M.; Gstir,~B.;
  Hanel,~G.; Denifl,~S.; Feketeova,~L.; Scheier,~P.; M\"{a}rk,~T.~D. {Inelastic
  Interactions of Protons and Electrons with Biologically Relevant Molecules}.
  \emph{Eur. Phys. J. D} \textbf{2002}, \emph{20}, 459--468\relax
\mciteBstWouldAddEndPuncttrue
\mciteSetBstMidEndSepPunct{\mcitedefaultmidpunct}
{\mcitedefaultendpunct}{\mcitedefaultseppunct}\relax
\EndOfBibitem
\bibitem[de~Vries \latin{et~al.}(2003)de~Vries, Hoekstra, Morgenstern, and
  Schlath\"{o}lter]{deVries_2003_PRL.91.053401}
de~Vries,~J.; Hoekstra,~R.; Morgenstern,~R.; Schlath\"{o}lter,~T. {Charge
  Driven Fragmentation of Nucleic Acid Bases}. \emph{Phys. Rev. Lett.}
  \textbf{2003}, \emph{91}, 053401\relax
\mciteBstWouldAddEndPuncttrue
\mciteSetBstMidEndSepPunct{\mcitedefaultmidpunct}
{\mcitedefaultendpunct}{\mcitedefaultseppunct}\relax
\EndOfBibitem
\bibitem[Hanel \latin{et~al.}(2003)Hanel, Gstir, Denifl, Scheier, Probst,
  Farizon, Farizon, Illenberger, and M\"{a}rk]{Hanel_2003_PRL.90.188104}
Hanel,~G.; Gstir,~B.; Denifl,~S.; Scheier,~P.; Probst,~M.; Farizon,~B.;
  Farizon,~M.; Illenberger,~E.; M\"{a}rk,~T.~D. {Electron Attachment to Uracil:
  Effective Destruction at Subexcitation Energies}. \emph{Phys. Rev. Lett.}
  \textbf{2003}, \emph{90}, 188104\relax
\mciteBstWouldAddEndPuncttrue
\mciteSetBstMidEndSepPunct{\mcitedefaultmidpunct}
{\mcitedefaultendpunct}{\mcitedefaultseppunct}\relax
\EndOfBibitem
\bibitem[Liu \latin{et~al.}(2006)Liu, Br{\o}ndsted~Nielsen, Hvelplund,
  Zettergren, Cederquist, Manil, and Huber]{Liu_2006_PRL.97.133401}
Liu,~B.; Br{\o}ndsted~Nielsen,~S.; Hvelplund,~P.; Zettergren,~H.;
  Cederquist,~H.; Manil,~B.; Huber,~B.~A. {Collision-Induced Dissociation of
  Hydrated Adenosine Monophosphate Nucleotide Ions: Protection of the Ion in
  Water Nanoclusters}. \emph{Phys. Rev. Lett.} \textbf{2006}, \emph{97},
  133401\relax
\mciteBstWouldAddEndPuncttrue
\mciteSetBstMidEndSepPunct{\mcitedefaultmidpunct}
{\mcitedefaultendpunct}{\mcitedefaultseppunct}\relax
\EndOfBibitem
\bibitem[Milosavljevi\'{c} \latin{et~al.}(2011)Milosavljevi\'{c}, Nicolas,
  Lemaire, Dehon, Thissen, Bizau, R\'{e}fr\'{e}giers, Nahon, and
  Giuliani]{Milosavljevic_2011_PCCP.13.15432}
Milosavljevi\'{c},~A.~R.; Nicolas,~C.; Lemaire,~J.; Dehon,~C.; Thissen,~R.;
  Bizau,~J.; R\'{e}fr\'{e}giers,~M.; Nahon,~L.; Giuliani,~A. {Photoionization
  of a Protein Isolated in Vacuo}. \emph{Phys. Chem. Chem. Phys.}
  \textbf{2011}, \emph{13}, 15432--15436\relax
\mciteBstWouldAddEndPuncttrue
\mciteSetBstMidEndSepPunct{\mcitedefaultmidpunct}
{\mcitedefaultendpunct}{\mcitedefaultseppunct}\relax
\EndOfBibitem
\bibitem[Gonz\'{a}lez-Maga\~{n}a \latin{et~al.}(2013)Gonz\'{a}lez-Maga\~{n}a,
  Tiemens, Reitsma, Boschman, Door, Bari, Hoekstra, Lahaie, Wagner, Huels, and
  Schlath\"{o}lter]{Gonzalez-Magana_2013_PRA}
Gonz\'{a}lez-Maga\~{n}a,~O.; Tiemens,~M.; Reitsma,~G.; Boschman,~L.; Door,~M.;
  Bari,~S.; Hoekstra,~R.; Lahaie,~P.~O.; Wagner,~J.~R.; Huels,~M.~A.;
  Schlath\"{o}lter,~T. {Fragmentation of Protonated Oligonucleotides by
  Energetic Photons, Protons, and C$^{q+}$ Ions}. \emph{Phys. Rev. A}
  \textbf{2013}, \emph{87}, 032702\relax
\mciteBstWouldAddEndPuncttrue
\mciteSetBstMidEndSepPunct{\mcitedefaultmidpunct}
{\mcitedefaultendpunct}{\mcitedefaultseppunct}\relax
\EndOfBibitem
\bibitem[Lalande \latin{et~al.}(2019)Lalande, Schwob, Vizcaino, Chirot,
  Dugourd, Schlath\"{o}lter, and Poully]{Lalande_2019_ChemBioChem}
Lalande,~M.; Schwob,~L.; Vizcaino,~V.; Chirot,~F.; Dugourd,~P.;
  Schlath\"{o}lter,~T.; Poully,~J.-C. {Direct Radiation Effects on the
  Structure and Stability of Collagen and Other Proteins}. \emph{ChemBioChem}
  \textbf{2019}, \emph{20}, 2972--2980\relax
\mciteBstWouldAddEndPuncttrue
\mciteSetBstMidEndSepPunct{\mcitedefaultmidpunct}
{\mcitedefaultendpunct}{\mcitedefaultseppunct}\relax
\EndOfBibitem
\bibitem[Maclot \latin{et~al.}(2016)Maclot, Delaunay, Piekarski, Domaracka,
  Huber, Adoui, Mart\'{i}n, Alcam\'{i}, Avaldi, Bolognesi, D\'{i}az-Tendero,
  and Rousseau]{Maclot_2016_PRL.117.073201}
Maclot,~S.; Delaunay,~R.; Piekarski,~D.~G.; Domaracka,~A.; Huber,~B.~A.;
  Adoui,~L.; Mart\'{i}n,~F.; Alcam\'{i},~M.; Avaldi,~L.; Bolognesi,~P.;
  D\'{i}az-Tendero,~S.; Rousseau,~P. {Determination of Energy-Transfer
  Distributions in Ionizing Ion-Molecule Collisions}. \emph{Phys. Rev. Lett.}
  \textbf{2016}, \emph{117}, 073201\relax
\mciteBstWouldAddEndPuncttrue
\mciteSetBstMidEndSepPunct{\mcitedefaultmidpunct}
{\mcitedefaultendpunct}{\mcitedefaultseppunct}\relax
\EndOfBibitem
\bibitem[Wang \latin{et~al.}(2022)Wang, Rathnachalam, Zamudio-Bayer, Bijlsma,
  Li, Hoekstra, Kubin, Timm, von Issendorff, Lau, Faraji, and
  Schlath\"{o}lter]{Wang_2022_PCCP.24.7815}
Wang,~X.; Rathnachalam,~S.; Zamudio-Bayer,~V.; Bijlsma,~K.; Li,~W.;
  Hoekstra,~R.; Kubin,~M.; Timm,~M.; von Issendorff,~B.; Lau,~J.~T.;
  Faraji,~S.; Schlath\"{o}lter,~T. {Intramolecular Hydrogen Transfer in DNA
  Induced by Site-Selective Resonant Core Excitation}. \emph{Phys. Chem. Chem.
  Phys.} \textbf{2022}, \emph{24}, 7815--7825\relax
\mciteBstWouldAddEndPuncttrue
\mciteSetBstMidEndSepPunct{\mcitedefaultmidpunct}
{\mcitedefaultendpunct}{\mcitedefaultseppunct}\relax
\EndOfBibitem
\bibitem[Palacios and Mart\'{i}n(2019)Palacios, and
  Mart\'{i}n]{Palacios_2019_CompMolSci}
Palacios,~A.; Mart\'{i}n,~F. {The Quantum Chemistry of Attosecond Molecular
  Science}. \emph{WIREs Comput. Mol. Sci.} \textbf{2019}, \emph{10},
  e1430\relax
\mciteBstWouldAddEndPuncttrue
\mciteSetBstMidEndSepPunct{\mcitedefaultmidpunct}
{\mcitedefaultendpunct}{\mcitedefaultseppunct}\relax
\EndOfBibitem
\bibitem[West \latin{et~al.}(2013)West, Womick, and
  Moran]{West_2013_JPCA.117.5865}
West,~B.~A.; Womick,~J.~M.; Moran,~A.~M. {Interplay between Vibrational Energy
  Transfer and Excited State Deactivation in DNA Components}. \emph{J. Phys.
  Chem. A} \textbf{2013}, \emph{117}, 5865--5874\relax
\mciteBstWouldAddEndPuncttrue
\mciteSetBstMidEndSepPunct{\mcitedefaultmidpunct}
{\mcitedefaultendpunct}{\mcitedefaultseppunct}\relax
\EndOfBibitem
\bibitem[Li \latin{et~al.}(2021)Li, Kavatsyuk, Douma, Wang, Hoekstra, Mayer,
  Robinson, G\"{u}hr, Lalande, Abdelmouleh, Ryszka, Poully, and
  Schlath\"{o}lter]{Li_2021_ChemSci.12.13177}
Li,~W.; Kavatsyuk,~O.; Douma,~W.; Wang,~X.; Hoekstra,~R.; Mayer,~D.;
  Robinson,~M.; G\"{u}hr,~M.; Lalande,~M.; Abdelmouleh,~M.; Ryszka,~M.;
  Poully,~J.~C.; Schlath\"{o}lter,~T. {Charge Reversing Multiple Electron
  Detachment Auger Decay of Inner-Shell Vacancies in Gas-Phase Deprotonated
  DNA}. \emph{Chem. Sci.} \textbf{2021}, \emph{12}, 13177--13186\relax
\mciteBstWouldAddEndPuncttrue
\mciteSetBstMidEndSepPunct{\mcitedefaultmidpunct}
{\mcitedefaultendpunct}{\mcitedefaultseppunct}\relax
\EndOfBibitem
\bibitem[Hu and Niemeyer(2019)Hu, and Niemeyer]{Hu_2019_AdvMater.31.1806294}
Hu,~Y.; Niemeyer,~C.~M. {From DNA Nanotechnology to Material Systems
  Engineering}. \emph{Adv. Mater.} \textbf{2019}, \emph{31}, 1806294\relax
\mciteBstWouldAddEndPuncttrue
\mciteSetBstMidEndSepPunct{\mcitedefaultmidpunct}
{\mcitedefaultendpunct}{\mcitedefaultseppunct}\relax
\EndOfBibitem
\bibitem[Rajendran \latin{et~al.}(2012)Rajendran, Endo, and
  Sugiyama]{Rajendran_2012_AngewChemIntEd.51.874}
Rajendran,~A.; Endo,~M.; Sugiyama,~H. {Single-Molecule Analysis Using DNA
  Origami}. \emph{Angew. Chem. Int. Ed.} \textbf{2012}, \emph{51},
  874--890\relax
\mciteBstWouldAddEndPuncttrue
\mciteSetBstMidEndSepPunct{\mcitedefaultmidpunct}
{\mcitedefaultendpunct}{\mcitedefaultseppunct}\relax
\EndOfBibitem
\bibitem[Keller \latin{et~al.}(2014)Keller, Rackwitz, Cau\"{e}t, Li\"{e}vin,
  K\"{o}rzd\"{o}rfer, Rotaru, Gothelf, Besenbacher, and
  Bald]{Keller_2014_SciRep.4.7391}
Keller,~A.; Rackwitz,~J.; Cau\"{e}t,~E.; Li\"{e}vin,~J.;
  K\"{o}rzd\"{o}rfer,~T.; Rotaru,~A.; Gothelf,~K.~V.; Besenbacher,~F.; Bald,~I.
  {Sequence Dependence of Electron-Induced DNA Strand Breakage Revealed by DNA
  Nanoarrays}. \emph{Sci. Rep.} \textbf{2014}, \emph{4}, 7391\relax
\mciteBstWouldAddEndPuncttrue
\mciteSetBstMidEndSepPunct{\mcitedefaultmidpunct}
{\mcitedefaultendpunct}{\mcitedefaultseppunct}\relax
\EndOfBibitem
\bibitem[Sala \latin{et~al.}(2021)Sala, Zerolov\'{a}, Rodriguez, Reimitz,
  Dav\'{i}dkov\'{a}, Ebel, Bald, and
  Ko\v{c}i\v{s}ek]{Sala_2021_Nanoscale.13.11197}
Sala,~L.; Zerolov\'{a},~A.; Rodriguez,~A.; Reimitz,~D.; Dav\'{i}dkov\'{a},~M.;
  Ebel,~K.; Bald,~I.; Ko\v{c}i\v{s}ek,~J. {Folding DNA into Origami
  Nanostructures Enhances Resistance to Ionizing Radiation}. \emph{Nanoscale}
  \textbf{2021}, \emph{13}, 11197--11203\relax
\mciteBstWouldAddEndPuncttrue
\mciteSetBstMidEndSepPunct{\mcitedefaultmidpunct}
{\mcitedefaultendpunct}{\mcitedefaultseppunct}\relax
\EndOfBibitem
\bibitem[Sala \latin{et~al.}(2022)Sala, Lyshchuk, \v{S}\'{a}chov\'{a},
  Chv\'{a}til, and Ko\v{c}i\v{s}ek]{Sala_2022_JPCL.13.3922}
Sala,~L.; Lyshchuk,~H.; \v{S}\'{a}chov\'{a},~J.; Chv\'{a}til,~D.;
  Ko\v{c}i\v{s}ek,~J. {Different Mechanisms of DNA Radiosensitization by
  8-Bromoadenosine and 2'-Deoxy-2'-Fluorocytidine Observed on DNA Origami
  Nanoframe Supports}. \emph{J. Phys. Chem. Lett.} \textbf{2022}, \emph{13},
  3922--3928\relax
\mciteBstWouldAddEndPuncttrue
\mciteSetBstMidEndSepPunct{\mcitedefaultmidpunct}
{\mcitedefaultendpunct}{\mcitedefaultseppunct}\relax
\EndOfBibitem
\bibitem[Fang \latin{et~al.}(2020)Fang, Xie, Hou, Liu, Zuo, Chao, Wang, Fan,
  Liu, and Wang]{Fang_2020_JACS.142.8782}
Fang,~W.; Xie,~M.; Hou,~X.; Liu,~X.; Zuo,~X.; Chao,~J.; Wang,~L.; Fan,~C.;
  Liu,~H.; Wang,~L. { DNA Origami Radiometers for Measuring Ultraviolet
  Exposure}. \emph{J. Am. Chem. Soc.} \textbf{2020}, \emph{142},
  8782--8789\relax
\mciteBstWouldAddEndPuncttrue
\mciteSetBstMidEndSepPunct{\mcitedefaultmidpunct}
{\mcitedefaultendpunct}{\mcitedefaultseppunct}\relax
\EndOfBibitem
\bibitem[Matsika(2021)]{Matsika_2021_ChemRev.121.9407}
Matsika,~S. {Electronic Structure Methods for the Description of Nonadiabatic
  Effects and Conical Intersections}. \emph{Chem. Rev.} \textbf{2021},
  \emph{121}, 9407--9449\relax
\mciteBstWouldAddEndPuncttrue
\mciteSetBstMidEndSepPunct{\mcitedefaultmidpunct}
{\mcitedefaultendpunct}{\mcitedefaultseppunct}\relax
\EndOfBibitem
\bibitem[Mart\'{i}nez-Fern\'{a}ndez and
  Franc\'{e}s-Monerris(2023)Mart\'{i}nez-Fern\'{a}ndez, and
  Franc\'{e}s-Monerris]{Martinez-Fernandez_2023_DNAphotostability}
Mart\'{i}nez-Fern\'{a}ndez,~L.; Franc\'{e}s-Monerris,~A. In \emph{Theoretical
  and Computational Photochemistry}; Garc\'{i}a-Iriepa,~C., Marazzi,~M., Eds.;
  Elsevier, 2023; pp 311--336\relax
\mciteBstWouldAddEndPuncttrue
\mciteSetBstMidEndSepPunct{\mcitedefaultmidpunct}
{\mcitedefaultendpunct}{\mcitedefaultseppunct}\relax
\EndOfBibitem
\bibitem[Westermayr \latin{et~al.}(2019)Westermayr, Gastegger, Menger, Mai,
  Gonz\'{a}lez, and Marquetand]{Westermayr_2019_ChemSci.10.8100}
Westermayr,~J.; Gastegger,~M.; Menger,~M. F. S.~J.; Mai,~S.; Gonz\'{a}lez,~L.;
  Marquetand,~P. {Machine Learning Enables Long Time Scale Molecular
  Photodynamics Simulations}. \emph{Chem. Sci.} \textbf{2019}, \emph{10},
  8100--8107\relax
\mciteBstWouldAddEndPuncttrue
\mciteSetBstMidEndSepPunct{\mcitedefaultmidpunct}
{\mcitedefaultendpunct}{\mcitedefaultseppunct}\relax
\EndOfBibitem
\bibitem[Poppleton \latin{et~al.}(2020)Poppleton, Bohlin, Matthies, Sharma,
  Zhang, and \v{S}ulc]{Poppleton_2020_NuclAcidsRes.48.e72}
Poppleton,~E.; Bohlin,~J.; Matthies,~M.; Sharma,~S.; Zhang,~F.; \v{S}ulc,~P.
  {Design, Optimization and Analysis of Large DNA and RNA Nanostructures
  through Interactive Visualization, Editing and Molecular Simulation}.
  \emph{Nucleic Acids Res.} \textbf{2020}, \emph{48}, e72\relax
\mciteBstWouldAddEndPuncttrue
\mciteSetBstMidEndSepPunct{\mcitedefaultmidpunct}
{\mcitedefaultendpunct}{\mcitedefaultseppunct}\relax
\EndOfBibitem
\bibitem[Huynh \latin{et~al.}(2020)Huynh, Hosny, Guthier, Bitterman, Petit,
  Haas-Kogan, Kann, Aerts, and Mak]{Huynh_2020_NatRevClinOncol.17.771}
Huynh,~E.; Hosny,~A.; Guthier,~C.; Bitterman,~D.~S.; Petit,~S.~F.;
  Haas-Kogan,~D.~A.; Kann,~B.; Aerts,~H. J. W.~L.; Mak,~R.~H. {Artificial
  Intelligence in Radiation Oncology}. \emph{Nat. Rev. Clin. Oncol.}
  \textbf{2020}, \emph{17}, 771--781\relax
\mciteBstWouldAddEndPuncttrue
\mciteSetBstMidEndSepPunct{\mcitedefaultmidpunct}
{\mcitedefaultendpunct}{\mcitedefaultseppunct}\relax
\EndOfBibitem
\bibitem[Singh \latin{et~al.}(2022)Singh, Sharma, Garg, Kumar, Baliyan, Rani,
  and Kumar]{Singh_2022_BiotechnolAdv.61.108052}
Singh,~M.; Sharma,~D.; Garg,~M.; Kumar,~A.; Baliyan,~A.; Rani,~R.; Kumar,~V.
  {Current Understanding of Biological Interactions and Processing of DNA
  Origami Nanostructures: Role of Machine Learning and Implications in Drug
  Delivery}. \emph{Biotechnol. Adv.} \textbf{2022}, \emph{61}, 108052\relax
\mciteBstWouldAddEndPuncttrue
\mciteSetBstMidEndSepPunct{\mcitedefaultmidpunct}
{\mcitedefaultendpunct}{\mcitedefaultseppunct}\relax
\EndOfBibitem
\bibitem[Rackwitz and Bald(2018)Rackwitz, and
  Bald]{Rackwitz_2018_ChemEurJ.24.4680}
Rackwitz,~J.; Bald,~I. {Low-Energy Electron-Induced Strand Breaks in
  Telomere-Derived DNA Sequences--Influence of DNA Sequence and Topology}.
  \emph{Chem. Eur. J.} \textbf{2018}, \emph{24}, 4680--4688\relax
\mciteBstWouldAddEndPuncttrue
\mciteSetBstMidEndSepPunct{\mcitedefaultmidpunct}
{\mcitedefaultendpunct}{\mcitedefaultseppunct}\relax
\EndOfBibitem
\bibitem[Berardinelli \latin{et~al.}(2017)Berardinelli, Coluzzi, Sgura, and
  Antoccia]{Berardinelli_2017_MutatResRev.773.204}
Berardinelli,~F.; Coluzzi,~E.; Sgura,~A.; Antoccia,~A. {Targeting Telomerase
  and Telomeres to Enhance Ionizing Radiation Effects in in Vitro and in Vivo
  Cancer Models}. \emph{Mutat. Res. Rev. Mutat. Res.} \textbf{2017},
  \emph{773}, 204--219\relax
\mciteBstWouldAddEndPuncttrue
\mciteSetBstMidEndSepPunct{\mcitedefaultmidpunct}
{\mcitedefaultendpunct}{\mcitedefaultseppunct}\relax
\EndOfBibitem
\bibitem[Keller and Linko(2020)Keller, and
  Linko]{Keller_2020_AngewChemIntEd.59.15818}
Keller,~A.; Linko,~V. {Challenges and Perspectives of DNA Nanostructures in
  Biomedicine}. \emph{Angew. Chem. Int. Ed.} \textbf{2020}, \emph{59},
  15818--15833\relax
\mciteBstWouldAddEndPuncttrue
\mciteSetBstMidEndSepPunct{\mcitedefaultmidpunct}
{\mcitedefaultendpunct}{\mcitedefaultseppunct}\relax
\EndOfBibitem
\bibitem[Paul \latin{et~al.}(2001)Paul, Toma, Breger, Mullot, Aug\'{e}, Balcou,
  Muller, and P.]{Paul_2001_Science.292.1689}
Paul,~P.~M.; Toma,~E.~S.; Breger,~P.; Mullot,~G.; Aug\'{e},~F.; Balcou,~P.;
  Muller,~H.~G.; P.,~A. {Observation of a Train of Attosecond Pulses from High
  Harmonic Generation}. \emph{Science} \textbf{2001}, \emph{292},
  1689--1692\relax
\mciteBstWouldAddEndPuncttrue
\mciteSetBstMidEndSepPunct{\mcitedefaultmidpunct}
{\mcitedefaultendpunct}{\mcitedefaultseppunct}\relax
\EndOfBibitem
\bibitem[Drescher \latin{et~al.}(2002)Drescher, Hentschel, Kienberger,
  Uiberacker, Yakovlev, Scrinzi, Westerwalbesloh, Kleineberg, U., and
  Krausz]{Drescher_2002_Nature.419.803}
Drescher,~M.; Hentschel,~M.; Kienberger,~R.; Uiberacker,~M.; Yakovlev,~V.;
  Scrinzi,~A.; Westerwalbesloh,~T.; Kleineberg,~U.; U.,~H.; Krausz,~F.
  {Time-Resolved Atomic Inner-Shell Spectroscopy}. \emph{Nature} \textbf{2002},
  \emph{419}, 803--807\relax
\mciteBstWouldAddEndPuncttrue
\mciteSetBstMidEndSepPunct{\mcitedefaultmidpunct}
{\mcitedefaultendpunct}{\mcitedefaultseppunct}\relax
\EndOfBibitem
\bibitem[Goulielmakis \latin{et~al.}(2010)Goulielmakis, Loh, Wirth, Santra,
  Rohringer, Yakovlev, Zherebtsov, Pfeifer, Azzeer, Kling, and
  et~al.]{Goulielmakis_2010_Nature.466.739}
Goulielmakis,~E.; Loh,~Z.-H.; Wirth,~A.; Santra,~R.; Rohringer,~N.;
  Yakovlev,~V.~S.; Zherebtsov,~S.; Pfeifer,~T.; Azzeer,~A.~M.; Kling,~M.~F.;
  et~al. {Real-Time Observation of Valence Electron Motion}. \emph{Nature}
  \textbf{2010}, \emph{466}, 739--743\relax
\mciteBstWouldAddEndPuncttrue
\mciteSetBstMidEndSepPunct{\mcitedefaultmidpunct}
{\mcitedefaultendpunct}{\mcitedefaultseppunct}\relax
\EndOfBibitem
\bibitem[Schultze \latin{et~al.}(2010)Schultze, Fies, Karpowicz, Gagnon,
  Korbman, Hofstetter, Neppl, Cavalieri, Komninos, Mercouris, and
  et~al.]{Schultze_2010_Science.328.1658}
Schultze,~M.; Fies,~M.; Karpowicz,~N.; Gagnon,~J.; Korbman,~M.; Hofstetter,~M.;
  Neppl,~S.; Cavalieri,~A.~L.; Komninos,~Y.; Mercouris,~T.; et~al. {Delay in
  Photoemission}. \emph{Science} \textbf{2010}, \emph{328}, 1658--1662\relax
\mciteBstWouldAddEndPuncttrue
\mciteSetBstMidEndSepPunct{\mcitedefaultmidpunct}
{\mcitedefaultendpunct}{\mcitedefaultseppunct}\relax
\EndOfBibitem
\bibitem[L\'{e}pine \latin{et~al.}(2014)L\'{e}pine, Ivanov, and
  Vrakking]{Lepine_2014_NatPhot.8.195}
L\'{e}pine,~F.; Ivanov,~M.~Y.; Vrakking,~M. J.~J. {Attosecond Molecular
  Dynamics: Fact or Fiction?} \emph{Nat. Photonics} \textbf{2014}, \emph{8},
  195--204\relax
\mciteBstWouldAddEndPuncttrue
\mciteSetBstMidEndSepPunct{\mcitedefaultmidpunct}
{\mcitedefaultendpunct}{\mcitedefaultseppunct}\relax
\EndOfBibitem
\bibitem[Sansone \latin{et~al.}(2010)Sansone, Kelkensberg, P\'{e}rez-Torres,
  Morales, Kling, Siu, Ghafur, Johnsson, Swoboda, Benedetti, and
  et~al.]{Sansone_2010_Nature.465.763}
Sansone,~G.; Kelkensberg,~F.; P\'{e}rez-Torres,~J.~F.; Morales,~F.;
  Kling,~M.~F.; Siu,~W.; Ghafur,~O.; Johnsson,~P.; Swoboda,~M.; Benedetti,~E.;
  et~al. {Electron Localization Following Attosecond Molecular
  Photoionization}. \emph{Nature} \textbf{2010}, \emph{465}, 763--766\relax
\mciteBstWouldAddEndPuncttrue
\mciteSetBstMidEndSepPunct{\mcitedefaultmidpunct}
{\mcitedefaultendpunct}{\mcitedefaultseppunct}\relax
\EndOfBibitem
\bibitem[Neidel \latin{et~al.}(2013)Neidel, Klei, Yang, Rouz\'{e}e, Vrakking,
  Kl\"{u}nder, Miranda, Arnold, Fordell, L'Huillier, and
  et~al.]{Neidel_2013_PRL.111.033001}
Neidel,~C.; Klei,~J.; Yang,~C.~H.; Rouz\'{e}e,~A.; Vrakking,~M. J.~J.;
  Kl\"{u}nder,~K.; Miranda,~M.; Arnold,~C.~L.; Fordell,~T.; L'Huillier,~A.;
  et~al. {Probing Time-Dependent Molecular Dipoles on the Attosecond Time
  Scale}. \emph{Phys. Rev. Lett.} \textbf{2013}, \emph{111}, 033001\relax
\mciteBstWouldAddEndPuncttrue
\mciteSetBstMidEndSepPunct{\mcitedefaultmidpunct}
{\mcitedefaultendpunct}{\mcitedefaultseppunct}\relax
\EndOfBibitem
\bibitem[Calegari \latin{et~al.}(2014)Calegari, Ayuso, Trabattoni, Belshaw,
  De~Camillis, Anumula, Frassetto, Poletto, Palacios, Decleva, and
  et~al.]{Calegari_2014_Science.346.336}
Calegari,~F.; Ayuso,~D.; Trabattoni,~A.; Belshaw,~L.; De~Camillis,~S.;
  Anumula,~S.; Frassetto,~F.; Poletto,~L.; Palacios,~A.; Decleva,~P.; et~al.
  {Ultrafast Electron Dynamics in Phenylalanine Initiated by Attosecond
  Pulses}. \emph{Science} \textbf{2014}, \emph{346}, 336--339\relax
\mciteBstWouldAddEndPuncttrue
\mciteSetBstMidEndSepPunct{\mcitedefaultmidpunct}
{\mcitedefaultendpunct}{\mcitedefaultseppunct}\relax
\EndOfBibitem
\bibitem[Herv\'{e} \latin{et~al.}(2022)Herv\'{e}, Boyer, Br\'{e}dy, Compagnon,
  and L\'{e}pine]{Herve_2022_AdvPhysX.7.2123283}
Herv\'{e},~M.; Boyer,~A.; Br\'{e}dy,~R.; Compagnon,~I.; L\'{e}pine,~F.
  {Ultrafast Dynamics in Molecular Ions Following UV and XUV Excitation: A
  Perspective}. \emph{Adv. Phys. X} \textbf{2022}, \emph{7}, 2123283\relax
\mciteBstWouldAddEndPuncttrue
\mciteSetBstMidEndSepPunct{\mcitedefaultmidpunct}
{\mcitedefaultendpunct}{\mcitedefaultseppunct}\relax
\EndOfBibitem
\bibitem[Herv\'{e} \latin{et~al.}(2022)Herv\'{e}, Boyer, Br\'{e}dy, Allouche,
  Compagnon, and L\'{e}pine]{Herve_2022_SciRep.12.13191}
Herv\'{e},~M.; Boyer,~A.; Br\'{e}dy,~R.; Allouche,~A.~R.; Compagnon,~I.;
  L\'{e}pine,~F. {On-the-fly Investigation of XUV Excited Large Molecular Ions
  Using a High Harmonic Generation Light Source}. \emph{Sci. Rep.}
  \textbf{2022}, \emph{12}, 13191\relax
\mciteBstWouldAddEndPuncttrue
\mciteSetBstMidEndSepPunct{\mcitedefaultmidpunct}
{\mcitedefaultendpunct}{\mcitedefaultseppunct}\relax
\EndOfBibitem
\bibitem[Vacher \latin{et~al.}(2017)Vacher, Bearpark, Robb, and
  Malhado]{Vacher_2017_PRL.118.083001}
Vacher,~M.; Bearpark,~M.~J.; Robb,~M.~A.; Malhado,~J.~P. {Electron Dynamics
  upon Ionization of Polyatomic Molecules: Coupling to Quantum Nuclear Motion
  and Decoherence}. \emph{Phys. Rev. Lett.} \textbf{2017}, \emph{118},
  083001\relax
\mciteBstWouldAddEndPuncttrue
\mciteSetBstMidEndSepPunct{\mcitedefaultmidpunct}
{\mcitedefaultendpunct}{\mcitedefaultseppunct}\relax
\EndOfBibitem
\bibitem[Despr\'{e} \latin{et~al.}(2018)Despr\'{e}, Golubev, and
  Kuleff]{Despre_2018_PRL.121.203002}
Despr\'{e},~V.; Golubev,~N.~V.; Kuleff,~A.~I. {Charge Migration in Propiolic
  Acid: A Full Quantum Dynamical Study}. \emph{Phys. Rev. Lett.} \textbf{2018},
  \emph{121}, 203002\relax
\mciteBstWouldAddEndPuncttrue
\mciteSetBstMidEndSepPunct{\mcitedefaultmidpunct}
{\mcitedefaultendpunct}{\mcitedefaultseppunct}\relax
\EndOfBibitem
\bibitem[Herv\'{e} \latin{et~al.}(2021)Herv\'{e}, Despr\'{e}, Castellanos~Nash,
  Loriot, Boyer, Scognamiglio, Karras, Br\'{e}dy, Constant, Tielens, and
  et~al.]{Herve_2021_NatPhys.17.327}
Herv\'{e},~M.; Despr\'{e},~V.; Castellanos~Nash,~P.; Loriot,~V.; Boyer,~A.;
  Scognamiglio,~A.; Karras,~G.; Br\'{e}dy,~R.; Constant,~E.; Tielens,~A. G.
  G.~M.; et~al. {Ultrafast Dynamics of Correlation Bands Following XUV
  Molecular Photoionization}. \emph{Nat. Phys.} \textbf{2021}, \emph{17},
  327--331\relax
\mciteBstWouldAddEndPuncttrue
\mciteSetBstMidEndSepPunct{\mcitedefaultmidpunct}
{\mcitedefaultendpunct}{\mcitedefaultseppunct}\relax
\EndOfBibitem
\bibitem[Berrah \latin{et~al.}(2019)Berrah, Sanchez-Gonzalez, Jurek, Obaid,
  Xiong, Squibb, Osipov, Lutman, Fang, Barillot, and
  et~al.]{Berrah_2019_NatPhys.15.1279}
Berrah,~N.; Sanchez-Gonzalez,~A.; Jurek,~Z.; Obaid,~R.; Xiong,~H.;
  Squibb,~R.~J.; Osipov,~T.; Lutman,~A.; Fang,~L.; Barillot,~T.; et~al.
  {Femtosecond-Resolved Observation of the Fragmentation of
  Buckminsterfullerene Following X-ray Multiphoton Ionization}. \emph{Nat.
  Phys.} \textbf{2019}, \emph{15}, 1279--1283\relax
\mciteBstWouldAddEndPuncttrue
\mciteSetBstMidEndSepPunct{\mcitedefaultmidpunct}
{\mcitedefaultendpunct}{\mcitedefaultseppunct}\relax
\EndOfBibitem
\bibitem[Wabnitz \latin{et~al.}(2002)Wabnitz, Bittner, de~Castro, D\"{o}hrmann,
  G\"{u}rtler, Laarmann, Laasch, Schulz, Swiderski, von Haeften, and
  et~al.]{Wabnitz_2002_Nature.420.482}
Wabnitz,~H.; Bittner,~L.; de~Castro,~A. R.~B.; D\"{o}hrmann,~R.;
  G\"{u}rtler,~P.; Laarmann,~T.; Laasch,~W.; Schulz,~J.; Swiderski,~A.; von
  Haeften,~K.; et~al. {Multiple Ionization of Atom Clusters by Intense Soft
  X-rays From a Free-Electron Laser}. \emph{Nature} \textbf{2002}, \emph{420},
  482--485\relax
\mciteBstWouldAddEndPuncttrue
\mciteSetBstMidEndSepPunct{\mcitedefaultmidpunct}
{\mcitedefaultendpunct}{\mcitedefaultseppunct}\relax
\EndOfBibitem
\bibitem[Baccarelli \latin{et~al.}(2011)Baccarelli, Bald, Gianturco,
  Illenberger, and Kopyra]{baccarelli11}
Baccarelli,~I.; Bald,~I.; Gianturco,~F.~A.; Illenberger,~E.; Kopyra,~J.
  {Electron-Induced Damage of DNA and Its Components: Experiments and
  Theoretical Models}. \emph{Phys. Rep.} \textbf{2011}, \emph{508}, 1--44\relax
\mciteBstWouldAddEndPuncttrue
\mciteSetBstMidEndSepPunct{\mcitedefaultmidpunct}
{\mcitedefaultendpunct}{\mcitedefaultseppunct}\relax
\EndOfBibitem
\bibitem[Fabrikant \latin{et~al.}(2017)Fabrikant, Eden, Mason, and
  Fedor]{fabrikant17}
Fabrikant,~I.~I.; Eden,~S.; Mason,~N.~J.; Fedor,~J. In \emph{Advances In
  Atomic, Molecular, and Optical Physics, Vol.~66}; Arimondo,~E., Lin,~C.~C.,
  Yelin,~S.~F., Eds.; Academic Press, Cambridge, MA, 2017; pp 545--657\relax
\mciteBstWouldAddEndPuncttrue
\mciteSetBstMidEndSepPunct{\mcitedefaultmidpunct}
{\mcitedefaultendpunct}{\mcitedefaultseppunct}\relax
\EndOfBibitem
\bibitem[Ko\v{c}i\v{s}ek \latin{et~al.}(2016)Ko\v{c}i\v{s}ek, Pysanenko,
  F{\'a}rn{\'{\i}}k, and Fedor]{kocisek16_uracil}
Ko\v{c}i\v{s}ek,~J.; Pysanenko,~A.; F{\'a}rn{\'{\i}}k,~M.; Fedor,~J.
  {Microhydration Prevents Fragmentation of Uracil and Thymine by Low-Energy
  Electrons}. \emph{J. Phys. Chem. Lett.} \textbf{2016}, \emph{7},
  3401--3405\relax
\mciteBstWouldAddEndPuncttrue
\mciteSetBstMidEndSepPunct{\mcitedefaultmidpunct}
{\mcitedefaultendpunct}{\mcitedefaultseppunct}\relax
\EndOfBibitem
\bibitem[Allan(2007)]{allan07_formic}
Allan,~M. {Electron Collisions with Formic Acid Monomer and Dimer}. \emph{Phys.
  Rev. Lett.} \textbf{2007}, \emph{98}, 123201\relax
\mciteBstWouldAddEndPuncttrue
\mciteSetBstMidEndSepPunct{\mcitedefaultmidpunct}
{\mcitedefaultendpunct}{\mcitedefaultseppunct}\relax
\EndOfBibitem
\bibitem[Ko\v{c}i\v{s}ek \latin{et~al.}(2018)Ko\v{c}i\v{s}ek, Sedmidubsk\'{a},
  Indrajith, F\'{a}rn\'{i}k, and Fedor]{kocisek18}
Ko\v{c}i\v{s}ek,~J.; Sedmidubsk\'{a},~B.; Indrajith,~S.; F\'{a}rn\'{i}k,~M.;
  Fedor,~J. {Electron Attachment to Microhydrated Deoxycytidine Monophosphate}.
  \emph{J. Phys. Chem. B} \textbf{2018}, \emph{122}, 5212--5217\relax
\mciteBstWouldAddEndPuncttrue
\mciteSetBstMidEndSepPunct{\mcitedefaultmidpunct}
{\mcitedefaultendpunct}{\mcitedefaultseppunct}\relax
\EndOfBibitem
\bibitem[Postler \latin{et~al.}(2015)Postler, Renzler, Kaiser, Huber, Probst,
  Scheier, and Ellis]{postler15}
Postler,~J.; Renzler,~M.; Kaiser,~A.; Huber,~S.~E.; Probst,~M.; Scheier,~P.;
  Ellis,~A.~M. {Electron-Induced Chemistry of Cobalt Tricarbonyl Nitrosyl
  (Co(CO)$_3$NO) in Liquid Helium Nanodroplets}. \emph{J. Phys. Chem. C}
  \textbf{2015}, \emph{119}, 20917--20922\relax
\mciteBstWouldAddEndPuncttrue
\mciteSetBstMidEndSepPunct{\mcitedefaultmidpunct}
{\mcitedefaultendpunct}{\mcitedefaultseppunct}\relax
\EndOfBibitem
\bibitem[Lengyel \latin{et~al.}(2017)Lengyel, Papp, Matej\v{c}\'{i}k,
  Ko\v{c}i\v{s}ek, F{\'a}rn{\'{\i}}k, and Fedor]{lengyel17}
Lengyel,~J.; Papp,~P.; Matej\v{c}\'{i}k,~{\v{S}}.; Ko\v{c}i\v{s}ek,~J.;
  F{\'a}rn{\'{\i}}k,~M.; Fedor,~J. {Suppression of Low-Energy Dissociative
  Electron Attachment in Fe(CO)$_5$ upon Clustering}. \emph{Beilstein J.
  Nanotechnol.} \textbf{2017}, \emph{8}, 2200--2207\relax
\mciteBstWouldAddEndPuncttrue
\mciteSetBstMidEndSepPunct{\mcitedefaultmidpunct}
{\mcitedefaultendpunct}{\mcitedefaultseppunct}\relax
\EndOfBibitem
\bibitem[Lengyel \latin{et~al.}(2016)Lengyel, Ko\v{c}i\v{s}ek,
  F{\'a}rn{\'{\i}}k, and Fedor]{lengyel16_scaveng}
Lengyel,~J.; Ko\v{c}i\v{s}ek,~J.; F{\'a}rn{\'{\i}}k,~M.; Fedor,~J.
  {Self-Scavenging of Electrons in Fe(CO)$_5$ Aggregates Deposited on Argon
  Nanoparticles}. \emph{J. Phys. Chem. C} \textbf{2016}, \emph{120},
  7397--7402\relax
\mciteBstWouldAddEndPuncttrue
\mciteSetBstMidEndSepPunct{\mcitedefaultmidpunct}
{\mcitedefaultendpunct}{\mcitedefaultseppunct}\relax
\EndOfBibitem
\bibitem[Landheer \latin{et~al.}(2011)Landheer, Rosenberg, Bernau, Swiderek,
  Utke, Hagen, and Fairbrother]{landheer11}
Landheer,~K.; Rosenberg,~S.~G.; Bernau,~L.; Swiderek,~P.; Utke,~I.;
  Hagen,~C.~W.; Fairbrother,~D.~H. {Low-Energy Electron-Induced Decomposition
  and Reactions of Adsorbed Tetrakis(trifluorophosphine)platinum
  [Pt(PF$_3$)$_4$]}. \emph{J. Phys. Chem. C} \textbf{2011}, \emph{115},
  17452--17463\relax
\mciteBstWouldAddEndPuncttrue
\mciteSetBstMidEndSepPunct{\mcitedefaultmidpunct}
{\mcitedefaultendpunct}{\mcitedefaultseppunct}\relax
\EndOfBibitem
\bibitem[F{\'a}rn{\'{\i}}k \latin{et~al.}(2021)F{\'a}rn{\'{\i}}k, Fedor,
  Ko\v{c}i\v{s}ek, Lengyel, Pluha\v{r}ov\'{a}, Poterya, and
  Pysanenko]{farnik21_persp}
F{\'a}rn{\'{\i}}k,~M.; Fedor,~J.; Ko\v{c}i\v{s}ek,~J.; Lengyel,~J.;
  Pluha\v{r}ov\'{a},~E.; Poterya,~V.; Pysanenko,~A. {Pickup and Reactions of
  Molecules on Clusters Relevant for Atmospheric and Interstellar Processes}.
  \emph{Phys. Chem. Chem. Phys.} \textbf{2021}, \emph{23}, 3195--3213\relax
\mciteBstWouldAddEndPuncttrue
\mciteSetBstMidEndSepPunct{\mcitedefaultmidpunct}
{\mcitedefaultendpunct}{\mcitedefaultseppunct}\relax
\EndOfBibitem
\bibitem[B\"{o}hler \latin{et~al.}(2013)B\"{o}hler, Warneke, and
  Swiderek]{bohler13}
B\"{o}hler,~E.; Warneke,~J.; Swiderek,~P. {Control of Chemical Reactions and
  Synthesis by Low-Energy Electrons}. \emph{Chem. Soc. Rev.} \textbf{2013},
  \emph{42}, 9219--9231\relax
\mciteBstWouldAddEndPuncttrue
\mciteSetBstMidEndSepPunct{\mcitedefaultmidpunct}
{\mcitedefaultendpunct}{\mcitedefaultseppunct}\relax
\EndOfBibitem
\bibitem[Arumainayagam \latin{et~al.}(2019)Arumainayagam, Garod, Boyer, Bao,
  Campbell, Wang, Nowak, Arumainayagam, and
  Hodge]{Arumainayagam_2019_ChemSocRev}
Arumainayagam,~C.~R.; Garod,~R.~T.; Boyer,~A.~K.,~M. C. an~Hay; Bao,~S.~T.;
  Campbell,~J.~S.; Wang,~J.; Nowak,~C.~M.; Arumainayagam,~M.~R.; Hodge,~P.~J.
  {Extraterrestrial Prebiotic Molecules: Photochemistry vs. Radiation Chemistry
  of Interstellar Ices}. \emph{Chem. Soc. Rev.} \textbf{2019}, \emph{48},
  2293--2314\relax
\mciteBstWouldAddEndPuncttrue
\mciteSetBstMidEndSepPunct{\mcitedefaultmidpunct}
{\mcitedefaultendpunct}{\mcitedefaultseppunct}\relax
\EndOfBibitem
\bibitem[Smyth \latin{et~al.}(2014)Smyth, Kohanoff, and Fabrikant]{smyth14}
Smyth,~M.; Kohanoff,~J.; Fabrikant,~I.~I. {Electron-Induced Hydrogen Loss in
  Uracil in a Water Cluster Environment}. \emph{J. Chem. Phys.} \textbf{2014},
  \emph{140}, 184313\relax
\mciteBstWouldAddEndPuncttrue
\mciteSetBstMidEndSepPunct{\mcitedefaultmidpunct}
{\mcitedefaultendpunct}{\mcitedefaultseppunct}\relax
\EndOfBibitem
\bibitem[Pysanenko \latin{et~al.}(2015)Pysanenko, Habartov\'{a},
  Svr\v{c}kov\'{a}, Lengyel, Poterya, Roeselov\'{a}, Fedor, and
  F{\'a}rn{\'{\i}}k]{pysanenko15}
Pysanenko,~A.; Habartov\'{a},~A.; Svr\v{c}kov\'{a},~P.; Lengyel,~J.;
  Poterya,~V.; Roeselov\'{a},~M.; Fedor,~J.; F{\'a}rn{\'{\i}}k,~M. {Lack of
  Aggregation of Molecules on Ice Nanoparticles}. \emph{J. Phys. Chem. A}
  \textbf{2015}, \emph{119}, 8991--8999\relax
\mciteBstWouldAddEndPuncttrue
\mciteSetBstMidEndSepPunct{\mcitedefaultmidpunct}
{\mcitedefaultendpunct}{\mcitedefaultseppunct}\relax
\EndOfBibitem
\bibitem[Chesnavich and Bowers(1977)Chesnavich, and
  Bowers]{Chesnavich_1977_JACS.99.1705}
Chesnavich,~W.~J.; Bowers,~M.~T. {Statistical Phase Space Theory of Polyatomic
  Systems. Application to the Unimolecular Reactions C$_6$H$_5$CN$\cdot^+
  \rightarrow$ C$_6$H$_4\cdot^+ +$ HCN and C$_4$H$_6\cdot^+ \rightarrow$
  C$_3$H$_3^+ + \cdot$CH$_3$}. \emph{J. Am. Chem. Soc.} \textbf{1977},
  \emph{99}, 1705--1711\relax
\mciteBstWouldAddEndPuncttrue
\mciteSetBstMidEndSepPunct{\mcitedefaultmidpunct}
{\mcitedefaultendpunct}{\mcitedefaultseppunct}\relax
\EndOfBibitem
\bibitem[Kat\={o}(1996)]{Kato_1996_JCP.105.9502}
Kat\={o},~T. {Phase Space Bottlenecks and Rates of No-Barrier Fragmentation
  Reactions into Polyatomic Molecules}. \emph{J. Chem. Phys.} \textbf{1996},
  \emph{105}, 9502--9508\relax
\mciteBstWouldAddEndPuncttrue
\mciteSetBstMidEndSepPunct{\mcitedefaultmidpunct}
{\mcitedefaultendpunct}{\mcitedefaultseppunct}\relax
\EndOfBibitem
\bibitem[Garrett and Truhlar(1979)Garrett, and
  Truhlar]{Garrett_1979_JPC.83.1052}
Garrett,~B.~C.; Truhlar,~D.~G. {Generalized Transition State Theory. Classical
  Mechanical Theory and Applications to Collinear Reactions of Hydrogen
  Molecules}. \emph{J. Phys. Chem.} \textbf{1979}, \emph{83}, 1052--1079\relax
\mciteBstWouldAddEndPuncttrue
\mciteSetBstMidEndSepPunct{\mcitedefaultmidpunct}
{\mcitedefaultendpunct}{\mcitedefaultseppunct}\relax
\EndOfBibitem
\bibitem[Ptasinska \latin{et~al.}(2005)Ptasinska, Denifl, Scheier, Illenberger,
  and Mark]{ptasinska05}
Ptasinska,~S.; Denifl,~S.; Scheier,~P.; Illenberger,~E.; Mark,~T.~D. {Bond- and
  Site-Selective Loss of H Atoms from Nucleobases by Very-Low-Energy Electrons
  ($<3$~eV)}. \emph{Angew. Chem. Int. Ed.} \textbf{2005}, \emph{44},
  6941--6943\relax
\mciteBstWouldAddEndPuncttrue
\mciteSetBstMidEndSepPunct{\mcitedefaultmidpunct}
{\mcitedefaultendpunct}{\mcitedefaultseppunct}\relax
\EndOfBibitem
\bibitem[McAllister \latin{et~al.}(2019)McAllister, Kazemigazestane, Henry, Gu,
  Fabrikant, Tribello, and Kohanoff]{mcallister19}
McAllister,~M.; Kazemigazestane,~N.; Henry,~L.~T.; Gu,~B.; Fabrikant,~I.;
  Tribello,~G.~A.; Kohanoff,~J. {Solvation Effects on Dissociative Electron
  Attachment to Thymine}. \emph{J. Phys. Chem. B} \textbf{2019}, \emph{123},
  1537--1544\relax
\mciteBstWouldAddEndPuncttrue
\mciteSetBstMidEndSepPunct{\mcitedefaultmidpunct}
{\mcitedefaultendpunct}{\mcitedefaultseppunct}\relax
\EndOfBibitem
\bibitem[Suchan \latin{et~al.}(2022)Suchan, Kolafa, and
  Slav\'i\v{c}ek]{suchan22}
Suchan,~J.; Kolafa,~J.; Slav\'i\v{c}ek,~P. {Electron-Induced Fragmentation of
  Water Droplets: Simulation Study}. \emph{J. Chem. Phys.} \textbf{2022},
  \emph{156}, 144303\relax
\mciteBstWouldAddEndPuncttrue
\mciteSetBstMidEndSepPunct{\mcitedefaultmidpunct}
{\mcitedefaultendpunct}{\mcitedefaultseppunct}\relax
\EndOfBibitem
\bibitem[Thorman \latin{et~al.}(2015)Thorman, P., Fairbrother, and
  Ing\'{o}lfsson]{thorman15}
Thorman,~R.~M.; P.,~R. K.~T.; Fairbrother,~D.~H.; Ing\'{o}lfsson,~O. {The Role
  of Low-Energy Electrons in Focused Electron Beam Induced Deposition: Four
  Case Studies of Representative Precursors}. \emph{Beilstein J. Nanotechnol.}
  \textbf{2015}, \emph{6}, 1904--1926\relax
\mciteBstWouldAddEndPuncttrue
\mciteSetBstMidEndSepPunct{\mcitedefaultmidpunct}
{\mcitedefaultendpunct}{\mcitedefaultseppunct}\relax
\EndOfBibitem
\bibitem[Johny \latin{et~al.}(2019)Johny, Onvlee, Kierspel, Bieker, Trippel,
  and K\"{u}pper]{KupperSelection}
Johny,~M.; Onvlee,~J.; Kierspel,~T.; Bieker,~H.; Trippel,~S.; K\"{u}pper,~J.
  {Spatial Separation of Pyrrole and Pyrrole-Water Clusters}. \emph{Chem. Phys.
  Lett.} \textbf{2019}, \emph{721}, 149--152\relax
\mciteBstWouldAddEndPuncttrue
\mciteSetBstMidEndSepPunct{\mcitedefaultmidpunct}
{\mcitedefaultendpunct}{\mcitedefaultseppunct}\relax
\EndOfBibitem
\bibitem[Mauracher \latin{et~al.}(2018)Mauracher, Echt, Ellis, Yang, Bohme,
  Postler, Kaiser, Denifl, and Scheier]{mauracher18}
Mauracher,~A.; Echt,~O.; Ellis,~A.~M.; Yang,~S.; Bohme,~D.~K.; Postler,~J.;
  Kaiser,~A.; Denifl,~S.; Scheier,~P. {Cold Physics and Chemistry: Collisions,
  Ionization and Reactions Inside Helium Nanodroplets Close to Zero K}.
  \emph{Phys. Rep.} \textbf{2018}, \emph{751}, 1--90\relax
\mciteBstWouldAddEndPuncttrue
\mciteSetBstMidEndSepPunct{\mcitedefaultmidpunct}
{\mcitedefaultendpunct}{\mcitedefaultseppunct}\relax
\EndOfBibitem
\bibitem[Dvo\v{r}\'{a}k \latin{et~al.}(2022)Dvo\v{r}\'{a}k, Rankovi\v{c},
  Houfek, Nag, \v{C}ur\'{i}k, Fedor, and \v{C}\'{i}\v{z}ek]{dvorak22_co2}
Dvo\v{r}\'{a}k,~J.; Rankovi\v{c},~M.; Houfek,~K.; Nag,~P.; \v{C}ur\'{i}k,~R.;
  Fedor,~J.; \v{C}\'{i}\v{z}ek,~M. {Vibronic Coupling through the Continuum in
  the $e + $CO$_2$ System}. \emph{Phys. Rev. Lett.} \textbf{2022}, \emph{129},
  013401\relax
\mciteBstWouldAddEndPuncttrue
\mciteSetBstMidEndSepPunct{\mcitedefaultmidpunct}
{\mcitedefaultendpunct}{\mcitedefaultseppunct}\relax
\EndOfBibitem
\bibitem[{Ragesh Kumar} \latin{et~al.}(2022){Ragesh Kumar}, Nag, Rankovi\v{c},
  Ko\v{c}i\v{s}ek, Ma\v{s}\'{i}n, and Fedor]{kumar_pyrole22}
{Ragesh Kumar},~T.~P.; Nag,~P.; Rankovi\v{c},~M.; Ko\v{c}i\v{s}ek,~J.;
  Ma\v{s}\'{i}n,~Z.; Fedor,~J. {Distant Symmetry Control in Electron-Induced
  Bond Cleavage}. \emph{J. Phys. Chem. Lett.} \textbf{2022}, \emph{13},
  11136--11142\relax
\mciteBstWouldAddEndPuncttrue
\mciteSetBstMidEndSepPunct{\mcitedefaultmidpunct}
{\mcitedefaultendpunct}{\mcitedefaultseppunct}\relax
\EndOfBibitem
\bibitem[Po\v{s}tulka \latin{et~al.}(2017)Po\v{s}tulka, Slav\'i\v{c}ek, Fedor,
  F\'{a}rn\'{i}k, and Ko\v{c}i\v{s}ek]{postulka17}
Po\v{s}tulka,~J.; Slav\'i\v{c}ek,~P.; Fedor,~J.; F\'{a}rn\'{i}k,~M.;
  Ko\v{c}i\v{s}ek,~J. {Energy Transfer in Microhydrated Uracil, 5-Fluorouracil,
  and 5-Bromouracil}. \emph{J. Phys. Chem. B} \textbf{2017}, \emph{121},
  8965--8974\relax
\mciteBstWouldAddEndPuncttrue
\mciteSetBstMidEndSepPunct{\mcitedefaultmidpunct}
{\mcitedefaultendpunct}{\mcitedefaultseppunct}\relax
\EndOfBibitem
\bibitem[Lin \latin{et~al.}(2021)Lin, Gao, Yang, Wu, Zhang, Feng, Dai, and
  Du]{Binwei_2021_FrontOncol.11.644400}
Lin,~B.; Gao,~F.; Yang,~Y.; Wu,~D.; Zhang,~Y.; Feng,~G.; Dai,~T.; Du,~X. {FLASH
  Radiotherapy: History and Future}. \emph{Front. Oncol.} \textbf{2021},
  \emph{11}, 644400\relax
\mciteBstWouldAddEndPuncttrue
\mciteSetBstMidEndSepPunct{\mcitedefaultmidpunct}
{\mcitedefaultendpunct}{\mcitedefaultseppunct}\relax
\EndOfBibitem
\bibitem[Zacheis \latin{et~al.}(1999)Zacheis, Gray, and
  Kamat]{Zacheis_1999_JPCB.103.2142}
Zacheis,~G.~A.; Gray,~K.~A.; Kamat,~P.~V. {Radiation-Induced Catalysis on Oxide
  Surfaces: Degradation of Hexachlorobenzene on $\gamma$-Irradiated Alumina
  Nanoparticles}. \emph{J. Phys. Chem. B} \textbf{1999}, \emph{103},
  2142--2150\relax
\mciteBstWouldAddEndPuncttrue
\mciteSetBstMidEndSepPunct{\mcitedefaultmidpunct}
{\mcitedefaultendpunct}{\mcitedefaultseppunct}\relax
\EndOfBibitem
\bibitem[Coekelbergs \latin{et~al.}(1962)Coekelbergs, Crucq, and
  Frennet]{Coekelbergs_1962_AdvCatal.13.55}
Coekelbergs,~R.; Crucq,~A.; Frennet,~A. {Radiation Catalysis}. \emph{Adv.
  Catal.} \textbf{1962}, \emph{13}, 55--136\relax
\mciteBstWouldAddEndPuncttrue
\mciteSetBstMidEndSepPunct{\mcitedefaultmidpunct}
{\mcitedefaultendpunct}{\mcitedefaultseppunct}\relax
\EndOfBibitem
\bibitem[Abedini \latin{et~al.}(2013)Abedini, Daud, Hamid, Othman, and
  Saion]{Abedini_2013_NanoscaleResLett.8.474}
Abedini,~A.; Daud,~A.~R.; Hamid,~M. A.~A.; Othman,~N.~K.; Saion,~E. {A Review
  on Radiation-Induced Nucleation and Growth of Colloidal Metallic
  Nanoparticles}. \emph{Nanoscale Res. Lett.} \textbf{2013}, \emph{8},
  474\relax
\mciteBstWouldAddEndPuncttrue
\mciteSetBstMidEndSepPunct{\mcitedefaultmidpunct}
{\mcitedefaultendpunct}{\mcitedefaultseppunct}\relax
\EndOfBibitem
\bibitem[Roy and Lahiri(2013)Roy, and Lahiri]{Roy_2013_AnalChem.80.7504}
Roy,~K.; Lahiri,~S. {In Situ $\gamma$-Radiation: One-Step Environmentally
  Benign Method to Produce Gold--Palladium Bimetallic Nanoparticles}.
  \emph{Anal. Chem.} \textbf{2013}, \emph{80}, 7504--7507\relax
\mciteBstWouldAddEndPuncttrue
\mciteSetBstMidEndSepPunct{\mcitedefaultmidpunct}
{\mcitedefaultendpunct}{\mcitedefaultseppunct}\relax
\EndOfBibitem
\bibitem[Zhang \latin{et~al.}(2023)Zhang, He, and
  Zhou]{Zhang_2023_AdvDrugDelivRev}
Zhang,~D.; He,~J.; Zhou,~M. {Radiation-Assisted Strategies Provide New
  Perspectives to Improve the Nanoparticle Delivery to Tumor}. \emph{Adv. Drug
  Deliv. Rev.} \textbf{2023}, \emph{193}, 114642\relax
\mciteBstWouldAddEndPuncttrue
\mciteSetBstMidEndSepPunct{\mcitedefaultmidpunct}
{\mcitedefaultendpunct}{\mcitedefaultseppunct}\relax
\EndOfBibitem
\bibitem[Gauduel \latin{et~al.}(2010)Gauduel, Glinec, Rousseau, Burgy, and
  Malka]{Gauduel_2010_EPJD.60.121}
Gauduel,~Y.~A.; Glinec,~Y.; Rousseau,~J.-P.; Burgy,~F.; Malka,~V. {High Energy
  Radiation Femtochemistry of Water Molecules: Early Electron-Radical Pairs
  Processes}. \emph{Eur. Phys. J. D} \textbf{2010}, \emph{60}, 121--135\relax
\mciteBstWouldAddEndPuncttrue
\mciteSetBstMidEndSepPunct{\mcitedefaultmidpunct}
{\mcitedefaultendpunct}{\mcitedefaultseppunct}\relax
\EndOfBibitem
\bibitem[Svoboda \latin{et~al.}(2020)Svoboda, Michiels, LaForge, Med,
  Stienkemeier, Slav\'{i}\v{c}ek, and W\"{o}rner]{Svoboda_2020_SciAdv.6.3}
Svoboda,~V.; Michiels,~R.; LaForge,~A.~C.; Med,~J.; Stienkemeier,~F.;
  Slav\'{i}\v{c}ek,~P.; W\"{o}rner,~H.~J. {Real-Time Observation of Water
  Radiolysis and Hydrated Electron Formation Induced by Extreme-Ultraviolet
  Pulses}. \emph{Sci. Adv.} \textbf{2020}, \emph{6}, eaaz0385\relax
\mciteBstWouldAddEndPuncttrue
\mciteSetBstMidEndSepPunct{\mcitedefaultmidpunct}
{\mcitedefaultendpunct}{\mcitedefaultseppunct}\relax
\EndOfBibitem
\bibitem[Baldacchino \latin{et~al.}(2004)Baldacchino, Vigneron, Renault, Pin,
  Abedinzadeh, Deycard, Balanzat, Bouffard, Gard\`{e}s-Albert, Hickel, and
  Mialocq]{Baldacchino_2004_CPL.385.66}
Baldacchino,~G.; Vigneron,~G.; Renault,~J.-P.; Pin,~S.; Abedinzadeh,~Z.;
  Deycard,~S.; Balanzat,~E.; Bouffard,~S.; Gard\`{e}s-Albert,~M.; Hickel,~B.;
  Mialocq,~J.-C. {A Nanosecond Pulse Radiolysis Study of the Hydrated Electron
  with High Energy Ions with a Narrow Velocity Distribution}. \emph{Chem. Phys.
  Lett.} \textbf{2004}, \emph{385}, 66--71\relax
\mciteBstWouldAddEndPuncttrue
\mciteSetBstMidEndSepPunct{\mcitedefaultmidpunct}
{\mcitedefaultendpunct}{\mcitedefaultseppunct}\relax
\EndOfBibitem
\bibitem[Dromey \latin{et~al.}(2016)Dromey, Coughlan, Senje, Taylor, Kuschel,
  Villagomez-Bernabe, Stefanuik, Nersisyan, Stella, Kohanoff, and
  et~al.]{Dromey_2016_NatCommun.7.10642}
Dromey,~B.; Coughlan,~M.; Senje,~L.; Taylor,~M.; Kuschel,~S.;
  Villagomez-Bernabe,~B.; Stefanuik,~R.; Nersisyan,~G.; Stella,~L.;
  Kohanoff,~J.; et~al. {Picosecond Metrology of Laser-Driven Proton Bursts}.
  \emph{Nat. Commun.} \textbf{2016}, \emph{7}, 10642\relax
\mciteBstWouldAddEndPuncttrue
\mciteSetBstMidEndSepPunct{\mcitedefaultmidpunct}
{\mcitedefaultendpunct}{\mcitedefaultseppunct}\relax
\EndOfBibitem
\bibitem[Prasselsperger \latin{et~al.}(2021)Prasselsperger, Coughlan, Breslin,
  Yeung, Arthur, Donnelly, White, Afshari, Speicher, Yang, and
  et~al.]{Prasselsberger_2021_PRL.127.186001}
Prasselsperger,~A.; Coughlan,~M.; Breslin,~N.; Yeung,~M.; Arthur,~C.;
  Donnelly,~H.; White,~S.; Afshari,~M.; Speicher,~M.; Yang,~R.; et~al.
  {Real-Time Electron Solvation Induced by Bursts of Laser-Accelerated Protons
  in Liquid Water}. \emph{Phys. Rev. Lett.} \textbf{2021}, \emph{127},
  186001\relax
\mciteBstWouldAddEndPuncttrue
\mciteSetBstMidEndSepPunct{\mcitedefaultmidpunct}
{\mcitedefaultendpunct}{\mcitedefaultseppunct}\relax
\EndOfBibitem
\bibitem[Coughlan \latin{et~al.}(2020)Coughlan, Donnelly, Breslin, Arthur,
  Nersisyan, Yeung, Villagomez-Bernabe, Afshari, Currell, Zepf, and
  Dromey]{Coughlan_2020_NJP.22.103023}
Coughlan,~M.; Donnelly,~H.; Breslin,~N.; Arthur,~C.; Nersisyan,~G.; Yeung,~M.;
  Villagomez-Bernabe,~B.; Afshari,~M.; Currell,~F.; Zepf,~M.; Dromey,~B.
  {Ultrafast Dynamics and Evolution of Ion-Induced Opacity in Transparent
  Dielectrics}. \emph{New J. Phys.} \textbf{2020}, \emph{22}, 103023\relax
\mciteBstWouldAddEndPuncttrue
\mciteSetBstMidEndSepPunct{\mcitedefaultmidpunct}
{\mcitedefaultendpunct}{\mcitedefaultseppunct}\relax
\EndOfBibitem
\bibitem[Ziegler(1999)]{Ziegler_1999_JAP.85.1249}
Ziegler,~J.~F. {Stopping of Energetic Light Ions in Elemental Matter}. \emph{J.
  Appl. Phys.} \textbf{1999}, \emph{85}, 1249--1272\relax
\mciteBstWouldAddEndPuncttrue
\mciteSetBstMidEndSepPunct{\mcitedefaultmidpunct}
{\mcitedefaultendpunct}{\mcitedefaultseppunct}\relax
\EndOfBibitem
\bibitem[Macchi \latin{et~al.}(2013)Macchi, Borghesi, and
  Passoni]{Macchi_2013_RMP.85.751}
Macchi,~A.; Borghesi,~M.; Passoni,~M. {Ion Acceleration by Superintense
  Laser-Plasma Interaction}. \emph{Rev. Mod. Phys.} \textbf{2013}, \emph{85},
  751--794\relax
\mciteBstWouldAddEndPuncttrue
\mciteSetBstMidEndSepPunct{\mcitedefaultmidpunct}
{\mcitedefaultendpunct}{\mcitedefaultseppunct}\relax
\EndOfBibitem
\bibitem[Kar \latin{et~al.}(2016)Kar, Ahmed, Prasad, Cerchez, Brauckmann,
  Aurand, Cantono, Hadjisolomou, Lewis, Macchi, and
  et~al.]{Kar_2016_NatCommun.7.10792}
Kar,~S.; Ahmed,~H.; Prasad,~R.; Cerchez,~M.; Brauckmann,~S.; Aurand,~B.;
  Cantono,~G.; Hadjisolomou,~P.; Lewis,~C. L.~S.; Macchi,~A.; et~al. {Guided
  Post-Acceleration of Laser-Driven Ions by a Miniature Modular Structure}.
  \emph{Nat. Commun.} \textbf{2016}, \emph{7}, 10792\relax
\mciteBstWouldAddEndPuncttrue
\mciteSetBstMidEndSepPunct{\mcitedefaultmidpunct}
{\mcitedefaultendpunct}{\mcitedefaultseppunct}\relax
\EndOfBibitem
\bibitem[Dromey \latin{et~al.}(2006)Dromey, Zepf, Gopal, Lancaster, Wei,
  Krushelnick, Tatarakis, Vakakis, Moustaizis, Kodama, and
  et~al.]{Dromey_2006_NatPhys.2.456}
Dromey,~B.; Zepf,~M.; Gopal,~A.; Lancaster,~K.; Wei,~M.~S.; Krushelnick,~K.;
  Tatarakis,~M.; Vakakis,~N.; Moustaizis,~S.; Kodama,~R.; et~al. {High Harmonic
  Generation in the Relativistic Limit}. \emph{Nat. Phys.} \textbf{2006},
  \emph{2}, 456--459\relax
\mciteBstWouldAddEndPuncttrue
\mciteSetBstMidEndSepPunct{\mcitedefaultmidpunct}
{\mcitedefaultendpunct}{\mcitedefaultseppunct}\relax
\EndOfBibitem
\bibitem[Ackland(2010)]{Ackland_2010_Science.327.1587}
Ackland,~G. {Controlling Radiation Damage}. \emph{Science} \textbf{2010},
  \emph{327}, 1587--1588\relax
\mciteBstWouldAddEndPuncttrue
\mciteSetBstMidEndSepPunct{\mcitedefaultmidpunct}
{\mcitedefaultendpunct}{\mcitedefaultseppunct}\relax
\EndOfBibitem
\bibitem[Amaldi and Kraft(2005)Amaldi, and
  Kraft]{Amaldi_2005_RepProgPhys.68.1861}
Amaldi,~U.; Kraft,~G. {Radiotherapy with Beams of Carbon Ions}. \emph{Rep.
  Prog. Phys.} \textbf{2005}, \emph{68}, 1861--1882\relax
\mciteBstWouldAddEndPuncttrue
\mciteSetBstMidEndSepPunct{\mcitedefaultmidpunct}
{\mcitedefaultendpunct}{\mcitedefaultseppunct}\relax
\EndOfBibitem
\bibitem[Schardt \latin{et~al.}(2010)Schardt, Els\"{a}sser, and
  Schulz-Ertner]{Schardt_2010_RMP.82.383}
Schardt,~D.; Els\"{a}sser,~T.; Schulz-Ertner,~D. {Heavy-Ion Tumor Therapy:
  Physical and Radiobiological Benefits}. \emph{Rev. Mod. Phys.} \textbf{2010},
  \emph{82}, 383--425\relax
\mciteBstWouldAddEndPuncttrue
\mciteSetBstMidEndSepPunct{\mcitedefaultmidpunct}
{\mcitedefaultendpunct}{\mcitedefaultseppunct}\relax
\EndOfBibitem
\bibitem[Durante and Cucinotta(2011)Durante, and
  Cucinotta]{Durante_2011_RMP.83.1245}
Durante,~M.; Cucinotta,~F.~A. {Physical Basis of Radiation Protection in Space
  Travel}. \emph{Rev. Mod. Phys.} \textbf{2011}, \emph{83}, 1245--1281\relax
\mciteBstWouldAddEndPuncttrue
\mciteSetBstMidEndSepPunct{\mcitedefaultmidpunct}
{\mcitedefaultendpunct}{\mcitedefaultseppunct}\relax
\EndOfBibitem
\bibitem[Kronenberg and Cucinotta(2012)Kronenberg, and
  Cucinotta]{Kronenberg_2012_HealthPhys.103.556}
Kronenberg,~A.; Cucinotta,~F.~A. {Space Radiation Protection Issues}.
  \emph{Health Phys.} \textbf{2012}, \emph{103}, 556--567\relax
\mciteBstWouldAddEndPuncttrue
\mciteSetBstMidEndSepPunct{\mcitedefaultmidpunct}
{\mcitedefaultendpunct}{\mcitedefaultseppunct}\relax
\EndOfBibitem
\bibitem[de~Vera \latin{et~al.}(2019)de~Vera, Surdutovich, and
  Solov'yov]{deVera_2019_CancerNano.10.5}
de~Vera,~P.; Surdutovich,~E.; Solov'yov,~A.~V. {The Role of Shock Waves on the
  Biodamage Induced by Ion Beam Radiation}. \emph{Cancer Nanotechnol.}
  \textbf{2019}, \emph{10}, 5\relax
\mciteBstWouldAddEndPuncttrue
\mciteSetBstMidEndSepPunct{\mcitedefaultmidpunct}
{\mcitedefaultendpunct}{\mcitedefaultseppunct}\relax
\EndOfBibitem
\bibitem[Bottl{\"a}nder \latin{et~al.}(2015)Bottl{\"a}nder, M{\"u}cksch, and
  Urbassek]{bottlander2015effect}
Bottl{\"a}nder,~D.; M{\"u}cksch,~C.; Urbassek,~H.~M. {Effect of Swift-Ion
  Irradiation on DNA Molecules: A Molecular Dynamics Study Using the REAX Force
  Field}. \emph{Nucl. Instrum. Meth. B} \textbf{2015}, \emph{365},
  622--625\relax
\mciteBstWouldAddEndPuncttrue
\mciteSetBstMidEndSepPunct{\mcitedefaultmidpunct}
{\mcitedefaultendpunct}{\mcitedefaultseppunct}\relax
\EndOfBibitem
\bibitem[Favaudon \latin{et~al.}(2014)Favaudon, Caplier, Monceau, Pouzoulet,
  Sayarath, Fouillade, Poupon, Brito, Hup\'{e}, Bourhis, and
  et~al.]{Favaudon_2014_SciTranslMed}
Favaudon,~V.; Caplier,~L.; Monceau,~V.; Pouzoulet,~F.; Sayarath,~M.;
  Fouillade,~C.; Poupon,~M.~F.; Brito,~I.; Hup\'{e},~P.; Bourhis,~J.; et~al.
  {Ultrahigh Dose-Rate FLASH Irradiation Increases the Differential Response
  between Normal and Tumor Tissue in Mice}. \emph{Sci. Transl. Med.}
  \textbf{2014}, \emph{6}, 245ra93\relax
\mciteBstWouldAddEndPuncttrue
\mciteSetBstMidEndSepPunct{\mcitedefaultmidpunct}
{\mcitedefaultendpunct}{\mcitedefaultseppunct}\relax
\EndOfBibitem
\bibitem[Malka \latin{et~al.}(2008)Malka, Faure, Gauduel, Lefebvre, Rousse, and
  Phuoc]{Malka_2008_NatPhys.4.447}
Malka,~V.; Faure,~J.; Gauduel,~Y.~A.; Lefebvre,~E.; Rousse,~A.; Phuoc,~K.~T.
  {Principles and Applications of Compact Laser-Plasma Accelerators}.
  \emph{Nat. Phys.} \textbf{2008}, \emph{4}, 447--453\relax
\mciteBstWouldAddEndPuncttrue
\mciteSetBstMidEndSepPunct{\mcitedefaultmidpunct}
{\mcitedefaultendpunct}{\mcitedefaultseppunct}\relax
\EndOfBibitem
\bibitem[Fuchs \latin{et~al.}(2009)Fuchs, Szymanowski, Oelfke, Glinec,
  Rechatin, Faure, and Malka]{Fuchs_2009_PMB.54.3315}
Fuchs,~T.; Szymanowski,~H.; Oelfke,~U.; Glinec,~Y.; Rechatin,~C.; Faure,~J.;
  Malka,~V. {Treatment Planning for Laser-Accelerated Very-High Energy
  Electrons}. \emph{Phys. Med. Biol.} \textbf{2009}, \emph{54},
  3315--3328\relax
\mciteBstWouldAddEndPuncttrue
\mciteSetBstMidEndSepPunct{\mcitedefaultmidpunct}
{\mcitedefaultendpunct}{\mcitedefaultseppunct}\relax
\EndOfBibitem
\bibitem[Malka \latin{et~al.}(2010)Malka, Faure, and
  Gauduel]{Malka_2010_MutatResRev.704.142}
Malka,~V.; Faure,~J.; Gauduel,~Y.~A. {Ultra-Short Electron Beams Based
  Spatio-Temporal Radiation Biology and Radiotherapy}. \emph{Mutat. Res. Rev.
  Mutat. Res.} \textbf{2010}, \emph{704}, 142--151\relax
\mciteBstWouldAddEndPuncttrue
\mciteSetBstMidEndSepPunct{\mcitedefaultmidpunct}
{\mcitedefaultendpunct}{\mcitedefaultseppunct}\relax
\EndOfBibitem
\bibitem[Ogawa(2016)]{Ogawa_2016_Cancers.8.28}
Ogawa,~Y. {Paradigm Shift in Radiation Biology/Radiation Oncology --
  Exploitation of the ``H$_2$O$_2$ Effect'' for Radiotherapy Using Low-LET
  (Linear Energy Transfer) Radiation Such as X-Rays and High-Energy Electrons}.
  \emph{Cancers} \textbf{2016}, \emph{8}, 28--40\relax
\mciteBstWouldAddEndPuncttrue
\mciteSetBstMidEndSepPunct{\mcitedefaultmidpunct}
{\mcitedefaultendpunct}{\mcitedefaultseppunct}\relax
\EndOfBibitem
\bibitem[Paganetti \latin{et~al.}(2002)Paganetti, Niemierko, Ancukiewicz,
  Gerweck, Goitein, Loeffler, and Suit]{Paganetti_2002_IJROBP.53.407}
Paganetti,~H.; Niemierko,~A.; Ancukiewicz,~M.; Gerweck,~L.~E.; Goitein,~M.;
  Loeffler,~J.~S.; Suit,~H.~D. {Relative Biological Effectiveness (RBE) Values
  for Proton Beam Therapy}. \emph{Int. J. Radiat. Oncol. Biol. Phys.}
  \textbf{2002}, \emph{53}, 407--421\relax
\mciteBstWouldAddEndPuncttrue
\mciteSetBstMidEndSepPunct{\mcitedefaultmidpunct}
{\mcitedefaultendpunct}{\mcitedefaultseppunct}\relax
\EndOfBibitem
\bibitem[Mazal \latin{et~al.}(2020)Mazal, Prezado, Ares, de~Marzi, Patriarca,
  Miralbell, and Favaudon]{Mazal_2020_BJR.93.20190807}
Mazal,~A.; Prezado,~Y.; Ares,~C.; de~Marzi,~L.; Patriarca,~A.; Miralbell,~R.;
  Favaudon,~V. {FLASH and Minibeams in Radiation Therapy: The Effect of
  Microstructures on Time and Space and Their Potential Application to
  Protontherapy}. \emph{Br. J. Radiol.} \textbf{2020}, \emph{93},
  20190807\relax
\mciteBstWouldAddEndPuncttrue
\mciteSetBstMidEndSepPunct{\mcitedefaultmidpunct}
{\mcitedefaultendpunct}{\mcitedefaultseppunct}\relax
\EndOfBibitem
\bibitem[Horendeck \latin{et~al.}(2021)Horendeck, Walsh, Hirakawa, Fujimori,
  Kitamura, and Kato]{Horendeck_2021_FrontOncol.11.690042}
Horendeck,~D.; Walsh,~K.~D.; Hirakawa,~H.; Fujimori,~A.; Kitamura,~H.;
  Kato,~T.~A. {High LET-Like Radiation Tracks at the Distal Side of Accelerated
  Proton Bragg Peak}. \emph{Front. Oncol.} \textbf{2021}, \emph{11},
  690042\relax
\mciteBstWouldAddEndPuncttrue
\mciteSetBstMidEndSepPunct{\mcitedefaultmidpunct}
{\mcitedefaultendpunct}{\mcitedefaultseppunct}\relax
\EndOfBibitem
\bibitem[Audouin \latin{et~al.}(2023)Audouin, Hofverberg, Ngono-Ravache,
  Desorgher, and Baldacchino]{Audouin_2023_SciRep}
Audouin,~J.; Hofverberg,~P.; Ngono-Ravache,~Y.; Desorgher,~L.; Baldacchino,~G.
  {Intermediate LET-like Effect in Distal Part of Proton Bragg Peak Revealed by
  Track-Ends Imaging During Super-Fricke Radiolysis}. \emph{Sci. Rep.}
  \textbf{2023}, \emph{13}, 15460\relax
\mciteBstWouldAddEndPuncttrue
\mciteSetBstMidEndSepPunct{\mcitedefaultmidpunct}
{\mcitedefaultendpunct}{\mcitedefaultseppunct}\relax
\EndOfBibitem
\bibitem[Bizzarri \latin{et~al.}(2020)Bizzarri, Naimark, Nieto-Villar, Fedeli,
  and Giuliani]{Bizzarri_2020_Entropy.22.885}
Bizzarri,~M.; Naimark,~O.; Nieto-Villar,~J.; Fedeli,~V.; Giuliani,~A.
  {Complexity in Biological Organization: Deconstruction (and Subsequent
  Restating) of Key Concepts. }. \emph{Entropy} \textbf{2020}, \emph{22},
  85\relax
\mciteBstWouldAddEndPuncttrue
\mciteSetBstMidEndSepPunct{\mcitedefaultmidpunct}
{\mcitedefaultendpunct}{\mcitedefaultseppunct}\relax
\EndOfBibitem
\bibitem[Erenpreisa \latin{et~al.}(2023)Erenpreisa, Giuliani, Yoshikawa, Falk,
  Hildenbrand, Salmina, Freivalds, Vainshelbaum, Weidner, Sievers, Pilarczyk,
  and Hausmann]{Erenpreisa_2023_IJMS.24.2658}
Erenpreisa,~J.; Giuliani,~A.; Yoshikawa,~K.; Falk,~M.; Hildenbrand,~G.;
  Salmina,~K.; Freivalds,~T.; Vainshelbaum,~N.; Weidner,~J.; Sievers,~A.;
  Pilarczyk,~G.; Hausmann,~M. {Spatial-Temporal Genome Regulation in
  Stress-Response and Cell-Fate Change}. \emph{Int. J. Mol. Sci.}
  \textbf{2023}, \emph{24}, 2658\relax
\mciteBstWouldAddEndPuncttrue
\mciteSetBstMidEndSepPunct{\mcitedefaultmidpunct}
{\mcitedefaultendpunct}{\mcitedefaultseppunct}\relax
\EndOfBibitem
\bibitem[Cremer \latin{et~al.}(2015)Cremer, Cremer, H\"{u}bner, Strickfaden,
  Smeets, Popken, Sterr, Markaki, Rippe, and
  Cremer]{Cremer_2015_FEBSLett.589.2931}
Cremer,~T.; Cremer,~M.; H\"{u}bner,~B.; Strickfaden,~H.; Smeets,~D.;
  Popken,~J.; Sterr,~M.; Markaki,~Y.; Rippe,~K.; Cremer,~C. {The 4D Nucleome:
  Evidence for a Dynamic Nuclear Landscape Based on Co-Aligned Active and
  Inactive Nuclear Compartments}. \emph{FEBS Lett.} \textbf{2015}, \emph{589},
  2931--2943\relax
\mciteBstWouldAddEndPuncttrue
\mciteSetBstMidEndSepPunct{\mcitedefaultmidpunct}
{\mcitedefaultendpunct}{\mcitedefaultseppunct}\relax
\EndOfBibitem
\bibitem[Sievers \latin{et~al.}(2021)Sievers, Sauer, Hausmann, and
  Hildenbrand]{Sievers_2021_Genes.12.1571}
Sievers,~A.; Sauer,~L.; Hausmann,~M.; Hildenbrand,~G. {Eukaryotic Genomes Show
  Strong Evolutionary Conservation of K-Mer Composition and Correlation
  Contributions between Introns and Intergenic Regions}. \emph{Genes}
  \textbf{2021}, \emph{12}, 1571\relax
\mciteBstWouldAddEndPuncttrue
\mciteSetBstMidEndSepPunct{\mcitedefaultmidpunct}
{\mcitedefaultendpunct}{\mcitedefaultseppunct}\relax
\EndOfBibitem
\bibitem[Sievers \latin{et~al.}(2023)Sievers, Sauer, Bisch, Sprengel, Hausmann,
  and Hildenbrand]{Sievers_2023_Genes.14.755}
Sievers,~A.; Sauer,~L.; Bisch,~M.; Sprengel,~J.; Hausmann,~M.; Hildenbrand,~G.
  {Moderation of Structural DNA Properties by Coupled Dinucleotide Contents in
  Eukaryotes. }. \emph{Genes} \textbf{2023}, \emph{14}, 755\relax
\mciteBstWouldAddEndPuncttrue
\mciteSetBstMidEndSepPunct{\mcitedefaultmidpunct}
{\mcitedefaultendpunct}{\mcitedefaultseppunct}\relax
\EndOfBibitem
\bibitem[Krigerts \latin{et~al.}(2021)Krigerts, Salmina, Freivalds, Zayakin,
  Rumnieks, Inashkina, Giuliani, Hausmann, and
  Erenpreisa]{Krigerts_2021_BiophysJ.120.711}
Krigerts,~J.; Salmina,~K.; Freivalds,~T.; Zayakin,~P.; Rumnieks,~F.;
  Inashkina,~I.; Giuliani,~A.; Hausmann,~M.; Erenpreisa,~J. {Differentiating
  Cancer Cells Reveal Early Large-Scale Genome Regulation by Pericentric
  Domains}. \emph{Biophys. J.} \textbf{2021}, \emph{120}, 711--724\relax
\mciteBstWouldAddEndPuncttrue
\mciteSetBstMidEndSepPunct{\mcitedefaultmidpunct}
{\mcitedefaultendpunct}{\mcitedefaultseppunct}\relax
\EndOfBibitem
\bibitem[Erenpreisa \latin{et~al.}(2021)Erenpreisa, Krigerts, Salmina,
  Gerashchenko, Freivalds, Kurg, Winter, Krufczik, Zayakin, Hausmann, and
  Giuliani]{Erenpreisa_2021_Cells.10.1582}
Erenpreisa,~J.; Krigerts,~J.; Salmina,~K.; Gerashchenko,~B.~I.; Freivalds,~T.;
  Kurg,~R.; Winter,~R.; Krufczik,~M.; Zayakin,~P.; Hausmann,~M.; Giuliani,~A.
  {Heterochromatin Networks: Topology, Dynamics, and Function (a Working
  Hypothesis)}. \emph{Cells} \textbf{2021}, \emph{10}, 1582\relax
\mciteBstWouldAddEndPuncttrue
\mciteSetBstMidEndSepPunct{\mcitedefaultmidpunct}
{\mcitedefaultendpunct}{\mcitedefaultseppunct}\relax
\EndOfBibitem
\bibitem[Hausmann \latin{et~al.}(2022)Hausmann, Hildenbrand, and
  Pilarczyk]{Hausmann2022networks}
Hausmann,~M.; Hildenbrand,~G.; Pilarczyk,~G. In \emph{Nuclear, Chromosomal, and
  Genomic Architecture in Biology and Medicine}; Kloc,~M., Kubiak,~J.~Z., Eds.;
  Springer International Publishing, Cham, 2022; pp 3--34\relax
\mciteBstWouldAddEndPuncttrue
\mciteSetBstMidEndSepPunct{\mcitedefaultmidpunct}
{\mcitedefaultendpunct}{\mcitedefaultseppunct}\relax
\EndOfBibitem
\bibitem[Lee(2021)]{Lee_Hausmann2021super-resolution}
Lee,~M.,~J.-H.;~Hausmann In \emph{DNA -- Damages and Repair Mechanisms};
  Behzadi,~P., Ed.; IntechOpen, 2021\relax
\mciteBstWouldAddEndPuncttrue
\mciteSetBstMidEndSepPunct{\mcitedefaultmidpunct}
{\mcitedefaultendpunct}{\mcitedefaultseppunct}\relax
\EndOfBibitem
\bibitem[Jezkova \latin{et~al.}(2018)Jezkova, Zadneprianetc, Kulikova,
  Smirnova, Bulanova, Depes, Falkova, Boreyko, Krasavin, Davidkova, Kozubek,
  Valentova, and Falk]{Jezkova_2018_Nanoscale.10.1162}
Jezkova,~L.; Zadneprianetc,~M.; Kulikova,~E.; Smirnova,~E.; Bulanova,~T.;
  Depes,~D.; Falkova,~I.; Boreyko,~A.; Krasavin,~E.; Davidkova,~M.;
  Kozubek,~S.; Valentova,~O.; Falk,~M. {Particles with Similar LET Values
  Generate DNA Breaks of Different Complexity and Reparability: A
  High-Resolution Microscopy Analysis of $\gamma$H2AX/53BP1 Foci}.
  \emph{Nanoscale} \textbf{2018}, \emph{10}, 1162--1179\relax
\mciteBstWouldAddEndPuncttrue
\mciteSetBstMidEndSepPunct{\mcitedefaultmidpunct}
{\mcitedefaultendpunct}{\mcitedefaultseppunct}\relax
\EndOfBibitem
\bibitem[Falk and Hausmann(2021)Falk, and Hausmann]{MF_MH_2021_Cancers.13.18}
Falk,~M.; Hausmann,~M. {A Paradigm Revolution or Just Better Resolution -- Will
  Newly Emerging Superresolution Techniques Identify Chromatin Architecture as
  a Key Factor in Radiation-Induced DNA Damage and Repair Regulation?}
  \emph{Cancers} \textbf{2021}, \emph{13}, 18\relax
\mciteBstWouldAddEndPuncttrue
\mciteSetBstMidEndSepPunct{\mcitedefaultmidpunct}
{\mcitedefaultendpunct}{\mcitedefaultseppunct}\relax
\EndOfBibitem
\bibitem[Bobkova \latin{et~al.}(2018)Bobkova, Depes, Lee, Jezkova, Falkova,
  Pagacova, Kopecna, Zadneprianetc, Bacikova, Kulikova, Smirnova, Bulanova,
  Boreyko, Krasavin, Wenz, Bestvater, Hildenbrand, Hausmann, and
  Falk]{Bobkova_2018_IJMS.19.3713}
Bobkova,~E. \latin{et~al.}  {Recruitment of 53BP1 Proteins for DNA Repair and
  Persistence of Repair Clusters Differ for Cell Types as Detected by Single
  Molecule Localization Microscopy}. \emph{Int. J. Mol. Sci.} \textbf{2018},
  \emph{19}, 3713\relax
\mciteBstWouldAddEndPuncttrue
\mciteSetBstMidEndSepPunct{\mcitedefaultmidpunct}
{\mcitedefaultendpunct}{\mcitedefaultseppunct}\relax
\EndOfBibitem
\bibitem[Scully \latin{et~al.}(2019)Scully, Panday, Elango, and
  Willis]{Scully_2019_NatRevMolCellBiol}
Scully,~R.; Panday,~A.; Elango,~R.; Willis,~N.~A. {DNA Double-Strand Break
  Repair-Pathway Choice in Somatic Mammalian Cells}. \emph{Nat. Rev. Mol. Cell
  Biol.} \textbf{2019}, \emph{20}, 698--714\relax
\mciteBstWouldAddEndPuncttrue
\mciteSetBstMidEndSepPunct{\mcitedefaultmidpunct}
{\mcitedefaultendpunct}{\mcitedefaultseppunct}\relax
\EndOfBibitem
\bibitem[Falk \latin{et~al.}(2014)Falk, Hausmann, Lukasova, Biswas,
  Hildenbrand, Davidkova, Krasavin, Kleibl, Falkova, Jezkova, Stefancikova,
  Sevcik, Hofer, Bacikova, Matula, Boreyko, Vachelova, Michaelidesova, and
  Kozubek]{Falk_2014_CritRev_PartA}
Falk,~M. \latin{et~al.}  {Determining Omics Spatiotemporal Dimensions Using
  Exciting New Nanoscopy Techniques to Assess Complex Cell Responses to DNA
  Damage: Part A -- Radiomics}. \emph{Crit. Rev. Eukaryot. Gene Expr.}
  \textbf{2014}, \emph{24}, 205--223\relax
\mciteBstWouldAddEndPuncttrue
\mciteSetBstMidEndSepPunct{\mcitedefaultmidpunct}
{\mcitedefaultendpunct}{\mcitedefaultseppunct}\relax
\EndOfBibitem
\bibitem[Falk \latin{et~al.}(2014)Falk, Hausmann, Lukasova, Biswas,
  Hildenbrand, Davidkova, Krasavin, Kleibl, Falkova, Jezkova, Stefancikova,
  Sevcik, Hofer, Bacikova, Matula, Boreyko, Vachelova, Michaelidesova, and
  Kozubek]{Falk_2014_CritRev_PartB}
Falk,~M. \latin{et~al.}  {Determining Omics Spatiotemporal Dimensions Using
  Exciting New Nanoscopy Techniques to Assess Complex Cell Responses to DNA
  Damage: Part B -- Structuromics}. \emph{Crit. Rev. Eukaryot. Gene Expr.}
  \textbf{2014}, \emph{24}, 225--247\relax
\mciteBstWouldAddEndPuncttrue
\mciteSetBstMidEndSepPunct{\mcitedefaultmidpunct}
{\mcitedefaultendpunct}{\mcitedefaultseppunct}\relax
\EndOfBibitem
\bibitem[Iliakis \latin{et~al.}(2019)Iliakis, Mladenov, and
  Mladenova]{Iliakis_2019_Cancers.11.1671}
Iliakis,~G.; Mladenov,~E.; Mladenova,~V. {Necessities in the Processing of DNA
  Double Strand Breaks and Their Effects on Genomic Instability and Cancer}.
  \emph{Cancers} \textbf{2019}, \emph{11}, 1671\relax
\mciteBstWouldAddEndPuncttrue
\mciteSetBstMidEndSepPunct{\mcitedefaultmidpunct}
{\mcitedefaultendpunct}{\mcitedefaultseppunct}\relax
\EndOfBibitem
\bibitem[Falk \latin{et~al.}(2007)Falk, Lukasova, Gabrielova, Ondrej, and
  Kozubek]{Falk_2007_BBA.1773.1534}
Falk,~M.; Lukasova,~E.; Gabrielova,~B.; Ondrej,~V.; Kozubek,~S. {Chromatin
  Dynamics during DSB Repair}. \emph{Biochim. Biophys. Acta} \textbf{2007},
  \emph{1773}, 1534--1545\relax
\mciteBstWouldAddEndPuncttrue
\mciteSetBstMidEndSepPunct{\mcitedefaultmidpunct}
{\mcitedefaultendpunct}{\mcitedefaultseppunct}\relax
\EndOfBibitem
\bibitem[Falk \latin{et~al.}(2008)Falk, Luk\'{a}sov\'{a}, and
  Kozubek]{Falk_2008_BBA.1783.2398}
Falk,~M.; Luk\'{a}sov\'{a},~E.; Kozubek,~S. {Chromatin Structure Influences the
  Sensitivity of DNA to Gamma-Radiation}. \emph{Biochim. Biophys. Acta}
  \textbf{2008}, \emph{1783}, 2398--2414\relax
\mciteBstWouldAddEndPuncttrue
\mciteSetBstMidEndSepPunct{\mcitedefaultmidpunct}
{\mcitedefaultendpunct}{\mcitedefaultseppunct}\relax
\EndOfBibitem
\bibitem[Falk \latin{et~al.}(2008)Falk, Lukasova, Gabrielova, Ondrej, and
  Kozubek]{Falk_2008_JPCS.101.012018}
Falk,~M.; Lukasova,~E.; Gabrielova,~B.; Ondrej,~V.; Kozubek,~S. {Local Changes
  of Higher-Order Chromatin Structure during DSB-Repair}. \emph{J. Phys. Conf.
  Ser.} \textbf{2008}, \emph{101}, 012018\relax
\mciteBstWouldAddEndPuncttrue
\mciteSetBstMidEndSepPunct{\mcitedefaultmidpunct}
{\mcitedefaultendpunct}{\mcitedefaultseppunct}\relax
\EndOfBibitem
\bibitem[Falk \latin{et~al.}(2010)Falk, Lukasova, and
  Kozubek]{Falk_2010_MutatRes.704.88}
Falk,~M.; Lukasova,~E.; Kozubek,~S. {Higher-Order Chromatin Structure in DSB
  Induction, Repair and Misrepair}. \emph{Mutat. Res.} \textbf{2010},
  \emph{704}, 88--100\relax
\mciteBstWouldAddEndPuncttrue
\mciteSetBstMidEndSepPunct{\mcitedefaultmidpunct}
{\mcitedefaultendpunct}{\mcitedefaultseppunct}\relax
\EndOfBibitem
\bibitem[Sievers \latin{et~al.}(2018)Sievers, Wenz, Hausmann, and
  Hildenbrand]{Sievers_2018_Genes.9.482}
Sievers,~A.; Wenz,~F.; Hausmann,~M.; Hildenbrand,~G. {Conservation of k-Mer
  Composition and Correlation Contribution between Introns and Intergenic
  Regions of Animalia Genomes}. \emph{Genes} \textbf{2018}, \emph{9}, 482\relax
\mciteBstWouldAddEndPuncttrue
\mciteSetBstMidEndSepPunct{\mcitedefaultmidpunct}
{\mcitedefaultendpunct}{\mcitedefaultseppunct}\relax
\EndOfBibitem
\bibitem[Weidner \latin{et~al.}(2023)Weidner, Neitzel, Gote, Deck,
  K\"{u}ntzelmann, Pilarczyk, Falk, and Hausmann]{Weidner2023advanced}
Weidner,~J.; Neitzel,~C.; Gote,~M.; Deck,~J.; K\"{u}ntzelmann,~K.;
  Pilarczyk,~G.; Falk,~M.; Hausmann,~M. {Advanced Image-Free Analysis of the
  Nano-Organization of Chromatin and Other Biomolecules by Single Molecule
  Localization Microscopy (SMLM)}. \emph{Computat. Struct. Biotechnol. J.}
  \textbf{2023}, \emph{21}, 2018--2034\relax
\mciteBstWouldAddEndPuncttrue
\mciteSetBstMidEndSepPunct{\mcitedefaultmidpunct}
{\mcitedefaultendpunct}{\mcitedefaultseppunct}\relax
\EndOfBibitem
\bibitem[Krufczik \latin{et~al.}(2017)Krufczik, Sievers, Hausmann, Lee,
  Hildenbrand, Schaufler, and Hausmann]{Krufczik_2017_IJMS.18.1005}
Krufczik,~M.; Sievers,~A.; Hausmann,~A.; Lee,~J.-H.; Hildenbrand,~G.;
  Schaufler,~W.; Hausmann,~M. {Combining Low Temperature Fluorescence
  DNA-Hybridization, Immunostaining, and Super-Resolution Localization
  Microscopy for Nano-Structure Analysis of ALU Elements and Their Influence on
  Chromatin Structure}. \emph{Int. J. Mol. Sci.} \textbf{2017}, \emph{18},
  1005\relax
\mciteBstWouldAddEndPuncttrue
\mciteSetBstMidEndSepPunct{\mcitedefaultmidpunct}
{\mcitedefaultendpunct}{\mcitedefaultseppunct}\relax
\EndOfBibitem
\bibitem[Zhang \latin{et~al.}(2015)Zhang, M\'{a}t\'{e}, M\"{u}ller,
  Hillebrandt, Krufczik, Bach, Kaufmann, Hausmann, and
  Heermann]{Zhang_2015_PLoSONE.10.e0128555}
Zhang,~Y.; M\'{a}t\'{e},~G.; M\"{u}ller,~P.; Hillebrandt,~S.; Krufczik,~M.;
  Bach,~M.; Kaufmann,~R.; Hausmann,~M.; Heermann,~D.~W. {Radiation Induced
  Chromatin Conformation Changes Analysed by Fluorescent Localization
  Microscopy, Statistical Physics, and Graph Theory}. \emph{PLoS ONE}
  \textbf{2015}, \emph{10}, e0128555\relax
\mciteBstWouldAddEndPuncttrue
\mciteSetBstMidEndSepPunct{\mcitedefaultmidpunct}
{\mcitedefaultendpunct}{\mcitedefaultseppunct}\relax
\EndOfBibitem
\bibitem[Zhang and Heermann(2014)Zhang, and Heermann]{Zhang_2014_Chromosoma}
Zhang,~Y.; Heermann,~D.~W. {DNA Double-Strand Breaks: Linking Gene Expression
  to Chromosome Morphology and Mobility}. \emph{Chromosoma} \textbf{2014},
  \emph{123}, 103--115\relax
\mciteBstWouldAddEndPuncttrue
\mciteSetBstMidEndSepPunct{\mcitedefaultmidpunct}
{\mcitedefaultendpunct}{\mcitedefaultseppunct}\relax
\EndOfBibitem
\bibitem[Hausmann \latin{et~al.}(2018)Hausmann, Wagner, Lee, Schrock,
  Schaufler, Krufczik, Papenfu{\ss}, Port, Bestvater, and
  Scherthan]{Hausmann_2018_Nanoscale.10.4320}
Hausmann,~M.; Wagner,~E.; Lee,~J.-H.; Schrock,~G.; Schaufler,~W.; Krufczik,~M.;
  Papenfu{\ss},~F.; Port,~M.; Bestvater,~F.; Scherthan,~H. {Super-Resolution
  Localization Microscopy of Radiation-Induced Histone H2AX-Phosphorylation in
  Relation to H3K9-Trimethylation in HeLa Cells}. \emph{Nanoscale}
  \textbf{2018}, \emph{10}, 4320--4331\relax
\mciteBstWouldAddEndPuncttrue
\mciteSetBstMidEndSepPunct{\mcitedefaultmidpunct}
{\mcitedefaultendpunct}{\mcitedefaultseppunct}\relax
\EndOfBibitem
\bibitem[Hausmann \latin{et~al.}(2021)Hausmann, Falk, Neitzel, Hofmann, Biswas,
  Gier, Falkova, Heermann, and Hildenbrand]{Hausmann_2021_IJMS.22.3636}
Hausmann,~M.; Falk,~M.; Neitzel,~C.; Hofmann,~A.; Biswas,~A.; Gier,~T.;
  Falkova,~I.; Heermann,~D.~W.; Hildenbrand,~G. {Elucidation of the Clustered
  Nano-Architecture of Radiation-Induced DNA Damage Sites and Surrounding
  Chromatin in Cancer Cells: A Single Molecule Localization Microscopy
  Approach}. \emph{Int. J. Mol. Sci.} \textbf{2021}, \emph{22}, 3636\relax
\mciteBstWouldAddEndPuncttrue
\mciteSetBstMidEndSepPunct{\mcitedefaultmidpunct}
{\mcitedefaultendpunct}{\mcitedefaultseppunct}\relax
\EndOfBibitem
\bibitem[Hofmann \latin{et~al.}(2018)Hofmann, Krufczik, Heermann, and
  Hausmann]{Hofmann_2018_IJMS.19.2263}
Hofmann,~A.; Krufczik,~M.; Heermann,~D.; Hausmann,~M. {Using Persistent
  Homology as a New Approach for Super-Resolution Localization Microscopy Data
  Analysis and Classification of $\gamma$H2AX Foci/Clusters}. \emph{Int. J.
  Mol. Sci.} \textbf{2018}, \emph{19}, 2263\relax
\mciteBstWouldAddEndPuncttrue
\mciteSetBstMidEndSepPunct{\mcitedefaultmidpunct}
{\mcitedefaultendpunct}{\mcitedefaultseppunct}\relax
\EndOfBibitem
\bibitem[Hahn \latin{et~al.}(2021)Hahn, Neitzel, Kope\v{c}n\'{a}, Heermann,
  Falk, and Hausmann]{Hahn_2021_Cancers.13.5561}
Hahn,~H.; Neitzel,~C.; Kope\v{c}n\'{a},~O.; Heermann,~D.~W.; Falk,~M.;
  Hausmann,~M. {Topological Analysis of $\gamma$H2AX and MRE11 Clusters
  Detected by Localization Microscopy during X-Ray-Induced DNA Double-Strand
  Break Repair}. \emph{Cancers} \textbf{2021}, \emph{13}, 5561\relax
\mciteBstWouldAddEndPuncttrue
\mciteSetBstMidEndSepPunct{\mcitedefaultmidpunct}
{\mcitedefaultendpunct}{\mcitedefaultseppunct}\relax
\EndOfBibitem
\bibitem[Scherthan \latin{et~al.}(2019)Scherthan, Lee, Maus, Schumann, Muhtadi,
  Chojowski, Port, Lassmann, Bestvater, and
  Hausmann]{Scherthan_2019_Cancers.11.1877}
Scherthan,~H.; Lee,~J.-H.; Maus,~E.; Schumann,~S.; Muhtadi,~R.; Chojowski,~R.;
  Port,~M.; Lassmann,~M.; Bestvater,~F.; Hausmann,~M. {Nanostructure of
  Clustered DNA Damage in Leukocytes after In-Solution Irradiation with the
  Alpha Emitter Ra-223}. \emph{Cancers} \textbf{2019}, \emph{11}, 1877\relax
\mciteBstWouldAddEndPuncttrue
\mciteSetBstMidEndSepPunct{\mcitedefaultmidpunct}
{\mcitedefaultendpunct}{\mcitedefaultseppunct}\relax
\EndOfBibitem
\bibitem[Morales \latin{et~al.}(2015)Morales, White, Streva, DeFreece, Hedges,
  and Deininger]{Morales_2015_PLoSGenet}
Morales,~M.~E.; White,~T.~B.; Streva,~V.~A.; DeFreece,~C.~B.; Hedges,~D.~J.;
  Deininger,~P.~L. {The Contribution of ALU Elements to Mutagenic DNA
  Double-Strand Break Repair}. \emph{PLoS Genet.} \textbf{2015}, \emph{11},
  e1005016\relax
\mciteBstWouldAddEndPuncttrue
\mciteSetBstMidEndSepPunct{\mcitedefaultmidpunct}
{\mcitedefaultendpunct}{\mcitedefaultseppunct}\relax
\EndOfBibitem
\bibitem[Hausmann \latin{et~al.}(2017)Hausmann, Ili\'{c}, Pilarczyk, Lee,
  Logeswaran, Borroni, Krufczik, Theda, Waltrich, Bestvater, Hildenbrand,
  Cremer, and Blank]{Hausmann_2017_IJMS.18.2066}
Hausmann,~M.; Ili\'{c},~N.; Pilarczyk,~G.; Lee,~J.-H.; Logeswaran,~A.;
  Borroni,~A.~P.; Krufczik,~M.; Theda,~F.; Waltrich,~N.; Bestvater,~F.;
  Hildenbrand,~G.; Cremer,~C.; Blank,~M. {Challenges for Super-Resolution
  Localization Microscopy and Biomolecular Fluorescent Nano-Probing in Cancer
  Research}. \emph{Int. J. Mol. Sci.} \textbf{2017}, \emph{18}, 2066\relax
\mciteBstWouldAddEndPuncttrue
\mciteSetBstMidEndSepPunct{\mcitedefaultmidpunct}
{\mcitedefaultendpunct}{\mcitedefaultseppunct}\relax
\EndOfBibitem
\bibitem[Vicar \latin{et~al.}(2021)Vicar, Gumulec, Kolar, Kopecna, Pagacova,
  Falkova, and Falk]{Vicar_2021_DeepFoci}
Vicar,~T.; Gumulec,~J.; Kolar,~R.; Kopecna,~O.; Pagacova,~E.; Falkova,~I.;
  Falk,~M. {DeepFoci: Deep Learning-Based Algorithm for Fast Automatic Analysis
  of DNA Double-Strand Break Ionizing Radiation-Induced Foci}. \emph{Comput.
  Struct. Biotechnol. J.} \textbf{2021}, \emph{19}, 6465--6480\relax
\mciteBstWouldAddEndPuncttrue
\mciteSetBstMidEndSepPunct{\mcitedefaultmidpunct}
{\mcitedefaultendpunct}{\mcitedefaultseppunct}\relax
\EndOfBibitem
\bibitem[Dobe\v{s}ov\'{a} \latin{et~al.}(2022)Dobe\v{s}ov\'{a}, Gier,
  Kope\v{c}n\'{a}, Pag\'{a}\v{c}ov\'{a}, Vi\v{c}ar, Bestvater, Toufar,
  Ba\v{c}\'{i}kov\'{a}, Kopel, Fedr, Hildenbrand, Falkov\'{a}, Falk, and
  Hausmann]{Dobesova_2022_Pharmaceutics}
Dobe\v{s}ov\'{a},~L.; Gier,~T.; Kope\v{c}n\'{a},~O.; Pag\'{a}\v{c}ov\'{a},~E.;
  Vi\v{c}ar,~T.; Bestvater,~F.; Toufar,~J.; Ba\v{c}\'{i}kov\'{a},~A.;
  Kopel,~P.; Fedr,~R.; Hildenbrand,~G.; Falkov\'{a},~I.; Falk,~M.; Hausmann,~M.
  {Incorporation of Low Concentrations of Gold Nanoparticles: Complex Effects
  on Radiation Response and Fate of Cancer Cells}. \emph{Pharmaceutics}
  \textbf{2022}, \emph{14}, 166\relax
\mciteBstWouldAddEndPuncttrue
\mciteSetBstMidEndSepPunct{\mcitedefaultmidpunct}
{\mcitedefaultendpunct}{\mcitedefaultseppunct}\relax
\EndOfBibitem
\bibitem[Ahmad \latin{et~al.}(2020)Ahmad, Schettino, Royle, Barry, Pankhurst,
  Tillement, Russell, and Ricketts]{Ahmad_2020_PartPartSystCharact}
Ahmad,~R.; Schettino,~G.; Royle,~G.; Barry,~M.; Pankhurst,~Q.~A.;
  Tillement,~O.; Russell,~B.; Ricketts,~K. {Radiobiological Implications of
  Nanoparticles Following Radiation Treatment}. \emph{Part. Part. Syst.
  Charact.} \textbf{2020}, \emph{37}, 1900411\relax
\mciteBstWouldAddEndPuncttrue
\mciteSetBstMidEndSepPunct{\mcitedefaultmidpunct}
{\mcitedefaultendpunct}{\mcitedefaultseppunct}\relax
\EndOfBibitem
\bibitem[Her \latin{et~al.}(2017)Her, Jaffray, and
  Allen]{Her_2017_AdvDrugDelivRev}
Her,~S.; Jaffray,~D.~A.; Allen,~C. {Gold Nanoparticles for Applications in
  Cancer Radiotherapy: Mechanisms and Recent Advancements}. \emph{Adv. Drug
  Deliv. Rev.} \textbf{2017}, \emph{109}, 84--101\relax
\mciteBstWouldAddEndPuncttrue
\mciteSetBstMidEndSepPunct{\mcitedefaultmidpunct}
{\mcitedefaultendpunct}{\mcitedefaultseppunct}\relax
\EndOfBibitem
\bibitem[Ricketts \latin{et~al.}(2018)Ricketts, Ahmad, Beaton, Cousins,
  Critchley, Davies, Evans, Fenuyi, Gavriilidis, Harmer, and
  et~al.]{Ricketts_2018_BJR.91.20180325}
Ricketts,~K.; Ahmad,~R.; Beaton,~L.; Cousins,~B.; Critchley,~K.; Davies,~M.;
  Evans,~S.; Fenuyi,~I.; Gavriilidis,~A.; Harmer,~Q.~J.; et~al.
  {Recommendations for Clinical Translation of Nanoparticle-Enhanced
  Radiotherapy}. \emph{Br. J. Radiol.} \textbf{2018}, \emph{91}, 20180325\relax
\mciteBstWouldAddEndPuncttrue
\mciteSetBstMidEndSepPunct{\mcitedefaultmidpunct}
{\mcitedefaultendpunct}{\mcitedefaultseppunct}\relax
\EndOfBibitem
\bibitem[Douglass \latin{et~al.}(2013)Douglass, Bezak, and
  Penfold]{Douglass_2013_MedPhys.40.071710}
Douglass,~M.; Bezak,~E.; Penfold,~S. {Monte Carlo Investigation of the
  Increased Radiation Deposition due to Gold Nanoparticles Using Kilovoltage
  and Megavoltage Photons in a 3D Randomized Cell Model}. \emph{Med. Phys.}
  \textbf{2013}, \emph{40}, 071710\relax
\mciteBstWouldAddEndPuncttrue
\mciteSetBstMidEndSepPunct{\mcitedefaultmidpunct}
{\mcitedefaultendpunct}{\mcitedefaultseppunct}\relax
\EndOfBibitem
\bibitem[Butterworth \latin{et~al.}(2012)Butterworth, McMahon, Currell, and
  Prise]{Butterworth_2012_Nanoscale.4.4830}
Butterworth,~K.~T.; McMahon,~S.~J.; Currell,~F.~J.; Prise,~K.~M. {Physical
  Basis and Biological Mechanisms of Gold Nanoparticle Radiosensitization}.
  \emph{Nanoscale} \textbf{2012}, \emph{4}, 4830\relax
\mciteBstWouldAddEndPuncttrue
\mciteSetBstMidEndSepPunct{\mcitedefaultmidpunct}
{\mcitedefaultendpunct}{\mcitedefaultseppunct}\relax
\EndOfBibitem
\bibitem[Rosa \latin{et~al.}(2017)Rosa, Connolly, Schettino, Butterworth, and
  Prise]{Rosa_2017_CancerNano.8.2}
Rosa,~S.; Connolly,~C.; Schettino,~G.; Butterworth,~K.~T.; Prise,~K.~M.
  {Biological Mechanisms of Gold Nanoparticle Radiosensitization}. \emph{Cancer
  Nanotechnol.} \textbf{2017}, \emph{8}, 2\relax
\mciteBstWouldAddEndPuncttrue
\mciteSetBstMidEndSepPunct{\mcitedefaultmidpunct}
{\mcitedefaultendpunct}{\mcitedefaultseppunct}\relax
\EndOfBibitem
\bibitem[Sicard-Roselli \latin{et~al.}(2014)Sicard-Roselli, Brun, Gilles,
  Baldacchino, Kelsey, McQuaid, Polin, Wardlow, and
  Currell]{Sicard-Roselli_2014_Small}
Sicard-Roselli,~C.; Brun,~E.; Gilles,~M.; Baldacchino,~G.; Kelsey,~C.;
  McQuaid,~H.; Polin,~C.; Wardlow,~N.; Currell,~F. {A New Mechanism for
  Hydroxyl Radical Production in Irradiated Nanoparticle Solutions}.
  \emph{Small} \textbf{2014}, \emph{10}, 3338--3346\relax
\mciteBstWouldAddEndPuncttrue
\mciteSetBstMidEndSepPunct{\mcitedefaultmidpunct}
{\mcitedefaultendpunct}{\mcitedefaultseppunct}\relax
\EndOfBibitem
\bibitem[Verkhovtsev \latin{et~al.}(2022)Verkhovtsev, Nichols, Mason, and
  Solov'yov]{Verkhovtsev_2022_JPCA.126.2170}
Verkhovtsev,~A.~V.; Nichols,~A.; Mason,~N.~J.; Solov'yov,~A.~V. {Molecular
  Dynamics Characterization of Radiosensitizing Coated Gold Nanoparticles in
  Aqueous Environment}. \emph{J. Phys. Chem. A} \textbf{2022}, \emph{126},
  2170--2184\relax
\mciteBstWouldAddEndPuncttrue
\mciteSetBstMidEndSepPunct{\mcitedefaultmidpunct}
{\mcitedefaultendpunct}{\mcitedefaultseppunct}\relax
\EndOfBibitem
\bibitem[MC_((accessed 2020-11-06))]{MC_Lecacy}
Radiotherapy: seizing the opportunity in cancer care. Maria Curie Legacy white
  paper. (accessed 2020-11-06);
  \url{https://www.cocir.org/uploads/media/Radiotherapy_-_Seizing_the_opportunity_in_cancer_care.pdf}\relax
\mciteBstWouldAddEndPuncttrue
\mciteSetBstMidEndSepPunct{\mcitedefaultmidpunct}
{\mcitedefaultendpunct}{\mcitedefaultseppunct}\relax
\EndOfBibitem
\bibitem[Gains \latin{et~al.}(2019)Gains, Beaton, Amos, and
  Sharma]{Gains_Amos_2019}
Gains,~J.; Beaton,~L.; Amos,~R.~A.; Sharma,~R.~A. In \emph{{Walter and Miller's
  Textbook of Radiotherapy: Radiation Physics, Therapy and Oncology}}, 8th ed.;
  Symonds,~P., Mills,~J.~A., Duxbury,~A., Eds.; Elsevier, 2019; pp
  579--588\relax
\mciteBstWouldAddEndPuncttrue
\mciteSetBstMidEndSepPunct{\mcitedefaultmidpunct}
{\mcitedefaultendpunct}{\mcitedefaultseppunct}\relax
\EndOfBibitem
\bibitem[Zeng \latin{et~al.}(2018)Zeng, Amos, Winey, Beltran, Saleh, Tochner,
  Kooy, and Both]{Zeng_Amos_2018}
Zeng,~C.; Amos,~R.~A.; Winey,~B.; Beltran,~C.; Saleh,~Z.; Tochner,~Z.;
  Kooy,~H.; Both,~S. In \emph{{Target Volume Delineation and Treatment Planning
  for Particle Therapy: A Practical Guide}}; Lee,~N.~Y., Leeman,~J.~E.,
  Cahlon,~O., Sine,~K., Jiang,~G., Lu,~J.~J., Both,~S., Eds.; Springer
  International Publishing, Cham, 2018; pp 45--105\relax
\mciteBstWouldAddEndPuncttrue
\mciteSetBstMidEndSepPunct{\mcitedefaultmidpunct}
{\mcitedefaultendpunct}{\mcitedefaultseppunct}\relax
\EndOfBibitem
\bibitem[Wang \latin{et~al.}(2013)Wang, Zhang, Li, Amos, Shaitelman, Hoffman,
  Howell, Salehpour, Zhang, Sun, and et~al.]{Wang_2013_BrJRadiol.86.20130176}
Wang,~X.; Zhang,~X.; Li,~X.; Amos,~R.~A.; Shaitelman,~S.~F.; Hoffman,~K.;
  Howell,~R.; Salehpour,~M.; Zhang,~S.~X.; Sun,~T.~L.; et~al. {Accelerated
  Partial-Breast Irradiation Using Intensity-Modulated Proton Radiotherapy: Do
  Uncertainties Outweigh Potential Benefits?} \emph{Br. J. Radiol.}
  \textbf{2013}, \emph{86}, 20130176\relax
\mciteBstWouldAddEndPuncttrue
\mciteSetBstMidEndSepPunct{\mcitedefaultmidpunct}
{\mcitedefaultendpunct}{\mcitedefaultseppunct}\relax
\EndOfBibitem
\bibitem[Kim \latin{et~al.}(2020)Kim, Kim, and Cho]{Kim_Cho_2020_ProgMedPhys}
Kim,~Y.; Kim,~J.; Cho,~S. {Review of the Existing Relative Biological
  Effectiveness Models for Carbon Ion Beam Therapy}. \emph{Prog. Med. Phys.}
  \textbf{2020}, \emph{31}, 1--7\relax
\mciteBstWouldAddEndPuncttrue
\mciteSetBstMidEndSepPunct{\mcitedefaultmidpunct}
{\mcitedefaultendpunct}{\mcitedefaultseppunct}\relax
\EndOfBibitem
\bibitem[Kelleter \latin{et~al.}(2019)Kelleter, Zhen-Hong~Tham, Saakyan,
  Griffiths, Amos, Jolly, and Gibson]{Kelleter_2019_MedPhys.46.3734}
Kelleter,~L.; Zhen-Hong~Tham,~B.; Saakyan,~R.; Griffiths,~J.; Amos,~R.;
  Jolly,~S.; Gibson,~A. {Technical Note: Simulation of Dose Buildup in Proton
  Pencil Beams}. \emph{Med. Phys.} \textbf{2019}, \emph{46}, 3734--3738\relax
\mciteBstWouldAddEndPuncttrue
\mciteSetBstMidEndSepPunct{\mcitedefaultmidpunct}
{\mcitedefaultendpunct}{\mcitedefaultseppunct}\relax
\EndOfBibitem
\bibitem[Ree and Redalen(2015)Ree, and Redalen]{Ree_2015_BrJRadiol.88.20150009}
Ree,~A.~H.; Redalen,~K.~R. {Personalized Radiotherapy: Concepts, Biomarkers and
  Trial Design}. \emph{Br. J. Radiol.} \textbf{2015}, \emph{88}, 20150009\relax
\mciteBstWouldAddEndPuncttrue
\mciteSetBstMidEndSepPunct{\mcitedefaultmidpunct}
{\mcitedefaultendpunct}{\mcitedefaultseppunct}\relax
\EndOfBibitem
\bibitem[Falls \latin{et~al.}(2018)Falls, Sharma, Lawrence, Amos, Advani,
  Ahmed, Vikram, Coleman, and Prasanna]{Falls_2018_RadRes.190.350}
Falls,~K.~C.; Sharma,~R.~A.; Lawrence,~Y.~R.; Amos,~R.~A.; Advani,~S.~J.;
  Ahmed,~M.~M.; Vikram,~B.; Coleman,~C.~N.; Prasanna,~P.~G. {Radiation-Drug
  Combinations to Improve Clinical Outcomes and Reduce Normal Tissue
  Toxicities: Current Challenges and New Approaches: Report of the Symposium
  Held at the 63rd Annual Meeting of the Radiation Research Society, 15-18
  October 2017; Cancun, Mexico}. \emph{Radiat. Res.} \textbf{2018}, \emph{190},
  350--360\relax
\mciteBstWouldAddEndPuncttrue
\mciteSetBstMidEndSepPunct{\mcitedefaultmidpunct}
{\mcitedefaultendpunct}{\mcitedefaultseppunct}\relax
\EndOfBibitem
\bibitem[Cho and Krishnan(2013)Cho, and
  Krishnan]{Cho_2017_CancerNanotechnology}
Cho,~S.~H., Krishnan,~S., Eds. \emph{{Cancer Nanotechnology: Principles and
  Application in Radiation Oncology}}; CRC Press: Boca Raton, 2013\relax
\mciteBstWouldAddEndPuncttrue
\mciteSetBstMidEndSepPunct{\mcitedefaultmidpunct}
{\mcitedefaultendpunct}{\mcitedefaultseppunct}\relax
\EndOfBibitem
\bibitem[Suzuki(2020)]{Suzuki_2020_IJClinOncol.25.43}
Suzuki,~M. {Boron Neutron Capture Therapy (BNCT): A Unique Role in Radiotherapy
  with a View to Entering the Accelerator-Based BNCT Era}. \emph{Int. J. Clin.
  Oncol.} \textbf{2020}, \emph{25}, 43--50\relax
\mciteBstWouldAddEndPuncttrue
\mciteSetBstMidEndSepPunct{\mcitedefaultmidpunct}
{\mcitedefaultendpunct}{\mcitedefaultseppunct}\relax
\EndOfBibitem
\bibitem[Porra \latin{et~al.}(2022)Porra, Sepp\"{a}l\"{a}, Wendland, Revitzer,
  Joensuu, Eide, Koivunoro, Smick, Smick, and
  Tenhunen]{Porra_2022_ActaOncol.61.269}
Porra,~L.; Sepp\"{a}l\"{a},~T.; Wendland,~L.; Revitzer,~H.; Joensuu,~H.;
  Eide,~P.; Koivunoro,~H.; Smick,~N.; Smick,~T.; Tenhunen,~M.
  {Accelerator-Based Boron Neutron Capture Therapy Facility at the Helsinki
  University Hospital}. \emph{Acta. Oncol.} \textbf{2022}, \emph{61},
  269--273\relax
\mciteBstWouldAddEndPuncttrue
\mciteSetBstMidEndSepPunct{\mcitedefaultmidpunct}
{\mcitedefaultendpunct}{\mcitedefaultseppunct}\relax
\EndOfBibitem
\bibitem[Favaudon \latin{et~al.}(2014)Favaudon, Caplier, Monceau, Pouzoulet,
  Sayarath, Fouillade, Poupon, Brito, Hup\'{e}, Bourhis, and
  et~al.]{Favaudon_2014_TransMed}
Favaudon,~V.; Caplier,~L.; Monceau,~V.; Pouzoulet,~F.; Sayarath,~M.;
  Fouillade,~C.; Poupon,~M.-F.; Brito,~I.; Hup\'{e},~P.; Bourhis,~J.; et~al.
  {Ultrahigh Dose-Rate FLASH Irradiation Increases the Differential Response
  Between Normal and Tumor Tissue in Mice}. \emph{Sci. Transl. Med.}
  \textbf{2014}, \emph{6}, 245ra93\relax
\mciteBstWouldAddEndPuncttrue
\mciteSetBstMidEndSepPunct{\mcitedefaultmidpunct}
{\mcitedefaultendpunct}{\mcitedefaultseppunct}\relax
\EndOfBibitem
\bibitem[Louren\c{c}o \latin{et~al.}(2023)Louren\c{c}o, Subiel, Lee, Flynn,
  Cotterill, Shipley, Romano, Speth, Lee, Zhang, and
  et~al.]{Lorenco_2023_SciRep.13.2054}
Louren\c{c}o,~A.; Subiel,~A.; Lee,~N.; Flynn,~S.; Cotterill,~J.; Shipley,~D.;
  Romano,~F.; Speth,~J.; Lee,~E.; Zhang,~Y.; et~al. {Absolute Dosimetry for
  FLASH Proton Pencil Beam Scanning Radiotherapy}. \emph{Sci. Rep.}
  \textbf{2023}, \emph{13}, 2054\relax
\mciteBstWouldAddEndPuncttrue
\mciteSetBstMidEndSepPunct{\mcitedefaultmidpunct}
{\mcitedefaultendpunct}{\mcitedefaultseppunct}\relax
\EndOfBibitem
\bibitem[Mascia \latin{et~al.}(2023)Mascia, Daugherty, Zhang, Lee, Xiao,
  Sertorio, Woo, Backus, McDonald, McCann, and et~al.]{Mascia_JAMAOncol.9.62}
Mascia,~A.~E.; Daugherty,~E.~C.; Zhang,~Y.; Lee,~E.; Xiao,~Z.; Sertorio,~M.;
  Woo,~J.; Backus,~L.~R.; McDonald,~J.~M.; McCann,~C.; et~al. {Proton FLASH
  Radiotherapy for the Treatment of Symptomatic Bone Metastases: The FAST-01
  Nonrandomized Trial}. \emph{JAMA Oncol.} \textbf{2023}, \emph{9},
  62--69\relax
\mciteBstWouldAddEndPuncttrue
\mciteSetBstMidEndSepPunct{\mcitedefaultmidpunct}
{\mcitedefaultendpunct}{\mcitedefaultseppunct}\relax
\EndOfBibitem
\bibitem[Prezado \latin{et~al.}(2019)Prezado, Jouvion, Guardiola, Gonzalez,
  Juchaux, Bergs, Nauraye, Labiod, De~Marzi, Pouzoulet, and
  et~al.]{Prezado_2019_IJROBP.104.266}
Prezado,~Y.; Jouvion,~G.; Guardiola,~C.; Gonzalez,~W.; Juchaux,~M.; Bergs,~J.;
  Nauraye,~C.; Labiod,~D.; De~Marzi,~L.; Pouzoulet,~F.; et~al. {Tumor Control
  in RG2 Glioma-Bearing Rats: A Comparison Between Proton Minibeam Therapy and
  Standard Proton Therapy}. \emph{Int. J. Radiat. Oncol. Biol. Phys.}
  \textbf{2019}, \emph{104}, 266--271\relax
\mciteBstWouldAddEndPuncttrue
\mciteSetBstMidEndSepPunct{\mcitedefaultmidpunct}
{\mcitedefaultendpunct}{\mcitedefaultseppunct}\relax
\EndOfBibitem
\bibitem[Strieth-Kalthoff \latin{et~al.}(2018)Strieth-Kalthoff, James, Teders,
  Pitzer, and Glorius]{Strieth-Kalthoff_2018}
Strieth-Kalthoff,~F.; James,~M.~J.; Teders,~M.; Pitzer,~L.; Glorius,~F. {Energy
  Transfer Catalysis Mediated by Visible Light: Principles, Applications,
  Directions}. \emph{Chem. Soc. Rev.} \textbf{2018}, \emph{47},
  7190--7202\relax
\mciteBstWouldAddEndPuncttrue
\mciteSetBstMidEndSepPunct{\mcitedefaultmidpunct}
{\mcitedefaultendpunct}{\mcitedefaultseppunct}\relax
\EndOfBibitem
\bibitem[Marzo \latin{et~al.}(2018)Marzo, Pagire, Reiser, and
  K\"{o}nig]{Marzo_2018_AngewChem.57.10034}
Marzo,~L.; Pagire,~S.~K.; Reiser,~O.; K\"{o}nig,~B. {Visible-Light
  Photocatalysis: Does It Make a Difference in Organic Synthesis?} \emph{Angew.
  Chem. Int. Ed.} \textbf{2018}, \emph{57}, 10034--10072\relax
\mciteBstWouldAddEndPuncttrue
\mciteSetBstMidEndSepPunct{\mcitedefaultmidpunct}
{\mcitedefaultendpunct}{\mcitedefaultseppunct}\relax
\EndOfBibitem
\bibitem[Zhan \latin{et~al.}(2018)Zhan, Chen, Yi, Li, Wu, and
  Tian]{Zhan_2018_NatRevChem.2.216}
Zhan,~C.; Chen,~X.-J.; Yi,~J.; Li,~J.-F.; Wu,~D.-Y.; Tian,~Z.-Q. {From
  Plasmon-Enhanced Molecular Spectroscopy to Plasmon-Mediated Chemical
  Reactions}. \emph{Nat. Rev. Chem.} \textbf{2018}, \emph{2}, 216--230\relax
\mciteBstWouldAddEndPuncttrue
\mciteSetBstMidEndSepPunct{\mcitedefaultmidpunct}
{\mcitedefaultendpunct}{\mcitedefaultseppunct}\relax
\EndOfBibitem
\bibitem[Liu \latin{et~al.}(2007)Liu, Atwater, Wang, and
  Huo]{Liu_2007_CollSurfB.58.3}
Liu,~X.; Atwater,~M.; Wang,~J.; Huo,~Q. {Extinction Coefficient of Gold
  Nanoparticles with Different Sizes and Different Capping Ligands}.
  \emph{Colloids Surf. B} \textbf{2007}, \emph{58}, 3--7\relax
\mciteBstWouldAddEndPuncttrue
\mciteSetBstMidEndSepPunct{\mcitedefaultmidpunct}
{\mcitedefaultendpunct}{\mcitedefaultseppunct}\relax
\EndOfBibitem
\bibitem[Gell\'{e} \latin{et~al.}(2019)Gell\'{e}, Jin, de~La~Garza, Price,
  Besteiro, and Moores]{Gelle_2019_ChemRev.120.986}
Gell\'{e},~A.; Jin,~T.; de~La~Garza,~L.; Price,~G.~D.; Besteiro,~L.~V.;
  Moores,~A. {Applications of Plasmon-Enhanced Nanocatalysis to Organic
  Transformations}. \emph{Chem. Rev.} \textbf{2019}, \emph{120},
  986--1041\relax
\mciteBstWouldAddEndPuncttrue
\mciteSetBstMidEndSepPunct{\mcitedefaultmidpunct}
{\mcitedefaultendpunct}{\mcitedefaultseppunct}\relax
\EndOfBibitem
\bibitem[Christopher \latin{et~al.}(2011)Christopher, Xin, and
  Linic]{Christopher_2011_NatChem.3.467}
Christopher,~P.; Xin,~H.; Linic,~S. {Visible-Light-Enhanced Catalytic Oxidation
  Reactions on Plasmonic Silver Nanostructures}. \emph{Nat. Chem.}
  \textbf{2011}, \emph{3}, 467--472\relax
\mciteBstWouldAddEndPuncttrue
\mciteSetBstMidEndSepPunct{\mcitedefaultmidpunct}
{\mcitedefaultendpunct}{\mcitedefaultseppunct}\relax
\EndOfBibitem
\bibitem[Dutta \latin{et~al.}(2021)Dutta, Sch\"{u}rmann, Kogikoski, Mueller,
  Reich, and Bald]{Dutta_2021_ACSCatal.11.8370}
Dutta,~A.; Sch\"{u}rmann,~R.; Kogikoski,~S.; Mueller,~N.~S.; Reich,~S.;
  Bald,~I. {Kinetics and Mechanism of Plasmon-Driven Dehalogenation Reaction of
  Brominated Purine Nucleobases on Ag and Au}. \emph{ACS Catal.} \textbf{2021},
  \emph{11}, 8370--8381\relax
\mciteBstWouldAddEndPuncttrue
\mciteSetBstMidEndSepPunct{\mcitedefaultmidpunct}
{\mcitedefaultendpunct}{\mcitedefaultseppunct}\relax
\EndOfBibitem
\bibitem[Kogikoski \latin{et~al.}(15)Kogikoski, Dutta, and
  Bald]{Kogikoski_2021_ACSNano.15.20562}
Kogikoski,~S.; Dutta,~A.; Bald,~I. {Spatial Separation of Plasmonic
  Hot-Electron Generation and a Hydrodehalogenation Reaction Center Using a DNA
  Wire}. \emph{ACS Nano} \textbf{15}, \emph{2021}, 20562--20573\relax
\mciteBstWouldAddEndPuncttrue
\mciteSetBstMidEndSepPunct{\mcitedefaultmidpunct}
{\mcitedefaultendpunct}{\mcitedefaultseppunct}\relax
\EndOfBibitem
\bibitem[Sch\"{u}rmann and Bald(2017)Sch\"{u}rmann, and
  Bald]{Schurmann_2017_Nanoscale.9.1951}
Sch\"{u}rmann,~R.; Bald,~I. {Real-Time Monitoring of Plasmon Induced
  Dissociative Electron Transfer to the Potential DNA Radiosensitizer
  8-Bromoadenine}. \emph{Nanoscale} \textbf{2017}, \emph{9}, 1951--1955\relax
\mciteBstWouldAddEndPuncttrue
\mciteSetBstMidEndSepPunct{\mcitedefaultmidpunct}
{\mcitedefaultendpunct}{\mcitedefaultseppunct}\relax
\EndOfBibitem
\bibitem[Zhou \latin{et~al.}(2018)Zhou, Swearer, Zhang, Robatjazi, Zhao,
  Henderson, Dong, Christopher, Carter, Nordlander, and
  Halas]{Zhou_2018_Science.362.69}
Zhou,~L.; Swearer,~D.~F.; Zhang,~C.; Robatjazi,~H.; Zhao,~H.; Henderson,~L.;
  Dong,~L.; Christopher,~P.; Carter,~E.~A.; Nordlander,~P.; Halas,~N.~J.
  {Quantifying Hot Carrier and Thermal Contributions in Plasmonic
  Photocatalysis}. \emph{Science} \textbf{2018}, \emph{362}, 69--72\relax
\mciteBstWouldAddEndPuncttrue
\mciteSetBstMidEndSepPunct{\mcitedefaultmidpunct}
{\mcitedefaultendpunct}{\mcitedefaultseppunct}\relax
\EndOfBibitem
\bibitem[Sivan \latin{et~al.}(2019)Sivan, Un, and Dubi]{Sivan_2019_Science.364}
Sivan,~J.,~Y. an~Baraban; Un,~I.~W.; Dubi,~Y. {Comment on ``Quantifying Hot
  Carrier and Thermal Contributions in Plasmonic Photocatalysis''}.
  \emph{Science} \textbf{2019}, \emph{364}, eaaw9367\relax
\mciteBstWouldAddEndPuncttrue
\mciteSetBstMidEndSepPunct{\mcitedefaultmidpunct}
{\mcitedefaultendpunct}{\mcitedefaultseppunct}\relax
\EndOfBibitem
\bibitem[Robatjazi \latin{et~al.}(2020)Robatjazi, Bao, Zhang, Zhou,
  Christopher, Carter, Nordlander, and Halas]{Robatjazi_2020_NatCatal.3.564}
Robatjazi,~H.; Bao,~J.~L.; Zhang,~M.; Zhou,~L.; Christopher,~P.; Carter,~E.~A.;
  Nordlander,~P.; Halas,~N.~J. {Plasmon-Driven Carbon--Fluorine (C($sp^3$)--F)
  Bond Activation with Mechanistic Insights into Hot-Carrier-Mediated
  Pathways}. \emph{Nat. Catal.} \textbf{2020}, \emph{3}, 564--573\relax
\mciteBstWouldAddEndPuncttrue
\mciteSetBstMidEndSepPunct{\mcitedefaultmidpunct}
{\mcitedefaultendpunct}{\mcitedefaultseppunct}\relax
\EndOfBibitem
\bibitem[Dubi \latin{et~al.}(2022)Dubi, Un, Baraban, and
  Sivan]{Dubi_2022_NatCatal.5.244}
Dubi,~Y.; Un,~I.~W.; Baraban,~J.~H.; Sivan,~Y. {Distinguishing Thermal From
  Non-Thermal Contributions to Plasmonic Hydrodefluorination}. \emph{Nat.
  Catal.} \textbf{2022}, \emph{5}, 244--246\relax
\mciteBstWouldAddEndPuncttrue
\mciteSetBstMidEndSepPunct{\mcitedefaultmidpunct}
{\mcitedefaultendpunct}{\mcitedefaultseppunct}\relax
\EndOfBibitem
\bibitem[Swaminathan \latin{et~al.}(2021)Swaminathan, Rao, Bera, and
  Chandra]{Swaminathan_221_AngewChem.60.12532}
Swaminathan,~S.; Rao,~V.~G.; Bera,~J.~K.; Chandra,~M. {The Pivotal Role of Hot
  Carriers in Plasmonic Catalysis of C--N Bond Forming Reaction of Amines}.
  \emph{Angew. Chem. Int. Ed.} \textbf{2021}, \emph{60}, 12532--12538\relax
\mciteBstWouldAddEndPuncttrue
\mciteSetBstMidEndSepPunct{\mcitedefaultmidpunct}
{\mcitedefaultendpunct}{\mcitedefaultseppunct}\relax
\EndOfBibitem
\bibitem[Baffou \latin{et~al.}(2020)Baffou, Bordacchini, Baldi, and
  Quidant]{Baffou_2020_LightSciAppl}
Baffou,~G.; Bordacchini,~I.; Baldi,~A.; Quidant,~R. {Simple Experimental
  Procedures to Distinguish Photothermal From Hot-Carrier Processes in
  Plasmonics}. \emph{Light Sci. Appl.} \textbf{2020}, \emph{9}, 108\relax
\mciteBstWouldAddEndPuncttrue
\mciteSetBstMidEndSepPunct{\mcitedefaultmidpunct}
{\mcitedefaultendpunct}{\mcitedefaultseppunct}\relax
\EndOfBibitem
\bibitem[Sch\"{u}rmann \latin{et~al.}(2022)Sch\"{u}rmann, Dutta, Ebel, Tapio,
  Milosavljevi\'{c}, and Bald]{Schurmann_2022_JCP.157.084708}
Sch\"{u}rmann,~R.; Dutta,~A.; Ebel,~K.; Tapio,~K.; Milosavljevi\'{c},~A.~R.;
  Bald,~I. {Plasmonic Reactivity of Halogen Thiophenols on Gold Nanoparticles
  Studied by SERS and XPS}. \emph{J. Chem. Phys.} \textbf{2022}, \emph{157},
  084708\relax
\mciteBstWouldAddEndPuncttrue
\mciteSetBstMidEndSepPunct{\mcitedefaultmidpunct}
{\mcitedefaultendpunct}{\mcitedefaultseppunct}\relax
\EndOfBibitem
\bibitem[Sch\"{u}rmann \latin{et~al.}(2019)Sch\"{u}rmann, Ebel, Nicolas,
  Milosavljevi\'{c}, and Bald]{Schurmann_2019_JPCL.10.3153}
Sch\"{u}rmann,~R.; Ebel,~K.; Nicolas,~C.; Milosavljevi\'{c},~A.~R.; Bald,~I.
  {Role of Valence Band States and Plasmonic Enhancement in
  Electron-Transfer-Induced Transformation of Nitrothiophenol}. \emph{J. Phys.
  Chem. Lett.} \textbf{2019}, \emph{10}, 3153--3158\relax
\mciteBstWouldAddEndPuncttrue
\mciteSetBstMidEndSepPunct{\mcitedefaultmidpunct}
{\mcitedefaultendpunct}{\mcitedefaultseppunct}\relax
\EndOfBibitem
\bibitem[Sprague-Klein \latin{et~al.}(2018)Sprague-Klein, Negru, Madison,
  Coste, Rugg, Felts, McAnally, Banik, Apkarian, Wasielewski, and
  et~al.]{Sprague-Klein_2018_JACS.140.10583}
Sprague-Klein,~E.~A.; Negru,~B.; Madison,~L.~R.; Coste,~S.~C.; Rugg,~B.~K.;
  Felts,~A.~M.; McAnally,~M.~O.; Banik,~M.; Apkarian,~V.~A.;
  Wasielewski,~M.~R.; et~al. {Photoinduced Plasmon-Driven Chemistry in
  trans-1,2-Bis(4-pyridyl)ethylene Gold Nanosphere Oligomers}. \emph{J. Am.
  Chem. Soc.} \textbf{2018}, \emph{140}, 10583--10592\relax
\mciteBstWouldAddEndPuncttrue
\mciteSetBstMidEndSepPunct{\mcitedefaultmidpunct}
{\mcitedefaultendpunct}{\mcitedefaultseppunct}\relax
\EndOfBibitem
\bibitem[Ding \latin{et~al.}(2017)Ding, Mertens, Lombardi, Scherman, and
  Baumberg]{Ding_2017_ACSPhot.4.1453}
Ding,~T.; Mertens,~J.; Lombardi,~A.; Scherman,~O.~A.; Baumberg,~J.~J.
  {Light-Directed Tuning of Plasmon Resonances via Plasmon-Induced
  Polymerization Using Hot Electrons}. \emph{ACS Photonics} \textbf{2017},
  \emph{4}, 1453--1458\relax
\mciteBstWouldAddEndPuncttrue
\mciteSetBstMidEndSepPunct{\mcitedefaultmidpunct}
{\mcitedefaultendpunct}{\mcitedefaultseppunct}\relax
\EndOfBibitem
\bibitem[Koopman \latin{et~al.}(2021)Koopman, Titov, Sarhan, Gaebel,
  Sch\"{u}rmann, Mostafa, Kogikoski, Milosavljevi\'{c}, Stete, Liebig, and
  et~al.]{Koopman_2021_AdvMaterInterf}
Koopman,~W.; Titov,~E.; Sarhan,~R.~M.; Gaebel,~T.; Sch\"{u}rmann,~R.;
  Mostafa,~A.; Kogikoski,~S.; Milosavljevi\'{c},~A.~R.; Stete,~F.; Liebig,~F.;
  et~al. {The Role of Structural Flexibility in Plasmon-Driven Coupling
  Reactions: Kinetic Limitations in the Dimerization of Nitro-Benzenes}.
  \emph{Adv. Mater. Interfaces} \textbf{2021}, \emph{8}, 2101344\relax
\mciteBstWouldAddEndPuncttrue
\mciteSetBstMidEndSepPunct{\mcitedefaultmidpunct}
{\mcitedefaultendpunct}{\mcitedefaultseppunct}\relax
\EndOfBibitem
\bibitem[Cheruvathoor~Poulose \latin{et~al.}(2022)Cheruvathoor~Poulose,
  Zoppellaro, Konidakis, Serpetzoglou, Stratakis, Tomanec, Beller,
  Bakandritsos, and Zbo\v{r}il]{Poulose_2022_NatNanotech.17.485}
Cheruvathoor~Poulose,~A.; Zoppellaro,~G.; Konidakis,~I.; Serpetzoglou,~E.;
  Stratakis,~E.; Tomanec,~O.; Beller,~M.; Bakandritsos,~A.; Zbo\v{r}il,~R.
  {Fast and Selective Reduction of Nitroarenes Under Visible Light With an
  Earth-Abundant Plasmonic Photocatalyst}. \emph{Nature Nanotechnol.}
  \textbf{2022}, \emph{17}, 485--492\relax
\mciteBstWouldAddEndPuncttrue
\mciteSetBstMidEndSepPunct{\mcitedefaultmidpunct}
{\mcitedefaultendpunct}{\mcitedefaultseppunct}\relax
\EndOfBibitem
\bibitem[Ezendam \latin{et~al.}(2022)Ezendam, Herran, Nan, Gruber, Kang,
  Gr\"{o}bmeyer, Lin, Gargiulo, Sousa-Castillo, and
  Cort\'{e}s]{Ezendam_2022_ACSEnergyLett}
Ezendam,~S.; Herran,~M.; Nan,~L.; Gruber,~C.; Kang,~Y.; Gr\"{o}bmeyer,~F.;
  Lin,~R.; Gargiulo,~J.; Sousa-Castillo,~A.; Cort\'{e}s,~E. {Hybrid Plasmonic
  Nanomaterials for Hydrogen Generation and Carbon Dioxide Reduction}.
  \emph{ACS Energy Lett.} \textbf{2022}, \emph{7}, 778--815\relax
\mciteBstWouldAddEndPuncttrue
\mciteSetBstMidEndSepPunct{\mcitedefaultmidpunct}
{\mcitedefaultendpunct}{\mcitedefaultseppunct}\relax
\EndOfBibitem
\bibitem[King \latin{et~al.}(2022)King, Wang, Fonseca~Guzman, and
  Ross]{King_2022_ChemCatal.2.1880}
King,~M.~E.; Wang,~C.; Fonseca~Guzman,~M.~V.; Ross,~M.~B. {Plasmonics for
  Environmental Remediation and Pollutant Degradation}. \emph{Chem. Catal.}
  \textbf{2022}, \emph{2}, 1880--1892\relax
\mciteBstWouldAddEndPuncttrue
\mciteSetBstMidEndSepPunct{\mcitedefaultmidpunct}
{\mcitedefaultendpunct}{\mcitedefaultseppunct}\relax
\EndOfBibitem
\bibitem[Takimoto \latin{et~al.}(2023)Takimoto, Toma, Suda, Shirokura, Tokura,
  Fukuda, Matsumoto, Imai, and Sugimoto]{Takimoto_2023_NatCommun.14.19}
Takimoto,~D.; Toma,~S.; Suda,~Y.; Shirokura,~T.; Tokura,~Y.; Fukuda,~K.;
  Matsumoto,~M.; Imai,~H.; Sugimoto,~W. {Platinum Nanosheets Synthesized via
  Topotactic Reduction of Single-Layer Platinum Oxide Nanosheets for
  Electrocatalysis}. \emph{Nat. Commun.} \textbf{2023}, \emph{14}, 19\relax
\mciteBstWouldAddEndPuncttrue
\mciteSetBstMidEndSepPunct{\mcitedefaultmidpunct}
{\mcitedefaultendpunct}{\mcitedefaultseppunct}\relax
\EndOfBibitem
\bibitem[Kodama \latin{et~al.}(2021)Kodama, Nagai, Kuwaki, Jinnouchi, and
  Morimoto]{Kodama_2021_NatNanotechnol.16.140}
Kodama,~K.; Nagai,~T.; Kuwaki,~A.; Jinnouchi,~R.; Morimoto,~Y. {Challenges in
  Applying Highly Active Pt-Based Nanostructured Catalysts for Oxygen Reduction
  Reactions to Fuel Cell Vehicles}. \emph{Nat. Nanotechnol.} \textbf{2021},
  \emph{16}, 140--147\relax
\mciteBstWouldAddEndPuncttrue
\mciteSetBstMidEndSepPunct{\mcitedefaultmidpunct}
{\mcitedefaultendpunct}{\mcitedefaultseppunct}\relax
\EndOfBibitem
\bibitem[Liu \latin{et~al.}(2020)Liu, Zhao, Peng, Duan, and
  Huang]{Liu_2020_JACS.142.17182}
Liu,~Z.; Zhao,~Z.; Peng,~B.; Duan,~X.; Huang,~Y. {Beyond Extended Surfaces:
  Understanding the Oxygen Reduction Reaction on Nanocatalysts}. \emph{J. Am.
  Chem. Soc.} \textbf{2020}, \emph{142}, 17812--17827\relax
\mciteBstWouldAddEndPuncttrue
\mciteSetBstMidEndSepPunct{\mcitedefaultmidpunct}
{\mcitedefaultendpunct}{\mcitedefaultseppunct}\relax
\EndOfBibitem
\bibitem[Xie \latin{et~al.}(2019)Xie, Niu, Kim, Li, and
  Yang]{Xie_2019_ChemRev.120.1184}
Xie,~C.; Niu,~Z.; Kim,~D.; Li,~M.; Yang,~P. {Surface and Interface Control in
  Nanoparticle Catalysis}. \emph{Chem. Rev.} \textbf{2019}, \emph{120},
  1184--1249\relax
\mciteBstWouldAddEndPuncttrue
\mciteSetBstMidEndSepPunct{\mcitedefaultmidpunct}
{\mcitedefaultendpunct}{\mcitedefaultseppunct}\relax
\EndOfBibitem
\bibitem[Jenkinson \latin{et~al.}(2020)Jenkinson, Wagner, Kornienko, Reisner,
  and Wheatley]{Jenkinson_2020_AdvFunctMater.30.2002633}
Jenkinson,~K.~J.; Wagner,~A.; Kornienko,~N.; Reisner,~E.; Wheatley,~A. E.~H. {A
  One-Pot Route to Faceted FePt-Fe$_3$O$_4$ Dumbbells: Probing
  Morphology--Catalytic Activity Effects in O$_2$ Reduction Catalysis}.
  \emph{Adv. Funct. Mater.} \textbf{2020}, \emph{30}, 2002633\relax
\mciteBstWouldAddEndPuncttrue
\mciteSetBstMidEndSepPunct{\mcitedefaultmidpunct}
{\mcitedefaultendpunct}{\mcitedefaultseppunct}\relax
\EndOfBibitem
\bibitem[Zhang \latin{et~al.}(2021)Zhang, Zhang, and
  Cui]{Zhang_2021_ChemCommun.5.11}
Zhang,~J.; Zhang,~L.; Cui,~Z. {Strategies to Enhance the Electrochemical
  Performances of Pt-Based Intermetallic Catalysts}. \emph{Chem. Commun.}
  \textbf{2021}, \emph{5}, 11--26\relax
\mciteBstWouldAddEndPuncttrue
\mciteSetBstMidEndSepPunct{\mcitedefaultmidpunct}
{\mcitedefaultendpunct}{\mcitedefaultseppunct}\relax
\EndOfBibitem
\bibitem[Lei \latin{et~al.}(2020)Lei, Li, He, Meng, Mu, Yu, Ross, and
  Yang]{Lei_2020_NanoRes.13.638}
Lei,~W.; Li,~M.; He,~L.; Meng,~X.; Mu,~Z.; Yu,~Y.; Ross,~F.~M.; Yang,~W. {A
  General Strategy for Bimetallic Pt-Based Nano-Branched Structures as Highly
  Active and Stable Oxygen Reduction and Methanol Oxidation Bifunctional
  Catalysts}. \emph{Nano Res.} \textbf{2020}, \emph{13}, 638--645\relax
\mciteBstWouldAddEndPuncttrue
\mciteSetBstMidEndSepPunct{\mcitedefaultmidpunct}
{\mcitedefaultendpunct}{\mcitedefaultseppunct}\relax
\EndOfBibitem
\bibitem[Mao \latin{et~al.}(2016)Mao, Chen, Pei, Wang, and
  Li]{Mao_2016_ChemCommun.52.5985}
Mao,~J.; Chen,~Y.; Pei,~J.; Wang,~D.; Li,~Y. {Pt--M (M = Cu, Fe, Zn, etc.)
  Bimetallic Nanomaterials with Abundant Surface Defects and Robust Catalytic
  Properties}. \emph{Chem. Commun.} \textbf{2016}, \emph{52}, 5985--5988\relax
\mciteBstWouldAddEndPuncttrue
\mciteSetBstMidEndSepPunct{\mcitedefaultmidpunct}
{\mcitedefaultendpunct}{\mcitedefaultseppunct}\relax
\EndOfBibitem
\bibitem[Tian \latin{et~al.}(2017)Tian, Xu, Zhang, Wu, Xia, and
  Wang]{Tian_2017_ACSEnergyLett.2.2035}
Tian,~X.~L.; Xu,~Y.~Y.; Zhang,~W.; Wu,~T.; Xia,~B.~Y.; Wang,~X. {Unsupported
  Platinum-Based Electrocatalysts for Oxygen Reduction Reaction}. \emph{ACS
  Energy Lett.} \textbf{2017}, \emph{2}, 2035--2043\relax
\mciteBstWouldAddEndPuncttrue
\mciteSetBstMidEndSepPunct{\mcitedefaultmidpunct}
{\mcitedefaultendpunct}{\mcitedefaultseppunct}\relax
\EndOfBibitem
\bibitem[Zhang \latin{et~al.}(2018)Zhang, Yang, Wang, Dou, and
  Liu]{Zhang_2018_AdvEnergyMater.8.1703597}
Zhang,~B.~W.; Yang,~H.~L.; Wang,~Y.~X.; Dou,~S.~X.; Liu,~H.~K. {A Comprehensive
  Review on Controlling Surface Composition of Pt-Based Bimetallic
  Electrocatalysts}. \emph{Adv. Energy Mater.} \textbf{2018}, \emph{8},
  1703597\relax
\mciteBstWouldAddEndPuncttrue
\mciteSetBstMidEndSepPunct{\mcitedefaultmidpunct}
{\mcitedefaultendpunct}{\mcitedefaultseppunct}\relax
\EndOfBibitem
\bibitem[Chaudhari \latin{et~al.}(2018)Chaudhari, Joo, Kwon, Kim, Kim, Joo, and
  Lee]{Chaudhari_2018_NanoRes.11.6111}
Chaudhari,~N.~K.; Joo,~J.; Kwon,~H.~B.; Kim,~B.; Kim,~H.~Y.; Joo,~S.~H.;
  Lee,~K. {Nanodendrites of Platinum-Group Metals for Electrocatalytic
  Applications}. \emph{Nano Res.} \textbf{2018}, \emph{11}, 6111--6140\relax
\mciteBstWouldAddEndPuncttrue
\mciteSetBstMidEndSepPunct{\mcitedefaultmidpunct}
{\mcitedefaultendpunct}{\mcitedefaultseppunct}\relax
\EndOfBibitem
\bibitem[Ming and Wheatley(223)Ming, and Wheatley]{Ming_2023_Nanoscale.15.8814}
Ming,~S.; Wheatley,~A. E.~H. {Manipulating Morphology and Composition in
  Colloidal Heterometallic Nanopods and Nanodendrites}. \emph{Nanoscale}
  \textbf{223}, \emph{15}, 8814--8824\relax
\mciteBstWouldAddEndPuncttrue
\mciteSetBstMidEndSepPunct{\mcitedefaultmidpunct}
{\mcitedefaultendpunct}{\mcitedefaultseppunct}\relax
\EndOfBibitem
\bibitem[Chen \latin{et~al.}(2022)Chen, Nguyen, and
  Xia]{Chen_2022_ChemNanoMater}
Chen,~R.; Nguyen,~Q.~N.; Xia,~Y. {Oriented Attachment: A Unique Mechanism for
  the Colloidal Synthesis of Metal Nanostructures}. \emph{ChemNanoMat}
  \textbf{2022}, \emph{8}, e202100474\relax
\mciteBstWouldAddEndPuncttrue
\mciteSetBstMidEndSepPunct{\mcitedefaultmidpunct}
{\mcitedefaultendpunct}{\mcitedefaultseppunct}\relax
\EndOfBibitem
\bibitem[Gao \latin{et~al.}(2023)Gao, Zhang, Nakajima, He, Liu, Zhang, Wang,
  and Wu]{Gao_2023_NatCommun.14.2640}
Gao,~R.~T.; Zhang,~J.; Nakajima,~T.; He,~J.; Liu,~X.; Zhang,~X.; Wang,~L.;
  Wu,~L. {Single-Atomic-Site Platinum Steers Photogenerated Charge Carrier
  Lifetime of Hematite Nanoflakes for Photoelectrochemical Water Splitting}.
  \emph{Nat. Commun.} \textbf{2023}, \emph{14}, 2640\relax
\mciteBstWouldAddEndPuncttrue
\mciteSetBstMidEndSepPunct{\mcitedefaultmidpunct}
{\mcitedefaultendpunct}{\mcitedefaultseppunct}\relax
\EndOfBibitem
\bibitem[Hughes \latin{et~al.}(2021)Hughes, Haque, Northey, and
  Giddey]{Hughes_2021_Resources.10.93}
Hughes,~A.~E.; Haque,~N.; Northey,~S.~A.; Giddey,~S. {Platinum Group Metals: A
  Review of Resources, Production and Usage with a Focus on Catalysts}.
  \emph{Resources} \textbf{2021}, \emph{10}, 93\relax
\mciteBstWouldAddEndPuncttrue
\mciteSetBstMidEndSepPunct{\mcitedefaultmidpunct}
{\mcitedefaultendpunct}{\mcitedefaultseppunct}\relax
\EndOfBibitem
\bibitem[Sushma \latin{et~al.}(2018)Sushma, Kumari, and
  Saroha]{Sushma2018performance}
Sushma; Kumari,~M.; Saroha,~A.~K. {Performance of Various Catalysts on
  Treatment of Refractory Pollutants in Industrial Wastewater by Catalytic Wet
  Air Oxidation: A Review}. \emph{J. Environ. Manage.} \textbf{2018},
  \emph{228}, 169--188\relax
\mciteBstWouldAddEndPuncttrue
\mciteSetBstMidEndSepPunct{\mcitedefaultmidpunct}
{\mcitedefaultendpunct}{\mcitedefaultseppunct}\relax
\EndOfBibitem
\bibitem[Fujiwara \latin{et~al.}(2017)Fujiwara, Okuyama, and
  Pratsinis]{Fujiwara_2017_EnvironSci}
Fujiwara,~K.; Okuyama,~K.; Pratsinis,~S.~E. {Metal--Support Interactions in
  Catalysts for Environmental Remediation}. \emph{Environ. Sci. Nano}
  \textbf{2017}, \emph{4}, 2076--2092\relax
\mciteBstWouldAddEndPuncttrue
\mciteSetBstMidEndSepPunct{\mcitedefaultmidpunct}
{\mcitedefaultendpunct}{\mcitedefaultseppunct}\relax
\EndOfBibitem
\bibitem[Huth \latin{et~al.}(2018)Huth, Porrati, and
  Dobrovolskiy]{Huth_2018_MicroelectronEng}
Huth,~M.; Porrati,~F.; Dobrovolskiy,~O.~V. {Focused Electron Beam Induced
  Deposition Meets Materials Science}. \emph{Microelectron. Eng.}
  \textbf{2018}, \emph{185-186}, 9--28\relax
\mciteBstWouldAddEndPuncttrue
\mciteSetBstMidEndSepPunct{\mcitedefaultmidpunct}
{\mcitedefaultendpunct}{\mcitedefaultseppunct}\relax
\EndOfBibitem
\bibitem[{De Teresa} \latin{et~al.}(2016){De Teresa}, Fern\'{a}ndez-Pacheco,
  C\'{o}rdoba, Serrano-Ram\'{o}n, Sangiao, and
  Ibarra]{DeTeresa_2016_JPD.49.243003}
{De Teresa},~J.~M.; Fern\'{a}ndez-Pacheco,~A.; C\'{o}rdoba,~R.;
  Serrano-Ram\'{o}n,~L.; Sangiao,~S.; Ibarra,~M.~R. {Review of Magnetic
  Nanostructures Grown by Focused Electron Beam Induced Deposition (FEBID)}.
  \emph{J. Phys. D: Appl. Phys.} \textbf{2016}, \emph{49}, 243003\relax
\mciteBstWouldAddEndPuncttrue
\mciteSetBstMidEndSepPunct{\mcitedefaultmidpunct}
{\mcitedefaultendpunct}{\mcitedefaultseppunct}\relax
\EndOfBibitem
\bibitem[Huth \latin{et~al.}(2021)Huth, Porrati, and
  Barth]{Huth_2021_JAP.130.170901}
Huth,~M.; Porrati,~F.; Barth,~S. {Living Up to Its Potential -- Direct-Write
  Nanofabrication with Focused Electron Beams}. \emph{J. Appl. Phys.}
  \textbf{2021}, \emph{130}, 170901\relax
\mciteBstWouldAddEndPuncttrue
\mciteSetBstMidEndSepPunct{\mcitedefaultmidpunct}
{\mcitedefaultendpunct}{\mcitedefaultseppunct}\relax
\EndOfBibitem
\bibitem[Reyntjens and Puers(2000)Reyntjens, and Puers]{Reyntjens_FIBID_2000}
Reyntjens,~S.; Puers,~R. {Focused Ion Beam Induced Deposition: Fabrication of
  Three-Dimensional Microstructures and Young's Modulus of the Deposited
  Material}. \emph{J. Micromech. Microeng.} \textbf{2000}, \emph{10},
  181--188\relax
\mciteBstWouldAddEndPuncttrue
\mciteSetBstMidEndSepPunct{\mcitedefaultmidpunct}
{\mcitedefaultendpunct}{\mcitedefaultseppunct}\relax
\EndOfBibitem
\bibitem[Cui \latin{et~al.}(2012)Cui, Li, Luo, Liu, and
  Gu]{Cui_2012_APL.100.143106}
Cui,~A.; Li,~W.; Luo,~Q.; Liu,~Z.; Gu,~C. {Freestanding Nanostructures for
  Three-Dimensional Superconducting Nanodevices}. \emph{Appl. Phys. Lett.}
  \textbf{2012}, \emph{100}, 143106\relax
\mciteBstWouldAddEndPuncttrue
\mciteSetBstMidEndSepPunct{\mcitedefaultmidpunct}
{\mcitedefaultendpunct}{\mcitedefaultseppunct}\relax
\EndOfBibitem
\bibitem[Nanda \latin{et~al.}(2015)Nanda, {van Veldhoven}, Maas, Sadeghian, and
  Alkemade]{Nanda_2015_JVSTB.33.06F503}
Nanda,~G.; {van Veldhoven},~E.; Maas,~D.; Sadeghian,~H.; Alkemade,~P. F.~A.
  {Helium Ion Beam Induced Growth of Hammerhead AFM Probes}. \emph{J. Vac. Sci.
  Technol. B} \textbf{2015}, \emph{33}, 06F503\relax
\mciteBstWouldAddEndPuncttrue
\mciteSetBstMidEndSepPunct{\mcitedefaultmidpunct}
{\mcitedefaultendpunct}{\mcitedefaultseppunct}\relax
\EndOfBibitem
\bibitem[C\'{o}rdoba \latin{et~al.}(2018)C\'{o}rdoba, Ibarra, Mailly, and {De
  Teresa}]{Cordoba_2018_NanoLett.18.1379}
C\'{o}rdoba,~R.; Ibarra,~A.; Mailly,~D.; {De Teresa},~J.~M. {Vertical Growth of
  Superconducting Crystalline Hollow Nanowires by He+ Focused Ion Beam Induced
  Deposition}. \emph{Nano Lett.} \textbf{2018}, \emph{18}, 1379--1386\relax
\mciteBstWouldAddEndPuncttrue
\mciteSetBstMidEndSepPunct{\mcitedefaultmidpunct}
{\mcitedefaultendpunct}{\mcitedefaultseppunct}\relax
\EndOfBibitem
\bibitem[Swiderek \latin{et~al.}(2018)Swiderek, Marbach, and
  Hagen]{Swiderek_2018_BJN.9.1317}
Swiderek,~P.; Marbach,~H.; Hagen,~C.~W. {Chemistry for Electron-Induced
  Nanofabrication}. \emph{Beilstein J. Nanotechnol.} \textbf{2018}, \emph{9},
  1317--1320\relax
\mciteBstWouldAddEndPuncttrue
\mciteSetBstMidEndSepPunct{\mcitedefaultmidpunct}
{\mcitedefaultendpunct}{\mcitedefaultseppunct}\relax
\EndOfBibitem
\bibitem[Botman \latin{et~al.}(2009)Botman, Mulders, and
  Hagen]{Botman_2009_Nanotechnol.20.372001}
Botman,~A.; Mulders,~J. J.~L.; Hagen,~C.~W. {Creating Pure Nanostructures From
  Electron-Beam-Induced Deposition Using Purification Techniques: A Technology
  Perspective}. \emph{Nanotechnology} \textbf{2009}, \emph{20}, 372001\relax
\mciteBstWouldAddEndPuncttrue
\mciteSetBstMidEndSepPunct{\mcitedefaultmidpunct}
{\mcitedefaultendpunct}{\mcitedefaultseppunct}\relax
\EndOfBibitem
\bibitem[Geier \latin{et~al.}(2014)Geier, Gspan, Winkler, Schmied, Fowlkes,
  Fitzek, Rauch, Rattenberger, Rack, and Plank]{Geier_2014_JPCC.118.14009}
Geier,~B.; Gspan,~C.; Winkler,~R.; Schmied,~R.; Fowlkes,~J.~D.; Fitzek,~H.;
  Rauch,~S.; Rattenberger,~J.; Rack,~P.~D.; Plank,~H. {Rapid and Highly Compact
  Purification for Focused Electron Beam Induced Deposits: A Low Temperature
  Approach Using Electron Stimulated H$_2$O Reactions}. \emph{J. Phys. Chem. C}
  \textbf{2014}, \emph{118}, 14009--14016\relax
\mciteBstWouldAddEndPuncttrue
\mciteSetBstMidEndSepPunct{\mcitedefaultmidpunct}
{\mcitedefaultendpunct}{\mcitedefaultseppunct}\relax
\EndOfBibitem
\bibitem[Fowlkes \latin{et~al.}(2015)Fowlkes, Geier, Lewis, Rack, Stanford,
  Winkler, and Plank]{Fowlkes_2015_PCCP.17.18294}
Fowlkes,~J.~D.; Geier,~B.; Lewis,~B.~B.; Rack,~P.~D.; Stanford,~M.~G.;
  Winkler,~R.; Plank,~H. {Electron Nanoprobe Induced Oxidation: A Simulation of
  Direct-Write Purification}. \emph{Phys. Chem. Chem. Phys.} \textbf{2015},
  \emph{17}, 18294--18304\relax
\mciteBstWouldAddEndPuncttrue
\mciteSetBstMidEndSepPunct{\mcitedefaultmidpunct}
{\mcitedefaultendpunct}{\mcitedefaultseppunct}\relax
\EndOfBibitem
\bibitem[Prosvetov \latin{et~al.}(2023)Prosvetov, Verkhovtsev, Sushko, and
  Solov'yov]{Prosvetov_2023_EPJD}
Prosvetov,~A.; Verkhovtsev,~A.~V.; Sushko,~G.; Solov'yov,~A.~V. {Atomistic
  Modeling of Thermal Effects in Focused Electron Beam-Induced Deposition of
  Me$_2$Au(tfac)}. \emph{Eur. Phys. J. D} \textbf{2023}, \emph{77}, 15\relax
\mciteBstWouldAddEndPuncttrue
\mciteSetBstMidEndSepPunct{\mcitedefaultmidpunct}
{\mcitedefaultendpunct}{\mcitedefaultseppunct}\relax
\EndOfBibitem
\bibitem[Huth \latin{et~al.}(2009)Huth, Klingenberger, Grimm, Porrati, and
  Sachser]{Huth_2009_NJP.11.033032}
Huth,~M.; Klingenberger,~D.; Grimm,~C.; Porrati,~F.; Sachser,~R. {Conductance
  Regimes of W-based Granular Metals Prepared by Electron Beam Induced
  Deposition}. \emph{New J. Phys.} \textbf{2009}, \emph{11}, 033032\relax
\mciteBstWouldAddEndPuncttrue
\mciteSetBstMidEndSepPunct{\mcitedefaultmidpunct}
{\mcitedefaultendpunct}{\mcitedefaultseppunct}\relax
\EndOfBibitem
\bibitem[Huth \latin{et~al.}(2020)Huth, Porrati, Gruszka, and
  Barth]{Huth_2020_Micromachines.11.28}
Huth,~M.; Porrati,~F.; Gruszka,~P.; Barth,~S. {Temperature-Dependent Growth
  Characteristics of Nb- and CoFe-Based Nanostructures by Direct-Write Using
  Focused Electron Beam-Induced Deposition}. \emph{Micromachines}
  \textbf{2020}, \emph{11}, 28\relax
\mciteBstWouldAddEndPuncttrue
\mciteSetBstMidEndSepPunct{\mcitedefaultmidpunct}
{\mcitedefaultendpunct}{\mcitedefaultseppunct}\relax
\EndOfBibitem
\bibitem[Stanford \latin{et~al.}(2016)Stanford, Mahady, Lewis, Fowlkes, Tan,
  Livengood, Magel, Moore, and Rack]{Stanford_2016_ACSAMI.8.29155}
Stanford,~M.~G.; Mahady,~K.; Lewis,~B.~B.; Fowlkes,~J.~D.; Tan,~S.;
  Livengood,~R.; Magel,~G.~A.; Moore,~T.~M.; Rack,~P.~D. {Laser-Assisted
  Focused He$^+$ Ion Beam Induced Etching with and without XeF$_2$ Gas Assist}.
  \emph{ACS Appl. Mater. Interfaces} \textbf{2016}, \emph{8},
  29155--29162\relax
\mciteBstWouldAddEndPuncttrue
\mciteSetBstMidEndSepPunct{\mcitedefaultmidpunct}
{\mcitedefaultendpunct}{\mcitedefaultseppunct}\relax
\EndOfBibitem
\bibitem[Seewald \latin{et~al.}(2021)Seewald, Winkler, Kothleitner, and
  Plank]{Seewald_2021_Micromachines.12.115}
Seewald,~L.~M.; Winkler,~R.; Kothleitner,~G.; Plank,~H. {Expanding 3D
  Nanoprinting Performance by Blurring the Electron Beam}. \emph{Micromachines}
  \textbf{2021}, \emph{12}, 115\relax
\mciteBstWouldAddEndPuncttrue
\mciteSetBstMidEndSepPunct{\mcitedefaultmidpunct}
{\mcitedefaultendpunct}{\mcitedefaultseppunct}\relax
\EndOfBibitem
\bibitem[Frabboni \latin{et~al.}(2006)Frabboni, Gazzadi, Felisari, and
  Spessot]{Frabboni_2006_APL.88.213116}
Frabboni,~S.; Gazzadi,~G.~C.; Felisari,~L.; Spessot,~A. {Fabrication by
  Electron Beam Induced Deposition and Transmission Electron Microscopic
  Characterization of Sub-10-nm Freestanding Pt Nanowires}. \emph{Appl. Phys.
  Lett.} \textbf{2006}, \emph{88}, 213116\relax
\mciteBstWouldAddEndPuncttrue
\mciteSetBstMidEndSepPunct{\mcitedefaultmidpunct}
{\mcitedefaultendpunct}{\mcitedefaultseppunct}\relax
\EndOfBibitem
\bibitem[Keller \latin{et~al.}(2018)Keller, {Al Mamoori}, Pieper, Gspan,
  Stockem, Schr\"{o}der, Barth, Winkler, Plank, Pohlit, M\"{u}ller, and
  Huth]{Keller_2018_SciRep.8.6160}
Keller,~L.; {Al Mamoori},~M. K.~I.; Pieper,~J.; Gspan,~C.; Stockem,~I.;
  Schr\"{o}der,~C.; Barth,~S.; Winkler,~R.; Plank,~H.; Pohlit,~M.;
  M\"{u}ller,~J.; Huth,~M. {Direct-Write of Free-Form Building Blocks for
  Artificial Magnetic 3D Lattices}. \emph{Sci. Rep.} \textbf{2018}, \emph{8},
  6160\relax
\mciteBstWouldAddEndPuncttrue
\mciteSetBstMidEndSepPunct{\mcitedefaultmidpunct}
{\mcitedefaultendpunct}{\mcitedefaultseppunct}\relax
\EndOfBibitem
\bibitem[Sanz-Hern\'{a}ndez \latin{et~al.}(2020)Sanz-Hern\'{a}ndez,
  Hierro-Rodriguez, Donnelly, Pablo-Navarro, Sorrentino, Pereiro, Mag\'{e}n,
  McVitie, {de Teresa}, Ferrer, Fischer, and
  Fern\'{a}ndez-Pacheco]{Sanz-Hernandez_2020_ACSNano.14.8084}
Sanz-Hern\'{a}ndez,~D.; Hierro-Rodriguez,~A.; Donnelly,~C.; Pablo-Navarro,~J.;
  Sorrentino,~A.; Pereiro,~E.; Mag\'{e}n,~C.; McVitie,~S.; {de Teresa},~J.~M.;
  Ferrer,~S.; Fischer,~P.; Fern\'{a}ndez-Pacheco,~A. {Artificial Double-Helix
  for Geometrical Control of Magnetic Chirality}. \emph{ACS Nano}
  \textbf{2020}, \emph{14}, 8084--8092\relax
\mciteBstWouldAddEndPuncttrue
\mciteSetBstMidEndSepPunct{\mcitedefaultmidpunct}
{\mcitedefaultendpunct}{\mcitedefaultseppunct}\relax
\EndOfBibitem
\bibitem[Passaseo \latin{et~al.}(2017)Passaseo, Esposito, Cuscun\`{a}, and
  Tasco]{Passaseo_2017_AdvOptMater.5.1601079}
Passaseo,~A.; Esposito,~M.; Cuscun\`{a},~M.; Tasco,~V. {Materials and 3D
  Designs of Helix Nanostructures for Chirality at Optical Frequencies}.
  \emph{Adv. Opt. Mater.} \textbf{2017}, \emph{5}, 1601079\relax
\mciteBstWouldAddEndPuncttrue
\mciteSetBstMidEndSepPunct{\mcitedefaultmidpunct}
{\mcitedefaultendpunct}{\mcitedefaultseppunct}\relax
\EndOfBibitem
\bibitem[Winkler \latin{et~al.}(2017)Winkler, Schmidt, Haselmann, Fowlkes,
  Lewis, Kothleitner, Rack, and Plank]{Winkler_2017_ACSApplMatInt.9.8233}
Winkler,~R.; Schmidt,~F.-P.; Haselmann,~U.; Fowlkes,~J.~D.; Lewis,~B.~B.;
  Kothleitner,~G.; Rack,~P.~D.; Plank,~H. {Direct-Write 3D Nanoprinting of
  Plasmonic Structures}. \emph{ACS Appl. Mater. Interfaces} \textbf{2017},
  \emph{9}, 8233--8240\relax
\mciteBstWouldAddEndPuncttrue
\mciteSetBstMidEndSepPunct{\mcitedefaultmidpunct}
{\mcitedefaultendpunct}{\mcitedefaultseppunct}\relax
\EndOfBibitem
\bibitem[Esposito \latin{et~al.}(2015)Esposito, Tasco, Cuscun\`{a}, Todisco,
  Benedetti, Tarantini, {De Giorgi}, Sanvitto, and
  Passaseo]{Esposito_2015_ACSPhoton.2.105}
Esposito,~M.; Tasco,~V.; Cuscun\`{a},~M.; Todisco,~F.; Benedetti,~A.;
  Tarantini,~I.; {De Giorgi},~M.; Sanvitto,~D.; Passaseo,~A. {Nanoscale 3D
  Chiral Plasmonic Helices with Circular Dichroism at Visible Frequencies}.
  \emph{ACS Photonics} \textbf{2015}, \emph{2}, 105--114\relax
\mciteBstWouldAddEndPuncttrue
\mciteSetBstMidEndSepPunct{\mcitedefaultmidpunct}
{\mcitedefaultendpunct}{\mcitedefaultseppunct}\relax
\EndOfBibitem
\bibitem[Beard and Gordeev(2010)Beard, and
  Gordeev]{Beard_2010_Nanotechnol.21.475702}
Beard,~J.~D.; Gordeev,~S.~N. {Large Flexibility of High Aspect Ratio Carbon
  Nanostructures Fabricated by Electron-Beam-Induced Deposition}.
  \emph{Nanotechnology} \textbf{2010}, \emph{21}, 475702\relax
\mciteBstWouldAddEndPuncttrue
\mciteSetBstMidEndSepPunct{\mcitedefaultmidpunct}
{\mcitedefaultendpunct}{\mcitedefaultseppunct}\relax
\EndOfBibitem
\bibitem[Burbridge and Gordeev(2009)Burbridge, and
  Gordeev]{Burbridge_2009_Nanotechnol.20.285308}
Burbridge,~D.~J.; Gordeev,~S.~N. {Proximity Effects in Free-Standing EBID
  Structures}. \emph{Nanotechnology} \textbf{2009}, \emph{20}, 285308\relax
\mciteBstWouldAddEndPuncttrue
\mciteSetBstMidEndSepPunct{\mcitedefaultmidpunct}
{\mcitedefaultendpunct}{\mcitedefaultseppunct}\relax
\EndOfBibitem
\bibitem[Mutunga \latin{et~al.}(2019)Mutunga, Winkler, Sattelkow, Rack, Plank,
  and Fowlkes]{Mutunga_2019_ACSNano.13.5198}
Mutunga,~E.; Winkler,~R.; Sattelkow,~J.; Rack,~P.~D.; Plank,~H.; Fowlkes,~J.~D.
  {Impact of Electron-Beam Heating during 3D Nanoprinting}. \emph{ACS Nano}
  \textbf{2019}, \emph{13}, 5198--5213\relax
\mciteBstWouldAddEndPuncttrue
\mciteSetBstMidEndSepPunct{\mcitedefaultmidpunct}
{\mcitedefaultendpunct}{\mcitedefaultseppunct}\relax
\EndOfBibitem
\bibitem[Fowlkes \latin{et~al.}(2020)Fowlkes, Winkler, Mutunga, Rack, and
  Plank]{Fowlkes_2020_Micromachines.11.8}
Fowlkes,~J.~D.; Winkler,~R.; Mutunga,~E.; Rack,~P.~D.; Plank,~H. {Simulation
  Informed CAD for 3D Nanoprinting}. \emph{Micromachines} \textbf{2020},
  \emph{11}, 8\relax
\mciteBstWouldAddEndPuncttrue
\mciteSetBstMidEndSepPunct{\mcitedefaultmidpunct}
{\mcitedefaultendpunct}{\mcitedefaultseppunct}\relax
\EndOfBibitem
\bibitem[Bret \latin{et~al.}(2004)Bret, Utke, Gaillard, and
  Hoffmann]{Bret_2004_JVSTB.22.2504}
Bret,~T.; Utke,~I.; Gaillard,~C.; Hoffmann,~P. {Periodic Structure Formation by
  Focused Electron-Beam-Induced Deposition}. \emph{J. Vac. Sci. Technol. B}
  \textbf{2004}, \emph{22}, 2504--2510\relax
\mciteBstWouldAddEndPuncttrue
\mciteSetBstMidEndSepPunct{\mcitedefaultmidpunct}
{\mcitedefaultendpunct}{\mcitedefaultseppunct}\relax
\EndOfBibitem
\bibitem[K.~M{\o}lhave \latin{et~al.}(2004)K.~M{\o}lhave, Dohn, and
  B{\o}ggild]{Molhave_2004_Nanotechnol.15.1047}
K.~M{\o}lhave,~D.~N.; Dohn,~S.; B{\o}ggild,~P. {Constructing, Connecting and
  Soldering Nanostructures by Environmental Electron Beam Deposition}.
  \emph{Nanotechnology} \textbf{2004}, \emph{15}, 1047--1053\relax
\mciteBstWouldAddEndPuncttrue
\mciteSetBstMidEndSepPunct{\mcitedefaultmidpunct}
{\mcitedefaultendpunct}{\mcitedefaultseppunct}\relax
\EndOfBibitem
\bibitem[Fowlkes \latin{et~al.}(2016)Fowlkes, Winkler, Lewis, Stanford, Plank,
  and Rack]{Fowlkes_2016_ACSNano.10.6163}
Fowlkes,~J.~D.; Winkler,~R.; Lewis,~B.~B.; Stanford,~M.~G.; Plank,~H.;
  Rack,~P.~D. {Simulation-Guided 3D Nanomanufacturing via Focused Electron Beam
  Induced Deposition}. \emph{ACS Nano} \textbf{2016}, \emph{10},
  6163--6172\relax
\mciteBstWouldAddEndPuncttrue
\mciteSetBstMidEndSepPunct{\mcitedefaultmidpunct}
{\mcitedefaultendpunct}{\mcitedefaultseppunct}\relax
\EndOfBibitem
\bibitem[Hirt \latin{et~al.}(2017)Hirt, Reiser, Spolenak, and
  Zambelli]{Hirt_2017_AdvMater.29.1604211}
Hirt,~L.; Reiser,~A.; Spolenak,~R.; Zambelli,~T. {Additive Manufacturing of
  Metal Structures at the Micrometer Scale}. \emph{Adv. Mater.} \textbf{2017},
  \emph{29}, 1604211\relax
\mciteBstWouldAddEndPuncttrue
\mciteSetBstMidEndSepPunct{\mcitedefaultmidpunct}
{\mcitedefaultendpunct}{\mcitedefaultseppunct}\relax
\EndOfBibitem
\bibitem[Sanz-Hern{\'{a}}ndez and
  Fern{\'{a}}ndez-Pacheco(2017)Sanz-Hern{\'{a}}ndez, and
  Fern{\'{a}}ndez-Pacheco]{Sanz-Hernandez_2017_BJN.8.2151}
Sanz-Hern{\'{a}}ndez,~D.; Fern{\'{a}}ndez-Pacheco,~A. {Modelling Focused
  Electron Beam Induced Deposition Beyond Langmuir Adsorption}. \emph{Beilstein
  J. Nanotechnol.} \textbf{2017}, \emph{8}, 2151--2161\relax
\mciteBstWouldAddEndPuncttrue
\mciteSetBstMidEndSepPunct{\mcitedefaultmidpunct}
{\mcitedefaultendpunct}{\mcitedefaultseppunct}\relax
\EndOfBibitem
\bibitem[Plank \latin{et~al.}(2008)Plank, Gspan, Dienstleder, Kothleitner, and
  Hofer]{Plank_2008_Nanotechnol.19.485302}
Plank,~H.; Gspan,~C.; Dienstleder,~M.; Kothleitner,~G.; Hofer,~F. {The
  Influence of Beam Defocus on Volume Growth Rates for Electron Beam Induced
  Platinum Deposition}. \emph{Nanotechnology} \textbf{2008}, \emph{19},
  485302\relax
\mciteBstWouldAddEndPuncttrue
\mciteSetBstMidEndSepPunct{\mcitedefaultmidpunct}
{\mcitedefaultendpunct}{\mcitedefaultseppunct}\relax
\EndOfBibitem
\bibitem[Kuhness \latin{et~al.}(2021)Kuhness, Gruber, Winkler, Sattelkow,
  Fitzek, Letofsky-Papst, Kothleitner, and
  Plank]{Kuhness_2021_ACSApplMatInt.13.1178}
Kuhness,~D.; Gruber,~A.; Winkler,~R.; Sattelkow,~J.; Fitzek,~H.;
  Letofsky-Papst,~I.; Kothleitner,~G.; Plank,~H. {High-Fidelity 3D Nanoprinting
  of Plasmonic Gold Nanoantennas}. \emph{ACS Appl. Mater. Interfaces}
  \textbf{2021}, \emph{13}, 1178--1191\relax
\mciteBstWouldAddEndPuncttrue
\mciteSetBstMidEndSepPunct{\mcitedefaultmidpunct}
{\mcitedefaultendpunct}{\mcitedefaultseppunct}\relax
\EndOfBibitem
\bibitem[Pablo-Navarro \latin{et~al.}(2019)Pablo-Navarro, Sangiao, Mag\'{e}n,
  and {Mar\'{i}a de Teresa}]{Pablo-Navarro_2019_Nanotechnol.30.505302}
Pablo-Navarro,~J.; Sangiao,~S.; Mag\'{e}n,~C.; {Mar\'{i}a de Teresa},~J.
  {Diameter Modulation of 3D Nanostructures in Focused Electron Beam Induced
  Deposition Using Local Electric Fields and Beam Defocus}.
  \emph{Nanotechnology} \textbf{2019}, \emph{30}, 505302\relax
\mciteBstWouldAddEndPuncttrue
\mciteSetBstMidEndSepPunct{\mcitedefaultmidpunct}
{\mcitedefaultendpunct}{\mcitedefaultseppunct}\relax
\EndOfBibitem
\bibitem[Winkler \latin{et~al.}(2018)Winkler, Lewis, Fowlkes, Rack, and
  Plank]{Winkler_2018_ACSApplNanoMater.1.1014}
Winkler,~R.; Lewis,~B.~B.; Fowlkes,~J.~D.; Rack,~P.~D.; Plank,~H.
  {High-Fidelity 3D-Nanoprinting via Focused Electron Beams: Growth
  Fundamentals}. \emph{ACS Appl. Nano Mater.} \textbf{2018}, \emph{1},
  1014--1027\relax
\mciteBstWouldAddEndPuncttrue
\mciteSetBstMidEndSepPunct{\mcitedefaultmidpunct}
{\mcitedefaultendpunct}{\mcitedefaultseppunct}\relax
\EndOfBibitem
\bibitem[Fowlkes \latin{et~al.}(2018)Fowlkes, Winkler, Lewis,
  Fern\'{a}ndez-Pacheco, Skoric, Sanz-Hern\'{a}ndez, Stanford, Mutunga, Rack,
  and Plank]{Fowlkes_2018_ACS_ApplNanoMater.1.1028}
Fowlkes,~J.~D.; Winkler,~R.; Lewis,~B.~B.; Fern\'{a}ndez-Pacheco,~A.;
  Skoric,~L.; Sanz-Hern\'{a}ndez,~D.; Stanford,~M.~G.; Mutunga,~E.;
  Rack,~P.~D.; Plank,~H. {High-Fidelity 3D-Nanoprinting via Focused Electron
  Beams: Computer-Aided Design (3BID)}. \emph{ACS Appl. Nano Mater.}
  \textbf{2018}, \emph{1}, 1028--1041\relax
\mciteBstWouldAddEndPuncttrue
\mciteSetBstMidEndSepPunct{\mcitedefaultmidpunct}
{\mcitedefaultendpunct}{\mcitedefaultseppunct}\relax
\EndOfBibitem
\bibitem[Toth \latin{et~al.}(2015)Toth, Lobo, Friedli, Szkudlarek, and
  Utke]{Toth_2015_BJN.6.1518}
Toth,~M.; Lobo,~C.; Friedli,~V.; Szkudlarek,~A.; Utke,~I. {Continuum Models of
  Focused Electron Beam Induced Processing}. \emph{Beilstein J. Nanotechnol.}
  \textbf{2015}, \emph{6}, 1518--1540\relax
\mciteBstWouldAddEndPuncttrue
\mciteSetBstMidEndSepPunct{\mcitedefaultmidpunct}
{\mcitedefaultendpunct}{\mcitedefaultseppunct}\relax
\EndOfBibitem
\bibitem[Guo \latin{et~al.}(2012)Guo, Kometani, Warisawa, and
  Ishihara]{Guo_2012_JpnJAP.51.065001}
Guo,~D.; Kometani,~R.; Warisawa,~S.; Ishihara,~S. {Three-Dimensional
  Nanostructure Fabrication by Controlling Downward Growth on Focused-Ion-Beam
  Chemical Vapor Deposition}. \emph{Jpn. J. Appl. Phys.} \textbf{2012},
  \emph{51}, 065001\relax
\mciteBstWouldAddEndPuncttrue
\mciteSetBstMidEndSepPunct{\mcitedefaultmidpunct}
{\mcitedefaultendpunct}{\mcitedefaultseppunct}\relax
\EndOfBibitem
\bibitem[Guo \latin{et~al.}(2013)Guo, Kometani, Warisawa, and
  Ishihara]{Guo_2013_JVSTB.31.061601}
Guo,~D.; Kometani,~R.; Warisawa,~S.; Ishihara,~S. {Growth of Ultra-Long
  Free-Space-Nanowire by the Real-Time Feedback Control of the Scanning Speed
  on Focused-Ion-Beam Chemical Vapor Deposition}. \emph{J. Vac. Sci. Technol.
  B} \textbf{2013}, \emph{31}, 061601\relax
\mciteBstWouldAddEndPuncttrue
\mciteSetBstMidEndSepPunct{\mcitedefaultmidpunct}
{\mcitedefaultendpunct}{\mcitedefaultseppunct}\relax
\EndOfBibitem
\bibitem[Winkler \latin{et~al.}(2014)Winkler, Fowlkes, Szkudlarek, Utke, Rack,
  and Plank]{Winkler_2014_ACS_ApplMatInt.6.2987}
Winkler,~R.; Fowlkes,~J.; Szkudlarek,~A.; Utke,~I.; Rack,~P.~D.; Plank,~H. {The
  Nanoscale Implications of a Molecular Gas Beam during Electron Beam Induced
  Deposition}. \emph{ACS Appl. Mater. Interfaces} \textbf{2014}, \emph{6},
  2987--2995\relax
\mciteBstWouldAddEndPuncttrue
\mciteSetBstMidEndSepPunct{\mcitedefaultmidpunct}
{\mcitedefaultendpunct}{\mcitedefaultseppunct}\relax
\EndOfBibitem
\bibitem[Olsen and Bohr(2010)Olsen, and Bohr]{Olsen_2010_TheorChemAcc.125.207}
Olsen,~K.; Bohr,~J. {The Generic Geometry of Helices and Their Close-Packed
  Structures}. \emph{Theor. Chem. Acc.} \textbf{2010}, \emph{125},
  207--215\relax
\mciteBstWouldAddEndPuncttrue
\mciteSetBstMidEndSepPunct{\mcitedefaultmidpunct}
{\mcitedefaultendpunct}{\mcitedefaultseppunct}\relax
\EndOfBibitem
\bibitem[Keller and Huth(2018)Keller, and Huth]{Keller_2018_BJN.9.2581}
Keller,~L.; Huth,~M. {Pattern Generation for Direct-Write Three-Dimensional
  Nanoscale Structures via Focused Electron Beam Induced Deposition}.
  \emph{Beilstein J. Nanotechnol.} \textbf{2018}, \emph{9}, 2581--2598\relax
\mciteBstWouldAddEndPuncttrue
\mciteSetBstMidEndSepPunct{\mcitedefaultmidpunct}
{\mcitedefaultendpunct}{\mcitedefaultseppunct}\relax
\EndOfBibitem
\bibitem[Schindelin \latin{et~al.}(2012)Schindelin, Arganda-Carreras, Frise,
  Kaynig, Longair, Pietzsch, Preibisch, Rueden, Saalfeld, Schmid, and
  \textit{et al.}]{Schindelin_2012_NatMeth.9.676}
Schindelin,~J.; Arganda-Carreras,~I.; Frise,~E.; Kaynig,~V.; Longair,~M.;
  Pietzsch,~T.; Preibisch,~S.; Rueden,~C.; Saalfeld,~S.; Schmid,~B.; \textit{et
  al.} {Fiji: an open-source platform for biological-image analysis}.
  \emph{Nat. Meth.} \textbf{2012}, \emph{9}, 676--682\relax
\mciteBstWouldAddEndPuncttrue
\mciteSetBstMidEndSepPunct{\mcitedefaultmidpunct}
{\mcitedefaultendpunct}{\mcitedefaultseppunct}\relax
\EndOfBibitem
\bibitem[Engmann \latin{et~al.}(2012)Engmann, Stano, \v{S}. Matej\v{c}\'{i}k,
  and Ing\'{o}lfsson]{Engmann_2012_PCCP.14.14611}
Engmann,~S.; Stano,~M.; \v{S}. Matej\v{c}\'{i}k; Ing\'{o}lfsson,~O. {Gas Phase
  Low Energy Electron Induced Decomposition of the Focused Electron Beam
  Induced Deposition (FEBID) Precursor Trimethyl (Methylcyclopentadienyl)
  Platinum(iv) (MeCpPtMe$_3$)}. \emph{Phys. Chem. Chem. Phys.} \textbf{2012},
  \emph{14}, 14611--14618\relax
\mciteBstWouldAddEndPuncttrue
\mciteSetBstMidEndSepPunct{\mcitedefaultmidpunct}
{\mcitedefaultendpunct}{\mcitedefaultseppunct}\relax
\EndOfBibitem
\bibitem[Wnuk \latin{et~al.}(2009)Wnuk, Gorham, Rosenberg, {van Dorp}, Madey,
  Hagen, and Fairbrother]{Wnuk_2009_JPCC.113.2487}
Wnuk,~J.~D.; Gorham,~J.~M.; Rosenberg,~S.~G.; {van Dorp},~W.~F.; Madey,~T.~E.;
  Hagen,~C.~W.; Fairbrother,~D.~H. {Electron Induced Surface Reactions of the
  Organometallic Precursor Trimethyl(methylcyclopentadienyl)platinum(IV)}.
  \emph{J. Phys. Chem. C} \textbf{2009}, \emph{113}, 2487--2496\relax
\mciteBstWouldAddEndPuncttrue
\mciteSetBstMidEndSepPunct{\mcitedefaultmidpunct}
{\mcitedefaultendpunct}{\mcitedefaultseppunct}\relax
\EndOfBibitem
\bibitem[Athanasopoulos(2019)]{Athanasopoulos2019_SpacePolicy}
Athanasopoulos,~H.~K. {The Moon Village and Space 4.0: The `Open Concept' as a
  New Way of Doing Space?} \emph{Space Policy} \textbf{2019}, \emph{49},
  101323\relax
\mciteBstWouldAddEndPuncttrue
\mciteSetBstMidEndSepPunct{\mcitedefaultmidpunct}
{\mcitedefaultendpunct}{\mcitedefaultseppunct}\relax
\EndOfBibitem
\bibitem[van Dishoeck(2014)]{vanDishoeck_2014_FadarayDisc}
van Dishoeck,~E.~F. {Astrochemistry of Dust, Ice and Gas: Introduction and
  Overview}. \emph{Faraday Discus.} \textbf{2014}, \emph{168}, 9--47\relax
\mciteBstWouldAddEndPuncttrue
\mciteSetBstMidEndSepPunct{\mcitedefaultmidpunct}
{\mcitedefaultendpunct}{\mcitedefaultseppunct}\relax
\EndOfBibitem
\bibitem[Ghosh \latin{et~al.}(2023)Ghosh, Hendy, Raush, and
  Momeni]{Ghosh_2023_Materials.16.383}
Ghosh,~M.; Hendy,~M.; Raush,~J.; Momeni,~K. {A Phase-Field Model for In-Space
  Manufacturing of Binary Alloys}. \emph{Materials} \textbf{2023}, \emph{16},
  383\relax
\mciteBstWouldAddEndPuncttrue
\mciteSetBstMidEndSepPunct{\mcitedefaultmidpunct}
{\mcitedefaultendpunct}{\mcitedefaultseppunct}\relax
\EndOfBibitem
\bibitem[Dohn\'{a}lek \latin{et~al.}(2003)Dohn\'{a}lek, Kimmel, Ayotte, Smith,
  and Kay]{Dohnalek_2003_JCP.118.364}
Dohn\'{a}lek,~Z.; Kimmel,~G.~A.; Ayotte,~P.; Smith,~R.~S.; Kay,~B.~D. {The
  Deposition Angle-Dependent Density of Amorphous Solid Water Films}. \emph{J.
  Chem. Phys.} \textbf{2003}, \emph{118}, 364--372\relax
\mciteBstWouldAddEndPuncttrue
\mciteSetBstMidEndSepPunct{\mcitedefaultmidpunct}
{\mcitedefaultendpunct}{\mcitedefaultseppunct}\relax
\EndOfBibitem
\bibitem[Kimmel \latin{et~al.}(2001)Kimmel, Stevenson, Dohn\'{a}lek, Smith, and
  Kay]{Kimmel_2001_JCP.114.5284}
Kimmel,~G.~A.; Stevenson,~K.~P.; Dohn\'{a}lek,~Z.; Smith,~R.~S.; Kay,~B.~D.
  {Control of Amorphous Solid Water Morphology Using Molecular Beams. I.
  Experimental Results}. \emph{J. Chem. Phys.} \textbf{2001}, \emph{114},
  5284--5294\relax
\mciteBstWouldAddEndPuncttrue
\mciteSetBstMidEndSepPunct{\mcitedefaultmidpunct}
{\mcitedefaultendpunct}{\mcitedefaultseppunct}\relax
\EndOfBibitem
\bibitem[Stevenson \latin{et~al.}(1999)Stevenson, Kimmel, Dohn\'{a}lek, Smith,
  and Kay]{Stevenson_1999_Science.283.1505}
Stevenson,~K.~P.; Kimmel,~G.~A.; Dohn\'{a}lek,~Z.; Smith,~R.~S.; Kay,~K.
  {Controlling the Morphology of Amorphous Solid Water}. \emph{Science}
  \textbf{1999}, \emph{283}, 1505--1507\relax
\mciteBstWouldAddEndPuncttrue
\mciteSetBstMidEndSepPunct{\mcitedefaultmidpunct}
{\mcitedefaultendpunct}{\mcitedefaultseppunct}\relax
\EndOfBibitem
\bibitem[Holtom \latin{et~al.}(2006)Holtom, Dawes, Mukerji, Davis, Webb,
  Hoffman, and Mason]{Holtom_2006_PCCP.8.714}
Holtom,~P.~D.; Dawes,~A.; Mukerji,~R.~J.; Davis,~M.~P.; Webb,~S.~M.;
  Hoffman,~S.~V.; Mason,~N.~J. {VUV Photoabsorption Spectroscopy of Sulfur
  Dioxide Ice}. \emph{Phys. Chem. Chem. Phys.} \textbf{2006}, \emph{8},
  714--718\relax
\mciteBstWouldAddEndPuncttrue
\mciteSetBstMidEndSepPunct{\mcitedefaultmidpunct}
{\mcitedefaultendpunct}{\mcitedefaultseppunct}\relax
\EndOfBibitem
\bibitem[Bacon and Osetsky(2002)Bacon, and Osetsky]{Bacon_2002_IntMaterRev}
Bacon,~D.~J.; Osetsky,~Y.~N. {Modelling Atomic Scale Radiation Damage Processes
  and Effects in Metals}. \emph{Int. Mater. Rev.} \textbf{2002}, \emph{47},
  233--241\relax
\mciteBstWouldAddEndPuncttrue
\mciteSetBstMidEndSepPunct{\mcitedefaultmidpunct}
{\mcitedefaultendpunct}{\mcitedefaultseppunct}\relax
\EndOfBibitem
\bibitem[Shingledecker and Herbst(2018)Shingledecker, and
  Herbst]{Shingledecker_2018_PCCP.20.5359}
Shingledecker,~C.~N.; Herbst,~E. {A General Method for the Inclusion of
  Radiation Chemistry in Astrochemical Models}. \emph{Phys. Chem. Chem. Phys.}
  \textbf{2018}, \emph{20}, 5359--5367\relax
\mciteBstWouldAddEndPuncttrue
\mciteSetBstMidEndSepPunct{\mcitedefaultmidpunct}
{\mcitedefaultendpunct}{\mcitedefaultseppunct}\relax
\EndOfBibitem
\bibitem[Shingledecker \latin{et~al.}(2018)Shingledecker, Tennis, Le~Gal, and
  Herbst]{Shingledecker_2018_AstrophysJ.861.20}
Shingledecker,~C.~N.; Tennis,~J.; Le~Gal,~R.; Herbst,~E. {On Cosmic-Ray-Driven
  Grain Chemistry in Cold Core Models}. \emph{Astrophys. J.} \textbf{2018},
  \emph{861}, 20\relax
\mciteBstWouldAddEndPuncttrue
\mciteSetBstMidEndSepPunct{\mcitedefaultmidpunct}
{\mcitedefaultendpunct}{\mcitedefaultseppunct}\relax
\EndOfBibitem
\bibitem[Sherwood(2019)]{Sherwood_2019_ActaAstron}
Sherwood,~B. {Principles for a Practical Moon Base}. \emph{Acta Astronautica}
  \textbf{2019}, \emph{160}, 116--124\relax
\mciteBstWouldAddEndPuncttrue
\mciteSetBstMidEndSepPunct{\mcitedefaultmidpunct}
{\mcitedefaultendpunct}{\mcitedefaultseppunct}\relax
\EndOfBibitem
\bibitem[Strigari \latin{et~al.}(2021)Strigari, Strolin, Morganti, and
  Bartoloni]{Strigari_2021_FrontPublHealth}
Strigari,~L.; Strolin,~S.; Morganti,~A.~G.; Bartoloni,~A. {Dose-Effects Models
  for Space Radiobiology: An Overview on Dose-Effect Relationships}.
  \emph{Front. Public Health} \textbf{2021}, \emph{9}, 733337\relax
\mciteBstWouldAddEndPuncttrue
\mciteSetBstMidEndSepPunct{\mcitedefaultmidpunct}
{\mcitedefaultendpunct}{\mcitedefaultseppunct}\relax
\EndOfBibitem
\bibitem[Mertens \latin{et~al.}(2018)Mertens, Slaba, and
  Hu]{Mertens_2018_SpaceWeather}
Mertens,~C.~J.; Slaba,~T.~C.; Hu,~S. {Active Dosimeter-Based Estimate of
  Astronaut Acute Radiation Risk for Real-Time Solar Energetic Particle
  Events}. \emph{Space Weather} \textbf{2018}, \emph{16}, 1291--1316\relax
\mciteBstWouldAddEndPuncttrue
\mciteSetBstMidEndSepPunct{\mcitedefaultmidpunct}
{\mcitedefaultendpunct}{\mcitedefaultseppunct}\relax
\EndOfBibitem
\bibitem[Zhu \latin{et~al.}(2020)Zhu, Chen, Weng, Yu, Wang, He, Sheng, McKenna,
  Jaroszynski, and Zhang]{Zhu_2020_SciAdv.6.eaaz7240}
Zhu,~X.-L.; Chen,~M.; Weng,~S.-M.; Yu,~T.-P.; Wang,~W.-M.; He,~F.;
  Sheng,~Z.-M.; McKenna,~P.; Jaroszynski,~D.~A.; Zhang,~J. {Extremely Brilliant
  GeV $\gamma$-Rays from a Two-Stage Laser-Plasma Accelerator}. \emph{Sci.
  Adv.} \textbf{2020}, \emph{6}, eaaz7240\relax
\mciteBstWouldAddEndPuncttrue
\mciteSetBstMidEndSepPunct{\mcitedefaultmidpunct}
{\mcitedefaultendpunct}{\mcitedefaultseppunct}\relax
\EndOfBibitem
\bibitem[Howell \latin{et~al.}(2022)Howell, Ahmed, Afanasev, Alesini, Annand,
  Aprahamian, Balabanski, Benson, Bernstein, Brune, and \textit{et
  al.}]{Howell_JPG_2022.49.010502}
Howell,~C.~R.; Ahmed,~M.~W.; Afanasev,~A.; Alesini,~D.; Annand,~J. R.~M.;
  Aprahamian,~A.; Balabanski,~D.~L.; Benson,~S.~V.; Bernstein,~A.;
  Brune,~C.~R.; \textit{et al.} {International Workshop on Next Generation
  Gamma-Ray Source}. \emph{J. Phys. G: Nucl. Part. Phys.} \textbf{2022},
  \emph{49}, 010502\relax
\mciteBstWouldAddEndPuncttrue
\mciteSetBstMidEndSepPunct{\mcitedefaultmidpunct}
{\mcitedefaultendpunct}{\mcitedefaultseppunct}\relax
\EndOfBibitem
\bibitem[Wu \latin{et~al.}(2006)Wu, Vinokurov, Mikhailov, Li, and
  Popov]{Wu_2006_PRL.96.224801}
Wu,~Y.~K.; Vinokurov,~N.~A.; Mikhailov,~S.; Li,~J.; Popov,~V. {High-Gain Lasing
  and Polarization Switch with a Distributed Optical-Klystron Free-Electron
  Laser}. \emph{Phys. Rev. Lett.} \textbf{2006}, \emph{96}, 224801\relax
\mciteBstWouldAddEndPuncttrue
\mciteSetBstMidEndSepPunct{\mcitedefaultmidpunct}
{\mcitedefaultendpunct}{\mcitedefaultseppunct}\relax
\EndOfBibitem
\bibitem[Doerr(2018)]{Doerr_2018_NatMeth.15.33}
Doerr,~A. {The New XFELs}. \emph{Nat. Meth.} \textbf{2018}, \emph{15}, 33\relax
\mciteBstWouldAddEndPuncttrue
\mciteSetBstMidEndSepPunct{\mcitedefaultmidpunct}
{\mcitedefaultendpunct}{\mcitedefaultseppunct}\relax
\EndOfBibitem
\bibitem[Seddon \latin{et~al.}(2017)Seddon, Clarke, Dunning, Masciovecchio,
  Milne, Parmigiani, Rugg, Spence, Thompson, Ueda, and \textit{et
  al.}]{Seddon_2017_RepProgPhys.80.115901}
Seddon,~E.~A.; Clarke,~J.~A.; Dunning,~D.~J.; Masciovecchio,~C.; Milne,~C.~J.;
  Parmigiani,~F.; Rugg,~D.; Spence,~J. C.~H.; Thompson,~N.~R.; Ueda,~K.;
  \textit{et al.} {Short-Wavelength Free-Electron Laser Sources and Science: A
  Review}. \emph{Rep. Prog. Phys.} \textbf{2017}, \emph{80}, 115901\relax
\mciteBstWouldAddEndPuncttrue
\mciteSetBstMidEndSepPunct{\mcitedefaultmidpunct}
{\mcitedefaultendpunct}{\mcitedefaultseppunct}\relax
\EndOfBibitem
\bibitem[Milne \latin{et~al.}(2017)Milne, Schietinger, Aiba, Alarcon, Alex,
  Anghel, Arsov, Beard, Beaud, Bettoni, and \textit{et
  al.}]{Milne_2017_ApplSci.7.720}
Milne,~C.~J.; Schietinger,~T.; Aiba,~M.; Alarcon,~A.; Alex,~J.; Anghel,~A.;
  Arsov,~V.; Beard,~C.; Beaud,~P.; Bettoni,~S.; \textit{et al.} {SwissFEL: The
  Swiss X-ray Free Electron Laser}. \emph{Appl. Sci.} \textbf{2017}, \emph{7},
  720\relax
\mciteBstWouldAddEndPuncttrue
\mciteSetBstMidEndSepPunct{\mcitedefaultmidpunct}
{\mcitedefaultendpunct}{\mcitedefaultseppunct}\relax
\EndOfBibitem
\bibitem[Bostedt \latin{et~al.}(2016)Bostedt, Boutet, Fritz, Huang, Lee, Lemke,
  Robert, Schlotter, Turner, and Williams]{Bostedt_2016_RMP.88.015007}
Bostedt,~C.; Boutet,~S.; Fritz,~D.~M.; Huang,~Z.; Lee,~H.~J.; Lemke,~H.~T.;
  Robert,~A.; Schlotter,~W.~F.; Turner,~J.~J.; Williams,~G.~J. {Linac Coherent
  Light Source: The First Five Years}. \emph{Rev. Mod. Phys.} \textbf{2016},
  \emph{88}, 015007\relax
\mciteBstWouldAddEndPuncttrue
\mciteSetBstMidEndSepPunct{\mcitedefaultmidpunct}
{\mcitedefaultendpunct}{\mcitedefaultseppunct}\relax
\EndOfBibitem
\bibitem[Couprie(2014)]{Couprie_2014_JESRP.196.3}
Couprie,~M.~E. {New Generation of Light Sources: Present and Future}. \emph{J.
  Electr. Spectrosc. Rel. Phenom.} \textbf{2014}, \emph{196}, 3--13\relax
\mciteBstWouldAddEndPuncttrue
\mciteSetBstMidEndSepPunct{\mcitedefaultmidpunct}
{\mcitedefaultendpunct}{\mcitedefaultseppunct}\relax
\EndOfBibitem
\bibitem[Tavares \latin{et~al.}(2014)Tavares, Leemann, Sj\"{o}str\"{o}m, and
  Andersson]{Tavares_2014_JSynchrRad.21.862}
Tavares,~P.~F.; Leemann,~S.~C.; Sj\"{o}str\"{o}m,~M.; Andersson,~{\AA}. {The
  MAX IV Storage Ring Project}. \emph{J. Synchrotron Rad.} \textbf{2014},
  \emph{21}, 862--877\relax
\mciteBstWouldAddEndPuncttrue
\mciteSetBstMidEndSepPunct{\mcitedefaultmidpunct}
{\mcitedefaultendpunct}{\mcitedefaultseppunct}\relax
\EndOfBibitem
\bibitem[Yabashi and Tanaka(2017)Yabashi, and
  Tanaka]{Yabashi_2017_NatPhoton.11.12}
Yabashi,~M.; Tanaka,~H. {The Next Ten Years of X-Ray Science}. \emph{Nat.
  Photonics} \textbf{2017}, \emph{11}, 12--14\relax
\mciteBstWouldAddEndPuncttrue
\mciteSetBstMidEndSepPunct{\mcitedefaultmidpunct}
{\mcitedefaultendpunct}{\mcitedefaultseppunct}\relax
\EndOfBibitem
\bibitem[Emma \latin{et~al.}(2010)Emma, Akre, Arthur, Bionta, Bostedt, Bozek,
  Brachmann, Bucksbaum, Coffee, Decker, and \textit{et
  al.}]{Emma-EtAl_NaturePhotonics_v4_015006_2010}
Emma,~P.; Akre,~R.; Arthur,~J.; Bionta,~R.; Bostedt,~C.; Bozek,~J.;
  Brachmann,~A.; Bucksbaum,~P.; Coffee,~R.; Decker,~F.-J.; \textit{et al.}
  {First Lasing and Operation of an {\AA}ngstrom-Wavelength Free-Electron
  Laser}. \emph{Nat. Photonics} \textbf{2010}, \emph{4}, 641--647\relax
\mciteBstWouldAddEndPuncttrue
\mciteSetBstMidEndSepPunct{\mcitedefaultmidpunct}
{\mcitedefaultendpunct}{\mcitedefaultseppunct}\relax
\EndOfBibitem
\bibitem[Korol \latin{et~al.}(1998)Korol, Solov'yov, and
  Greiner]{AVK_AVS_WG_1998_JPG.24.L45}
Korol,~A.~V.; Solov'yov,~A.~V.; Greiner,~W. {Coherent Radiation of an
  Ultrarelativistic Charged Particle Channelled in a Periodically Bent
  Crystal}. \emph{J. Phys. G: Nucl. Part. Phys.} \textbf{1998}, \emph{24},
  L45--L53\relax
\mciteBstWouldAddEndPuncttrue
\mciteSetBstMidEndSepPunct{\mcitedefaultmidpunct}
{\mcitedefaultendpunct}{\mcitedefaultseppunct}\relax
\EndOfBibitem
\bibitem[Ayvazyan \latin{et~al.}(2002)Ayvazyan, Baboi, Bohnet, Brinkmann,
  Castellano, Castro, Catani, Choroba, Cianchi, Dohlus, and \textit{et
  al.}]{Ayvazyan_2002_EPJD.20.149}
Ayvazyan,~V.; Baboi,~N.; Bohnet,~I.; Brinkmann,~R.; Castellano,~M.; Castro,~P.;
  Catani,~L.; Choroba,~S.; Cianchi,~A.; Dohlus,~M.; \textit{et al.} {A New
  Powerful Source for Coherent VUV Radiation: Demonstration of Exponential
  Growth and Saturation at the TTF Free-Electron Laser}. \emph{Eur. Phys. J. D}
  \textbf{2002}, \emph{20}, 149--156\relax
\mciteBstWouldAddEndPuncttrue
\mciteSetBstMidEndSepPunct{\mcitedefaultmidpunct}
{\mcitedefaultendpunct}{\mcitedefaultseppunct}\relax
\EndOfBibitem
\bibitem[Schm\"{u}ser \latin{et~al.}(2009)Schm\"{u}ser, Dohlus, and
  Rossbach]{Schmueser_FELs_book}
Schm\"{u}ser,~P.; Dohlus,~M.; Rossbach,~J. \emph{{Ultraviolet and Soft X-Ray
  Free-Electron Lasers}}; Springer Berlin, Heidelberg, 2009\relax
\mciteBstWouldAddEndPuncttrue
\mciteSetBstMidEndSepPunct{\mcitedefaultmidpunct}
{\mcitedefaultendpunct}{\mcitedefaultseppunct}\relax
\EndOfBibitem
\bibitem[Bessonov(1986)]{Bessonov_SovJQuantElectr_1986}
Bessonov,~E.~G. {Theory of Parametric Free-Electron Lasers}. \emph{Sov. J.
  Quantum Electron.} \textbf{1986}, \emph{16}, 1056--1063\relax
\mciteBstWouldAddEndPuncttrue
\mciteSetBstMidEndSepPunct{\mcitedefaultmidpunct}
{\mcitedefaultendpunct}{\mcitedefaultseppunct}\relax
\EndOfBibitem
\bibitem[McNeil and Thompson(2010)McNeil, and
  Thompson]{McNeil_2010_NatPhoton.4.814}
McNeil,~B. W.~J.; Thompson,~N.~R. {X-Ray Free-Electron Lasers}. \emph{Nat.
  Photonics} \textbf{2010}, \emph{4}, 814--821\relax
\mciteBstWouldAddEndPuncttrue
\mciteSetBstMidEndSepPunct{\mcitedefaultmidpunct}
{\mcitedefaultendpunct}{\mcitedefaultseppunct}\relax
\EndOfBibitem
\bibitem[Gover \latin{et~al.}(2019)Gover, Ianconescu, Friedman, Emma, Sudar,
  Musumeci, and Pellegrini]{Gover_2019_RMP.91.035003}
Gover,~A.; Ianconescu,~R.; Friedman,~A.; Emma,~C.; Sudar,~N.; Musumeci,~P.;
  Pellegrini,~C. {Superradiant and Stimulated-Superradiant Emission of Bunched
  Electron Beams}. \emph{Rev. Mod. Phys.} \textbf{2019}, \emph{91},
  035003\relax
\mciteBstWouldAddEndPuncttrue
\mciteSetBstMidEndSepPunct{\mcitedefaultmidpunct}
{\mcitedefaultendpunct}{\mcitedefaultseppunct}\relax
\EndOfBibitem
\bibitem[Greiner \latin{et~al.}(2010)Greiner, Korol, Kostyuk, and
  Solov'yov]{CU_LS_patent}
Greiner,~W.; Korol,~A.~V.; Kostyuk,~A.; Solov'yov,~A.~V. {Vorrichtung und
  Verfahren zur Erzeugung electromagnetischer Strahlung}. 2010; Application for
  German patent, June 14, Ref.: 10 2010 023 632.2\relax
\mciteBstWouldAddEndPuncttrue
\mciteSetBstMidEndSepPunct{\mcitedefaultmidpunct}
{\mcitedefaultendpunct}{\mcitedefaultseppunct}\relax
\EndOfBibitem
\bibitem[Kostyuk \latin{et~al.}(2010)Kostyuk, Korol, Solov'yov, and
  Greiner]{Kostyuk_2010_JPB.43.151001}
Kostyuk,~A.; Korol,~A.~V.; Solov'yov,~A.~V.; Greiner,~W. {Stable propagation of
  a modulated positron beam in a bent crystal channel}. \emph{J. Phys. B: At.
  Mol. Opt. Phys.} \textbf{2010}, \emph{43}, 151001\relax
\mciteBstWouldAddEndPuncttrue
\mciteSetBstMidEndSepPunct{\mcitedefaultmidpunct}
{\mcitedefaultendpunct}{\mcitedefaultseppunct}\relax
\EndOfBibitem
\bibitem[Ledingham \latin{et~al.}(2003)Ledingham, McKenna, and
  Singhal]{Ledingham_2003_Science.300.1107}
Ledingham,~K. W.~D.; McKenna,~P.; Singhal,~R.~P. {Applications for Nuclear
  Phenomena Generated by Ultra-Intense Lasers}. \emph{Science} \textbf{2003},
  \emph{300}, 1107--1111\relax
\mciteBstWouldAddEndPuncttrue
\mciteSetBstMidEndSepPunct{\mcitedefaultmidpunct}
{\mcitedefaultendpunct}{\mcitedefaultseppunct}\relax
\EndOfBibitem
\bibitem[{ur Rehman} \latin{et~al.}(2017){ur Rehman}, Lee, and
  Kim]{urRehman_2017_AnnNuclEn.105.150}
{ur Rehman},~H.; Lee,~J.; Kim,~Y. {Optimization of the Laser-Compton Scattering
  Spectrum for the Transmutation of High-Toxicity and Long-Living Nuclear
  Waste}. \emph{Ann. Nucl. Energy} \textbf{2017}, \emph{105}, 150--160\relax
\mciteBstWouldAddEndPuncttrue
\mciteSetBstMidEndSepPunct{\mcitedefaultmidpunct}
{\mcitedefaultendpunct}{\mcitedefaultseppunct}\relax
\EndOfBibitem
\bibitem[{ur Rehman} \latin{et~al.}(2018){ur Rehman}, Lee, and
  Kim]{urRehman_2018_IntJEnergyRes.42.236}
{ur Rehman},~H.; Lee,~J.; Kim,~Y. {Comparison of the Laser-Compton Scattering
  and the Conventional Bremsstrahlung X-Rays for Photonuclear Transmutation}.
  \emph{Int. J. Energy Res.} \textbf{2018}, \emph{42}, 236--244\relax
\mciteBstWouldAddEndPuncttrue
\mciteSetBstMidEndSepPunct{\mcitedefaultmidpunct}
{\mcitedefaultendpunct}{\mcitedefaultseppunct}\relax
\EndOfBibitem
\bibitem[Weon \latin{et~al.}(2008)Weon, Je, Hwu, and
  Margaritondo]{Weon_2008_PRL.100.217403}
Weon,~B.~M.; Je,~J.~H.; Hwu,~Y.; Margaritondo,~G. {Decreased Surface Tension of
  Water by Hard-X-Ray Irradiation}. \emph{Phys. Rev. Lett.} \textbf{2008},
  \emph{100}, 217403\relax
\mciteBstWouldAddEndPuncttrue
\mciteSetBstMidEndSepPunct{\mcitedefaultmidpunct}
{\mcitedefaultendpunct}{\mcitedefaultseppunct}\relax
\EndOfBibitem
\bibitem[Vanraes \latin{et~al.}(2021)Vanraes, Venugopalan, and
  Bogaerts]{Vanraes2021}
Vanraes,~P.; Venugopalan,~S.~P.; Bogaerts,~A. {Multiscale Modeling of
  Plasma--Surface Interaction -- General Picture and a Case Study of Si and
  SiO$_2$ Etching by Fluorocarbon-Based Plasmas}. \emph{Appl. Phys. Rev.}
  \textbf{2021}, \emph{8}\relax
\mciteBstWouldAddEndPuncttrue
\mciteSetBstMidEndSepPunct{\mcitedefaultmidpunct}
{\mcitedefaultendpunct}{\mcitedefaultseppunct}\relax
\EndOfBibitem
\bibitem[Bonitz \latin{et~al.}(2019)Bonitz, Filinov, Abraham, Balzer,
  K\"{a}hlert, Pehlke, Bronold, Pamperin, Becker, Loffhagen, and
  Fehske]{Bonitz2019}
Bonitz,~M.; Filinov,~A.; Abraham,~J.-W.; Balzer,~K.; K\"{a}hlert,~H.;
  Pehlke,~E.; Bronold,~F.~X.; Pamperin,~M.; Becker,~M.; Loffhagen,~D.;
  Fehske,~H. {Towards an Integrated Modeling of the Plasma-Solid Interface}.
  \emph{Front. Chem. Sci. Eng.} \textbf{2019}, \emph{13}, 201--237\relax
\mciteBstWouldAddEndPuncttrue
\mciteSetBstMidEndSepPunct{\mcitedefaultmidpunct}
{\mcitedefaultendpunct}{\mcitedefaultseppunct}\relax
\EndOfBibitem
\bibitem[Ebert \latin{et~al.}(2006)Ebert, Montijn, Briels, Hundsdorfer,
  Meulenbroek, Rocco, and {van Veldhuizen}]{Ebert2006}
Ebert,~U.; Montijn,~C.; Briels,~T. M.~P.; Hundsdorfer,~W.; Meulenbroek,~B.;
  Rocco,~A.; {van Veldhuizen},~E.~M. {The Multiscale Nature of Streamers}.
  \emph{Plasma Sources Sci. Technol.} \textbf{2006}, \emph{15},
  S118--S129\relax
\mciteBstWouldAddEndPuncttrue
\mciteSetBstMidEndSepPunct{\mcitedefaultmidpunct}
{\mcitedefaultendpunct}{\mcitedefaultseppunct}\relax
\EndOfBibitem
\bibitem[Brault(2018)]{Brault2018}
Brault,~P. {Multiscale Molecular Dynamics Simulation of Plasma Processing:
  Application to Plasma Sputtering}. \emph{Front. Phys.} \textbf{2018},
  \emph{6}\relax
\mciteBstWouldAddEndPuncttrue
\mciteSetBstMidEndSepPunct{\mcitedefaultmidpunct}
{\mcitedefaultendpunct}{\mcitedefaultseppunct}\relax
\EndOfBibitem
\bibitem[Crose \latin{et~al.}(2018)Crose, Zhang, Tran, and
  Christofides]{Crose2018}
Crose,~M.; Zhang,~W.; Tran,~A.; Christofides,~P.~D. {Multiscale
  Three-Dimensional CFD Modeling for PECVD of Amorphous Silicon Thin Films}.
  \emph{Comput. Chem. Eng.} \textbf{2018}, \emph{113}, 184--195\relax
\mciteBstWouldAddEndPuncttrue
\mciteSetBstMidEndSepPunct{\mcitedefaultmidpunct}
{\mcitedefaultendpunct}{\mcitedefaultseppunct}\relax
\EndOfBibitem
\bibitem[Zhu \latin{et~al.}(2023)Zhu, Han, Xiao, and Gan]{Zhu2023}
Zhu,~G.; Han,~M.; Xiao,~B.; Gan,~Z. {Influence of Sputtering Pressure on the
  Micro-Topography of Sputtered Cu/Si Films: Integrated Multiscale Simulation}.
  \emph{Processes} \textbf{2023}, \emph{11}, 1649\relax
\mciteBstWouldAddEndPuncttrue
\mciteSetBstMidEndSepPunct{\mcitedefaultmidpunct}
{\mcitedefaultendpunct}{\mcitedefaultseppunct}\relax
\EndOfBibitem
\bibitem[Adamovich \latin{et~al.}(2022)Adamovich, Agarwal, Ahedo, Alves,
  Baalrud, Babaeva, Bogaerts, Bourdon, Bruggeman, Canal, and \textit{et
  al.}]{Adamovich2022}
Adamovich,~I.; Agarwal,~S.; Ahedo,~E.; Alves,~L.~L.; Baalrud,~S.; Babaeva,~N.;
  Bogaerts,~A.; Bourdon,~A.; Bruggeman,~P.~J.; Canal,~C.; \textit{et al.} {The
  2022 Plasma Roadmap: Low Temperature Plasma Science and Technology}. \emph{J.
  Phys. D: Appl. Phys.} \textbf{2022}, \emph{55}, 373001\relax
\mciteBstWouldAddEndPuncttrue
\mciteSetBstMidEndSepPunct{\mcitedefaultmidpunct}
{\mcitedefaultendpunct}{\mcitedefaultseppunct}\relax
\EndOfBibitem
\bibitem[Dollet(2004)]{Dollet2004}
Dollet,~A. {Multiscale Modeling of CVD Film Growth -- A Review of Recent
  Works}. \emph{Surf. Coat. Technol.} \textbf{2004}, \emph{177-178},
  245--251\relax
\mciteBstWouldAddEndPuncttrue
\mciteSetBstMidEndSepPunct{\mcitedefaultmidpunct}
{\mcitedefaultendpunct}{\mcitedefaultseppunct}\relax
\EndOfBibitem
\bibitem[Schleder \latin{et~al.}(2019)Schleder, Padilha, {Mera Acosta}, Costa,
  and Fazzio]{Schleder2019}
Schleder,~G.~R.; Padilha,~A. C.~M.; {Mera Acosta},~C.; Costa,~M.; Fazzio,~A.
  {From DFT to Machine Learning: Recent Approaches to Materials Science -- A
  Review}. \emph{J. Phys.: Materials} \textbf{2019}, \emph{2}, 032001\relax
\mciteBstWouldAddEndPuncttrue
\mciteSetBstMidEndSepPunct{\mcitedefaultmidpunct}
{\mcitedefaultendpunct}{\mcitedefaultseppunct}\relax
\EndOfBibitem
\bibitem[Neyts and Brault(2017)Neyts, and Brault]{Neyts2017}
Neyts,~E.~C.; Brault,~P. {Molecular Dynamics Simulations for Plasma-Surface
  Interactions}. \emph{Plasma Proc. Polym.} \textbf{2017}, \emph{14},
  1600145\relax
\mciteBstWouldAddEndPuncttrue
\mciteSetBstMidEndSepPunct{\mcitedefaultmidpunct}
{\mcitedefaultendpunct}{\mcitedefaultseppunct}\relax
\EndOfBibitem
\bibitem[Brault \latin{et~al.}(2023)Brault, Thomann, and Cavarroc]{Brault2023}
Brault,~P.; Thomann,~A.-L.; Cavarroc,~M. {Theory and molecular simulations of
  plasma sputtering, transport and deposition processes}. \emph{Eur. Phys. J.
  D} \textbf{2023}, \emph{77}, 19\relax
\mciteBstWouldAddEndPuncttrue
\mciteSetBstMidEndSepPunct{\mcitedefaultmidpunct}
{\mcitedefaultendpunct}{\mcitedefaultseppunct}\relax
\EndOfBibitem
\bibitem[Yang \latin{et~al.}(2015)Yang, Lively, and Allain]{Yang2015}
Yang,~Z.; Lively,~M.~A.; Allain,~J.~P. {Kinetic Monte Carlo simulation of
  self-organized pattern formation induced by ion beam sputtering using crater
  functions}. \emph{Phys. Rev. B} \textbf{2015}, \emph{91}, 075427\relax
\mciteBstWouldAddEndPuncttrue
\mciteSetBstMidEndSepPunct{\mcitedefaultmidpunct}
{\mcitedefaultendpunct}{\mcitedefaultseppunct}\relax
\EndOfBibitem
\bibitem[Verboncoeur(2005)]{Verboncoeur2005}
Verboncoeur,~J.~P. {Particle Simulation of Plasmas: Review and Advances}.
  \emph{Plasma Phys. Controlled Fusion} \textbf{2005}, \emph{47},
  A231--A260\relax
\mciteBstWouldAddEndPuncttrue
\mciteSetBstMidEndSepPunct{\mcitedefaultmidpunct}
{\mcitedefaultendpunct}{\mcitedefaultseppunct}\relax
\EndOfBibitem
\bibitem[Benilov(2020)]{Benilov2020}
Benilov,~M.~S. {Modeling the Physics of Interaction of High-Pressure Arcs with
  Their Electrodes: Advances and Challenges}. \emph{J. Phys. D: Appl. Phys.}
  \textbf{2020}, \emph{53}, 013002\relax
\mciteBstWouldAddEndPuncttrue
\mciteSetBstMidEndSepPunct{\mcitedefaultmidpunct}
{\mcitedefaultendpunct}{\mcitedefaultseppunct}\relax
\EndOfBibitem
\bibitem[Murphy and Park(2017)Murphy, and Park]{Murphy2017}
Murphy,~A.~B.; Park,~H. {Modeling of Thermal Plasma Processes: The Importance
  of Two-Way Plasma-Surface Interactions}. \emph{Plasma Proc. Polym.}
  \textbf{2017}, \emph{14}, 1600177\relax
\mciteBstWouldAddEndPuncttrue
\mciteSetBstMidEndSepPunct{\mcitedefaultmidpunct}
{\mcitedefaultendpunct}{\mcitedefaultseppunct}\relax
\EndOfBibitem
\bibitem[Murphy \latin{et~al.}(2008)Murphy, Boulos, Colombo, Fauchais, Ghedini,
  Gleizes, Proulx, and Schram]{Murphy2008}
Murphy,~A.~B.; Boulos,~M.~I.; Colombo,~V.; Fauchais,~P.; Ghedini,~E.;
  Gleizes,~A.; Proulx,~P.; Schram,~D.~C. {Avanced Thermal Plasma Modelling}.
  \emph{High Temp. Mater. Proc.} \textbf{2008}, \emph{12}, 255--336\relax
\mciteBstWouldAddEndPuncttrue
\mciteSetBstMidEndSepPunct{\mcitedefaultmidpunct}
{\mcitedefaultendpunct}{\mcitedefaultseppunct}\relax
\EndOfBibitem
\bibitem[Trelles(2018)]{Trelles2018}
Trelles,~J.~P. {Advances and Challenges in Computational Fluid Dynamics of
  Atmospheric Pressure Plasmas}. \emph{Plasma Sources Sci. Technol.}
  \textbf{2018}, \emph{27}, 093001\relax
\mciteBstWouldAddEndPuncttrue
\mciteSetBstMidEndSepPunct{\mcitedefaultmidpunct}
{\mcitedefaultendpunct}{\mcitedefaultseppunct}\relax
\EndOfBibitem
\bibitem[Kadlec(2007)]{Kadlec2007}
Kadlec,~S. {Simulation of Neutral Particle Flow During High Power Magnetron
  Impulse}. \emph{Plasma Proc. Polym.} \textbf{2007}, \emph{4},
  S419--S423\relax
\mciteBstWouldAddEndPuncttrue
\mciteSetBstMidEndSepPunct{\mcitedefaultmidpunct}
{\mcitedefaultendpunct}{\mcitedefaultseppunct}\relax
\EndOfBibitem
\bibitem[Kushner(2009)]{Kushner2009}
Kushner,~M.~J. {Hybrid Modelling of Low Temperature Plasmas for Fundamental
  Investigations and Equipment Design}. \emph{J. Phys. D: Appl. Phys.}
  \textbf{2009}, \emph{42}, 194013\relax
\mciteBstWouldAddEndPuncttrue
\mciteSetBstMidEndSepPunct{\mcitedefaultmidpunct}
{\mcitedefaultendpunct}{\mcitedefaultseppunct}\relax
\EndOfBibitem
\bibitem[Kim \latin{et~al.}(2005)Kim, Iza, Yang, Radmilovi\'{c}-Radjenovi\'{c},
  and Lee]{Kim2005}
Kim,~H.~C.; Iza,~F.; Yang,~S.~S.; Radmilovi\'{c}-Radjenovi\'{c},~M.; Lee,~J.~K.
  {Particle and Fluid Simulations of Low-Temperature Plasma Discharges:
  Benchmarks and Kinetic Effects}. \emph{J. Phys. D: Appl. Phys.}
  \textbf{2005}, \emph{38}, R283--R301\relax
\mciteBstWouldAddEndPuncttrue
\mciteSetBstMidEndSepPunct{\mcitedefaultmidpunct}
{\mcitedefaultendpunct}{\mcitedefaultseppunct}\relax
\EndOfBibitem
\bibitem[Economou(2017)]{Economou2017}
Economou,~D.~J. {Hybrid Simulation of Low Temperature Plasmas: A Brief
  Tutorial}. \emph{Plasma Proc. Polym.} \textbf{2017}, \emph{14}, 1600152\relax
\mciteBstWouldAddEndPuncttrue
\mciteSetBstMidEndSepPunct{\mcitedefaultmidpunct}
{\mcitedefaultendpunct}{\mcitedefaultseppunct}\relax
\EndOfBibitem
\bibitem[Nijdam \latin{et~al.}(2020)Nijdam, Teunissen, and Ebert]{Nijdam2020}
Nijdam,~S.; Teunissen,~J.; Ebert,~U. {The Physics of Streamer Discharge
  Phenomena}. \emph{Plasma Sources Sci. Technol.} \textbf{2020}, \emph{29},
  103001\relax
\mciteBstWouldAddEndPuncttrue
\mciteSetBstMidEndSepPunct{\mcitedefaultmidpunct}
{\mcitedefaultendpunct}{\mcitedefaultseppunct}\relax
\EndOfBibitem
\bibitem[Ebert and Sentman(2008)Ebert, and Sentman]{Ebert2008}
Ebert,~U.; Sentman,~D.~D. {Streamers, Sprites, Leaders, Lightning: From Micro-
  to Macroscales}. \emph{J. Phys. D: Appl. Phys.} \textbf{2008}, \emph{41},
  230301\relax
\mciteBstWouldAddEndPuncttrue
\mciteSetBstMidEndSepPunct{\mcitedefaultmidpunct}
{\mcitedefaultendpunct}{\mcitedefaultseppunct}\relax
\EndOfBibitem
\bibitem[Ebert \latin{et~al.}(2010)Ebert, Nijdam, Li, Luque, Briels, and {van
  Veldhuizen}]{Ebert2010}
Ebert,~U.; Nijdam,~S.; Li,~C.; Luque,~A.; Briels,~T.; {van Veldhuizen},~E.
  {Review of Recent Results on Streamer Discharges and Discussion of Their
  Relevance for Sprites and Lightning}. \emph{J. Geophys. Res. Space Phys.}
  \textbf{2010}, \emph{115}, A00E43\relax
\mciteBstWouldAddEndPuncttrue
\mciteSetBstMidEndSepPunct{\mcitedefaultmidpunct}
{\mcitedefaultendpunct}{\mcitedefaultseppunct}\relax
\EndOfBibitem
\bibitem[Jimenez and Dew(2012)Jimenez, and Dew]{Jimenez2012}
Jimenez,~F.~J.; Dew,~S.~K. {Comprehensive Computer Model for Magnetron
  Sputtering. I. Gas Heating and Rarefaction}. \emph{J. Vac. Sci. Technol. A}
  \textbf{2012}, \emph{30}, 041302\relax
\mciteBstWouldAddEndPuncttrue
\mciteSetBstMidEndSepPunct{\mcitedefaultmidpunct}
{\mcitedefaultendpunct}{\mcitedefaultseppunct}\relax
\EndOfBibitem
\bibitem[Gudmundsson(2020)]{Gudmundsson2020}
Gudmundsson,~J.~T. {Physics and Technology of Magnetron Sputtering Discharges}.
  \emph{Plasma Sources Sci. Technol.} \textbf{2020}, \emph{29}, 113001\relax
\mciteBstWouldAddEndPuncttrue
\mciteSetBstMidEndSepPunct{\mcitedefaultmidpunct}
{\mcitedefaultendpunct}{\mcitedefaultseppunct}\relax
\EndOfBibitem
\bibitem[Anders(2014)]{Anders2014}
Anders,~A. {A Review Comparing Cathodic Arcs and High Power Impulse Magnetron
  Sputtering (HiPIMS)}. \emph{Surf. Coat. Technol.} \textbf{2014}, \emph{257},
  308--325\relax
\mciteBstWouldAddEndPuncttrue
\mciteSetBstMidEndSepPunct{\mcitedefaultmidpunct}
{\mcitedefaultendpunct}{\mcitedefaultseppunct}\relax
\EndOfBibitem
\bibitem[Brenning \latin{et~al.}(2013)Brenning, Lundin, Minea, Costin, and
  Vitelaru]{Brenning2013}
Brenning,~N.; Lundin,~D.; Minea,~T.; Costin,~C.; Vitelaru,~C. {Spokes and
  Charged Particle Transport in HiPIMS Magnetrons}. \emph{J. Phys. D: Appl.
  Phys.} \textbf{2013}, \emph{46}, 084005\relax
\mciteBstWouldAddEndPuncttrue
\mciteSetBstMidEndSepPunct{\mcitedefaultmidpunct}
{\mcitedefaultendpunct}{\mcitedefaultseppunct}\relax
\EndOfBibitem
\bibitem[Kadlec and \v{C}apek(2017)Kadlec, and \v{C}apek]{Kadlec2017}
Kadlec,~S.; \v{C}apek,~J. {Return of Target Material Ions Leads to a Reduced
  Hysteresis in Reactive High Power Impulse Magnetron Sputtering: Model}.
  \emph{J. Appl. Phys.} \textbf{2017}, \emph{121}, 171910\relax
\mciteBstWouldAddEndPuncttrue
\mciteSetBstMidEndSepPunct{\mcitedefaultmidpunct}
{\mcitedefaultendpunct}{\mcitedefaultseppunct}\relax
\EndOfBibitem
\bibitem[Anders(2008)]{Anders2008}
Anders,~A. \emph{{Cathodic Arcs Cathodic Arcs From Fractal Spots to Energetic
  Condensation}}; Springer New York, NY, 2008\relax
\mciteBstWouldAddEndPuncttrue
\mciteSetBstMidEndSepPunct{\mcitedefaultmidpunct}
{\mcitedefaultendpunct}{\mcitedefaultseppunct}\relax
\EndOfBibitem
\bibitem[Schneider(2006)]{Schneider2006}
Schneider,~R. {Plasma--Wall Interaction: A Multiscale Problem}. \emph{Phys.
  Scr.} \textbf{2006}, \emph{T124}, 76--79\relax
\mciteBstWouldAddEndPuncttrue
\mciteSetBstMidEndSepPunct{\mcitedefaultmidpunct}
{\mcitedefaultendpunct}{\mcitedefaultseppunct}\relax
\EndOfBibitem
\bibitem[Cheimarios \latin{et~al.}(2021)Cheimarios, Kokkoris, and
  Boudouvis]{Cheimarios2021}
Cheimarios,~N.; Kokkoris,~G.; Boudouvis,~A.~G. {Multiscale Modeling in Chemical
  Vapor Deposition Processes: Models and Methodologies}. \emph{Arch. Comput.
  Methods Eng.} \textbf{2021}, \emph{28}, 637--672\relax
\mciteBstWouldAddEndPuncttrue
\mciteSetBstMidEndSepPunct{\mcitedefaultmidpunct}
{\mcitedefaultendpunct}{\mcitedefaultseppunct}\relax
\EndOfBibitem
\bibitem[Kambara \latin{et~al.}(2023)Kambara, Kawaguchi, Lee, Ikuse, Hamaguchi,
  Ohmori, and Ishikawa]{Kambara2023}
Kambara,~M.; Kawaguchi,~S.; Lee,~H.~J.; Ikuse,~K.; Hamaguchi,~S.; Ohmori,~T.;
  Ishikawa,~K. {Science-Based, Data-Driven Developments in Plasma Processing
  for Material Synthesis and Device-Integration Technologies}. \emph{Jpn. J.
  Appl. Phys.} \textbf{2023}, \emph{62}, SA0803\relax
\mciteBstWouldAddEndPuncttrue
\mciteSetBstMidEndSepPunct{\mcitedefaultmidpunct}
{\mcitedefaultendpunct}{\mcitedefaultseppunct}\relax
\EndOfBibitem
\bibitem[Gunasegaram \latin{et~al.}(2021)Gunasegaram, Murphy, Barnard, DebRoy,
  Matthews, Ladani, and Gu]{Gunasegaram2021}
Gunasegaram,~D.~R.; Murphy,~A.~B.; Barnard,~A.; DebRoy,~T.; Matthews,~M.~J.;
  Ladani,~L.; Gu,~D. {Towards Developing Multiscale-Multiphysics Models and
  Their Surrogates for Digital Twins of Metal Additive Manufacturing}.
  \emph{Addit. Manuf.} \textbf{2021}, \emph{46}, 102089\relax
\mciteBstWouldAddEndPuncttrue
\mciteSetBstMidEndSepPunct{\mcitedefaultmidpunct}
{\mcitedefaultendpunct}{\mcitedefaultseppunct}\relax
\EndOfBibitem
\bibitem[Jetly and Chaudhury(2021)Jetly, and
  Chaudhury]{Jetly_2021_MLST.2.035025}
Jetly,~V.; Chaudhury,~B. {Extracting Electron Scattering Cross Sections from
  Swarm Data using Deep Neural Networks}. \emph{Mach. Learn.: Sci. Technol.}
  \textbf{2021}, \emph{2}, 035025\relax
\mciteBstWouldAddEndPuncttrue
\mciteSetBstMidEndSepPunct{\mcitedefaultmidpunct}
{\mcitedefaultendpunct}{\mcitedefaultseppunct}\relax
\EndOfBibitem
\bibitem[Nam \latin{et~al.}(2021)Nam, Yong, Hwang, and
  Choi]{Nam_2021_PLA.387.127005}
Nam,~J.; Yong,~H.; Hwang,~J.; Choi,~J. {Training an Artificial Neural Network
  for Recognizing Electron Collision Patterns}. \emph{Phys. Lett. A}
  \textbf{2021}, \emph{387}, 127005\relax
\mciteBstWouldAddEndPuncttrue
\mciteSetBstMidEndSepPunct{\mcitedefaultmidpunct}
{\mcitedefaultendpunct}{\mcitedefaultseppunct}\relax
\EndOfBibitem
\bibitem[Kr\"{u}ger \latin{et~al.}(2019)Kr\"{u}ger, Gergs, and
  Trieschmann]{Krueger2019}
Kr\"{u}ger,~F.; Gergs,~T.; Trieschmann,~J. {Machine Learning Plasma-Surface
  Interface for Coupling Sputtering and Gas-Phase Transport Simulations}.
  \emph{Plasma Sources Sci. Technol.} \textbf{2019}, \emph{28}, 035002\relax
\mciteBstWouldAddEndPuncttrue
\mciteSetBstMidEndSepPunct{\mcitedefaultmidpunct}
{\mcitedefaultendpunct}{\mcitedefaultseppunct}\relax
\EndOfBibitem
\bibitem[Raissi \latin{et~al.}(2019)Raissi, Perdikaris, and
  Karniadakis]{Raissi2019}
Raissi,~M.; Perdikaris,~P.; Karniadakis,~G. {Physics-Informed Neural Networks:
  A Deep Learning Framework for Solving Forward and Inverse Problems Involving
  Nonlinear Partial Differential Equations}. \emph{J. Comput. Phys.}
  \textbf{2019}, \emph{378}, 686--707\relax
\mciteBstWouldAddEndPuncttrue
\mciteSetBstMidEndSepPunct{\mcitedefaultmidpunct}
{\mcitedefaultendpunct}{\mcitedefaultseppunct}\relax
\EndOfBibitem
\bibitem[Karniadakis \latin{et~al.}(2021)Karniadakis, Kevrekidis, Lu,
  Perdikaris, Wang, and Yang]{Karniadakis2021}
Karniadakis,~G.~E.; Kevrekidis,~I.~G.; Lu,~L.; Perdikaris,~P.; Wang,~S.;
  Yang,~L. {Physics-Informed Machine Learning}. \emph{Nat. Rev. Phys.}
  \textbf{2021}, \emph{3}, 422--440\relax
\mciteBstWouldAddEndPuncttrue
\mciteSetBstMidEndSepPunct{\mcitedefaultmidpunct}
{\mcitedefaultendpunct}{\mcitedefaultseppunct}\relax
\EndOfBibitem
\bibitem[Spears \latin{et~al.}(2018)Spears, Brase, Bremer, Chen, Field,
  Gaffney, Kruse, Langer, Lewis, Nora, Peterson, Thiagarajan, {Van Essen}, and
  Humbird]{Spears2018}
Spears,~B.~K.; Brase,~J.; Bremer,~P.-T.; Chen,~B.; Field,~J.; Gaffney,~J.;
  Kruse,~M.; Langer,~S.; Lewis,~K.; Nora,~R.; Peterson,~J.~L.;
  Thiagarajan,~J.~J.; {Van Essen},~B.; Humbird,~K. {Deep Learning: A Guide for
  Practitioners in the Physical Sciences}. \emph{Phys. Plasmas} \textbf{2018},
  \emph{25}, 080901\relax
\mciteBstWouldAddEndPuncttrue
\mciteSetBstMidEndSepPunct{\mcitedefaultmidpunct}
{\mcitedefaultendpunct}{\mcitedefaultseppunct}\relax
\EndOfBibitem
\bibitem[Cuomo \latin{et~al.}(2022)Cuomo, {Schiano Di Cola}, Giampaolo, Rozza,
  Raissi, and Piccialli]{Cuomo2022}
Cuomo,~S.; {Schiano Di Cola},~V.; Giampaolo,~F.; Rozza,~G.; Raissi,~M.;
  Piccialli,~F. {Scientific Machine Learning Through Physics -- Informed Neural
  Networks: Where we are and What's Next}. \emph{J. Sci. Comput.}
  \textbf{2022}, \emph{92}, 88\relax
\mciteBstWouldAddEndPuncttrue
\mciteSetBstMidEndSepPunct{\mcitedefaultmidpunct}
{\mcitedefaultendpunct}{\mcitedefaultseppunct}\relax
\EndOfBibitem
\bibitem[Markidis(2021)]{Markidis2021}
Markidis,~S. {The Old and the New: Can Physics-Informed Deep-Learning Replace
  Traditional Linear Solvers?} \emph{Front. Big Data} \textbf{2021}, \emph{4},
  669097\relax
\mciteBstWouldAddEndPuncttrue
\mciteSetBstMidEndSepPunct{\mcitedefaultmidpunct}
{\mcitedefaultendpunct}{\mcitedefaultseppunct}\relax
\EndOfBibitem
\bibitem[Park \latin{et~al.}(2020)Park, Jang, Cha, Noh, Choi, Lee, Seong, Kim,
  Cho, Park, Seo, Yang, and Kim]{Park2020}
Park,~S.; Jang,~Y.; Cha,~T.; Noh,~Y.; Choi,~Y.; Lee,~J.; Seong,~J.; Kim,~B.;
  Cho,~T.; Park,~Y.; Seo,~R.; Yang,~J.-H.; Kim,~G.-H. {Predictive Control of
  the Plasma Processes in the OLED Display Mass Production Referring to the
  Discontinuity Qualifying PI-VM}. \emph{Phys. Plasmas} \textbf{2020},
  \emph{27}, 083507\relax
\mciteBstWouldAddEndPuncttrue
\mciteSetBstMidEndSepPunct{\mcitedefaultmidpunct}
{\mcitedefaultendpunct}{\mcitedefaultseppunct}\relax
\EndOfBibitem
\bibitem[Bonzanini \latin{et~al.}(2023)Bonzanini, Shao, Graves, Hamaguchi, and
  Mesbah]{Bonzanini2023}
Bonzanini,~A.~D.; Shao,~K.; Graves,~D.~B.; Hamaguchi,~S.; Mesbah,~A.
  {Foundations of Machine Learning for Low-Temperature Plasmas: Methods and
  Case Studies}. \emph{Plasma Sources Sci. Technol.} \textbf{2023}, \emph{32},
  024003\relax
\mciteBstWouldAddEndPuncttrue
\mciteSetBstMidEndSepPunct{\mcitedefaultmidpunct}
{\mcitedefaultendpunct}{\mcitedefaultseppunct}\relax
\EndOfBibitem
\bibitem[Kawaguchi and Murakami(2022)Kawaguchi, and Murakami]{Kawaguchi2022}
Kawaguchi,~S.; Murakami,~T. {Physics-Informed Neural Networks for Solving the
  Boltzmann Equation of the Electron Velocity Distribution Function in Weakly
  Ionized Plasmas}. \emph{Jpn. J. Appl. Phys.} \textbf{2022}, \emph{61},
  086002\relax
\mciteBstWouldAddEndPuncttrue
\mciteSetBstMidEndSepPunct{\mcitedefaultmidpunct}
{\mcitedefaultendpunct}{\mcitedefaultseppunct}\relax
\EndOfBibitem
\bibitem[Cai \latin{et~al.}(2021)Cai, Wang, Wang, Perdikaris, and
  Karniadakis]{Cai2021}
Cai,~S.; Wang,~Z.; Wang,~S.; Perdikaris,~P.; Karniadakis,~G.~E.
  {Physics-Informed Neural Networks for Heat Transfer Problems}. \emph{J. Heat
  Transfer} \textbf{2021}, \emph{143}, 060801\relax
\mciteBstWouldAddEndPuncttrue
\mciteSetBstMidEndSepPunct{\mcitedefaultmidpunct}
{\mcitedefaultendpunct}{\mcitedefaultseppunct}\relax
\EndOfBibitem
\bibitem[Zeng \latin{et~al.}(2023)Zeng, Zhang, Ren, and Shao]{Zeng2023}
Zeng,~X.; Zhang,~S.; Ren,~C.; Shao,~T. {Physics Informed Neural Networks for
  Electric Field Distribution Characteristics Analysis}. \emph{J. Phys. D:
  Appl. Phys.} \textbf{2023}, \emph{56}, 165202\relax
\mciteBstWouldAddEndPuncttrue
\mciteSetBstMidEndSepPunct{\mcitedefaultmidpunct}
{\mcitedefaultendpunct}{\mcitedefaultseppunct}\relax
\EndOfBibitem
\bibitem[Zhong \latin{et~al.}(2023)Zhong, Wu, and Wang]{Zhong2023}
Zhong,~L.; Wu,~B.; Wang,~Y. {Accelerating Physics-Informed Neural Network Based
  1D Arc Simulation by Meta Learning}. \emph{J. Phys. D: Appl. Phys.}
  \textbf{2023}, \emph{56}, 074006\relax
\mciteBstWouldAddEndPuncttrue
\mciteSetBstMidEndSepPunct{\mcitedefaultmidpunct}
{\mcitedefaultendpunct}{\mcitedefaultseppunct}\relax
\EndOfBibitem
\bibitem[Carbone \latin{et~al.}(2021)Carbone, Graef, Hagelaar, Boer, Hopkins,
  Stephens, Yee, Pancheshnyi, {van Dijk}, and Pitchford]{Carbone2021}
Carbone,~E.; Graef,~W.; Hagelaar,~G.; Boer,~D.; Hopkins,~M.~M.;
  Stephens,~J.~C.; Yee,~B.~T.; Pancheshnyi,~S.; {van Dijk},~J.; Pitchford,~L.
  {Data Needs for Modeling Low-Temperature Non-Equilibrium Plasmas: The LXCat
  Project, History, Perspectives and a Tutorial}. \emph{Atoms} \textbf{2021},
  \emph{9}, 16\relax
\mciteBstWouldAddEndPuncttrue
\mciteSetBstMidEndSepPunct{\mcitedefaultmidpunct}
{\mcitedefaultendpunct}{\mcitedefaultseppunct}\relax
\EndOfBibitem
\bibitem[HIT((accessed 2023-10-18))]{HITRAN_database}
HITRAN (HIgh-resolution TRANsmission molecular absorption) database. (accessed
  2023-10-18); \url{https://hitran.org/}\relax
\mciteBstWouldAddEndPuncttrue
\mciteSetBstMidEndSepPunct{\mcitedefaultmidpunct}
{\mcitedefaultendpunct}{\mcitedefaultseppunct}\relax
\EndOfBibitem
\bibitem[Buehler \latin{et~al.}(2022)Buehler, Brath, O.~Lemke, Pincus,
  Eriksson, Gordon, and Larsson]{Buehler_2022_JAMES.14}
Buehler,~S.~A.; Brath,~M.; O.~Lemke,~{\O}.~H.; Pincus,~R.; Eriksson,~P.;
  Gordon,~I.; Larsson,~R. {A New Halocarbon Absorption Model Based on HITRAN
  Cross-Section Data and New Estimates of Halocarbon Instantaneous Clear-Sky
  Radiative Forcing}. \emph{J. Adv. Model. Earth Syst.} \textbf{2022},
  \emph{14}, e2022MS003239\relax
\mciteBstWouldAddEndPuncttrue
\mciteSetBstMidEndSepPunct{\mcitedefaultmidpunct}
{\mcitedefaultendpunct}{\mcitedefaultseppunct}\relax
\EndOfBibitem
\bibitem[ALA((accessed 2023-10-18))]{ALADDIN_database}
ALADDIN, a database of atomic, molecular and plasma-material interactions data
  for fusion research. (accessed 2023-10-18);
  \url{https://www.iaea.org/resources/databases/aladdin}\relax
\mciteBstWouldAddEndPuncttrue
\mciteSetBstMidEndSepPunct{\mcitedefaultmidpunct}
{\mcitedefaultendpunct}{\mcitedefaultseppunct}\relax
\EndOfBibitem
\bibitem[Celiberto \latin{et~al.}(2016)Celiberto, Armenise, Cacciatore,
  Capitelli, Esposito, Gamallo, Janev, Lagan\`{a}, Laporta, Laricchiuta, and
  \textit{et al.}]{Celiberto_2016_PSST.25.033004}
Celiberto,~R.; Armenise,~I.; Cacciatore,~M.; Capitelli,~M.; Esposito,~F.;
  Gamallo,~P.; Janev,~R.~K.; Lagan\`{a},~A.; Laporta,~V.; Laricchiuta,~A.;
  \textit{et al.} {Atomic and Molecular Data for Spacecraft Re-entry Plasmas}.
  \emph{Plasma Sources Sci. Technol.} \textbf{2016}, \emph{25}, 033004\relax
\mciteBstWouldAddEndPuncttrue
\mciteSetBstMidEndSepPunct{\mcitedefaultmidpunct}
{\mcitedefaultendpunct}{\mcitedefaultseppunct}\relax
\EndOfBibitem
\bibitem[Song \latin{et~al.}(2012)Song, Kwon, Jhang, Kwang, Park, Kang, and
  Yoon]{Industrial_Appl_Plasmas}
Song,~M.-Y.; Kwon,~D.-C.; Jhang,~W.-S.; Kwang,~S.-H.; Park,~J.-H.; Kang,~Y.-K.;
  Yoon,~J.-S. In \emph{Atomic Processes in Basic and Applied Physics. Springer
  Series on Atomic, Optical, and Plasma Physics, vol 68.}; Shevelko,~V.,
  Tawara,~H., Eds.; Springer, Berlin, Heidelberg, 2012; pp 357--391\relax
\mciteBstWouldAddEndPuncttrue
\mciteSetBstMidEndSepPunct{\mcitedefaultmidpunct}
{\mcitedefaultendpunct}{\mcitedefaultseppunct}\relax
\EndOfBibitem
\bibitem[Samukawa \latin{et~al.}(2012)Samukawa, Hori, Rauf, Tachibana,
  Bruggeman, Kroesen, Whitehead, Murphy, Gutsol, Starikovskaia, and \textit{et
  al.}]{Plasma_Roadmap_JPD_2012}
Samukawa,~S.; Hori,~M.; Rauf,~S.; Tachibana,~K.; Bruggeman,~P.; Kroesen,~G.;
  Whitehead,~J.~C.; Murphy,~A.~B.; Gutsol,~A.~F.; Starikovskaia,~S.; \textit{et
  al.} {The 2012 Plasma Roadmap}. \emph{J. Phys. D: Appl. Phys.} \textbf{2012},
  \emph{45}, 253001\relax
\mciteBstWouldAddEndPuncttrue
\mciteSetBstMidEndSepPunct{\mcitedefaultmidpunct}
{\mcitedefaultendpunct}{\mcitedefaultseppunct}\relax
\EndOfBibitem
\bibitem[Adamovich \latin{et~al.}(2017)Adamovich, Baalrud, Bogaerts, Bruggeman,
  Cappelli, Colombo, Czarnetzki, Ebert, Eden, Favia, and \textit{et
  al.}]{Plasma_Roadmap_JPD_2017}
Adamovich,~I.; Baalrud,~S.~D.; Bogaerts,~A.; Bruggeman,~P.~J.; Cappelli,~M.;
  Colombo,~V.; Czarnetzki,~U.; Ebert,~U.; Eden,~J.~G.; Favia,~P.; \textit{et
  al.} {The 2017 Plasma Roadmap: Low temperature plasma science and
  technology}. \emph{J. Phys. D: Appl. Phys.} \textbf{2017}, \emph{50},
  323001\relax
\mciteBstWouldAddEndPuncttrue
\mciteSetBstMidEndSepPunct{\mcitedefaultmidpunct}
{\mcitedefaultendpunct}{\mcitedefaultseppunct}\relax
\EndOfBibitem
\bibitem[Anirudh \latin{et~al.}(2023)Anirudh, Archibald, Asif, Becker,
  Benkadda, Bremer, Bud\'{e}, Chang, Chen, Churchill, and \textit{et
  al.}]{Anirudh_PlasmaScience_2022}
Anirudh,~R.; Archibald,~R.; Asif,~M.~S.; Becker,~M.~M.; Benkadda,~S.;
  Bremer,~P.-T.; Bud\'{e},~R. H.~S.; Chang,~C.~S.; Chen,~L.; Churchill,~R.~M.;
  \textit{et al.} {2022 Review of Data-Driven Plasma Science}. \emph{IEEE
  Trans. Plasma Sci.} \textbf{2023}, \emph{51}, 1750--1838\relax
\mciteBstWouldAddEndPuncttrue
\mciteSetBstMidEndSepPunct{\mcitedefaultmidpunct}
{\mcitedefaultendpunct}{\mcitedefaultseppunct}\relax
\EndOfBibitem
\bibitem[Wei((accessed 2023-10-18))]{Weizmann_PlasmaLab_DBs}
Databases for Atomic and Plasma Physics. (accessed 2023-10-18);
  \url{https://plasma-gate.weizmann.ac.il/directories/databases}\relax
\mciteBstWouldAddEndPuncttrue
\mciteSetBstMidEndSepPunct{\mcitedefaultmidpunct}
{\mcitedefaultendpunct}{\mcitedefaultseppunct}\relax
\EndOfBibitem
\bibitem[Exo((accessed 2023-10-18))]{ExoMol_database}
ExoMol -- High-temperature molecular line lists for modeling exoplanet
  atmospheres. (accessed 2023-10-18); \url{https://www.exomol.com/}\relax
\mciteBstWouldAddEndPuncttrue
\mciteSetBstMidEndSepPunct{\mcitedefaultmidpunct}
{\mcitedefaultendpunct}{\mcitedefaultseppunct}\relax
\EndOfBibitem
\bibitem[RAD((accessed 2023-10-18))]{RADAM_portal}
RADAM (RAdiation DAMage) database portal. (accessed 2023-10-18);
  \url{https://radamdb.mbnresearch.com/}\relax
\mciteBstWouldAddEndPuncttrue
\mciteSetBstMidEndSepPunct{\mcitedefaultmidpunct}
{\mcitedefaultendpunct}{\mcitedefaultseppunct}\relax
\EndOfBibitem
\bibitem[Denifl \latin{et~al.}(2013)Denifl, Garcia, Huber, Marinkovi\'{c},
  Mason, Postler, Rabus, Rixon, Solov'yov, Suraud, and
  Yakubovich]{RADAM_DB_Denifl_JPCS}
Denifl,~S.; Garcia,~G.; Huber,~B.~A.; Marinkovi\'{c},~B.~P.; Mason,~N.~J.;
  Postler,~J.; Rabus,~H.; Rixon,~G.; Solov'yov,~A.~V.; Suraud,~E.;
  Yakubovich,~A.~V. {Radiation Damage of Biomolecules (RADAM) Database
  Development: Current Status}. \emph{J. Phys.: Conf. Ser.} \textbf{2013},
  \emph{438}, 012016\relax
\mciteBstWouldAddEndPuncttrue
\mciteSetBstMidEndSepPunct{\mcitedefaultmidpunct}
{\mcitedefaultendpunct}{\mcitedefaultseppunct}\relax
\EndOfBibitem
\bibitem[Cha((accessed 2023-10-18))]{ChannelingDB_portal}
Channeling database portal. (accessed 2023-10-18);
  \url{https://mbnresearch.com/databases}\relax
\mciteBstWouldAddEndPuncttrue
\mciteSetBstMidEndSepPunct{\mcitedefaultmidpunct}
{\mcitedefaultendpunct}{\mcitedefaultseppunct}\relax
\EndOfBibitem
\bibitem[Qua((accessed 2023-10-18))]{Quantemol_website}
Quantemol software tools. (accessed 2023-10-18);
  \url{http://www.quantemol.com/}\relax
\mciteBstWouldAddEndPuncttrue
\mciteSetBstMidEndSepPunct{\mcitedefaultmidpunct}
{\mcitedefaultendpunct}{\mcitedefaultseppunct}\relax
\EndOfBibitem
\bibitem[Joshipura and Mason(2018)Joshipura, and Mason]{Joshipura_Mason_book}
Joshipura,~K.~N.; Mason,~N.~J. \emph{{Atomic--Molecular Ionization by Electron
  Scattering: Theory and Applications}}; Cambridge University Press, Cambridge,
  2018\relax
\mciteBstWouldAddEndPuncttrue
\mciteSetBstMidEndSepPunct{\mcitedefaultmidpunct}
{\mcitedefaultendpunct}{\mcitedefaultseppunct}\relax
\EndOfBibitem
\bibitem[{Del Zanna} \latin{et~al.}(2019){Del Zanna}, Fern\'{a}ndez-Menchero,
  and Badnell]{DelZanna_2019_MNRAS.484.4754}
{Del Zanna},~G.; Fern\'{a}ndez-Menchero,~L.; Badnell,~N.~R. {Uncertainties on
  Atomic data. A case study: N IV}. \emph{Mon. Notices Royal Astron. Soc.}
  \textbf{2019}, \emph{484}, 4754--4759\relax
\mciteBstWouldAddEndPuncttrue
\mciteSetBstMidEndSepPunct{\mcitedefaultmidpunct}
{\mcitedefaultendpunct}{\mcitedefaultseppunct}\relax
\EndOfBibitem
\bibitem[Chung \latin{et~al.}(2016)Chung, Braams, Bartschat, Cs\'{a}sz\'{a}r,
  Drake, Kirchner, Kokoouline, and Tennyson]{Chung_2016_JPD.49.363002}
Chung,~H.-K.; Braams,~B.~J.; Bartschat,~K.; Cs\'{a}sz\'{a}r,~A.~G.; Drake,~G.
  W.~F.; Kirchner,~T.; Kokoouline,~V.; Tennyson,~J. {Uncertainty Estimates for
  Theoretical Atomic and Molecular Data}. \emph{J. Phys. D: Appl. Phys.}
  \textbf{2016}, \emph{49}, 363002\relax
\mciteBstWouldAddEndPuncttrue
\mciteSetBstMidEndSepPunct{\mcitedefaultmidpunct}
{\mcitedefaultendpunct}{\mcitedefaultseppunct}\relax
\EndOfBibitem
\bibitem[F\'{a}rn\'{i}k(2023)]{Farnik_2023_JPCL.14.287}
F\'{a}rn\'{i}k,~M. {Bridging Gaps between Clusters in Molecular-Beam
  Experiments and Aerosol Nanoclusters}. \emph{J. Phys. Chem. Lett.}
  \textbf{2023}, \emph{14}, 287--294\relax
\mciteBstWouldAddEndPuncttrue
\mciteSetBstMidEndSepPunct{\mcitedefaultmidpunct}
{\mcitedefaultendpunct}{\mcitedefaultseppunct}\relax
\EndOfBibitem
\bibitem[Mason \latin{et~al.}(2008)Mason, Drage, Webb, Dawes, McPheat, and
  Hayes]{Mason_2008_FaradayDisc.137.367}
Mason,~N.~J.; Drage,~E.~A.; Webb,~S.~M.; Dawes,~A.; McPheat,~R.; Hayes,~G. {The
  Spectroscopy and Chemical Dynamics of Microparticles Explored Using an
  Ultrasonic Trap}. \emph{Faraday Discuss.} \textbf{2008}, \emph{137},
  367--376\relax
\mciteBstWouldAddEndPuncttrue
\mciteSetBstMidEndSepPunct{\mcitedefaultmidpunct}
{\mcitedefaultendpunct}{\mcitedefaultseppunct}\relax
\EndOfBibitem
\bibitem[Dangi and Dickerson(2021)Dangi, and
  Dickerson]{Dangi_2021_ACSOmega.6.10447}
Dangi,~B.~B.; Dickerson,~D.~J. {Design and Performance of an Acoustic Levitator
  System Coupled with a Tunable Monochromatic Light Source and a Raman
  Spectrometer for In Situ Reaction Monitoring}. \emph{ACS Omega}
  \textbf{2021}, \emph{6}, 10447--10453\relax
\mciteBstWouldAddEndPuncttrue
\mciteSetBstMidEndSepPunct{\mcitedefaultmidpunct}
{\mcitedefaultendpunct}{\mcitedefaultseppunct}\relax
\EndOfBibitem
\bibitem[Rafferty \latin{et~al.}(2023)Rafferty, Vennes, Bain, and
  Preston]{Rafferty_2023_PCCP.25.7066}
Rafferty,~A.; Vennes,~B.; Bain,~A.; Preston,~T.~C. {Optical Trapping and Light
  Scattering in Atmospheric Aerosol Science}. \emph{Phys. Chem. Chem. Phys.}
  \textbf{2023}, \emph{25}, 7066--7089\relax
\mciteBstWouldAddEndPuncttrue
\mciteSetBstMidEndSepPunct{\mcitedefaultmidpunct}
{\mcitedefaultendpunct}{\mcitedefaultseppunct}\relax
\EndOfBibitem
\bibitem[Nomura \latin{et~al.}(2017)Nomura, Tsuchida, Kajiwara, Yoshida,
  Majima, and Saito]{Nomura_2017_JCP.147.225103}
Nomura,~S.; Tsuchida,~H.; Kajiwara,~A.; Yoshida,~S.; Majima,~T.; Saito,~M.
  {Dissociation of Biomolecules in Liquid Environments During Fast Heavy-Ion
  Irradiation}. \emph{J. Chem. Phys.} \textbf{2017}, \emph{2017}, 225103\relax
\mciteBstWouldAddEndPuncttrue
\mciteSetBstMidEndSepPunct{\mcitedefaultmidpunct}
{\mcitedefaultendpunct}{\mcitedefaultseppunct}\relax
\EndOfBibitem
\bibitem[Haume \latin{et~al.}(2016)Haume, Rosa, Grellet, \'{S}mia{\l}ek,
  Butterworth, Solov'yov, Prise, Golding, and Mason]{Haume_2016_CNano.7.8}
Haume,~K.; Rosa,~S.; Grellet,~S.; \'{S}mia{\l}ek,~M.~A.; Butterworth,~K.~T.;
  Solov'yov,~A.~V.; Prise,~K.~M.; Golding,~J.; Mason,~N.~J. {Gold Nanoparticles
  for Cancer Radiotherapy: A Review}. \emph{Cancer Nanotechnol.} \textbf{2016},
  \emph{7}, 8\relax
\mciteBstWouldAddEndPuncttrue
\mciteSetBstMidEndSepPunct{\mcitedefaultmidpunct}
{\mcitedefaultendpunct}{\mcitedefaultseppunct}\relax
\EndOfBibitem
\bibitem[CA_((accessed 2023-11-15))]{CA_MultIChem_website}
COST Action ``Multiscale Irradiation and Chemistry Driven Processes and Related
  Technologies'' (MultIChem). (accessed 2023-11-15);
  \url{http://mbnresearch.com/ca20129-multichem/main}\relax
\mciteBstWouldAddEndPuncttrue
\mciteSetBstMidEndSepPunct{\mcitedefaultmidpunct}
{\mcitedefaultendpunct}{\mcitedefaultseppunct}\relax
\EndOfBibitem
\bibitem[H20((accessed 2023-11-15))]{H2020_RISE_RADON_website}
Horizon 2020 Research and Innovation Staff Exchange (RISE) Project
  ``Irradiation Driven Nanofabrication: Computational Modelling Versus
  Experiment'' (RADON). (accessed 2023-11-15);
  \url{http://mbnresearch.com/radon/main}\relax
\mciteBstWouldAddEndPuncttrue
\mciteSetBstMidEndSepPunct{\mcitedefaultmidpunct}
{\mcitedefaultendpunct}{\mcitedefaultseppunct}\relax
\EndOfBibitem
\bibitem[H20((accessed 2023-11-15))]{H2020_RISE_N-LIGHT_website}
Horizon 2020 Research and Innovation Staff Exchange (RISE) Project ``Novel
  Light Sources: Theory and Experiment'' (N-LIGHT). (accessed 2023-11-15);
  \url{http://mbnresearch.com/N-Light/main}\relax
\mciteBstWouldAddEndPuncttrue
\mciteSetBstMidEndSepPunct{\mcitedefaultmidpunct}
{\mcitedefaultendpunct}{\mcitedefaultseppunct}\relax
\EndOfBibitem
\bibitem[Gra((accessed 2023-11-15))]{Graphene_Flagship_website}
The Graphene Flagship, a European Union scientific research initiative.
  (accessed 2023-11-15); \url{https://graphene-flagship.eu/}\relax
\mciteBstWouldAddEndPuncttrue
\mciteSetBstMidEndSepPunct{\mcitedefaultmidpunct}
{\mcitedefaultendpunct}{\mcitedefaultseppunct}\relax
\EndOfBibitem
\bibitem[Hum((accessed 2023-11-15))]{Human-Brain-Project_website}
Human Brain Project, a European Union scientific research initiative. (accessed
  2023-11-15); \url{https://www.humanbrainproject.eu/en/}\relax
\mciteBstWouldAddEndPuncttrue
\mciteSetBstMidEndSepPunct{\mcitedefaultmidpunct}
{\mcitedefaultendpunct}{\mcitedefaultseppunct}\relax
\EndOfBibitem
\bibitem[Qua((accessed 2023-11-15))]{Quantum_Flagship_EU_website}
Quantum Technologies Flagship, a European Union scientific research initiative.
  (accessed 2023-11-15);
  \url{https://digital-strategy.ec.europa.eu/en/policies/quantum-technologies-flagship}\relax
\mciteBstWouldAddEndPuncttrue
\mciteSetBstMidEndSepPunct{\mcitedefaultmidpunct}
{\mcitedefaultendpunct}{\mcitedefaultseppunct}\relax
\EndOfBibitem
\bibitem[Qua((accessed 2023-11-15))]{Quantum_Flagship_website}
Quantum Technologies Flagship. (accessed 2023-11-15);
  \url{https://qt.eu/}\relax
\mciteBstWouldAddEndPuncttrue
\mciteSetBstMidEndSepPunct{\mcitedefaultmidpunct}
{\mcitedefaultendpunct}{\mcitedefaultseppunct}\relax
\EndOfBibitem
\end{mcitethebibliography}

\end{document}